%% file: thesis.tex
\documentclass[british]{mimosis}
\usepackage{setspace}
\usepackage[T1]{fontenc}
\usepackage[square,sort&compress,numbers]{natbib}
\usepackage{amssymb}
\usepackage{mathtools}
\usepackage{isomath}
\usepackage[Symbolsmallscale]{upgreek}
\usepackage[nf,p]{ebgaramond}
\usepackage[utopia,varg,slantedGreek]{newtxmath}
\usepackage[cleanlook,printdayon]{isodate}
\usepackage[protrusion,tracking,kerning,spacing,final,babel]{microtype}
\usepackage{physics}
\usepackage[final,unicode,hyperfootnotes=false]{hyperref}
\hypersetup{%
    colorlinks = true,
    citecolor  = NavyBlue,
    linkcolor  = NavyBlue,
    urlcolor   = NavyBlue,
}
\usepackage{cleveref}
\usepackage{tikz}
\usepackage{scalerel}
\usepackage{anyfontsize}
\usepackage{tensor}
\usepackage{epigraph}
\usepackage[title,titletoc]{appendix}
\usepackage{xfrac}
\usepackage[flushmargin]{footmisc}
\usepackage[format=plain,indention=.5cm,labelfont=bf]{caption}
\usepackage{floatpag}
\usepackage[paper=a4paper,hmargin=20mm,vmargin=28.29mm,bindingoffset=20mm,foot=1cm]{geometry}

\DeclareMathOperator{\Var}{Var}
\DeclareMathOperator{\Cov}{Cov}

\newacronym{AGWB}{AGWB}{Astrophysical gravitational-wave background}
\newacronym{AION}{AION}{Atom Interferometer Observatory and Network}
\newacronym{BAO}{BAO}{Baryon acoustic oscillations}
\newacronym{BBH}{BBH}{Binary black hole}
\newacronym{BBN}{BBN}{Big-bang nucleosynthesis}
\newacronym{BH}{BH}{Black hole}
\newacronym{BHNS}{BHNS}{Black hole-neutron star}
\newacronym{BNS}{BNS}{Binary neutron star}
\newacronym{CBC}{CBC}{Compact binary coalescence}
\newacronym{CDM}{CDM}{Cold dark matter}
\newacronym{CE}{CE}{Cosmic Explorer}
\newacronym{CGF}{CGF}{Cumulant-generating function}
\newacronym{CP}{CP}{Common process}
\newacronym{DM}{DM}{Dark matter}
\newacronym{EFE}{EFE}{Einstein field equation}
\newacronym{EM}{EM}{Electromagnetic}
\newacronym{EMRI}{EMRI}{Extreme mass-ratio inspiral}
\newacronym{EoM}{EoM}{Equation of motion}
\newacronym{EPTA}{EPTA}{European Pulsar Timing Array}
\newacronym{ET}{ET}{Einstein Telescope}
\newacronym{FLRW}{FLRW}{Friedmann-Lema\^itre-Robertson-Walker}
\newacronym{FOPT}{FOPT}{First-order phase transition}
\newacronym{FPE}{FPE}{Fokker-Planck equation}
\newacronym{GR}{GR}{General relativity}
\newacronym{GUT}{GUT}{Grand unified theory}
\newacronym{GW}{GW}{Gravitational wave}
\newacronym{GWB}{GWB}{Gravitational-wave background}
\newacronym{GWTC}{GWTC}{Gravitational-wave transient catalogue}
\newacronym{IFO}{IFO}{Interferometer}
\newacronym{i.i.d.}{i.i.d.}{Independent and identically distributed}
\newacronym{IPTA}{IPTA}{International Pulsar Timing Array}
\newacronym{ISCO}{ISCO}{Innermost stable circular orbit}
\newacronym{ISW}{ISW}{Integrated Sachs-Wolfe}
\newacronym{KAGRA}{KAGRA}{Kamioka Gravitational-Wave Detector}
\newacronym{KM}{KM}{Kramers-Moyal}
\newacronym{LHC}{LHC}{Large Hadron Collider}
\newacronym{LIGO}{LIGO}{Laser Interferometer Gravitational-Wave Observatory}
\newacronym{LISA}{LISA}{Laser Interferometer Space Antenna}
\newacronym{LLR}{LLR}{Lunar laser ranging}
\newacronym{LR}{LR}{Laser ranging}
\newacronym{LSS}{LSS}{Large-scale structure}
\newacronym{LVK}{LVK}{LIGO/Virgo/KAGRA}
\newacronym{MAGIS}{MAGIS}{Matter-wave Atomic Gradiometer Interferometric Sensor}
\newacronym{MHD}{MHD}{Magnetohydrodynamic}
\newacronym{MSP}{MSP}{Millisecond pulsar}
\newacronym{MVUE}{MVUE}{Minimum-variance unbiased estimator}
\newacronym{NANOGrav}{NANOGrav}{North American Nanohertz Observatory for Gravitational Waves}
\newacronym{NS}{NS}{Neutron star}
\newacronym{ODE}{ODE}{Ordinary differential equation}
\newacronym{ORF}{ORF}{Overlap reduction function}
\newacronym{PBH}{PBH}{Primordial black hole}
\newacronym{PDE}{PDE}{Partial differential equation}
\newacronym{PDF}{PDF}{Probability density function}
\newacronym{PI}{PI}{Power-law integrated}
\newacronym{PN}{PN}{Post-Newtonian}
\newacronym{PPTA}{PPTA}{Parkes Pulsar Timing Array}
\newacronym{PSD}{PSD}{Power spectral density}
\newacronym{PTA}{PTA}{Pulsar timing array}
\newacronym{QNM}{QNM}{Quasinormal mode}
\newacronym{rms}{rms}{Root-mean-square}
\newacronym{SFR}{SFR}{Star formation rate}
\newacronym{SGWB}{SGWB}{Stochastic gravitational-wave background}
\newacronym{SHC}{SHC}{Spherical harmonic component}
\newacronym{SKA}{SKA}{Square Kilometre Array}
\newacronym{SLR}{SLR}{Satellite laser ranging}
\newacronym{SNR}{SNR}{Signal-to-noise ratio}
\newacronym{SW}{SW}{Sachs-Wolfe}
\newacronym{ToA}{ToA}{Time of arrival}
\newacronym{TT}{TT}{Transverse-traceless}
\newacronym{VEV}{VEV}{Vacuum expectation value}
\newacronym{WD}{WD}{White dwarf}
\newacronym{ZFL}{ZFL}{Zero-frequency limit}

\makeindex
\makeglossaries

\title{Cosmology and Fundamental Physics in the Era of Gravitational-Wave Astronomy}
\author{Alexander C. Jenkins}

\begin{document}


\newcommand*{\rme}{\mathrm{e}}
\newcommand*{\rmi}{\mathrm{i}}
\newcommand*{\rmc}{\mathrm{c}}
\newcommand*{\rmk}{\mathrm{k}}
\newcommand*{\rmb}{\mathrm{b}}
\newcommand*{\Pl}{\mathrm{Pl}}
\newcommand*{\eps}{\varepsilon}
\newcommand*{\obs}{\mathrm{obs}}
\newcommand*{\gw}{\mathrm{gw}}

\newcommand*{\wasyfamily}{\fontencoding{U}\fontfamily{wasy}\selectfont}
\newcommand*\fakeslant[1]{\pdfliteral{1 0 0.167 1 0 0 cm}#1\pdfliteral{1 0 -0.167 1 0 0 cm}}
\newcommand*{\asc}{{\scaleobj{0.8}{\fakeslant{\textbf{\wasyfamily\char19}}}}}

\newcommand*{\dalemb}{{\scaleobj{1.3}{\square}}}

\renewcommand*{\partial}{\uppartial}

\renewcommand*\bibpreamble{%
    \epigraph{\enquote{There's more to life than books you know,\\but not much more.}}{Morrissey}}


\frontmatter
    \include{chapters/title}
    \include{chapters/abstract}
    \include{publications}
    \include{chapters/acknowledgements}
    \null\newpage
    \include{chapters/notation}
    \tableofcontents

\mainmatter

    \include{chapters/introduction}
    \include{chapters/anisotropies}
    \include{chapters/cosmic-strings}
    \include{chapters/binary-resonance}
    \include{chapters/conclusion}
    \include{chapters/appendices}

\backmatter

    \begingroup
        \let\clearpage\relax
        \glsaddall
        \printglossary[type=\acronymtype]
    \endgroup

    \printindex
    \bibliography{thesis}

\end{document}

%% file: chapters/title.tex
\begin{titlepage}
  \vspace*{2cm}
  \makeatletter
  \begin{center}
    \begin{Huge}
      \@title
    \end{Huge}\\[0.1cm]
    \begin{Large}
        \emph{by}\\
        \@author
    \end{Large}\\
    \vfill
    \begin{center}
        \includegraphics[width=0.3\textwidth]{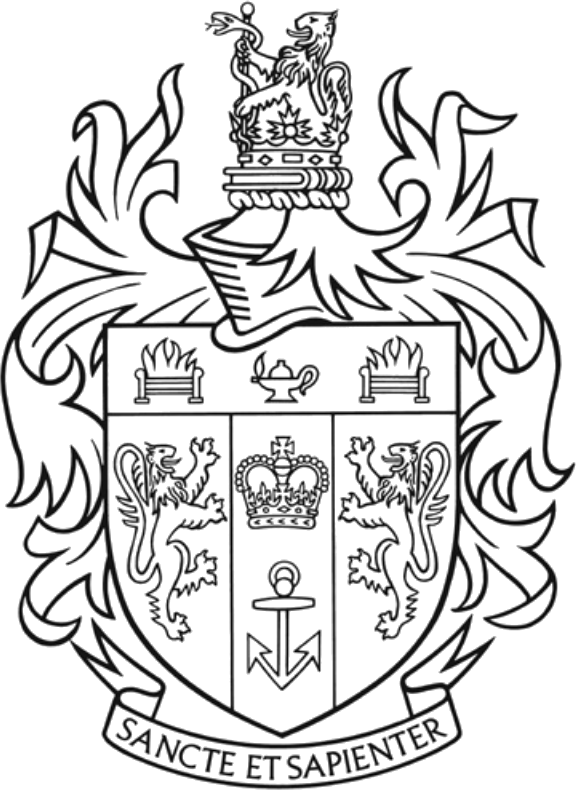}
    \end{center}
    \vfill
    A thesis presented for the degree of\\
    \emph{Doctor of Philosophy}\\
    at\\
    \textsc{King's College London}\\
    Department of Physics\\
    \printdate{27/1/2022}
  \end{center}
  \makeatother
  \vspace*{3cm}
\end{titlepage}

\newpage
\null
\thispagestyle{empty}
\newpage

%% file: chapters/abstract.tex
\begin{center}
  \textsc{Abstract}
\end{center}

\noindent

The advent of gravitational-wave (GW) astronomy has presented us with a completely new means for observing the Universe, allowing us to probe its structure and evolution like never before.
In this thesis, we explore three distinct but complementary avenues for using GW observations to gain new insights into cosmology and fundamental physics.

In chapter~\ref{chap:anisotropies}, we study the astrophysical GW background (AGWB): the cumulative GW signal arising from a large number of compact binary coalescences (CBCs) throughout the Universe.
Since these compact binaries reside in galaxies, the AGWB contains \emph{anisotropies} (i.e., intensity fluctuations on the sky) that trace out the large-scale structure of the cosmic matter distribution.
Despite their intrinsic interest as a source of novel cosmological information, these anisotropies have been neglected in most studies of the AGWB until quite recently.
Applying tools and concepts from other cosmological observables such as galaxy surveys and the cosmic microwave background (CMB), we investigate the angular power spectrum of the AGWB, with the goal of developing predictions that can be confronted with current and future directional AGWB searches.
Our key result is a simulated full-sky map of the AGWB anisotropies that we construct using data from the Millennium $N$-body simulation.
We find that these anisotropies are much larger in amplitude than in early-Universe observables such as the CMB, and that the lowest few multipoles of the angular power spectrum are likely to be observed by third-generation GW observatories.
We also highlight the issue of \emph{shot noise} due to the relatively low rate of CBCs in our frequency band of interest, and develop an optimal data-analysis strategy for estimating the true angular power spectrum in the presence of this shot noise.

In chapter~\ref{chap:cosmic-strings}, we investigate the \emph{nonlinear GW memory effect}, a fascinating prediction of general relativity in the dynamical, nonlinear regime, in which essentially all GW emission is accompanied by a hereditary, monotonically-increasing GW strain sourced by the energy of the escaping gravitons.
Essentially all of the literature on this effect has focused on the memory signals associated with CBCs; we broaden this scope by calculating, for the first time, the nonlinear memory emitted by cusps and kinks on cosmic string loops, which are among the most promising \emph{cosmological} sources of GWs.
Working in the Nambu-Goto approximation, we obtain simple analytical waveforms for the memory emitted by cusps and kinks, as well as the \enquote{memory of the memory} and other higher-order effects.
Summing over all of these contributions, we show that, surprisingly, the combined cusp memory signal \emph{diverges} for sufficiently large loops, indicating a breakdown in the validity of the weak-field description of the cusp.
We trace this divergence back to the high-frequency behaviour of the original cusp waveform, which gives rise to a trans-Planckian energy flux in the direction of the cusp's motion.
We then present one tentative possible solution to this divergence, in which the portion of the string surrounding the cusp collapses to form a primordial black hole (PBH).
We investigate the observational predictions of this scenario, and show that these PBHs could act as a \enquote{smoking gun} signature of cosmic strings.

Finally, in chapter~\ref{chap:binary-resonance} we develop a powerful new method for GW detection based on precision measurements of the orbits of binary systems.
In the presence of a stochastic GW background (GWB) the trajectories of the binary's components are perturbed, giving rise to a random walk in the system's orbital parameters over time.
By searching for this stochastic orbital evolution, we can infer the presence or absence of a GWB, turning the binary into a dynamical GW detector.
We develop here a novel Fokker-Planck formalism for calculating the expected evolution in all six orbital elements.
We then apply this formalism to two observational probes: timing of binary millisecond pulsars, and laser ranging of the Moon and artificial satellites.
We use a Fisher-forecasting approach to estimate the sensitivity of each of these probes to the GWB, and show that present data are already sensitive enough to place the strongest constraints to date in the $\upmu$Hz frequency band.
This band lies between the frequencies probed by pulsar timing arrays and by future space-based interferometers such as LISA, and is therefore an extremely attractive observational target, which could contain numerous cosmological GW signals.
As an example, we consider the GWB sourced by a cosmological first-order phase transition (FOPT), and show that the binary resonance searches we propose (in particular, with lunar laser ranging) will be sensitive to a region of the FOPT parameter space that no other current or near-future GW experiment can reach.

%% file: publications.tex
\begin{center}
  \textsc{Publications}
\end{center}

\noindent
The contents of this thesis are the result of my own work, except where specific reference is made to the work of others.
This work was originally presented in the publications listed below.

\noindent
Chapter~\ref{chap:anisotropies} is based on:
\begin{itemize}
    \item \textbf{ACJ} and M.~Sakellariadou, \emph{Anisotropies in the stochastic gravitational-wave background: Formalism and the cosmic string case}, \href{https://doi.org/10.1103/PhysRevD.98.063509}{Phys. Rev. D \textbf{98} (2018), 063509}, \href{https://arxiv.org/abs/1802.06046}{arXiv:1802.06046 [astro-ph.CO]}, \cite{Jenkins:2018nty}
    \item \textbf{ACJ}, M.~Sakellariadou, T.~Regimbau, and E.~Slezak, \emph{Anisotropies in the astrophysical gravitational-wave background: Predictions for the detection of compact binaries by LIGO and Virgo}, \href{https://doi.org/10.1103/PhysRevD.98.063501}{Phys. Rev. D \textbf{98} (2018), 063501}, \href{https://arxiv.org/abs/1806.01718}{arXiv:1806.01718 [astro-ph.CO]}, \cite{Jenkins:2018uac}
    \item \textbf{ACJ}, R.~O’Shaughnessy, M.~Sakellariadou, and D.~Wysocki, \emph{Anisotropies in the astrophysical gravi-tational-wave background: The impact of black hole distributions}, \href{https://doi.org/10.1103/PhysRevLett.122.111101}{Phys. Rev. Lett. \textbf{122} (2019), 111101}, \href{https://arxiv.org/abs/1810.13435}{arXiv:1810.13435 [astro-ph.CO]}, \cite{Jenkins:2018kxc}
    \item \textbf{ACJ} and M.~Sakellariadou, \emph{Shot noise in the astrophysical gravitational-wave background}, \href{https://doi.org/10.1103/PhysRevD.100.063508}{Phys. Rev. D \textbf{100} (2019), 063508}, \href{https://arxiv.org/abs/1902.07719}{arXiv:1902.07719 [astro-ph.CO]}, \cite{Jenkins:2019uzp}
    \item \textbf{ACJ}, J.~D.~Romano, and M.~Sakellariadou, \emph{Estimating the angular power spectrum of the gravi-tational-wave background in the presence of shot noise}, \href{https://doi.org/10.1103/PhysRevD.100.083501}{Phys. Rev. D \textbf{100} (2019), 083501}, \href{https://arxiv.org/abs/1907.06642}{arXiv: 1907.06642 [astro-ph.CO]}, \cite{Jenkins:2019nks}
\end{itemize}
Chapter~\ref{chap:cosmic-strings} is based on:
\begin{itemize}
    \item \textbf{ACJ} and M.~Sakellariadou, \emph{Primordial black holes from cusp collapse on cosmic strings} (2020), \href{https://arxiv.org/abs/2006.16249}{arXiv:2006.16249 [astro-ph.CO]}, \cite{Jenkins:2020ctp}
    \item \textbf{ACJ} and M.~Sakellariadou, \emph{Nonlinear gravitational-wave memory from cusps and kinks on cosmic strings}, \href{https://doi.org/10.1088/1361-6382/ac1084}{Class. Quant. Grav. \textbf{38} (2021), 165004}, \href{https://arxiv.org/abs/2102.12487}{arXiv:2102.12487 [gr-qc]}, \cite{Jenkins:2021kcj}
\end{itemize}
Chapter~\ref{chap:binary-resonance} is based on:
\begin{itemize}
    \item D.~Blas and \textbf{ACJ}, \emph{Detecting stochastic gravitational waves with binary resonance}, \href{https://doi.org/10.1103/PhysRevD.105.064021}{Phys. Rev. D \textbf{105} (2022), 064021}, \href{https://arxiv.org/abs/2107.04063}{arXiv: 2107.04063 [gr-qc]}, \cite{Blas:2021mpc}
    \item D.~Blas and \textbf{ACJ}, \emph{Bridging the $\mu$Hz gap in the gravitational-wave landscape with binary resonance}, \href{https://doi.org/10.1103/PhysRevLett.128.101103}{Phys. Rev. Lett. \textbf{128} (2022), 101103}, \href{https://arxiv.org/abs/2107.04601}{arXiv:2107.04601 [astro-ph.CO]}, \cite{Blas:2021mqw}
\end{itemize}
The following publications, of which I am also an author, are beyond the scope of this thesis:
\begin{itemize}
    \item \textbf{ACJ}, A.~G.~A.~Pithis, and M.~Sakellariadou, \emph{Can we detect quantum gravity with compact binary inspirals?}, \href{https://doi.org/10.1103/PhysRevD.98.104032}{Phys. Rev. D \textbf{98} (2018), 104032}, \href{https://arxiv.org/abs/1809.06275}{arXiv:1809.06275 [gr-qc]}, \cite{Jenkins:2018ysa}
    \item P.~Auclair, J.-J.~Blanco-Pillado, D.~G.~Figueroa, \textbf{ACJ}, M.~Lewicki, M.~Sakellariadou, S.~Sanidas, L.~Sousa, D.~A.~Steer, J.~M.~Wachter, and S.~Kuroyanagi, \emph{Probing the gravitational wave background from cosmic strings with LISA}, \href{https://doi.org/10.1088/1475-7516/2020/04/034}{JCAP \textbf{04} (2020), 034}, \href{https://arxiv.org/abs/1909.00819}{arXiv:1909.00819 [astro-ph.CO]}, \cite{Auclair:2019wcv}
    \item D.~Bertacca, A.~Ricciardone, N.~Bellomo, \textbf{ACJ}, S.~Matarrese, A.~Raccanelli, T.~Regimbau, and M.~Sakellariadou, \emph{Projection effects on the observed angular spectrum of the astrophysical stochastic gravitational wave background}, \href{https://doi.org/10.1103/PhysRevD.101.103513}{Phys. Rev. D \textbf{101} (2020), 103513}, \href{https://arxiv.org/abs/1909.11627}{arXiv:1909.11627 [astro-ph.CO]}, \cite{Bertacca:2019fnt}
    \item G.~Boileau, \textbf{ACJ}, M.~Sakellariadou, R.~Meyer, and N.~Christensen, \emph{Ability of LISA to detect a gravitational-wave background of cosmological origin: the cosmic string case}, \href{https://doi.org/10.1103/PhysRevD.105.023510}{Phys. Rev. D \textbf{105} (2022), 023510}, \href{https://arxiv.org/abs/2109.06552}{arXiv:2109.06552 [gr-qc]}, \cite{Boileau:2021gbr}
    \item N.~Bellomo, D.~Bertacca, \textbf{ACJ}, S.~Matarrese, A.~Raccanelli, T.~Regimbau, A.~Ricciardone, and M.~Sakellariadou, \emph{\texttt{CLASS\_GWB}: robust modeling of the astrophysical gravitational wave background anisotropies}, \href{https://arxiv.org/abs/2110.15059}{arXiv:2110.15059 [gr-qc]}, \cite{Bellomo:2021mer}
    \item N.~Bartolo, D.~Bertacca, R.~Caldwell, C.~R.~Contaldi, G.~Cusin, V.~De~Luca, E.~Dimastrogiovanni, M.~Fasiello, D.~G.~Figueroa, G.~Franciolini, \textbf{ACJ}, M.~Peloso, M.~Pieroni, A.~Renzini, A.~Ricciardone, A.~Riotto, M.~Sakellariadou, L.~Sorbo, G.~Tasinato, J.~Torrado, S.~Clesse, and S.~Kuroyanagi, \emph{Probing Anisotropies of the Stochastic Gravitational Wave Background with LISA}, \href{https://arxiv.org/abs/2201.08782}{arXiv:2201.08782 [astro-ph.CO]}, \cite{Bartolo:2022pez}
    \item A.~Renzini, B.~Goncharov, \textbf{ACJ}, and P.~M.~Meyers, \emph{Stochastic Gravitational-Wave Backgrounds: Current Detection Efforts and Future Prospects}, \href{https://doi.org/10.3390/galaxies10010034}{Galaxies \textbf{10} (2022), 34}, \href{https://arxiv.org/abs/2202.00178}{arXiv:2202.00178 [gr-qc]}, \cite{Renzini:2022alw}
\end{itemize}

%% file: chapters/acknowledgements.tex
\begin{center}
  \textsc{Acknowledgements}
\end{center}

\noindent

First of all, thank you to Mairi Sakellariadou---my supervisor, collaborator, mentor, and friend---for giving me the chance to go on this journey in the first place, and for expertly guiding me through the ups and downs of the past four years.
Thank you for having faith in me, for always being so generous with your time and ideas, and for helping me to grow as a physicist and a researcher.

\vspace{0.4cm}
I'm acutely aware of how privileged I am to be able to think about gravitational waves all day rather than doing a \enquote{real job}, so thank you to the Faculty of Natural and Mathematical Sciences at King's for enabling this by supporting me financially throughout my PhD.

\vspace{0.4cm}
Thank you to the whole TPPC group for making King's such a fun and stimulating place to do research.
I'm particularly grateful to Diego Blas, Malcolm Fairbairn, and Eugene Lim for their support and advice, and for all the physics I've learnt from each of them over the years.
Thank you also to all my colleagues in the LVK Stochastic group who I've had the pleasure of interacting and collaborating with, and who have shaped my understanding of many of the topics discussed in this thesis, particularly Joe Romano and Nelson Christensen.

\vspace{0.4cm}
Thank you to my examiners, David Wands and Stephen Fairhurst, for their insightful questions and feedback on this thesis, and for making the viva a genuinely enjoyable experience.

\vspace{0.4cm}
Thank you to all the great friends who've made the past four years so much fun: Nick, Gavi, Robert, Jenni, Joshua, and Lauren, for helping make London feel like home, as well as Josu, Ali, Thomas, Steph, Katarina, Manya, James, Matt, Giuseppe, Louis, Patrick, Andreas, and all the other past and present PhD students at King's, for distracting me from physics when I needed it most, and for helping me explore the Strand's many drinking establishments.

\vspace{0.4cm}
Special thanks to Stuart, Bryony, Niki, and Nigel, whose company kept my spirits high and my sanity intact through two successive lockdowns.

\vspace{0.4cm}
Thank you to Mum and Dad for their endless love and support, for nurturing my interest in science and maths from the very beginning, and for always encouraging me to be the best that I can be.
I'd never have got this far without you, and there's so much I owe you both.
Thank you for everything.

\vspace{0.4cm}
Finally, my biggest thanks go to Saskia---the centre of my Universe---for all the countless ways you've helped me and encouraged me through writing this thesis, and all the way through three different degrees in two different cities over the past eight years.
Thank you for making the highs feel even higher, and the lows feel not so low.
Every single day is better with you by my side.

%% file: chapters/notation.tex
\begin{center}
  \textsc{Notation}
\end{center}

\noindent
\begin{itemize}
    \item We generally use $\sim$ to mean equality to within an order of magnitude, $\approx$ to mean approximate equality, $\simeq$ to mean asymptotic equality (e.g., $\rme^x\simeq1+x$), and $\propto$ to mean proportionality.
    The symbol $\equiv$ denotes equality by definition.
    \item We use units such that the speed of light in vacuum and the Boltzmann constant are both equal to unity, $c=k_\mathrm{B}=1$, but keep Newton's constant $G$ and the reduced Planck constant $\hbar$ explicit throughout.
        Unless otherwise specified, we set the Hubble constant to the \emph{Planck} 2018 value of $H_0\approx67.7\,\mathrm{km}\,\mathrm{s}^{-1}\,\mathrm{Mpc}^{-1}\approx2.19\times10^{-18}\,\mathrm{Hz}$~\cite{Aghanim:2018vyg}.
    \item Spacetime indices running over $(0,1,2,3)$ are indicated by lowercase Greek letters, while Latin letters indicate either spatial indices running over $(1,2,3)$ or, where specified, some other set of positive integers $(1,2,\dots)$.
        We also use $A,A'$ to denote GW polarisation indices running over $(+,\times)$.
        Round brackets around pairs of indices indicate a symmetrisation operation, e.g.,
        \begin{equation*}
            X_{(\mu\nu)}\equiv\frac{1}{2}(X_{\mu\nu}+X_{\nu\mu}).
        \end{equation*}
    \item Our conventions for general relativity match those of Misner, Thorne, and Wheeler~\cite{Misner:1974qy} and Wald~\cite{Wald:1984rg}.
    The spacetime metric $g_{\mu\nu}$, which is used to raise and lower indices, has signature $({-}\,{+}\,{+}\,{+})$, and defines the Christoffel symbols
        \begin{equation*}
            \Gamma^\mu_{\alpha\beta}\equiv\frac{1}{2}g^{\mu\nu}(\partial_\alpha g_{\beta\nu}+\partial_\beta g_{\alpha\nu}-\partial_\nu g_{\alpha\beta}).
        \end{equation*}
    The Riemann curvature is given by
        \begin{equation*}
            \tensor{R}{^\mu_{\nu\alpha\beta}}\equiv\partial_\alpha\Gamma^\mu_{\nu\beta}-\partial_\beta\Gamma^\mu_{\nu\alpha}+\Gamma^\mu_{\rho\alpha}\Gamma^\rho_{\nu\beta}-\Gamma^\mu_{\rho\beta}\Gamma^\rho_{\nu\alpha},
        \end{equation*}
        which defines the Ricci curvature, $R_{\mu\nu}\equiv\tensor{R}{^\alpha_{\mu\alpha\nu}}$, and Ricci scalar, $R\equiv\tensor{R}{^\mu_\mu}$.
    We \emph{do not} use commas and semicolons to denote partial and covariant derivatives.
    \item Spatial 3-vectors (and Euclidean vectors more generally) are written in boldface, $\vb*x$, with the corresponding letter used to denote their length, $x=|\vb*x|$.
    Unit-length 3-vectors are written with a hat, $\vu*x=\vb*x/x$.
    \item Complex conjugates are denoted with an asterisk, $z^*$.
    \item Matrices are denoted by uppercase sans-serif Latin letters, $\mathsf{M}$, with their transpose and Hermitian conjugate written as $\mathsf{M}^\mathsf{T}$ and $\mathsf{M}^\dagger$, respectively.
    \item We denote integrals over the real line $\mathbb{R}=(-\infty,+\infty)$ and over 3D Euclidean space $\mathbb{R}^3$ by
        \begin{equation*}
            \int_\mathbb{R}\dd{x}\equiv\int_{-\infty}^{+\infty}\dd{x},\qquad\int_{\mathbb{R}^3}\dd[3]{\vb*x}\equiv\int_{-\infty}^{+\infty}\dd{x_1}\int_{-\infty}^{+\infty}\dd{x_2}\int_{-\infty}^{+\infty}\dd{x_3},
        \end{equation*}
        while integrals over the 2-sphere $S^2$ are written as
        \begin{equation*}
            \int_{S^2}\dd[2]{\vu*r}\equiv\int_{-1}^{+1}\dd{(\cos\theta)}\int_0^{2\uppi}\dd{\phi}=\int_0^\uppi\dd{\theta}\sin\theta\int_0^{2\uppi}\dd{\phi},
        \end{equation*}
        where $(\theta,\phi)$ are the usual polar coordinates.
    \item We define the Fourier transform of a function $h(t)$ by
        \begin{equation*}
            \tilde{h}(f)=\mathcal{F}[h](f)\equiv\int_\mathbb{R}\dd{t}\rme^{-2\uppi\rmi ft}h(t),\qquad h(t)=\int_\mathbb{R}\dd{f}\rme^{2\uppi\rmi ft}\tilde{h}(f).
        \end{equation*}
\end{itemize}

%% file: chapters/introduction.tex
\chapter[Introduction: the dawn of gravitational-wave astronomy]{Introduction:\\the dawn of gravitational-wave astronomy}
\label{chap:intro}

\epigraph{%
    \enquote{When I ask myself \enquote{What are the great things we got from the Renaissance?}, it's the great art, the great music, the science insights of Leonardo da Vinci.
    Two hundred years from now when you ask \enquote{What are the great things that came from this era?}, I think it’s going to be an understanding of the Universe around us.}
    }{Kip Thorne}


\noindent
The era of gravitational-wave (GW) astronomy has begun.
On \printdate{14/09/2015}, almost a century after Einstein published his General Theory of Relativity (GR)~\cite{Einstein:1916vd,Einstein:1918btx}, one of the most important and controversial predictions of his theory was verified in spectacular fashion by the Advanced LIGO interferometers~\cite{Harry:2010zz,Aasi:2014jea} with the detection of GWs from two merging black holes (BHs)~\cite{Abbott:2016blz}.
This event---dubbed GW150914---helped dispel decades of scepticism around the physical existence of GWs due to theoretical misconceptions (including by Einstein himself)~\cite{Kennefick:2007zz,Cervantes-Cota:2016zjc} and spurious observational claims (most notably by Weber and his collaborators)~\cite{Weber:1969bz,Collins:2004jj}, and was immediately heralded as a major breakthrough by the scientific community.
In the six years since this first detection, LIGO and its European counterpart Virgo~\cite{Acernese:2014hva} have discovered dozens more GW signals~\cite{Abbott:2018mvr,Abbott:2020niy,Abbott:2021usb,Abbott:2021djp}, firmly establishing GW astronomy as a new observational science for the 21$^\text{st}$ century.

While testing the predictions of GR is a laudable goal in itself, it is far from the only motivation for GW science.
Much of the appeal of GWs lies in their r\^ole as an astronomical \emph{messenger}, akin to the photon or the neutrino: a new and fundamentally different tool with which to observe the Universe.
The power of combining this new messenger with more traditional astronomical observations has already been forcefully demonstrated by the neutron star (NS) merger event GW170817~\cite{Abbott:2017qsa,Abbott:2017lvd,Abbott:2017mdv}, whose GW signal, combined with observations across the electromagnetic (EM) spectrum, has eliminated a large class of modified gravity theories~\cite{Creminelli:2017sry,Sakstein:2017xjx,Ezquiaga:2017ekz,Baker:2017hug}, probed the properties of nuclear matter under extreme conditions~\cite{Margalit:2017dij,Radice:2017lry,Abbott:2018exr}, confirmed NS mergers as key sites of $r$-process nucleosynthesis~\cite{Drout:2017ijr,Chornock:2017sdf}, and provided a novel and independent measurement of the local expansion rate of the Universe~\cite{Schutz:1986gp,Abbott:2017xzu}.

Even more exciting, in my view, is the opportunity to probe objects and phenomena that \emph{only} GWs can access: \enquote{dark} parts of the Universe that are invisible to EM astronomy.
This is important because we know that luminous matter only constitutes a small fraction of the cosmic matter inventory, with the majority consisting of non-baryonic \enquote{dark matter} (DM) whose nature is still mysterious~\cite{Bertone:2004pz}.
Black holes (which may or may not account for most or all of the DM, see section~\ref{sec:primordial-black-holes}) are a quintessential example of a dark astronomical object, made not of hot radiating gas like the Sun and other stars, but formed instead from pure spacetime geometry.
Until September 2015, all observational evidence for BHs was \emph{indirect}, mostly due to their gravitational influence on surrounding luminous matter (e.g., in quasars~\cite{Silk:1997xw} or X-ray binaries~\cite{Remillard:2006fc}).
With the advent of GW astronomy, we can now observe BHs \emph{directly}, and LIGO/Virgo are already giving us fascinating new insights into their properties and origins as a result~\cite{Rodriguez:2015oxa,Antonini:2016gqe,deMink:2016vkw,Belczynski:2017gds,Abbott:2018jsj,Abbott:2020mjq,Abbott:2020gyp,Tiwari:2020otp}.

Going back further, beyond the myriad dark objects in our cosmic neighbourhood, we reach an entire epoch in the Universe's history that is inaccessible to EM observations.
For the first 370,000 years or so after the Big Bang, the Universe was so hot that the electrons were unbound from their atomic nuclei, forming a dense ionised plasma that was impenetrable to EM radiation~\cite{Weinberg:2008zzc}.
The light emitted at the end of this epoch, which we observe today as the Cosmic Microwave Background (CMB)~\cite{Dicke:1965zz,Penzias:1965wn}, represents the earliest point in the Universe's history that can be probed with EM observations.
Since the CMB has been mapped out in exquisite detail by the Planck satellite~\cite{Aghanim:2018vks} and other missions~\cite{Smoot:1992td,deBernardis:2000sbo,Hanany:2000qf,Bennett:2012zja,Ade:2015tva}, we have already, in a sense, hit the limit of EM astronomy's reach into the early Universe.
GWs, on the other hand, face no such restriction; since they couple so weakly to matter, they pass through the hot pre-recombination plasma unimpeded, carrying signals from the Universe's earliest moments.
Thus, while EM observations can tell us about conditions in the Universe at energies below $\sim0.3\,\mathrm{eV}$ (i.e. $\sim3000\,\mathrm{K}$, the temperature at which photons decoupled from baryons), GWs can reach far further, far beyond the $\sim10^{13}\,\mathrm{eV}$ energies probed by terrestrial experiments such as the Large Hadron Collider (LHC)~\cite{Evans:2008zzb}, all the way up to the Planck scale $m_\Pl\sim10^{28}\,\mathrm{eV}$ and into the realm of quantum gravity.

This ability to probe the hidden side of the Universe, from BHs to the Big Bang, makes GWs the theoretical physicist's most powerful new tool, opening up countless new avenues for studying the nature of the cosmos and the fundamental laws that govern it.
The goal of this thesis is to explore a few of these avenues.
In chapter~\ref{chap:anisotropies}, we look at ways in which GW observations can be used to probe the large-scale structure (LSS) of the Universe by studying \emph{anisotropies} in the cosmic GW background, analogous to those in the CMB.
In chapter~\ref{chap:cosmic-strings}, we study novel GW signatures and phenomenology of \emph{cosmic strings}, which are one of the most the promising ways in which GWs can probe the pre-recombination era.
Finally, in chapter~\ref{chap:binary-resonance}, we explore the possibility of using binary systems as \emph{dynamical detectors} of GWs, and investigate the constraints that such detectors can place on cosmological first-order phase transitions.

In the remainder of this chapter, we set the stage by giving a self-contained introduction to GW astronomy.
We begin in section~\ref{sec:what-are-gws} with a discussion of GWs themselves, in which we derive their basic physical properties in the framework of linearised GR.
Then, in section~\ref{sec:gwb-intro}, we introduce a GW signal which is of particular importance from the point of view of cosmology: the stochastic GW background (GWB).
In section~\ref{sec:gw-sources} we give an overview of some of the various GW sources which contribute to the GWB, focusing on those which are the most promising as probes of cosmology and fundamental physics.
Finally, in section~\ref{sec:gw-detection} we describe the basic principles and current status of observational searches for GWs.

\section{What are gravitational waves?}
\label{sec:what-are-gws}

\begin{figure}[t!]
    \begin{center}
        \includegraphics[width=0.6\textwidth]{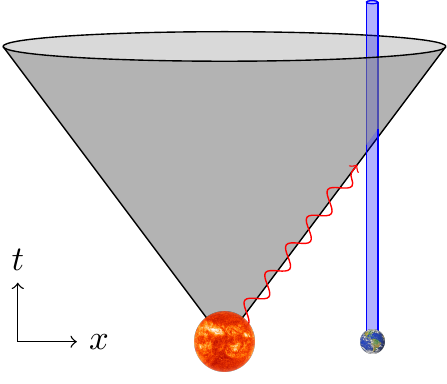}
    \end{center}
    \caption{%
    In Einsteinian gravity, disturbances in the Sun's gravitational field (shown in red) travel along the lightcone (grey), reaching the Earth's worldline (blue) some finite time after they originated.
    We call these propagating disturbances \emph{gravitational waves}.
    }
    \label{fig:lightcone}
\end{figure}

The existence of GWs arises very naturally from the basic premise of GR, which is to formulate a \emph{relativistic} theory of the gravitational field; i.e., one in which all observers agree on the speed of light in vacuum, $c\approx3\times10^8\,\mathrm{m}\,\mathrm{s}^{-1}$.
This requirement imposes a particular kind of causal structure on the theory, in which no physical degree of freedom can propagate faster than light, and the consequences of any event can therefore only be felt within its \enquote{lightcone}: the region of spacetime that can be reached by particles travelling at speeds $\le c$ (see figure~\ref{fig:lightcone}).
Classical electrodynamics is a key example of such a theory.

This causal structure is in marked contrast with that of Newtonian gravity, in which changes in the gravitational field are felt \emph{simultaneously} everywhere, no matter how far from the source.
If, for example, the Sun were to spontaneously disappear, then according to Newton the Earth would immediately deviate from its orbit.
Meanwhile, the last photons emitted by the Sun, obeying the relativistic laws of electrodynamics, would travel along the lightcone, and it would take $\sim8.3$ minutes for the Sun to cease shining in the sky (as measured by an observer comoving with the Earth).
This instantaneous Newtonian propagation is nonsensical from the perspective of a relativistic theory, as different observers cannot agree on what is meant by \enquote{simultaneous}---some would even see the Earth's orbit change \emph{before} the Sun disappeared, making the causal relationship between the two events ambiguous.
However, all observers would agree that the sunlight ceased to reach the Earth \emph{after} the Sun disappeared, as the speed-of-light propagation of this signal has a special status in a relativistic theory.

Clearly this situation would make much more sense if changes to the gravitational field also propagated at the speed of light.
As we discuss below, this is indeed the situation in GR.\footnote{%
    In fact, this idea was discussed decades before Einstein's formulation of GR, most notably by Heaviside~\cite{Heaviside:1893gw} and Poincar\'e~\cite{Poincare:1906el}.}
This simple example illustrates the most general and straightforward way to understand what GWs are: they are degrees of freedom of the gravitational field which propagate at the speed of light, thereby endowing gravitational interactions with a relativistic causal structure akin to that of electrodynamics.\footnote{%
    Note however that GWs do not \emph{have} to propagate at the speed of light in a generic relativistic theory~\cite{Carballo-Rubio:2020ttr,deRham:2020zyh}; this is merely a particularly appealing and well-motivated option.
In practice, joint GW and EM observations of GW170817 have demonstrated that any differences between the speed of light and the speed of GWs must be incredibly small: less than one part in $10^{15}$~\cite{Creminelli:2017sry,Sakstein:2017xjx,Ezquiaga:2017ekz,Baker:2017hug}.}

In the remainder of this section, we explore the physical properties of these light-like excitations of the gravitational field.
After a very brief recap of GR, we introduce the linearised form of the theory and use this to show that there are in fact \emph{two} degrees of freedom, before discussing how these interact with test masses, how they are sourced by the motion of matter, and how we can assign energy and momentum to them.
We end with a discussion of the \emph{GW memory effect}, a fascinating prediction of GR which is further explored in chapter~\ref{chap:cosmic-strings} of this thesis.
Throughout we use units in which $c=1$.

\subsection{A lightning review of general relativity}

We have referred several times above to \enquote{the gravitational field}.
In GR, this means the \emph{spacetime metric} $g_{\mu\nu}(x)$: a symmetric rank-$(0,2)$ tensor field, whose components in any coordinate chart $\{x^\mu\}$ determine the spacetime interval $\dd{s}$ between two infinitesimally separated events,
    \begin{equation}
        \dd{s}^2=g_{\mu\nu}\dd{x}^\mu\dd{x}^\nu.
    \end{equation}
The resulting spacetime geometry is generically \emph{curved}, and this curvature influences the trajectories of particles through spacetime.
In the absence of non-gravitational interactions, these trajectories are solutions of the geodesic equation
    \begin{equation}
    \label{eq:geodesic}
        \dv{U^\mu}{\lambda}+\Gamma^\mu_{\alpha\beta}U^\alpha U^\beta=0,
    \end{equation}
    where $U^\mu(\lambda)$ is the 4-velocity of the particle along a trajectory with affine parameter $\lambda$, and $\Gamma^\mu_{\alpha\beta}$ are the Christoffel symbols associated with the metric $g_{\mu\nu}$.
The metric, in turn, evolves dynamically according to the Einstein field equation (EFE),\footnote{%
    This is sometimes written with an additional cosmological constant term, $R_{\mu\nu}-\tfrac{1}{2}Rg_{\mu\nu}+\Lambda g_{\mu\nu}=8\uppi GT_{\mu\nu}$.
    Here we choose instead to absorb $\Lambda$ into $T_{\mu\nu}$, treating it as a homogeneous matter field with negative pressure, $p=-\rho$.}
    \begin{equation}
    \label{eq:efe}
        R_{\mu\nu}-\frac{1}{2}Rg_{\mu\nu}=8\uppi GT_{\mu\nu},
    \end{equation}
    where $R_{\mu\nu}$ is the Ricci tensor corresponding to $g_{\mu\nu}$ and $R$ is the Ricci scalar, while the \emph{energy-momentum tensor} $T_{\mu\nu}$ describes how matter (i.e., all other fields that exist in spacetime) acts as a source of spacetime curvature.
This beautiful reciprocal interplay between the geometry of spacetime and the matter fields that inhabit it is succinctly captured by Wheeler~\cite{Wheeler:1998vs}: \enquote{Spacetime tells matter how to move; matter tells spacetime how to curve.}

The EFE can also be derived from an action principle, writing
    \begin{equation}
    \label{eq:einstein-hilbert-action}
        S=S_\mathrm{EH}+S_\mathrm{mat},\qquad S_\mathrm{EH}\equiv\frac{1}{16\uppi G}\int\dd[4]{x}\sqrt{-g}R,\qquad T^{\mu\nu}\equiv\frac{2}{\sqrt{-g}}\frac{\updelta S_\mathrm{mat}}{\updelta g_{\mu\nu}},
    \end{equation}
    where $g\equiv\det g_{\mu\nu}$ is the metric determinant, $S_\mathrm{EH}$ is the \emph{Einstein-Hilbert action}, and $S_\mathrm{mat}$ is the action of the matter fields.
The metric which extremises the total action $S$ then obeys equation~\eqref{eq:efe}.

One of the most important features of the EFE~\eqref{eq:efe} is that it holds in any coordinate chart; we are free to arbitrarily redefine our coordinates $x\to x'(x)$ (subject to differentiability conditions) by appropriately redefining the metric,
    \begin{equation}
    \label{eq:diffeo}
        g_{\mu\nu}(x)\to g'_{\mu\nu}(x')=\pdv{x^\alpha}{x'^\mu}\pdv{x^\beta}{x'^\nu}g_{\alpha\beta}(x),
    \end{equation}
    and similarly redefining the energy-momentum tensor.
Such a transformation is called a \emph{diffeomorphism}, and the indifference of GR to diffeomorphisms is called the principle of \emph{general covariance}.
This principle reflects the notion that spacetime coordinates are merely artificial constructs that we introduce to facilitate calculations, and should play no fundamental r\^ole in the laws of physics.
The resulting freedom to choose coordinates in GR can thus be thought of as a gauge freedom, analogous to the choice of potential in electrodynamics.
Practically, this huge redundancy in description means that we must be careful to separate the physical degrees of freedom from gauge degrees of freedom; historically it was this difficulty that caused much of the early confusion over the existence of GWs in GR~\cite{Cervantes-Cota:2016zjc}.

\subsection{Linearised general relativity}

The EFE~\eqref{eq:efe}, while famous for its elegance and conceptual simplicity, is also infamously difficult to solve.
While many exact solutions do exist~\cite{Stephani:2003tm}, only a handful represent physically relevant spacetimes, and those that do often assume a high degree of symmetry that is all but guaranteed to be broken in reality.
Chief among these are the flat Minkowski spacetime, the Friedmann-Lema\^itre-Roberson-Walker (FLRW) solutions used in cosmology, and the Kerr-Newman family of BH spacetimes.

Despite the relative simplicity of these solutions, there are many physical situations where differences between the true spacetime metric $g_{\mu\nu}$ and that of some exact solution $g^{(0)}_{\mu\nu}$ are \enquote{small} in some sense.
We can then make progress by using \emph{perturbation theory}, defining the small quantity $h_{\mu\nu}\equiv g_{\mu\nu}-g^{(0)}_{\mu\nu}$, and writing the relevant equations as series expansions in $h$.
While these methods are necessarily approximate, due to the need to truncate the expansion at some order in $h$, they are often incredibly powerful: for example, cosmological perturbation theory (where $g^{(0)}_{\mu\nu}$ belongs to the FLRW family) provides an excellent description of the anisotropies in the CMB~\cite{Weinberg:2008zzc} as well as many other cosmological observables at lower redshift~\cite{Bernardeau:2001qr}, while BH perturbation theory (where $g^{(0)}_{\mu\nu}$ belongs to the Kerr-Newman family) allows us to compute the quasinormal \enquote{ringing} of BHs formed in mergers~\cite{Kokkotas:1999bd,Berti:2009kk} and the orbits of small bodies around supermassive BHs through the \enquote{self-force} formalism~\cite{Barack:2018yvs}.

The case where $g^{(0)}_{\mu\nu}$ is the Minkowski metric $\eta_{\mu\nu}=\mathrm{diag}(-1,+1,+1,+1)$ is by far the simplest, describing a situation where there are no strongly-gravitating objects (such as black holes and neutron stars) and where cosmological expansion is negligible.
Despite its simplicity, however, this flat-space perturbation theory is sufficient to capture many of the most salient features of GWs.
In fact, for much of the discussion we can simplify things even further by considering only the first-order (linear) term in the series expansion.
The resulting theory is called \emph{linearised} GR.\footnote{%
    It is important to note that linearised GR is not the only tool for studying GWs, and that it is not necessary to assume that GWs are perturbatively small.
    For example, one can use the Bondi-Sachs formalism~\cite{Bondi:1962px,Sachs:1962wk,dInverno:1992gxs} to study the large-distance behaviour of GWs of arbitrary amplitude emitted by an isolated source, and there are also numerous exact solutions describing arbitrarily strong gravitational plane waves~\cite{dInverno:1992gxs,Stephani:2003tm}.}

Inserting the perturbed flat-space metric
    \begin{equation}
    \label{eq:perturbed-flat-space}
        g_{\mu\nu}=\eta_{\mu\nu}+h_{\mu\nu},\qquad|h_{\mu\nu}|\ll1,
    \end{equation}
    into the EFE~\eqref{eq:efe} and keeping only the linear-order terms, we obtain~\cite{Maggiore:2007zz}
    \begin{equation}
    \label{eq:efe-linear}
        \dalemb\bar{h}_{\mu\nu}+\eta_{\mu\nu}\partial^\alpha\partial^\beta\bar{h}_{\alpha\beta}-2\partial_{(\mu}\partial^\alpha\bar{h}_{\nu)\alpha}=-16\uppi GT_{\mu\nu},
    \end{equation}
    which we have written in terms of the trace-reversed GW perturbation,
    \begin{equation}
    \label{eq:trace-reversed}
        \bar{h}_{\mu\nu}\equiv h_{\mu\nu}-\frac{1}{2}\tensor{h}{^\alpha_\alpha}\eta_{\mu\nu}.
    \end{equation}
The resulting equation~\eqref{eq:efe-linear} is a hyperbolic partial differential equation (PDE), i.e., one which has wavelike solutions which propagate at a characteristic finite velocity; in this case, the speed of light, as suggested by the presence of the (flat-space) d'Alembertian operator $\dalemb\equiv\partial^\mu\partial_\mu=-\partial_0^2+\partial^i\partial_i$, and the absence of any mass term $\sim m^2\bar{h}_{\mu\nu}$.
The other derivative terms in equation~\eqref{eq:efe-linear} do not spoil this behaviour, and in fact we see below that we can safely set them to zero with an appropriate gauge choice.

\subsection{Fixing the gauge}

Since $\bar{h}_{\mu\nu}$ has two symmetric spacetime indices, the linearised EFE~\eqref{eq:efe-linear} has, in principle, ten independent degrees of freedom.
However, due to general covariance, we must be extremely cautious in identifying which of these degrees of freedom are physical (if any), and which can be removed by an appropriate gauge choice.

Writing a general diffeomorphism as $x^\mu\to x^\mu+\xi^\mu$, and using equation~\eqref{eq:diffeo}, we find that the metric perturbation transforms as
    \begin{equation}
    \label{eq:h-diffeo}
        h_{\mu\nu}\to h_{\mu\nu}-2\partial_{(\mu}\xi_{\nu)}.
    \end{equation}
We thus see that specifying an almost-flat metric~\eqref{eq:perturbed-flat-space} has already ruled out a large class of diffeomorphisms, as the only remaining gauge transformations consistent with this ansatz are those in which $|\partial\xi|\ll1$.
Nonetheless, the $\partial\xi$ terms are still permitted to be of the same size as $h$, so they must still be carefully separated from the physical degrees of freedom.

We saw above that the interpretation of the linearised EFE~\eqref{eq:efe-linear} as a wave equation was obscured somewhat by the presence of terms like $\partial^\mu\bar{h}_{\mu\nu}$, and mentioned that these can be removed by a gauge transformation.
Indeed, from equations~\eqref{eq:trace-reversed} and~\eqref{eq:h-diffeo} we find that $\partial^\mu\bar{h}_{\mu\nu}\to\partial^\mu\bar{h}_{\mu\nu}-\dalemb\xi_\nu$, which shows that it is always possible to set
    \begin{equation}
    \label{eq:harmonic}
        \partial^\mu\bar{h}_{\mu\nu}=0
    \end{equation}
    with an appropriate choice of $\xi$, as the latter can always be constructed using the Green's function for the d'Alembertian.
This is the \emph{harmonic} gauge condition,\footnote{%
    This name comes from the fact that the spacetime coordinates are then solutions of the curved-space wave equation, $\nabla^\mu\nabla_\mu x^\nu=0$.
    Other names include the de Donder gauge, the Hilbert gauge, and in the context of linearised GR specifically, the Lorenz or (erroneously) Lorentz gauge~\cite{Jackson:2001ia}, by analogy with electrodynamics.}
    in which the EFE becomes
    \begin{equation}
    \label{eq:efe-linear-harmonic-gauge}
        \dalemb\bar{h}_{\mu\nu}=-16\uppi GT_{\mu\nu},
    \end{equation}
    making its interpretation as a massless relativistic wave equation much clearer.

Setting $\partial^\mu\bar{h}_{\mu\nu}=0$ removes four gauge degrees of freedom from the ten that we started with: one for each value of the free spacetime index $\nu$.
This does not fully specify $\xi$ however, as it is still possible to perform a smaller set of gauge transformations that obey $\dalemb\xi_\mu=0$; i.e., we are free to choose four harmonic functions.
If we are in a region of spacetime with no matter sources, $T_{\mu\nu}=0$, then this allows us to set certain components of $\bar{h}_{\mu\nu}$ to zero.\footnote{%
    We cannot set components of $\bar{h}_{\mu\nu}$ to zero inside a matter source by subtracting only harmonic functions obeying $\dalemb\xi_\mu=0$, because the flat-space d'Alembertian commutes with partial derivatives to give
        $\dalemb\bar{h}_{\mu\nu}\to\dalemb(\bar{h}_{\mu\nu}-2\partial_{(\mu}\xi_{\nu)}+\eta_{\mu\nu}\partial^\alpha\xi_\alpha)=\dalemb\bar{h}_{\mu\nu}\ne0$.}
A particularly appealing choice is to set
    \begin{equation}
    \label{eq:transverse}
        \bar{h}_{0i}=0,
    \end{equation}
    which is called the \emph{transverse} condition.
To see why, consider a plane-wave solution to the vacuum EFE, $\dalemb\bar{h}_{\mu\nu}=0$, given by
    \begin{equation}
    \label{eq:plane-wave}
        \bar{h}_{\mu\nu}(t,\vb*x)=\rme^{2\uppi\rmi f(t-\vu*r\vdot\vb*x)}\bar{H}_{\mu\nu},
    \end{equation}
    where $f$ is the frequency of the wave, $\vu*r$ is a unit vector pointing in the propagation direction, and the components of $\bar{H}_{\mu\nu}$ are constants describing the magnitude of the trace-reversed metric perturbation.
Applying the transverse condition~\eqref{eq:transverse} in the harmonic gauge~\eqref{eq:harmonic} gives us
    \begin{equation}
        \partial^\mu\bar{h}_{\mu i}=\partial^j\bar{h}_{ji}=0\qquad\Longrightarrow\qquad\hat{r}^j\bar{H}_{ji}=0,
    \end{equation}
    such that the projection of the plane wave onto the propagation direction is zero; i.e., there is no perturbation along the longitudinal direction, only in the transverse directions.

The transverse condition~\eqref{eq:transverse} uses up three of our four free harmonic functions (one for each of the three values of the spatial index $i$).
We can use the final one to additionally enforce the \emph{traceless} condition,
    \begin{equation}
    \label{eq:traceless}
        \bar{h}\tensor{}{^\mu_\mu}=0,
    \end{equation}
    which erases the distinction between the metric perturbation and its trace-reversed form, $h_{\mu\nu}=\bar{h}_{\mu\nu}$.\footnote{%
    Note that while we have set $h_{0i}=0$, we have kept $h_{00}$ thus far.
    This component must be constant since $\partial^0h_{00}=\partial^\mu h_{\mu0}=0$ by the harmonic condition, such that the vacuum EFE reduces to the Laplace equation, $\partial^i\partial_ih_{00}=0$, and thus does not exhibit wavelike behaviour.
    In fact, in the appropriate weak-field, slow-motion limit of GR, this component is related to the Newtonian gravitational potential, $\phi(\vb*x)=-\frac{1}{2}h_{00}$~\cite{Misner:1974qy}.
    We can thus neglect it if we are sufficiently far from any massive objects, leaving just the spatial components $h_{ij}$ as the nonzero parts of the metric perturbation.
    The traceless condition is then just $\tensor{h}{^i_i}=0$.}
Since the coordinate volume element in our perturbed flat spacetime is proportional to $\sqrt{-g}=\sqrt{1+\tensor{h}{^\mu_\mu}}$, equation~\eqref{eq:traceless} can be interpreted as choosing the gauge such that the coordinate volume of a spacetime region is not affected by a passing GW.

Combining the three conditions in equations~\eqref{eq:harmonic}, \eqref{eq:transverse}, and~\eqref{eq:traceless}, we have exhausted our possibilities for specifying $\xi$, and have thus removed all gauge freedom from the vacuum linearised EFE.
We call this the \emph{transverse-traceless} (TT) gauge.\footnote{\label{ft:TT}%
    In constructing the TT gauge, we have chosen our coordinates such that the metric perturbation is transverse and traceless.
    However, it is useful to note that this is not strictly necessary in practice.
    Given a generic coordinate system in linearised GR, it is always possible to extract the \emph{TT part} of the metric perturbation with an appropriate projection operation.
    So long as the coordinate system is in the harmonic gauge (so that equation~\eqref{eq:efe-linear-harmonic-gauge} holds), one can show that this TT part is equal to the value of the metric perturbation in the TT gauge, allowing us to identify this part as representing GWs even when we are not in the TT gauge.}
Of our initial ten degrees of freedom, we have removed four by choosing the harmonic gauge, and a further four with the transverse and traceless conditions, leaving \emph{two} physical degrees of freedom.
We explore these two GW modes below.

\subsection{The two polarisation modes}
\label{sec:polarisation-modes}

Let us now return to the plane wave~\eqref{eq:plane-wave} and apply the TT gauge.
Without loss of generality, we can choose our spatial coordinates $(x,y,z)$ such that the GW is propagating in the $\vu*z$ direction.
The transverse condition then becomes $h_{zi}=0$, leaving just four nonzero components: two on-diagonal ($h_{xx}$ and $h_{yy}$) and two off-diagonal ($h_{xy}$ and $h_{yx}$).
The former must be the same up to a minus sign to ensure that the trace vanishes, and the latter must be equal to each other to ensure $h_{ij}$ is symmetric, so we see that only two components can be freely chosen; these are the two radiative degrees of freedom of the vacuum gravitational field in GR, which we call the \emph{plus} and \emph{cross polarisations}, $h_+$ and $h_\times$, for reasons that will become apparent in section~\ref{sec:gw-test-masses}.

We can thus write the $\vu*z$-pointing plane wave as
    \begin{equation}
    \label{eq:plane-wave-z}
        h_{ij}(t,z)=\rme^{2\uppi\rmi f(t-z)}
        \begin{pmatrix}
            h_+ & h_\times & 0 \\
            h_\times & -h_+ & 0 \\
            0 & 0 & 0
        \end{pmatrix}
        =\sum_{A=+,\times}\rme^{2\uppi\rmi f(t-z)}e^A_{ij}h_A,
    \end{equation}
    where in the second equality we have defined the TT \emph{polarisation tensors},
    \begin{equation}
        e^+_{ij}=\hat{x}_i\hat{x}_j-\hat{y}_i\hat{y}_j,\qquad e^\times_{ij}=2\hat{x}_{(i}\hat{y}_{j)},
    \end{equation}
    which encapsulate the geometric pattern of each of the two polarisations.
We can also generalise to an arbitrary propagation direction $\vu*r$, which we specify using the standard spherical polar angles $\theta\in[0,\uppi]$ and $\phi\in[0,2\uppi)$; we then have
    \begin{equation}
        h_{ij}(t,\vb*x)=\sum_{A=+,\times}\rme^{2\uppi\rmi f(t-\vu*r\vdot\vb*x)}e^A_{ij}(\vu*r)h_A,
    \end{equation}
    with the polarisation tensors given by
    \begin{equation}
    \label{eq:polarisation-tensors}
        e^+_{ij}(\vu*r)=\hat{\theta}_i\hat{\theta}_j-\hat{\phi}_i\hat{\phi}_j,\qquad e^\times_{ij}(\vu*r)=2\hat{\theta}_{(i}\hat{\phi}_{j)},
    \end{equation}
    where the propagation direction $\vu*r$ and two transverse directions $\vu*\theta$ and $\vu*\phi$ are given by
    \begin{align}
    \begin{split}
        \vu*r&=\sin\theta\cos\phi\,\vu*x+\sin\theta\sin\phi\,\vu*y+\cos\theta\,\vu*z,\\
        \vu*\theta&=\cos\theta\cos\phi\,\vu*x+\cos\theta\sin\phi\,\vu*y-\sin\theta\,\vu*z,\\
        \vu*\phi&=-\sin\phi\,\vu*x+\cos\phi\,\vu*y.
    \end{split}
    \end{align}
Note that these are normalised such that
    \begin{equation}
    \label{eq:polarisation-tensor-normalisation}
        e^A_{ij}(\vu*r)e^{A',ij}(\vu*r)=2\delta_{AA'},
    \end{equation}
    where $A,A'$ are polarisation indices running over $+,\times$, and such that
    \begin{equation}
        a_{ij}=\frac{1}{2}e^A_{ij}e^{A,kl}a_{kl}
    \end{equation}
    for any symmetric TT rank-$(0,2)$ tensor $a_{\mu\nu}$, with implicit summation over the polarisation index.
We are free to redefine the two polarisation modes by rotating $\vu*\theta$ and $\vu*\phi$ around the $\vu*r$-axis by an arbitrary angle $\psi$; this gives
    \begin{equation}
    \label{eq:polarisation-rotation}
        \begin{pmatrix}
            e^+_{ij} \\
            e^\times_{ij}
        \end{pmatrix}
        \to
        \begin{pmatrix}
            \cos2\psi & -\sin2\psi \\
            \sin2\psi & \cos2\psi
        \end{pmatrix}
        \begin{pmatrix}
            e^+_{ij} \\
            e^\times_{ij}
        \end{pmatrix}
        =
        \begin{pmatrix}
            \cos2\psi e^+_{ij}-\sin2\psi e^\times_{ij} \\
            \sin2\psi e^+_{ij}+\cos2\psi e^\times_{ij}
        \end{pmatrix},
    \end{equation}
    mixing the amplitudes of the two polarisations.
We refer to $\psi$ as the \enquote{polarisation angle}.

For a generic plane wave propagating in the $\vu*r$-direction, it is often convenient to define
    \begin{equation}
    \label{eq:complex-strain}
        h(t,\vb*x)\equiv\frac{1}{2}\qty[e^{+,ij}(\vu*r)-\rmi e^{\times,ij}(\vu*r)]h_{ij}(t,\vb*x)=h_+(t,\vb*x)-\rmi h_\times(t,\vb*x),
    \end{equation}
    thereby encoding the two real TT degrees of freedom as a single complex variable, whose amplitude is
    \begin{equation}
        |h|^2=hh^*=h_+^2+h_\times^2=\frac{1}{2}h^{ij}h_{ij}.
    \end{equation}
Under rotations of the polarisation angle, this complex strain transforms like
    \begin{equation}
    \label{eq:complex-polarisation-rotation}
        h\to\rme^{-2\rmi\psi}h,
    \end{equation}
    which is the transformation law for a massless spin-2 particle.

Since we are working in linearised GR, any interactions between different plane waves (which correspond to terms of order $\sim h^2$) are neglected.
This allows us to trivially superpose as many plane-wave solutions as we like, each with arbitrary frequency, propagation direction, and plus- and cross-mode amplitudes (so long as the latter are both much smaller than unity).
In fact, the set of all plane waves serves as a basis in the space of solutions to the linearised vacuum EFE.
We can thus write an arbitrary vacuum GW solution as a superposition of plane waves,
    \begin{equation}
    \label{eq:plane-wave-expansion}
        h_{ij}(t,\vb*x)=\int_\mathbb{R}\dd{f}\int_{S^2}\dd[2]{\vu*r}\rme^{2\uppi\rmi f(t-\vu*r\vdot\vb*x)}e^A_{ij}(\vu*r)\tilde{h}_A(f,\vu*r),
    \end{equation}
    with the contribution from each polarisation, frequency, and propagation direction described by the Fourier components $\tilde{h}_A(f,\vu*r)$, and with implicit summation over the polarisation index.
We can also invert this to write
    \begin{equation}
    \label{eq:strain-fourier-components}
        \tilde{h}_A(f,\vu*r)=\frac{1}{2}f^2\int_{\mathbb{R}^3}\dd[3]{\vb*x}\rme^{-2\uppi\rmi f(t-\vu*r\vdot\vb*x)}e^{A,ij}(\vu*r)h_{ij}(t,\vb*x).
    \end{equation}
This plane-wave expansion will prove useful when we discuss stochastic GW backgrounds in section~\ref{sec:gwb-intro}.

\subsection{Interaction of gravitational waves with test masses}
\label{sec:gw-test-masses}

Since GWs are distortions in the curvature of spacetime, they affect the solutions to the geodesic equation~\eqref{eq:geodesic}, and can thus influence the trajectories of particles.
(Of course, there is no absolute reference frame in GR, so we are really interested in how GWs affect the pairwise separation between different geodesics.)
These effects are obscured if we work in TT coordinates, as the relevant Christoffel symbols for a particle initially at rest ($U^i=0$) are
    \begin{equation}
        \Gamma^i_{00}=\tensor{\partial}{_0}\tensor{h}{_0^i}-\frac{1}{2}\tensor{\partial}{^i}\tensor{h}{_{00}},
    \end{equation}
    which vanish by construction in the TT frame.
As a result, the \emph{coordinate} distance between two initially stationary freely-falling particles is unaffected by GWs in this frame at linear order.
However, the \emph{proper} distance between them (i.e., as measured using the spacetime metric) is affected, due to the time-varying metric perturbation $h_{ij}$.
Since proper distances are diffeomorphism-invariant, we see that this must be a physical effect and not a coordinate artefact.

In order to analyse the response of a system of test masses to an impinging GW, it is therefore preferable to choose an alternative coordinate frame in which coordinate distances more closely follow the behaviour of proper distances.
We can do this by constructing the \emph{proper detector frame} (also called \emph{Fermi coordinates}), in which the metric is locally flat with vanishing first derivatives along the trajectory of the system, with no rotation of the coordinates along this trajectory (in the sense that the coordinate spin axes of freely-spinning gyroscopes do not change over time).
Since the metric is locally flat, coordinate distances are then equal to proper distances if we work sufficiently close to the origin.
It is straightforward to show that, in this frame, the separation vector $\xi^i$ between two nearby geodesics obeys $\ddot{\xi}^i=\tensor{R}{^i_{00j}}\xi^j$~\cite{Maggiore:2007zz}, where dots denote derivatives with respect to the coordinate time (which is locally equivalent to the system's proper time).

\begin{figure}[t!]
    \includegraphics[width=\textwidth]{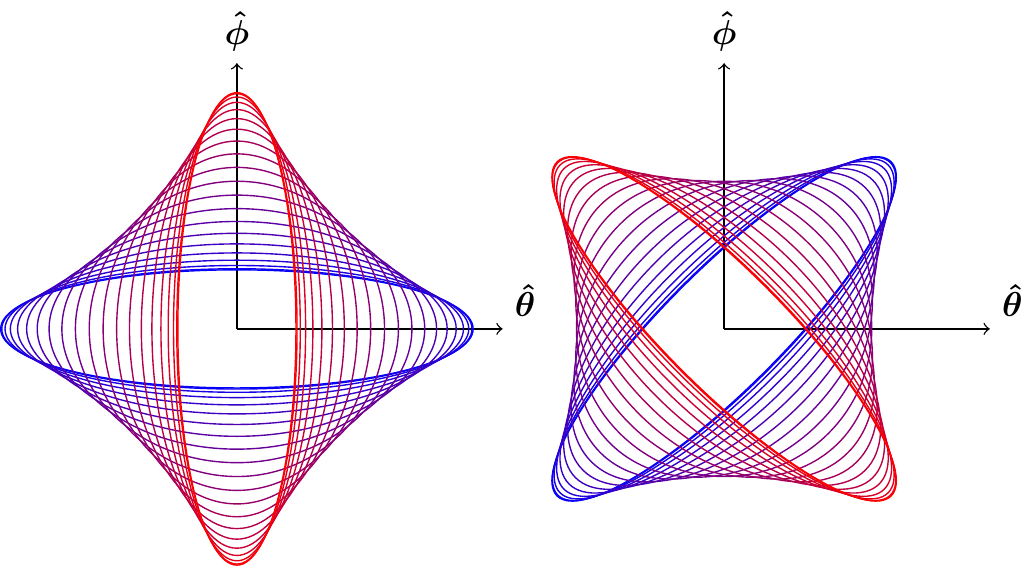}
    \caption{%
    Deformation of a circular ring of test particles by a plane GW propagating out of the page.
    The left panel shows the pure $+$-polarised case, while the right panel shows the $\times$-polarised case.
    The colour gradient (from blue to red) shows how the deformation evolves over a time interval equal to half the GW oscillation period.
    (Note that the amplitude of the deformation has been greatly exaggerated here for clarity.)
    }
    \label{fig:plus-cross}
\end{figure}

At this point, we could evaluate the Riemann tensor as a function of the metric perturbation in the proper detector frame.
However, it turns out that the components of the Riemann tensor are \emph{invariant} under gauge transformations in linearised GR (note this is a stronger statement than the tensor obeying general covariance), so we are free to evaluate it in the TT frame, making the link to the physical GW perturbation clearer.
We therefore find that
    \begin{equation}
    \label{eq:test-mass-response}
        \ddot{\xi}^i=\frac{1}{2}\ddot{h}^{ij}\xi_j,
    \end{equation}
    where $\xi^i$ is the coordinate separation in the proper detector frame as before, and $h_{ij}$ is the metric perturbation in the TT frame.

Substituting in the $\vu*z$-pointing plane wave solution from equation~\eqref{eq:plane-wave-z}, we find that equation~\eqref{eq:test-mass-response} has solutions
    \begin{equation}
        \vb*\xi=\frac{1}{2}\rme^{2\uppi\rmi f(t-z)}\qty[h_+(x_0\vu*x-y_0\vu*y)+h_\times(y_0\vu*x+x_0\vu*y)],
    \end{equation}
    where $(x_0,y_0)$ are the initial coordinates of the separation vector (we can treat these as constant on the right-hand side here, as their variation over time only gives corrections of order $\sim h^2$).
The resulting deformations of a circular ring of test particles are shown in figure~\ref{fig:plus-cross}, clearly demonstrating where the plus and cross polarisations get their names from.
We see that $\vb*\xi\vdot\vu*z=0$, so the GW is transverse as expected.

When we say that equation~\eqref{eq:test-mass-response} holds for pairs of \enquote{nearby} geodesics, we mean that the size of the separation vector $\xi$ is much less than the lengthscale over which the spacetime curvature varies---in this case, the wavelength of the GW, $\lambda\sim1/f$.
This situation, in which $\xi\ll\lambda$, is called the \emph{small-antenna limit} in the context of GW detectors.
As we discuss in section~\ref{sec:gw-detection}, this limit holds for some GW experiments (e.g., ground-based interferometers such as LIGO and Virgo), but not for others (e.g., pulsar timing arrays).
In the latter case, the right-hand side of equation~\eqref{eq:test-mass-response} picks up higher-order terms in $\xi$, as well as a nontrivial dependence on the frequency of the GW.

\subsection{Generation of gravitational waves}
\label{sec:gw-generation}

We have focused thus far on the propagation of GWs in vacuum.
However, in order to use GWs as \emph{messengers} which tell us about objects in the cosmos, we need to understand how said objects generate GWs in the first place.
To do so, we abandon the TT gauge momentarily (since our freedom to set certain metric components to zero is inhibited by the presence of a source term), and return to the linearised EFE in the harmonic gauge, equation~\eqref{eq:efe-linear-harmonic-gauge}.
We can invert the flat-space d'Alembertian using the retarded Green's function,
    \begin{equation}
        \mathcal{G}(x,x')=-\frac{\delta(t-t'-|\vb*x-\vb*x'|)}{4\uppi|\vb*x-\vb*x'|},\qquad\dalemb\mathcal{G}(x,x')=\delta^{(4)}(x-x').
    \end{equation}
The corresponding solution for the trace-reversed metric perturbation is then given by
    \begin{equation}
        \bar{h}_{ij}(t,\vb*x)=-16\uppi G\int\dd[4]{x'}\mathcal{G}(x,x')T_{ij}(x')=4G\int\frac{\dd[3]{\vb*x'}}{|\vb*x-\vb*x'|}T_{ij}(t-|\vb*x-\vb*x'|,\vb*x'),
    \end{equation}
    where the integral runs over a constant-time hyperslice.
Outside of the source we are once again free to impose the TT gauge, writing
    \begin{equation}
    \label{eq:gw-source-solution}
        h_{ij}(t,\vb*x)=4G\int\frac{\dd[3]{\vb*x'}}{|\vb*x-\vb*x'|}\qty[T_{ij}(t-|\vb*x-\vb*x'|,\vb*x')]^\mathrm{TT},
    \end{equation}
    where the \enquote{TT} superscript indicates that we have extracted the transverse-traceless part of the energy-momentum tensor (with the transverse directions defined relative to the propagation from the source to $\vb*x$), so that the quantity on the left-hand side corresponds to a gauge-invariant GW propagating in vacuum (see footnote~\ref{ft:TT}).

In many physical situations, the motion of the source is more straightforwardly described in the Fourier domain.
We thus rewrite the energy-momentum tensor in terms of its spacetime Fourier transform,
    \begin{equation}
        T_{ij}(t,\vb*x)=\int\dd{f}\int\dd[3]{\vb*k}\rme^{2\uppi\rmi(ft-\vb*k\vdot\vb*x)}\tilde{T}_{ij}(f,\vb*k).
    \end{equation}
Inserting this into equation~\eqref{eq:gw-source-solution}, we find that the GW solution at distances much greater than the size of the source is given by
    \begin{equation}
        h_{ij}(t,\vb*x)=\frac{4G}{r}\int\dd{f}\rme^{2\uppi\rmi f(t-r)}\qty[\tilde{T}_{ij}(f,f\vu*r)]^\mathrm{TT},\qquad r\gg\max|\vb*x'|,
    \end{equation}
    where $\vb*r$ is a 3-vector pointing from the centre of the source to $\vb*x$.
Equivalently, we can write this in terms of the plane wave components $\tilde{h}_A$ using equation~\eqref{eq:strain-fourier-components},
    \begin{equation}
        \tilde{h}_A(f,\vu*n)=\frac{2G}{r}\rme^{-2\uppi\rmi fr}e^{A,ij}(\vu*r)\tilde{T}_{ij}(f,f\vu*r)\delta^{(2)}(\vu*r,\vu*n).
    \end{equation}
Here the exponential phase factor reflects the time taken for the signal to reach distance $r$ (this can be removed by translating the time coordinate used in the Fourier transforms), while the Dirac delta tells us that the only nonzero plane wave component is that which points in the direction of the source.\footnote{Here $\delta^{(2)}(\vu*r,\vu*r')\equiv\delta(\cos\theta-\cos\theta')\delta(\phi-\phi')$ is the Dirac delta function on the 2-sphere, defined such that we have $\int_{S^2}\dd[2]{\vu*r'}f(\vu*r')\delta^{(2)}(\vu*r,\vu*r')=f(\vu*r)$ for test functions $f:S^2\to\mathbb{R}$.}
(Note that the \enquote{TT} subscript is no longer needed, as we have projected onto the TT polarisation tensors.)

Thus far we have made no assumptions about the source other than requiring it to be weakly-gravitating, so as to be consistent with linearised GR.
However, we can greatly simplify our description of GW generation if we consider sources with internal velocities much smaller than the speed of light,\footnote{%
    Note that in virialised self-gravitating systems, $v^2$ is of the same order of magnitude as the gravitational potential $GM/R$, where $M$ and $R$ are the mass and radius of the system; any self-gravitating system which is consistent with the linearised GR requirement that $GM/R\ll1$ must therefore also be slowly-moving, $v\ll1$.} $v\ll1$.
In this limit, the resulting GW signal is dominated by the mass quadrupole of the source,
    \begin{equation}
        h_{ij}(t,\vb*x)\simeq\frac{2G}{r}\qty[\ddot{\mathcal{Q}}_{ij}(t-r)]^\mathrm{TT},\qquad\mathcal{Q}_{ij}(t)\equiv\int\dd[3]{\vb*x'}\rho(t,\vb*x')x'_ix'_j,
    \end{equation}
    where $\rho\equiv T_{00}$ is the mass density.
As a simple example, consider a system of two slowly-moving point masses.
In a frame where the centre of mass of the system is fixed at the origin, the quadrupole moment is given by
    \begin{equation}
        \mathcal{Q}^{ij}=m_1x_1^ix_1^j+m_2x_2^ix_2^j=\mu R^iR^j
    \end{equation}
    (ignoring terms of order $v^2$), where $\vb*R\equiv\vb*x_1-\vb*x_2$ is a 3-vector pointing from the position of one mass to the other, and $\mu\equiv m_1m_2/(m_1+m_2)$ is the reduced mass of the system.
The emitted GW signal is thus
    \begin{equation}
    \label{eq:gw-signal-2-body}
        h_{ij}\simeq\frac{4G\mu}{r}\qty[\dot{R}_i\dot{R}_j+R_{(i}\ddot{R}_{j)}]^\mathrm{TT},
    \end{equation}
    where the right-hand side is evaluated at the retarded time $t-r$.

\subsection{Energy and momentum of gravitational waves}
\label{sec:gw-energy}

The notions of energy and momentum are extremely intuitive and useful in understanding the physics of GWs.
However, they are also notoriously thorny issues in GR~\cite{Szabados:2009eka}, as general covariance makes it impossible to find a local measure (i.e., one defined at each point in spacetime) of gravitational energy-momentum that all observers can agree on.
To see this, note that the energy density of the gravitational field in Newtonian gravity goes like $\sim(\nabla\phi)^2$ (where $\phi$ is the Newtonian potential), so to ensure consistency in the Newtonian limit we would expect any analogous quantity in GR to be quadratic in the first derivatives of the metric, $\sim(\partial g)^2$.
The problem is that it is always possible to transform to a local inertial frame in which these first derivatives vanish, and we can therefore always set any such quantity to zero at any point in spacetime.
(This problem arises in both linearised GR and the full nonlinear theory, and is just as much of an obstacle for defining, e.g., the mass density of a BH as it is for defining the energy density of GWs.)

Nonetheless, we can still define a sensible \emph{quasi-local} energy-momentum tensor for the gravitational field by averaging over a region of spacetime, rather than focusing on a single point.
The intuition here is that, while $\partial g$ can be set to zero at any individual point, it cannot be simultaneously set to zero at every point in a sufficiently \enquote{large} region of spacetime in any one coordinate frame.

There are two key approaches for constructing a quasi-local energy-momentum tensor for GWs.\footnote{%
    A detailed derivation of equation~\eqref{eq:gw-energy-momentum} using both approaches described here can be found in \citet{Maggiore:2007zz}.}
One is to write down a Lagrangian for linearised GR and use Noether's theorem to identify the conserved currents associated with the invariance of this Lagrangian under spacetime translations (this is the usual procedure for defining energy and momentum in other field theories).
The other is to work directly with the EFE and study how GWs \emph{backreact} on spacetime, acting as a source for further perturbations to the metric.\footnote{%
    This approach requires us to go beyond the flat-background perturbation theory we have adopted so far, as we must allow the background spacetime to evolve dynamically to study how it responds to the energy-momentum carried by the GWs.
    In order for the distinction between the background and the GWs to be unambiguous, we require a separation of scales (either spatial or temporal) between the low-frequency/long-wavelength variations in the background metric and the high-frequency/short-wavelength perturbations due to the GWs.
    One can then develop an effective theory of the metric on large scales by \enquote{integrating out} the GW perturbations, with equation~\eqref{eq:gw-energy-momentum} representing a renormalisation correction to the total energy-momentum tensor due to this coarse-graining.}
Both approaches yield the same answer:
    \begin{equation}
    \label{eq:gw-energy-momentum}
        T_\gw^{\mu\nu}=\frac{1}{32\uppi G}\ev{\partial^\mu h^{\alpha\beta}\partial^\nu h_{\alpha\beta}},
    \end{equation}
    where the angle brackets denote a quasi-local spatial average over a region much larger than the wavelength of the GW (or equivalently, a temporal average over many periods).
In particular, the quasi-local GW energy density is
    \begin{equation}
    \label{eq:gw-energy-density}
        \rho_\gw=\frac{1}{32\uppi G}\ev{\dot{h}^{\alpha\beta}\dot{h}_{\alpha\beta}}.
    \end{equation}
These expressions apply so long as we are in the harmonic gauge, and are invariant under the remaining gauge transformations by virtue of the averaging process.

Using equation~\eqref{eq:gw-energy-momentum}, we can thus calculate the total GW energy flux radiated by a generic isolated source in a given direction $\vu*r$ over all time,
    \begin{equation}
    \label{eq:gw-energy-flux}
        \frac{\dd{E_\gw}}{\dd[2]{\vu*r}}=\frac{r^2}{32\uppi G}\int_\mathbb{R}\dd{t}\dot{h}^{ij}\dot{h}_{ij}=\frac{r^2}{16\uppi G}\int_\mathbb{R}\dd{t}|\dot{h}|^2=\frac{\uppi r^2}{2G}\sum_{A=+,\times}\int_0^\infty\dd{f}f^2|\tilde{h}_A|^2.
    \end{equation}
Here we assume that the flux is measured at a radius $r$ much larger than both the size of the source and the wavelengths of interest,\footnote{The leading-order piece of the GW strain we are interested in goes as $\sim1/r$ at large distances, and is thus compensated by the factor of $r^2$ in equation~\eqref{eq:gw-energy-flux}. To get an invariant expression for the energy flux we must therefore choose $r$ large enough that the next-to-leading order $\sim1/r^2$ piece of the strain is negligible. Taking $r$ much larger than the GW wavelength and the size of the source achieves this.} yet small enough that cosmological expansion can be neglected.
We have assumed we are in vacuum, such that we are free to select the TT gauge and replace $\alpha$ and $\beta$ by purely spatial indices.
We have also removed the angle brackets, since the time integral averages over many GW periods for us.
The final equality uses Parseval's theorem to convert the time integral into a frequency integral.

\subsection{Gravitational-wave memory}
\label{sec:gw-memory}

\begin{figure}[t!]
    \begin{center}
        \includegraphics[width=\textwidth]{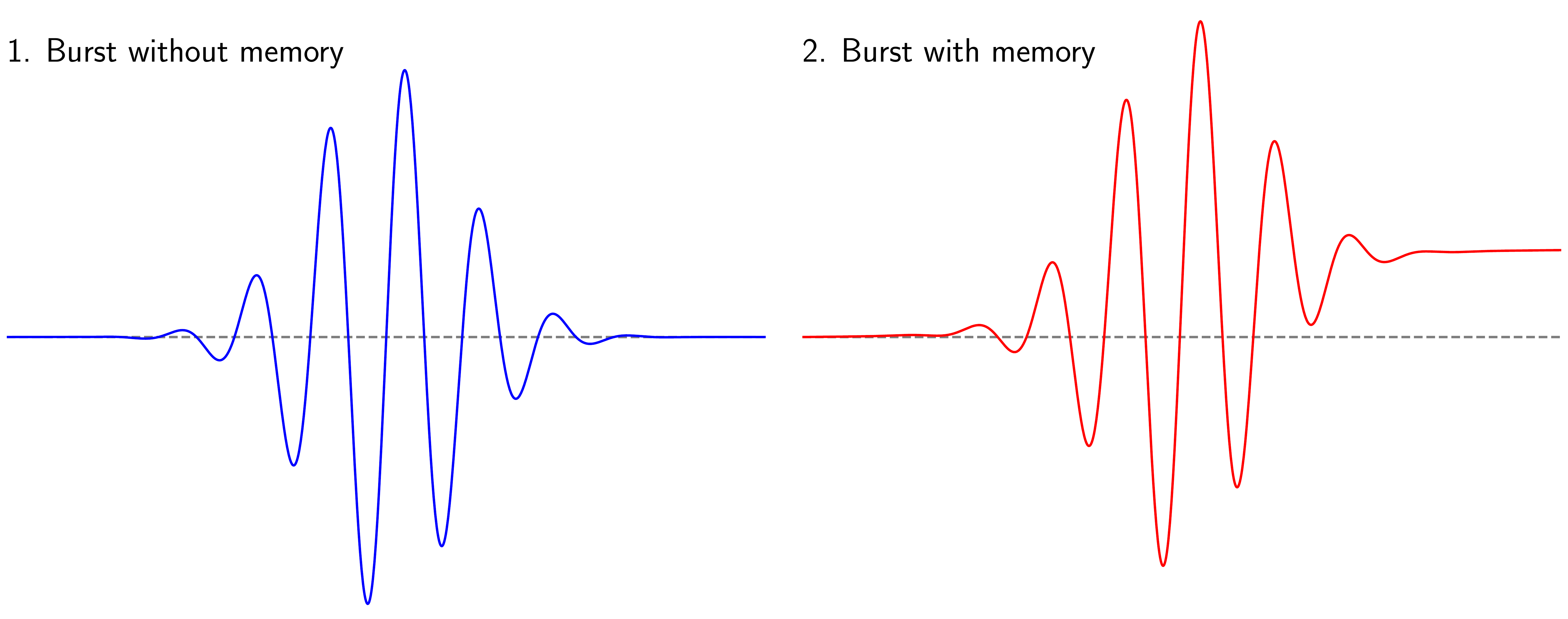}
    \end{center}
    \caption{%
    A cartoon illustration of the memory effect.
    In the left panel, a gravitational-wave burst passes and the strain returns to its initial value.
    In the right panel, a gravitational-wave burst passes and leaves a nonzero memory offset.
    }
    \label{fig:memory}
\end{figure}

Much of our basic intuition for the behaviour of GWs comes from examples of wavelike systems we are familiar with in our everyday lives, such as the propagation of ripples on the surface of a pond.
These analogies encourage us to visualise GWs as shown in the left panel of figure~\ref{fig:memory}: an incoming wave packet causes some initially quiescent quantity (e.g., the vertical displacement of a leaf floating on the pond, or the proper distance between two freely-falling test masses) to oscillate until it eventually returns to its starting value.
Remarkably, this intuitive picture is often wrong in the case of GWs, as there are many situations in which the GW strain does \emph{not} return to its initial value, but instead settles down with some fixed offset from its starting point (as illustrated in the right panel of figure~\ref{fig:memory}), i.e.,
    \begin{equation}
        \Updelta h_{ij}\equiv\lim_{t\to+\infty}h_{ij}(t)-\lim_{t\to-\infty}h_{ij}(t)\ne0.
    \end{equation}
This corresponds to a permanent change in the GW measurement apparatus, and is therefore called the \emph{GW memory effect}~\cite{Braginsky:1986ia,Braginsky:1987gw,Favata:2010zu}, as the apparatus \enquote{remembers} the passage of the wave packet.

Though surprising at first, it is straightforward to derive this effect by solving the linearised EFE for sources whose energy-momentum tensor undergoes a permanent change between early and late times.
For example, consider the GW signal~\eqref{eq:gw-signal-2-body} that we derived above for two slowly-moving test masses.
We see immediately that if the relative velocity or acceleration of the two bodies is different at early and late times, then the resulting GW signal will exhibit a memory effect like that shown in figure~\ref{fig:memory}.
Such situations typically occur when the system is gravitationally unbound,\footnote{%
    Gravitationally-bound systems such as compact binaries do generate \emph{some} linear memory if the merger product has a non-zero recoil velocity, but this linear memory signal is likely far too weak to be detected~\cite{Favata:2008ti}, even for systems with maximal recoil velocities (the so-called \enquote{super-kick} configuration)~\cite{Campanelli:2007cga}.}
     either in \enquote{scattering-like} events where the masses follow a hyperbolic orbit, undergoing a permanent change in their velocities due to their gravitational interaction, or in \enquote{explosion-like} events where the masses are initially bound, but become unbound due to some injection of energy and escape to infinity with nonzero velocities.
These examples are both simplified descriptions of real astrophysical GW sources, such as stellar scattering~\cite{Zeldovich:1974gvh,Smarr:1977fy,Turner:1977hyp,Turner:1978zz,Kovacs:1978eu,Bontz:1979zfl}, core-collapse supernovae~\cite{Epstein:1978dv,Turner:1978jj,Burrows:1995bb,Kotake:2005zn,Ott:2008wt}, and gamma-ray bursts~\cite{Segalis:2001ns,Sago:2004pn,Birnholtz:2013bea,Akiba:2013qwa}, as well as more idealised theoretical setups such as high-energy particle decay~\cite{Tolish:2014bka,Tolish:2014oda,Allen:2019hnd}.
In both cases the GW memory associated with the signal~\eqref{eq:gw-signal-2-body} is given by
    \begin{equation}
        \Updelta h_{ij}\simeq\Updelta\frac{4G}{r}\mu(v_iv_j)^\mathrm{TT},
    \end{equation}
    where $\vb*v\ll1$ is the relative velocity, and we have neglected the acceleration term, which vanishes at $t\to\pm\infty$ in both examples.
This expression is readily generalised to relativistic velocities and more than two particles, giving~\cite{Thorne:1992sdb}
    \begin{equation}
    \label{eq:linear-memory}
        \Updelta h_{ij}=\Updelta\frac{4G}{r}\sum_am_a\gamma_a\frac{(v_{a,i}v_{a,j})^\mathrm{TT}}{1-\vu*r\vdot\vb*v_a},
    \end{equation}
    where the index $a$ labels the particles, $\gamma_a\equiv(1-v_a^2)^{-1/2}$ is the Lorentz factor of particle $a$, and $\vu*r$ is the GW propagation direction.

While GW memory effects associated with unbound systems have been studied since the 1970s~\cite{Zeldovich:1974gvh,Smarr:1977fy,Turner:1977hyp,Kovacs:1978eu,Epstein:1978dv,Bontz:1979zfl,Turner:1978zz,Turner:1978jj}, it was not until 1991 that \citet{Christodoulou:1991cr} pointed out a much more generic memory effect which accompanies essentially \emph{any} source of GWs,\footnote{%
    The only exception is when the integrand of equation~\eqref{eq:nonlinear-memory} has vanishing TT component; e.g., when the radiated GW flux is exactly isotropic.}
    called the \emph{nonlinear} memory (or sometimes, \enquote{Christodoulou memory}),
    \begin{equation}
    \label{eq:nonlinear-memory}
        \Updelta h_{ij}=\frac{4G}{r}\int_{S^2}\dd[2]{\vu*r'}\frac{\dd{E_\gw}}{\dd[2]{\vu*r'}}\frac{(\hat{r}'_i\hat{r}'_j)^\mathrm{TT}}{1-\vu*r\vdot\vu*r'}.
    \end{equation}
(In contrast, the effect encapsulated by equation~\eqref{eq:linear-memory} is called the \emph{linear} memory.)
The reason the nonlinear memory effect went unnoticed for so long is that it is absent in the linearised theory, and was only discovered by Christodoulou using rigorous asymptotic methods~\cite{Christodoulou:1991cr} developed in his proof (with Klainerman) of the global nonlinear stability of Minkowski space~\cite{Christodoulou:1993uv}.
(The same effect was also discovered soon after by Blanchet and Damour~\cite{Blanchet:1992br} in the very different context of post-Minkowskian theory.)
We see that the nonlinear memory is sourced not by the momenta of unbound particles escaping to infinity as in equation~\eqref{eq:linear-memory}, but instead by the GW energy radiated by the source.
In fact, as \citet{Thorne:1992sdb} pointed out soon after Christodoulou's result, we can straightforwardly interpret equation~\eqref{eq:nonlinear-memory} as the \emph{linear} memory~\eqref{eq:linear-memory} associated with the gravitons radiated by the source.
To see this, we need only make three small modifications to equation~\eqref{eq:linear-memory}:
    \begin{enumerate}
        \item replace $m_a\gamma_a$ by $E_a$, the energy of each graviton (as measured in the observer's rest frame);
        \item let $\vb*v_a$ have unit magnitude (since gravitons propagate at the speed of light);
        \item take the limit of infinitely many particles (i.e., the classical limit), so that the sum $\sum_a$ is replaced by an integral over a smooth angular distribution.
    \end{enumerate}
(The $\Updelta$ symbol is also unnecessary on the right-hand side of equation~\eqref{eq:nonlinear-memory}, since we assume the GW flux from the source is zero in the distant past, $t\to-\infty$.)
This interpretation makes clear why such an effect is only present in the full, nonlinear theory: it is due to the backreaction of gravitational perturbations on spacetime, a phenomenon which is neglected when working with a fixed, non-dynamical background.

While the memory effect was first discovered in the context of classical GR, it has been recognised in recent years that analogous memory effects are a generic feature of field theories with massless degrees of freedom~\cite{Bieri:2013hqa,Tolish:2014bka,Tolish:2014oda,Susskind:2015hpa,Pate:2017vwa}.
Explicit examples that have been studied include \enquote{electromagnetic memory} in electrodynamics~\cite{Bieri:2013hqa,Kapec:2015vwa,Susskind:2015hpa,Campoleoni:2019ptc} and \enquote{colour memory} in Yang-Mills theory~\cite{Pate:2017vwa,Ball:2018prg,Campoleoni:2019ptc,Jokela:2019apz,Jokela:2020ibz}, with the photon and the gluon playing the r\^ole of the graviton, respectively.
Related work (principally conducted by Strominger and his collaborators) has shown that these memory effects are intimately related to both the asymptotic symmetries of the fields in each theory and to \enquote{soft theorems} which govern the production of low-energy massless particles in the corresponding quantum theory~\cite{He:2014laa,Strominger:2014pwa,Pasterski:2015tva,Pasterski:2015zua,Kapec:2015vwa,Flanagan:2015pxa,Strominger:2017zoo}.
These deep theoretical links add to the appeal of GW memory as a tool for understanding GR and probing the true nature of the gravitational field.

\section{The gravitational-wave background}
\label{sec:gwb-intro}

We have seen in section~\ref{sec:what-are-gws} that GR predicts the existence of two massless propagating degrees of freedom associated with TT perturbations to the gravitational field.
These are generated by the motion of matter fields, meaning that the GW signals we receive on Earth can be used to infer the properties of distant objects in the Universe.
In this section we turn our attention to a GW signal that is of particular interest to cosmologists: the stochastic GW background (SGWB)~\cite{Allen:1996vm,Maggiore:1999vm,Regimbau:2011rp,Romano:2016dpx,Christensen:2018iqi,Caprini:2018mtu}.
We begin by discussing what distinguishes this signal from the compact binary signals detected thus far by LIGO/Virgo, before covering its statistical properties, how to model its expected amplitude, and how best to search for it in noisy data.

\begin{figure}[t!]
    \begin{center}
        \includegraphics[width=\textwidth]{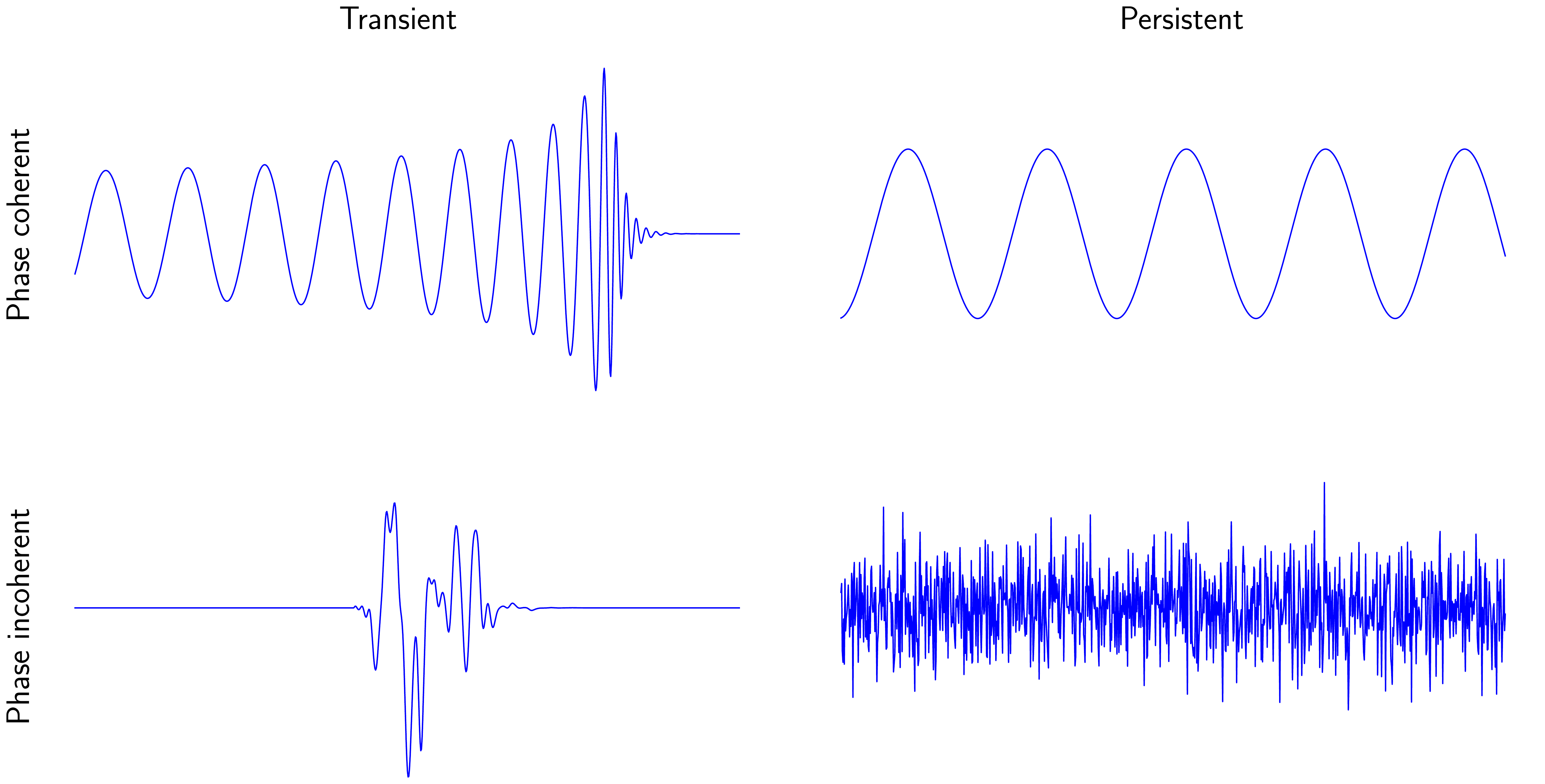}
    \end{center}
    \caption{%
    A taxonomy of four key gravitational-wave signal morphologies: phase-coherent transients such as compact binaries; persistent phase-coherent signals such as the \enquote{continuous waves} emitted by spinning neutron stars; incoherent transients (i.e., bursts), such as those emitted by core-collapse supernovae; and persistent, incoherent signals---namely, the stochastic background.}
    \label{fig:taxonomy}
\end{figure}

\subsection{What is the stochastic background?}

Imagine we have constructed a GW detector (a few examples of which are described in section~\ref{sec:gw-detection}), and we switch it on.
What kinds of signals might we expect to see?
This simple question is incredibly important for GW astronomy, due to the fact that GWs are typically extremely faint and therefore difficult to distinguish from instrumental noise.
Unlike in EM astronomy, where the signal is often very easily distinguishable from the noise (even instruments with very high noise levels, such as the naked eye, can easily detect many EM signals in the night sky), signal detection in GW astronomy relies on sophisticated statistical methods, with the best approach depending strongly on the morphology of the signal.

Of all the ways we might characterise a GW signal, there are two questions that are perhaps the most important:
    \begin{enumerate}
        \item Is the signal \emph{transient}, or is it \emph{persistent}?
        i.e., does the signal only appear in the detector for a relatively short time, or is it \enquote{always on}?\footnote{%
        Note that \enquote{relatively short} means relative to a typical observational timescale.
        A signal lasting for, say, several centuries, while certainly transient on astronomical timescales, would be treated as persistent for our purposes here.}
        \item Is the signal \emph{coherent}, or is it \emph{incoherent}?
        i.e., are we able to deterministically model the phase of the signal, or does our lack of knowledge about the source force us to treat the phase as random?\footnote{%
        It is interesting to note that in EM astronomy, essentially \emph{all} signals are incoherent; one rarely attempts to measure the phase of electromagnetic waves, only their intensity.
        This is because astronomical EM radiation is typically associated with the random, thermal motion of charges on microscopic scales, which is impossible to model in a phase-coherent way.
        Viewed through this lens, we see that the use of matched filtering in GW astronomy relies on the large-scale, coherent motion of macroscopic massive bodies (such as the components of a compact binary) to generate signals whose phase can be modelled accurately.
        One can see this as a consequence of the universally-attractive nature of gravity; an analogous system of coherently oscillating macroscopic charges would be very difficult to realise in the Universe due to the cancellation of positive and negative charges on smaller scales.}
    \end{enumerate}
These two pairs of categories define four different kinds of GW signal (as illustrated in figure~\ref{fig:taxonomy}), each of which has its own set of specialised search methods.\footnote{%
    As a result of this, these four categories correspond exactly to the four main data analysis working groups in the LIGO/Virgo/KAGRA Collaboration.
    For an introduction to some of the methods used by each group, see \citet{Creighton:2011zz} or \citet{Maggiore:2007zz}.}

Phase-coherent signals are by far the easiest to search for, since one can use the technique of \emph{matched filtering} (convolving the data with a template which encapsulates our knowledge of the signal); assuming a perfect template and Gaussian noise, this returns the maximum signal-to-noise ratio (SNR) of any possible search~\cite{Wainstein:1962vrq}.
In the case of incoherent signals it is, by definition, impossible to construct a template, so this option is not available to us.
Transient signals also present a key search advantage compared to persistent signals, as their arrival represents a change in the state of the detector; persistent signals, on the other hand, are harder to disentangle from the instrumental noise, which is also \enquote{always on}.\footnote{On the other hand, of course, there are many advantages to searching for a signal one knows is always present. With transient searches, there is always a risk of missing rare and exciting signals due to changes in the state of the detector.}
It is unsurprising, then, that all of the GWs detected by LIGO/Virgo thus far originated from compact binaries; as we show in section~\ref{sec:compact-binaries}, these are prime examples of transient, phase-coherent signals.

\begin{figure}[t!]
    \begin{center}
        \includegraphics[width=0.55\textwidth]{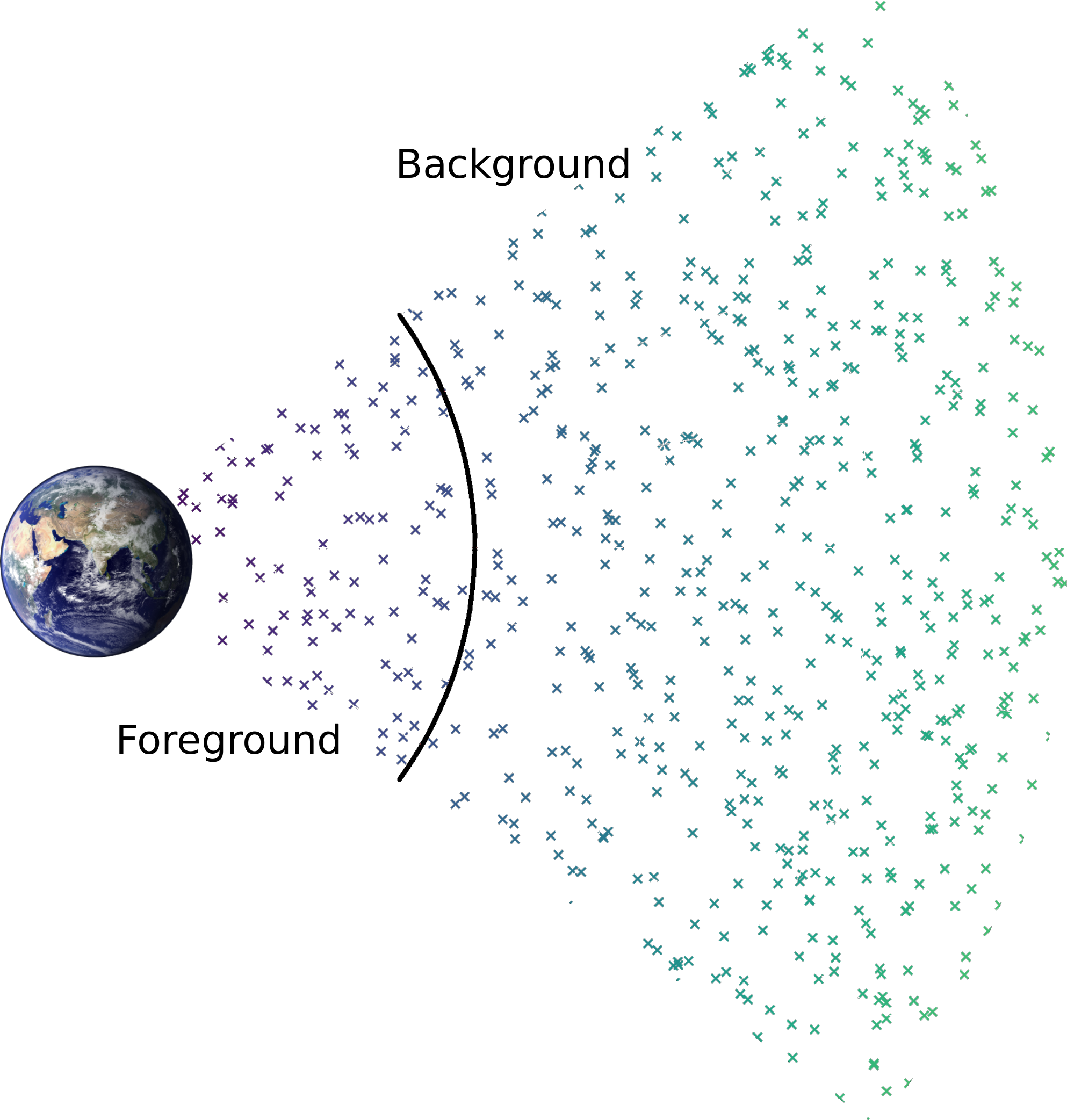}
    \end{center}
    \caption{%
    A cosmic population of gravitational-wave sources gives typically gives rise to a small number of signals that are individually resolvable (the \enquote{foreground}), and a much larger number that are not (the \enquote{background}).
    The combined emission from the latter gives rise to a persistent, incoherent signal that we call the stochastic background.}
    \label{fig:foreground-background}
\end{figure}

Consider, however, the situation depicted in figure~\ref{fig:foreground-background}, in which we have a cosmic population of sources distributed throughout the Universe; these could be compact binaries, or any other population of GW sources.\footnote{%
    In some contexts it makes sense to replace the notion of \enquote{individual sources} here with \enquote{individual Hubble patches} in which some cosmological GW production process takes place; e.g., the GWs generated by a first-order phase transition, which we discuss below in section~\ref{sec:phase-transitions}.
    The rest of the discussion here still holds in cases like this, with the caveat that typically \emph{none} of the \enquote{sources} are individually resolvable.}
Since the strain signal from each of these sources falls off like $\sim1/r$, any instrument will inevitably have a \enquote{detection horizon} beyond which the signals are too faint to be distinguished from noise.
We can also see in figure~\ref{fig:foreground-background} that, so long as the number density of sources doesn't fall off too quickly, the total number of sources grows rapidly as a function of distance (since the size of a spherical shell of radius $r$ grows like $\sim r^2$).
As a result, we might have a situation where it is impossible to distinguish individual sources from each other past a certain distance due to the \enquote{confusion noise} from many overlapping signals, even if these are all inside the detection horizon.

In both situations, matched filtering is no longer effective at detecting individual sources.
But the GWs emitted by these sources are still present in the detector, and can still be searched for using other methods.
Instead of focusing on individual sources, it becomes more useful to consider the entire population of sources together, treating their combined GW emission as one aggregate signal.
Since these sources are typically very widely separated from each other, we expect them to be causally disconnected, meaning that there is no discernible pattern in the arrival times of each individual signal.
This makes it impossible to deterministically model the phase of the aggregate signal, even if we have a perfect template for each individual signal, as we cannot predict how each of their independent phases will combine.
It also means that we should expect the signal to be time-translation invariant (at least in a statistical sense), since there is no reason to expect that all of these distant, independent sources should conspire to emit their GWs at the same time.
We therefore have a \emph{persistent}, \emph{incoherent} signal associated with essentially any cosmic population of GW sources.
We call this signal a \emph{stochastic gravitational-wave background} (SGWB);\footnote{%
    We will often refer to \enquote{a} stochastic background, meaning the persistent, incoherent signal associated with one particular population of sources (e.g., compact binaries), as opposed to \enquote{the} stochastic background, which refers to the superposition of all such stochastic signals, which is what we see in a detector.
    The question of how best to distinguish between multiple stochastic signals which might be simultaneously measured by a detector is an important open problem in GW astronomy~\cite{Harms:2008xv,Regimbau:2016ike,Caprini:2019pxz,Parida:2019ybm,Boileau:2020rpg,Flauger:2020qyi,Sachdev:2020bkk,Sharma:2020btq,Boileau:2021sni,Martinovic:2020hru,Suresh:2021rsn}, which we do not discuss further in this thesis.
    } \enquote{stochastic} because of its random, nondeterministic phase evolution, and \enquote{background} because of its association with a large number of distant sources.
In contrast, the small number of nearby, individually resolvable sources form what we might call a \enquote{foreground}.\footnote{%
    In some contexts the term \enquote{foreground} has negative connotations, and implies that the signal in question is a nuisance factor which one must mitigate in order to observe the underlying target signal (e.g., polarised dust emission from the Milky Way acts as a nuisance foreground for CMB observatories~\cite{Aghanim:2018vks}).
    The situation is different here, as both foreground and background are extremely scientifically valuable in GW astronomy.}

As discussed above, persistent and incoherent signals like the SGWB are the most difficult to search for, since our ignorance of the phase evolution means we are unable to use matched filtering, and the lack of a distinct arrival time makes it hard to differentiate between the signal and the detector noise.
So why should we bother?
The answer can be seen in figure~\ref{fig:foreground-background}: being associated with distant sources, the SGWB can give us a glimpse of the Universe at \emph{high redshift} and on \emph{large scales}, revealing cosmological information that is inaccessible with individual nearby detections alone.
We therefore see that the SGWB is an indispensable tool in our effort to probe cosmology and fundamental physics with GW observations.

In practice, the distinction between \enquote{phase-coherent} and \enquote{phase-incoherent} signals is somewhat blurry, particularly when the incoherence comes from a finite number of overlapping signals (e.g., in the compact binary case)~\cite{Cornish:2015pda}.
If the number of such signals is sufficiently small, and if the phase evolution of each individual signal is well-known, then it is possible to use semi-coherent search methods that leverage this knowledge and outperform traditional stochastic searches---see, in particular, \citet{Smith:2017vfk} for a Bayesian methodology that relies on the known phase evolution of BBH signals, as well as the relatively low expected rate of stellar-mass BBHs in the Universe.
In recognition of this ambiguity, we will often drop the word \enquote{stochastic} below, and simply refer to the GW signal we have described above as the \emph{gravitational-wave background} (GWB).
Our philosophy is that the words \enquote{deterministic/stochastic}, \enquote{coherent/incoherent}, etc., are best applied to GW \emph{search methods}, and that different search methods with varying degrees of phase coherence may be suited to different GWB signals from different populations of sources.
Notwithstanding this distinction, the remainder of this section (and indeed, of this thesis) is primarily concerned with phase-incoherent GWB searches, as these are simpler conceptually, and have thus far been applied much more successfully in practice due to computational challenges associated with semi-coherent Bayesian searches.

\subsection{Statistical properties of the gravitational-wave background}
\label{sec:gwb-statistics}

Assuming we are unable to deterministically model the phase evolution of the GWB, the best we can hope to do is specify the \emph{statistical} properties of the strain.
Performing the plane wave expansion as in equation~\eqref{eq:plane-wave-expansion}, we therefore treat the components $\tilde{h}_A(f,\vu*r)$ as random variables.
In this section, we \enquote{derive} a list of standard assumptions that are made about the probability distribution governing these random strain components.

The fact that the GWB is, by definition, composed of a large number of statistically independent sources allows us to invoke the central limit theorem and treat $\tilde{h}_A(f,\vu*r)$ as Gaussian.
This reduces the problem of specifying an entire probability density function down to just specifying the first two moments,
    \begin{equation}
        \ev{\tilde{h}_A(f,\vu*r)},\qquad\ev{\tilde{h}_A(f,\vu*r)\tilde{h}^*_{A'}(f',\vu*r')},
    \end{equation}
    where $\ev{\cdots}$ here denotes an expectation value.
This follows from Isserlis' theorem, which states that all higher-order moments are trivial for a set of Gaussian random variables.

Since $\tilde{h}_A(f,\vu*r)$ is a complex number, we can write it in terms of an amplitude and a complex phase (both of which are functions of frequency, propagation direction, and polarisation mode),
    \begin{equation}
        \tilde{h}_A(f,\vu*r)=\mathcal{A}_A(f,\vu*r)\rme^{\rmi\varphi_A(f,\vu*r)}.
    \end{equation}
It is not difficult to convince oneself that the phase $\varphi_A(f,\vu*r)$ should be uniformly distributed on $[0,2\uppi)$, and should be statistically independent of the amplitude $\mathcal{A}_A(f,\vu*r)$.
Indeed, this follows immediately from the assumption that the signal is statistically invariant under time translations $t\to t-\tau$ (i.e., the signal is \emph{stationary}), since the definition of the Fourier transform implies that such a translation gives $\varphi_A(f,\vu*r)\to\varphi_A(f,\vu*r)-2\uppi f\tau$ while leaving the amplitude constant.
We see that the probability distribution of $\varphi_A(f,\vu*r)$ must be identical to that of $\varphi_A(f,\vu*r)+c$ (modulo $2\uppi$) for an arbitrary constant shift $c$, which can only be satisfied if the distribution is uniform.
We also see that the amplitude is invariant even though the phase is shifted, meaning that the two must be statistically independent of each other.
As a result, we see that the first moment of the GWB strain must vanish,
    \begin{equation}
        \ev{\tilde{h}_A(f,\vu*r)}=\ev{\mathcal{A}_A(f,\vu*r)\rme^{\rmi\varphi_A(f,\vu*r)}}=\ev{\mathcal{A}_A(f,\vu*r)}\ev{\rme^{\rmi\varphi_A(f,\vu*r)}}=0,
    \end{equation}
    since $\ev*{\rme^{\rmi\varphi}}$ vanishes if $\varphi$ is uniform.

In order to evaluate the second moment, we need to know about the two-point statistics of $\mathcal{A}_A(f,\vu*r)$ and $\varphi_A(f,\vu*r)$.
Here we can use the distinguishing feature of the stochastic background, which is that the arrival times of GWs from each individual source are statistically independent due to each source being causally disconnected from the others.
(Recall that this is the fundamental reason why we are unable to use matched filtering for stochastic signals.)
This implies that the phases of any two plane-wave components are statistically independent from each other, with the only exception being when they have the same frequency, propagation direction, and polarisation state (since they are then just two copies of the same random variable),\footnote{%
    One might object that cosmological sources of GWs \emph{can} be causally connected if the Universe underwent a period of inflation at early times which shrank the comoving size of the cosmological horizon; this is the leading explanation for why different Hubble patches in the CMB have such similar temperatures.
    (Indeed, this \enquote{horizon problem} was one of the original theoretical motivations for inflation~\cite{Weinberg:2008zzc}.)
    However, \citet{Margalit:2020sxp} have shown that even if phase-coherent correlations do exist in the GW signal at early times, this coherence is lost as the GWs propagate across the Universe, as each GW experiences a large phase shift $\updelta\varphi\gg2\uppi$ due to inhomogeneities in the gravitational potential along its line of sight.
    This argument holds for GW frequencies $f\gtrsim10^{-12}\,\mathrm{Hz}$, which covers the entirety of the frequency range we are interested in here, meaning that we can safely assume zero phase correlation between different lines of sight.}
    \begin{equation}
        \Cov\!\qty[\rme^{\rmi\varphi_A(f,\vu*r)},\rme^{\rmi\varphi_{A'}(f',\vu*r')}]=\ev{\rme^{\rmi[\varphi_A(f,\vu*r)-\varphi_{A'}(f',\vu*r')]}}\propto\delta_{AA'}\delta(f-f')\delta^{(2)}(\vu*r,\vu*r').
    \end{equation}
As a result, we have
    \begin{align}
    \begin{split}
    \label{eq:second-moment-propto}
        \ev{\tilde{h}_A(f,\vu*r)\tilde{h}^*_{A'}(f',\vu*r')}&=\ev{\mathcal{A}_A(f,\vu*r)\mathcal{A}_{A'}(f',\vu*r')}\ev{\rme^{\rmi[\varphi_A(f,\vu*r)-\varphi_{A'}(f',\vu*r')]}}\\
        &\propto\ev{\qty[\mathcal{A}_A(f,\vu*r)]^2}\delta_{AA'}\delta(f-f')\delta^{(2)}(\vu*r,\vu*r').
    \end{split}
    \end{align}
(Note that we have made no assumptions about whether the amplitudes of different sources are statistically independent or not.)

The two Dirac delta functions in equation~\eqref{eq:second-moment-propto} are somewhat surprising at first, as they seem to imply that the variance of each individual plane wave component, $\ev{|\tilde{h}_A(f,\vu*r)|^2}$, is infinite.
In fact, this is only true in the idealised situation where we have infinitely fine frequency resolution $\updelta f$ and angular resolution $\updelta \vu*r$.
In practice, we are only able to resolve a finite set of discrete frequency bins $f_i$ and sky pixels $\vu*r_i$, so that the Dirac deltas are replaced by Kronecker deltas,
    \begin{equation}
        \delta(f_i-f_j)\to\frac{\delta_{ij}}{\updelta f},\qquad\delta^{(2)}(\vu*r_i,\vu*r_j)\to\frac{\delta_{ij}}{\updelta\vu*r}.
    \end{equation}
However, since we are usually only interested in integrals over the second moment~\eqref{eq:second-moment-propto} (e.g., in equation~\eqref{eq:omega_gw-derivation} below) we can safely keep the Dirac deltas in equation~\eqref{eq:second-moment-propto}, with the understanding that they represent an idealised continuum limit.

We have thus determined the second moment up to some unknown deterministic function of frequency, sky direction, and polarisation,
    \begin{equation}
        \ev{\tilde{h}_A(f,\vu*r)\tilde{h}^*_{A'}(f',\vu*r')}=S_A(f,\vu*r)\delta_{AA'}\delta(f-f')\delta^{(2)}(\vu*r,\vu*r').
    \end{equation}
At this point it is useful to recall from section~\ref{sec:polarisation-modes} that the amplitudes of the two polarisation modes are not fixed for a given plane wave, but undergo mixing when ones rotates around the line of sight, as in equation~\eqref{eq:polarisation-rotation}.
If we assume the polarisation angle of each source, $\psi(f,\vu*r)$, is uniformly distributed, then by a very similar argument to that above we conclude that this unknown function must be equal for each polarisation mode,
    \begin{equation}
    \label{eq:unpolarised}
        S(f,\vu*r)\equiv S_+(f,\vu*r)=S_\times(f,\vu*r),
    \end{equation}
    since we can arbitrarily rotate $+$-modes into $\times$-modes and vice versa.
This assumption of a uniform distribution in $\psi(f,\vu*r)$ is fairly robust, particularly for astrophysical sources, since there is generally no reason to expect sources to be preferentially aligned along any particular direction around the line of sight.
Note, however, that there are several models of the early Universe that give rise to a GWB with nontrivial polarisation content, for example through couplings to parity-violating matter fields~\cite{Alexander:2004us,Barnaby:2011qe,Adshead:2013qp,Adshead:2013nka,Dimastrogiovanni:2016fuu,Machado:2018nqk}, the addition of parity-violating terms to the gravitational action~\cite{Satoh:2007gn,Contaldi:2008yz,Takahashi:2009wc,Alexander:2009tp,Wang:2012fi,Bartolo:2017szm}, or the generation of helical magnetohydrodynamic (MHD) turbulence in the primordial plasma~\cite{Kahniashvili:2005qi,Caprini:2014mja,Kisslinger:2015hua,Ellis:2019tjf}.
Specialised search methods have been developed for polarised backgrounds such as these~\cite{Seto:2006hf,Seto:2006dz,Seto:2007tn,Seto:2008sr,Smith:2016jqs,Domcke:2019zls,Ellis:2020uid,Martinovic:2021hzy}; however, we will not discuss these further here, since the unpolarised assumption~\eqref{eq:unpolarised} is expected to be valid for all of the GW sources we investigate in this thesis.

\begin{figure}[t!]
    \begin{center}
        \includegraphics[width=0.49\textwidth]{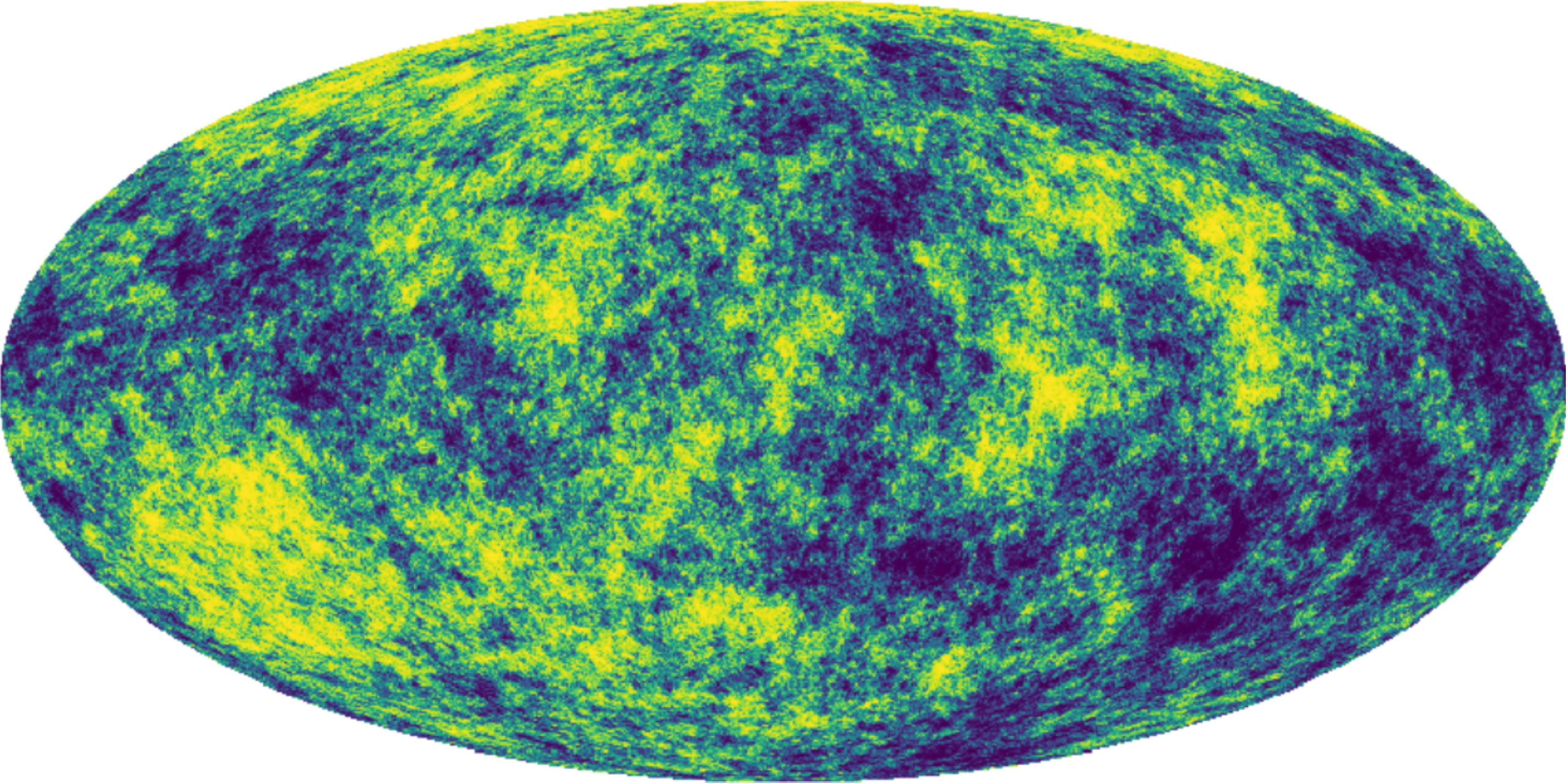}
        \includegraphics[width=0.49\textwidth]{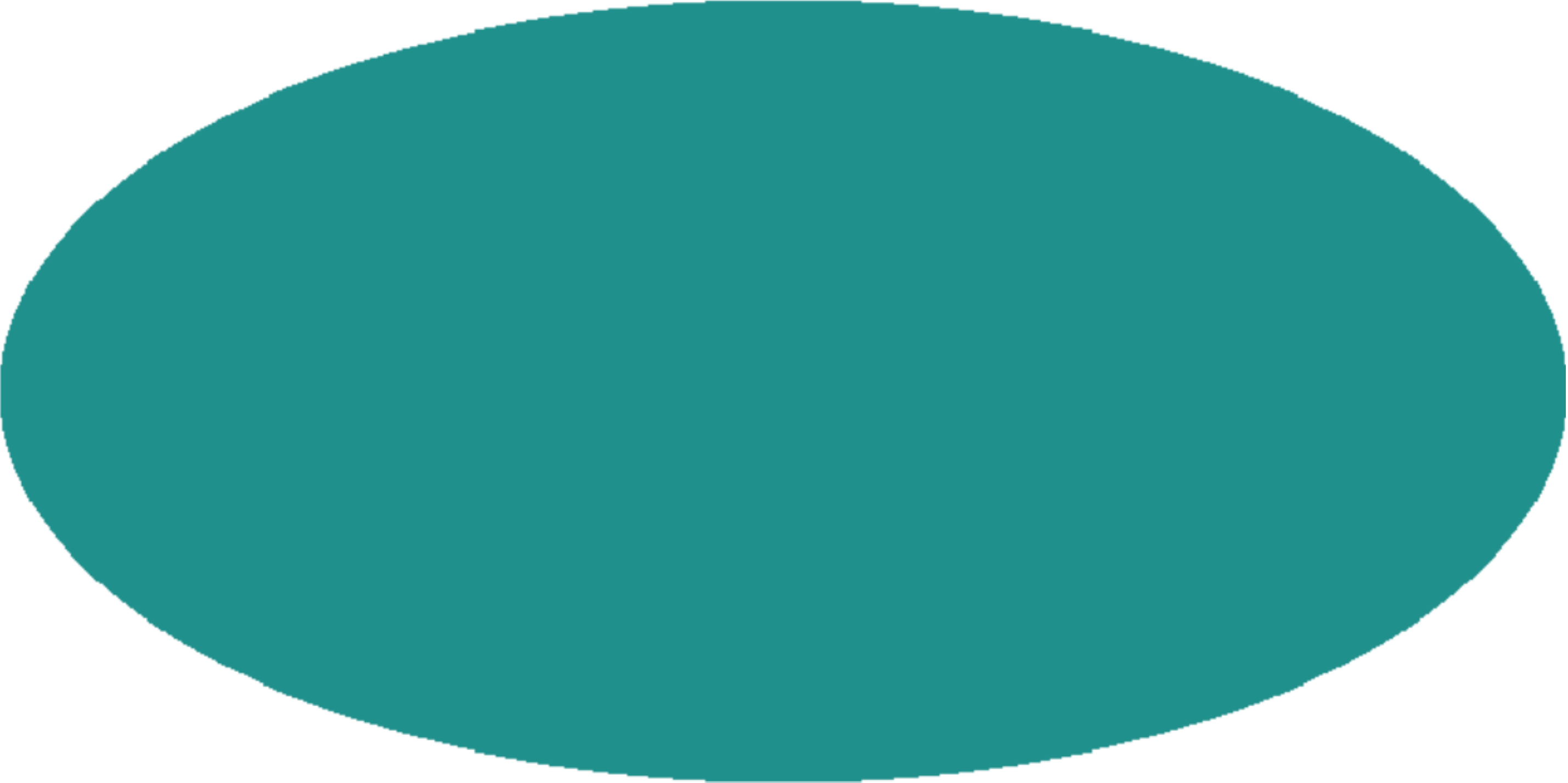}
    \end{center}
    \caption{%
    Left panel: a statistically isotropic field on the sphere.
    Right panel: an \emph{exactly} isotropic field on the sphere.
    Most stochastic background searches assume the latter.}
    \label{fig:statistical-isotropy}
\end{figure}

The final assumption that is usually made in describing the statistics of the GWB is that its intensity is \emph{isotropic}, i.e., the same in all directions on the sky.
This implies that the function $S(f,\vu*r)$ characterising the second moment of the GWB strain is independent of sky direction, and therefore depends only on frequency.
The motivation for this assumption comes from the statistical isotropy of many other cosmological observables; the CMB, for instance, varies in intensity by only a few parts in $10^5$ across the sky.
It is important to point out, however, that \emph{statistical} isotropy (i.e., invariance of expectation values of the observed field under sky rotations) is not the same as \emph{exact} isotropy (i.e., invariance of \emph{the field itself} under rotations), which is being assumed here; see figure~\ref{fig:statistical-isotropy}.
Indeed, modern cosmology has forcefully demonstrated that, even if departures from exact isotropy and homogeneity are small, they are incredibly important tools for understanding the structure and evolution of the Universe.
In this sense, isotropy is the least well-justified of all the assumptions we have made thus far.
In chapter~\ref{chap:anisotropies} of this thesis we revisit this assumption, and investigate what can be learned by considering \emph{anisotropies} in the GWB.

\subsection{The gravitational-wave density parameter}
\label{sec:gw-density-parameter}

To summarise, we have shown that by assuming the GWB to be Gaussian, stationary, unpolarised, and isotropic, with no nontrivial phase correlations, we can write
    \begin{equation}
    \label{eq:gwb-moments}
        \ev{\tilde{h}_A(f,\vu*r)}=0,\qquad\ev{\tilde{h}_A(f,\vu*r)\tilde{h}^*_{A'}(f',\vu*r')}=S(f)\delta_{AA'}\delta(f-f')\delta^{(2)}(\vu*r,\vu*r'),
    \end{equation}
    such that the strain statistics are fully characterised by a single function of frequency, $S(f)$.
The fact that $h_{ij}(t,\vb*x)$ is real implies that $S(f)$ is real and even-valued, $S(f)=S^*(f)=S(-f)$.

This function is not arbitrary; it is directly related to the energy spectrum of the GWB.
To see this, we insert the plane-wave expansion~\eqref{eq:plane-wave-expansion} and the GWB moments~\eqref{eq:gwb-moments} into the expression for the quasi-local GW energy density we obtained previously, equation~\eqref{eq:gw-energy-density}.
Placing our detector at the origin $\vb*x=\vb*0$ (which we assume to be effectively in vacuum, such that we can work in the TT gauge), this gives
    \begin{align}
    \begin{split}
    \label{eq:omega_gw-derivation}
        \rho_\gw&=\frac{1}{32\uppi G}\ev{\dot{h}^{ij}\dot{h}_{ij}}\\
        &=\frac{\uppi}{8G}\int_\mathbb{R}\dd{f}\int_\mathbb{R}\dd{f'}\int_{S^2}\dd[2]{\vu*r}\int_{S^2}\dd[2]{\vu*r'}\rme^{2\uppi\rmi t(f-f')}e^{A,ij}(\vu*r)e^{A'}_{ij}(\vu*r')ff'\ev{\tilde{h}_A(f,\vu*r)\tilde{h}^*_{A'}(f',\vu*r')}\\
        &=\frac{\uppi}{8G}\int_\mathbb{R}\dd{f}\int_{S^2}\dd[2]{\vu*r}e^{A,ij}(\vu*r)e^A_{ij}(\vu*r)f^2S(f)=\frac{4\uppi^2}{G}\int_\mathbb{R}\dd{(\ln f)}f^3S(f),
    \end{split}
    \end{align}
    where in the last line we have used equation~\eqref{eq:polarisation-tensor-normalisation}, and have restricted ourselves to positive frequencies (without loss of generality), so that $\ln f$ is well-defined.\footnote{\label{ft:ergodicity}%
    Note that we have indulged in a slight abuse of notation here: we have treated the angle brackets $\ev{\cdots}$ in equation~\eqref{eq:gw-energy-density} identically to those in equation~\eqref{eq:gwb-moments}, even though the former represent a spatial or temporal average of a single GWB realisation over a region of spacetime, while the latter represent an ensemble average over many random realisations of the GWB at a single point in spacetime.
    We can interchange these two averaging processes if we assume that the GWB is \emph{ergodic} such that, given sufficient time or a large enough spatial volume, the number of occurrences of a given value of the GWB strain is proportional to the probability assigned to that value by the distribution in equation~\eqref{eq:gwb-moments}.
    Ergodic processes are ubiquitous in physics whenever we have a configuration space of finite size---as is the case here, since the GWB strain cannot become arbitrarily large.}
We therefore find
    \begin{equation}
        S(f)=\frac{G}{4\uppi^2f^3}\dv{\rho_\gw}{(\ln f)},
    \end{equation}
    so that the strain statistics of the GWB are uniquely determined by the frequency spectrum of the GW energy density.

Since the GWB is a cosmological observable, it is often convenient to normalise its energy density with respect to the present-day critical energy density of the Universe (i.e., the energy density required for the Universe to be spatially flat).
We therefore define the dimensionless \emph{GWB density parameter} (or simply \emph{GWB spectrum}),
    \begin{equation}
        \Omega_\gw(f)\equiv\frac{1}{\rho_\mathrm{c}}\dv{\rho_\gw}{(\ln f)},\qquad\rho_\mathrm{c}\equiv\frac{3H_0^2}{8\uppi G},
    \end{equation}
    with the total normalised energy density in GWs being given by $\int_\mathbb{R}\dd{(\ln f)}\Omega_\gw(f)$.
This brings the GWB in line with how cosmologists describe other sources of energy density in the Universe; see, e.g., table~\ref{tab:density-parameters}.
(When there is no chance of confusion with other density parameters we will omit the \enquote{gw} subscript, and simply write $\Omega(f)$.)
Equation~\eqref{eq:gwb-moments} thus becomes
    \begin{align}
    \begin{split}
    \label{eq:gwb-moments-omega}
        \ev{\tilde{h}_A(f,\vu*r)}=0,\qquad\ev{\tilde{h}_A(f,\vu*r)\tilde{h}^*_{A'}(f',\vu*r')}&=\frac{3H_0^2\Omega_\gw(f)}{32\uppi^3|f|^3}\delta_{AA'}\delta(f-f')\delta^{(2)}(\vu*r,\vu*r').
    \end{split}
    \end{align}

\begin{table}[t!]
    \begin{center}
        \begin{tabular}{l l}
            Source of energy density, $X$ & $\phantom{1}$Density parameter, $\Omega_X\equiv\rho_X/\rho_\mathrm{c}$\\
            \hline
            Dark energy ($\Lambda$) & $\phantom{1}0.689$\\
            Cold dark matter (CDM) & $\phantom{1}0.261$\\
            Baryonic matter & $\phantom{1}0.0490$\\
            Photons ($\gamma$) & $\phantom{1}0.0000543$\\
            Neutrinos ($\nu$) & $\phantom{1}0.0000375$\\
            \hline
            All matter (baryons + CDM) & $\phantom{1}0.311$\\
            All radiation ($\gamma$ + $\nu$) & $\phantom{1}0.0000918$\\
            \hline
            Total & $\phantom{0}1.00$
        \end{tabular}
    \end{center}
    \caption{%
        Present-day cosmological density parameters of the main components of the $\Lambda$CDM model, as determined by \emph{Planck} 2018, TTTEEE+lowE+lensing+BAO~\cite{Aghanim:2018vyg}.
        We assume the Standard Model value of the effective number of neutrino species, $N_\mathrm{eff}=3.045$~\cite{deSalas:2016ztq}.}
    \label{tab:density-parameters}
\end{table}

As we will see below, the values of $\Omega_\gw(f)$ predicted by most models of the GWB (and indeed, the values allowed by current observations) are much smaller than any of those in table~\ref{tab:density-parameters}, even those of photons and neutrinos.
This shows that GW cosmology is \enquote{doubly hard}: not only are GWs difficult to detect due to how weakly they couple to our detectors, but they are also not very abundant in the Universe, making up a minuscule fraction of the cosmic energy budget.
However, as we will see in subsequent chapters, the challenges involved in detecting the GWB are more than compensated by the immense scientific value such a detection would have.

It is worth noting briefly that the energy density we have chosen to normalise against, $\rho_\mathrm{c}$, depends on a quantity whose value has historically been somewhat poorly determined: the Hubble constant, $H_0$.
For this reason, some authors characterise the GWB in terms of $h^2\Omega_\gw$ rather than just $\Omega_\gw$, where $h$ is defined by $H_0=h\times100\,\mathrm{km}\,\mathrm{s}^{-1}\,\mathrm{Mpc}^{-1}$; this then has the advantage of being insensitive to uncertainties in $H_0$.
We choose not to do this here.
Instead, we assume the \emph{Planck} 2018 value $h=0.677$ whenever quoting values for $\Omega_\gw$.
While this is discrepant with the value $h=0.732$ inferred from local distance ladder measurements~\cite{Riess:2020fzl} (the infamous \enquote{Hubble tension}---see, e.g., \citet{Verde:2019ivm}), the resulting change in $\Omega_\gw$ is only a factor of $\approx1.17$, which is negligible in most situations.

\subsection{Modelling gravitational-wave backgrounds}
\label{sec:modelling-stochastic-backgrounds}

Given a model for a particular cosmic population of GW sources, we can calculate the corresponding density parameter $\Omega(f)$ as follows.
First, consider the GW energy density we observe today due to a single source, which is equal to the rate of GW energy passing through a spatial 2-surface surrounding us per unit surface area.
In a Euclidean Universe, we can write this as
    \begin{equation}
    \label{eq:rho_gw-individual-source}
        \rho=\frac{1}{4\uppi r^2}\dv{E}{t},
    \end{equation}
    where $r$ is the distance to the source, and $\dv*{E}{t}$ is the rate at which it emits energy in GWs (which we have implicitly averaged over inclination).
In a FLRW Universe, this expression still holds so long as we replace the Euclidean distance $r$ by the luminosity distance $d_L$, and replace the global time $t$ by the time in the source's rest frame, $t_\mathrm{s}$.
Both $d_L$ and $t_\mathrm{s}$ are functions of redshift, $z$.

For an individual signal, $\dv*{E}{t_\mathrm{s}}$ is an arbitrary function of time.
However, when we combine a large number of signals, we expect this energy emission rate to reduce to its time-averaged value (due to ergodicity, see footnote~\ref{ft:ergodicity}), which is just the product of the mean rate of signals\footnote{%
    In using this language, we appear to be restricting ourselves to transient signals.
    However, we can easily accommodate individual sources that are persistent (e.g., spinning neutron stars) by letting $E$ be the total energy radiated by a persistent source in some fixed time interval $\tau$, and setting $R$ equal to the number of such sources divided by $\tau$.}
    per unit source time, $R$, and the total energy emitted by each signal, $E$, so that $\dv*{E}{t_\mathrm{s}}\to RE$.
If we focus on just the GWs emitted in a particular logarithmic frequency bin, then \eqref{eq:rho_gw-individual-source} becomes
    \begin{equation}
        \dv{\rho}{(\ln f)}=\frac{R}{4\uppi d_L^2}\dv{E}{(\ln f_\mathrm{s})},
    \end{equation}
    where we distinguish between the observed frequency $f$ and the source-frame frequency $f_\mathrm{s}=(1+z)\times f$.
(Note that we would pick up a factor of $1+z$ here if we were using linear frequency bins, but redshifting does not change the size of a logarithmic bin.)

Now we go from the time-averaged contribution from a single source to the total contribution from all sources at all redshifts.
Since the signal rate $R$ will, in general, vary over cosmological timescales, it is more useful to think in terms of the \emph{comoving rate density} $\mathcal{R}(z)$, which is the rate of signals per unit comoving volume.
We can then encapsulate the contributions from all sources by computing a volume integral,
    \begin{equation}
    \label{eq:rho_gw-volume-integral}
        \dv{\rho}{(\ln f)}=\int\dd{\mathcal{V}}\frac{\mathcal{R}(z)}{4\uppi d_L^2(z)}\dv{E}{(\ln f_\mathrm{s})},
    \end{equation}
    where $\dd{\mathcal{V}}$ is the comoving volume element, which can also be written as $\dd{\mathcal{V}}=4\uppi r^2\dd{z}/H(z)$ with $r(z)$ the comoving distance.
Inserting this into equation~\eqref{eq:rho_gw-volume-integral}, and using the fact that $d_L/r=1+z$, we therefore find
    \begin{equation}
    \label{eq:phinney-formula}
        \Omega(f)=\frac{1}{\rho_\mathrm{c}}\int_0^\infty\frac{\dd{z}}{(1+z)^2}\frac{\mathcal{R}(z)}{H(z)}\dv{E}{(\ln f_\mathrm{s})}.
    \end{equation}

This result (sometimes called the \enquote{Phinney formula}~\cite{Phinney:2001di}) allows us to calculate the expected GWB spectrum from a cosmic population of sources.
All we need to know is the energy spectrum radiated by each source, the comoving rate density of sources, and the expansion rate of the cosmological background.
Often the first two of these three quantities will depend on some set of parameters $\vb*\zeta$ describing the sources (e.g., the masses of compact binaries), in which case the formula becomes
    \begin{equation}
    \label{eq:phinney-formula-parameterised}
        \Omega(f)=\frac{1}{\rho_\mathrm{c}}\int_0^\infty\frac{\dd{z}}{(1+z)^2}\int\dd{\vb*\zeta}\frac{\mathcal{R}(z,\vb*\zeta)}{H(z)}\dv{E(\vb*\zeta)}{(\ln f_\mathrm{s})}.
    \end{equation}

\subsection{Stochastic search methods}
\label{sec:stochastic-searches}

As mentioned above, persistent and incoherent signals such as the GWB are arguably the most challenging to search for, since we cannot use matched filtering, and there is no distinct arrival time to distinguish the signal from instrumental noise.
Faced with these limitations, how does one go about searching for the GWB?

\subsubsection{Cross-correlation searches}
The most powerful approach that has been developed thus far is to conduct \emph{cross-correlation} searches between two or more detectors~\cite{Christensen:1992wi,Flanagan:1993ix,Allen:1997ad,Romano:2016dpx}.
The key insight here is that even if the GWB is indistinguishable from noise in any individual detector, two different detectors will always \enquote{see} the same GWB signal, while ideally having statistically independent sources of noise.
We can therefore use the output of one detector as a (noisy) template for the signal we expect to see in the other detector.

More quantitatively, let us write the data stream $d(t)$ in each detector as a linear sum of the GWB strain $h(t)$ and some random noise process $n(t)$.
The expectation value of the cross-correlation statistic $d_1d_2$ can then be written as a sum of four terms,
    \begin{equation}
        \ev{d_1d_2}=\ev{(h_1+n_1)(h_2+n_2)}=\ev{h_1h_2}+\ev{h_1n_2}+\ev{n_1h_2}+\ev{n_1n_2},
    \end{equation}
    with the subscripts labelling the two detectors.
We expect the second and third of these terms to vanish, since there is no reason for the detector noise to have any coherence with the GW signal.
Similarly, the fourth term should vanish if each detector is subject to statistically-independent sources of noise.\footnote{%
    Note that this is not always the case.
    In particular, any pair of ground-based interferometers, no matter how widely separated, are expected to possess some level of correlated noise due to \emph{Schumann resonances}~\cite{Christensen:1992wi,Thrane:2013npa,Meyers:2020qrb}: coherent electromagnetic fields in the cavity between the Earth's surface and the ionosphere that are continuously excited by lightning strikes.}
This leaves just the first term, which, heuristically, is proportional to the GWB density parameter, $\ev{d_1d_2}=\ev{h_1h_2}\propto\Omega$.

Of course, we don't have observational access to ensemble-averaged quantities, only to individual random realisations.
However, since both the signal and noise are ergodic (see footnote~\ref{ft:ergodicity}), we can approximate the ensemble average by summing over many successive measurements of the cross-correlation statistic made at different times.
As the number of measurements, $N_t$, goes to infinity, we recover exactly the ensemble average,
    \begin{equation}
        \lim_{N_t\to\infty}\frac{1}{N_t}\sum_{i=1}^{N_t}d_1(t_i)d_2(t_i)=\ev{d_1d_2}=\ev{h_1h_2}.
    \end{equation}
Since $N_t$ is finite in practice, the cross-correlation statistic will have some random scatter around its expectation value; this is the fundamental limitation on our ability to measure the GWB.
Assuming that the signal and the noise are both Gaussian, and that measurements at different times are independent and identically distributed (i.i.d.), this scatter is given by
    \begin{align}
    \begin{split}
    \label{eq:cross-correlation-variance}
        \Var&\qty[\frac{1}{N_t}\sum_{i=1}^{N_t}d_1(t_i)d_2(t_i)]=-\ev{d_1d_2}^2+\frac{1}{N_t^2}\sum_{i=1}^{N_t}\sum_{j=1}^{N_t}\ev{d_1(t_i)d_2(t_i)d_1(t_j)d_2(t_j)}\\
        &=-\ev{h_1h_2}^2+\frac{1}{N_t^2}\sum_{i=1}^{N_t}\sum_{j=1}^{N_t}\bigg[\ev{d_1(t_i)d_2(t_i)}\ev{d_1(t_j)d_2(t_j)}+\ev{d_1(t_i)d_1(t_j)}\ev{d_2(t_i)d_2(t_j)}\\
        &\qquad\qquad\qquad\qquad\qquad\qquad\qquad\qquad\qquad+\ev{d_1(t_i)d_2(t_j)}\ev{d_1(t_j)d_2(t_i)}\bigg]\\
        &=-\ev{h_1h_2}^2+\frac{1}{N_t^2}\sum_{i=1}^{N_t}\sum_{j=1}^{N_t}\qty[\ev{h_1h_2}^2+\delta_{ij}\qty(\ev{n_1^2}+\ev{h_1^2})\qty(\ev{n_2^2}+\ev{h_2^2})+\delta_{ij}\ev{h_1h_2}^2]\\
        &=\frac{1}{N_t}\qty[\qty(\ev{n_1^2}+\ev{h_1^2})\qty(\ev{n_2^2}+\ev{h_2^2})+\ev{h_1h_2}^2]\simeq\frac{1}{N_t}\ev{n_1^2}\ev{n_2^2},
    \end{split}
    \end{align}
    where we have used Isserlis' theorem to rewrite the fourth moment in terms of products of second moments.
In the final line we have assumed we are in the \enquote{weak-signal} regime such that $|h|\ll|n|$, as is the case for essentially all GW experiments.
We can then calculate a signal-to-noise ratio (SNR) for the cross-correlation search by dividing the GWB term by the square-root of equation~\eqref{eq:cross-correlation-variance},
    \begin{equation}
    \label{eq:cross-correlation-snr}
        \mathrm{SNR}\simeq\sqrt{N_t}\frac{\ev{h_1h_2}}{\sqrt{\ev{n_1^2}\ev{n_2^2}}}.
    \end{equation}

There are two comments to make about equation~\eqref{eq:cross-correlation-snr}.
First, we see that the cross-correlated GWB power is divided by the geometric mean of the auto-correlated noise power in each detector.
In the weak-signal regime, this factor is very small.
However, since $N_t$ is proportional to the total observing time $T_\obs$, we see that $\mathrm{SNR}\propto\sqrt{T_\obs}$.
This is the key reason that cross-correlation searches are so useful: in principle, \emph{any} pair of detectors can eventually measure the GWB with arbitrarily large SNR, so long as we observe for a long enough time.\footnote{%
    In practice, if the noise power is too large compared to the GWB, then the requisite time might be far too long for such a measurement to be feasible.
    For this reason, any time spent upgrading the sensitivity of the detector is often more than worth the consequent loss of observing time.}

\subsubsection{The overlap reduction function}
There are many complications that we have swept under the carpet here (see \citet{Romano:2016dpx} for an authoritative treatment), but one which is important to highlight is the precise relationship between the GW strain cross-correlation $\ev{h_1h_2}$ and the GW density parameter $\Omega(f)$.
If the two detectors are co-located, then these quantities really are proportional to each other,\footnote{%
    This quantity isn't really what is measured in cross-correlation searches (which usually work with frequency-domain data $\tilde{d}(f)$ instead), and even if it was, we're missing a factor in the integrand to account for the loss of sensitivity of the instrument outside of the small-antenna regime, where equation~\eqref{eq:pattern-functions} breaks down.
    However, these details are unimportant for our purpose here, which is to illustrate the origin of the overlap reduction function.}
    \begin{equation}
    \label{eq:gwb-time-domain-amplitude}
        \ev{h_1(t)h_2(t)}=\frac{3H_0^2}{4\uppi^2}\int_0^\infty\frac{\dd{f}}{f^3}\Gamma_{12}\Omega(f).
    \end{equation}
Here $\Gamma_{12}$ is a proportionality constant which encodes the coupling of each detector's strain readout $h(t)$ to the actual GW strain $h_{ij}(t)$.
Defining the \emph{detector response functions} $\mathcal{D}^A(\vu*r)$ through the response of a detector to each polarisation mode of a GW plane wave propagating in the $\vu*r$-direction (and assuming we are in the small-antenna limit),
    \begin{equation}
    \label{eq:pattern-functions}
        h(t)\equiv\mathcal{D}^A(\vu*r)h_A(t,\vu*r),
    \end{equation}
    we find that the proportionality constant is
    \begin{equation}
        \Gamma_{12}=\frac{1}{8\uppi}\int_{S^2}\dd[2]{\vu*r}\mathcal{D}^A_1(\vu*r)\mathcal{D}^A_2(\vu*r).
    \end{equation}
For the simplest case of two identical, equal-arm, co-located and co-aligned interferometers, this is given by $\Gamma_{12}=\sin^2\beta/5$, where $\beta$ is the opening angle between the arms of each interferometer.

\begin{figure}[t!]
    \begin{center}
        \includegraphics[width=0.8\textwidth]{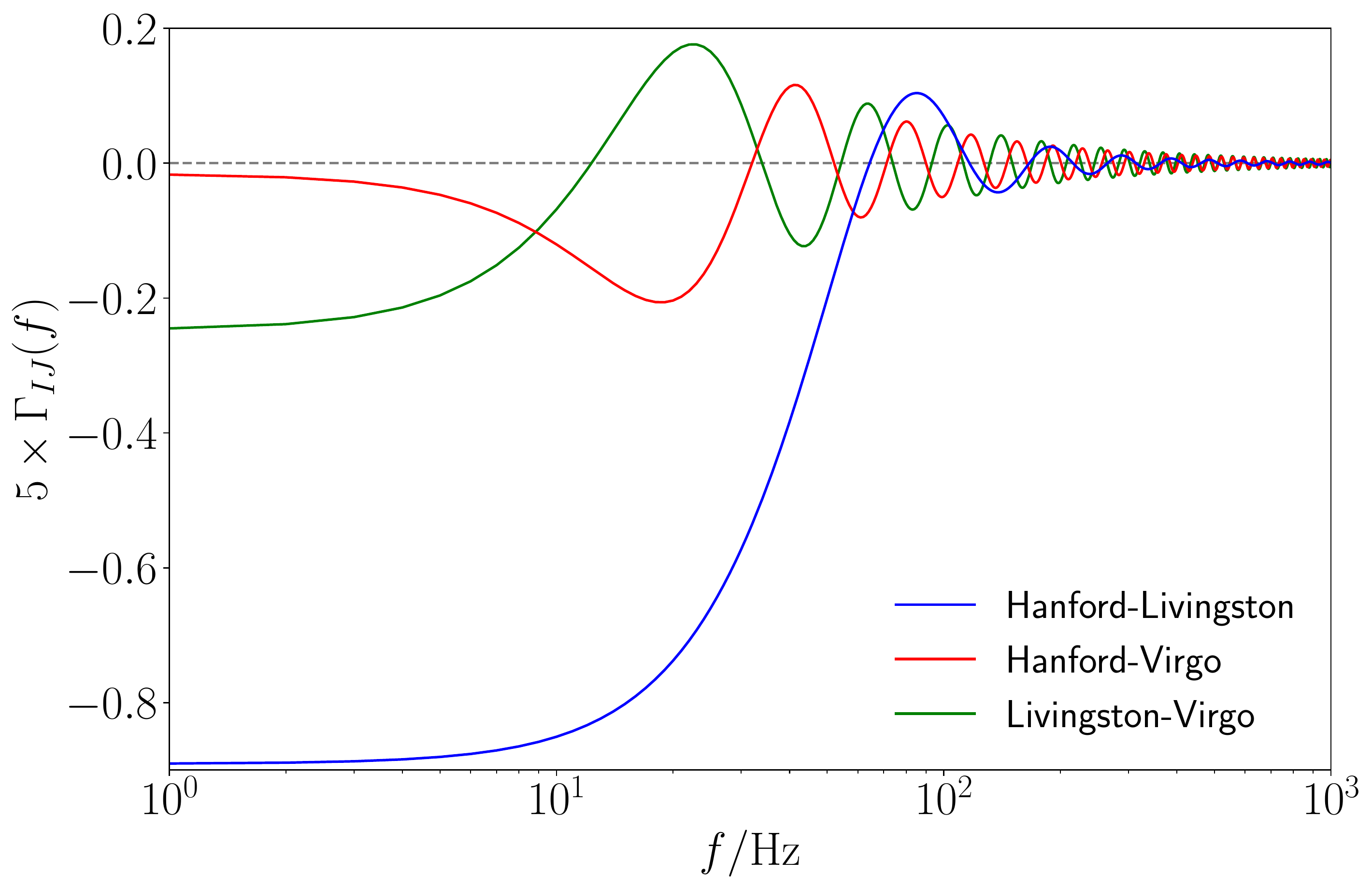}
    \end{center}
    \caption{%
    Overlap reduction functions for the three detectors that make up the LIGO/Virgo network.
    The factor of $5$ ensures that $5\times\Gamma_{IJ}(f)=1$ for two co-located, co-aligned detectors in the small-antenna limit.
    At low frequencies the Hanford-Livingston ORF is approximately $5\times\Gamma_\mathrm{HL}\approx-0.89$; this is close to $-1$, due to the two LIGO interferometers being nearly co-aligned up to a $90^\circ$ rotation (there is a slight misalignment due to the curvature of the Earth between the two detector sites, which causes $|5\times\Gamma_\mathrm{HL}|$ to drop below unity).
    Each pair of detectors has a series of \enquote{blind spots} set by their separation, with longer baselines corresponding to lower-frequency blind spots.}
    \label{fig:orf}
\end{figure}

The issue with cross-correlating between two co-located detectors is that they both experience the same sources of environmental noise, making it extremely difficult to ensure that $\ev{n_1n_2}$ vanishes.\footnote{%
    Nonetheless, GWB searches have been conducted with co-located detectors in the past; in particular, \citet{Aasi:2014sej} carried out such a search using two interferometers at the LIGO Hanford site, H1 and H2.
    (H2 has since been decommissioned.)}
We are therefore usually interested in cross-correlating between widely-separated pairs of interferometers such as LIGO Hanford and LIGO Livingston, which are situated about $3000\,\mathrm{km}$ apart.
In this case, however, $\Gamma_{12}$ becomes a frequency-dependent transfer function,
    \begin{equation}
    \label{eq:orf}
        \Gamma_{12}(f)=\frac{1}{8\uppi}\int_{S^2}\dd[2]{\vu*r}\mathcal{D}^A_1(\vu*r)\mathcal{D}^A_2(\vu*r)\cos[2\uppi f\vu*r\vdot(\vb*x_1-\vb*x_2)],
    \end{equation}
    called the \emph{overlap reduction function} (ORF).
The cosine in the integrand here means that this function is generally smaller than the constant $\Gamma_{12}$ we found for the co-located case.
This is because the two detectors, being at different locations, are generally out of phase in their response to a given plane wave, and this phase difference causes a loss of coherence between their strain readouts.
In fact, $\Gamma_{12}(f)$ usually oscillates between positive and negative values as a function of $f$, meaning that there are a series of frequencies $f_*$ set by the inverse separation $|\vb*x_1-\vb*x_2|^{-1}$ where there is, on average, \emph{no} coherence between the two detectors, $\Gamma_{12}(f_*)=0$ (see figure~\ref{fig:orf}).
Since we can only measure the combination $\Gamma_{12}(f)\Omega(f)$, these frequencies represent \enquote{blind spots} in the search, at which it is impossible to measure $\Omega(f)$.
(Note however that this can be mitigated by combining multiple pairs of detectors, each with different pairwise separations and therefore different blind spots.)

\subsubsection{Excess-power searches}
What if we only have access to a single detector?
In principle, we can still carry out a GWB search if we have some prior information about the amplitude or spectral shape of the detector noise.
If the measured amplitude is greater than expected, then we can try to attribute the excess power to a GWB signal.
This is the exact method that \citet{Penzias:1965wn} used in their groundbreaking first detection of the CMB, which relied on ruling out all other possible sources of noise in their radio antenna.
This method is very difficult to apply in GWB searches, however, since a detection would require us to determine the noise spectrum of our detector with an uncertainty much smaller than $\ev*{h^2}$.
Since $|h|\ll|n|$, this is practically impossible in most cases.
One important exception is the future space-based interferometer LISA, for which it will be possible to construct a \emph{null channel} which is (approximately) insensitive to GWs, giving a direct measurement of the noise power~\cite{Adams:2010vc}.

\subsubsection{Sensitivity curves}
Whether one is conducting a cross-correlation search in a network of detectors, or an excess-power search in a single detector, the ultimate goal is to reconstruct the GWB spectrum $\Omega(f)$ within some frequency band set by the sensitivity of the experiment.
In both cases, however, this is often a very difficult task; for cross-correlation searches, we have seen that the ORF generally causes \enquote{blind spots} which make it difficult to reconstruct certain frequencies, while for excess-power searches a lack of knowledge about the spectral shape of the GWB makes it even harder to distinguish signal power from noise power.
Fortunately, in many situations the GWB spectrum is sufficiently slowly-varying across the frequency range of the search that we can model it with a simple power law,
    \begin{equation}
    \label{eq:power-law-gwb}
        \Omega(f)=\Omega_\alpha(f/f_\mathrm{ref})^\alpha,
    \end{equation}
    where $\alpha$ is the spectral index and $\Omega_\alpha$ is the amplitude at some reference frequency $f_\mathrm{ref}$.
The case $\alpha=0$ corresponds to a scale-invariant spectrum, while $\alpha=3$ corresponds to white noise (since $\ev*{h^2(t)}$ then receives equal contributions from each frequency bin; c.f. equation~\eqref{eq:gwb-time-domain-amplitude}).
As we will see in section~\ref{sec:gw-sources} below, the GWB from inspiralling compact binaries is described by $\alpha=\sfrac{2}{3}$, while cosmic strings have $\alpha=0$ at high frequencies, and first-order phase transitions are typically described by a \emph{broken} power law that interpolates between $\alpha=3$ and $\alpha=-4$.

Using equation~\eqref{eq:power-law-gwb} massively reduces the complexity of the search, as we are now attempting to infer just two parameters for the whole spectrum, rather than one free parameter for every frequency bin.
We can even reduce this further to just one parameter by assuming a particular value of $\alpha$ and attempting to infer the corresponding value of $\Omega_\alpha$; this approach forms the basis of the stochastic searches conducted by the LIGO/Virgo/KAGRA Collaboration, which typically focus on $\alpha=0$, $\sfrac{2}{3}$, and $3$.
Since the GWB has yet to be detected, this yields upper limits on $\Omega_\alpha$ for each of the spectral indices that are searched for.

\begin{figure}[t!]
    \begin{center}
        \includegraphics[width=0.8\textwidth]{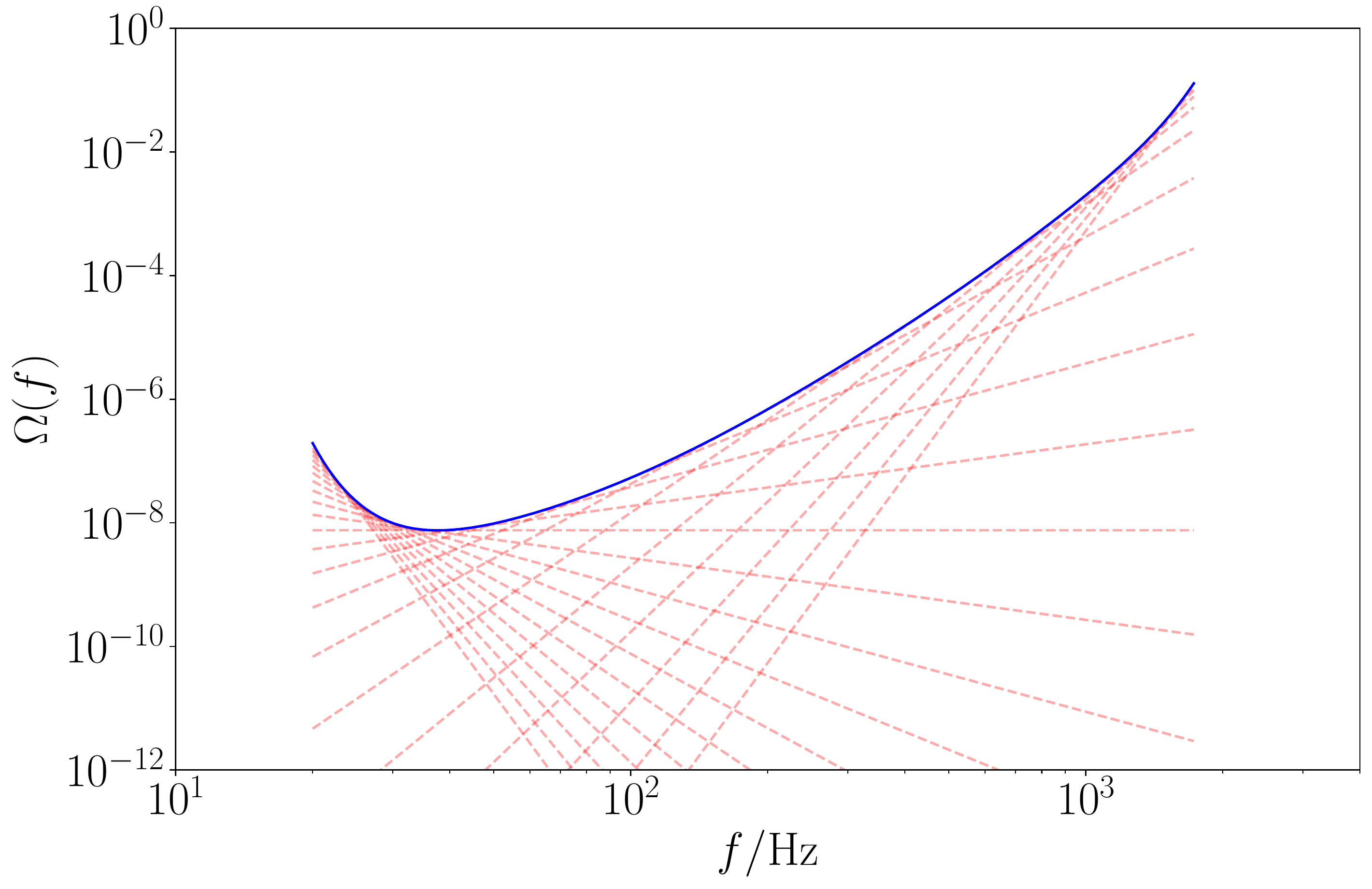}
    \end{center}
    \caption{%
    The power-law integrated sensitivity curve for LIGO/Virgo's first three observing runs~\cite{Abbott:2021pi,Abbott:2021kbb} (solid blue curve), along with the individual power-law upper limits for spectral indices $\alpha=-10,-9,\ldots,+10$ (dashed red lines).}
    \label{fig:pi-curve}
\end{figure}

Taking this power-law search method to its logical conclusion, we can construct a \emph{power-law integrated sensitivity curve} (PI curve)~\cite{Thrane:2013oya} by scanning over a large set of spectral indices $A=\{\alpha_0,\alpha_1,\dots\}$ and combining the individual power-law constraints to give
    \begin{equation}
        \Omega_\mathrm{PI}(f)\equiv\max_{\alpha\in A}\Omega_\alpha(f/f_\mathrm{ref})^\alpha.
    \end{equation}
Under the assumption that the signal can be described by a power law with index $\alpha\in A$, this PI curve has the property that any signal which intersects the curve, or lies tangent to it at any point, will be detected by the search.
In principle, we could imagine having a very sharply peaked signal (e.g., a delta function), which could intersect the PI curve without being detected.
In practice, however, there are few models that predict such sharply peaked signals, and the PI curve therefore represents an extremely useful summary of the sensitivity of a given GW experiment to the GWB.

\section{Sources of gravitational waves}
\label{sec:gw-sources}

As mentioned at the beginning of this chapter, our primary goal is to use GWs (and in particular, the GWB) as \emph{messengers}, allowing us to probe exotic sources in the early and late Universe---particularly those \enquote{dark} sources that are invisible to EM astronomy.
In this section we give a brief introduction to a few of these, focusing primarily on \emph{cosmological} sources, whose GW emission can teach us about the structure of the Universe on the largest scales and the fundamental laws that govern it.

We begin, however, with a quintessentially \emph{astrophysical} source: compact binaries.
Despite not being cosmological in origin, these sources are still extremely important for GW cosmology, as they give rise to a strong stochastic background; in fact, this signal is likely to be the dominant component of the GWB across a very broad frequency range.
As we will see in chapter~\ref{chap:anisotropies} below, compact binaries can also be used as tracers of the cosmic matter distribution, allowing us to use anisotropies in the GWB as a novel probe of the large-scale structure of the Universe.

There is one key class of cosmological GWs that we will not discuss here in any detail, which is the primordial tensor spectrum that is generated during inflation through the amplification of quantum fluctuations~\cite{Grishchuk:1974ny,Starobinsky:1979ty}.
Though these primordial GWs are of fundamental importance in modern cosmology, and are doubtless a fascinating observational target, their spectrum is so weak at the frequencies we are interested in ($f\gtrsim10^{-10}\,\mathrm{Hz}$) that they are not likely to be directly detected by GW observatories in the foreseeable future.\footnote{%
    At least, this is the case in the standard inflationary scenario of a single slowly-rolling inflaton field.
    For a summary of some alternative scenarios in which the inflationary GWB is enhanced at high frequencies, see \citet{Bartolo:2016ami}, as well as \citet{Iacconi:2019vgc,Iacconi:2020yxn} for more recent work on this.}
Instead, they are much more likely to be probed indirectly by future CMB missions, through the \enquote{$B$-mode} polarisation patterns that they imprint on the CMB~\cite{Kamionkowski:1996zd,Seljak:1996gy,Ade:2015tva}.

\subsection{Compact binaries}
\label{sec:compact-binaries}

We saw in section~\ref{sec:gw-generation} that in the weak-field, slow-motion limit, GW generation is driven by the second time derivative of the mass quadrupole of the source.
In order to maximise the strength of the GW signal, we thus require a large, high-mass system, with rapid (but non-spherically symmetric) internal velocities.
Perhaps the most efficient mechanism nature has invented for achieving this is the self-gravitating motion of \emph{binary stars}.
Binaries are ubiquitous in astronomy; roughly half of solar-type stars, and a majority of more massive stars, are in binaries or systems of higher multiplicity (triples, etc.)~\cite{Duchene:2013cba}.
However, as we show below, for the purposes of GW detection our attention is restricted to \emph{compact} binaries, whose components are not stars but dense stellar remnants such as white dwarfs (WDs), neutron stars (NSs), and black holes (BHs).

\subsubsection{Circular Newtonian binaries}
As a first approximation, we can model a binary system as two point masses interacting via Newtonian gravity, whose separation vector $\vb*R$ thus obeys
    \begin{equation}
    \label{eq:newtonian-eom}
        \ddot{\vb*R}+\frac{GM}{R^3}\vb*R=\vb*0,
    \end{equation}
    where $M\equiv m_1+m_2$ is the total mass of the system.
This is a good approximation when $GM/R\ll1$, or equivalently, when the orbital velocity of the binary is small, $v\ll1$.
Assuming for now that the binary is on a circular orbit, equation~\eqref{eq:newtonian-eom} fixes the orbital period $P$ of the binary such that
    \begin{equation}
    \label{eq:kepler-circular}
        \frac{GM}{R^3}=\qty(\frac{2\uppi}{P})^2.
    \end{equation}
(The same equation holds for more general elliptical orbits if one replaces $R$---which is no longer constant---with $a$, the semi-major axis of the ellipse; this equation is then called Kepler's third law.)
It is straightforward to substitute equations~\eqref{eq:newtonian-eom} and~\eqref{eq:kepler-circular} into the GW signal we derived for a system of two point masses~\eqref{eq:gw-signal-2-body}, giving\footnote{%
    Here the time $t$ on the right-hand side should properly be interpreted as the retarded time $t-r$ at the observer's position.
    However, if the observer is at a fixed distance from the binary (as is usually a good approximation), this distinction is unimportant, as the two are equivalent up to a time translation.}
    \begin{equation}
    \label{eq:hij-newtonian-binary}
        h_{ij}=-\frac{4G\mu}{r}\frac{GM}{R}
        \begin{pmatrix}
            \cos(4\uppi t/P) & \sin(4\uppi t/P) & 0\\
            \sin(4\uppi t/P) & -\cos(4\uppi t/P) & 0\\
            0 & 0 & 0
        \end{pmatrix}^\mathrm{TT},
    \end{equation}
    where we have, without loss of generality, chosen our coordinates such that the binary is confined to the $z=0$ plane, and such that $\vu*R=\vu*x$ at time $t=0$.

We see that the GWs emitted by a Newtonian binary are monochromatic, with a frequency which is twice that of the binary's orbital frequency,
    \begin{equation}
    \label{eq:binary-gw-frequency}
        f=\frac{2}{P}=\frac{1}{\uppi}\sqrt{\frac{GM}{R^3}}.
    \end{equation}
One way of understanding this intuitively is to notice that after half an orbital period the angular positions of the two binary components are interchanged, resulting in a configuration which is equivalent from the point of view of GW production; this halving of the period corresponds to a doubling of the frequency.

\begin{table}[t!]
    \begin{center}
    \begin{tabular}{c c c c}
        Object & $m$ & $d$ & $Gm/d$\\
        \hline
        Solar-type star & $\sim m_\odot$ & $\sim R_\odot$ & $\sim10^{-6}$\\
        White dwarf & $\sim0.6\,m_\odot$ & $\sim10^{-2}\,R_\odot$ & $\sim10^{-4}$\\
        Neutron star & $\sim1.4\,m_\odot$ & $\sim10^{-5}\,R_\odot$ & $\sim0.2$\\
        Black hole & $\dots$ & $2Gm$ & $1/2$
    \end{tabular}
    \end{center}
    \caption{%
    Typical mass, radius, and compactness of various astronomical objects.
    The compactness of a (nonspinning) BH is always $1/2$ regardless of its mass, as the Schwarzschild radius is always $2Gm$.}
    \label{tab:compactness}
\end{table}

We also see that equation~\eqref{eq:hij-newtonian-binary} contains a factor of $GM/R$ as a result of Kepler's third law.
This dimensionless factor is called the \emph{compactness} of the binary, and is a measure of how strongly-gravitating the system is.
For a binary composed of objects of mass $m$ and size $d$, the compactness of the system is bounded from above by the compactness of the objects, $Gm/d$, since $R\gtrsim d$.
We therefore see that the maximum amplitude of the signal (and thus the prospect of detection) is greatly enhanced if the binary is composed of highly compact objects such as WDs, NSs, or BHs.\footnote{%
    Of course, the weak-field, slow-motion approach we have adopted here breaks down when $GM/R\sim1$, so it is not strictly legitimate to apply equation~\eqref{eq:hij-newtonian-binary} here.
    In principle, one must account for relativistic effects to say anything about this regime, either by going to much higher order in perturbation theory (e.g. in a post-Newtonian framework), or by using numerical simulations of full general relativity.
    However, in practice the simple Newtonian description of the signal is sufficient to illustrate its most important features, once one accounts for radiation reaction as discussed below.}
In particular, we see from table~\ref{tab:compactness} that the maximum strain from binary black hole (BBH) and binary neutron star (BNS) systems is about six orders of magnitude greater than that of a binary composed of solar-type stars.
This is perhaps the simplest way of understanding why compact binaries, particularly those composed of BHs and NSs, are the primary observational target for current and planned GW observatories.

Eliminating the binary separation $R$ in favour of the GW frequency $f$ using equation~\eqref{eq:binary-gw-frequency}, and projecting onto the polarisation tensors to obtain the complex strain~\eqref{eq:complex-strain}, we can rewrite the GW signal emitted by a Newtonian binary as
    \begin{equation}
    \label{eq:newtonian-binary-complex-strain}
        h=-\frac{2G\mathcal{M}}{r}(\uppi G\mathcal{M}f)^{2/3}\qty[(1+\cos^2\theta)\cos(2\uppi ft-2\phi)+2\rmi\cos\theta\sin(2\uppi ft-2\phi)].
    \end{equation}
This is determined by a peculiar combination of $m_1$ and $m_2$ called the \emph{chirp mass},
    \begin{equation}
        \mathcal{M}\equiv\mu^{3/5}M^{2/5}=\eta^{3/5}M=\frac{(m_1m_2)^{3/5}}{(m_1+m_2)^{1/5}}.
    \end{equation}
Here $\eta\equiv\mu/M$ is a dimensionless quantity called the \emph{symmetric mass ratio}, which varies between $\eta=1/4$ for equal-mass binaries, and $\eta\to0$ for \enquote{extreme mass-ratio} binaries in which one object is much more massive than the other.
Since $\eta\le1/4$, we see that the chirp mass is always smaller than the total mass by a factor of $\le(1/4)^{3/5}\approx0.44$; for example, GW150914 had a total mass of $M\approx66\,m_\odot$, but a chirp mass of only $\mathcal{M}\approx29\,m_\odot$.
The angles $(\theta,\phi)$ in equation~\eqref{eq:newtonian-binary-complex-strain} specify the observer's line of sight with respect to the binary reference frame used in equation~\eqref{eq:hij-newtonian-binary}; from the observer's point of view $\theta$ is thus interpreted as the inclination of the binary with respect to this line of sight, and we will therefore denote it by $I$.
The factor $\uppi G\mathcal{M}f$ that appears in equation~\eqref{eq:newtonian-binary-complex-strain}---which will also crop up in many of the expressions below---is another measure of how strongly-gravitating the system is, and is related to the compactness $GM/r$ and orbital velocity $v$ of the binary by $(\uppi G\mathcal{M}f)^{2/3}=\eta^{2/5}GM/R=\eta^{2/5}v^2$.

\begin{figure}[t!]
    \begin{center}
        \includegraphics[width=0.9\textwidth]{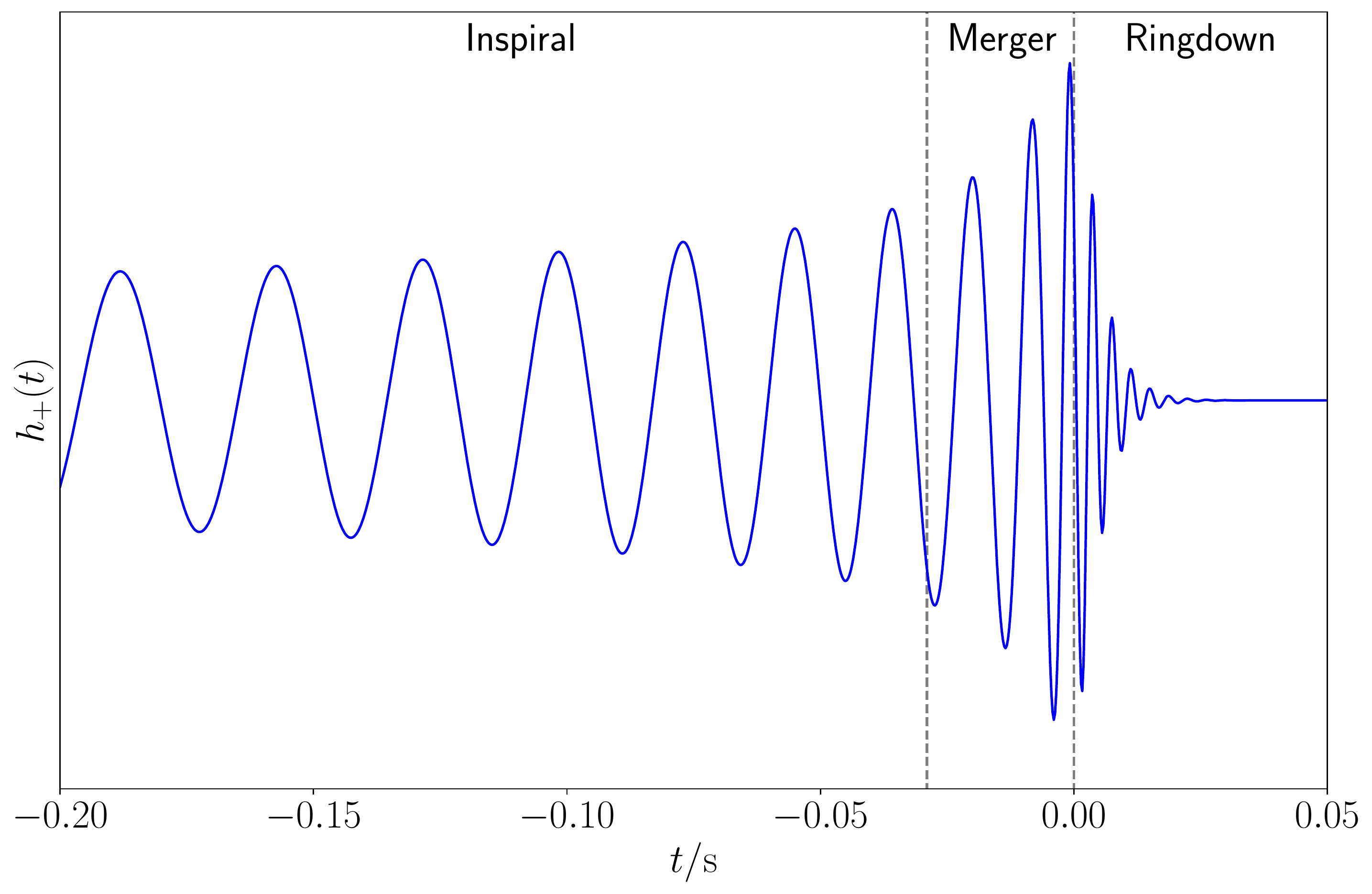}
    \end{center}
    \caption{%
    Gravitational-wave strain signal from the final stages of a compact binary coalescence.
    We show the $+$-polarisation (in arbitrary units) for an observer at inclination $I=0$.
    This waveform was generated using the \texttt{SEOBNRv4\_opt} approximant~\cite{Bohe:2016gbl} (as implemented in the Python package PyCBC~\cite{alex_nitz_2021_4849433,Usman:2015kfa}) for a GW150914-like system with $m_1=35.6\,m_\odot$ and $m_2=30.6\,m_\odot$~\cite{Abbott:2018mvr}.
    }
    \label{fig:chirp}
\end{figure}

\subsubsection{Radiation reaction}
If we believe the Newtonian result~\eqref{eq:hij-newtonian-binary}, then a binary system will continuously emit GWs at the same fixed frequency.
However, we know from section~\ref{sec:gw-energy} that these GWs carry away energy, and there must therefore be a corresponding loss of energy from the dynamics of the system.
Inserting the complex strain~\eqref{eq:newtonian-binary-complex-strain} into equation~\eqref{eq:gw-energy-flux}, we find that the binary loses energy at a rate
    \begin{equation}
    \label{eq:newtonian-binary-gw-power}
        \dv{E_\gw}{t}=\int_{S^2}\dd[2]{\vu*r}\frac{\ev{|r\dot{h}|^2}}{16\uppi G}=\frac{32}{5G}(\uppi G\mathcal{M}f)^{10/3}.
    \end{equation}
Assuming a circular orbit, the binary's orbital energy (kinetic plus potential) is
    \begin{equation}
    \label{eq:orbital-energy}
        E_\mathrm{orb}=-\frac{GM\mu}{2R}=-\frac{1}{2}\mathcal{M}(\uppi G\mathcal{M}f)^{2/3}.
    \end{equation}
Taking the time derivative of this and setting $\dot{E}_\mathrm{orb}+\dot{E}_\gw=0$, we obtain
    \begin{equation}
    \label{eq:circular-binary-fdot}
        \dot{f}=\frac{96\uppi}{5}f^2(\uppi G\mathcal{M}f)^{5/3},
    \end{equation}
    or equivalently, in terms of the orbital period $P=2/f$,
    \begin{equation}
    \label{eq:circular-period-decay}
        \dot{P}=-\frac{192\uppi}{5}\qty(\frac{2\uppi G\mathcal{M}}{P})^{5/3}.
    \end{equation}
We thus see that energy loss due to GW emission causes a circular binary to spiral inwards, orbiting at an ever-increasing rate.\footnote{%
    Strictly speaking the orbit is no longer circular, since the separation $R$ shrinks as the binary period decays.
    However, this process happens on much longer timescales than that of the orbit so long as $|\dot{P}|\ll1$, which is equivalent to our earlier requirement that $GM/R\ll1$.
    We can thus think of these orbits as \enquote{quasi-circular}, since the separation is effectively constant over the course of a single period.}
It was this effect which provided the first indirect evidence for the existence of GWs, thanks to observations of the binary pulsar B1913+16 by \citet{Hulse:1974eb}.
We call this process \enquote{radiation reaction}, in analogy with the damping force experienced by radiating charges in classical electromagnetism~\cite{Jackson:1998nia}.

Note that, while our focus here is on circular orbits, equation~\eqref{eq:circular-period-decay} is readily extended to elliptical orbits, giving~\cite{Peters:1963ux}
    \begin{equation}
    \label{eq:period-decay-radiation-reaction}
        \dot{P}=-\frac{192\uppi}{5}\qty(\frac{2\uppi G\mathcal{M}}{P})^{5/3}\frac{1+\tfrac{73}{24}e^2+\tfrac{37}{96}e^4}{(1-e^2)^{7/2}},
    \end{equation}
    where $e\in[0,1)$ is the eccentricity of the ellipse.
In this case, there is also a radiation reaction effect on the eccentricity itself due to the radiation of angular momentum~\cite{Peters:1963ux},
    \begin{equation}
    \label{eq:eccentricity-decay-radiation-reaction}
        \dot{e}=-\frac{608\uppi}{15P}\qty(\frac{2\uppi G\mathcal{M}}{P})^{5/3}\frac{e+\tfrac{121}{304}e^3}{(1-e^2)^{5/2}}.
    \end{equation}
Since the right-hand side above is always negative, we see that GW emission causes eccentric binaries to circularise over time, at a rate which increases drastically as the binary's period decays.
This justifies somewhat our focus on circular binaries here; indeed all of the GW signals detected thus far by LIGO/Virgo are consistent with zero eccentricity~\cite{Abbott:2020niy} (see however \citet{Gayathri:2020coq}).

Returning to the circular case, we see from equation~\eqref{eq:newtonian-binary-complex-strain} that as the binary's period decays the amplitude of the GW strain grows, and as a result, so does the rate at which the system loses energy~\eqref{eq:newtonian-binary-gw-power}.
The resulting \enquote{chirp} signal, whose amplitude and frequency grow at an accelerating rate over time, is highly distinctive (see figure~\ref{fig:chirp}), and reflects the inherent gravitational instability of binary systems.\footnote{%
    \label{ft:thermo}
    We can understand this instability from a thermodynamical point of view.
    Equation~\eqref{eq:orbital-energy} implies that a decrease in the energy of the system corresponds to an \emph{increase} in its \enquote{temperature} (whether this is defined in terms of the energy radiated by the system or in terms of the velocities of its components).
    The system therefore has a \emph{negative heat capacity}, meaning that it is unable to reach equilibrium, and is thus inherently unstable.
    This behaviour is not restricted to binaries, but is a generic feature of finite-size self-gravitating systems, including galaxies~\cite{Binney:2008gd} and black holes~\cite{Wald:1984rg}.}
The runaway inspiral associated with this instability only stops once the binary's components are so close to each other that they merge to become a single object; to a first approximation, we expect this to happen once the separation reaches $R=6GM$, which is the radius of the innermost stable circular orbit (ISCO) around a Schwarzschild BH of mass $M$.
Inserting this radius into equation~\eqref{eq:binary-gw-frequency}, we see that the merger frequency is roughly
    \begin{equation}
    \label{eq:isco-frequency}
        f_\mathrm{ISCO}=\frac{1}{6^{3/2}\uppi GM}\approx63\,\mathrm{Hz}\times\qty(\frac{M}{70\,m_\odot})^{-1}.
    \end{equation}
We can find the time taken for the binary starting at some initial frequency $f_0\ll f_\mathrm{ISCO}$ to reach merger by integrating equation~\eqref{eq:circular-binary-fdot},
    \begin{equation}
    \label{eq:time-to-merger}
        t_\mathrm{merge}\simeq\frac{5G\mathcal{M}}{256}(\uppi G\mathcal{M}f_0)^{-8/3}\approx4.8\,\mathrm{s}\times\qty(\frac{\mathcal{M}}{30\,m_\odot})^{-5/3}\qty(\frac{f_0}{10\,\mathrm{Hz}})^{-8/3}.
    \end{equation}

The resulting merger product is initially highly excited, and undergoes a period of further GW emission before settling down to a quiescent state.
If the final object is a BH (as is always the case for BBHs, and also occurs for sufficiently massive BNSs~\cite{Baiotti:2016qnr}), this post-merger emission is described by a superposition of damped sinusoids at a fixed set of quasi-normal mode (QNM) frequencies determined by the structure of the BH spacetime; the resulting signal is called the \enquote{ringdown}.
The overall inspiral-merger-ringdown process is called a \emph{compact binary coalescence} (CBC).

\subsubsection{Frequency spectrum and the stochastic background}
So far our discussion has been purely in the time domain, where the evolution of the binary is clearer.
However, it is often more convenient to work in the frequency domain, particularly when studying stochastic backgrounds.
In general, deriving the frequency-domain GW signal from a binary is rather involved, as we must account for both the rapid phase oscillations in the signal and the more gradual increase in the orbital frequency when taking the Fourier transform.
Fortunately, for quasicircular Newtonian binaries in the quadrupole approximation there is a simple trick we can use to obtain the frequency spectrum of the radiated GW energy~\cite{Romano:2019yrj}.
Taking advantage of the fact that the GW signal in this approximation is emitted at a single frequency which can be written as a function of time, $f(t)$, by integrating equation~\eqref{eq:circular-binary-fdot}, we write the GW energy spectrum as
    \begin{equation}
    \label{eq:cbc-energy-spectrum}
        \dv{E_\gw}{(\ln f)}=\frac{f}{\dot{f}}\dv{E_\gw}{t}=\frac{1}{3}\mathcal{M}(\uppi G\mathcal{M}f)^{2/3}=\frac{1}{3}\eta Mv^2.
    \end{equation}
This agrees exactly with what one finds from a more careful analysis, using the stationary-phase approximation to evaluate the Fourier transform of the strain~\cite{Maggiore:2007zz}.
Note that this expression is identical to the orbital energy~\eqref{eq:orbital-energy} up to an $\order{1}$ constant; this tells us that GW emission is highly efficient in extracting energy from the orbit.
Indeed, if we integrate equation~\eqref{eq:cbc-energy-spectrum} up to the ISCO frequency, we find that the total energy radiated is $E_\gw=\mu/12$, or just over $2\%$ of the total rest-mass of the system for equal-mass binaries.
(In comparison, the equivalent fraction for Hydrogen burning in stars is roughly $0.7\%$.)
In reality, the true fraction is even higher due to radiation emitted in the merger-ringdown phase.

Inserting equation~\eqref{eq:cbc-energy-spectrum} into the Phinney formula~\eqref{eq:phinney-formula}, we see that the GWB signal from a cosmic population of CBCs has a characteristic $\sim f^{2/3}$ spectral slope, the amplitude of which is determined by the CBC rate density and mass distribution.
Note however that this $\sim f^{2/3}$ power law does not continue indefinitely; since the spectrum in equation~\eqref{eq:cbc-energy-spectrum} is truncated when the two objects merge around the ISCO frequency~\eqref{eq:isco-frequency}, the GWB spectrum will gradually reach a peak and then drop off at higher frequencies, as fewer and fewer CBCs contribute.

Ignoring this for now, we can combine equations~\eqref{eq:cbc-energy-spectrum} and~\eqref{eq:phinney-formula} to find
    \begin{equation}
        \Omega(f)=\frac{8f^{2/3}}{9H_0^2}\int_0^\infty\frac{\dd{z}}{(1+z)^{4/3}}\frac{\mathcal{R}(z)}{H(z)}\int\dd{\mathcal{M}}p(\mathcal{M}|z)(\uppi G\mathcal{M})^{5/3},
    \end{equation}
    where $p(\mathcal{M}|z)$ is the chirp mass distribution as a function of redshift.
We can obtain a crude, order-of-magnitude estimate for the GWB from CBCs by setting the redshift integral equal to the value of its integrand at redshift zero.
For the LIGO/Virgo frequency band, this gives
    \begin{equation}
        \Omega(f)\sim10^{-9}\times\qty(\frac{f}{100\,\mathrm{Hz}})^{2/3}\qty(\frac{\ev{\mathcal{M}}}{30\,m_\odot})^{5/3}\frac{\mathcal{R}_0}{10\,\mathrm{Gpc}^{-3}\,\mathrm{yr}^{-1}},
    \end{equation}
    where $\ev{\mathcal{M}}$ is the mean chirp mass (which we have normalised to the chirp mass of GW150914) and $\mathcal{R}_0$ is the BBH merger rate at $z=0$ (which we have normalised to the value reported in \citet{Abbott:2018mvr}).
This simple estimate is actually remarkably close to the answer we find from a much more detailed calculation in chapter~\ref{chap:anisotropies}.

\subsection{First-order phase transitions}
\label{sec:phase-transitions}

We now turn to look at GWs of cosmological origin.
Although we have yet to detect GWs from any source other than CBCs, the rise of GW astronomy has prompted many cosmologists and particle theorists to investigate mechanisms for producing GWs in the early Universe.
Few such possibilities have excited the interest of the community as much as \emph{cosmological first-order phase transitions}, which present an extremely promising way to search for new physics beyond the Standard Model~\cite{Kamionkowski:1993fg,Caprini:2015zlo,Caprini:2018mtu,Caprini:2019egz}.

\begin{figure}[t!]
    \begin{center}
        \includegraphics[width=0.9\textwidth]{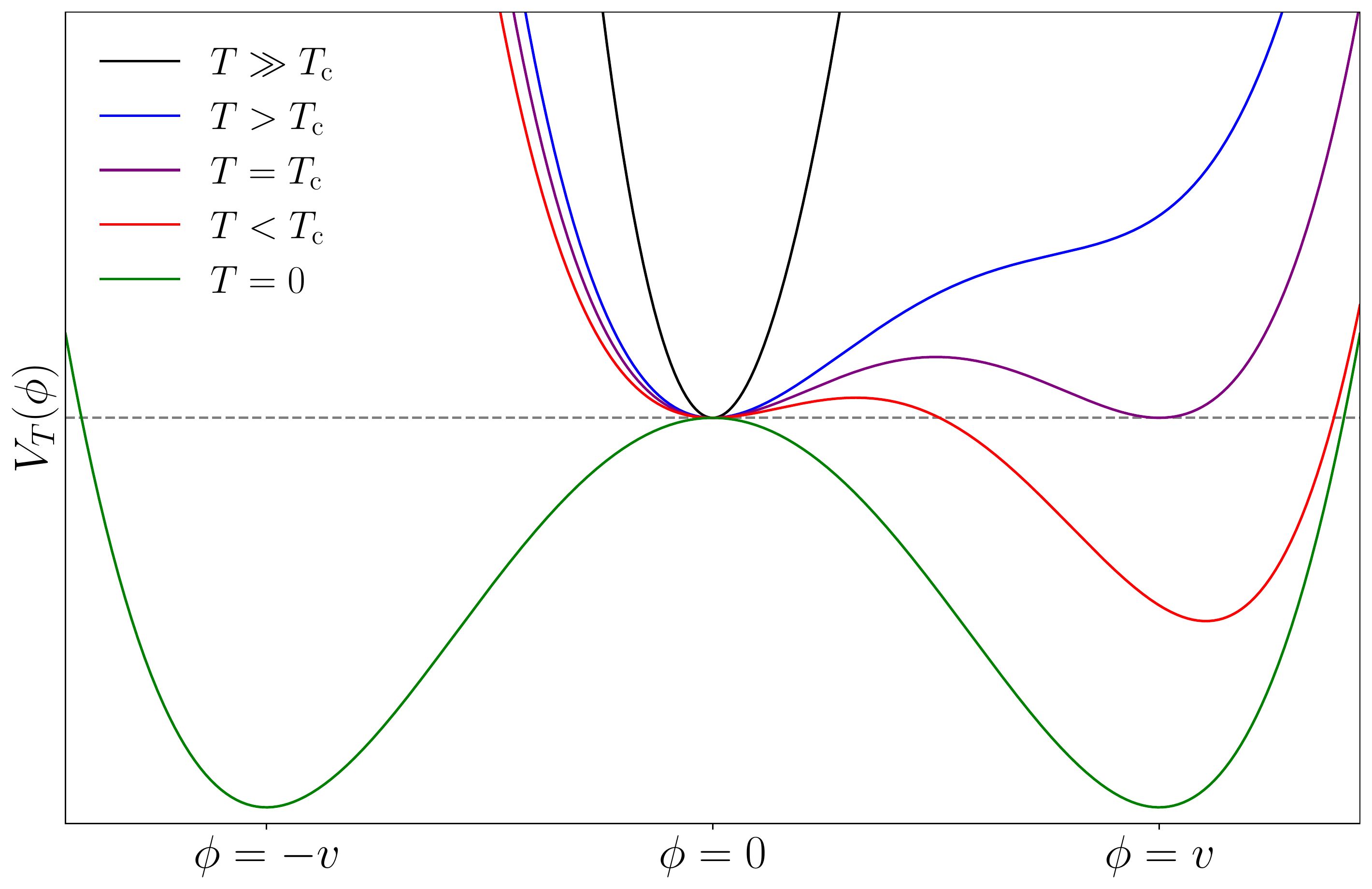}
    \end{center}
    \caption{%
    The thermal effective potential~\eqref{eq:thermal-effective-potential} for a scalar field $\phi$ that undergoes a first-order phase transition.
    The field is initially localised around $\phi=0$, which is a stable vacuum state at high temperatures.
    As the Universe cools below a critical temperature $T_\rmc$ due to cosmic expansion, $\phi=0$ becomes a metastable false vacuum, separated from the true vacuum by a potential barrier.
    }
    \label{fig:fopt-potential}
\end{figure}

The basic picture is as follows: at some high-temperature epoch in the early Universe, we have a scalar field $\phi$ whose dynamics are described by a thermal effective potential of the form~\cite{Hindmarsh:2020hop}
    \begin{equation}
    \label{eq:thermal-effective-potential}
        \hbar^3V_T(\phi)\simeq A(T^2-T_0^2)\phi^2-BT\phi^3+C\phi^4,
    \end{equation}
    where $A$, $B$, and $C$ are dimensionless constants, $T$ is the temperature of the Universe, and $T_0$ is some fixed energy scale.
(Recall that we use units with $k_\mathrm{B}=1$ but $\hbar\ne1$.
The factor of $\hbar^3$ here ensures that the potential has dimensions of mass per unit volume, while $T$ and $\phi$ both have dimensions of mass.)
For example, $\phi$ could be the Higgs field, in which case $T_0$ is related to the Higgs mass, and the values of $A,B,C$ are set by the Higgs' couplings to other particles.
As the Universe expands and cools, $T$ decreases, changing the shape of the potential and therefore the dynamics of the scalar field.
At zero temperature and at very high temperatures, there is a discrete $\mathbb{Z}_2$ symmetry under $\phi\to-\phi$.

For appropriate values of $A,B,C$, the potential exhibits the behaviour shown in figure~\ref{fig:fopt-potential} where, as well as possessing a local minimum at $\phi=0$ at high temperatures, a second minimum develops at some $\phi>0$ once the temperature falls below a certain threshold.
As the Universe continues to cool, it passes a critical temperature $T_\rmc$ where this point becomes a \emph{global} minimum, and it becomes energetically favourable for the field to fall into this state and acquire a nonzero vacuum expectation value (VEV), $v\equiv\ev{\phi}>0$.
We call this state the \emph{broken phase}, as it breaks the $\phi\to-\phi$ symmetry, in contrast with the \emph{symmetric phase} at $\phi=0$.

For a broad class of values of $A,B,C$, the potential has a barrier between $\phi=0$ and $\phi=v$ at temperatures $T\lesssim T_\rmc$, which prevents the field from straightforwardly rolling down into the broken phase.
The symmetric phase thus becomes a metastable \emph{false vacuum} (as opposed to the \emph{true vacuum} at the bottom of the potential well) which can persist for some time after the critical temperature is reached.
Eventually, however, the field escapes this metastable state, either through thermal fluctuations \enquote{over the barrier}, or through quantum tunnelling \enquote{under the barrier}.
These random decay events occur at localised points in spacetime, resulting in the nucleation of \enquote{bubbles} of true vacuum which then expand outwards at relativistic velocities and collide with each other until the entire Universe has transitioned to the broken phase.
This process is called a \emph{first-order phase transition} (FOPT).

While we have motivated this discussion in terms of a Higgs-like scalar field $\phi$, it turns out that the masses and couplings of the heaviest particles in the Standard Model are such that the Higgs potential does \emph{not} have a barrier, and instead of a FOPT there is a simple cross-over transition which occurs homogeneously throughout the Universe in thermal equilibrium~\cite{Kajantie:1996mn}.
However, in alternative particle physics models which couple new fields to the Higgs, or which have new Higgs-like scalar fields, the potential generically acquires a barrier between the symmetric and broken phases, meaning that FOPTs are a very generic consequence of such modifications.
As such, any detection of a FOPT in the early Universe would be incredibly exciting, as it would provide compelling evidence for physics beyond the Standard Model.

GW astronomy provides the most promising means of making such a detection.
This is because false vacuum decay represents a violent and highly energetic departure from equilibrium in the early-Universe plasma, with the resulting dynamics of the matter fields leading to copious production of GWs.
Modelling the resulting GWB signal is a considerable theoretical and numerical challenge, and is currently a very active area of research.
However, the general picture is that there are three key processes that contribute to the GWB~\cite{Caprini:2015zlo}:
    \begin{enumerate}
        \item the dynamics of the scalar field during collisions of true-vacuum bubbles;
        \item the bulk motion of the plasma due to sound waves;
        \item the chaotic motion of the plasma due to MHD turbulence.
    \end{enumerate}

While all three of these mechanisms depend in a complicated way on the details of the underlying particle physics model, the sound-wave contribution dominates the combined GWB spectrum in most cases.
This contribution is usually modelled as a broken power law,
    \begin{equation}
    \label{eq:fopt-gwb-spectrum}
        \Omega(f)=\Omega(f_*)\times(f/f_*)^3\qty[\frac{7}{4+3(f/f_*)^2}]^{7/2},
    \end{equation}
    which grows like $\sim f^3$ at low frequencies and falls off as $\sim f^{-4}$ at high frequencies.
The changeover between these two regimes occurs at a peak frequency~\cite{Caprini:2015zlo,Schmitz:2020syl}
    \begin{equation}
    \label{eq:fopt-peak-freq}
        f_*\approx 19\,\upmu\mathrm{Hz}\times\frac{T_*}{100\,\mathrm{GeV}}\frac{\beta/H_*}{v_\mathrm{w}}\qty(\frac{g_*}{106.75})^{1/6},
    \end{equation}
    which depends on a few key parameters describing the phase transition: $T_*$, the temperature at which the transition takes place (which we have normalised here relative to the electroweak scale); $\beta$, the inverse duration of the transition (normalised to the Hubble rate at the transition epoch, $H_*$); and $v_\mathrm{w}$, the velocity of the bubble walls.
(Here $g_*$ is the number of relativistic degrees of freedom in the plasma, which we have normalised to the value it takes in the Standard Model at early times.)
The peak amplitude of the spectrum is approximated by
    \begin{equation}
    \label{eq:fopt-peak-intensity}
        \Omega(f_*)\approx5.7\times10^{-6}\times\frac{v_\mathrm{w}}{\beta/H_*}\qty(\frac{\kappa\alpha}{1+\alpha})^2\qty(\frac{g_*}{106.75})^{-1/3}\qty[1-\qty(1+2\tau_\mathrm{sw}H_*)^{-1/2}],
    \end{equation}
    where $\alpha$ is the energy density released by the transition, in units of the energy density of the plasma at that epoch; $\kappa$ is an efficiency parameter determined by $\alpha$ and $v_\mathrm{w}$; and $\tau_\mathrm{sw}$ is the lifetime of the sound-wave source, which is a function of $\alpha$, $\beta$, and $v_\mathrm{w}$.

We can understand the strongly-peaked nature of this spectrum as being a consequence of the transient nature of the FOPT source: the GW production is centred around a single epoch in the Universe's history (set by the transition temperature $T_*$), and as a result this imprints a characteristic frequency scale $f_*$ on the spectrum.
This makes it very challenging to probe the entire FOPT parameter space $(T_*,\alpha,\beta/H_*,v_\mathrm{w})$, as it is possible for the peak of the spectrum to lie between the sensitive frequency bands of various GW experiments.
We will revisit this problem in chapter~\ref{chap:binary-resonance}, and show how the novel GW detection method we develop there can be used to probe a unique region of the FOPT parameter space.

\subsection{Cosmic strings}
\label{sec:cosmic-strings}

\begin{figure}[t!]
    \begin{center}
        \includegraphics[width=\textwidth]{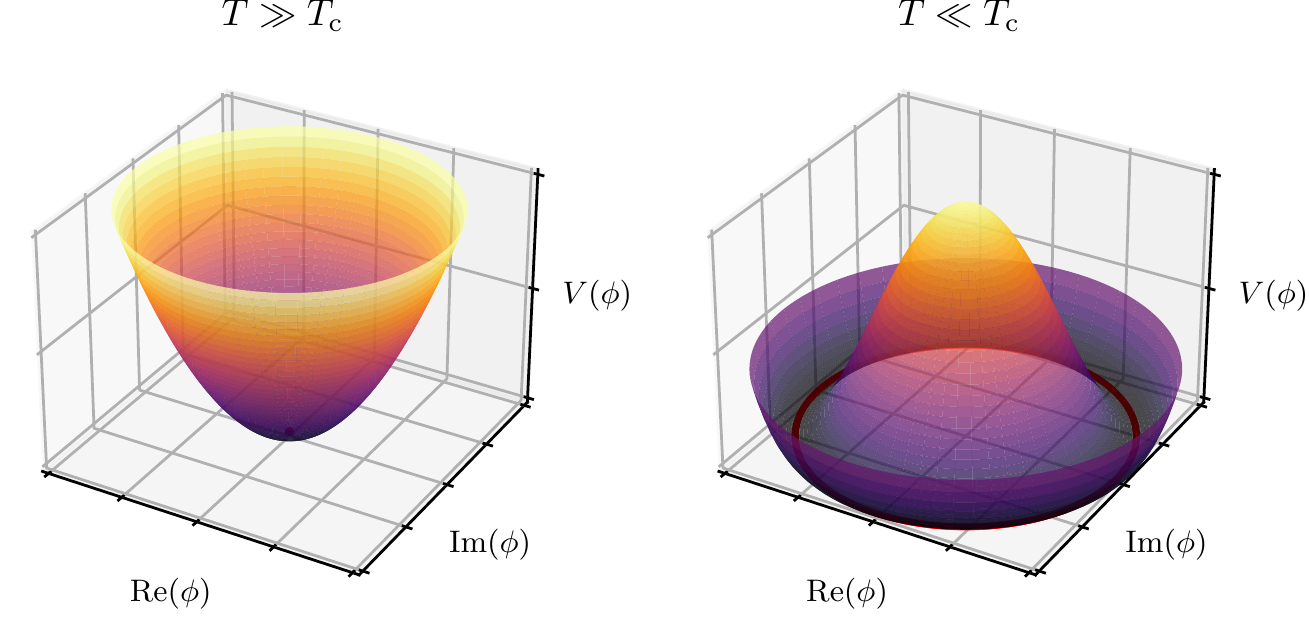}
    \end{center}
    \caption{%
    Effective potential for the complex scalar field $\phi$.
    At high temperatures (left panel), this is simply $\sim|\phi|^2T^2$, and the symmetric phase $\phi=0$ is the unique vacuum state.
    At low temperatures (right panel), the potential transitions to the form in equation~\eqref{eq:complex-phi-potential}; the symmetric phase then becomes unstable, and the field decays into one of a continuous family of vacua given by $|\phi|=v$.
    }
    \label{fig:mexican-hat}
\end{figure}

Consider now what happens if we replace the real scalar field $\phi$ of the previous section with a \emph{complex} scalar field, whose effective potential at low temperatures $T\ll T_\rmc$ is given by
    \begin{equation}
    \label{eq:complex-phi-potential}
        \hbar^3V(\phi)\simeq C(|\phi|^2-v^2)^2,
    \end{equation}
    which is minimised by setting $|\phi|=v$.
(This is the same potential as equation~\eqref{eq:thermal-effective-potential} at $T\ll T_0$, up to a constant shift.)
Rather than having two disconnected vacua at $\phi=\pm v$ related by a broken $\mathbb{Z}_2$ symmetry, we now have a connected circle of vacua at $\phi=\rme^{\rmi\vartheta}v$ that are related by a broken $U(1)$ symmetry, as illustrated in figure~\ref{fig:mexican-hat}; we label each of these vacua by a phase angle $\vartheta$.

\begin{figure}[p!]
    \begin{center}
        \includegraphics[width=0.78\textwidth]{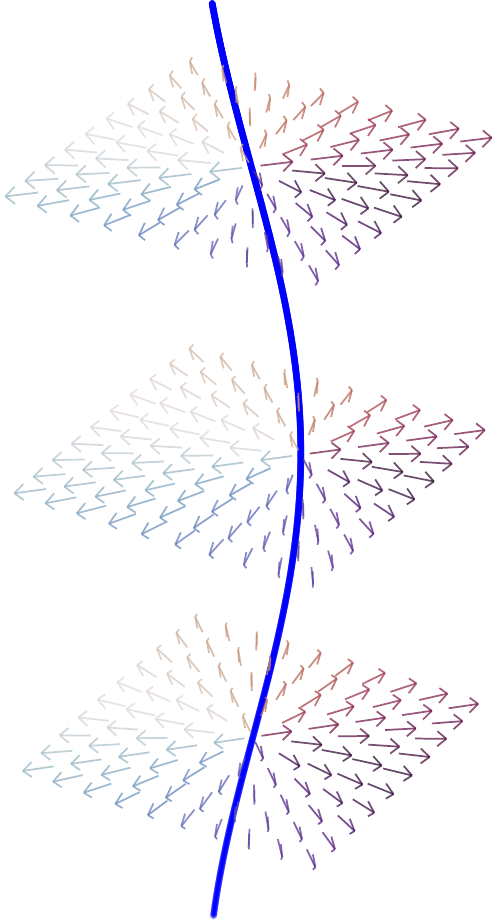}
    \end{center}
    \caption{%
    Configuration of the complex scalar field $\phi$ on three spatial slices intersecting a cosmic string (shown in blue).
    The colour and direction of the arrows at each point on these slices indicate the phase angle $\vartheta$ of the vacuum state at that point in space.
    If we travel around a closed spatial loop that encloses the string, we find that this phase has a nonzero winding number $n$ around the vacuum circle, $\vartheta\to\vartheta+2\uppi n$.
    }
    \label{fig:cosmic-string-slice}
\end{figure}

As the Universe cools, different spatial regions decay from the symmetric phase $\phi=0$ to a different, randomly-selected point on the vacuum circle, resulting in a phase $\vartheta$ that is a smoothly-varying field on spacetime.
This process generically gives rise to configurations such as the one shown in figure~\ref{fig:cosmic-string-slice}, where $\vartheta$ executes a nonzero winding around the vacuum circle.
As we approach the centre of such a configuration, we see that it is impossible for the scalar field $\phi$ to reside in any of the vacua without causing a discontinuity; $\phi$ is therefore forced into the higher-energy symmetric phase at this point.
Topologically, it is impossible to have $\phi=0$ at just this single isolated point.
Instead, we necessarily obtain a linelike object of nonzero energy density by stacking many of these points together, as illustrated in figure~\ref{fig:cosmic-string-slice}.
We call this object a \emph{cosmic string}~\cite{Kibble:1976sj,Vilenkin:1984ib,Hindmarsh:1994re,Vilenkin:2000jqa}.
As with FOPTs, cosmic strings do not occur in the Standard Model, but are a generic consequence of many high-energy modifications to the Standard Model~\cite{Jeannerot:2003qv}, and are therefore an extremely well-motivated means of searching for new physics in the early Universe.

Once formed, cosmic strings are protected from decay by the topological stability of the underlying field configuration.
This stability is much more robust than the metastability of the false vacuum state we encountered in the previous section, which was vulnerable to decay via any single point in space tunnelling to the true vacuum.
Due to the continuity of the phase angle $\vartheta$ throughout space, it is impossible to \enquote{unwind} the scalar field at any single point along the string: one would have to perform this unwinding simultaneously \emph{everywhere} along the entire length of the string.
The probability of the field tunnelling to such a globally unwound configuration is vanishingly small.
As a result, cosmic strings can survive over cosmological timescales, acting as \emph{relics} of their early formation epoch that we can, in principle, observe at late times.

We can understand the importance of cosmic strings as GW sources by estimating the amount of mass/energy they carry per unit length.
To calculate this, we need to know ($i$) the energy density $\rho$ of the scalar field near the core of the string, and ($ii$) the characteristic width $\delta$ of the string, i.e., the lengthscale over which $\phi$ decays from the symmetric phase to one of the vacuum states as we travel away from the core of the string.
We can then approximate the linear mass density, $\mu$, as the product of the cross-sectional area of the string and the energy density in its core,
    \begin{equation}
        \mu\approx\uppi\delta^2\times\rho.
    \end{equation}
It is natural to interpret $\rho$ as the difference in the potential~\eqref{eq:complex-phi-potential} between the symmetric phase and the broken phase, which gives us $\rho=V(0)-V(v)\sim\hbar^{-3}v^4$.
For the string width $\delta$, we notice that since the scalar field VEV, $v$, is the only dimensionful quantity in equation~\eqref{eq:complex-phi-potential}, the only lengthscale that we can construct is the associated Compton wavelength, $\delta\sim\hbar/v$.
Neglecting $\order{1}$ constants, we therefore find that the linear density must be
    \begin{equation}
    \label{eq:string-tension}
        G\mu\sim(v/m_\Pl)^2,
    \end{equation}
    which we have made dimensionless by normalising relative to the Planckian linear density $m_\Pl/\ell_\Pl=m_\Pl^2/\hbar=1/G$.
Equivalently, this means we can write the string width $\delta$ as
    \begin{equation}
    \label{eq:string-width}
        \delta\sim\sqrt{\hbar/\mu}=\ell_\Pl/\sqrt{G\mu}.
    \end{equation}
As with any string, the linear density $\mu$ is related to the string tension, $\mathcal{T}$, and the propagation speed of waves along the string, $u$, by the equation $\mu=\mathcal{T}/u^2$.
As we will see in chapter~\ref{chap:cosmic-strings}, the motion of cosmic strings can be written as a \emph{relativistic} wave equation with $u=1$, so this equation tells us we are free to interchange their tension and linear density.
We therefore refer to $G\mu$ as the \emph{(dimensionless) string tension} of a cosmic string.

Equation~\eqref{eq:string-tension} shows that strings formed at early epochs in the Universe's history will have enormous string tensions compared to objects from our everyday experience.
For example, a string formed at the grand unified theory (GUT) scale, $v\sim 10^{13}\,\mathrm{TeV}\sim10^{-3}m_\Pl$, would have a dimensionless string tension of $G\mu\sim10^{-6}$.
(Compare this with, e.g., a steel guitar string, which has $G\mu\approx5\times10^{-30}$.)
This means that a GUT-scale string spanning the size of the Earth's orbit around the Sun would have a mass of $\mu\times\mathrm{AU}\sim100\,m_\odot$, far greater than the total mass of the Solar System.
Such objects would clearly exert considerable gravitational influence over their surroundings.
The fact that they also oscillate at relativistic velocities tells us that cosmic strings must be incredibly powerful sources of GWs.

Since cosmic strings are topologically forbidden from having endpoints, they are forced to form either closed loops~\cite{Kibble:1982cb,Turok:1984cn} or open-ended \enquote{long} strings (also called \enquote{infinite} strings) which stretch across an entire Hubble volume.
Any given Hubble patch will typically have $\order{1}$ long strings passing through it~\cite{Vilenkin:2000jqa}, which are formed at the epoch of the symmetry-breaking phase transition.
These long strings are free to oscillate on sub-horizon scales, and generically intersect themselves to chop off closed loops, which produce copious GWs and decay to smaller sizes by trading their length for radiated energy (since their linear density $\mu$ is a constant fixed by the underlying field theory, any loss of energy necessarily translates into a decay in the loop's size).
This process (combined with loop-loop intersections which also chop off smaller loops) causes a downwards cascade, eventually filling the Hubble patch with a network of loops with a continuous spectrum of sizes, through which the energy in the long strings is gradually dissipated into GWs~\cite{Bennett:1987vf}.

Since loop production persists over cosmological timescales, the GW emission from the loop network accumulates over time to give a strong GWB spectrum~\cite{Auclair:2019wcv}.
Unlike the peaked spectrum~\eqref{eq:fopt-gwb-spectrum} we saw in the FOPT case, which encoded the single epoch of GW production, this collective emission from cosmic strings of all sizes and at all epochs leads to a GWB which is flat over a very broad frequency band (although this plateau is usually also accompanied by a peak at lower frequencies).
As a result, cosmic strings allow us to probe much higher energy scales than those accessible with FOPT signals.
For example, the GWB from a FOPT occurring at the GUT scale would peak at $f\sim10^{10}\,\mathrm{Hz}$, far beyond the sensitive frequencies of current or future GW experiments, while the signal from a cosmic string network formed at that epoch would be observable at much lower frequencies.

GW searches for cosmic strings with LIGO/Virgo~\cite{Abbott:2009rr,Abbott:2017mem,Abbott:2019vic,Abbott:2021kbb,Abbott:2021nrg} and pulsar timing arrays~\cite{Lasky:2015lej,Blanco-Pillado:2017rnf,Yonemaru:2020bmr} have so far returned only null results, allowing us to place a conservative upper limit on the string tension of $G\mu\lesssim10^{-11}$ (there is some uncertainty due to the modelling of the cosmic string loop network---the most stringent constraints are at the level of $G\mu\lesssim10^{-15}$~\cite{Abbott:2021nrg}).
These constraints are orders of magnitude better than those derived from CMB observations, which are of the order $G\mu\lesssim10^{-7}$~\cite{Kaiser:1984iv,Ade:2013xla,McEwen:2016aob}.

In chapter~\ref{chap:cosmic-strings} we will describe the GW emission from loops in more detail, focusing on three key emission mechanisms called \emph{cusps}, \emph{kinks}, and \emph{kink-kink collisions}.
We will derive, for the first time, the nonlinear GW memory~\eqref{eq:nonlinear-memory} associated with these mechanisms, and will show that this has surprising implications for the gravitational dynamics of cusps.

\subsection{Primordial black holes}
\label{sec:primordial-black-holes}

We saw in section~\ref{sec:compact-binaries} that compact binaries, and binary black holes in particular, are vitally important sources of GWs.
Previously, we characterised these sources as \emph{astrophysical} rather than cosmological, as the main formation channel for BHs and NSs is thought to be through the gravitational collapse of stars that have exhausted their nuclear fuel.
However, it is also possible that BHs may have formed through \emph{cosmological} processes in the early Universe, meaning that the BBH signals observed by LIGO/Virgo could be a hint of something much more exotic.
Efforts to understand the formation and observational signatures of these \emph{primordial} black holes (PBHs)~\cite{Zeldovich:1967aa,Hawking:1971ei,Carr:1974nx,Carr:1975qj} have formed a major strand of research in cosmology for the past fifty years.

While there are various ways of producing PBHs in the early Universe that have been studied in the literature, the most well-established mechanism is the gravitational collapse of horizon-sized density perturbations.
Since a given Hubble patch must be significantly more massive than average to undergo such a collapse, this mechanism requires the production of very large overdensities during inflation, e.g., due to a phase transition between two distinct inflationary phases~\cite{Garcia-Bellido:1996mdl}, or the presence of a feature in the inflationary potential~\cite{Ivanov:1994pa}.
These large perturbations are \enquote{frozen out} by the inflationary expansion, and only become dynamical again and collapse once they re-enter the horizon during the radiation-dominated era.
We can estimate the masses of the resulting PBHs by noticing that, since they correspond to the collapse of an entire Hubble patch, their Schwarzschild radius $2Gm_\mathrm{pbh}$ should be of the same order of magnitude as the particle horizon at the time of horizon re-entry.
In a radiation-dominated Universe this is given by $r_\mathrm{hor}=2t_\mathrm{age}$, where $t_\mathrm{age}$ is the age of the Universe at that epoch.
We therefore have the rough estimate
    \begin{equation}
    \label{eq:pbh-horizon-mass}
        m_\mathrm{pbh}\sim\frac{t_\mathrm{age}}{G}\sim m_\Pl\times\qty(\frac{t_\mathrm{age}}{t_\Pl})\sim 10^6\,m_\odot\times\qty(\frac{t_\mathrm{age}}{10\,\mathrm{s}}).
    \end{equation}
This shows that the masses of horizon-collapse PBHs can span an enormous range of physical scales, from tiny Planckian objects formed in the Universe's earliest moments, all the way up to supermassive BHs comparable to those residing at the centres of galaxies, formed shortly before the epoch of big-bang nucleosynthesis.
PBHs that are sufficiently massive can produce detectable GW signals at late times by forming compact binaries~\cite{Bird:2016dcv,Clesse:2016vqa,Sasaki:2016jop,Raidal:2017mfl,Sasaki:2018dmp,Raidal:2018bbj,DeLuca:2020qqa}, as described in section~\ref{sec:compact-binaries}.
They can also produce a powerful GWB signal during the collapse process itself, with the dynamics of the large scalar overdensities leading to the generation of sizeable tensor perturbations~\cite{Ananda:2006af}.

One of the most exciting features of equation~\eqref{eq:pbh-horizon-mass} is that it provides a mechanism for producing BHs with masses less than that of the Sun, $m_\mathrm{pbh}<m_\odot$, which is impossible to achieve through the standard astrophysical process of stellar collapse.
This means that any detection of a subsolar-mass BH, whether through GWs or otherwise, would provide powerful evidence for PBH formation in the early Universe, and could substantially advance our understanding of inflation and other areas of fundamental physics.
Aside from their masses, PBHs may also possess other features which could help to distinguish them from astrophysical BHs.
For example, the radiation-era formation process described above is expected to lead to the PBHs having negligibly small spins, due to the angular momentum of the collapsing region being dissipated by radiation pressure during the collapse; this is in contrast with astrophysical BHs, which can have quite sizeable spins due to angular momentum conservation during their collapse.

\begin{figure}[t!]
    \begin{center}
        \includegraphics[width=\textwidth]{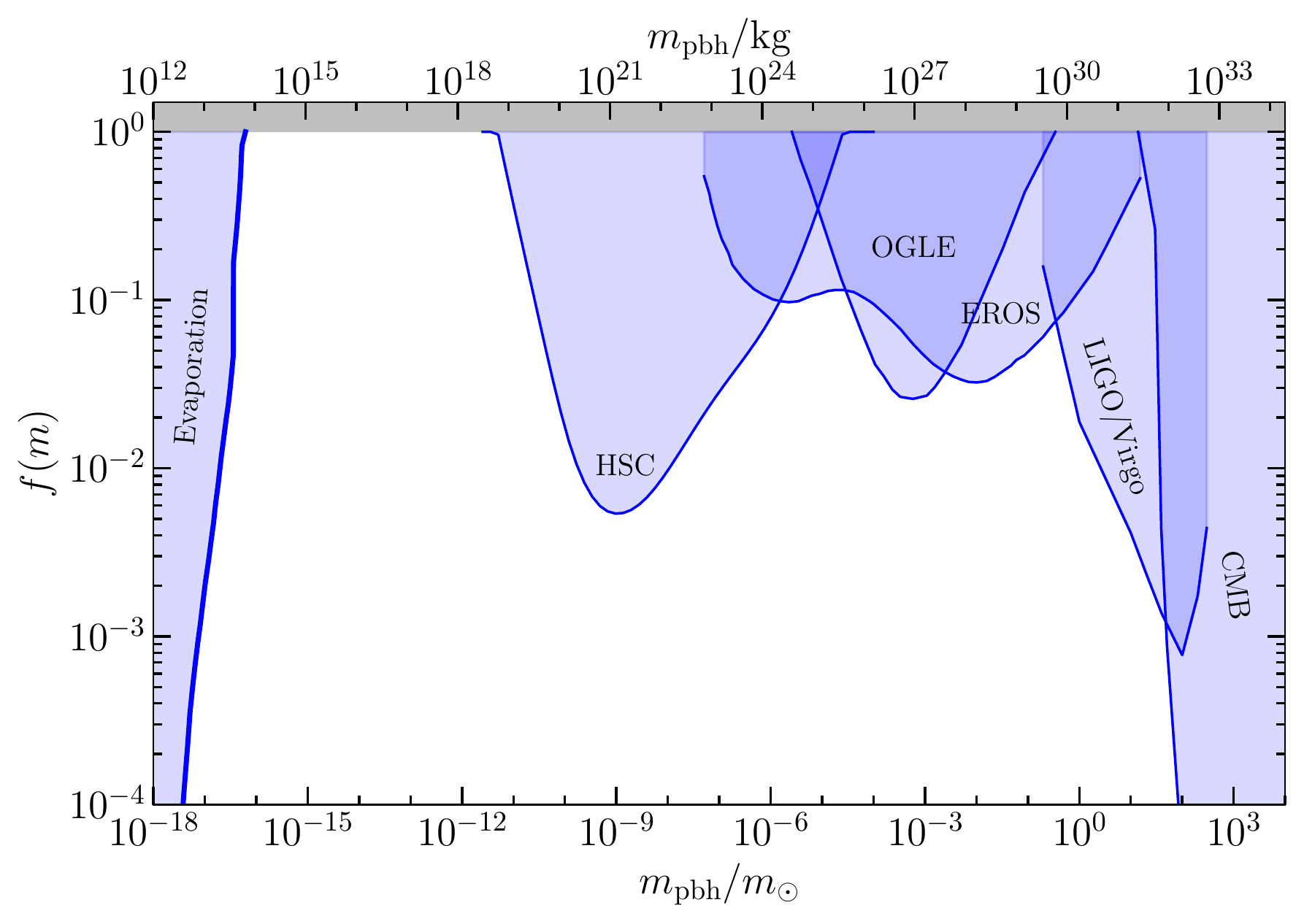}
    \end{center}
    \caption{%
    Current observational bounds on the PBH mass spectrum.
    Very light PBHs ($m_\mathrm{pbh}\lesssim10^{-16}\,m_\odot$) are constrained by the non-observation of high-energy particles (e.g. gamma rays) emitted during evaporation~\cite{Carr:2009jm,Clark:2016nst,Poulin:2016anj,Clark:2018ghm,Boudaud:2018hqb,Laha:2019ssq,DeRocco:2019fjq,Kim:2020ngi}, while very massive PBHs ($m_\mathrm{pbh}\gtrsim10\,m_\odot$) are constrained by the sensitivity of the CMB anisotropies to PBH accretion at the epoch of recombination~\cite{Serpico:2020ehh}.
    PBHs at intermediate masses ($10^{-11}\,m_\odot\lesssim m_\mathrm{pbh}\lesssim100\,m_\odot$) are constrained by stellar microlensing experiments such as OGLE~\cite{Niikura:2019kqi}, EROS~\cite{Tisserand:2006ryy}, and Subaru-HSC~\cite{Croon:2020ouk}, and by the BBH event rate observed by LIGO/Virgo~\cite{Abbott:2019qbw,Kavanagh:2018ggo}.
    PBHs in the \enquote{sublunar} mass range ($10^{-16}\,m_\odot\lesssim m_\mathrm{pbh}\lesssim10^{-12}\,m_\odot$) could account for the entirety of the Universe's CDM budget.
    This figure was produced using the Python package \texttt{PBH-bounds}~\cite{bradley_j_kavanagh_2019_3538999,Green:2020jor}.
    }
    \label{fig:pbh-bounds}
\end{figure}

As we go towards lower end of the mass range predicted by equation~\eqref{eq:pbh-horizon-mass}, it becomes increasingly important to treat the PBHs as quantum-mechanical objects.
By considering the behaviour of quantum fields on BH spacetimes, \citet{Hawking:1974sw} famously showed that BHs emit thermal blackbody radiation with a temperature\footnote{Strictly speaking, this expression is only correct for Schwarzschild BHs. For Kerr BHs one must multiply this by a factor of $2\sqrt{1-\chi^2}/(1+\sqrt{1-\chi^2})$, where $\chi\equiv J/(Gm^2)\in[0,1]$ is the BH's dimensionless spin.} $T_\mathrm{BH}=\hbar/(8\uppi Gm)$, causing them to radiate away their mass and evaporate on a timescale~\cite{Hawking:1974rv,Page:1976df,Page:1976ki}
    \begin{equation}
    \label{eq:pbh-evaporation-timescale}
        t_\mathrm{evap}\sim t_\Pl\times\qty(\frac{m}{m_\Pl})^3.
    \end{equation}
As a result, PBHs with masses less than
    \begin{equation}
    \label{eq:evaporated-pbh-mass}
        m_*\approx3\times10^{19}m_\Pl\approx5\times10^{11}\,\mathrm{kg}\approx3\times10^{-19}\,m_\odot
    \end{equation}
    are expected to have evaporated completely in the time since their formation.
Equation~\eqref{eq:evaporated-pbh-mass} therefore provides a lower bound on the masses of PBHs surviving to the present day.

As we mentioned near the beginning of this chapter, non-evaporating PBHs are interesting not just as GW sources, but also as a very well-motivated candidate for dark matter~\cite{Chapline:1975ojl,Carr:2016drx}.
Since they are massive, nonbaryonic, have nonrelativistic velocities, and interact only through gravity, PBHs fulfil all of the properties that we require for a DM particle.
For this reason, constraints on their cosmological abundance are usually phrased in terms of their \emph{CDM mass fraction},
    \begin{equation}
    \label{eq:cdm-mass-fraction}
        f(m)\equiv\frac{1}{\rho_\mathrm{cdm}}\dv{\rho_\mathrm{pbh}}{(\ln m)}=\frac{m^2n_\mathrm{pbh}(m)}{\rho_\mathrm{cdm}},
    \end{equation}
    which tells us how much of the observed CDM mass density in the Universe is made up of PBHs of mass $m$.
(We will also refer to this quantity as the \emph{PBH mass spectrum}.)
Here $\rho_\mathrm{pbh}$ and $n_\mathrm{pbh}$ are the PBH mass density and number density respectively, while $\rho_\mathrm{cdm}\approx0.261\,\rho_\mathrm{c}\approx30\,m_\odot\,\mathrm{kpc}^{-3}$ (c.f. table~\ref{tab:density-parameters}).
To successfully account for the entirety of the DM, a given PBH formation scenario must give $\int\dd{(\ln m)}f(m)=1$; however, consistency with various observational probes places upper bounds on $f(m)$ over a broad range of masses, as shown in figure~\ref{fig:pbh-bounds}.

We will encounter PBHs again in chapter~\ref{chap:cosmic-strings}, in which we investigate the possibility of them being formed through the gravitational collapse of cusps on cosmic string loops.

\section[Gravitational-wave detection: basic principles and current results]{Gravitational-wave detection:\\basic principles and current results}
\label{sec:gw-detection}

Having familiarised ourselves with a few important sources of gravitational waves in section~\ref{sec:gw-sources}, the obvious question now is: how do we go about detecting these signals?

The key idea is to use the results of section~\ref{sec:gw-test-masses}, in which we found that the passage of a GW causes an oscillation in the proper distance between two freely-falling test masses.
In particular, this results in an oscillation in the light-travel time $T$ between the two masses~\cite{Estabrook:1975dop},
    \begin{equation}
    \label{eq:light-travel-time}
        \updelta T=\frac{1}{2}\int_0^T\dd{t}\hat{u}^i\hat{u}^jh_{ij}(t,\vb*x(t)),
    \end{equation}
    with $\vu*u$ the propagation direction of the light pulse.
If one knows the expected arrival time of the pulse with sufficient accuracy, then searching for offsets from this expected arrival time allows us to probe the GW strain $h_{ij}$.
The two leading methods for GW detection, laser interferometry and pulsar timing, are both fundamentally based on this principle, each using different pairs of objects as test masses and different means of calibrating the expected arrival time.
We give a brief overview of each of these methods and their current observational results in sections~\ref{sec:laser-interferometers} and~\ref{sec:pulsar-timing} respectively.
We then describe a completely different class of indirect cosmological bounds in section~\ref{sec:cosmological-gwb-bounds}, before briefly touching on a variety of other GW searches in section~\ref{sec:other-gw-searches}.

\subsection{Laser interferometers}
\label{sec:laser-interferometers}

One approach to extracting the perturbations~\eqref{eq:light-travel-time} in the round-trip time of a light pulse is to use an \emph{interferometer} (IFO), sending two laser pulses of the same wavelength along different paths through space before recombining them and measuring their relative phase.
The power of this method is that it removes the need to calibrate the light travel time along each path; instead, we can perform a \emph{differential} measurement between the two paths.
If each path is affected differently by the passage of a GW, then this results in a perturbation to the relative phase.
For example, suppose we use a beam-splitter to send half of our laser photons along the $\vu*x$-axis and the other half along the $\vu*y$-axis, each being reflected by a test mass and returning to the origin.
The resulting differential signal is then
    \begin{equation}
    \label{eq:ifo-phase-difference}
        T_x(t)-T_y(t)=\frac{1}{2}\int_0^T\dd{t}(\hat{x}^i\hat{x}^j-\hat{y}^i\hat{y}^j)h_{ij}(t,\vb*x(t))\simeq\frac{T}{2}(\hat{x}^i\hat{x}^j-\hat{y}^i\hat{y}^j)h_{ij}(t,\vb*0),
    \end{equation}
    where in the second equality we have taken the \emph{small-antenna limit} in which the arms of the IFO are much smaller than the wavelength of the GW.

In order to measure this signal, we require the test masses at the ends of each arm of the IFO to be free to move along the axis of the corresponding arm, so that their response to an incoming GW corresponds to that of a freely-falling object along that axis.
However, this freedom means that there are countless sources of instrumental noise that can shake the test masses and therefore generate a phase difference~\eqref{eq:ifo-phase-difference}, limiting the sensitivity to GW signals.
The two key sources of noise for ground-based IFOs are seismic noise at low frequencies and photon shot noise in the laser at high frequencies~\cite{Maggiore:2007zz,Saulson:1995zi}; as a result, the best sensitivity comes at a \enquote{sweet spot} between these two noise floors at about $10$--$1000\,\mathrm{Hz}$.

Several ground-based IFOs have been used to conduct GW searches since this idea was first proposed by \citet{Weiss:1972} in 1972.
Of these, by far the most successful to date are the Laser Interferometer Gravitational-Wave Observatory~\cite{Harry:2010zz,Aasi:2014jea} (LIGO, which consists of two instruments in Hanford, Washington and Livingston, Louisiana, USA), and its European counterpart Virgo~\cite{Acernese:2014hva} (which is a single instrument in Cascina, Italy).
These three IFOs, operating together as a global GW observatory network, have jointly detected dozens of CBC signals~\cite{Abbott:2018mvr,Abbott:2020niy,Abbott:2021usb}, primarily from BBHs, but also including several BNSs and, more recently, two systems which have been interpreted as black hole-neutron star binaries (BHNS)~\cite{Abbott:2021qlt}.
These detections have revealed fascinating insights into the population properties of BHs and NSs in the local Universe~\cite{Abbott:2020gyp}, and have been used to perform stringent new tests of GR~\cite{Abbott:2016src,Yunes:2016jcc,Abbott:2020tif}.
They have also included a number of \enquote{exceptional events} with unexpected properties~\cite{Abbott:2020aai,Abbott:2020zkf,Abbott:2020iuh,Abbott:2020mjq} that challenge our understanding of the astrophysical formation channels for these systems.

\begin{figure}[t!]
    \includegraphics[width=\textwidth]{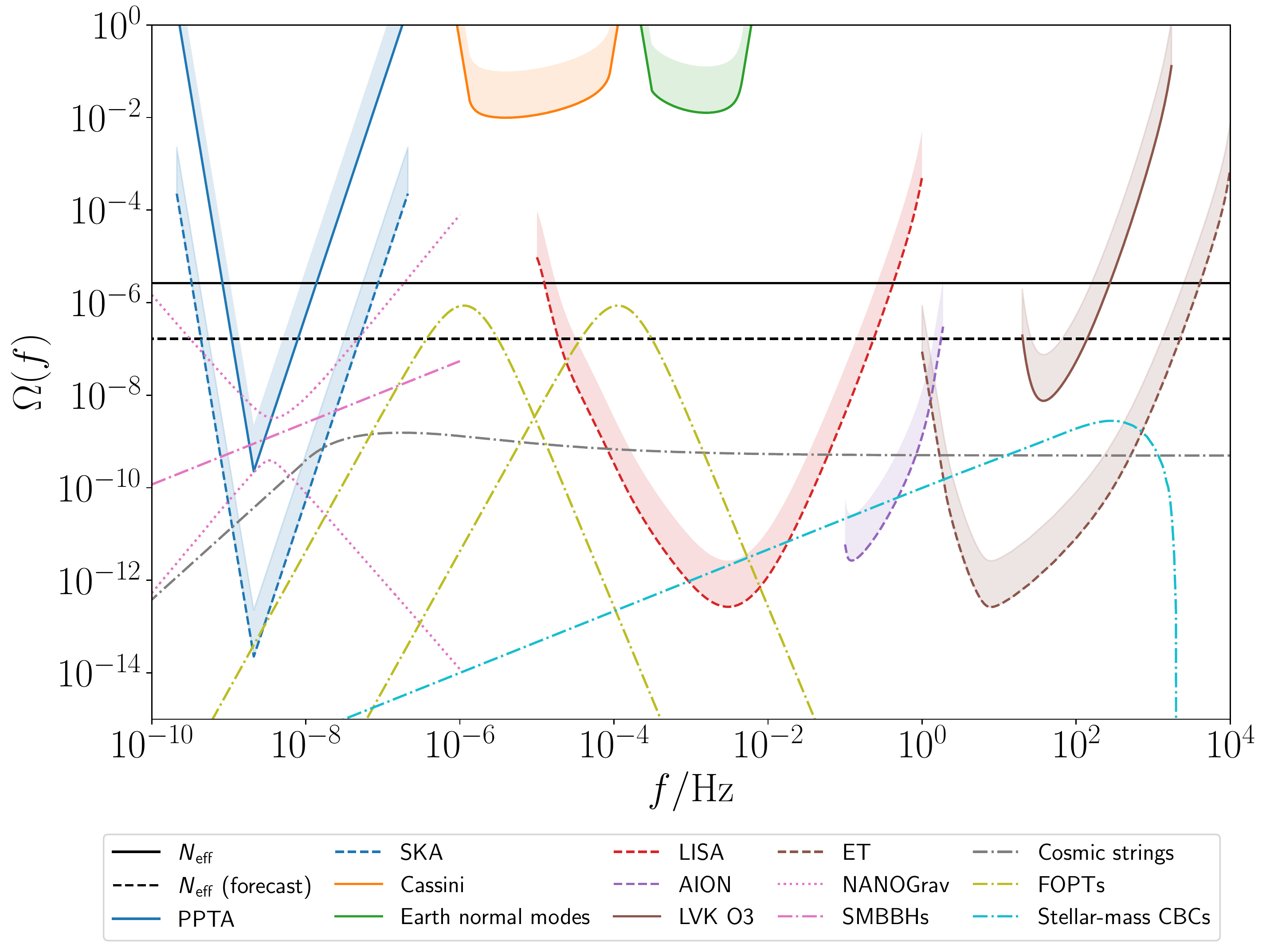}
    \caption{%
    An overview of current and future observational bounds on the stochastic gravitational-wave background.
    Solid curves indicate existing results from the LIGO/Virgo/KAGRA Collaboration (LVK)~\cite{Abbott:2021kbb}, pulsar timing by the Parkes PTA~\cite{Lasky:2015lej}, CMB+BBN constraints from $N_\mathrm{eff}$~\cite{Pagano:2015hma}, gravimeter monitoring of the Earth's normal modes~\cite{Coughlin:2014xua}, and Doppler tracking of the Cassini spacecraft~\cite{Armstrong:2003ay}.
    Dashed curves indicate forecast bounds from Einstein Telescope (ET)~\cite{Punturo:2010zz}, LISA~\cite{Amaro-Seoane:2017drz}, the Square Kilometre Array (SKA)~\cite{Janssen:2014dka}, and the proposed km-scale atom interferometer AION~\cite{Badurina:2019hst}, as well as improved $N_\mathrm{eff}$ constraints~\cite{Pagano:2015hma}.
    Dot-dashed curves show various potential GWB signals.
    The light blue curve shows the expected spectrum from stellar-mass CBCs (based on our current knowledge of the cosmic BBH, BNS, and BHNS populations), which we discuss in chapter~\ref{chap:anisotropies}.
    The grey curve shows an example spectrum from Nambu-Goto cosmic string loops (using \enquote{model 2} of the loop network, with string tension $G\mu=10^{-11}$), which we discuss in chapter~\ref{chap:cosmic-strings}.
    The yellow curves show two first-order phase transition (FOPT) spectra at temperatures $T_*=2\,\mathrm{GeV}$ and $200\,\mathrm{GeV}$, peaking at $f\approx1\,\upmu\mathrm{Hz}$ and $\approx100\,\upmu\mathrm{Hz}$ respectively.
    The pink dotted curves indicate a range of possible signals associated with the common process (CP) detected by NANOGrav~\cite{Arzoumanian:2020vkk}, while the overlaid dot-dashed curve shows the median inferred amplitude for the NANOGrav CP when assuming a $\Omega\sim f^{2/3}$ spectrum, as expected for inspiralling supermassive binary black holes (SMBBHs).
    }
    \label{fig:gwb-constraints}
\end{figure}

Aside from their success in detecting CBCs, LIGO/Virgo have also conducted some of the most sensitive GWB searches of any experiment to date~\cite{Abbott:2016jlg,Abbott:2019vic,Abbott:2021kbb}.
In figure~\ref{fig:gwb-constraints} we show the PI curve from LIGO/Virgo's third observing run, which reaches a maximum sensitivity of $\Omega(f)\approx7.5\times10^{-9}$ at $f\approx38\,\mathrm{Hz}$ for an isotropic background~\cite{Abbott:2021kbb}.
LIGO/Virgo have also carried out searches for anisotropic backgrounds~\cite{Abbott:2016nwa,Abbott:2019gaw,Abbott:2021tss}, backgrounds with non-GR polarisation modes~\cite{Abbott:2018czr,Callister:2017ocg}, and modelled searches for the GWB from cosmic strings~\cite{Abbott:2017mem,Abbott:2021nrg}.

Looking to the future, the recent addition of the Japanese IFO KAGRA~\cite{Abbott:2020npa,Akutsu:2018axf} and the future addition of LIGO India~\cite{Unnikrishnan:2013qwa} will extend the capabilities of the GW observatory network, while planned upgrades to the existing instruments will further enhance their sensitivity.
This network is expected to detect the GWB signal from compact binaries once LIGO and Virgo reach their design sensitivities~\cite{Abbott:2017zlf}.
By that time, we can expect to have detected many more individual CBC signals, which in turn will give us a much firmer understanding of the expected amplitude and spectral shape of the stochastic CBC signal.

By the late 2030s, these experiments will likely be replaced by so-called \enquote{third generation} IFOs such as Einstein Telescope (ET)~\cite{Punturo:2010zz} and Cosmic Explorer (CE)~\cite{Hall:2020dps,Reitze:2019iox}.
These observatories will be so sensitive that their BBH detection horizons will reach all the way out to the earliest epochs of star formation at $z\approx20$ and beyond~\cite{Maggiore:2019uih} (the BNS and BHNS detection horizons are at later epochs, due to the lower masses of these systems), allowing them to individually resolve more than $99.9\%$ of all BBH signals in the Universe~\cite{Regimbau:2016ike} (assuming that they are not primordial in origin), as well as enormous numbers of BNS and BHNS signals.
As a result, it might not even make sense to talk about a stochastic CBC signal for ET and CE; instead, the goal will likely be to coherently model and subtract the majority of individual CBCs, and to use stochastic methods to search for cosmological GWs beneath this astrophysical foreground~\cite{Regimbau:2016ike,Sachdev:2020bkk,Sharma:2020btq}.

By this time, we also expect to have data from the first \emph{space-based} GW observatory, the Laser Interferometer Space Antenna (LISA)~\cite{Amaro-Seoane:2017drz}.
Freed from the seismic noise that limits the sensitivity of ground-based IFOs below $\sim10\,\mathrm{Hz}$, LISA will be able to search for GWs at much lower frequencies, with a sensitivity peaking in the mHz regime.
This will allow LISA probe a completely different population of CBCs, with masses much greater than those of the stellar-mass binaries detected by LIGO/Virgo.
(Recall that CBC signals are truncated around the ISCO frequency~\eqref{eq:isco-frequency}, so that binaries with masses greater than $\sim200\,m_\odot$ merge below the LIGO/Virgo frequency band.)
In particular, these more massive sources are expected to include extreme mass-ratio inspirals (EMRIs)~\cite{Amaro-Seoane:2007osp,Babak:2017tow} in which a stellar-mass object orbits a supermassive black hole (SMBH) with mass $\gtrsim10^6\,m_\odot$, as well as supermassive binary black holes (SMBBHs)~\cite{Klein:2015hvg} that are expected to form when two galaxies merge.
LISA will also prove incredibly useful in searching for cosmological signals, probing cosmic strings with tensions of $G\mu\gtrsim10^{-17}$~\cite{Auclair:2019wcv}, and first-order phase transitions at temperatures $T_*\sim1\,\mathrm{GeV}\text{--}10^3\,\mathrm{TeV}$~\cite{Caprini:2015zlo,Caprini:2019egz} (note that this includes transitions at the electroweak scale $\sim200\,\mathrm{GeV}$, which are particularly well-motivated from the point of view of particle physics models).

\subsection{Pulsar timing}
\label{sec:pulsar-timing}

The other main GW detection technique that has been explored to date relies on \emph{millisecond pulsars}~\cite{Backer:1982msp,Lorimer:2008se}: highly-spinning neutron stars that emit strong beams of EM radiation (particularly radio emission) from their magnetic poles.
Since these beams are usually misaligned with the pulsar's rotation axis, the EM emission acts like the beam of a lighthouse, only reaching the Earth once per spin period.
Pulsar spin periods are incredibly stable over time, and in many cases have been measured to within one part per trillion, allowing us to predict with very high precision the times of arrival (ToAs) of subsequent pulses.
Subtracting this \enquote{timing formula} from the observed ToAs then gives us a set of \emph{timing residuals} that directly measure the GW-induced perturbations to the light-travel time between the pulsar and the Earth.
These residuals are subject to noise in the radio antenna (which causes a certain level of uncertainty in the reconstruction of each pulse, and therefore in determining the ToA), as well as some level of intrinsic noise in the pulse profiles and emission times from each pulsar, the nature of which is not yet fully understood.

\begin{figure}[t!]
    \begin{center}
        \includegraphics[width=0.85\textwidth]{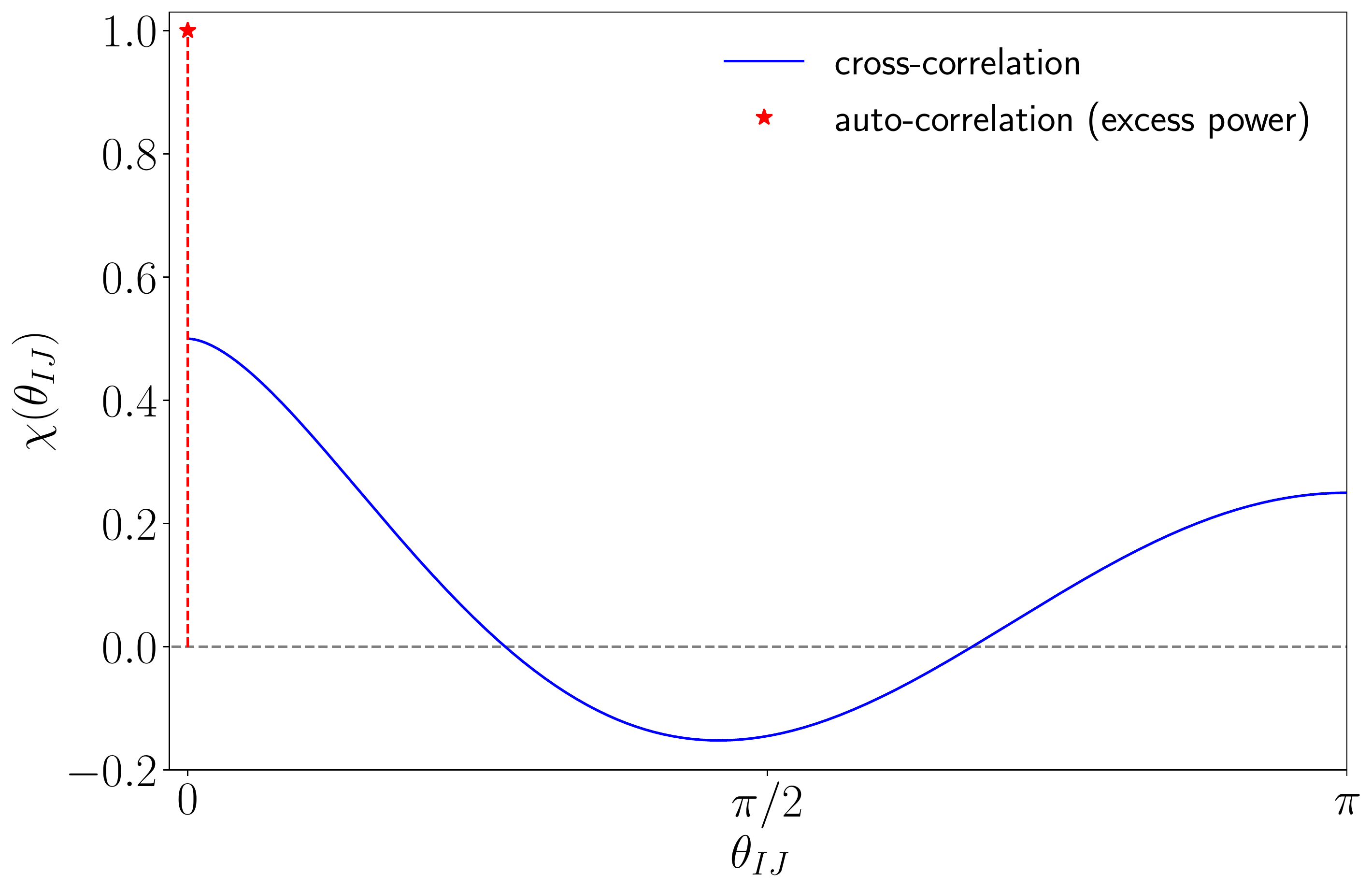}
    \end{center}
    \caption{%
    The Hellings-Downs function (blue curve) for cross-correlation GWB searches as a function of the angle $\theta_{IJ}$ between the two pulsars.
    The red point emphasises that the signal power is doubled for auto-correlation (i.e., excess power) searches with individual pulsars.
    }
    \label{fig:hellings-downs}
\end{figure}

We saw in section~\ref{sec:stochastic-searches} that the strongest GWB searches typically come from cross-correlating data between multiple detectors.
In the context of pulsar timing, this corresponds to measuring the ToAs of multiple pulsars and cross-correlating the timing residuals between them~\cite{Detweiler:1979wn}.
Such an experiment is called a \emph{pulsar timing array} (PTA).
The ORF in this case has a simple analytical expression, and is proportional to the \emph{Hellings-Downs curve}~\cite{Hellings:1983fr},
    \begin{equation}
    \label{eq:hellings-downs-curve}
        \chi(\theta_{IJ})\equiv\frac{1+\delta_{IJ}}{2}-\frac{1-\cos\theta_{IJ}}{2}\qty[\frac{1}{4}-\frac{3}{2}\ln\qty(\frac{1-\cos\theta_{IJ}}{2})],
    \end{equation}
    with $I$ and $J$ labelling different pulsars in the array.
This depends only on the angle $\theta_{IJ}$ between the two pulsars on the sky, and is maximised for pulsars in the same sky direction, $\theta_{IJ}=0$ (see figure~\ref{fig:hellings-downs}).
It is also possible to perform excess-power searches for each individual pulsar in the array by looking at the auto-correlation of their timing residuals; we see from the Kronecker delta term in equation~\eqref{eq:hellings-downs-curve} that this auto-correlation signal is a factor of two stronger than the cross-correlation signal for two different pulsars at the same sky location, $\theta_{IJ}=0$, as a single set of timing residuals is guaranteed to be perfectly in phase with itself.

In equation~\eqref{eq:ifo-phase-difference}, we used the small-antenna approximation to simplify the phase difference measured by an IFO.
One important difference between IFOs and PTAs is that the latter are \emph{never} in the small-antenna regime.
Since the distances to known pulsars are on the order of $d\sim0.1\text{--}100\,\mathrm{kpc}$, this regime would correspond to frequencies much lower than $1/d\sim10^{-10}\text{--}10^{-13}\,\mathrm{Hz}$; but GWs with these frequencies would cause much less than a single oscillation in the timing residuals over observational timescales, making them extremely challenging to measure.
Essentially, the issue is that the light-travel time across the \enquote{detector} is much longer than the observation time.
PTAs are therefore forced to measure GWs with frequencies $f\gg1/d$, which undergo many oscillations during the light-travel time from the pulsar to the Earth, causing the total GW perturbation~\eqref{eq:light-travel-time} to average to a smaller value.
As a result, PTAs are most sensitive at the lower end of their frequency band, giving a sharp PI curve which grows like $\sim f^5$~\cite{Thrane:2013oya} (see figure~\ref{fig:gwb-constraints}).
This curve is truncated at a minimum frequency corresponding to the inverse of the total observation time, $f_\mathrm{min}=1/T_\obs$, which for decade-long pulsar timing campaigns can reach below a few nHz.
The key observational target in this frequency band is the GWB from inspiralling SMBBHs~\cite{Jaffe:2002rt}, which has a characteristic $\sim f^{2/3}$ spectrum, just like the GWB from stellar-mass CBCs in the LIGO/Virgo band.
However, PTAs have also proven extremely useful in probing cosmological GW sources---particularly cosmic strings~\cite{Blanco-Pillado:2017rnf}, whose GWB spectrum typically contains a peak around this frequency range.

There are three key PTAs currently operating, all of which have been taking data for more than a decade: the Parkes Pulsar Timing Array (PPTA)~\cite{Hobbs:2013aka}, the European Pulsar Timing Array (EPTA)~\cite{Kramer:2013kea}, and the North American Nanohertz Observatory for Gravitational Waves (NANOGrav)~\cite{McLaughlin:2013ira}.
As well as conducting independent GW searches, these three collaborations also perform joint analyses as the International Pulsar Timing Array (IPTA)~\cite{Manchester:2013lea}.
The strongest reported upper limit on the nHz GWB comes from a 2015 PPTA analysis, setting $\Omega(f)<2.3\times10^{-10}$ at a reference frequency of $f=\mathrm{yr}^{-1}\approx32\,\mathrm{nHz}$~\cite{Shannon:2015ect,Lasky:2015lej} (see figure~\ref{fig:gwb-constraints}).
However, NANOGrav have recently claimed a detection of \emph{excess power} in the auto-correlation spectra of their timing residuals, with an amplitude and spectral shape that is consistent across all of the pulsars in the array~\cite{Arzoumanian:2020vkk}.
This \enquote{common process} signal is completely consistent with what one would expect from a GWB, leading several authors to explore how a spectrum with this amplitude and spectral tilt could be generated by various exotic cosmological GW sources~\cite{Ellis:2020ena,Blasi:2020mfx,Vaskonen:2020lbd,DeLuca:2020agl,Buchmuller:2020lbh,Ratzinger:2020koh,Vagnozzi:2020gtf,Neronov:2020qrl,Kuroyanagi:2020sfw,Bian:2020urb,Blanco-Pillado:2021ygr} (though the spectrum is also perfectly consistent with the GWB from SMBBHs~\cite{Middleton:2020asl}).
Consistent findings have very recently been reported by PPTA and EPTA in a re-analysis of their data~\cite{Goncharov:2021oub,Chen:2021rqp}, as well as in a joint analysis by the IPTA~\cite{Antoniadis:2022pcn}.

Confirming the NANOGrav common process signal as being due to the GWB will require a detection of the corresponding \emph{cross-correlation} signal; if the measured cross-correlation spectra are consistent with the distinctive shape of the Hellings-Downs curve as a function of the inter-pulsar angle $\theta_{IJ}$, then this will provide extremely strong evidence for GWs.
We know from equation~\eqref{eq:hellings-downs-curve} that the cross-correlation signal is a factor of two smaller than the auto-correlation signal detected by NANOGrav, meaning that the present situation is unsurprising, and also that the cross-correlation signal will likely be within reach once NANOGrav have collected a few more years of data.
We have therefore reached a very exciting point in the development of GW astronomy, where NANOGrav and other PTAs appear to be on the brink of making the first-ever detection of the GWB.
If this signal is genuine, then future PTA observations such as those planned for the Square Kilometre Array (SKA)~\cite{Janssen:2014dka} will detect it with extremely large SNR, allowing highly detailed reconstruction of the GWB spectrum in the nHz band.

\subsection{Cosmological bounds}
\label{sec:cosmological-gwb-bounds}

We saw in section~\ref{sec:gw-density-parameter} that the energy density of GWs in the Universe can be thought of in the same terms as the other key components of the cosmic energy inventory: matter, radiation, and dark energy.
In fact, since GWs redshift in exactly the same way as photons and other relativistic particles, the GW density parameter $\Omega_\gw(f)$ should really be treated as just one part of the total radiation density, in much the same way that CDM and baryons each contribute to the total matter density.
This radiation density parameter is crucially important for early-Universe physics, as it determines the expansion rate of the Universe at redshifts $z\gtrsim3400$.
As a result, if the GW energy density $\Omega_\gw(f)$ is large enough at early times, it can contribute significantly to the Hubble rate, and thereby leave an imprint on early-Universe observables such as the light element abundances predicted by big-bang nucleosynthesis (BBN), and the temperature and polarisation anisotropies of the CMB~\cite{Maggiore:1999vm}.
Of course, this only holds for primordial GWs that were already present at these early epochs, such as those from FOPTs; astrophysical sources have no impact on these probes, and even some cosmological sources do not contribute (e.g. the GWs emitted by cosmic string loops or PBH binaries at later times).

The resulting constraints are usually phrased in terms of the effective number of neutrino species, $N_\mathrm{eff}$, which is a key model parameter for both BBN and the CMB.
Since neutrinos have extremely small masses, they are relativistic at these early epochs, and therefore redshift in the same way as photons and GWs.
BBN and CMB analyses typically do not account for the GW energy density, meaning that a given GWB spectrum $\Omega_\gw(f)$ is effectively absorbed into the measured value of $N_\mathrm{eff}$, giving an extra contribution~\cite{Maggiore:1999vm} relative to the Standard Model prediction of $N_\mathrm{eff}^{(\mathrm{SM})}\approx3.045$~\cite{deSalas:2016ztq},
    \begin{equation}
    \label{eq:n_eff-constraints}
        N_\mathrm{eff}-N_\mathrm{eff}^{(\mathrm{SM})}=\frac{16}{7\Omega_\gamma}\qty[\frac{g_S(T_*)}{g_S(T_0)}]^{4/3}\int_{\ln f_*}^\infty\dd{(\ln f)}\Omega_\gw(f)\approx2.50\times10^5\int_{\ln f_*}^\infty\dd{(\ln f)}\Omega_\gw(f).
    \end{equation}
Here $\Omega_\gamma\approx5.43\times10^{-5}$ is the present-day photon density, and $g_S(T)$ is the number of entropic degrees of freedom in the Universe as a function of photon temperature $T$, with $T_*$ and $T_0$ the temperatures at the relevant early-Universe epoch and at the present day, respectively.

We see from equation~\eqref{eq:n_eff-constraints} that, rather than constraining the GWB spectrum in a certain frequency band like the IFOs and PTAs we have discussed above, $N_\mathrm{eff}$ measurements constrain the \emph{integrated} GW energy density across a very broad frequency range.
(This range is cut off at a minimum frequency $f_*\approx10^{-15}\,\mathrm{Hz}$ corresponding to the horizon size at the epoch of BBN, since super-horizon GWs are non-dynamical and thus do not contribute to the GW energy density.)
As a result, if we were to detect the GWB through its contribution to $N_\mathrm{eff}$, we would have no information about its frequency content, making it extremely difficult to infer the source of the signal.
In fact, it would be very difficult to attribute such a detection to the GWB in the first place, as there are numerous other mechanisms for changing the value of $N_\mathrm{eff}$ (e.g., by adding further relativistic particle species to the Standard Model).

Despite these caveats, $N_\mathrm{eff}$ measurements still provide very useful constraints on the GWB.
The most up-to-date analysis, which combines the Planck 2015 temperature and polarisation angular power spectra with information from BBN Deuterium abundances, CMB lensing, and baryon acoustic oscillation (BAO) data, gives~\cite{Pagano:2015hma}
    \begin{equation}
        \int_{\ln f_*}^\infty\dd{(\ln f)}\Omega_\gw(f)<2.6\times10^{-6},
    \end{equation}
Forecasts by the same authors predict this bound will improve to $1.7\times10^{-7}$ with future CMB and BAO measurements by COrE and EUCLID.
Both bounds are shown in figure~\ref{fig:gwb-constraints}, along with the PI curves from various IFOs and PTAs, and the predicted GWB spectra from a variety of astrophysical and cosmological sources.

There are several other cosmological constraints on the GWB that we do not mention here, primarily coming from CMB observations (e.g. through searches for $B$-mode polarisation patterns~\cite{Kamionkowski:1996zd,Seljak:1996gy,Ade:2015tva}, or GW-induced spectral distortions~\cite{Kite:2020uix}).
However, $N_\mathrm{eff}$ measurements provide the only cosmological constraint on GWs at the frequencies $f\gtrsim10^{-10}\,\mathrm{Hz}$ that we are interested in, which is why we have focused on them here.

\subsection{Other searches}
\label{sec:other-gw-searches}

There are many GW experiments that we have omitted from the discussion above.
In this section we very briefly highlight a few that are of particular interest, all of which are shown in figure~\ref{fig:gwb-constraints}.

The oldest method of GW detection, pioneered by \citet{Weber:1960zz,Weber:1969bz} in the 1960s with his \enquote{resonant bar} experiments, is to monitor the resonant frequencies of some macroscopic test object.
When acted upon by a GW with a frequency matching one of the object's normal modes, the stretching and squeezing action they induce excites vibrations in these modes which can be amplified and detected.
Numerous such resonant-mass experiments have been operated since the 1960s~\cite{Aguiar:2010kn}, but none has had sufficient sensitivity to place meaningful bounds on the GWB below the $\Omega(f)<1$ level required to prevent GWs from over-closing the Universe.
However, the same idea has been applied with success on a much larger scale, using the Earth itself as a resonant mass.
By monitoring the Earth's normal mode frequencies using gravimeter data, \citet{Coughlin:2014xua} placed an upper limit on the GWB in the mHz frequency band.
The corresponding PI curve has a minimum of $\Omega(f)\approx1.3\times10^{-2}$ at $f\approx1.6\,\mathrm{mHz}$.
Recently, a trio of similar experiments have been proposed in which gravimeters installed on the Lunar surface could be used to monitor the Moon's normal modes, providing a cleaner probe of GWs in the mHz frequency band~\cite{Harms:2020mma,Jani:2020gnz,Katsanevas:2020lsga}.

Another GW experiment which has successfully constrained the GWB below the $\Omega(f)<1$ level is the work of \citet{Armstrong:2003ay}, who carried out Doppler-tracking observations of the Cassini spacecraft.
The principle here is more similar to IFOs and PTAs in that it tracks the trajectories of photons in the presence of GWs; however, rather than measuring changes in the light-travel time, one attempts to measure changes in the photon frequency due to GW-induced Doppler shifts between the two test masses (this is essentially the time derivative of the perturbation to the light-travel time).
This analysis probed GWs in the $1\text{--}100\,\upmu\mathrm{Hz}$ band, with the corresponding PI curve reaching a minimum of $\Omega(f)\approx9.9\times10^{-3}$ at $f\approx3.9\,\upmu\mathrm{Hz}$.

Finally, a GW detection technique which is subject to growing interest in the community is \emph{atom interferometry}, in which, rather than using interference between photons to measure the phase difference between two spatial paths, one exploits the wavelike behaviour of matter on the quantum scale to measure interference between atoms.
While first-generation atom interferometry experiments such as AION~\cite{Badurina:2019hst} and MAGIS~\cite{Graham:2017pmn,Abe:2021ksx} are in their early stages and are not expected to have significant sensitivity to GWs, proposals for future km-scale experiments are forecast to provide extremely useful GWB constraints in the $0.1\text{--}1\,\mathrm{Hz}$ band.
In particular, the AION-km proposal has a forecast sensitivity of $\Omega(f)\approx2.7\times10^{-12}$ at $f\approx0.12\,\mathrm{Hz}$.


%% file: chapters/anisotropies.tex
\chapter{Anisotropies in the gravitational-wave background}
\label{chap:anisotropies}

In section~\ref{sec:gwb-intro} we introduced the \emph{gravitational-wave background}, and argued that, by giving us access to physics at high redshift, this provides one of our most powerful tools for probing cosmology with GW observations.
We also discussed the standard set of statistical assumptions that are usually made about the GWB: that it has no phase correlations between different sky directions, and that it is Gaussian, stationary, unpolarised, and isotropic.
This last assumption, however, is particularly restrictive and unrealistic in the context of the inhomogeneous Universe we inhabit, with its beautiful and fascinating array of structures on a vast range of scales.
In order to put the GWB on an equal footing with other cosmological observables, we must treat it as \emph{anisotropic} (though for extragalactic sources it is still \emph{statistically} isotropic---see section~\ref{sec:characterising-anisotropies} below).

Aside from providing a more faithful description of reality, one of the key motivations for treating the GWB as anisotropic is the \emph{information content} of the anisotropies, which is generally complementary to that of the isotropic component.
As we saw with the \enquote{Phinney formula}~\eqref{eq:phinney-formula} in section~\ref{sec:modelling-stochastic-backgrounds}, the isotropic GWB spectrum encodes the average rate, number density, and energy spectra of GW sources in the Universe, as well as the isotropic expansion rate of the FLRW background.
Meanwhile, the anisotropies, as well as depending on all of these ingredients, also encode the \emph{spatial clustering} of GW sources, and the inhomogeneities in the spacetime metric that the GWs encounter as they propagate towards us.
This situation is entirely analogous to other cosmological observables such as the CMB: while the isotropic component (i.e., the mean CMB temperature $T_0\approx2.73\,\mathrm{K}$) is sensitive to the expansion history of the homogeneous background, the anisotropies reveal a treasure trove of information about the origins and dynamical evolution of perturbations in the Universe's matter content and spacetime metric.

Of course, given the historical headstart electromagnetic astronomy has had relative to GW astronomy, the observational status of EM probes such as the CMB is vastly more advanced than that of the GWB.
As emphasised by \citet{Romano:2019yrj}, it took half a century for CMB observations to progress from Penzias and Wilson's initial discovery of the isotropic component to the exquisitely detailed maps of the anisotropies provided by Planck, whereas in 2021 we're yet to even detect the isotropic component of the GWB.
It therefore seems unlikely that studies of the GWB anisotropies will be able to compete with EM probes any time soon in terms of, say, constraining the values of cosmological parameters like the Hubble constant.
So why study GWB anisotropies?

The key motivation for GWB searches, in my view, lies in their \emph{discovery potential}.
While EM observables each probe a known set of sources described by well-understood physics, the GWB has the potential to unveil previously-unknown sources across an enormous redshift range, thereby giving us the opportunity to probe exotic new physics.
The sources we discussed in section~\ref{sec:gw-sources}---first-order phase transitions, cosmic strings, and primordial black holes---are just a few of the most prominent possibilities studied in the literature, any of which would revolutionise our understanding of fundamental physics if detected.
Studying the anisotropies associated with each of these GWB signals will likely be crucial in distinguishing and characterising the underlying sources, enabling us to extract as much cosmological information as possible from our GW data.
Another key motivation is that searches for GWB anisotropies are already underway, with LIGO/Virgo in particular providing impressively strong directional upper limits on the GWB intensity~\cite{Abbott:2016nwa,Abbott:2019gaw,Abbott:2021tss}.
As these searches become increasingly sensitive, it is important for us to understand what they should expect to see, so that the results can be properly interpreted.

For these reasons, there has been a recent surge of interest in modelling the expected spectrum of anisotropies in the GWB~\cite{Alba:2015cms,Bartolo:2019oiq,Bartolo:2019yeu,Bartolo:2019zvb,Bertacca:2019fnt,Capurri:2021zli,Conneely:2018wis,Contaldi:2016koz,Cusin:2017fwz,Cusin:2017mjm,Cusin:2018rsq,Cusin:2019jhg,Cusin:2019jpv,Geller:2018mwu,Jenkins:2018nty,Jenkins:2018uac,Jenkins:2018kxc,Jenkins:2019uzp,Kumar:2021ffi,Liu:2020mru,Pitrou:2019rjz,ValbusaDallArmi:2020ifo,Wang:2021djr,Bartolo:2022pez}, and their cross-correlations with EM observables~\cite{Adshead:2020bji,Braglia:2021fxn,Cusin:2017fwz,Cusin:2018rsq,Canas-Herrera:2019npr,Malhotra:2020ket,Ricciardone:2021kel}, as well as in developing associated data analysis techniques~\cite{Ain:2018zvo,Ali-Haimoud:2020iyz,Ali-Haimoud:2020ozu,Alonso:2020mva,Alonso:2020rar,Banagiri:2021ovv,Contaldi:2020rht,Hotinli:2019tpc,Jenkins:2019nks,Mentasti:2020yyd,Panda:2019hyg,Renzini:2021iim,Renzini:2018nee,Renzini:2018vkx,Renzini:2019vmt,Suresh:2020khz,Suresh:2021rsn,Yang:2020usq}.
While this body of work has included studies of the anisotropies associated with inflation~\cite{Alba:2015cms,Bartolo:2019oiq,Bartolo:2019yeu}, FOPTs~\cite{Geller:2018mwu,Kumar:2021ffi}, cosmic strings~\cite{Olmez:2011cg,Kuroyanagi:2016ugi,Jenkins:2018nty}, and PBHs~\cite{Bartolo:2019zvb,Wang:2021djr}, the primary focus has been on the \emph{astrophysical} GW background (AGWB) from compact binary coalescences.
As I see it, there are three key reasons why this signal is interesting:
    \begin{enumerate}
        \item Since the black holes and neutron stars that give rise to the AGWB are created via stellar evolution, we expect them to reside in galaxies and other star-forming environments.
            As a result, these CBCs act as \emph{tracers} of the inhomogeneous distribution of galaxies, and therefore of the cosmic matter distribution.
            By studying the statistical clustering of this background signal on the sky, we can therefore probe the large-scale structure of the Universe at late times, and gain insights into the nonlinear gravitational dynamics that governs these structures.
            While this LSS is already being measured by late-Universe EM observations such as galaxy surveys, there are several features of GWB searches which might allow us to access new and complementary information compared to these existing probes: namely that they automatically include the full sky and extend to arbitrarily high redshifts (in contrast with galaxy surveys, which typically cover only a fraction of the sky, and are inevitably limited in their depth), as well as using a different population of tracers of the matter distribution (i.e., CBC-hosting galaxies), whose clustering properties will, in general, differ from those used by existing cosmological probes.
        \item Aside from providing novel cosmological information, the GWB from CBCs also has the potential to reveal interesting \emph{astrophysical} information, principally about the rates and masses of CBCs and how these evolve throughout cosmic history.
            This could help us answer some of the many open questions about these systems; for example, what fraction of them originate from isolated binary evolution, and what fraction are dynamically assembled in dense stellar environments~\cite{Rodriguez:2015oxa}?
            Are there other important formation channels, e.g., in AGN discs~\cite{Bartos:2016dgn}?
            What are the mass distributions of BHs and NSs, and how do these depend on the stellar physics of their progenitors~\cite{Giacobbo:2018etu}?
            What are the properties of the host galaxies of these CBCs~\cite{Adhikari:2020wpn}, and how do these impact the CBCs themselves?
            While individually-resolved CBCs can help us try to pin down these questions at low redshifts, the properties of the AGWB and its anisotropies could help us paint a much more complete picture, covering the full history of star formation in the Universe.
        \item Perhaps most importantly of all, the AGWB is likely to be the loudest stochastic signal across a broad range of frequencies, and is therefore likely to the first component of the GWB that we detect.
            In order to dig out the cosmological signals which may be lurking beneath it, and thereby exploit the GWB to its full potential, we are thus obliged to understand the AGWB as thoroughly as possible, including a complete characterisation of its anisotropies.
    \end{enumerate}

With these motivations in mind, the goal of this chapter is to investigate the anisotropies in the AGWB, focusing on the background from stellar-mass CBCs that is the target of ground-based interferometers such as LIGO/Virgo.
We begin in section~\ref{sec:characterising-anisotropies} by introducing some of the necessary tools for studying the GWB as a random field on the sphere; this is almost entirely a review of pre-existing results, though the presentation is somewhat different to that which one usually encounters in the GWB literature.
In section~\ref{sec:agwb} we calculate predictions for the AGWB spectrum, starting with the isotropic component, before constructing a simulated map of the AGWB's anisotropies using data from the Millennium $N$-body simulation.
In section~\ref{sec:shot-noise} we examine the issue of \emph{shot noise} as an obstacle for AGWB searches, deriving the expected amplitude of this noise, before developing an optimal statistical methodology for mitigating its impact on measurements of the angular power spectrum.
We summarise our results in section~\ref{sec:anisotropies-summary}, and discuss some of the ways forward for future studies of GWB anisotropies.

\section{Characterising the anisotropic background}
\label{sec:characterising-anisotropies}

In section~\ref{sec:gwb-statistics}, we saw that the plane wave components of the GWB strain, $\tilde{h}_A(f,\vu*r)$, must be treated as random variables due to our ignorance of their incoherent phases, and that under the standard set of assumptions the statistics of these components are fully characterised by a single function of frequency.
In section~\ref{sec:gw-density-parameter}, we identified this function with the GW density parameter $\Omega(f)$.
Motivated by the discussion above, we now relax the assumption of perfect isotropy, so that the GWB strain moments become
    \begin{align}
    \begin{split}
    \label{eq:gwb-strain-moments-anisotropic}
        \ev{\tilde{h}_A(f,\vu*r)}=0,\qquad\ev{\tilde{h}_A(f,\vu*r)\tilde{h}^*_{A'}(f',\vu*r')}&=\frac{3\uppi H_0^2}{(2\uppi|f|)^3}\Omega(f,\vu*r)\delta_{AA'}\delta(f-f')\delta^{(2)}(\vu*r,\vu*r'),
    \end{split}
    \end{align}
    where we have promoted the GWB intensity to a field on the 2-sphere,
    \begin{equation}
        \Omega(f,\vu*r)\equiv\frac{1}{\rho_\mathrm{c}}\frac{\dd{\rho_\gw}}{\dd{(\ln f)}\dd[2]{\vu*r}}.
    \end{equation}
This has units of inverse steradians, and when integrated gives the isotropic GWB spectrum from before,
    \begin{equation}
        \Omega(f)=\int_{S^2}\dd[2]{\vu*r}\Omega(f,\vu*r).
    \end{equation}

\subsection{The gravitational-wave background as a random field}

Previously, we treated $\Omega(f)$ as a deterministic (i.e., non-random) quantity, fixed by the average properties of the cosmic population of GW sources.
However, when studying the distribution of GW intensity on the sky, $\Omega(f,\vu*r)$, we are faced once again with a fundamental degree of uncertainty, similar to our ignorance about the phases of the strain components $\tilde{h}_A(f,\vu*r)$.
In this case, the uncertainty is due to our lack of knowledge about the angular positions of individual GW sources on the sky.
For example, while we can attempt to deterministically model the average rate of CBC signals that contribute to the GWB, there is no hope of modelling where on the sky these signals should come from without a complete knowledge of the positions of all the BBHs and BNSs in the Universe.
We are therefore forced to treat $\Omega(f,\vu*r)$ itself as essentially random.\footnote{%
    Note that we now have two distinct levels of randomness in the GWB strain: the randomness of the incoherent phases, and the randomness of the intensity as a function of sky direction.
    It is important to stress that these are two logically distinct sources of randomness, with the former being tied to the random GW emission times of different sources in the Universe, and the latter tied to the spatial clustering of those sources due to LSS.
    We will return to this point in section~\ref{sec:shot}.}

Of course, the above argument doesn't hold for GW sources which inhabit a preferred set of directions on the sky.
For example, the stochastic signal from white dwarf binaries that will be observed by LISA will come primarily from the galactic plane, and it should therefore be possible to model the angular distribution of this signal using the known distribution of stars within the galaxy.
However, we are interested here in \emph{extragalactic} sources, for which there are no preferred directions.
(Or at least, if there are preferred directions, these depend on information we do not have access to, such as the locations of CBCs throughout the Universe.)
We therefore assume that the GWB is \emph{statistically isotropic}, such that all ensemble-averaged functions of $\Omega(f,\vu*r)$ are invariant under sky rotations.
In particular, this implies that the mean of the GWB intensity field is independent of direction, and that the covariance between any two field points depends only on the invariant angle between those points,
    \begin{equation}
    \label{eq:gwb-moments-frequency}
        \ev{\Omega(f,\vu*r)}\equiv \bar{\Omega}(f),\qquad\Cov[\Omega(f,\vu*r),\Omega(f,\vu*r')]\equiv C(f,\vu*r\vdot\vu*r').
    \end{equation}
If the GWB is a Gaussian random field, then these two moments are sufficient to fully describe its statistics.

Note that we haven't written down the most general covariance function here, which would specify what happens when we cross-correlate two different frequencies, $f$ and $f'$, as well as two different sky directions.
In fact, this is not necessary, since we generally expect that it is possible to factorise the GWB into a frequency spectrum (which is deterministic) and an angular intensity map (which is a random field),
    \begin{equation}
    \label{eq:factorisation-assumption}
        \Omega(f,\vu*r)=\mathcal{H}(f)\Omega(\vu*r).
    \end{equation}
To see why this might be a reasonable assumption, it is useful to focus for a moment on the AGWB in the LIGO/Virgo frequency band.
In this case, the frequency spectrum encodes the characteristic $\sim f^{2/3}$ emission of each of the individual CBCs as they \enquote{chirp} up and approach merger, while the angular intensity encodes the distribution of CBCs throughout the Universe.
Since each CBC spans the entire frequency range of interest while it is in-band, there is no distinction between the angular distribution of \enquote{high-frequency CBCs} and \enquote{low-frequency CBCs}---they are just the same sources, with the same distribution on the sky, and we can therefore capture this distribution by factorising out the frequency spectrum as in equation~\eqref{eq:factorisation-assumption}.\footnote{%
    This argument does not hold in all settings.
    For example, many of the CBCs that will be observed by LISA will undergo negligible frequency evolution over observational timescales (c.f. equation~\eqref{eq:time-to-merger}), meaning that there \emph{is} then a distinction between the CBC populations emitting at different frequencies, and this could in principle lead to interesting frequency-dependent effects in the two-point statistics of the GWB.
    We do not investigate this further here.}
We therefore suppress the frequency arguments in equation~\eqref{eq:gwb-moments-frequency}, and focus on the angular intensity $\Omega(\vu*r)$ at some reference frequency.

\subsection{Spherical harmonics}

The statistical invariance of the GWB under rotations means that it's not actually very useful to think about $\Omega$ as a function of sky direction, since its value at any given direction $\vu*r$ is entirely dependent on our arbitrary choice of spherical coordinates.
This is exactly the same reason why it's not useful to describe the GWB strain in the time domain, as statistical time-translation invariance (i.e., stationarity) means that specifying the value of the strain at any given time $t$ is not very informative.
Instead, we treat the strain in the frequency domain, projecting $h_A(t,\vu*r)$ onto the Fourier modes $\rme^{2\uppi\rmi ft}$, which provide a complete and orthogonal set of basis functions on the real line $\mathbb{R}$.
Stationarity then means that the phases of the Fourier components $\tilde{h}_A(f,\vu*r)$ are uninformative, but their amplitudes \emph{are} useful, allowing us to describe the statistics of the GWB strain over different timescales (i.e., different frequencies) in terms of the power spectrum (which is proportional to $\Omega(f,\vu*r)$).
Ideally we would like to do something analogous here, projecting $\Omega(\vu*r)$ onto some set of basis functions on the sphere $S^2$, and thereby describing the statistics of the field on different \emph{angular} scales.

\begin{figure}[t!]
    \begin{center}
        \includegraphics[height=0.667\textwidth,angle=-90]{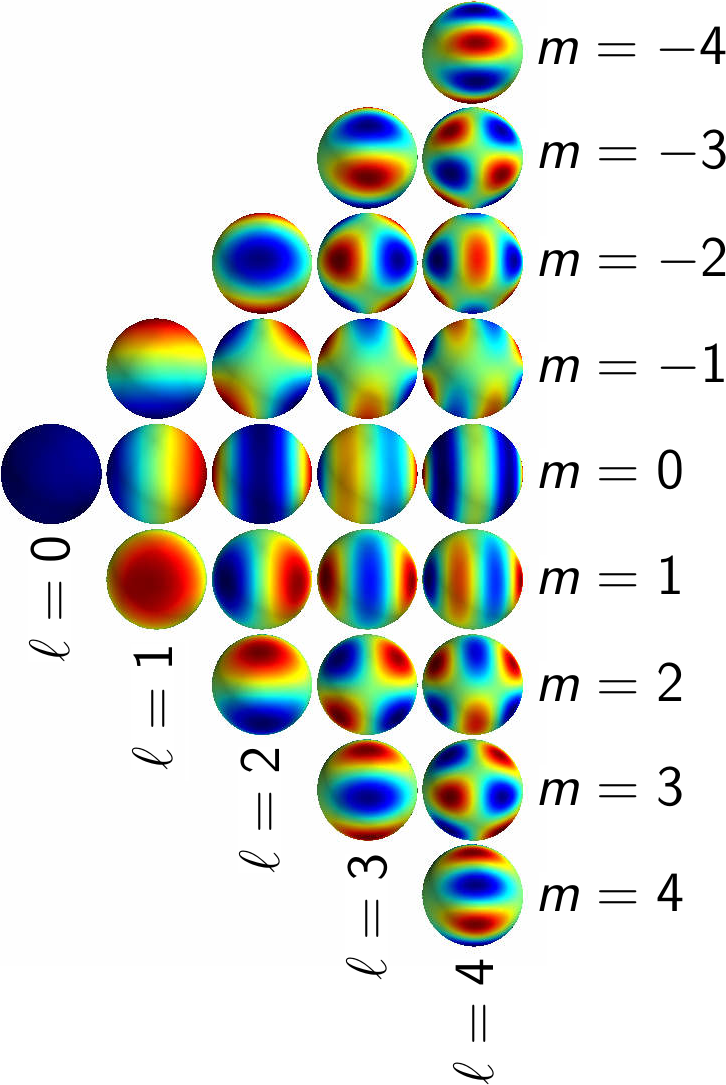}
    \end{center}
    \caption{%
    Visualisations of the real parts of the first few spherical harmonics, $\Re{Y_{\ell m}(\vu*r)}$, for multipoles $\ell\le4$.
    The polar axis ($\theta=0,\uppi$) is oriented vertically.
    }
    \label{fig:spherical-harmonics}
\end{figure}

The obvious candidate for these basis functions is the set of \emph{spherical harmonics} $Y_{\ell m}$ (see figure~\ref{fig:spherical-harmonics}), which are defined as solutions to the equation $[\nabla^2+\ell(\ell+1)]g(\vu*r)=0$, where $\ell$ is any non-negative integer.
Note that this is analogous to the equation $(\partial_t^2+\omega^2)g(t)=0$ obeyed by the Fourier modes $\rme^{\rmi\omega t}$, with $\ell$ here playing the r\^ole of \enquote{angular frequency}.
One important difference that comes with the higher dimensionality is an increase in the number of independent solutions of the same frequency: for any given $\ell$ there are $2\ell+1$ linearly independent harmonics $Y_{\ell m}(\vu*r)$, which we label with the index $m\in\{-\ell,-\ell+1,\ldots,+\ell\}$ such that $|m|\le\ell$.
These form a complete basis on the 2-sphere $S^2$, and are orthonormal,
    \begin{equation}
    \label{eq:spherical-harmonics-orthonormality}
        \int_{S^2}\dd[2]{\vu*r}Y_{\ell m}(\vu*r)Y^*_{\ell'm'}(\vu*r)=\delta_{\ell\ell'}\delta_{mm'}.
    \end{equation}
The first few spherical harmonics are
    \begin{align}
    \begin{split}
    \label{eq:first-few-spherical-harmonics}
        Y_{00}(\vu*r)&=\sqrt{\frac{1}{4\uppi}},\\
        Y_{10}(\vu*r)&=\sqrt{\frac{3}{4\uppi}}\cos\theta,\qquad Y_{11}(\vu*r)=-\sqrt{\frac{3}{8\uppi}}\sin\theta\rme^{\rmi\phi},\\
        Y_{20}(\vu*r)&=\sqrt{\frac{5}{16\uppi}}(3\cos^2-1),\qquad Y_{21}(\vu*r)=-\sqrt{\frac{15}{8\uppi}}\sin\theta\cos\theta\rme^{\rmi\phi},\qquad Y_{22}(\vu*r)=\sqrt{\frac{15}{32\uppi}}\sin^2\theta\rme^{2\rmi\phi},
    \end{split}
    \end{align}
    with the expressions for negative values of $m$ given by the identity
    \begin{equation}
    \label{eq:spherical-harmonic-complex-conjugate}
        Y^*_{\ell m}(\vu*r)=(-)^mY_{\ell,-m}(\vu*r).
    \end{equation}

We therefore decompose the GWB intensity into spherical harmonics,
    \begin{equation}
    \label{eq:shcs-definition}
        \Omega_{\ell m}\equiv\int_{S^2}\dd[2]{\vu*r}Y^*_{\ell m}(\vu*r)\Omega(\vu*r),
    \end{equation}
    allowing us to replace a real random field $\Omega:S^2\to\mathbb{R}$ with an infinite set of complex random numbers $\Omega_{\ell m}\in\mathbb{C}$, which we call the spherical harmonic components (SHCs) of the field.
Note that the fact that $\Omega(\vu*r)$ is real, combined with equation~\eqref{eq:spherical-harmonic-complex-conjugate}, implies that $\Omega^*_{\ell m}=(-)^m\Omega_{\ell,-m}$.
We can invert equation~\eqref{eq:shcs-definition} to write
    \begin{equation}
    \label{eq:inverse-shc-transform}
        \Omega(\vu*r)=\sum_{\ell m}Y_{\ell m}(\vu*r)\Omega_{\ell m},
    \end{equation}
    where we use the shorthand notation $\sum_{\ell m}\equiv\sum_{\ell=0}^\infty\sum_{m=-\ell}^{+\ell}$ for brevity.
Equations~\eqref{eq:shcs-definition} and~\eqref{eq:inverse-shc-transform} are the spherical analogues of the Fourier transform and its inverse.
One important difference with frequency spectra on $\mathbb{R}$ is that the spherical harmonic spectrum~\eqref{eq:shcs-definition} is discrete rather than continuous, which we can understand as a consequence of $S^2$ being compact (i.e., having finite volume).

\subsection{The angular power spectrum}

Now that we have decomposed the GWB into spherical harmonics, we would like to replace the field moments in equation~\eqref{eq:gwb-moments-frequency} with the first and second moments of the SHCs, writing these in terms of the constant $\bar{\Omega}$ and the function $C(\vu*r\vdot\vu*r')$.
We can obtain the first moment pretty much immediately by writing
    \begin{equation}
        \ev{\Omega_{\ell m}}=\bar{\Omega}\int_{S^2}\dd[2]{\vu*r}Y^*_{\ell m}(\vu*r)=\sqrt{4\uppi}\bar{\Omega}\delta_{\ell0}\delta_{m0},
    \end{equation}
    where in the second equality we have used the orthonormality condition~\eqref{eq:spherical-harmonics-orthonormality} with the $\ell=m=0$ harmonic from equation~\eqref{eq:first-few-spherical-harmonics}.
This tells us that all of the SHCs have zero mean, except for the monopole $\Omega_{00}$.

For the second moment, it is useful to note that since the dot product $\vu*r\vdot\vu*r'$ always lies in the interval $[-1,1]$, the function $C(\vu*r\vdot\vu*r')$ can be decomposed into Legendre polynomials $P_\ell(\vu*r\vdot\vu*r')$, which form a complete basis for functions on this interval.
We therefore write
    \begin{equation}
    \label{eq:covariance-multipole-expansion}
        C(\vu*r\vdot\vu*r')\equiv\sum_{\ell=0}^\infty\frac{2\ell+1}{4\uppi}C_\ell P_\ell(\vu*r\vdot\vu*r'),
    \end{equation}
    where the coefficients $C_\ell$ are a set of real numbers describing the decomposition of the covariance function in this basis, and the normalisation factor $(2\ell+1)/(4\uppi)$ is chosen such that we can write the inverse expression as
    \begin{equation}
    \label{eq:C_ell-multipole-integral}
        C_\ell=\int_{S^2}\dd[2]{\vu*r'}P_\ell(\vu*r\vdot\vu*r')C(\vu*r\vdot\vu*r'),
    \end{equation}
    using the orthogonality relation for the Legendre polynomials, $\int_{-1}^{+1}\dd{x}P_\ell(x)P_{\ell'}(x)=2\delta_{\ell\ell'}/(2\ell+1)$.

The Legendre polynomials are related to the spherical harmonics of the same angular scale $\ell$ by the \emph{addition theorem}~\cite{Durrer:2008eom}
    \begin{equation}
        P_\ell(\vu*r\vdot\vu*r')=\sum_{m=-\ell}^{+\ell}\frac{4\uppi}{2\ell+1}Y_{\ell m}(\vu*r)Y^*_{\ell m}(\vu*r'),
    \end{equation}
    which allows us to write equation~\eqref{eq:covariance-multipole-expansion} as
    \begin{equation}
        C(\vu*r\vdot\vu*r')=\sum_{\ell m}C_\ell Y_{\ell m}(\vu*r)Y_{\ell m}(\vu*r').
    \end{equation}
We can now derive the second moment of the SHCs,\footnote{%
    Note that here, and throughout, we define the covariance of two complex random variables with a complex conjugate on the second argument, i.e., $\Cov[X,Y]\equiv\ev{XY^*}-\ev{X}\ev{Y^*}=\Cov[Y,X]^*$.
    This ensures that the variance is always real, since $\Var[X]\equiv\Cov[X,X]=\ev*{|X|^2}-|\ev{X}|^2$.}
    \begin{align}
    \begin{split}
    \label{eq:shc-covariance-derivation}
        \Cov[\Omega_{\ell m},\Omega_{\ell'm'}]&=\int_{S^2}\dd[2]{\vu*r}\int_{S^2}\dd[2]{\vu*r'}Y_{\ell m}(\vu*r)Y^*_{\ell'm'}(\vu*r')C(\vu*r\vdot\vu*r')\\
        &=\sum_{LM}C_L\qty[\int_{S^2}\dd[2]{\vu*r}Y_{\ell m}(\vu*r)Y^*_{LM}(\vu*r)]\qty[\int_{S^2}\dd[2]{\vu*r'}Y^*_{\ell'm'}(\vu*r')Y_{LM}(\vu*r')]\\
        &=\sum_{LM}C_L\delta_{\ell L}\delta_{mM}\delta_{\ell'L}\delta_{m'M}=C_\ell\delta_{\ell\ell'}\delta_{mm'},
    \end{split}
    \end{align}
    which shows us that the SHCs $\Omega_{\ell m}$ are all uncorrelated from each other (which, if they are Gaussian, further implies that they are statistically independent).
This illustrates the usefulness of going to spherical harmonic space, since having a diagonal covariance matrix like equation~\eqref{eq:shc-covariance-derivation} is extremely convenient for both theory and data analysis.
Compare with the covariance matrix in field space (also called \enquote{pixel space}, since in practice the sky is usually partitioned into a finite number of pixels), $\mathsf{C}_{ij}\equiv C(\vu*r_i\vdot\vu*r_j)$, in which the off-diagonal terms are generically nonzero (otherwise the field would have no spatial correlations at all), making it much more cumbersome to work with.
Again, the analogy with Fourier analysis is clear: one of the key strengths of working in Fourier space is that modes of different frequency are usually uncorrelated, giving a diagonal covariance matrix.

We have thus shown that we can replace equation~\eqref{eq:gwb-moments-frequency} with\footnote{%
    Here we replace $\delta_{\ell0}\delta_{m0}$ with just $\delta_{\ell0}$.
    We can always do this, since having $\ell=0$ necessarily implies that $m=0$, in order to satisfy $|m|\le\ell$.}
    \begin{equation}
    \label{eq:shc-moments}
        \ev{\Omega_{\ell m}}=\sqrt{4\uppi}\bar{\Omega}\delta_{\ell0},\qquad\Cov[\Omega_{\ell m},\Omega_{\ell'm'}]=C_\ell\delta_{\ell\ell'}\delta_{mm'},
    \end{equation}
    which is analogous to equation~\eqref{eq:gwb-strain-moments-anisotropic} for the Fourier components of the GWB strain.
In particular, the set of numbers $C_\ell$ is analogous to the Fourier power spectrum; we call it the \emph{angular power spectrum}.
One important difference is that the monopole $\Omega_{00}$ has a nonzero mean value, as opposed to the Fourier components, which are all zero-mean.
This is because, while we can set the mean value of the strain $\ev*{h_{ij}(t)}$ to zero by a coordinate transformation, the mean GWB intensity $\bar{\Omega}$ must always be positive---otherwise there would be no GWB to observe.

Intuitively, the angular power spectrum $C_\ell$ tells us about the amplitude of statistical fluctuations in the intensity field on angular scales $\theta\sim\uppi/\ell$.
We can make this notion more precise by using equation~\eqref{eq:covariance-multipole-expansion} to write the variance of a single field point $\vu*r$ as\footnote{%
    At first glance, it seems like there should be a factor of $\ell(\ell+1/2)$ here rather than $\ell(\ell+1)$.
    However, one can show that $(\ell+1/2)/[\ln(\ell+1)-\ln\ell]\simeq\ell(\ell+1)+\order*{1}$ for $\ell\gg1$, so this is actually the consistent choice when approximating the sum as an integral.
    In any case, the different is negligible in the $\ell\gg1$ regime where this approximation holds.}
    \begin{equation}
        \Var[\Omega(\vu*r)]\equiv C(\vu*r\vdot\vu*r)=\sum_{\ell=0}^\infty\frac{2\ell+1}{4\uppi}C_\ell\approx\int\dd{(\ln\ell)}\frac{\ell(\ell+1)}{2\uppi}C_\ell,
    \end{equation}
    where we have used the fact that $P_\ell(1)=1$.
The quantity $\ell(\ell+1)C_\ell/(2\uppi)$ thus approximately describes the field variance associated with a given logarithmic bin in $\ell$, and is therefore the quantity that is usually presented in, e.g., CMB analyses, rather than the bare angular power spectrum $C_\ell$.
Comparing with the GWB strain once again, $\ell(\ell+1)C_\ell/(2\uppi)$ is analogous to the density parameter $\Omega(f)$, while $C_\ell$ is analogous to the strain power spectrum $S(f)\sim f^{-3}\Omega(f)$.
Having $\ell(\ell+1)C_\ell=\mathrm{constant}$ corresponds to a scale-invariant angular power spectrum, while having $C_\ell=\mathrm{constant}$ corresponds to a white spectrum.

\subsection{Estimating the angular power spectrum}
\label{sec:estimating-C_ells}

Suppose we detect the GWB and measure some set of SHCs, $\Omega_{\ell m}$.
What can we say about the angular power spectrum based on these measurements?
(We ignore here the problem of actually inferring the SHCs from noisy strain data, and the associated uncertainties in their values---see, e.g., \citet{Thrane:2009fp} or \citet{Romano:2016dpx} for thorough treatments of these topics.)

The SHCs that we observe represent just a single random realisation from the distribution parameterised by the angular power spectrum.
As such, we cannot measure these distribution parameters directly; the best we can do is construct some function of the observed SHCs whose ensemble-averaged value is related to the parameters we're interested in.
We call such a function an \emph{estimator}.
If the ensemble average of the estimator is equal to the parameter we're trying to infer, then we call it \emph{unbiased}.

From equation~\eqref{eq:shc-moments}, we see immediately that we can define the naive estimator $\hat{C}_\ell^{(m)}\equiv|\Omega_{\ell m}|^2$, which is unbiased for all $\ell>0$,
    \begin{equation}
    \label{eq:naive-estimator}
        \ev{\hat{C}_\ell^{(m)}}=\Var[\Omega_{\ell m}]+\qty|\ev{\Omega_{\ell m}}|^2=C_\ell+4\uppi\bar{\Omega}^2\delta_{\ell0}.
    \end{equation}
If we knew the value of $\bar{\Omega}$, we could make this unbiased for all $\ell$ by defining $\hat{C}_\ell^{(m)}\equiv|\Omega_{\ell m}-\sqrt{4\uppi}\bar{\Omega}\delta_{\ell0}|^2$; however, $\bar{\Omega}$ is not a known quantity, but another parameter to be estimated.

For any given set of measured SHCs, the estimator $\hat{C}_\ell^{(m)}$ will not be exactly equal to its average value, but will be subject to some random scatter.
We can quantify this by calculating the variance of the estimator,
    \begin{align}
    \begin{split}
    \label{eq:naive-estimator-variance}
        \Var[\hat{C}_\ell^{(m)}]&=\ev{|\Omega_{\ell m}|^4}-\ev{|\Omega_{\ell m}|^2}^2=\qty[2\ev{|\Omega_{\ell m}|^2}^2+\ev{(\Omega_{\ell m})^2}\ev{(\Omega^*_{\ell m})^2}]-\ev{|\Omega_{\ell m}|^2}^2\\
        &=\ev{|\Omega_{\ell m}|^2}^2+(-)^{2m}\ev{\Omega_{\ell m}\Omega^*_{\ell,-m}}\ev{\Omega_{\ell,-m}\Omega^*_{\ell m}}=(1+\delta_{m0})C_\ell^2,
    \end{split}
    \end{align}
    where we have assumed Gaussianity in order to write the fourth moment in terms of second moments.
We see that this is a rather poor estimator, as its standard deviation is equal to (or greater than, for $m=0$) the mean value we want to extract, meaning that we can typically expect our estimate to be off by $100\%$.

Normally with a problem like this we would try to make as many independent measurements as possible, and combine all of these to create an estimator with lower variance.
More explicitly, if we have $N$ independent and identically distributed (i.i.d.) unbiased estimators $\hat{X}_i$, $i=1,2,\ldots,N$ for some parameter $X$, each with some variance $\sigma^2$, then we can define an unbiased mean estimator $\hat{X}\equiv(1/N)\sum_{i=1}^N\hat{X}_i$ whose variance is reduced by a factor of $1/N$,
    \begin{equation}
        \Var[\hat{X}]=\frac{1}{N^2}\sum_{i=1}^N\sum_{j=1}^N\Cov[\hat{X}_i,\hat{X}_j]=\frac{1}{N^2}\sum_{i=1}^N\sum_{j=1}^N\sigma^2\delta_{ij}=\frac{\sigma^2}{N}.
    \end{equation}
In cosmological contexts, taking more measurements is not an option, as nature presents us with a single random realisation of LSS.
This results in a certain irreducible uncertainty when trying to estimate ensemble-averaged quantities like the angular power spectrum from this single realisation.
We call this uncertainty \emph{cosmic variance}.

Despite this fundamental limitation, however, we can still do significantly better than the naive estimator~\eqref{eq:naive-estimator}.
The key here is to note that, while we only have one random realisation of each SHC $\Omega_{\ell m}$, the $(2\ell+1)$ different $m$ indices for a given $\ell$ index are i.i.d., and therefore act as independent measurements of the same $C_\ell$ multipole in the angular power spectrum.
We can therefore average over these to define an improved estimator,
    \begin{equation}
    \label{eq:standard-estimator}
        \hat{C}_\ell\equiv\frac{1}{2\ell+1}\sum_m|\Omega_{\ell m}|^2,
    \end{equation}
    whose variance is
    \begin{align}
    \begin{split}
    \label{eq:standard-estimator-variance}
        \Var[\hat{C}_\ell]&=\frac{1}{(2\ell+1)^2}\sum_m\sum_{m'}\Cov[|\Omega_{\ell m}|^2,|\Omega_{\ell m'}|^2]\\
        &=\frac{1}{(2\ell+1)^2}\sum_m\sum_{m'}\qty[\ev{\Omega_{\ell m}\Omega^*_{\ell m'}}\ev{\Omega^*_{\ell m}\Omega_{\ell m'}}+\ev{\Omega_{\ell m}\Omega_{\ell m'}}\ev{\Omega^*_{\ell m}\Omega^*_{\ell m'}}]\\
        &=\frac{1}{(2\ell+1)^2}\sum_m\sum_{m'}C_\ell^2(\delta_{mm'}+\delta_{m,-m'})=\frac{2}{2\ell+1}C_\ell^2.
    \end{split}
    \end{align}

This is significantly better than the variance in equation~\eqref{eq:naive-estimator-variance}, particularly at higher multipoles $\ell\gg1$.
In fact, if we assume the intensity field to be Gaussian, we can prove that the improved estimator~\eqref{eq:standard-estimator} is the best that we can construct, in the sense that it is the minimum-variance unbiased estimator (MVUE) for the angular power spectrum.
To show this, we use the fact that the variance of any unbiased estimator $\hat{\theta}(x)$ for a parameter $\theta$ of a probability distribution $p(x|\theta)$ obeys the \emph{Cram\'er-Rao bound}~\cite{Kay:1993},
    \begin{equation}
    \label{eq:cramer-rao-bound}
        \Var[\hat{\theta}]\ge-\ev{\pdv[2]{\mathcal{L}}{\theta}}^{-1},
    \end{equation}
    where $\mathcal{L}(x|\theta)\equiv\ln p(x|\theta)$ is the log-likelihood for the distribution.
In our case, the Gaussian log-likelihood for the SHCs is
    \begin{equation}
    \label{eq:gaussian-likelihood}
        \mathcal{L}(\Omega_{\ell m}|\bar{\Omega},C_\ell)=-\frac{1}{2}\sum_{\ell m}\qty[\ln(2\uppi C_\ell)+\frac{|\Omega_{\ell m}-\sqrt{4\uppi}\bar{\Omega}\delta_{\ell0}|^2}{C_\ell}],
    \end{equation}
    so equation~\eqref{eq:cramer-rao-bound} becomes
    \begin{align}
    \begin{split}
        \Var[\hat{C}_\ell]\ge\ev{\sum_m\qty[\frac{|\Omega_{\ell m}-\sqrt{4\uppi}\bar{\Omega}\delta_{\ell0}|^2}{C_\ell^3}-\frac{1}{2C_\ell^2}]}^{-1}=\qty(\sum_m\frac{1}{2C_\ell^2})^{-1}=\frac{2}{2\ell+1}C_\ell^2,
    \end{split}
    \end{align}
    which is identical to what we found in equation~\eqref{eq:standard-estimator-variance}.
This corresponds to the irreducible uncertainty on the angular power spectrum due to cosmic variance.
We thus see that the estimator~\eqref{eq:standard-estimator} saturates this bound, meaning that it must be the MVUE for all $\ell>0$.

For the monopole $\ell=0$ we still have the issue of the estimator being biased, due to our inability to subtract off the unknown mean intensity $\bar{\Omega}$.
All that we can do with the monopole is construct an estimator for the mean intensity, $\hat{\Omega}\equiv\Omega_{00}/\sqrt{4\uppi}$; this is unbiased, $\ev*{\hat{\Omega}}=\bar{\Omega}$, and has variance $\Var[\hat{\Omega}]=C_0/(4\uppi)$.
The problem here is that we are attempting to infer both the mean and the variance of a population (the ensemble of all possible realisations of the GWB monopole) with only a single sample (the GWB monopole that we observe).
This is impossible to do, since we have no way of knowing whether the single realisation of the monopole that we observe is near to the cosmic mean value, or whether it is out in the tails of the distribution (which could be the case if, e.g., we live near the centre of a cosmological under-/over-density of GW sources).
We therefore neglect $C_0$ in what follows, focusing on the higher multipoles $\ell>0$.

\subsection{The gravitational-wave background in an inhomogeneous Universe}
\label{sec:inhomogeneous}

We have now discussed in some detail how to treat the anisotropic GWB intensity as a random field on the sphere, but so far we are lacking an anisotropic version of the Phinney formula~\eqref{eq:phinney-formula} to predict what this field should look like for a given population of GW sources.

To write down such a formula, we consider three sources of anisotropies in the GWB.
First, we treat the comoving number density of sources, $n$, as inhomogeneous, writing $n(t,\vb*x)=\bar{n}(t)[1+\delta_n(t,\vb*x)]$, where $\delta_n$ is the number density contrast, and $\bar{n}$ is the homogeneous mean number density.
(Since the comoving rate density $\mathcal{R}$ is proportional to $n$, we therefore also have $\mathcal{R}=\bar{\mathcal{R}}(1+\delta_n)$, where $\bar{\mathcal{R}}$ is the homogeneous mean rate density.)
Second, we assume the emitted GWs propagate through a FLRW spacetime with scalar perturbations, whose line element in the Newtonian gauge is
    \begin{equation}
        \dd{s}^2=a^2(\eta)\qty[-(1+2\psi(\eta,\vb*x))\dd{\eta}^2+(1-2\phi(\eta,\vb*x))\dd{\vb*x}\vdot\dd{\vb*x}],
    \end{equation}
    where $\psi$ and $\phi$ are the Bardeen potentials~\cite{Bardeen:1980kt}, and $\eta$ is conformal time, $\dd{\eta}\equiv\dd{t}/a(t)$.
Finally, we allow the sources and the observer to have peculiar velocities with respect to the cosmic rest frame, which follow some cosmological velocity field $\vb*v(\eta,\vb*x)$.
The GWB intensity is then given to first order in the perturbations by\footnote{
    See \citet{Bertacca:2019fnt} for a thorough derivation of a more complete version of this formula, as well as \citet{Bellomo:2021mer} for a numerical implementation of this.}~\cite{Contaldi:2016koz,Cusin:2017fwz,Jenkins:2018nty}
    \begin{align}
    \begin{split}
    \label{eq:anisotropic-phinney-formula}
        \Omega(f,\vu*r)=\frac{2G}{3H_0^2}\int^{\eta_0}_0\dd{\eta}a^2(\eta)&\int\dd{\vb*\zeta}\bar{\mathcal{R}}(\eta,\vb*\zeta)\dv{E}{(\ln f_\mathrm{s})}\\
        &\times\qty[1+\delta_n-4\psi_0+5\psi+\vu*r\vdot(4\vb*v_0-3\vb*v)+2\int^{\eta_0}_\eta\dd{\eta'}\partial_\eta(\psi+\phi)],
    \end{split}
    \end{align}
    where \enquote{$0$} subscripts indicate quantities evaluated at the observer's position $\vb*x_0$, while all other perturbations are evaluated along the line of sight, $\vb*x(\eta,\vu*r)=\vb*x_0+(\eta_0-\eta)\vu*r$.
We recall that $\vb*\zeta$ denotes the set of parameters characterising each CBC signal (masses, spins, etc.), and that $\dv*{E}{(\ln f_\mathrm{s})}$ is the dimensionless GW energy spectrum as a function of the source-frame frequency $f_\mathrm{s}$, which in this context is related to the observed frequency $f$ by
    \begin{equation}
    \label{eq:source-frame-frequency}
        f_\mathrm{s}=\frac{f}{a}\,\qty[1+\psi_0-\psi-\vu*r\vdot(\vb*v_0-\vb*v)+\int^{\eta_0}_\eta\dd{\eta'}\partial_\eta(\psi+\phi)].
    \end{equation}

Equation~\eqref{eq:anisotropic-phinney-formula} is analogous to the standard approach for calculating the scalar perturbations in the CMB temperature field, with the same three key contributions~\cite{Sachs:1967er,Durrer:2008eom}: a Sachs-Wolfe (SW) term due to gravitational redshifting of the photons/gravitons, expressed in terms of the values of the Bardeen potential $\psi$ at the source and at the observer; a Doppler term written in terms of the peculiar velocities of the source and the observer projected along the line of sight; and an integrated Sachs-Wolfe (ISW) term encoding the time-evolution of both Bardeen potentials, $\psi$ and $\phi$.
An important difference arises, however, in the density contrast term $\delta_n$.
While the CMB is emitted from a last-scattering surface which fills the entire sky, the GWB is in many cases (including the astrophysical background that is the focus of this chapter) emitted from a discrete population of sources, whose number density fluctuations are far and away the most important contribution to the angular power spectrum.
As we will see in section~\ref{sec:agwb-sky} below, the resulting anisotropies are significantly larger than those in the CMB from the SW, Doppler, and ISW effects.
(Indeed, \citet{Bertacca:2019fnt} found that these other terms contribute at most $\sim10\%$ of the total angular power spectrum at a reference frequency of $f=50\,\mathrm{Hz}$.)
In the remainder of this chapter, we therefore neglect the metric perturbations $\psi$, $\phi$, and consider an inhomogeneous population of discrete sources in a homogeneous FLRW background spacetime.

One other effect which is important to include, aside from the source density contrast, is the effect of the observer's peculiar velocity $\vb*v_0$.
Using equations~\eqref{eq:anisotropic-phinney-formula} and~\eqref{eq:source-frame-frequency}, we see that this induces a \emph{kinematic dipole} along the $\vb*v_0$ direction,
    \begin{equation}
        \Omega(\vu*r)=\bar{\Omega}+\mathcal{D}\,\vu*v_0\vdot\vu*r+\cdots,\qquad\mathcal{D}\equiv v_0\left.\qty(4\bar{\Omega}-\pdv{\bar{\Omega}}{(\ln f_\mathrm{s})})\right|_{f_\mathrm{s}=f/a},
    \end{equation}
    or equivalently, in spherical harmonics,
    \begin{equation}
        \Omega_{10}=\sqrt{\frac{4\uppi}{3}}\cos\theta_v\mathcal{D},\qquad\Omega_{11}=-\sqrt{\frac{2\uppi}{3}}\sin\theta_v\rme^{-\rmi\phi_v}\mathcal{D},\qquad C_1=\frac{|\Omega_{10}|^2+2|\Omega_{11}|^2}{3}=\frac{4\uppi}{9}\mathcal{D}^2,
    \end{equation}
    where $\vu*v_0=(\theta_v,\phi_v)$ and $v_0\equiv|\vb*v_0|$ are the direction and magnitude of the observer's peculiar velocity, with the latter being measured by Planck as $v_0\approx1.23\times10^{-3}$ for the Solar System barycentre~\cite{Aghanim:2018vks}.
We see that the coefficient $\mathcal{D}$ of the dipole term depends on the tilts of the GW energy spectra of the sources; the intuition here is that for a fixed observer-frame frequency $f$, the corresponding source-frame frequency $f_\mathrm{s}$ is lower for sources in the Doppler-boosted direction $\vb*v_0$, meaning that sources with red-tilted spectra give rise to a larger coefficient, and vice versa.
(We have been slightly lazy with our notation for this term---the derivative should in fact be taken inside of both integrals in equation~\eqref{eq:anisotropic-phinney-formula}, so that the spectral tilt is averaged over the source population.)
For a population of inspiralling CBCs this tilt is simply $\sim f_\mathrm{s}^{2/3}$, so we have $\mathcal{D}=(10/3)v_0\bar{\Omega}$.
This provides an extremely useful consistency relationship between the GWB monopole and kinematic dipole; if the dipole were observed to deviate from this value, this could provide interesting evidence for some other population of sources contributing to the GWB.

Before we move on, it is important to point out that the statistics of the density contrast $\delta_n$ that determine the AGWB angular power spectrum are usually \emph{non-Gaussian}, particularly for late-Universe sources such as CBCs.
While the primordial perturbations that act as seeds for these inhomogeneities are extremely well-described by Gaussian statistics (with CMB observations and other probes providing strong bounds on primordial non-Gaussianity), the nonlinear dynamics of gravitational clustering are known to cause significant non-Gaussianity in the cosmic matter distribution at late times~\cite{Bernardeau:2001qr,Peebles:1980lss}.
This implies that our treatment in terms of the angular power spectrum does not capture the full statistical information present in the AGWB anisotropies, and that it may be interesting to consider higher-order statistics such as the bispectrum, trispectrum, etc.
We leave this for future work.

\section{Modelling the astrophysical background and its anisotropies}
\label{sec:agwb}

The goal of this section is to compute the angular power spectrum of the AGWB, encapsulating the clustering statistics of CBC host galaxies.
To accomplish this, we use an all-sky mock lightcone galaxy catalogue constructed from the Millennium $N$-body simulation.
There are two key advantages of this approach, compared to other calculations based on CMB-inspired linear Boltzmann codes~\cite{Cusin:2018rsq}: first, we are able to capture the full nonlinear gravitational dynamics of the cosmological matter distribution at late times; and second, we are able to extract detailed astrophysical information about the galaxies that form at the peaks of this distribution, which is necessary to calculate the GW emission from CBCs hosted in each galaxy.

Before we come to this calculation, it is first necessary to construct a model for the isotropic component of the AGWB, $\bar{\Omega}$, as this depends on the CBC rate density and population properties that will determine the contribution of each galaxy in the mock catalogue to our AGWB sky map.

\subsection{The isotropic component}

Averaging all cosmological perturbations to zero, we can rewrite equation~\eqref{eq:anisotropic-phinney-formula} as
    \begin{equation}
    \label{eq:agwb-monopole}
        \bar{\Omega}(f)=\sum_i\frac{2G}{3H_0^2}\int_0^\infty\frac{\dd{z}}{(1+z)^2H(z)}\int\dd{\vb*\zeta}\bar{\mathcal{R}}_i(z,\vb*\zeta)\dv{E_i}{(\ln f_\mathrm{s})},
    \end{equation}
    which we have written as a sum over the three key CBC populations: BBHs, BNSs, and BHNSs.
(In principle we could include other astrophysical sources here too, e.g. supernovae.
However, the contributions from these other sources are expected to be negligible compared to CBCs~\cite{Regimbau:2011rp}.)
There are three ingredients we must specify in order to compute equation~\eqref{eq:agwb-monopole}:
    \begin{itemize}
        \item the Hubble rate $H(z)$ of the FLRW background;
        \item the isotropic GW energy spectra $\dv*{E_i}{(\ln f_\mathrm{s})}$ of the CBCs;
        \item the mean comoving rate density $\bar{\mathcal{R}}_i(z,\vb*\zeta)$ of each population, as a function of both the galaxy parameters and the CBC parameters.
    \end{itemize}

The first ingredient is the easiest; we assume a standard $\Lambda$CDM Universe with Planck 2018 parameters,
    \begin{equation}
        H(z)\simeq H_0\sqrt{\Omega_\Lambda+\Omega_m(1+z)^3},\qquad H_0\approx67.7\,\mathrm{km}\,\mathrm{s}^{-1}\,\mathrm{Mpc}^{-1},\qquad\Omega_\mathrm{m}\approx0.311,\qquad\Omega_\Lambda=1-\Omega_\mathrm{m}.
    \end{equation}

\begin{figure}[t!]
    \begin{center}
        \includegraphics[width=0.85\textwidth]{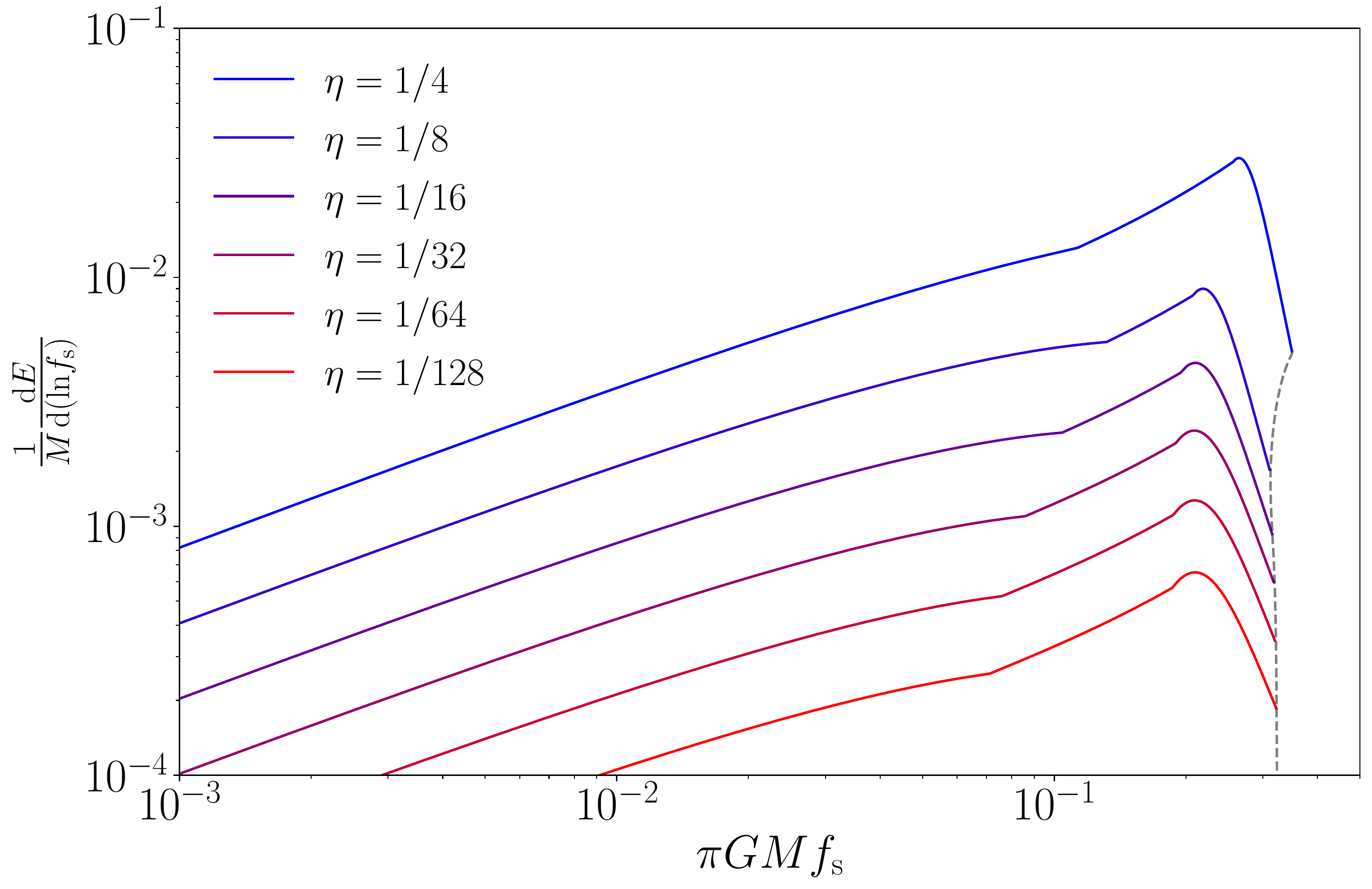}
    \end{center}
    \caption{%
    The hybrid energy spectrum~\eqref{eq:bbh-spectrum} for non-spinning BBHs from \citet{Ajith:2007kx,Ajith:2009bn}, shown here as a fraction of the total mass $M$ of the system, and as a function of the dimensionless frequency $\uppi GMf_\mathrm{s}=v_\mathrm{s}^3$.
    Written in this form, the only remaining variable is the dimensionless mass ratio $\eta\equiv m_1m_2/(m_1+m_2)^2\le1/4$.
    The black dashed line shows the high-frequency cutoff at $\uppi GMf_4$ as a function of $\eta$.
    }
    \label{fig:cbc-spectrum}
\end{figure}

For the second ingredient, we could use the $\sim f^{2/3}$ energy spectrum we computed in section~\ref{sec:compact-binaries} if we were interested purely in the inspiral regime, and were happy to neglect all post-Newtonian effects.
While this might be an adequate approximation for BNSs in the LIGO/Virgo frequency band, many of the more massive BBHs in our population have their merger and ringdown in this frequency band, meaning that we must adopt a more complete description of the energy spectrum.
To this end, we use the hybrid waveform models developed by \citet{Ajith:2007kx,Ajith:2009bn} (as shown in figure~\ref{fig:cbc-spectrum}), which combine post-Newtonian results in the inspiral regime with data from numerical relativity simulations to fit a parameterised function covering the full inspiral-merger-ringdown spectrum.
The hybrid BBH energy spectrum is written as a piecewise function,
    \begin{equation}
    \label{eq:bbh-spectrum}
        \dv{E}{(\ln f_\mathrm{s})}\approx\frac{1}{3}\eta Mv_\mathrm{s}^2\times
        \begin{cases}
            \qty[1-\qty(\tfrac{323}{224}-\tfrac{451}{168}\eta)v_\mathrm{s}^2]^2, & f_\mathrm{s}<f_1\\
            c_1(f_\mathrm{s}/f_1)\qty(1-1.8897\,v_\mathrm{s}+1.6557\,v_\mathrm{s}^2)^2, & f_1\le f_\mathrm{s}<f_2\\
            c_2(f_\mathrm{s}/f_2)^{7/3}\mathcal{L}^2(f_\mathrm{s}|f_2,f_3), & f_2\le f_\mathrm{s}<f_4
        \end{cases},
    \end{equation}
    with the three pieces corresponding to the inspiral, merger, and ringdown respectively.
Here $v_\mathrm{s}\equiv(\uppi GMf_\mathrm{s})^{1/3}$ is the binary's relative velocity, $c_1$ and $c_2$ are numerical constants enforcing continuity of the spectrum, and $\mathcal{L}(x|x_0,\sigma)\equiv(\sigma/\uppi)/[\sigma^2+(x-x_0)^2]$ is the Lorentzian function.
(Note that this reduces to the simple Newtonian result~\eqref{eq:cbc-energy-spectrum} in the limit $v_\mathrm{s}\ll1$, as expected.)
The frequencies $\{f_1,f_2,f_3,f_4\}$ are given by
    \begin{align}
    \begin{split}
        \uppi GMf_1&\approx0.0660+0.6437\,\eta-0.05822\,\eta^2-7.092\,\eta^3,\\
        \uppi GMf_2&\approx0.185+0.1469\,\eta-0.0249\,\eta^2+2.325\,\eta^3,\\
        \uppi GMf_3&\approx0.0925-0.4098\,\eta+1.829\,\eta^2-2.87\,\eta^3,\\
        \uppi GMf_4&\approx0.3236-0.1331\,\eta-0.2714\,\eta^2+4.922\,\eta^3,
    \end{split}
    \end{align}
    and correspond to the ISCO frequency, QNM frequency, inverse QNM damping time, and high-frequency cutoff, respectively.
We assume here that the constituents of the binary are non-spinning; while there is evidence for some large BH spins amongst the CBCs detected by LIGO/Virgo~\cite{Abbott:2020niy}, modelling the distribution of these spins greatly complicates our calculations by significantly increasing the dimensionality of the parameter space, and has negligible effect on the final GWB spectrum~\cite{Abbott:2021kbb,Jenkins:2018kxc}, so we choose to ignore them here.

\begin{figure}[t!]
    \begin{center}
        \includegraphics[width=0.85\textwidth]{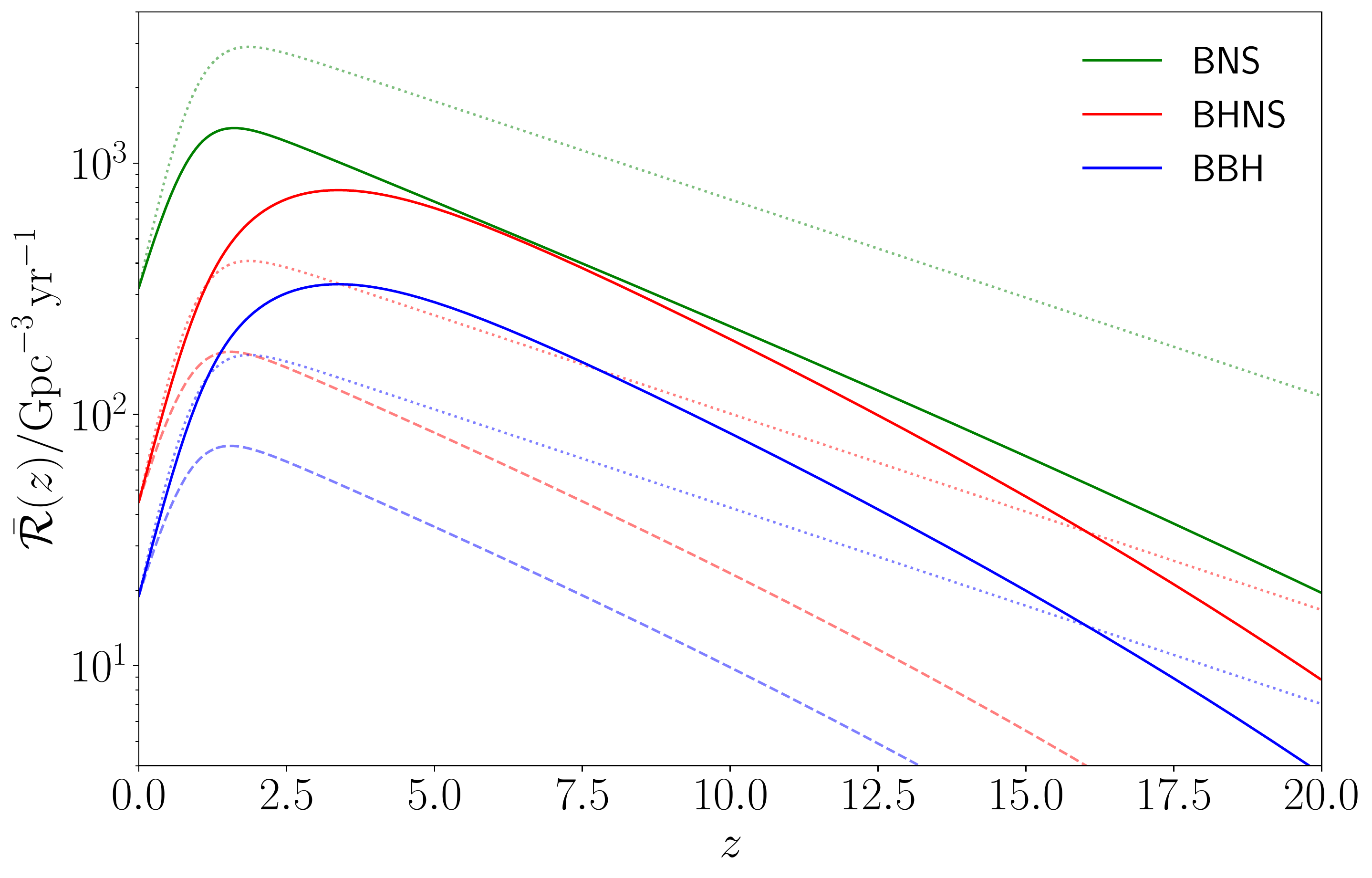}
    \end{center}
    \caption{%
    Comoving merger rate densities as functions of redshift for each of the three CBC populations we consider: BBHs, BHNSs, and BNSs.
    The dotted curves, given by equation~\eqref{eq:rate-density-direct-sfr}, are directly proportional to the cosmic mean SFR.
    The dashed curves are proportional to the \emph{delayed} SFR~\eqref{eq:sfr-delayed}, with delay time distributions as described in the text.
    The solid curves additionally account for the suppression in BH formation from high-metallicity stellar progenitors using equation~\eqref{eq:sfr-metallicity}.
    (For BNSs, the latter two curves are identical.)
    In each case, the overall amplitude is fixed by the redshift-zero merger rates~\eqref{eq:local-rates} inferred from the direct detections made by LIGO/Virgo; this has the slightly counter-intuitive effect of boosting the merger rate at higher redshifts when the metallicity suppression is taken into account, since this suppression is strongest at redshift zero.
    }
    \label{fig:merger-rates}
\end{figure}

Equation~\eqref{eq:bbh-spectrum} is not valid for BNSs and BHNSs, as in these cases the NS matter leaves imprints on the spectrum at frequencies $f_\mathrm{s}\gtrsim f_1$.
However, the lower masses of these systems mean that they typically merge outside of the LIGO/Virgo frequency band.
We can therefore conservatively use equation~\eqref{eq:bbh-spectrum} for BNSs and BHNSs too, but truncate the spectrum at the ISCO frequency $f_1$.

For the final ingredient, the comoving rate density, we need to capture the evolution in the number density of CBCs of different masses over cosmological timescales.
Since all CBCs are products of stellar evolution, one simple choice is to tie the CBC rate density to the star formation rate (SFR) density,
    \begin{equation}
    \label{eq:rate-density-direct-sfr}
        \bar{\mathcal{R}}_i(z,\vb*\zeta)\propto\bar{\psi}(z)p_i(\vb*\zeta),
    \end{equation}
    where $\bar{\psi}$ is the mean SFR density as a function of redshift, and $p_i(\vb*\zeta)$ is the probability density function (PDF) of the parameters of each population of CBCs (in particular, their masses), with $i$ running over the different CBC populations as before.
The unknown proportionality factor here encodes the efficiency with which newly-formed stars are converted into merging compact objects.
This factor is challenging to model directly; but fortunately, we don't need to model it.
Instead, we can fix the proportionality using the local (i.e., $z=0$) CBC rate density, $\bar{\mathcal{R}}_i^{(0)}$, inferred from the individual detections made by LIGO/Virgo,
    \begin{equation}
        \bar{\mathcal{R}}_i(z,\vb*\zeta)=\bar{\mathcal{R}}^{(0)}_ip_i(\vb*\zeta)\frac{\bar{\psi}(z)}{\bar{\psi}(0)}.
    \end{equation}
(Here we assume that the PDFs are normalised, $\int\dd{\vb*\zeta}p_i(\vb*\zeta)=1$.)

One major problem with this model is that it implicitly assumes that newly-formed stars are \emph{immediately} converted into merging CBCs.
In reality, the stars must first go through their main-sequence evolution before eventually becoming compact objects, and then must gradually inspiral from large initial separations until they reach the LIGO/Virgo band and merge.
This entire process gives rise to a significant \emph{delay time} between star formation and the associated CBC mergers, typically several Myr.
We therefore convolve the SFR density at each redshift with some distribution of delay times,
    \begin{equation}
    \label{eq:sfr-delayed}
        \bar{\psi}_{\mathrm{d},i}(z)=\int\dd{t_\mathrm{d}}p_i(t_\mathrm{d})\bar{\psi}(z_\mathrm{f}(z,t_\mathrm{d})),
    \end{equation}
    where $z_\mathrm{f}$ is the redshift at which the stars were formed, which occurs at a lookback time $t_\mathrm{d}$ before the redshift $z$ at which the binary merges.
By replacing $\bar{\psi}$ with $\bar{\psi}_{\mathrm{d},i}$ in equation~\eqref{eq:rate-density-direct-sfr}, we obtain an appropriately delayed estimate for the merger rate density.

So far we have assumed that the proportionality in equation~\eqref{eq:rate-density-direct-sfr} holds regardless of the properties of the binary and of the host galaxy.
One exception to this which is important to capture here is that BH formation is suppressed for stellar progenitors with large metallicities.
Following \citet{Abbott:2021kbb}, we model this metallicity dependence with a sharp cutoff on all BH formation for metallicities $Z>0.1\,Z_\odot$, where $Z_\odot\approx0.02$ is the solar metallicity.
For BBHs and BHNSs, we therefore re-weight the rate density by the fraction of SFR occurring at metallicities below this threshold,
    \begin{equation}
    \label{eq:sfr-metallicity}
        \bar{\psi}_{\mathrm{d},i}\to\bar{\psi}^{(Z)}_{\mathrm{d},i}\equiv\int\dd{t_\mathrm{d}}p_i(t_\mathrm{d})f_i(z_\mathrm{f})\bar{\psi}(z_\mathrm{f}),\qquad f_i(z)\equiv
        \begin{cases}
            1, & i=\mathrm{BNS},\\
            \int_0^{0.1\,Z_\odot}\dd{Z}p(Z|z), & i=\mathrm{BBH},\mathrm{BHNS}.
        \end{cases}
    \end{equation}
This is our model for the CBC rate density.
See figure~\ref{fig:merger-rates} for a comparison between this and the more simplistic models described above.

We are now ready to flesh out each of the individual parts that make up this model; many of the details here agree with the \enquote{fiducial} model adopted by the LVK in, e.g., \citet{Abbott:2021kbb}.
For the cosmic mean SFR density, we use the fitting function from \citet{Vangioni:2014axa},
    \begin{equation}
    \label{eq:sfrd-vangioni-et-al}
        \bar{\psi}(z)=\bar{\psi}_\mathrm{peak}\frac{\alpha\exp[\beta(z-z_\mathrm{peak})]}{\alpha-\beta+\beta\exp[\alpha(z-z_\mathrm{peak})]},
    \end{equation}
    which has a maximum value of $\bar{\psi}_\mathrm{peak}\approx0.145\,m_\odot\,\mathrm{yr}^{-1}\,\mathrm{Mpc}^{-3}$ at redshift $z_\mathrm{peak}\approx1.86$, with redshift scaling either side of this peak set by the dimensionless constants $\alpha=2.80$ and $\beta=2.62$.
This SFR density determines the rate at which stars produce metals and return these to the interstellar medium.
Equation~\eqref{eq:sfrd-vangioni-et-al} therefore sets the cosmic mean metallicity, using an expression which we take from \citet{Belczynski:2016obo},
    \begin{equation}
        \bar{Z}(z)=\frac{H_0}{\bar{\psi}_\mathrm{norm}}\int_z^{20}\dd{z'}\frac{\bar{\psi}(z')}{(1+z')H(z')},\qquad\bar{\psi}_\mathrm{norm}\approx8.90\,m_\odot\,\mathrm{yr}^{-1}\,\mathrm{Mpc}^{-3}.
    \end{equation}
We assume that $\log_{10}Z$ follows a Gaussian distribution around this mean with standard deviation $\sigma=1/2$.
This distribution is cut off at $\log_{10}Z=0$ (since $Z\le1$ by definition), such that its PDF is
    \begin{equation}
        p(\log_{10}Z|z)=\sqrt{8/\uppi}\,\frac{\exp[-2\qty(\log_{10}Z-\log_{10}\bar{Z}(z))^2]}{\mathrm{erfc}\qty[\sqrt{2}\log_{10}\bar{Z}(z)]},
    \end{equation}
    and the corresponding metallicity correction factor is
    \begin{equation}
        f_i(z)=
        \begin{cases}
            1, & i=\mathrm{BNS},\\
            \dfrac{\mathrm{erfc}\qty[\sqrt{2}\qty(\log_{10}\bar{Z}(z)-\log_{10}0.1\,Z_\odot)]}{\mathrm{erfc}\qty[\sqrt{2}\log_{10}\bar{Z}(z)]}, & i=\mathrm{BBH},\mathrm{BHNS},
        \end{cases}
    \end{equation}
    where $\mathrm{erfc}(x)\equiv1-\int_0^x\dd{t}\rme^{-t^2}$ is the complementary error function.

\begin{figure}[t!]
    \begin{center}
        \includegraphics[width=0.85\textwidth]{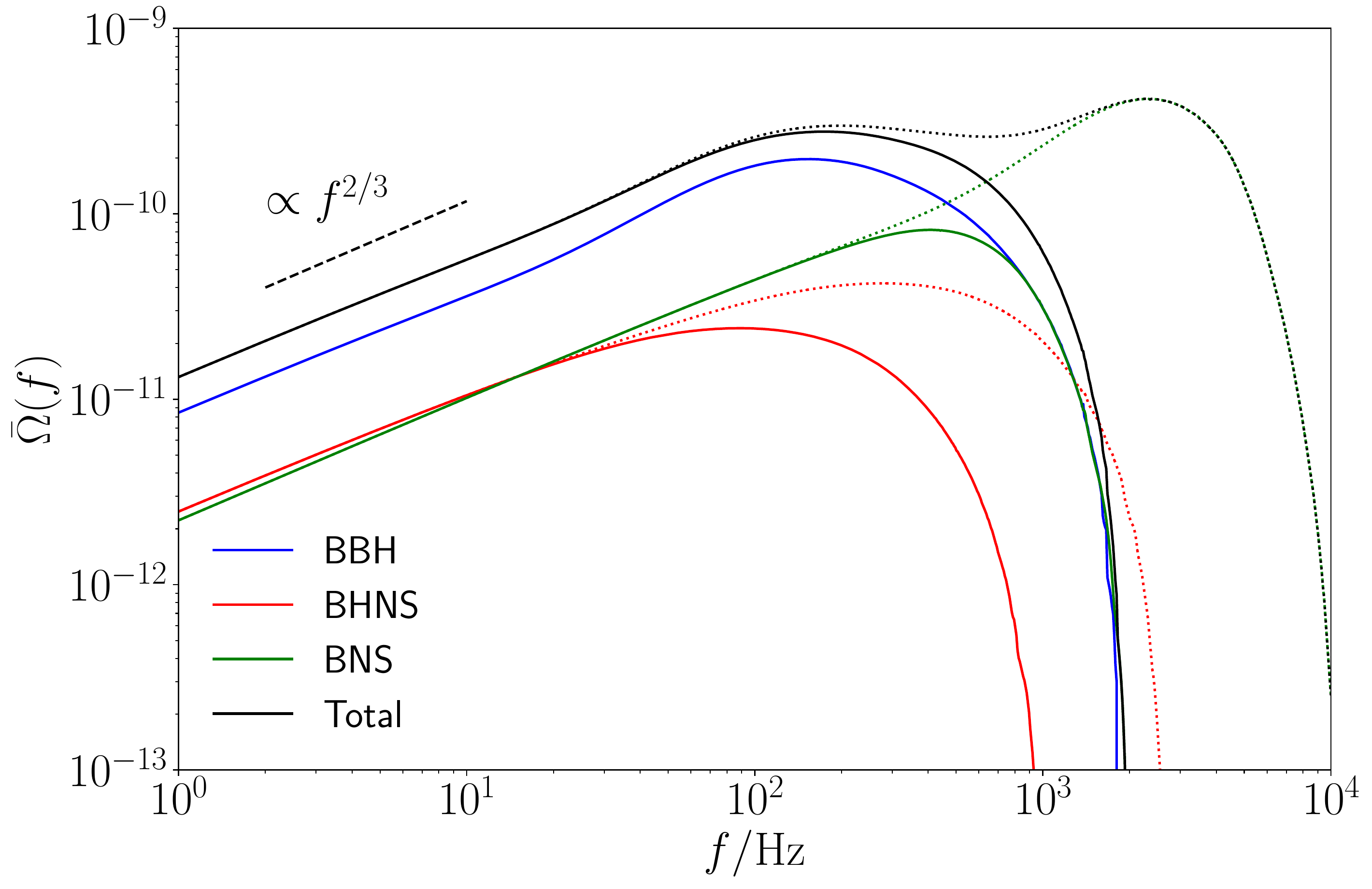}
    \end{center}
    \caption{%
    The mean GWB intensity $\bar{\Omega}(f)$ from stellar-mass CBCs in the LIGO/Virgo band, as predicted by our model.
    (Note that this has units of inverse steradians, and is therefore smaller than the isotropic GWB spectrum $\Omega(f)$ discussed in chapter~\ref{chap:intro} by a factor of $4\uppi$.)
    Solid curves show our default model, in which the GW emission from BNSs and BHNSs is truncated at the ISCO frequency, while dotted curves show the case where all binaries use the BBH waveform~\eqref{eq:bbh-spectrum}.
    The true GWB spectrum will likely lie somewhere between these two extremes; however, note that LIGO/Virgo stochastic searches are insensitive to this frequency range.
    }
    \label{fig:Omega-bar-cbcs}
\end{figure}

For the delay time distribution, we take $p_i(t_\mathrm{d})\propto1/t_\mathrm{d}$ between a minimum value $t_\mathrm{min}$ of $20\,\mathrm{Myr}$ for BNSs and $50\,\mathrm{Myr}$ for BBHs and BHNSs, and a maximum value equal to the age of the Universe at that redshift, which for our cosmological model is
    \begin{equation}
        t_\mathrm{age}(z)=\frac{2}{3}\Omega_\Lambda^{-1/2}H_0^{-1}\mathrm{sinh}^{-1}\qty(\sqrt{\frac{\Omega_\Lambda}{\Omega_\mathrm{m}(1+z)^3}}).
    \end{equation}
The delayed, metallicity-corrected SFR density can thus be written as
    \begin{equation}
        \bar{\psi}^{(Z)}_{\mathrm{d},i}=\frac{1}{\ln(t_\mathrm{age}(z)/t_{\mathrm{min},i})}\int_{\ln t_{\mathrm{min},i}}^{\ln t_\mathrm{age}(z)}\dd{(\ln t_\mathrm{d})}f_i(z_\mathrm{f}(z,t_\mathrm{d}))\bar{\psi}(z_\mathrm{f}(z,t_\mathrm{d})),
    \end{equation}
    where the formation redshift $z_\mathrm{f}$ is given by
    \begin{equation}
        \frac{1+z_\mathrm{f}}{1+z}=\qty[\cosh\qty(\frac{3}{2}\Omega_\Lambda^{1/2}H_0t_\mathrm{d})-\frac{H(z)}{\Omega_\Lambda^{1/2}H_0}\sinh\qty(\frac{3}{2}\Omega_\Lambda^{1/2}H_0t_\mathrm{d})]^{-2/3}.
    \end{equation}

\begin{figure}[t!]
    \begin{center}
        \includegraphics[width=0.8\textwidth]{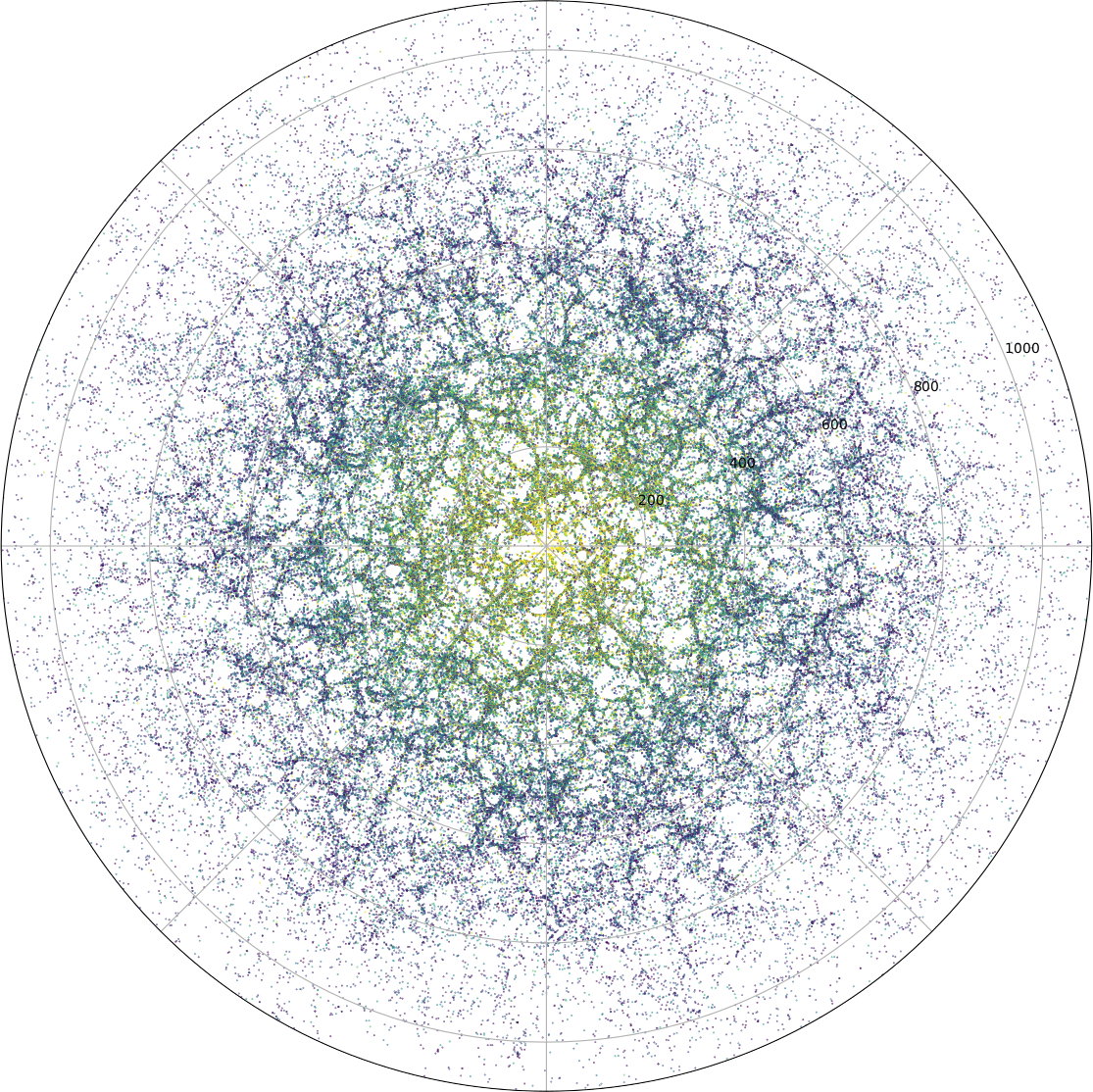}
    \end{center}
    \caption{%
    Equatorial slice through the three-dimensional distribution of galaxies in the Millennium mock galaxy catalogue.
    Each galaxy is coloured according to its contribution to the AGWB intensity, with nearby galaxies dominating due to the $\sim1/r^2$ fall-off in the GW flux.
    The numbered radial rings indicate comoving distances in Mpc.
    }
    \label{fig:polar-millennium}
\end{figure}

Finally, we need to specify the local rates and mass distributions of each CBC population.
For BBHs, following \citet{Abbott:2021kbb}, we assume the primary mass $m_1$ follows a broken power law distribution (as inferred from the second GW transient catalogue, GWTC-2),
    \begin{equation}
        p(m_1)\propto
        \begin{cases}
            m_1^{-1.58}, & 5.875\,m_\odot<m_1<40.82\,m_\odot,\\
            m_1^{-5.59}, & 40.82\,m_\odot<m_1<87.14\,m_\odot,
        \end{cases}
    \end{equation}
    with the secondary mass $m_2\le m_1$ given by a power-law distribution in the mass ratio $q\equiv m_2/m_1$, $p(q|m_1)\propto q^{1.4}$.
For BNSs, we assume that both NS masses are uniformly distributed between $1\,m_\odot$ and $2.5\,m_\odot$, while for BHNSs we assume the NS mass follows this same distribution, while the BH mass follows the same distribution as the primary mass $m_1$ of the BBHs.
We set the local BBH and BNS rates equal to those inferred for these models~\cite{Abbott:2021kbb}, while for BHNSs we use the recent value inferred in \citet{Abbott:2021qlt},
    \begin{equation}
    \label{eq:local-rates}
        \bar{\mathcal{R}}_\mathrm{BBH}^{(0)}\approx19\,\mathrm{Gpc}^{-3}\,\mathrm{yr}^{-1},\qquad\bar{\mathcal{R}}_\mathrm{BHNS}^{(0)}\approx45\,\mathrm{Gpc}^{-3}\,\mathrm{yr}^{-1},\qquad\bar{\mathcal{R}}_\mathrm{BNS}^{(0)}\approx320\,\mathrm{Gpc}^{-3}\,\mathrm{yr}^{-1}.
    \end{equation}

Putting all of these ingredients together, we find the AGWB spectrum shown in figure~\ref{fig:Omega-bar-cbcs}.
We see that, as expected, the spectrum follows a $\sim f^{2/3}$ power law at low frequencies, before peaking and falling off rapidly above $\sim1\,\mathrm{kHz}$ due to the post-merger truncation of the CBC energy spectrum.
This fall-off occurs at higher frequencies for BBHs and BNSs than for BHNSs, since the BBH spectrum includes the merger-ringdown emission from equation~\eqref{eq:bbh-spectrum}, while the BNS spectrum is dominated by much lower total masses, and therefore much higher ISCO frequencies.
The dotted curves show the hypothetical case in which BNSs and BHNSs have the same merger-ringdown emission as BBHs; we see that this only changes the overall spectrum at frequencies $f\gtrsim1\,\mathrm{kHz}$, and is therefore beyond the reach of LIGO/Virgo stochastic searches.
As such, we can safely neglect post-merger emission from these systems.

\begin{figure}[t!]
    \begin{center}
        \includegraphics[width=0.9\textwidth]{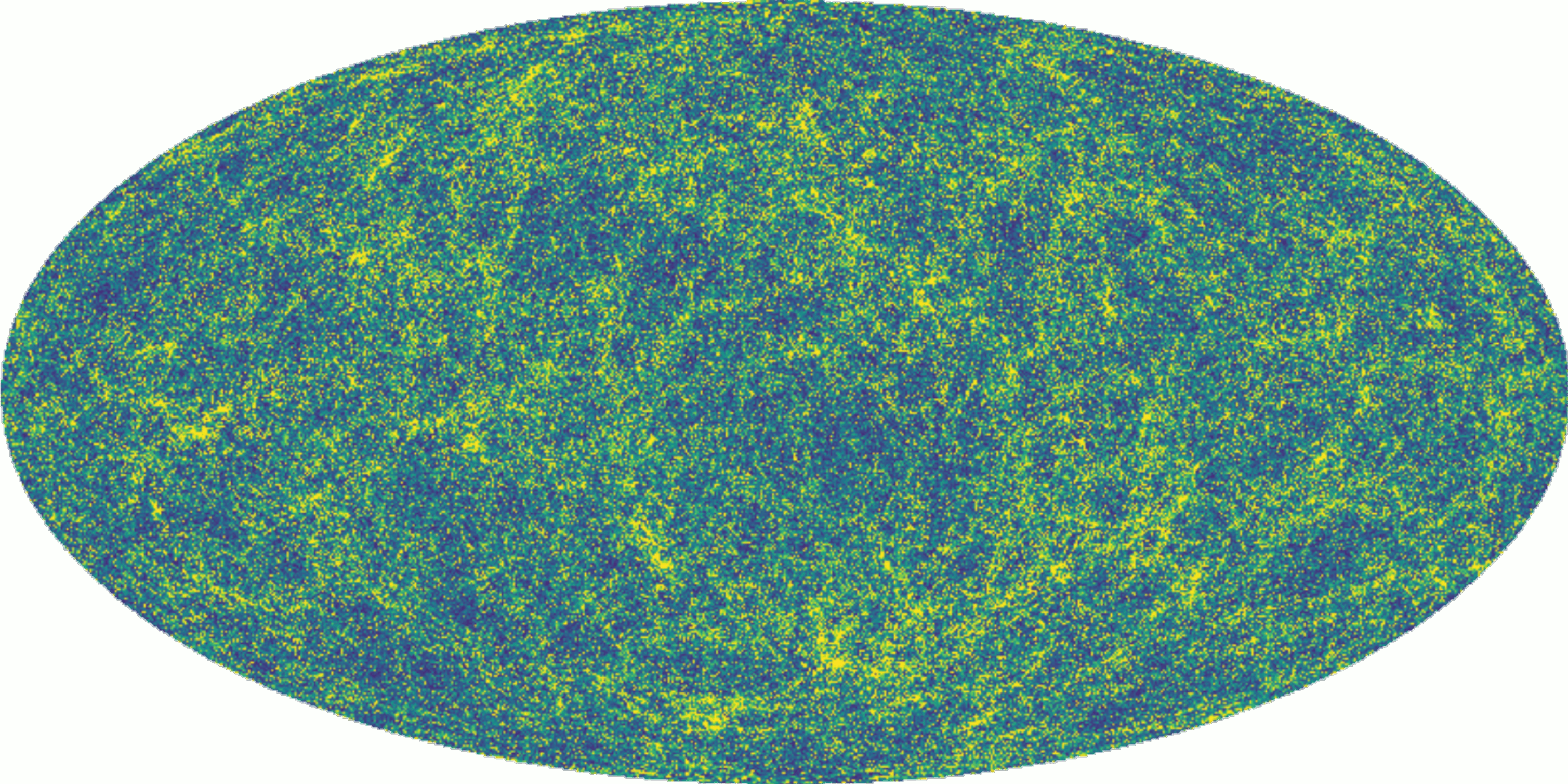}
    \end{center}
    \caption{%
    Mollweide projection of our full-sky mock HEALPix map of the AGWB, based on data from the Millennium simulation as described in the main text.
    Bright/dark pixels indicate over-/under-densities in the AGWB intensity field at a reference frequency of $25\,\mathrm{Hz}$.
    }
    \label{fig:map}
\end{figure}

\begin{figure}[t!]
    \begin{center}
        \includegraphics[width=0.9\textwidth]{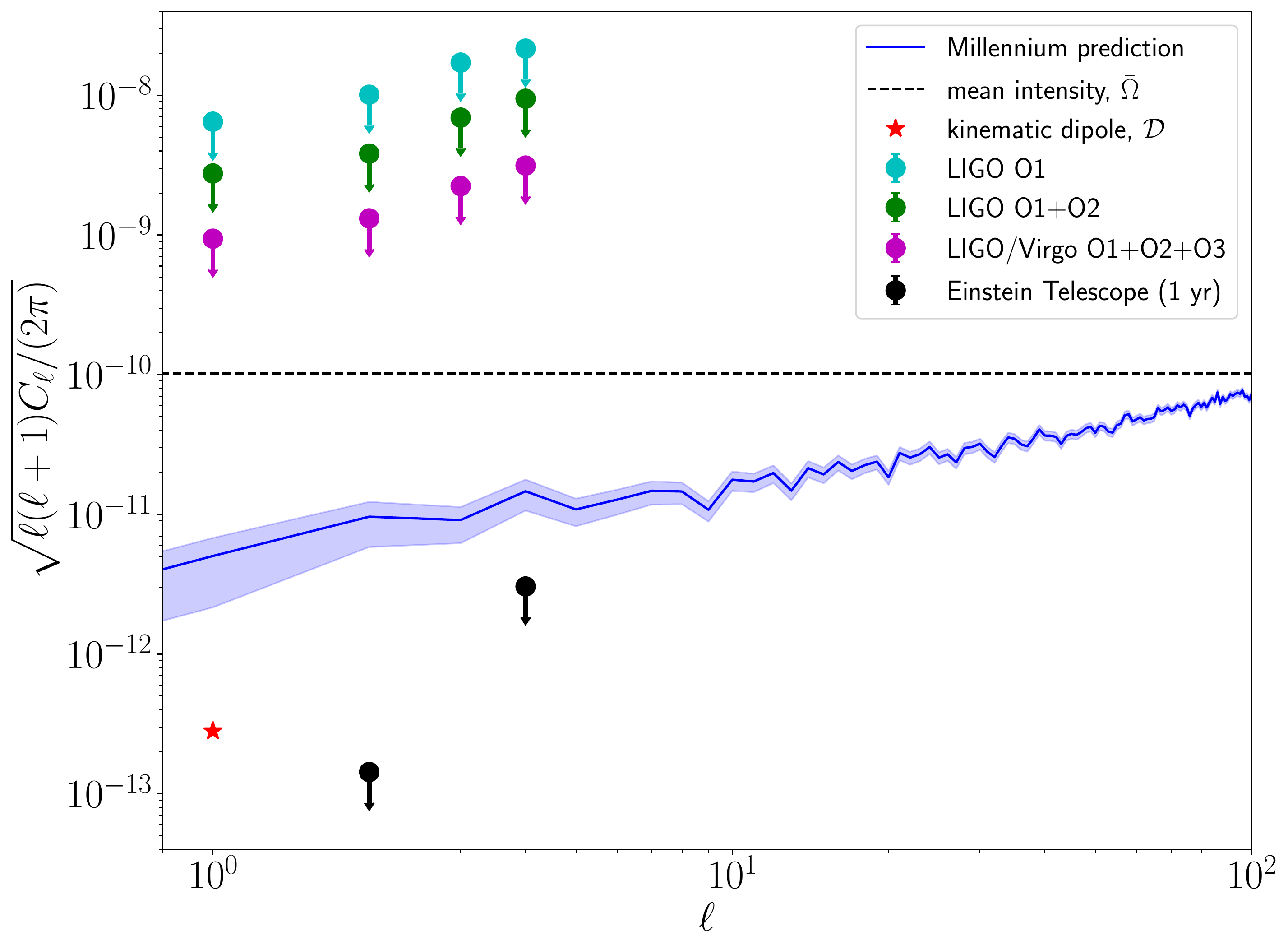}
    \end{center}
    \caption{%
    The AGWB angular power spectrum computed from the map shown in figure~\ref{fig:map}.
    The quantity plotted here, $\sqrt{\ell(\ell+1)C_\ell/(2\uppi)}$, is roughly the contribution of each angular multipole $\ell$ to the standard deviation in the intensity field fluctuations.
    This has units of inverse steradians, and can be directly compared with the mean intensity $\bar{\Omega}$, which is shown as a horizontal black dashed line.
    The shaded blue region shows the $1\sigma$ uncertainty in the angular power spectrum due to cosmic variance.
    The red point shows the kinematic dipole, $\sqrt{\ell(\ell+1)C_\ell/(2\uppi)}=(2/3)\mathcal{D}\simeq(20/9)v_0\bar{\Omega}$.
    The other coloured points with arrows show current upper limits (95~\% confidence) on the AGWB angular power spectrum from LIGO/Virgo, as well as forecast upper limits from the Einstein Telescope (as computed in appendix~\ref{sec:et}).
    All of the values shown here assume a reference frequency of $25\,\mathrm{Hz}$.
    }
    \label{fig:C_ell-cbcs}
\end{figure}

\subsection{The full AGWB sky}
\label{sec:agwb-sky}

We now construct a full AGWB sky map by applying the recipes described above to a mock light cone galaxy catalogue~\cite{Blaizot:2003av}.
This catalogue consists of 5,715,694 galaxies, and was built using a random tiling technique applied to 64 post-processed snapshots saved during the Millennium $N$-body simulation~\cite{Springel:2005nw}, with a time step of roughly $100\,\mathrm{Myr}$ and a box size of $500\,h^{-1}\,\mathrm{Mpc}$.
These snapshots of the CDM distribution were analysed using the \enquote{L-galaxies} semi-analytic model~\cite{DeLucia:2006szx} to populate the simulation box with galaxies and capture the baryonic physics of each of these galaxies (star formation, metallicity, etc.), which is crucial for our purposes.
The resulting large-scale galaxy distribution is illustrated in figure~\ref{fig:polar-millennium}.

In order to compute the contribution of each galaxy to the AGWB, we query the catalogue database~\cite{Lemson:2006ee} to extract its sky location, redshift, star formation history, metallicity history, and peculiar velocity (the last of which has a small effect on the galaxy's total GW energy emission and source frame frequency through equations~\eqref{eq:anisotropic-phinney-formula} and~\eqref{eq:source-frame-frequency}.)
Computing the galaxy's \emph{delayed} SFR, as well as the metallicity at the time of star formation, requires us to extract information that is not on the lightcone, so for each galaxy we query the full Millennium simulation to extract its SFR and metallicity at each of the previous snapshots.
This process is made more complicated by the fact that the lightcone galaxies are the result of a sequence of mergers of smaller galaxies, each with its own independent star formation history.
Tracing all of the branches of these merger trees for all 5,715,694 lightcone galaxies results in a set of 973,224,532 measurements of SFR, metallicity, and redshift, which we process to calculate the CBC merger rate of each galaxy on the lightcone.

\begin{figure}[t!]
    \begin{center}
        \includegraphics[width=0.85\textwidth]{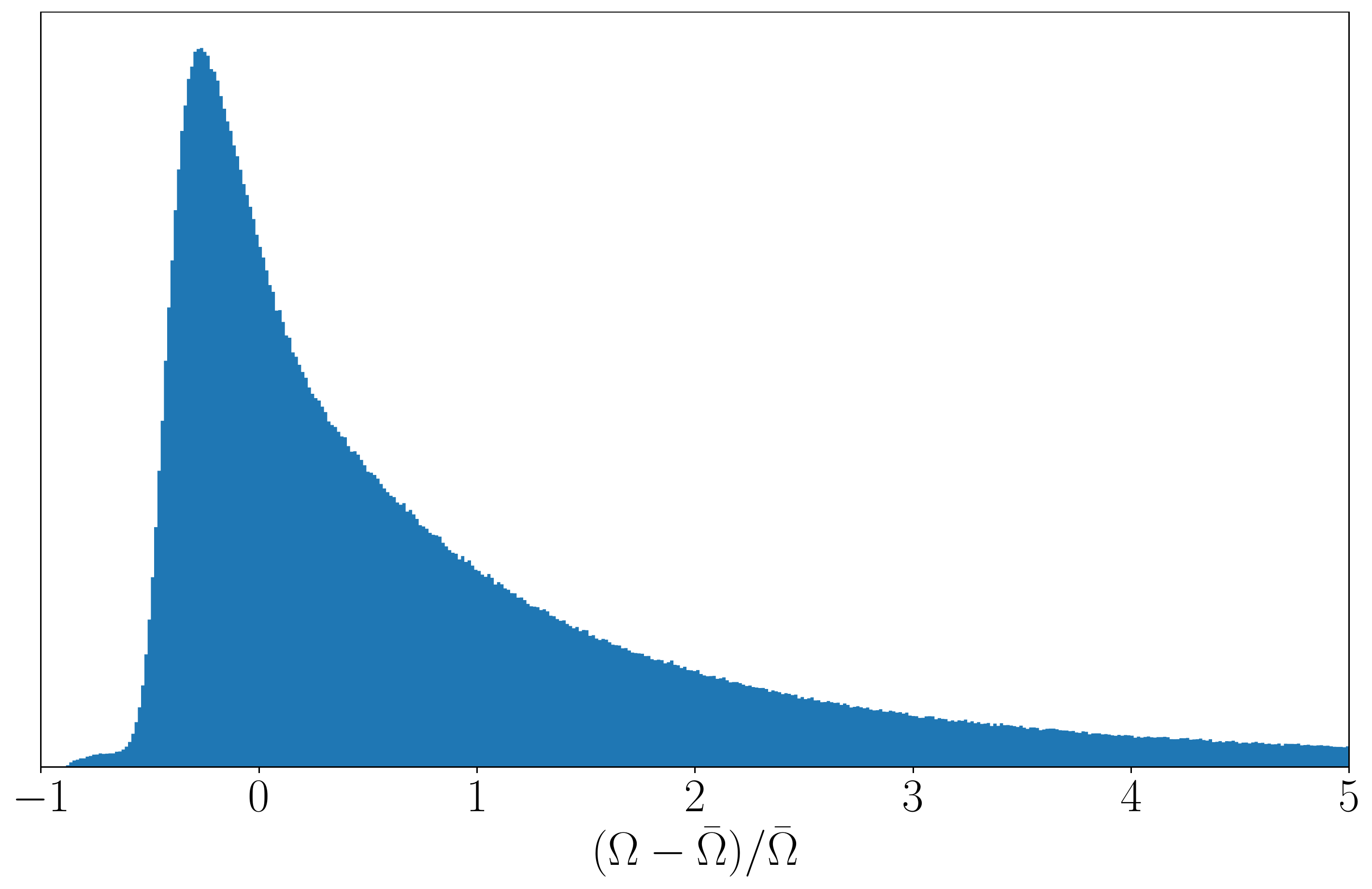}
    \end{center}
    \caption{%
    Histogram showing the distribution in GW intensity for each of the pixels in the map shown in figure~\ref{fig:map}, normalised to the mean intensity $\bar{\Omega}$.
    We see that the distribution is strongly non-Gaussian, with a heavy tail due to numerous nonlinear overdensities.
    }
    \label{fig:agwb-1-point}
\end{figure}

This results in a list of sky locations and GW intensities for each of the galaxies, which we combine to produce the full-sky AGWB map shown in figure~\ref{fig:map}.
This map uses the HEALPix\footnote{\url{http://healpix.sourceforge.net}} pixelisation scheme to partition the sky into equal-area curvilinear quadrilaterals, whose configuration is chosen to optimise the computation of spherical harmonics~\cite{Gorski:2004by}.
We estimate the angular power spectrum $C_\ell$ of this map using HEALPix's \enquote{\texttt{anafast}} routine; the result is shown in figure~\ref{fig:C_ell-cbcs}.
Note that this spectrum corresponds to a single random realisation of LSS,\footnote{%
In principle, we could reduce the cosmic variance here by averaging the angular power spectra from multiple statistically-independent mock catalogues; at the time of writing, however, the catalogue we use is the only full-sky lightcone which is publicly available and is sufficiently large for our purposes.
Furthermore, it is not possible to re-sample from this one catalogue without introducing correlations that would interfere with the goal of reducing cosmic variance.} such that each of the $C_\ell$ multipoles shown here should be interpreted as random draws from a distribution centred on the true angular power spectrum, with width set by the cosmic variance we calculated in equation~\eqref{eq:standard-estimator-variance}, $\sigma_\ell\equiv C_\ell/\sqrt{\ell+\sfrac{1}{2}}$.
We see in figure~\ref{fig:C_ell-cbcs} that the anisotropies are very large compared to those in early-Universe observables such as the CMB, with fluctuations on the order of a few percent at the largest angular scales $\ell\sim1$, rising to fluctuations of order unity at smaller angular scales $\ell\sim100$.
The resulting dipole component $C_1$ is significantly larger than the kinematic dipole due to our peculiar motion, which is denoted with a red star in figure~\ref{fig:C_ell-cbcs}.
These large fluctuations reflect the nonlinear gravitational clustering of matter at late times, and retroactively justify our decision to focus on the density contrast term in equation~\eqref{eq:anisotropic-phinney-formula}, neglecting the much smaller contributions from the SW and ISW terms.

The impact of this nonlinearity can also be seen in the one-point statistics of the AGWB map.
In figure~\ref{fig:agwb-1-point} we show a histogram of the GW intensity $\Omega(\vu*r_i)$ in each of the pixels $\vu*r_i$ in our map, normalised to the mean value $\bar{\Omega}$.
The distribution is strongly skewed towards large, nonlinear overdensities, and is clearly in very poor agreement with a Gaussian distribution (as expected based on our discussion in section~\ref{sec:inhomogeneous}).

In figure~\ref{fig:C_ell-cbcs} we also compare our predicted angular power spectrum with existing upper limits from LIGO/Virgo searches, as well as sensitivity forecasts for Einstein Telescope~\cite{Punturo:2010zz}.
These upper limits focus on the lowest few multipoles (i.e., the largest angular scales), as these are the ones that the current detector network is most sensitive to.
We see that, despite steady improvement from O1 onward, LIGO/Virgo are still several orders of magnitude away from detecting the AGWB anisotropies---which is unsurprising, given that the isotropic component has still not yet been detected, and this is louder by a factor of $\sim100$.
However, looking forward to the late 2030s, we see that ET is likely to make a confident detection of the AGWB anisotropies within just one year of observing time.
This forecast assumes a cross-correlation search between the three co-located interferometers that make up the ET proposal (one at each corner of an equilateral triangle).
As we show in appendix~\ref{sec:et}, the geometry of this setup is such that ET on its own is only sensitive to the $\ell=2$ and $\ell=4$ multipoles (as well as the monopole).
However, adding further third-generation ground-based interferometers to the network, such as the US-based Cosmic Explorer proposal~\cite{Hall:2020dps,Reitze:2019iox}, would break this degeneracy and provide sensitivity for a broader range of multipoles, as well as improving the sensitivity to the AGWB overall.
It is important to note that these third-generation interferometers will be able to individually resolve the vast majority of BBH events in the Universe (though only a smaller fraction of the BNS and BHNS events)~\cite{Regimbau:2016ike,Sachdev:2020bkk,Sharma:2020btq,Maggiore:2019uih}.
As such, it is not clear how stochastic searches for the AGWB will change in the third-generation era, or indeed whether it will still be useful to think about this signal as being inherently stochastic, rather than attempting some kind of semi-coherent search along the lines of \citet{Smith:2017vfk}.
We leave a deeper investigation of these questions, and how they impact upon cosmological studies using CBCs as tracers of the cosmic matter distribution, for future work.

\subsection{Estimating the impact of population uncertainties}
\label{sec:cbc-populations}

\begin{figure}[t!]
    \begin{center}
        \includegraphics[width=0.85\textwidth]{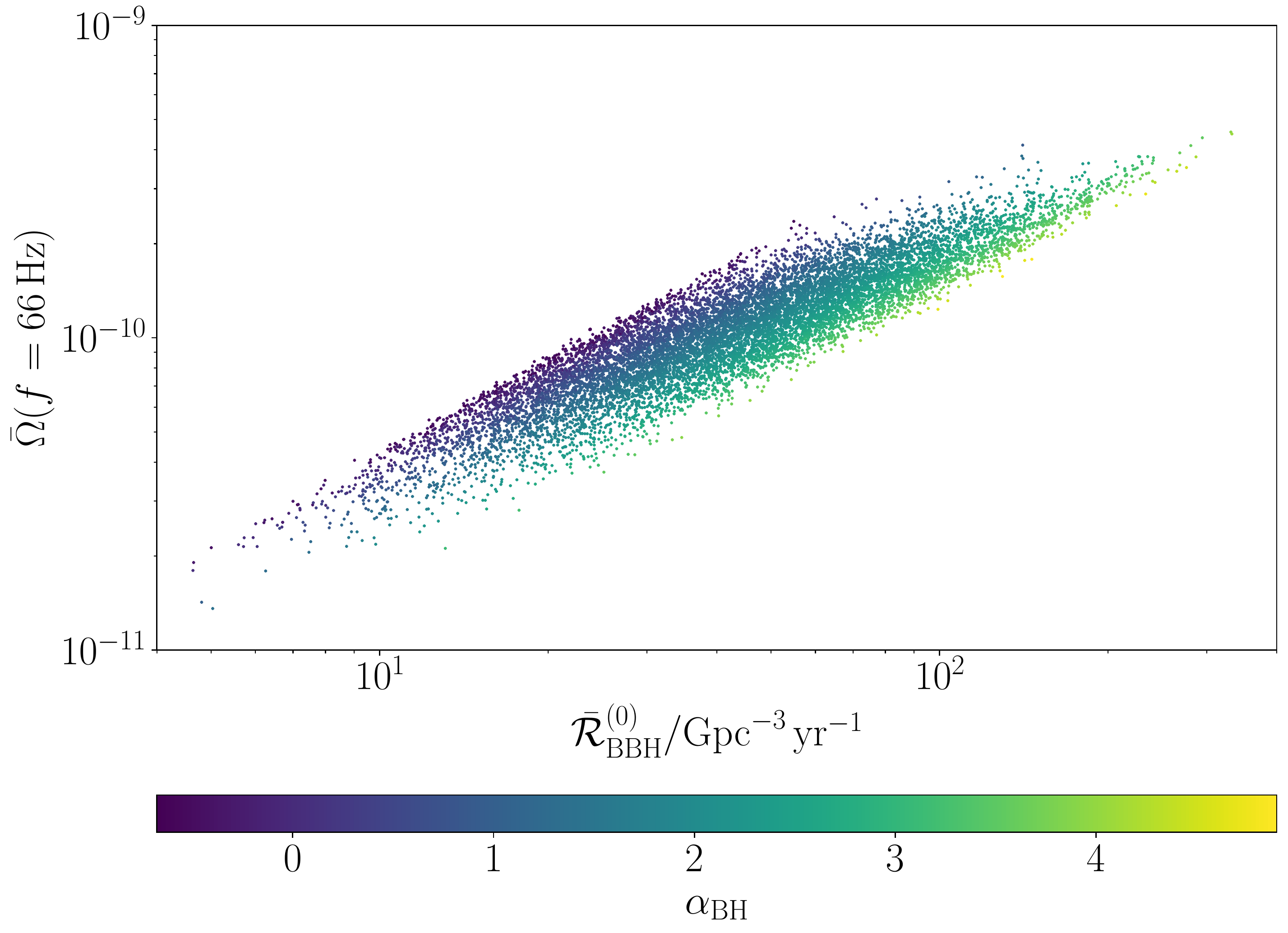}
    \end{center}
    \caption{%
    Scatter plot of the mean AGWB intensity (at a reference frequency of $66\,\mathrm{Hz}$) and local BBH rate density $\bar{\mathcal{R}}_\mathrm{BBH}^{(0)}$ in $\sim10^4$ samples from the distribution of possible BBH populations inferred from the first GW transient catalogue.
    Each sample is coloured according to its value of the BH mass power law index $\alpha_\mathrm{BH}$, with larger values corresponding to steeper fall-offs at large BH masses.
    }
    \label{fig:omega_bar-vs-rate-colour}
\end{figure}

There are clearly many modelling assumptions which go into our model for the AGWB and its angular power spectrum, of which perhaps the least well-determined at present are the rates and mass distributions we assume for the CBC populations.
In order to investigate the impact of these assumptions on our results, we can re-calculate the mean AGWB intensity and the angular power spectrum for a range of alternative rates and mass distributions, and see how much variation this causes in the final results.

\begin{figure}[t!]
    \begin{center}
        \includegraphics[width=0.85\textwidth]{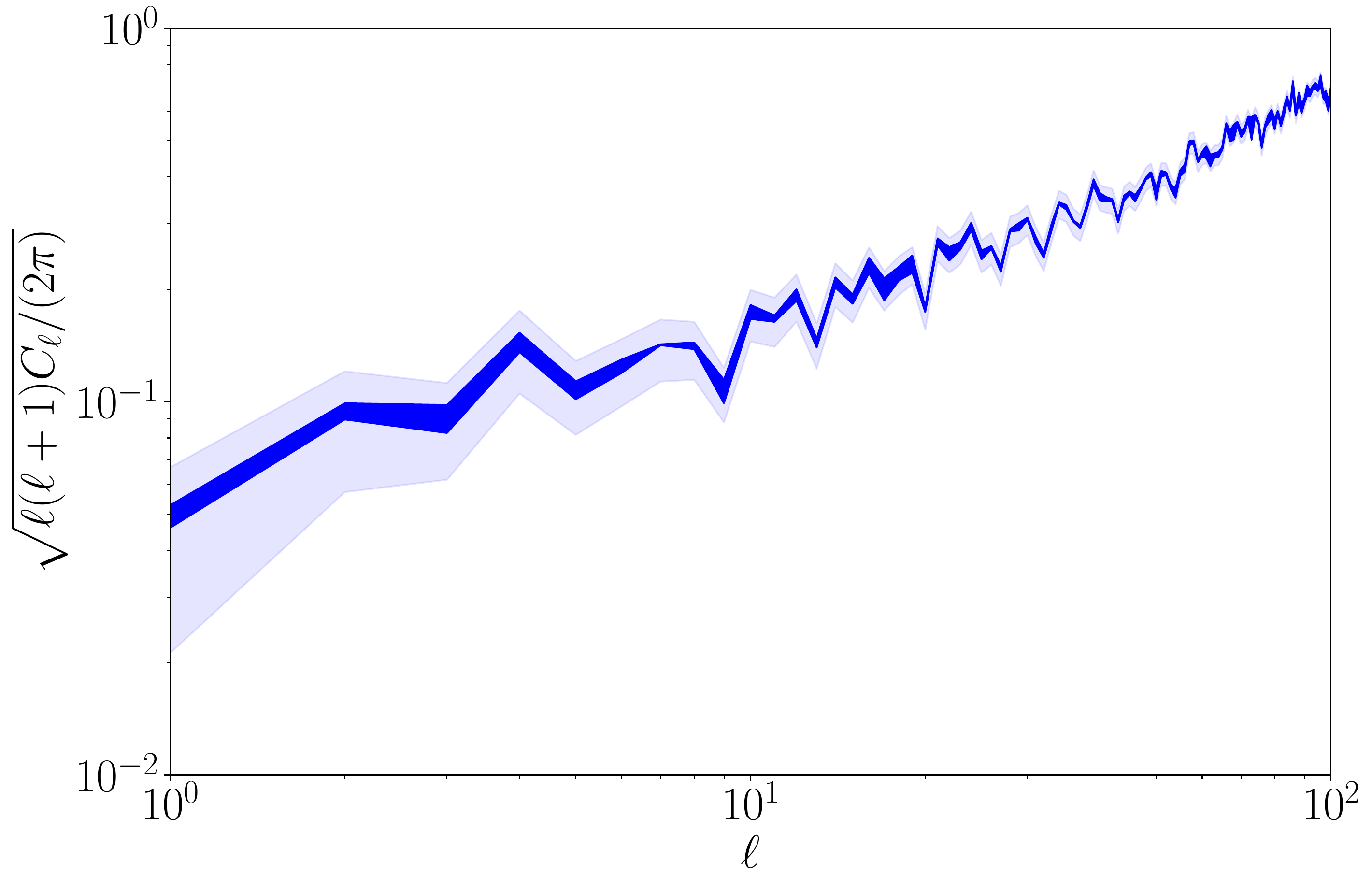}
    \end{center}
    \caption{%
    The AGWB angular power spectrum, normalised to the mean intensity $\bar{\Omega}$.
    While the light shaded region shows $1\sigma$ cosmic variance as before, the dark shaded region now shows the $3\sigma$ uncertainty region corresponding to the variation in the BBH population described in the main text.
    We see that this population uncertainty is always much smaller than cosmic variance (at least for the range of BBH population models considered here).
    }
    \label{fig:C_ell-astro-sigma}
\end{figure}

In figures~\ref{fig:omega_bar-vs-rate-colour} and~\ref{fig:C_ell-astro-sigma} we perform a scan over $\sim10^4$ different models for the BBH distribution, computing the isotropic component and angular power spectrum for each model.
Each set of model parameters is drawn from a posterior hyperparameter distribution inferred from the first five confident BBH detections reported by LIGO/Virgo (GW150914~\cite{Abbott:2016blz}, GW151226~\cite{Abbott:2016sjg}, GW170104~\cite{Abbott:2017bnn}, GW170608~\cite{Abbott:2017vox}, and GW170814~\cite{Abbott:2017ycc}),\footnote{This was the most complete study possible in Autumn 2018, when this work was first performed. Many more BBHs have been detected since that time~\cite{Abbott:2018mvr,Abbott:2020niy,Abbott:2021usb}.} using the methodology developed by \citet{Wysocki:2018mpo}.
For simplicity we assume a straightforward power-law distribution for the primary BH mass, $p(m_1)\propto m_1^{-\alpha_\mathrm{BH}}$, between a minimum mass of $5\,m_\odot$ and a maximum mass $m_\mathrm{max}$ that is allowed to vary.
The secondary BH mass is assumed to be uniformly distributed between $5\,m_\odot$ and $m_1$.
The parameters $\alpha_\mathrm{BH}$ and $m_\mathrm{max}$, along with the local BBH rate density $\bar{\mathcal{R}}_\mathrm{BBH}^{(0)}$, therefore fully specify the BBH distribution needed for our calculations, with all other details (star formation rate, metallicity threshold, etc.) held fixed.

In figure~\ref{fig:omega_bar-vs-rate-colour}, we see that this variation in the BBH population causes the mean AGWB intensity $\bar{\Omega}$ to vary by roughly an order of magnitude.
While there is a clear (and intuitively sensible) trend of $\bar{\Omega}$ increasing in line with the local rate density $\bar{\mathcal{R}}_\mathrm{BBH}^{(0)}$, there is still a significant amount of scatter around this trend.
Much of this scatter is captured by changes in the BH mass power law index $\alpha_\mathrm{BH}$, with shallower indices giving rise to stronger AGWB spectra.
Again, this makes intuitive sense, as decreasing $\alpha_\mathrm{BH}$ while holding the local rate density fixed effectively increases the relative importance of more massive BBHs, and these provide a much larger contribution in terms of GW energy radiated.
This interesting structure highlights the fact that, once detected, the isotropic component of the AGWB could provide interesting and useful astrophysical information about the CBC populations from which it is emitted.

In contrast with this large variation encoded in the isotropic component, we see in figure~\ref{fig:C_ell-astro-sigma} that, provided we normalise with respect to the isotropic amplitude, the AGWB angular power spectrum varies surprisingly little over the range of BBH population models we consider.
This shows us that slightly re-weighting different galaxies in terms of their BBH populations does little to change the overall clustering statistics of the signal.
As a result, our prediction for the angular power spectrum appears reasonably robust to population uncertainties, once one factors out the variation in the overall isotropic amplitude.

In the time since this work was first performed, the number of detected BBH events has grown significantly, and the GW astronomy community has begun to use ever more sophisticated models to capture their mass spectrum~\cite{Abbott:2020niy}.
As such, it would be interesting to repeat this exercise again in future work, to see how our evolving knowledge of BBHs (and other CBCs) affects our expectations for the AGWB and its anisotropies.
It would also be interesting to explore the impact of varying other model assumptions, particularly in ways which could have more influence on the relative contributions of different galaxy populations with different clustering properties.
For example, \citet{Adhikari:2020wpn} showed that if CBCs have longer delay times, then the GW events we observe today are concentrated in galaxies with larger stellar mass (since these dominate the star formation at higher redshifts), which live in more massive dark matter haloes and are therefore more tightly clustered.

\section{Shot noise in the astrophysical background}
\label{sec:shot-noise}

In sections~\ref{sec:characterising-anisotropies} and~\ref{sec:agwb}, we characterised the anisotropies in the AGWB in terms of their angular power spectrum components $C_\ell$.
These are statistical quantities, describing the expected angular correlation of the AGWB after averaging the signal in at least three distinct ways:
    \begin{enumerate}
        \item averaging over random realisations of LSS (i.e., over an \enquote{ensemble of Universes});
        \item averaging over the discrete positions of galaxies within the matter distribution to give a continuous galaxy number density field;
        \item averaging over the merger times of CBCs within each galaxy to give a mean merger rate.
    \end{enumerate}
However, the AGWB we observe corresponds to a single Universe, with some discrete number of galaxies, each with some discrete number of CBCs.
In practice, therefore, we are unable to perform the averaging process described above, leading to random fluctuations in the observed $C_\ell$'s, which must be accounted for when comparing to theoretical predictions.
What's more, our theoretical predictions in the previous section are themselves based on a simulated galaxy catalogue with a single realisation of LSS and a finite number of galaxies, so it is doubly important to understand these effects.

We have already accounted for the first point above: the uncertainty due to our observation of a single realisation of LSS, i.e., cosmic variance.
In section~\ref{sec:estimating-C_ells} we saw that, for a Gaussian AGWB, this uncertainty is given by $\Var[C_\ell]=C_\ell^2/(\ell+\sfrac{1}{2})$.
Besides cosmic variance, we must account for the fact that the AGWB is emitted from a finite number of galaxies (sampled from the underlying density field), each hosting a finite number of CBCs (sampled from the mean merger rate).
These two sampling processes, corresponding to the second and third points in the list above, follow Poissonian statistics and introduce \emph{shot noise} to the observed angular power spectrum.
This is a very important effect, which has been studied for decades in contexts such as galaxy redshift surveys~\cite{Feldman:1993ky,Hamilton:2005kz} and the cosmic infrared background~\cite{Kashlinsky:2018mnu}.
In the context of the AGWB, \citet{Meacher:2014aca} used numerical simulations to study the effects of shot noise on the isotropic component, but until now the effects of shot noise on the AGWB anisotropies have been ignored.

In this section, we derive expressions for the AGWB angular power spectrum in the presence of shot noise, and calculate the size of these shot-noise effects for our model of the AGWB in the LIGO/Virgo frequency band.
We then develop an optimal data-analysis method for inferring the \emph{true}, cosmological angular power spectrum in the presence of shot noise.

\subsection{The angular power spectrum, with and without shot noise}

We now show how the inclusion of generic shot-noise effects in the underlying statistics of the AGWB leads to an additional term in the angular power spectrum.
For later convenience, we write the intensity field as an integral over comoving distance,
    \begin{equation}
    \label{eq:Omega-integral-omega}
        \Omega(f,\vu*r)=\int\dd{r}r^2\omega(f,\vb*r),
    \end{equation}
    where $\omega$ is an \enquote{AGWB density} with dimensions $[\mathrm{length}]^{-3}$.
Comparing with equation~\eqref{eq:anisotropic-phinney-formula}, we see that
    \begin{equation}
    \label{eq:omega-definition}
        \omega(f,\vb*r)=\frac{2G}{3(1+z)^2}\qty(\frac{r_H}{r})^2\int\dd{\vb*\zeta}\mathcal{R}(\vb*r,\vb*\zeta)\dv{E}{(\ln f_\mathrm{s})},
    \end{equation}
    where $r_H\equiv1/H_0$ is the Hubble radius.
Note that the rate density $\mathcal{R}$ is now a random field in three dimensions, with mean equal to the isotropic but redshift-dependent value $\bar{\mathcal{R}}$ that we discussed in the previous section.
The SHCs are given in terms of the density $\omega$ by
    \begin{equation}
    \label{eq:spherical-harmonics}
        \Omega_{\ell m}\equiv\int_{S^2}\dd[2]{\vu*r}Y^*_{\ell m}(\vu*r)\Omega(\vu*r)=\int\dd[3]{\vb*r}Y^*_{\ell m}(\vu*r)\omega(\vb*r).
    \end{equation}

We calculate the angular power spectrum by specifying the second moment of the $\omega$ field.
Neglecting shot noise, this is simply
    \begin{equation}
    \label{eq:cov-lss}
        \Cov[\omega(\vb*r),\omega(\vb*r')]_\mathrm{LSS}=\bar{\omega}(r)\bar{\omega}(r')\xi(r,r',\theta),
    \end{equation}
    where $\xi$ is the two-point correlation function of $\omega$, describing the probability in excess of random of similar values of the field being clustered together.
Due to statistical isotropy, $\xi$ only depends on the angular positions $\vu*r,\vu*r'$ of the two points through their separation $\theta\equiv\cos^{-1}(\vu*r\vdot\vu*r')$.
(It does, however, depend on their radial distances, as these influence the intensity of the GW flux.)
Using equations~\eqref{eq:Omega-integral-omega},~\eqref{eq:C_ell-multipole-integral}, and~\eqref{eq:cov-lss}, we find the $C_\ell$'s in the absence of shot noise,
    \begin{equation}
    \label{eq:C_ell-lss}
        C_\ell^\mathrm{LSS}=2\uppi\int_{-1}^{+1}\dd{\cos\theta}P_\ell(\cos\theta)\int\dd{r}r^2\int\dd{r'}r'^2\bar{\omega}(r)\bar{\omega}(r')\xi(r,r',\theta).
    \end{equation}

We now introduce a shot-noise term that encompasses both the galaxy sampling and the CBC rate sampling.
Assuming these effects can jointly be treated as a local Poisson process that is independent of LSS, we have
    \begin{equation}
    \label{eq:cov}
        \Cov[\omega(\vb*r),\omega(\vb*r')]=\bar{\omega}(r)\bar{\omega}(r')\xi(r,r',\theta)+\mathcal{V}(r)\delta^3(\vb*r-\vb*r'),
    \end{equation}
    where $\mathcal{V}$ is some function describing the variance due to the finite sample, which is independent of direction due to statistical isotropy.
The form of this new shot-noise term (in particular, the fact that it is proportional to $\delta^3(\vb*r-\vb*r')$), is motivated by the equivalent expression for galaxy surveys (see, e.g., appendix~A of \citet{Feldman:1993ky}, or section~2.4 of \citet{Hamilton:2005kz}).
It is quite simple to convince oneself that the modification due to shot noise should take this form: there should be an extra term added to the covariance, as shot-noise fluctuations increase the variations in the measured values of $\omega$ throughout space, above the intrinsic variance due to the clustering of GW sources.
However, since the shot-noise fluctuations at one point in space are causally disconnected from those at any other point, the fluctuations at any two points are statistically independent, and the extra term in the covariance should vanish except when the two points are coincident, leading to the delta function.
Equation~\eqref{eq:cov} thus arises naturally from the superposition of independent Poisson processes at each point in space.
We flesh this argument out more quantitatively in section~\ref{sec:calc-shot-power} below.

Using equations~\eqref{eq:Omega-integral-omega},~\eqref{eq:C_ell-multipole-integral},~\eqref{eq:C_ell-lss}, and~\eqref{eq:cov}, as well as the fact that $P_\ell(1)=1$ for all $\ell$, the full angular power spectrum is then
    \begin{equation}
    \label{eq:lss+W}
        C_\ell=C_\ell^\mathrm{LSS}+\mathcal{W},\qquad\mathcal{W}\equiv\int\dd{r}r^2\mathcal{V}(r).
    \end{equation}
We therefore see that the shot noise generates a spectrally white (i.e., independent of $\ell$) contribution to the $C_\ell$'s.

\begin{figure}[p!]
    \begin{center}
    \includegraphics[width=0.667\textwidth]{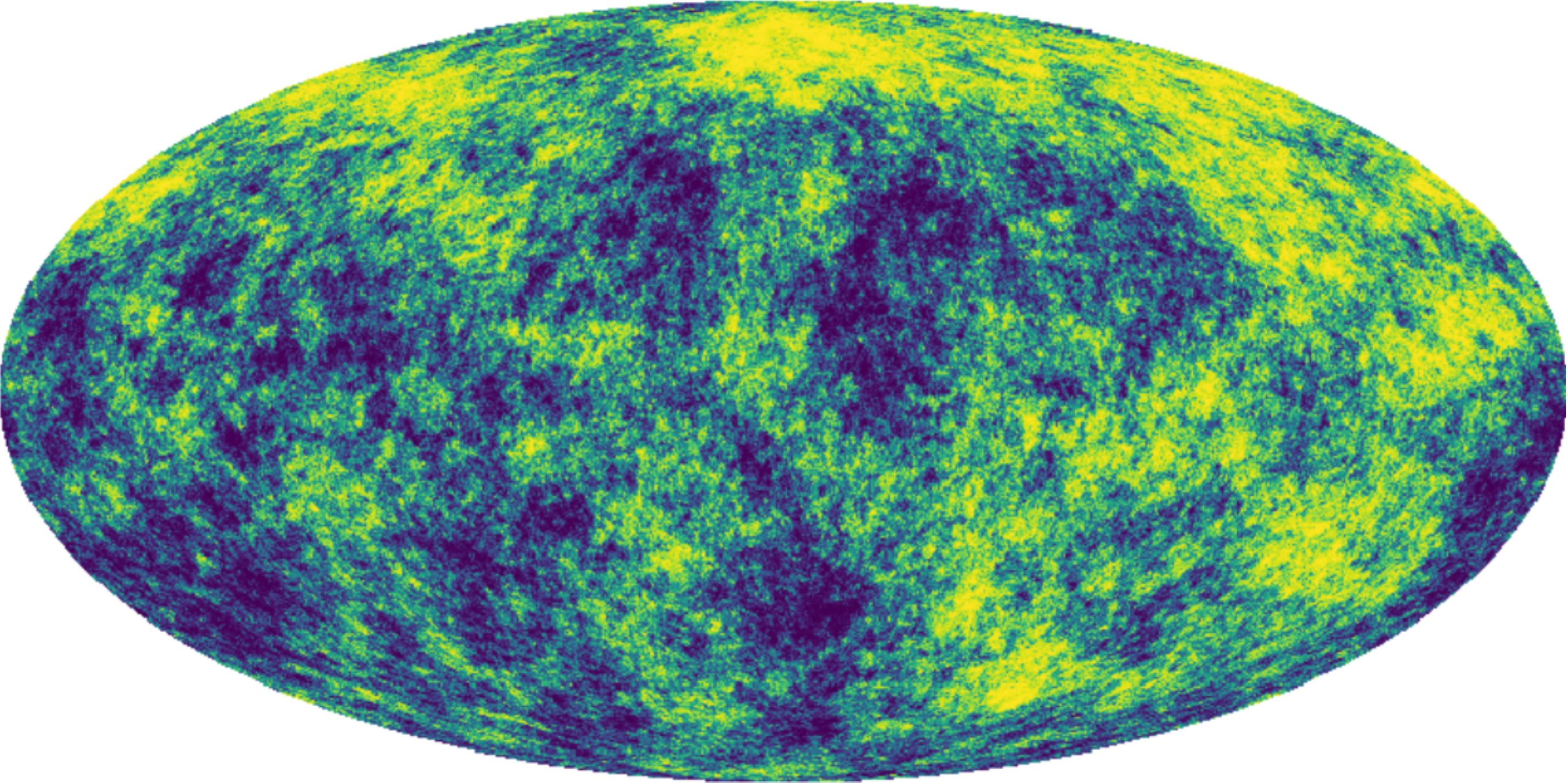}\\
    \includegraphics[width=0.667\textwidth]{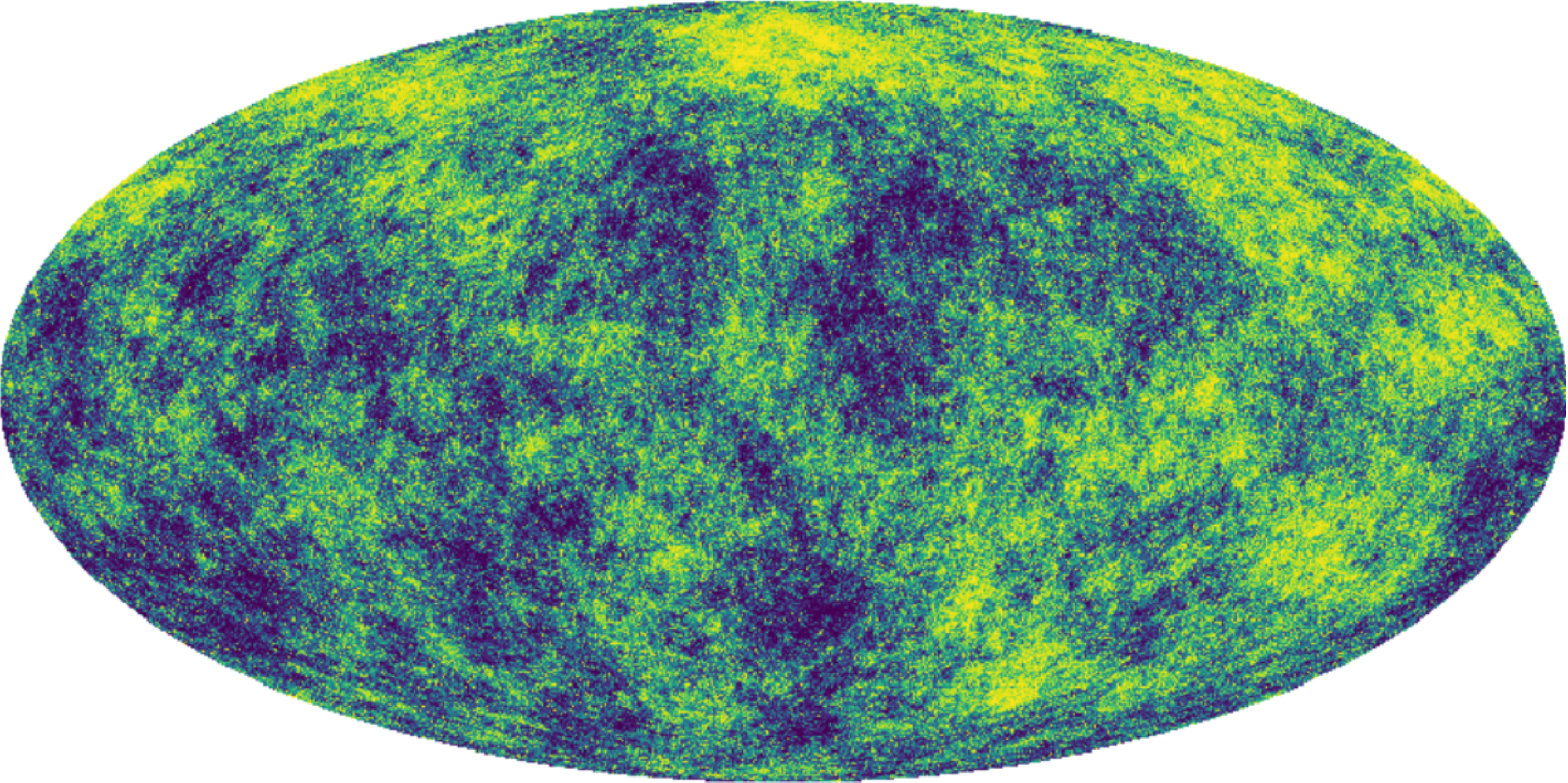}\\
    \includegraphics[width=0.667\textwidth]{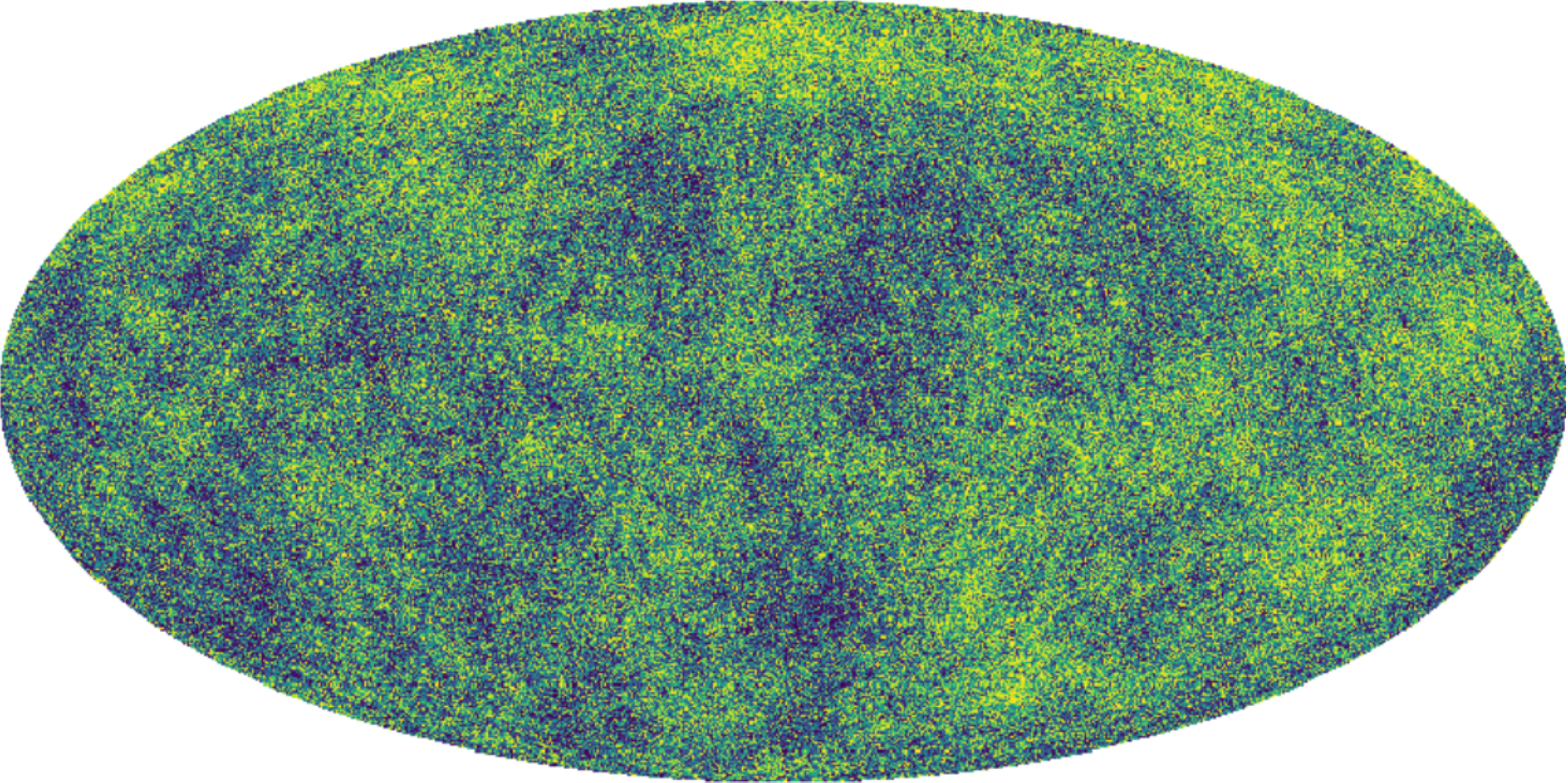}\\
    \includegraphics[width=0.667\textwidth]{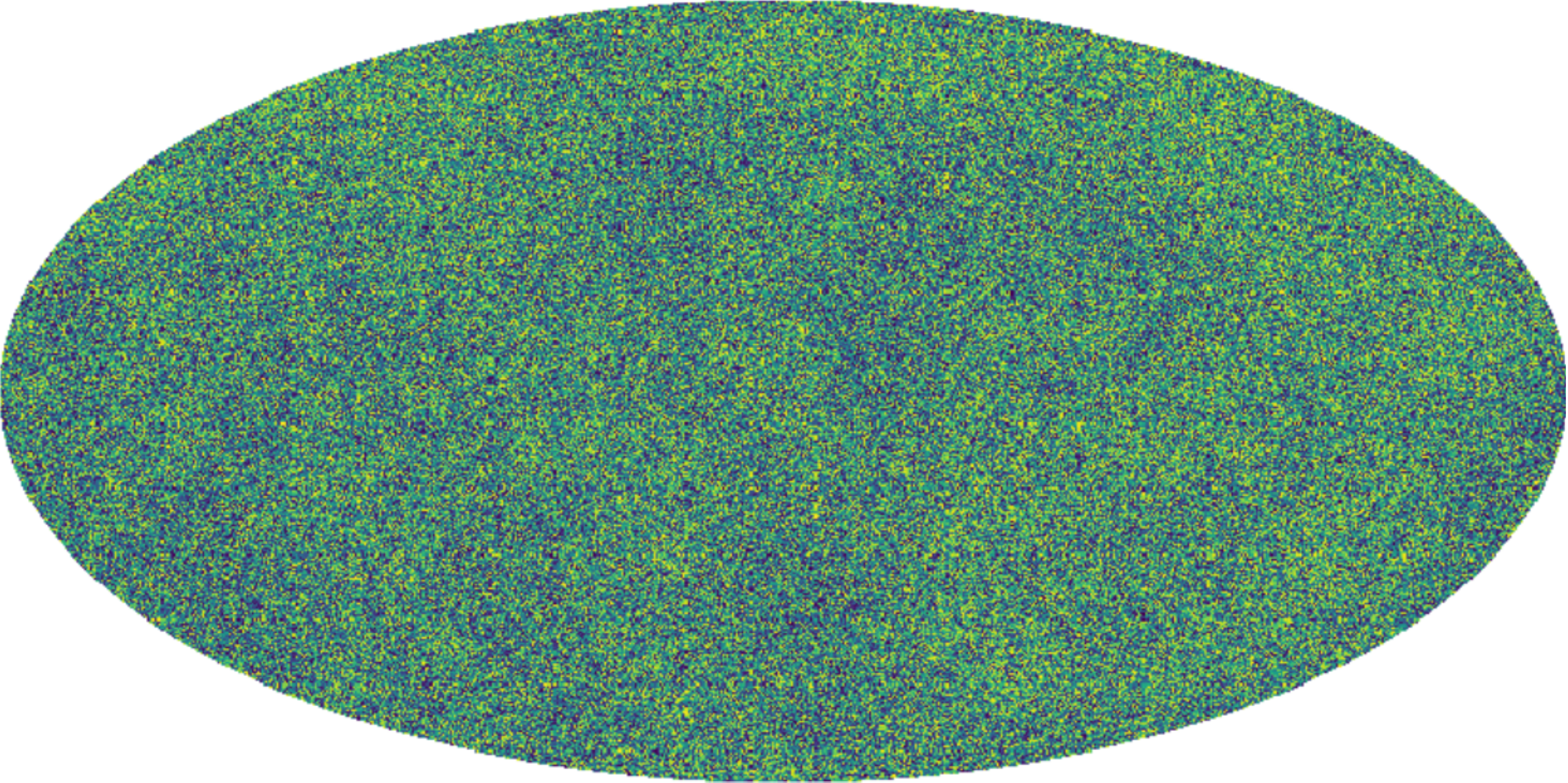}
    \end{center}
    \caption{%
    A toy-model depiction of shot noise.
    All four images are HEALPix maps with scale-invariant angular power spectra $\ell(\ell+1)C_\ell=\mathrm{constant}$, plus varying degrees of shot noise power.
    From top to bottom, the shot noise power is equal to $\mathcal{W}=0$, $\mathcal{W}=10^{-5}\bar{\Omega}^2$, $\mathcal{W}=10^{-4}\bar{\Omega}^2$, and $\mathcal{W}=10^{-3}\bar{\Omega}^2$.
    Physically, these represent different observations of the AGWB, with different observation time intervals, leading to different levels of shot noise power.
    All four maps have the same underlying random realisation of LSS, which is why the same large-scale features can be recognised in each of them.
    However, increasing the amount of shot noise leads to much stronger anisotropies on small scales, making it harder to discern the relatively subtle large-scale features.
    }%
    \label{fig:maps}
\end{figure}

\subsection{Calculating the shot-noise power}
\label{sec:calc-shot-power}

In order to evaluate equation~\eqref{eq:lss+W}, we must derive an expression for the local Poisson variance function $\mathcal{V}$, accounting for the random sampling of both the galaxy number density $n(\vb*r)$ and the CBC event rate per galaxy $R$, whose product gives the CBC rate density $\mathcal{R}=nR$.

Consider a volume element $\updelta V$ at position $\vb*r$.
We treat the number of galaxies in this region as a Poisson random variable,\footnote{%
    We stress that this is only an approximation.
    A more sophisticated approach would use the halo model of LSS~\cite{Cooray:2002dia}, accounting for the statistical properties of dark matter haloes and of different populations of galaxies within them.
    However, this approximation is sufficiently accurate for our purposes (particularly as the galaxy-number contribution to the shot noise is much less than the CBC rate contribution---see below).
    }%
     $N\sim\mathrm{Pois}[\updelta Vn(\vb*r)]$.
Assuming for now that these galaxies and the CBCs in them all have the same parameter values $\vb*\zeta$, the CBC event counts for each galaxy in a given source-frame time interval $T_\mathrm{s}$ are i.i.d. Poisson random variables, $\lambda_i\sim\mathrm{Pois}[RT_\mathrm{s}]$.
The total CBC event count from $\updelta V$ is then $\Lambda=\sum_{i=1}^N\lambda_i$.
This quantity---a sum over Poisson random variables, with a number of terms that is itself a Poisson random variable---follows a \emph{compound} Poisson distribution.
In appendix~\ref{sec:compound-poisson}, we show that the variance of this distribution is
    \begin{equation}
    \label{eq:var-Lambda}
        \Var[\Lambda]=\updelta V\bar{n}\qty[RT_\mathrm{s}+(RT_\mathrm{s})^2].
    \end{equation}

Taking the sampling as independent for different spatial volumes and for different parameter values, we therefore find
    \begin{equation}
        \Cov[\Lambda(\vb*r_i,\vb*\zeta),\Lambda(\vb*r_j,\vb*\zeta')]=\updelta V\bar{n}\qty[RT_\mathrm{s}+(RT_\mathrm{s})^2]\delta_{ij}\delta(\vb*\zeta,\vb*\zeta'),
    \end{equation}
    where we have introduced a Kronecker delta over the volume cells and a Dirac delta over parameter space to enforce this independence, and have averaged over LSS so that $n(\vb*r)\to\bar{n}(r)$.
Replacing $\Lambda$ with the comoving rate density, $\mathcal{R}=nR\simeq\Lambda/(\updelta VT_\mathrm{s})$, and taking $\updelta V\to0$, we are left with
    \begin{equation}
        \Cov[\mathcal{R},\mathcal{R}']_\mathrm{shot}=\qty(\frac{\bar{\mathcal{R}}}{T_\mathrm{s}}+\frac{\bar{\mathcal{R}}^2}{\bar{n}})\delta^{(3)}(\vb*r-\vb*r')\delta(\vb*\zeta,\vb*\zeta'),
    \end{equation}
    where $\mathcal{R}'$ is shorthand for $\mathcal{R}(\vb*r',\vb*\zeta')$, and where we have set $\delta_{ij}/\updelta V\to\delta^{(3)}(\vb*r-\vb*r')$.
The relationship between the source-frame time $T_\mathrm{s}$ and the observer-frame time $T$ will generally depend on cosmological metric perturbations and peculiar velocities at the location of the source (c.f. equation~\eqref{eq:source-frame-frequency} for the source-frame GW frequency), but to leading order it is simply $T_\mathrm{s}=T/(1+z)$.

Using equations~\eqref{eq:omega-definition} and~\eqref{eq:cov}, the shot-noise power is therefore
    \begin{equation}
    \label{eq:W-main}
        \mathcal{W}=\frac{4G^2}{9H_0^2}\int\frac{\dd{r}}{(1+z)^4}\qty(\frac{r_H}{r})^2\int\dd{\vb*\zeta}\qty(\frac{\bar{\mathcal{R}}}{T_\mathrm{s}}+\frac{\bar{\mathcal{R}}^2}{\bar{n}})\qty[\dv{E}{(\ln f_\mathrm{s})}]^2
    \end{equation}
Equation~\eqref{eq:W-main} is our main result, and has two distinct applications: ($i$) interpreting future observations of the AGWB, and ($ii$) improving our theoretical models.
For the first application, we can use equation~\eqref{eq:W-main} to calculate the expected shot noise in the observed $C_\ell$ spectrum as a function of observing time, given a model for the galaxy number density and CBC merger rate.
The $\bar{\mathcal{R}}/T_\mathrm{s}$ term dominates over the $\bar{\mathcal{R}}^2/\bar{n}$ term, since the latter is suppressed by a factor of $\bar{\mathcal{R}}T_\mathrm{s}/\bar{n}=RT_\mathrm{s}$, which is $\sim10^{-6}$ for a typical galaxy (i.e., the CBC rate per galaxy is on the order of $\sim1\,\mathrm{Myr}^{-1}$).
For the second application, we can use equation~\eqref{eq:W-main} to calculate the error inherent to our theoretically-predicted spectrum from the previous section due to the finite galaxy sampling in the simulated galaxy catalogue.
Since these predictions do not involve simulating the time of arrival of discrete GW signals, but average over the CBC rate, they exclude the shot noise due to sampling of this rate.
This is identical to taking the limit $T\to\infty$, meaning that the catalogue predictions contain shot noise due to the $\bar{\mathcal{R}}^2/\bar{n}$ term only.
We therefore distinguish between the \enquote{observational shot noise} and the \enquote{catalogue shot noise},
    \begin{align}
    \label{eq:W-obs-cat}
    \begin{split}
        \mathcal{W}_\mathrm{obs}&=\frac{4G^2}{9H_0^2T}\int\frac{\dd{r}}{(1+z)^3}\qty(\frac{r_H}{r})^2\int\dd{\vb*\zeta}\bar{\mathcal{R}}\qty[\dv{E}{(\ln f_\mathrm{s})}]^2,\\
        \mathcal{W}_\mathrm{cat}&=\frac{4G^2}{9H_0^2}\int\frac{\dd{r}}{(1+z)^4}\qty(\frac{r_H}{r})^2\int\dd{\vb*\zeta}\frac{\bar{\mathcal{R}}^2}{\bar{n}_\mathrm{cat}}\qty[\dv{E}{(\ln f_\mathrm{s})}]^2.
    \end{split}
    \end{align}
Here, $\bar{n}_\mathrm{cat}$ represents the galaxy number density in the catalogue, which is significantly less than the true galaxy number density $\bar{n}$ (this is because only galaxies brighter than a certain magnitude are included).

\subsection{Removing the foreground}

Inspecting equation~\eqref{eq:W-main}, we see that the integrand diverges as $r\to0$.
This is to be expected, for two reasons.
First, the Poisson statistics become progressively worse at small distances, as we are looking at smaller spherical shells that contain fewer galaxies, so the notion of a smooth number density $\bar{n}$ breaks down as $r\to0$.
Second, the contribution of a single CBC to the total AGWB flux becomes much larger at smaller distances, so a CBC that is arbitrarily close can bias the power spectrum by an arbitrarily large amount.

In order to regulate this divergence, it is necessary to introduce a cutoff distance $r_*$, below which we remove any CBC signals and do not consider them part of the AGWB.
(This is similar to what is done with, e.g.,~the cosmic infrared background~\cite{Kashlinsky:2018mnu}.)
We are free to choose the value of $r_*$, with larger values helping to reduce the shot noise as much as possible.
However, the choice of $r_*$ will also affect $C_\ell^\mathrm{LSS}$, and this must be accounted for when making theoretical predictions.

In principle, one can implement this foreground cut by removing from the GW strain time series $h(t)$ any intervals in which an individual CBC with a comoving distance $r<r_*$ is identified.
The CBC chirp signal encodes the luminosity distance $d_L$, which can easily be converted to the comoving distance $r$ by assuming a fiducial cosmology.
However, $d_L$ cannot be measured with arbitrary precision, particularly as it is degenerate with the inclination of the binary.
What's more, the detectability of CBCs at a given distance is also a function of sky position, due to the anisotropic beam pattern of the detectors.
Removing nearby sources based on a signal-to-noise threshold (as in \citet{Meacher:2014aca}) thus risks biasing the power spectrum by favouring particular kinds of CBC events and particular regions of the sky.
The alternative is to set $r_*$ small enough that \emph{all} CBCs at distances $r<r_*$ are detectable, so the foreground removal can be implemented in an unbiased and isotropic manner, without selection effects.
This would mean including some number of CBCs at $r>r_*$ in the stochastic analysis despite the fact that they can be individually resolved.
This proposal presents a serious data analysis challenge, and it is unclear whether it will be possible in practice to remove the foreground in a way which does not introduce directional biases.
Looking forward to third generation interferometers such as Einstein Telescope, we expect that essentially all BBHs in the observable Universe will be individually resolvable~\cite{Regimbau:2016ike,Sachdev:2020bkk,Sharma:2020btq,Maggiore:2019uih}, meaning that our framing here in terms of a resolvable foreground and a stochastic target signal will likely have to be modified.
Indeed, it will likely become preferable to study anisotropies in the distribution of resolved events, although some care will have to be taken in combining this with the large number of BNS and BHNS events that will remain unresolvable.
We leave a detailed examination of these issues for future work.

\begin{figure}[t!]
    \begin{center}
        \includegraphics[width=0.75\textwidth]{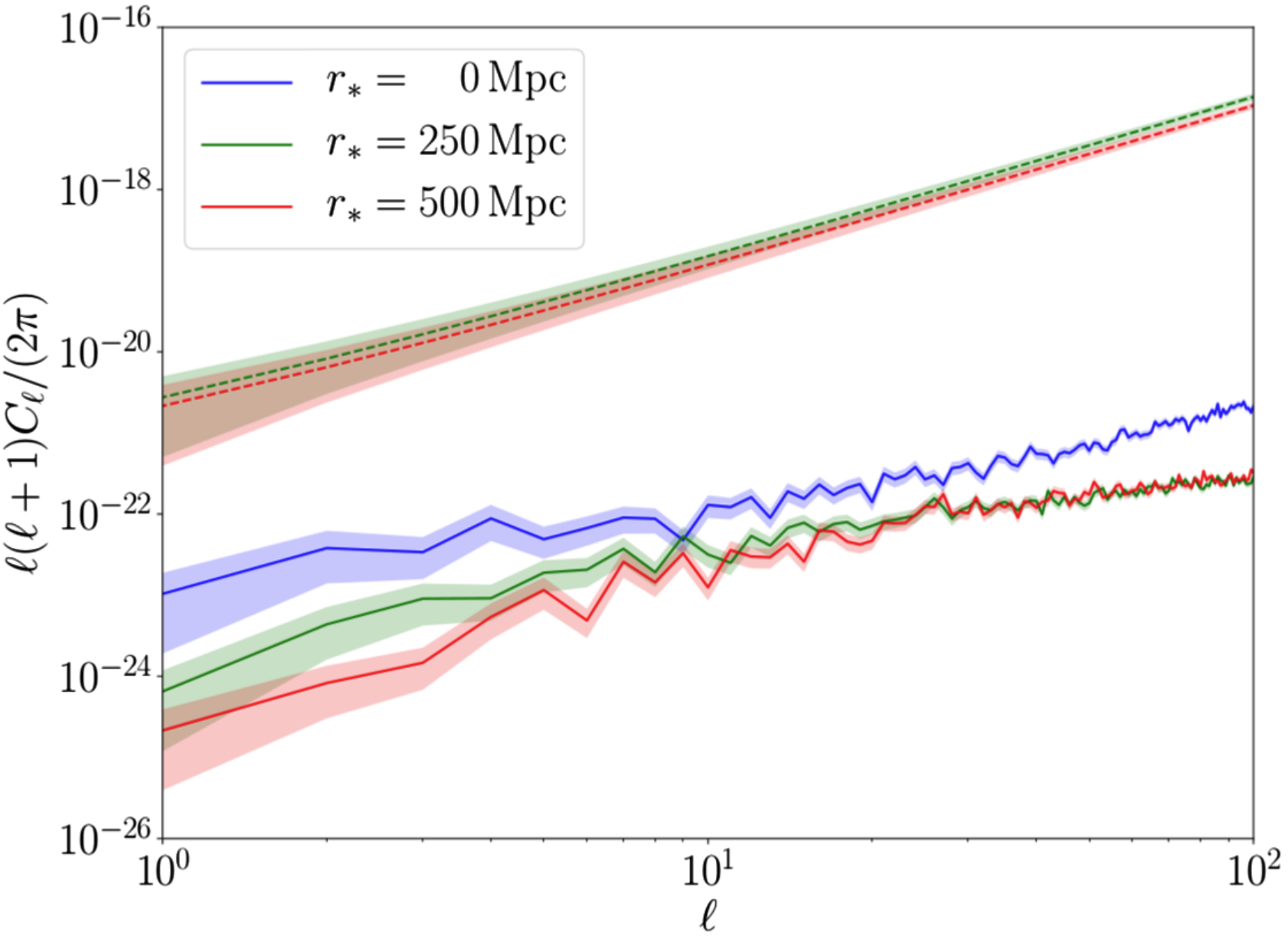}
    \end{center}
    \caption{%
    The AGWB angular power spectrum, plotted for different values of the cutoff distance $r_*$.
    The solid curves are calculated using a mock galaxy catalogue from the Millennium Simulation (with the catalogue error $\mathcal{W}_\mathrm{cat}$ removed, though this has negligible effect), and represent the cosmological power spectrum $C_\ell^\mathrm{LSS}$.
    The dashed curves include the observational shot noise after $T=1\,\mathrm{yr}$, giving the full power spectrum $C_\ell=C_\ell^\mathrm{LSS}+\mathcal{W}_\mathrm{obs}$.
    The shaded regions indicate cosmic variance.
    $\mathcal{W}_\mathrm{obs}$ is divergent in the case $r_*=0$, and so is not plotted.
    }
    \label{fig:C_ells_r_min}
\end{figure}

\subsection{Results and discussion}

Using the AGWB model described above, we can calculate the size of the observational and catalogue shot-noise terms from equation~\eqref{eq:W-obs-cat} for different values of the foreground cutoff distance $r_*$.
We find that the catalogue shot noise is at most $\mathcal{W}_\mathrm{cat}\approx3\times10^{-29}$, and is typically several orders of magnitude smaller than this for larger values of $r_*$.
Since the monopole is $\bar{\Omega}\approx10^{-10}$, this represents shot-noise fluctuations of $\lesssim0.01\%$.
This is negligible compared to the true power spectrum $C_\ell^\mathrm{LSS}$, so our earlier predictions based on the Millennium simulation are completely unaffected.

On the other hand, we find that the observational shot noise $\mathcal{W}_\mathrm{obs}$ is generically several orders of magnitude larger than the true angular power spectrum $C_\ell^\mathrm{LSS}$ for any reasonable values of $r_*$ and $T$.
(In fact, $\sqrt{\mathcal{W}_\mathrm{obs}}\approx\bar{\Omega}$, so the shot-noise fluctuations are typically as large as the monopole itself.)
This is illustrated in figure~\ref{fig:C_ells_r_min} for $r_*=250\,\mathrm{Mpc}$ and $r_*=500\,\mathrm{Mpc}$.
Note that $C_\ell^\mathrm{LSS}$ also changes with $r_*$, partly because the total GW emission is reduced, causing an overall reduction in the spectrum, and partly because the nearest galaxies contribute most strongly to the AGWB, so that their removal changes the shape of the spectrum.
Calculations of $C_\ell^\mathrm{LSS}$ using the catalogue are not reliable for values of $r_*$ significantly larger than $\sim500\,\mathrm{Mpc}$, due to the incompleteness of the catalogue at high redshift.
However, our calculations indicate that even increasing $r_*$ to $2\,\mathrm{Gpc}$ can only reduce $\mathcal{W}_\mathrm{obs}$ by less than an order of magnitude, so this seems unlikely to solve the problem.

The numerical values given in figure~\ref{fig:C_ells_r_min} depend on the details of the AGWB model, and include a multitude of random and systematic uncertainties in, e.g., the populations of astrophysical sources that contribute, their emission rates, and the nature of their clustering.
However, we stress that the main result, equation~\eqref{eq:lss+W}, is generic, and is grounded in simple and realistic physical principles.
Any finite population of sources will have random Poissonian fluctuations.
If these fluctuations are statistically independent at different spatial locations (i.e., if the shot-noise fluctuations are causally disconnected), then the angular power spectrum generically gains an extra white-noise component, $\mathcal{W}$.
If these sources are finite in time, then basic Poisson statistics dictates that this noise decays as the inverse of the observation time, $\mathcal{W}\propto1/T$.

\subsection{Temporal shot noise and hierarchical averaging}
\label{sec:shot}

Now that we have diagnosed the issue of shot noise, we turn our attention to possible ways in which it can be mitigated when carrying out searches for AGWB anisotropies.
In order to do this, it is helpful to first think more carefully about the statistics of the observed AGWB intensity under shot noise fluctuations, and how these statistics are distinct from those associated with LSS.

We have shown above that, for the AGWB, the \emph{temporal} shot noise associated with finite CBCs per observation time is significantly larger than the true $C_\ell$ spectrum, while the \emph{spatial} shot noise associated with a finite galaxy number density is significantly smaller; we therefore focus exclusively on the former in this analysis.
This allows us to exploit the fact that we have observational access to multiple realisations of the temporal shot noise: for each successive observation interval $\tau$, we can obtain a set of SHCs with shot noise power $\mathcal{W}_\tau\propto1/\tau$.
The shot-noise fluctuations in each set of SHCs are associated with a different set of CBCs in a different set of galaxies, so it is immediately clear that each successive shot noise realisation is statistically independent of the others.
This is in contrast with the spatial shot noise, for which we can only observe a single realisation (in other words, the random positions of galaxies are a \emph{persistent} random field, while the random sky locations of CBCs form a \emph{transient} random field).

\begin{figure}[t!]
    \includegraphics[width=\textwidth]{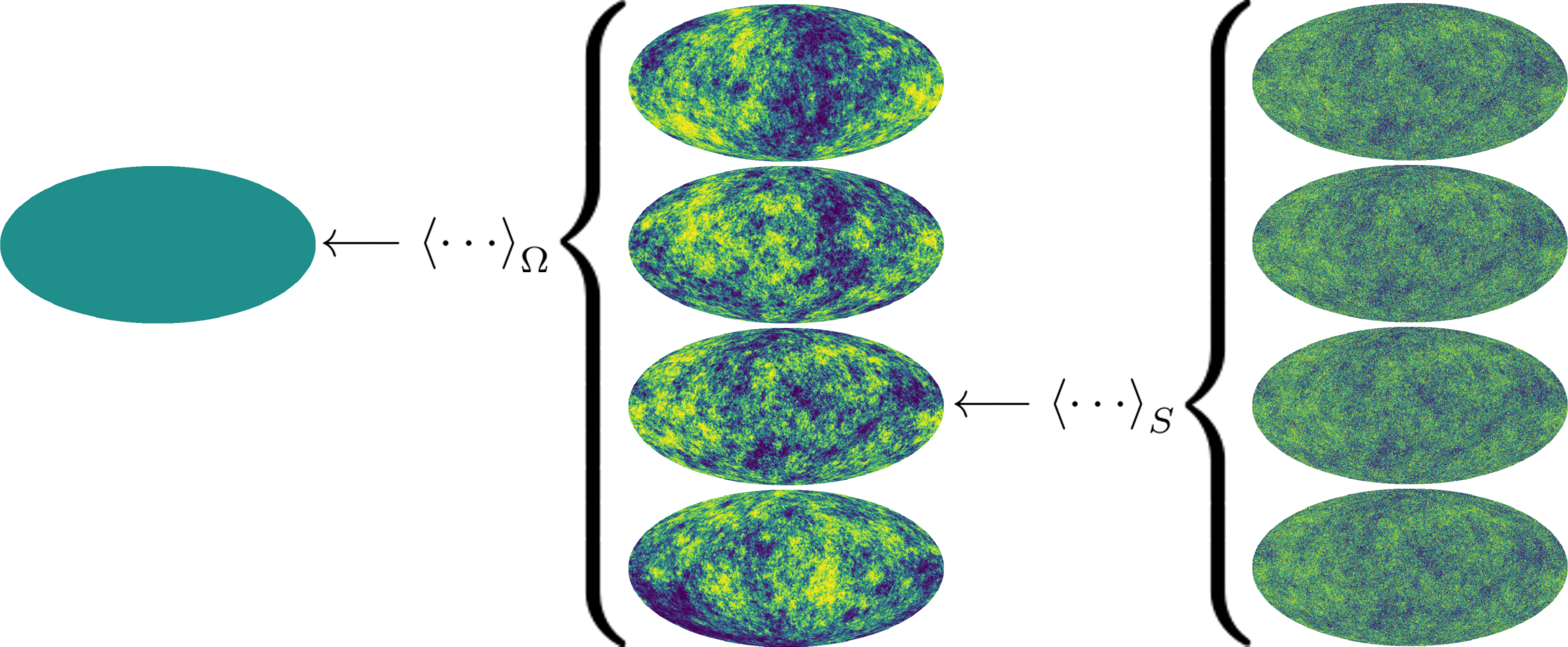}
    \caption{%
    An illustration of the two averaging procedures introduced in Sec.~\ref{sec:shot}.
    The right column shows four independent realisations of shot noise (at a level of $\mathcal{W}=10^{-4}\bar{\Omega}^2$) for a single given realisation of LSS.
    Averaging over many such realisations corresponds to the $\ev{\cdots}_S$ operation.
    Similarly, the central column shows four independent realisations of LSS, with zero shot noise.
    Averaging over many such realisations corresponds to the $\ev{\cdots}_\Omega$ operation, and results in a perfectly uniform field (i.e. all SHCs equal to zero, except the monopole).
    }%
    \label{fig:ensembles}
\end{figure}

There are two logically distinct random processes that govern the observed SHCs: the distribution of matter on large scales, and the emission of a finite number of GW signals from this matter distribution in a given observation period.
We model these processes, and their corresponding ensemble averages, in a hierarchical manner.
    \begin{enumerate}
        \item The true $\Omega_{\ell m}$ are drawn from Gaussian distributions with variance $C_\ell$.
        This process is associated with the cosmological averaging operation from section~\ref{sec:characterising-anisotropies},
            \begin{equation}
                \ev{\cdots}_\Omega\equiv\text{cosmological average},
            \end{equation}
            which can be thought of as an average over an \enquote{ensemble of Universes}, with each Universe having a distinct random realisation of LSS.
        (Of course, we only have access to a single such realisation, and this gives rise to cosmic variance.)
        \item The true $\Omega_{\ell m}$ are modulated by shot noise, so that a set of \enquote{noisy} SHCs $\Omega^i_{\ell m}$ is drawn from a distribution with the true components $\Omega_{\ell m}$ as its mean.
        This draw is independent for each time interval, with different intervals being labelled by the index $i$.
        We write the associated average over shot noise realisations as
            \begin{equation}
                \ev{\cdots}_S\equiv\text{shot noise average}.
            \end{equation}
        Physically, this reflects the fact that a given cosmological distribution of galaxies can correspond to many different GWB realisations, as the number and times-of-arrival of transient GW signals from each galaxy are essentially random.
    \end{enumerate}
Figure~\ref{fig:ensembles} is a schematic representation of the two averaging procedures.

More quantitatively, the first two moments of the noisy SHCs under the shot noise average for a fixed realisation of LSS are
    \begin{equation}
        \ev{\Omega^i_{\ell m}}_S=\Omega_{\ell m},\qquad\Cov[\Omega^i_{\ell m},\Omega^j_{\ell'm'}]_S=\delta_{\ell\ell'}\delta_{mm'}\delta_{ij}\mathcal{W}_\tau.
    \end{equation}
The first equality holds by definition, while the second states that each SHC in each time interval has equal shot noise power, and is uncorrelated with all the others.\footnote{%
    The lack of correlation between time intervals is due to the acausal relationship between distant GW sources, while the lack of correlation between different $\ell m$ is due to statistical isotropy.}
Averaging also over realisations of LSS, we find
    \begin{equation}
    \label{eq:cross-spectrum}
        \ev{\Omega^i_{\ell m}}_{S,\Omega}=0,\qquad\Cov[\Omega^i_{\ell m},\Omega^j_{\ell'm'}]_{S,\Omega}=\delta_{\ell\ell'}\delta_{mm'}\qty(C_\ell+\delta_{ij}\mathcal{W}_\tau),
    \end{equation}
    where we have used equation~\eqref{eq:shc-moments}, and have introduced the shorthand $\ev{\cdots}_{S,\Omega}\equiv\ev{\ev{\cdots}_S}_\Omega$.\footnote{%
    Note that $\Cov[X,Y]_{S,\Omega}\equiv\ev{XY^*}_{S,\Omega}-\ev{X}_{S,\Omega}\ev{Y^*}_{S,\Omega}$, which is \emph{not} equal to $\ev{\Cov[X,Y]_S}_\Omega$.}
(Recall that we are focusing on $\ell>0$, so the SHCs all have zero mean.)

\subsection{Mitigating the shot noise}
\label{sec:estimator}

In this section, we use \eqref{eq:cross-spectrum} to define a function of the noisy SHCs $\Omega_{\ell m}^i$ that is an unbiased estimator of the true angular power spectrum $C_\ell$ in the presence of shot noise.
We then show that (in the appropriate limit) this is the minimum-variance unbiased estimator (MVUE) for the true angular power spectrum.

We start by modifying the standard autocorrelation estimator \eqref{eq:standard-estimator}, forming a set of cross-correlations between different time intervals,
    \begin{equation}
        \hat{C}_\ell^{ij}\equiv\frac{1}{2\ell+1}\sum_{m=-\ell}^{+\ell}\Omega_{\ell m}^i\Omega_{\ell m}^{j*}.
    \end{equation}
These are unbiased if and only if $i\ne j$,
    \begin{equation}
        \ev{\hat{C}_\ell^{ij}}_{S,\Omega}=C_\ell+\delta_{ij}\mathcal{W}_\tau.
    \end{equation}
We can combine these estimators in much the same way that we combine multiple \enquote{naive} estimators \eqref{eq:naive-estimator} to form the standard estimator \eqref{eq:standard-estimator}.
Suppose that our total observing time is $T$, so that there are $N_\tau\equiv T/\tau$ time segments.
Then we have $N_\tau\qty(N_\tau-1)/2$ pairs $(i,j)$ for which $i\ne j$.
Summing over these, we define the combined estimator
    \begin{equation}
    \label{eq:new-estimator}
        \hat{C}_\ell\equiv\frac{2}{N_\tau\qty(N_\tau-1)}\sum_{ij}\hat{C}_\ell^{ij},
    \end{equation}
    where we introduce the shorthand $\sum_{ij}\equiv\sum_{i=1}^{N_\tau}\sum_{j=i+1}^{N_\tau}$.
This is unbiased, $\ev*{\hat{C}_\ell}_{S,\Omega}=C_\ell$, with variance given by
    \begin{equation}
    \label{eq:var-4th-moment}
        \Var[\hat{C}_\ell]_{S,\Omega}=\qty[\frac{2}{N_\tau\qty(N_\tau-1)\qty(2\ell+1)}]^2\sum_m\sum_{m'}\sum_{ij}\sum_{i'j'}\Cov\qty[\Omega_{\ell m}^i\Omega_{\ell m}^{j*},\Omega_{\ell m'}^{i'}\Omega_{\ell m'}^{j'*}]_{S,\Omega}.
    \end{equation}

Evaluating \eqref{eq:var-4th-moment} requires us to evaluate the fourth moment of the noisy SHCs.
This would be trivial if the SHCs were all Gaussian, but we must account for the Poisson-like nature of the shot noise.
In appendix~\ref{sec:compound-poisson} we calculate the fourth moment using the same statistical model for the CBC rate density as in section~\ref{sec:calc-shot-power}; this results in
    \begin{equation}
    \label{eq:var-main-result}
        \Var[\hat{C}_\ell]_{S,\Omega}=\frac{2}{2\ell+1}\qty[C_\ell^2+\frac{2\mathcal{W}_\tau C_\ell}{N_\tau}+\frac{\mathcal{W}_\tau^2}{N_\tau\qty(N_\tau-1)}].
    \end{equation}
This expression is tied to the fact that we have excluded the on-diagonal terms $i=j$ when constructing \eqref{eq:new-estimator}; otherwise, there would be additional contributions to the variance (the estimator would also no longer be unbiased).
Note that since $\mathcal{W}_\tau\propto1/\tau$, we have $\mathcal{W}_\tau/N_\tau\propto1/T$.
This means that we can't \enquote{win} by decreasing the length of the data segments $\tau$, only by increasing the total observing time $T$.
In fact, writing $\mathcal{W}_T=\mathcal{W}_\tau\qty(\tau/T)=\mathcal{W}_\tau/N_\tau$, we see that in the limit where $N_\tau\gg1$ (i.e., the limit in which we minimise the contribution of auto-power from each data segment), \eqref{eq:var-main-result} becomes
    \begin{equation}
    \label{eq:var-limit}
        \Var[\hat{C}_\ell]_{S,\Omega}\simeq\frac{2}{2\ell+1}\qty(C_\ell+\mathcal{W}_T)^2.
    \end{equation}
This is exactly the standard cosmic variance expression from \eqref{eq:standard-estimator-variance}, but with $C_\ell$ replaced by $C_\ell+\mathcal{W}_T$.
In section~\ref{sec:min-var} below, we show that this is in fact the minimum possible variance of any unbiased estimator for the $C_\ell$'s in the presence of shot noise, saturating the Cram\'er-Rao bound.
The estimator \eqref{eq:new-estimator} is therefore the MVUE in the limit $N_\tau\gg1$.

\begin{figure}[t!]
    \includegraphics[width=\textwidth]{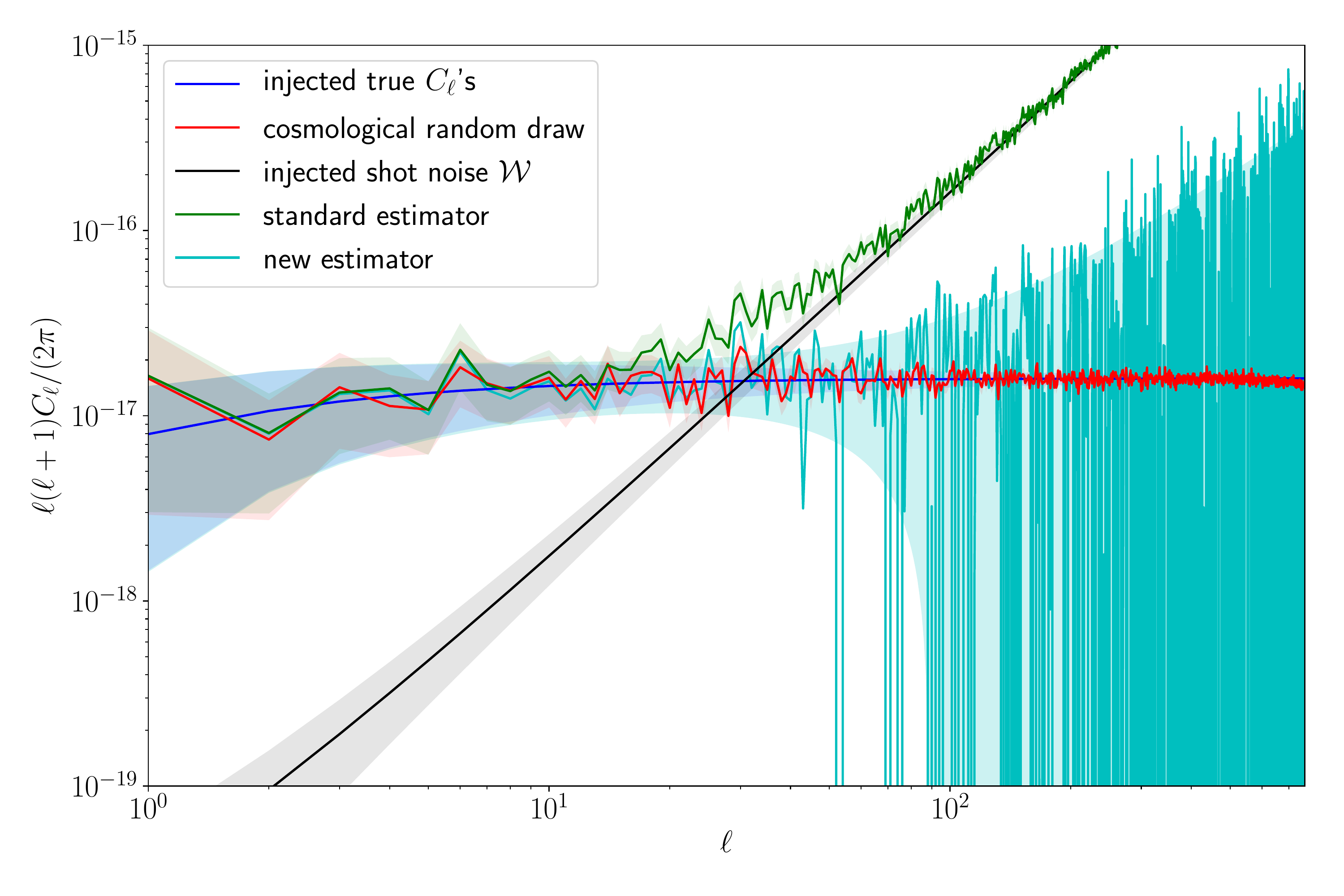}
    \caption{%
    Simulated angular power spectra using the standard estimator~\eqref{eq:standard-estimator} and the new estimator~\eqref{eq:new-estimator}.
    The dark blue line is the chosen \enquote{true} spectrum to be estimated, which is here taken as scale-invariant for simplicity, $\ell\qty(\ell+1)C_\ell\approx\mathrm{constant}$.
    The red line is the spectrum of a single random cosmological realisation of the AGWB (i.e. a single Universe), distributed around the dark blue line according to cosmic variance.
    The black line is the shot-noise power $\mathcal{W}$, here set to $10^{-3}$ times the monopole.
    The green line is the spectrum resulting from the standard estimator \eqref{eq:standard-estimator} for a single random realisation from the shot noise ensemble, which follows the sum of the true spectrum and the shot-noise power, $C_\ell+\mathcal{W}$.
    The cyan line is the spectrum resulting from the new estimator \eqref{eq:new-estimator}, for the same shot noise realisation, subdivided into $N_\tau=10$ independent segments.
    The shaded regions in all cases show the $1\sigma$ uncertainty, which for the cyan line is given by \eqref{eq:var-main-result}.
    }%
    \label{fig:new-estimator}
\end{figure}

At the opposite extreme, for the minimum number of segments, $N_\tau=2$, the variance is nearly twice as large (taking $\mathcal{W}_T\gg C_\ell$),
    \begin{equation}
        \Var[\hat{C}_\ell]_{S,\Omega}\approx\frac{4}{2\ell+1}\mathcal{W}_T^2.
    \end{equation}
An illustrative example for $N_\tau=10$ is shown in figure~\ref{fig:new-estimator}.

The term $\mathcal{W}_T$ in \eqref{eq:var-limit} is the same as that appearing in the mean of the standard estimator, $\ev*{C_\ell^{(\mathrm{std})}}_{S,\Omega}=C_\ell+\mathcal{W}_T$, so this new optimal estimator is still affected by the presence of shot noise; the crucial improvement is that the shot noise only adds to the variance of the estimator, and does not bias the spectrum as in the standard case.

\subsection{Minimum-variance estimation of the angular power spectrum}
\label{sec:min-var}

Recall that in section~\ref{sec:estimating-C_ells} we used the Gaussian likelihood~\eqref{eq:gaussian-likelihood} to show that the standard cosmic variance expression~\eqref{eq:standard-estimator-variance} saturates the Cram\'er-Rao bound, making it the MVUE for the angular power spectrum in the absence of shot noise.
Now we include shot noise, and consider the noisy SHCs, $\Omega_{\ell m}^i$.
Though we know these are not Gaussian (see appendix~\ref{sec:compound-poisson}), the Gaussian case is by far the most tractable, so we consider it first.
The joint Gaussian log-likelihood is fully specified by \eqref{eq:cross-spectrum},
    \begin{align}
    \label{eq:log-likelihood}
        \mathcal{L}=-\frac{1}{2}\sum_{\ell m}\qty[\ln(\mathrm{det}\qty(2\uppi\mathsf{C}_\ell))+\vb*\Omega_{\ell m}^\dagger\mathsf{C}_\ell^{-1}\vb*\Omega_{\ell m}],
    \end{align}
    where $\vb*\Omega_{\ell m}=(\Omega^1_{\ell m},\ldots,\Omega^{N_\tau}_{\ell m})$ is a vector of the noisy SHCs for a given $\ell m$, and $\mathsf{C}_\ell$ is the corresponding $N_\tau\times N_\tau$ covariance matrix,
    \begin{equation}
        \mathsf{C}_\ell=
        \begin{pmatrix}
            C_\ell+\mathcal{W}_\tau & C_\ell & \cdots & C_\ell \\
            C_\ell & C_\ell+\mathcal{W}_\tau & \cdots & C_\ell \\
            \vdots & \vdots & \ddots & \vdots \\
            C_\ell & C_\ell & \cdots & C_\ell+\mathcal{W}_\tau
        \end{pmatrix}.
    \end{equation}
One can show that this has determinant $\mathrm{det}\,\mathsf{C}_\ell=\qty(N_\tau C_\ell+\mathcal{W}_\tau)\mathcal{W}_\tau^{N_\tau-1}$, and inverse
    \begin{equation}
        \mathsf{C}_\ell^{-1}=\frac{1}{\qty(N_\tau C_\ell+\mathcal{W}_\tau)\mathcal{W}_\tau}
        \begin{pmatrix}
            \qty(N_\tau-1)C_\ell+\mathcal{W}_\tau & -C_\ell & \cdots & -C_\ell \\
            -C_\ell & \qty(N_\tau-1)C_\ell+\mathcal{W}_\tau & \cdots & -C_\ell \\
            \vdots & \vdots & \ddots & \vdots \\
            -C_\ell & -C_\ell & \cdots & \qty(N_\tau-1)C_\ell+\mathcal{W}_\tau
        \end{pmatrix},
    \end{equation}
    using the matrix determinant lemma and the Sherman-Morrison formula, respectively.
Taking the second derivative with respect to $C_\ell$, we therefore find
    \begin{align}
    \begin{split}
    \label{eq:cramer-rao-shot-noise}
        \Var[\hat{C}_\ell]_{S,\Omega}&\ge-\ev{\pdv[2]{\mathcal{L}}{C_\ell}}^{-1}_{S,\Omega}=-\ev{\sum_{m=-\ell}^{+\ell}\frac{1}{2\qty(C_\ell+\mathcal{W}_T)^2}-\frac{1}{N_\tau^2\qty(C_\ell+\mathcal{W}_T)^3}\sum_{i=1}^{N_\tau}\sum_{j=1}^{N_\tau}\Omega^{i*}_{\ell m}\Omega^{j}_{\ell m}}^{-1}_{S,\Omega}\\
        &=\qty[\sum_{m=-\ell}^{+\ell}-\frac{1}{2\qty(C_\ell+\mathcal{W}_T)^2}+\frac{1}{N_\tau^2\qty(C_\ell+\mathcal{W}_T)^3}\sum_{i=1}^{N_\tau}\sum_{j=1}^{N_\tau}\qty(C_\ell+\delta_{ij}\mathcal{W}_\tau)]^{-1}\\
        &=\qty[\sum_{m=-\ell}^{+\ell}\frac{1}{2\qty(C_\ell+\mathcal{W}_T)^2}]^{-1}=\frac{2}{2\ell+1}\qty(C_\ell+\mathcal{W}_T)^2,
    \end{split}
    \end{align}
    so the Cram\'er-Rao bound is the same as before, but with $C_\ell\to C_\ell+\mathcal{W}_T$.
This is exactly the variance we derived for our estimator in \eqref{eq:var-limit}.

Equation \eqref{eq:cramer-rao-shot-noise} was derived using the Gaussian log-likelihood \eqref{eq:log-likelihood}, and one may worry about whether it holds in the case we are interested in, given that the noisy SHCs do not follow a Gaussian distribution.
However, we have seen that our estimator saturates this bound in the limit $N_\tau\gg1$.
As shown in, e.g., \citet{Jaynes:2003}, the only probability distribution that saturates the Cram\'er-Rao bound under a given set of constraints (e.g., the constraints on the first two moments in \eqref{eq:cross-spectrum}) is that which maximises the entropy under those constraints.
The maximum-entropy distribution with fixed mean and variance is Gaussian~\cite{Jaynes:2003}, so this shows that our estimator must be Gaussian\footnote{At least, under the shot noise distribution---the LSS distribution is non-Gaussian as discussed above, but we ignore this complication for now.} in the limit $N_\tau\gg1$, and that the calculations above are valid in that limit.
(The approach to Gaussianity for $N_\tau\gg1$ can also be shown using the central limit theorem.)

\section{Summary and outlook}
\label{sec:anisotropies-summary}

In this chapter, we have investigated the angular power spectrum of the astrophysical gravitational-wave background from stellar-mass compact binary coalescences in the LIGO/Virgo frequency band, with the goal of using this signal as a novel probe of late-Universe cosmology.
Our key result is the predicted spectrum shown in figure~\ref{fig:C_ell-cbcs}, which is constructed using a full-sky mock lightcone galaxy catalogue based on the Millennium simulation.
Applying simple astrophysical recipes, we are able to model the GW emission from CBCs in all $\sim5.7\times10^6$ galaxies in this catalogue, allowing us to construct a full-sky simulated map of the AGWB (shown in figure~\ref{fig:map}), from which we calculate the corresponding angular power spectrum (shown in figure~\ref{fig:C_ell-cbcs}).
We find that the predicted anisotropies are much larger than in early-Universe observables such as the CMB, and that the one-point statistics of the AGWB map are strongly non-Gaussian, both of which can be understood in terms of the nonlinear gravitational dynamics that generate the clustering signal.
We have also calculated the kinematic dipole anisotropy associated with the peculiar motion of our detectors with respect to the cosmic rest frame; however, this is an order of magnitude smaller than the \enquote{intrinsic} clustering signal.
While our predicted $C_\ell$ spectrum is several orders of magnitude below the current sensitivity of LIGO/Virgo, we saw in figure~\ref{fig:C_ell-cbcs} that it should be detected by third-generation ground-based interferometers such as Einstein Telescope.

In section~\ref{sec:cbc-populations}, we considered the impact of CBC population uncertainties on our results.
This is an important factor to account for, as the rates and mass distributions of each class of CBC (BBH, BNS, and BHNS) are still rather poorly-determined from present observations.
We have therefore calculated the isotropic intensity and angular power spectrum of the AGWB using $\sim10^4$ different possible sets of BBH population hyperparameters, as inferred from the first five BBH events detected by LIGO/Virgo.
We found that the isotropic component of the AGWB is very sensitive to the underlying BBH distribution, varying by more than an order of magnitude, and encoding some interesting information about the shape of the BH mass function.
The angular power spectrum, on the other hand, is fairly insensitive to these changes in the BBH population once we normalise with respect to the isotropic amplitude, meaning that our predicted spectrum is robust to these uncertainties.
In future work, it would be interesting to carry out a more thorough and up-to-date study of the dependence of the AGWB and its anisotropies on the underlying astrophysical modelling, both in terms of the rates and mass distributions of CBCs, and also in terms of other modelled quantities like the delay-time distribution or the metallicity threshold for BH formation, which could have more of an impact on the angular power spectrum.

In section~\ref{sec:shot-noise}, we have pointed out a serious obstacle to using AGWB anisotropies as a cosmological probe: \emph{shot noise}.
This is an additional white-noise component of the observed angular power spectrum caused by the finite sampling of the CBC event rate in each galaxy.
Since a typical galaxy has a CBC rate of $\sim\mathrm{Myr}^{-1}$, one has to observe for an extremely long time to fully sample the large-scale clustering of galaxies using only GW observations.
(There is also an analogous contribution to the shot noise from the finite number of galaxies in a given spatial volume, but this is many orders of magnitude smaller than both the temporal shot noise associated with the CBC event rate and the \enquote{true} angular power spectrum, so we can safely neglect it.)
We have developed a novel data-analysis method for mitigating the impact of shot noise on measurements of the $C_\ell$ spectrum, constructing an estimator which provides an unbiased measurement of the true spectrum, and which attains the minimum possible variance of any such estimator.
However, while this method optimally removes the shot noise offset in our $C_\ell$ measurements, the presence of shot noise still greatly increases the \emph{uncertainty} in these measurements, far beyond the level of cosmic variance.

One promising avenue for future work is to focus on cross-correlating the AGWB anisotropies with other probes of LSS, such as galaxy surveys, rather than simply auto-correlating the AGWB with itself.
There are two key advantages to this: first, since the galaxy clustering signal has already been measured with very high SNR in galaxy surveys, the cross-correlation search can act as a \enquote{template} for the AGWB anisotropies, making a detection much easier; second, since galaxy surveys have no temporal shot noise, the total shot noise associated with the cross-correlation spectrum is drastically reduced compared to that of the auto-correlation spectrum, making it much easier (in principle) to extract interesting new information about LSS.

%% file: chapters/cosmic-strings.tex
\chapter{Nonlinear gravitational-wave memory from cosmic strings}
\label{chap:cosmic-strings}

In section~\ref{sec:gw-memory}, we introduced the nonlinear gravitational-wave memory effect---a fascinating prediction of general relativity in which oscillatory GW signals are generically accompanied by a permanent strain offset~\cite{Christodoulou:1991cr}, as illustrated in figure~\ref{fig:memory}.
This effect is sourced by the energy-momentum of the radiated gravitons~\cite{Thorne:1992sdb}, thus allowing us to directly probe the \enquote{ability of gravity to gravitate} (in the words of \citet{Favata:2009ii}).
The nonlinear memory effect also has surprising and illuminating links to other fundamental aspects of GR in the infrared regime~\cite{Strominger:2017zoo}, as well as analogous memory effects in other field theories~\cite{Bieri:2013hqa,Pate:2017vwa}, and therefore presents a powerful tool for advancing our theoretical understanding of gravitational physics.

With these motivations in mind, there has been a significant effort in recent years to calculate~\cite{Wiseman:1991ss,Favata:2008yd,Favata:2009ii,Favata:2010zu,Pollney:2010hs,Favata:2011qi,Nichols:2017rqr,Talbot:2018sgr,Khera:2020mcz} and search for~\cite{Kennefick:1994nw,vanHaasteren:2009fy,Seto:2009nv,Cordes:2012zz,Madison:2014vca,Wang:2014zls,Arzoumanian:2015cxr,Lasky:2016knh,Yang:2018ceq,Johnson:2018xly,Islo:2019qht,Divakarla:2019zjj,Aggarwal:2019ypr,Hubner:2019sly,Boersma:2020gxx,Ebersold:2020zah,Burko:2020gse} the nonlinear memory associated with compact binary coalescences.
This focus on CBCs is unsurprising, as we have seen in section~\ref{sec:gw-detection} that these are the primary observational target of both interferometer experiments and pulsar timing arrays (PTAs), and are predicted to be abundant sources of nonlinear memory.
However, CBCs are far from the only source of interest in GW astronomy, and in particular, it would be interesting to expand our scope to study the nonlinear memory emitted by \emph{cosmological} sources of GWs.
On the one hand, these sources might lead to memory observables that are distinct from those associated with CBCs, giving us novel signals to search for.
On the other hand, understanding the memory radiated by cosmological sources of GWs might help us to sharpen our theoretical understanding of these sources.

In this chapter, we calculate the nonlinear GW memory emitted by \emph{cosmic strings}, which, as we saw in section~\ref{sec:cosmic-strings}, are one of the most important and well-motivated cosmological sources of GWs.
We focus in particular on the memory associated with \emph{cusps} and \emph{kinks}, which are the two main mechanisms for GW emission from cosmic string loops.
There are at least two reasons to expect \emph{a priori} that cusps and kinks could be important sources of nonlinear memory.\footnote{%
    We note that cosmic string loops can also source significant amounts of \emph{linear} memory by emitting radiation in the underlying matter fields they are made from.
    These effects are ignored by the Nambu-Goto approximation we adopt here, and can only be resolved by field-theory simulations.
    In reality, we expect loops to generate a combination of linear and nonlinear memory, by radiating both matter and GWs.
    This has recently been demonstrated for collapsing circular cosmic string loops, using numerical-relativity/field-theory simulations~\cite{Aurrekoetxea:2020tuw}.}
    \begin{enumerate}
        \item GW memory waveforms tend to be associated with lower frequencies than the primary GW emission that sources them~\cite{Favata:2010zu}.
        This is because the memory grows monotonically over a period of order the duration of the primary signal, which is typically much longer than the oscillation period of the primary signal.
        This means that, e.g., massive stellar binary black holes which merge near the bottom end of the LIGO/Virgo frequency band produce memory signals which are shifted to lower frequencies, and are thus challenging to detect with LIGO/Virgo.
        Cosmic string signals, on the other hand, have durations which are comparable to their oscillation period (i.e. they have a \enquote{burst-like} morphology, which is very different to an inspiralling compact binary), meaning that we should expect the resulting memory signals to have power at similar frequencies to the original signal.
        What's more, the cosmic string signals we consider also have significant power at very high frequencies, and it is possible that the hereditary nature of the memory effect could transfer some of this power, enhancing the amplitude of the signal at observable frequencies.
        (A similar idea of \enquote{orphan} signals---where the memory emission is detectable even though the primary signal is not---was studied by \citet{McNeill:2017uvq}.)
        \item The angular pattern of the GW memory signal on the sky is typically different to that of the primary GW emission that sources it.
        (We can see this from the angular integral in equation~\eqref{eq:nonlinear-memory}.)
        As we will see in section~\ref{sec:cusps-kinks}, cosmic string cusps and kinks emit GWs in narrow beams, meaning that only a very small fraction of all cusps and kinks are oriented such that their GWs can be observed.
        However, if the associated GW memory signal is more broadly distributed on the sphere, this might allow us to observe some of the many cusps and kinks whose beams are not oriented towards us.
    \end{enumerate}

We find that both expectations are borne out by our calculations below: the GW memory from cusps and kinks is indeed emitted in a much broader range of directions than the initial beam, and the memory signal does indeed have a similar frequency profile to the primary signal, with the high-frequency behaviour of the primary GWs playing an important r\^ole in determining the strength of the memory effect.
In fact, we show that these two ingredients lead to a \emph{divergence} in the memory signal from cusps on sufficiently large loops, potentially signifying a breakdown of the standard weak-field approach for calculating the GW signal from cusps.
We attempt to clarify the root cause of this breakdown, and suggest one tentative possible resolution in which the cusp collapses to form a primordial black hole, \enquote{trapping} the high-frequency GW emission behind a horizon and thereby preventing the memory from diverging.
Other resolutions are possible however, and ultimately a fully general-relativistic treatment will be required to understand the true behaviour of cusps.

The remainder of this chapter is structured as follows.
We begin in section~\ref{sec:loops-intro} by introducing the dynamics of cosmic string loops in the Nambu-Goto approximation in the context of linearised GR, deriving the standard GW waveforms for cusps and kinks, and briefly discussing how these are used to calculate the GW background emitted by the cosmic string loop network.
This first section is a review of existing work, mainly following the exposition in chapter~6 of \citet{Vilenkin:2000jqa}, as well as the derivation of the cusp and kink waveforms by \citet{Damour:2001bk}.
The remainder of the chapter consists of original work.
In section~\ref{sec:nonlinear-memory} we derive some useful formulae for nonlinear memory, building on equation~\eqref{eq:nonlinear-memory} to find expressions for the late-time memory and Fourier-domain memory waveform in terms of the primary strain signal.
In section~\ref{sec:cusps}, we calculate the nonlinear GW memory signal associated with cusps, obtaining the simple frequency-domain waveform~\eqref{eq:cusp-result}; we show that the total GW energy radiated by this memory signal diverges for Nambu-Goto strings, and regularise this divergence by imposing a cutoff at the scale of the string width $\delta$; we then go on to consider higher-order memory effects (the nonlinear GW memory sourced by the memory GWs themselves) and show that accounting for all such contributions leads to a divergence for large loops, which persists even after applying the string-width regularisation; finally, we sketch how this divergence is resolved in the cusp-collapse scenario.
In section~\ref{sec:kinks} we repeat the memory calculation for kinks, obtaining the leading-order waveform~\eqref{eq:kink-result}; unlike in the cusp case, the memory signal is strongly suppressed at high frequencies due to interference effects, and no divergence occurs.
In section~\ref{sec:cusp-collapse} we flesh out the cusp-collapse proposal, showing how it arises naturally from arguments based on the hoop conjecture, and derive some of the properties of the resulting PBHs.
In section~\ref{sec:observational-consequences} we study the observable consequences of our results.
We find that, in the scenario where the memory divergence is cured by cusp collapse, the GW emission associated with the memory is strongly suppressed, and is beyond the reach of current or planned GW searches.
However, we show that the PBHs formed in this scenario form a unique population compared to other astrophysical and primordial formation mechanisms, making them a \enquote{smoking gun} signature of cosmic strings.
Finally, we summarise our results in section~\ref{sec:cosmic-strings-summary}.
In appendix~\ref{app:angular-integrals} we argue that the nonlinear memory divergence is associated with a trans-Planckian GW flux, as well as giving some technical details of the angular distribution of the memory radiation from cusps and kinks.

\section{Cosmic string loops: dynamics and gravitational-wave emission}
\label{sec:loops-intro}

The formation of cosmic strings in the early Universe naturally leads to the production of cosmic string \emph{loops} through the self-intersection of long strings.
The lengths of these loops span an enormous range of physical scales; initially created with lengths comparable to the Hubble scale, the emission of GWs causes loops to gradually shrink to microscopic sizes, eventually unwinding and dispersing once their length is comparable to the string width scale.
In order to understand the GWs emitted through this process, it is vital to understand the dynamics of individual loops.
In principle, we should do this by solving the equations of motion for the matter fields that constitute the loop, which usually involves running highly expensive large-scale lattice simulations.
However, in practice we can make significant progress analytically by effacing the microscopic dynamics of the underlying fields, and focusing on the effective, macroscopic equations of motion for the loops themselves.
We do this using the \emph{Nambu-Goto approximation}, which we introduce below, following sections~6.1 and~6.2 of \citet{Vilenkin:2000jqa}.

\subsection{The Nambu-Goto approximation}
\label{sec:nambu-goto}

We saw in section~\ref{sec:cosmic-strings} that the width of a cosmic string is roughly $\delta\sim\sqrt{\hbar/\mu}$, where $\mu$ is the string tension.
Since cosmic strings are formed at such high energies, this is a \emph{tiny} lengthscale: some seventeen orders of magnitude smaller than the radius of a proton for GUT-scale strings.
By comparison, the lengths $\ell$ of the loops we are interested in are astronomically large, $\ell\sim\mathrm{pc}\text{--}\mathrm{Gpc}$.
It is therefore natural for us to treat these loops as purely one-dimensional objects, setting their width $\delta$ to zero.
(Note that this is equivalent to taking $\hbar\to0$, so we can think of this as a classical limit in which we neglect the microphysical degrees of freedom.)
This allows us to describe their motion in terms of a two-dimensional surface in spacetime called the \emph{worldsheet}.

We cover this surface with coordinates $\zeta^I$, ($I=0,1$), and describe the location of a given point on the worldsheet in four-dimensional spacetime in terms of four functions, $X^\alpha(\zeta)$.
The metric that determines the spacetime interval between points on the worldsheet, $\gamma_{IJ}$, is given by restricting the metric of the full spacetime, $g_{\alpha\beta}$, down to this surface,
    \begin{equation}
        \gamma_{IJ}\equiv g_{\alpha\beta}\pdv{X^\alpha}{\zeta^I}\pdv{X^\beta}{\zeta^J}.
    \end{equation}

If the strings are formed through the breaking of a \emph{local} symmetry, then we should be able to encapsulate their equations of motion with a local worldsheet action\footnote{%
    In contrast, strings associated with \emph{global} symmetries are characterised by long-range non-gravitational interactions, and have a very distinct phenomenology compared to local strings as a result.}
    of the form $S=\int\dd[2]{\zeta}\sqrt{-\gamma}\mathcal{L}$, where $\gamma\equiv\det\gamma_{IJ}$.
Since $S$ has dimensions of $(\mathrm{mass}\times\mathrm{length})$, the Lagrangian density $\mathcal{L}$ has dimensions of $(\mathrm{mass}/\mathrm{length})$, just like the string tension $\mu$.
One obvious choice is therefore to take $\mathcal{L}=A\mu$, where $A$ is some unknown dimensionless constant.
We can fix the appropriate value of $A$ by evaluating the energy-momentum tensor of a string with this action, using the procedure in equation~\eqref{eq:einstein-hilbert-action} to obtain
    \begin{equation}
    \label{eq:nambu-goto-energy-momentum}
        T^{\alpha\beta}(x)=\frac{2}{\sqrt{-g}}\frac{\updelta S}{\updelta g_{\alpha\beta}}=\frac{A\mu}{\sqrt{-g}}\int\dd[2]{\zeta}\sqrt{-\gamma}\gamma^{IJ}\pdv{X^\alpha}{\zeta^I}\pdv{X^\beta}{\zeta^J}\delta^{(4)}(x-X),
    \end{equation}
    where we have used the standard identity
    \begin{equation}
    \label{eq:metric-determinant-derivative}
        \updelta\sqrt{-\gamma}=\frac{1}{2}\sqrt{-\gamma}\gamma^{IJ}\updelta\gamma_{IJ}.
    \end{equation}
Evaluating this for the simplest possible case of an infinite straight string (aligned with the $x^1$ axis without loss of generality) on a flat background spacetime, and choosing $\zeta^0=x^0$, $\zeta^1=x^1$, we find the energy density
    \begin{equation}
        T_{00}(x)=-A\mu\delta(x^2-X^2)\delta(x^3-X^3).
    \end{equation}
We see that if we set $A=-1$, then this agrees exactly with the energy density of an infinitely thin string of linear density $\mu$.
We thus obtain the \emph{Nambu-Goto action}
    \begin{equation}
    \label{eq:nambu-goto-action}
        S_\mathrm{NG}=-\mu\int\dd[2]{\zeta}\sqrt{-\gamma}.
    \end{equation}
Since $\sqrt{-\gamma}$ is the area element on the worldsheet, we can interpret this as the action that minimises the total worldsheet area.

Note that $\mu$ is not the only quantity we can use to construct the Lagrangian.
On dimensional grounds, we could write down the more general expression
    \begin{equation}
    \label{eq:worldsheet-lagrangian}
        \mathcal{L}=-\mu+B\hbar\kappa+C\frac{\hbar^2\kappa^2}{\mu}+\cdots,
    \end{equation}
    where the $\kappa$'s represent worldsheet curvature invariants, with indices suppressed.
However, since a loop of length $\ell$ will generally have curvature of order $\kappa\sim\ell^{-2}$, we see that $\hbar\kappa/\mu\sim(\delta/\ell)^2\ll1$, so we can safely discard all but the leading term, leaving the Nambu-Goto action~\eqref{eq:nambu-goto-action}.

This approximation breaks down if we consider small enough loops.
In particular, a more detailed analysis shows that field-theoretic effects become important for loops smaller than
    \begin{equation}
    \label{eq:ell-min}
        \ell_\mathrm{min}\equiv\frac{\delta}{G\mu}\approx\frac{\ell_\Pl}{(G\mu)^{3/2}}\approx5.1\times10^{-19}\,\mathrm{m}\,\times\qty(\frac{G\mu}{10^{-11}})^{-3/2}.
    \end{equation}
(Recall from section~\ref{sec:cosmic-strings} that GWB searches by LIGO/Virgo~\cite{Abbott:2009rr,Abbott:2017mem,Abbott:2019vic,Abbott:2021kbb,Abbott:2021nrg} and PTAs~\cite{Lasky:2015lej,Blanco-Pillado:2017rnf,Yonemaru:2020bmr} constrain the string tension to be $G\mu\lesssim10^{-11}$.
We therefore use $G\mu=10^{-11}$ as a representative value throughout this chapter, and use this to give numerical values of key physical quantities where relevant.)
These loops are expected to rapidly lose their energy through radiation in the underlying matter fields~\cite{Srednicki:1986xg,Hindmarsh:1994re,Matsunami:2019fss} or through topological unwinding and dispersion~\cite{Helfer:2018qgv,Aurrekoetxea:2020tuw}, and are therefore uninteresting for our purposes.\footnote{%
    There is a long-standing debate within the cosmic string community as to whether matter radiation can be safely neglected for loops much larger than $\ell_\mathrm{min}$.
    Field theory simulations of Abelian-Higgs loop networks appear to show that matter radiation dominates over GW emission on large scales~\cite{Vincent:1997cx,Hindmarsh:2008dw,Hindmarsh:2017qff}, causing the network to decay rapidly into the matter fields, thus significantly suppressing the expected GWB signal.
    These results are in tension with theoretical arguments and with simulations of individual loops, which support the idea that loops much larger than $\ell_\mathrm{min}$ can safely be treated in the Nambu-Goto approximation.
    It has been suggested that the anomalous matter radiation rates observed in the field-theory network simulations may be a numerical artefact, tied to the limited dynamical range of these simulations compared to the enormous separation of scales in the cosmological scenarios of interest.
    However, this remains an open question.
    Throughout this chapter, we assume that field-theory effects are negligible for loops much larger than $\ell_\mathrm{min}$ (though we will return to this point in section~\ref{sec:cusp-collapse-memory}).
    We refer the interested reader to, e.g., section 3.4 of \citet{Auclair:2019wcv} or section 9.2 of \citet{Caprini:2018mtu} for further discussion of this issue.
}

\subsection{The flat-space equations of motion}
\label{sec:flat-space-eoms}

We can find the equations of motion for a Nambu-Goto string by extremising the action~\eqref{eq:nambu-goto-action} with respect to the spacetime coordinates $X^\alpha(\zeta)$ which, from the worldsheet perspective, are just a set of four scalar fields.
Since our goal is ultimately to use the EoMs to calculate GW signals emitted by loops in linearised GR, we can specialise to a flat background spacetime, such that
    \begin{equation}
    \label{eq:flat-space-worldsheet-metric}
        \gamma_{IJ}=\eta_{\alpha\beta}\pdv{X^\alpha}{\zeta^I}\pdv{X^\beta}{\zeta^J}.
    \end{equation}
Varying the action then gives
    \begin{align}
    \begin{split}
        \updelta S&=-\frac{\mu}{2}\int\dd[2]{\zeta}\sqrt{-\gamma}\gamma^{IJ}\updelta\gamma_{IJ}=-\mu\eta_{\alpha\beta}\int\dd[2]{\zeta}\sqrt{-\gamma}\gamma^{IJ}\pdv{X^\alpha}{\zeta^I}\pdv{\zeta^J}\updelta X^\beta\\
        &=\mu\int\dd[2]{\zeta}\updelta X_\alpha\pdv{\zeta^J}\qty(\sqrt{-\gamma}\gamma^{IJ}\pdv{X^\alpha}{\zeta^I}),
    \end{split}
    \end{align}
    where we have used equation~\eqref{eq:metric-determinant-derivative} again, and have ignored the boundary term when integrating by parts.\footnote{%
    If we had kept a general spacetime metric $g_{\alpha\beta}$, then we would have picked up an additional term here containing the four-dimensional Christoffel symbols $\Gamma^\nu_{\alpha\beta}$.
    This correction vanishes on a flat spacetime.}
In order for this to vanish, the coordinates $X^\alpha$ must each obey a wave equation,
    \begin{equation}
    \label{eq:worldsheet-wave-equation}
        {}^{(2)}\dalemb X^\alpha\equiv\frac{1}{\sqrt{-\gamma}}\pdv{\zeta^I}\qty(\sqrt{-\gamma}\gamma^{IJ}\pdv{X^\alpha}{\zeta^J})=0,
    \end{equation}
    where ${}^{(2)}\dalemb$ is the two-dimensional covariant d'Alembertian operator on the worldsheet.
The lack of a mass term here shows us that the loop's degrees of freedom propagate at the speed of light.

Thus far, we have kept our worldsheet coordinates completely general, but as with the gauge freedom in the four-dimensional spacetime coordinates that we dealt with in section~\ref{sec:what-are-gws}, there is a great deal of redundancy here.
We can simplifying things considerably with a judicious choice of gauge.
One particularly convenient choice is to select $\zeta$ such that the worldsheet metric is conformally flat, $\gamma_{IJ}=\sqrt{-\gamma}\eta_{IJ}$; this is called the \enquote{conformal gauge}.
Equations~\eqref{eq:flat-space-worldsheet-metric} and~\eqref{eq:worldsheet-wave-equation} then become
    \begin{equation}
        \ddot{X}^\alpha-{X^\alpha}''=0,\qquad\dot{X}^\alpha\dot{X}_\alpha+{X^\alpha}'X'_\alpha=0,\qquad\dot{X}^\alpha X'_\alpha=0,
    \end{equation}
    where dots and primes indicate derivatives with respect to $\zeta^0$ and $\zeta^1$, respectively.
These equations are simplified even further if we now choose the timelike coordinate such that $\zeta^0=X^0\equiv t$.
The loop's trajectory is then fully described by a 3-vector $\vb*X(t,\sigma)$, where $\sigma\equiv\zeta^1$ is the spacelike coordinate in this gauge, and the EoMs become
    \begin{align}
        \label{eq:flat-space-eoms-1}
        \ddot{\vb*X}-\vb*X''&=\vb*0,\\
        \label{eq:flat-space-eoms-2}
        |\dot{\vb*X}|^2+|\vb*X'|^2&=1,\\
        \label{eq:flat-space-eoms-3}
        \dot{\vb*X}\vdot\vb*X'&=0.
    \end{align}

Each of these three equations has an intuitive physical interpretation.
Since $\vb*X'$ is the tangent vector pointing along the string, equation~\eqref{eq:flat-space-eoms-3} tells us that $\dot{\vb*X}$ should be interpreted as the \emph{transverse} velocity of the string; we cannot observe velocities along the tangential direction.
Rearranging equation~\eqref{eq:flat-space-eoms-2} shows us that intervals in the worldsheet coordinate $\sigma$ are related to intervals in spatial distance $|\vb*X|$ by
    \begin{equation}
    \label{eq:sigma-propto-energy}
        \dd{\sigma}=\frac{\qty|\dd{\vb*X}|}{|\vb*X'|}=\frac{\qty|\dd{\vb*X}|}{\sqrt{1-|\dot{\vb*X}|^2}}=\gamma\qty|\dd{\vb*X}|=\frac{\dd{E}}{\mu},
    \end{equation}
    where $\gamma=1/\sqrt{1-|\dot{\vb*X}|^2}$ is the Lorentz factor of the string segment (not to be confused with the determinant of the worldsheet metric $\gamma_{IJ}$), and $\dd{E}=\gamma\mu\qty|\dd{\vb*X}|$ is the energy (kinetic plus rest mass) of the string segment.\footnote{%
    This expression for the energy accords with our intuition from point particles in flat space, but we also could have derived it from the energy momentum tensor~\eqref{eq:nambu-goto-energy-momentum}.}
This shows us that by choosing this gauge, we have scaled $\sigma$ such that it is proportional to the energy at every point along the string, $\mu\dd{\sigma}=\dd{E}$.
Finally, equation~\eqref{eq:flat-space-eoms-1} tells us that the degrees of freedom propagating tangentially along the string obey a wave equation and propagate at the speed of light.
We can make this more explicit by decomposing $\vb*X$ into left- and right-moving modes,
    \begin{equation}
    \label{eq:left-right-decomposition}
        \vb*X(t,\sigma)=\frac{1}{2}\qty[\vb*X_+(\sigma_+)+\vb*X_-(\sigma_-)],\qquad\sigma_\pm\equiv t\pm\sigma.
    \end{equation}
This decomposition is always possible for any solution to equation~\eqref{eq:flat-space-eoms-1}, and also obeys equations~\eqref{eq:flat-space-eoms-2} and~\eqref{eq:flat-space-eoms-3} if we set
    \begin{equation}
    \label{eq:left-right-unit-velocity}
        |\dot{\vb*X}_+|=|\dot{\vb*X}_-|=1,
    \end{equation}
    showing that both modes do indeed propagate at the speed of light.
(Note that differentiating $\vb*X_\pm$ with respect to time gives the same result as differentiating with respect to their argument, $\sigma_\pm$.)

Since we are interested in cosmic string \emph{loops}, $\vb*X$ must be periodic in $\sigma$, with some finite periodicity length $\ell$.
From equation~\eqref{eq:sigma-propto-energy}, we see that this length is set by the total energy of the loop, since $E=\mu\int_0^\ell\dd{\sigma}=\mu\ell$.
This implies that $\ell$ must be constant over time (neglecting for now energy loss from GW radiation), so we call it the \emph{invariant length}.
Note that this differs from the coordinate length of the loop,
    \begin{equation}
        L=\int_0^\ell\dd{\sigma}|\vb*X'|=\int_0^\ell\dd{\sigma}\sqrt{1-|\dot{\vb*X}^2|}\le\ell,
    \end{equation}
    which fluctuates over time, undergoing relativistic length contraction due to the motion of the loop.

The left-right decomposition~\eqref{eq:left-right-decomposition} makes it clear that $\vb*X_\pm$ must also be periodic in $\sigma_\pm\in[0,\ell)$, which in turn implies that $\vb*X$ is periodic in \emph{time} as well as space.
In fact, the period of the loop's motion over time is $\ell/2$ rather than $\ell$, since equation~\eqref{eq:left-right-decomposition} gives $\vb*X(t+\ell/2,\sigma+\ell/2)=\vb*X(t,\sigma)$.
This shows us that it's not just modes moving along the loop that are relativistic, but that the loop itself must oscillate at relativistic speeds.
Indeed, using equations~\eqref{eq:flat-space-eoms-1} and~\eqref{eq:flat-space-eoms-2} we see that the loop's rms velocity, averaged over its length and over an oscillation period, is given by
    \begin{align}
    \begin{split}
    \label{eq:loop-v-rms}
        v_\mathrm{rms}^2&\equiv\ev{|\dot{\vb*X}|^2}_\ell=\ev{\vb*X\vdot\ddot{\vb*X}}_\ell=\ev{\vb*X\vdot\vb*X''}_\ell=\ev{|\vb*X'|^2}_\ell=\ev{1-|\dot{\vb*X}|^2}_\ell=1-v_\mathrm{rms}^2,\\
        \ev{\cdots}_\ell&\equiv\int_0^\ell\frac{\dd{\sigma}}{\ell}\int_0^{\ell/2}\frac{\dd{t}}{\ell/2}(\cdots),
    \end{split}
    \end{align}
    where periodicity allows us to drop the boundary terms when integrating by parts.
We therefore have $v_\mathrm{rms}=1/\sqrt{2}$; i.e., the loop moves at $\approx71\%$ the speed of light on average.

\subsubsection{Cusps and kinks}
Another way of understanding equation~\eqref{eq:loop-v-rms} is to think about the net velocity of the loop as the average of the velocity vectors of the left- and right-moving modes, $\dot{\vb*X}=(\dot{\vb*X}_++\dot{\vb*X}_-)/2$.
Since we saw in equation~\eqref{eq:left-right-unit-velocity} that these each have unit magnitude at all times and at all points on the loop, the net velocity depends only on the \emph{relative} velocity of the two modes.
If we define $\theta_\pm\equiv\cos^{-1}(\dot{\vb*X}_+\vdot\dot{\vb*X}_-)$ as the angle between the two velocity vectors, we see that the net velocity is given by
    \begin{equation}
    \label{eq:loop-net-velocity}
        |\dot{\vb*X}|^2=\frac{1}{4}\qty(|\dot{\vb*X}_+|^2+2\cos\theta_\pm+|\dot{\vb*X}_-|^2)=\frac{1+\cos\theta_\pm}{2}.
    \end{equation}
We thus reproduce equation~\eqref{eq:loop-v-rms} if $\cos\theta_\pm$ averages to zero over an oscillation period (i.e., the two mode velocities are just as likely to be aligned as anti-aligned).

\begin{figure}[t!]
    \begin{center}
        \includegraphics[width=0.55\textwidth]{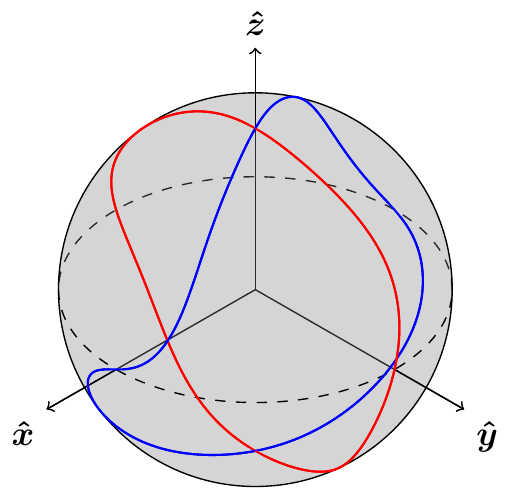}
    \end{center}
    \caption{%
    The left- and right-moving mode velocities $\dot{\vb*X}_\pm$ for a given loop configuration can be drawn as two curves on the Kibble-Turok sphere.
    These curves generically intersect each other, giving rise to cusps.
    Note that the curves drawn here are continuous, and therefore correspond to loops which have no kinks.
    }
    \label{fig:kibble-turok}
\end{figure}

From equation~\eqref{eq:loop-net-velocity}, we see that it is possible for the loop itself to move at the speed of light if we can arrange for both the left- and right-moving modes to be moving in the same direction, $\dot{\vb*X}_+=\dot{\vb*X}_-$ (since then $\theta_\pm=0$).
A point on the worldsheet where such a situation occurs is called a \emph{cusp}~\cite{Turok:1984cn,Vilenkin:1984ib,Hindmarsh:1994re,Vilenkin:2000jqa}.
At first glance, cusps seem finely-tuned and unlikely to occur for generic loop trajectories.
However, while it is true that a randomly-selected point on the worldsheet is unlikely to be a cusp, the occurrence of a cusp \emph{somewhere} on the loop during the course of an oscillation period is actually very generic.
To see this, consider the possible forms that the functions $\dot{\vb*X}_\pm$ can take.
These are both periodic 3-vector functions with unit magnitude, so we can draw them as closed curves on the unit 2-sphere (as illustrated in figure~\ref{fig:kibble-turok}; this is often referred to as the Kibble-Turok sphere).
Since $\int_0^\ell\dd{\sigma_\pm}\dot{\vb*X}_\pm=\vb*X_\pm(\ell)-\vb*X_\pm(0)=\vb*0$, neither of these curves can reside entirely in one hemisphere, as the integral $\int_0^\ell\dd{\sigma_\pm}\dot{\vb*X}_\pm$ would then be nonzero.
As a result, it is quite hard for the two curves to avoid each other completely, and there is generically at least one point at which they intersect, with each intersection corresponding to a cusp.

One way in which the two curves can more easily avoid intersecting is if they possess discontinuities; i.e., if there are values of $\sigma_\pm$ at which $\dot{\vb*X}_\pm$ jumps between two disconnected points on the unit sphere.
Physically this corresponds to a discontinuity in the loop's tangent vector $\vb*X'$, at which two smooth string segments are joined at a sharp angle.\footnote{%
    Of course, such a sharp feature is only possible in our idealised Nambu-Goto setup, where the string has zero width.
    In the full field-theoretic description, kinks correspond to points where the loop's tangent vector changes significantly over an interval of order the string width $\delta$, but in such a way that the matter fields are everywhere continuous.}
These features are called \emph{kinks}, and occur very naturally due to the string self-intersections from which loops are created.
Each kink is associated with a particular fixed value of $\sigma_+$ or $\sigma_-$, meaning that the kink propagates around the loop at the speed of light, with the direction of travel depending on whether the discontinuity is associated with a left- or a right-moving mode.

In the following section, we show that cusps and kinks are the two main mechanisms through which loops radiate GWs.

\subsection{Gravitational-wave emission from loops}
\label{sec:cusps-kinks}

We now turn to the loop's GW emission, following the derivation of the cusp and kink waveforms first given by \citet{Damour:2001bk}.
Using the results from section~\ref{sec:gw-generation}, we can write the Fourier transform of the complex strain waveform $h=h_+-\rmi h_\times$ emitted by the loop in direction $\vu*r$ as
    \begin{equation}
        \tilde{h}(f,\vu*r)=\frac{2G}{r}\qty[e^{+,ij}(\vu*r)-\rmi e^{\times,ij}(\vu*r)]\tilde{T}_{ij}(f,f\vu*r),
    \end{equation}
    where $\tilde{T}_{ij}$ is the spacetime Fourier transform of the energy-momentum tensor,
    \begin{equation}
        \tilde{T}_{ij}(f,\vb*k)=\int\dd{t}\int\dd[3]{\vb*x}\rme^{-2\uppi\rmi(ft-\vb*k\vdot\vb*x)}T_{ij}(t,\vb*x).
    \end{equation}
In our case, this is given by evaluating equation~\eqref{eq:nambu-goto-energy-momentum} in the conformal gauge,
    \begin{align}
    \begin{split}
    \label{eq:conformal-gauge-energy-momentum}
        T_{ij}(t,\vb*x)&=\mu\int_0^\ell\dd{\sigma}\delta^{(3)}(\vb*x-\vb*X)\qty(\dot{X}_i\dot{X}_j-X'_iX'_j),\\
        \tilde{T}_{ij}(f,\vb*k)&=\mu\int_0^{\ell/2}\dd{t}\int_0^\ell\dd{\sigma}\rme^{-2\uppi\rmi(ft-\vb*k\vdot\vb*X)}\qty(\dot{X}_i\dot{X}_j-X'_iX'_j),
    \end{split}
    \end{align}
    where we have restricted the time integral to $t\in[0,\ell/2]$, as we are interested in the GW emission from a single oscillation period.
We can simplify this by writing everything in terms of the left- and right-moving modes defined in equation~\eqref{eq:left-right-decomposition}.
Using $\vb*X_\pm'=\pm\dot{\vb*X}_\pm$, and changing the area element to $\dd{\sigma_+}\dd{\sigma_-}=2\dd{t}\dd{\sigma}$, we see that the energy-momentum tensor can be factorised into two separate integrals corresponding to the left- and right-moving parts, such that the waveform is given by
    \begin{equation}
    \label{eq:gw-strain-left-right}
        \tilde{h}(f,\vu*r)=\frac{G\mu}{r}(e^+_{ij}-\rmi e^\times_{ij})\mathcal{I}^i_+\mathcal{I}^j_-,\qquad\mathcal{I}^i_\pm(f,\vu*r)\equiv\int_{-\ell/2}^{+\ell/2}\dd{\sigma_\pm}\rme^{-\uppi\rmi f(\sigma_\pm-\vu*r\vdot\vb*X_\pm)}\dot{X}^i_\pm.
    \end{equation}

Since the loop's motion is periodic, the GW emission described by equation~\eqref{eq:gw-strain-left-right} is not a continuous spectrum, but is concentrated at the discrete set of frequencies $2n/\ell$, where $n$ is a nonzero integer, and $2/\ell$ is the fundamental mode of the loop.
This distinction is unimportant in most cases however, as the fundamental mode corresponds to a very low frequency for the loop sizes $\ell\sim\mathrm{pc}\text{--}\mathrm{Gpc}$ we are interested in, $2/\ell\approx2\times10^{-8}\,\mathrm{Hz}\times(\ell/\mathrm{pc})^{-1}$, meaning that we are typically interested in very high harmonics $|n|\gg1$, particularly for interferometers like LIGO/Virgo.
In this high-frequency regime, the emission modes of the loop are so closely spaced that they effectively form a continuous spectrum.
The downside of this is that the integrals $\mathcal{I}^i_\pm$ are highly oscillatory in this regime, meaning that the GW emission typically falls off exponentially with frequency.

There are two key ways in which this exponential suppression can be avoided, giving a strong high-frequency GW signal~\cite{Damour:2001bk}:
    \begin{enumerate}
        \item If there is a \emph{stationary point} in the phase, such that $\pdv*{\sigma_\pm}(\sigma_\pm-\vu*r\vdot\vb*X_\pm)=0$ for some value of $\sigma_\pm$.
        This corresponds to having $\dot{\vb*X}_\pm=\vu*r$, which occurs whenever the left-/right-moving mode in question is propagating directly towards the observer.
        \item If there is a \emph{discontinuity} in the integrand or one of its derivatives.
        Since $\vb*X_\pm$ itself must be continuous (recall that cosmic strings are topologically forbidden from having open ends), the strongest discontinuity that we can allow is a jump in the first derivative $\dot{\vb*X}_\pm$, i.e., a kink.
    \end{enumerate}
In order to prevent the GW strain~\eqref{eq:gw-strain-left-right} from decaying exponentially, we need one or the other of these conditions to be satisfied for \emph{both} the left- and right-moving modes.
There are three possible situations that achieve this:
    \begin{description}
        \item[Cusp:] Both integrals $\mathcal{I}^i_\pm$ possess a stationary point along the same propagation direction $\vu*r_\rmc$.
        This corresponds to having $\vu*r_\rmc=\dot{\vb*X}_+=\dot{\vb*X}_-$; in other words, a cusp pointing in the $\vu*r_\rmc$ direction.
        \item[Kink:] One integral has a stationary point along the GW propagation direction $\vu*r$, and the other has a kink.
        This corresponds to a point in the loop's oscillation at which the kink is travelling in the $\vu*r$ direction.
        \item[Kink-kink collision:] Both integrals have kinks.
        We can then interpret the emitted GWs as being due to a \emph{kink-kink collision}.
        Since neither kink is associated with a particular direction $\vu*r$ (as there is no stationary-phase requirement), the resulting GW emission must be isotropic.
    \end{description}
We calculate the corresponding waveform for each of these cases below.

\subsubsection{Cusps}
\begin{figure}[t!]
    \begin{center}
        \includegraphics[width=0.5\textwidth]{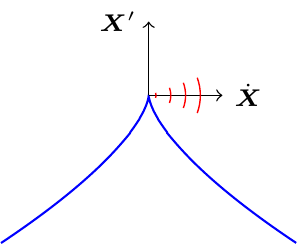}
    \end{center}
    \caption{%
    Diagram of a string segment (blue) near a cusp, showing the tangent vector $\vb*X'$ and velocity $\dot{\vb*X}$ at the cusp.
    Note the $y\propto x^{2/3}$ shape of this curve (where $x$ and $y$ are the distances of the loop from the cusp in the directions parallel to and transverse to the cusp velocity, respectively), which is characteristic of all cusps.
    As we show below, the emitted GWs (red) are beamed along the direction of the cusp velocity.
    }
    \label{fig:cusp}
\end{figure}

Let's look at the cusp case first.
Here we can use the stationary-phase approximation to evaluate the integrals $\mathcal{I}^i_\pm$, focusing on a small interval in $\sigma_\pm$ near the cusp.
This is useful because the gauge conditions~\eqref{eq:left-right-unit-velocity} constrain the form of the functions $\vb*X_\pm(\sigma_\pm)$ in a small region near the cusp, allowing us to derive results that are generic to \emph{all} cusps.
Choosing our coordinates so that the cusp occurs at $\sigma_\pm=0$, we can write the position and velocity of the left- and right-moving modes as Taylor expansions around this point,
    \begin{equation}
    \label{eq:cusp-xpm-taylor-expansions}
        \vb*X_\pm=\vu*r_\rmc\sigma_\pm+\frac{1}{2}\ddot{\vb*X}_\pm\sigma_\pm^2+\frac{1}{6}\dddot{\vb*X}_\pm\sigma_\pm^3+\cdots,\qquad\dot{\vb*X}_\pm=\vu*r_\rmc+\ddot{\vb*X}_\pm\sigma_\pm+\frac{1}{2}\dddot{\vb*X}_\pm\sigma_\pm^2+\cdots,
    \end{equation}
    with all the derivatives evaluated at $\sigma_\pm=0$.
We can then apply the gauge conditions~\eqref{eq:left-right-unit-velocity} order-by-order to find
    \begin{equation}
    \label{eq:left-right-constraints-at-cusp}
        \vu*r_\rmc\vdot\ddot{\vb*X}_\pm=0,\qquad\vu*r_\rmc\vdot\dddot{\vb*X}_\pm=-|\ddot{\vb*X}_\pm|^2.
    \end{equation}
As an example of how this determines the properties of the loop near the cusp, consider the loop's position at time $t=0$,
    \begin{equation}
        \vb*X(0,\sigma)=\frac{1}{2}[\vb*X_+(\sigma)+\vb*X_-(-\sigma)]=\frac{1}{4}(\ddot{\vb*X}_++\ddot{\vb*X}_-)\sigma^2+\frac{1}{12}(\dddot{\vb*X}_+-\dddot{\vb*X}_-)\sigma^3+\cdots.
    \end{equation}
Using the constraints in equation~\eqref{eq:left-right-constraints-at-cusp}, we see that the loop's distance from the cusp in the direction parallel to $\dot{\vb*X}$ (call this $x$) is related to the distance in the transverse direction (call this $y$) by $y\propto x^{2/3}$.
This gives rise to a universal shape for the string segment near the cusp, which is illustrated in figure~\ref{fig:cusp}.
The proportionality constant relating $x$ and $y$ depends on the values of the derivatives $\ddot{\vb*X}_\pm$, $\dddot{\vb*X}_\pm$ at the cusp, but we can generically expect it to include a factor of $\ell^{1/3}$ on dimensional grounds, as the loop length $\ell$ is the only relevant lengthscale in the problem (here it is hidden in the values of the derivatives, as we discuss below when estimating $|\ddot{\vb*X}_\pm|$).
In other words, this universal cusp shape is scaled up and down according to the total length of the loop; a slightly counterintuitive idea, given that the cusp is a small, localised feature which one might not expect to \enquote{know} about the rest of the loop, but one which is important later in this chapter.

Returning now to the GW signal emitted by a cusp, we see that equation~\eqref{eq:left-right-constraints-at-cusp} implies that the leading contribution to the phase factor in equation~\eqref{eq:gw-strain-left-right} is of order $\sigma_\pm^3$,
    \begin{equation}
    \label{eq:phase-taylor-expansion}
        \sigma_\pm-\vu*r_\rmc\vdot\vb*X_\pm=-\frac{1}{2}\vu*r_\rmc\vdot\ddot{\vb*X}_\pm\sigma_\pm^2-\frac{1}{6}\vu*r_\rmc\vdot\dddot{\vb*X}_\pm\sigma_\pm^3+\cdots=\frac{1}{6}|\ddot{\vb*X}_\pm|^2\sigma_\pm^3+\cdots.
    \end{equation}
We therefore write
    \begin{equation}
        \mathcal{I}^i_\pm(f,\vu*r_\rmc)\simeq\int_{-\ell/2}^{+\ell/2}\dd{\sigma_\pm}\qty(\hat{r}_*^i+\ddot{X}_\pm^i\sigma_\pm)\exp(-\frac{\uppi\rmi}{6}f|\ddot{\vb*X}_\pm|^2\sigma_\pm^3).
    \end{equation}
Notice that the leading-order term here is directed along the line of sight, $\mathcal{I}^i_\pm\propto\hat{r}_*^i$, and therefore vanishes when contracted with the TT polarisation tensors $e^A_{ij}$.
This means that actual leading-order contribution to the GW signal is given by
    \begin{equation}
    \label{eq:stationary-phase-leading-order}
        \mathcal{I}^i_\pm(f,\vu*r_\rmc)\simeq\ddot{X}_\pm^i\int_{-\ell/2}^{+\ell/2}\dd{\sigma_\pm}\sigma_\pm\exp(-\frac{\uppi\rmi}{6}f|\ddot{\vb*X}_\pm|^2\sigma_\pm^3)\simeq\frac{\ddot{X}_\pm^i}{|\ddot{\vb*X}_\pm|^{4/3}}(\uppi f/6)^{-2/3}\int\dd{u}u\,\rme^{-\rmi u^3},
    \end{equation}
    where we have rewritten the integral in terms of the dimensionless variable $u\equiv\sigma_\pm(\uppi f|\ddot{\vb*X}_\pm|^2/6)^{1/3}$.
We have assumed here that the GW frequency is positive; for negative frequencies, we can replace $f$ with $|f|$ and take the complex conjugate of the integral.

We can estimate $|\ddot{\vb*X}_\pm|$ by noting that generic solutions for $\vb*X_\pm$ can be written as a sum of Fourier modes, $\vb*X_\pm=\sum_{n=1}^\infty\vb*X_\pm^{(n)}\exp(2\uppi\rmi n\sigma_\pm/\ell)$.
Consider first the unrealistic case of a solution with a single mode, $n$.
Since $|\dot{\vb*X}_\pm|=1$, we would then have $|\ddot{\vb*X}_\pm|=2\uppi n/\ell$.
For a more realistic solution, there are cross-terms from various different modes, but in general we can write $|\ddot{\vb*X}_\pm|=2\uppi\bar{n}_\pm/\ell$, where the \enquote{effective mode number} $\bar{n}_\pm$ is of order unity for smooth strings, and becomes larger for very wiggly strings.
(One generally expects gravitational backreaction to dampen higher-order modes, which would dynamically drive $\bar{n}_\pm$ towards smaller values over time.)
We therefore write $u=\sigma_\pm\uppi(\bar{n}_\pm/\ell)^{2/3}(2f/3)^{1/3}$.

Note that we neglected the limits of the $u$ integral in equation~\eqref{eq:stationary-phase-leading-order}.
These are given by $|u|=\uppi\bar{n}_\pm^{2/3}(f\ell/12)^{1/3}$, and are thus much greater than unity, since we are interested in the high-frequency regime $f\ell\gg1$.
As such, we can safely extend the limits of the integral to infinity, since the final value is dominated by the region $|u|\lesssim1$ anyway.
This allows us to evaluate the integral analytically, giving
    \begin{equation}
        \int_{-\infty}^{+\infty}\dd{u}u\,\rme^{-\rmi u^3}=-\frac{2\uppi\rmi}{3\Gamma(\sfrac{1}{3})},
    \end{equation}
    where $\Gamma(z)$ is the Euler Gamma function.

Going back to equation~\eqref{eq:gw-strain-left-right}, we therefore find that the cusp waveform is given at high frequencies $|f|\ell\gg1$ by\footnote{%
    Note that the value of $A_\rmc$ given here is a factor of two smaller than the value of $0.8507$ quoted by \citet{Damour:2001bk}.
    This is because we work in terms of the complex strain $h$ instead of the metric perturbation $h_{ij}$, with the missing factor of two being carried by the normalisation of the polarisation tensors $e^A_{ij}$.}
    \begin{equation}
    \label{eq:cusp-waveform-beaming-direction}
        \tilde{h}_\rmc(f,\vu*r_\rmc)\simeq-A_\rmc\frac{G\mu\ell^{2/3}}{r|f|^{4/3}}(e^+_{ij}-\rmi e^\times_{ij})\hat{\tau}^i_+\hat{\tau}^j_-,\qquad A_\rmc\equiv\frac{4\times(2/3)^{2/3}}{(\bar{n}_+\bar{n}_-)^{1/3}\Gamma^2(\sfrac{1}{3})}\lesssim0.4253,
    \end{equation}
    where $\vu*\tau_\pm\equiv\ddot{\vb*X}_\pm/|\ddot{\vb*X}_\pm|$ are two unit vectors transverse to the GW propagation direction.
Each of these is defined by some angle $\theta_\pm$ in the transverse plane, and it is straightforward to show that
    \begin{equation}
        (e^+_{ij}-\rmi e^\times_{ij})\hat{\tau}^i_+\hat{\tau}^j_-=\rme^{-2\rmi(\theta_++\theta_-)},
    \end{equation}
    so that this factor can be absorbed into our choice of coordinates by performing a polarisation rotation, c.f. equation~\eqref{eq:complex-polarisation-rotation}.
The fact that we can do this indicates that the cusp's GW emission is \emph{linearly polarised}.

Now that we know the GW signal emitted along the direction of the cusp, $\vu*r_\rmc$, we can ask how this emission is modified if there is some nonzero angle $\theta$ between this direction and the observer's line of sight $\vu*r$.
If $\theta$ is too large, then the phase in $\mathcal{I}^i_\pm$ will no longer be stationary, and the GW emission will be exponentially suppressed at high frequencies.
The question then is, how small does this angle have to be for the observer to still see the much slower $f^{-4/3}$ power law decay?
We can obtain a rough answer to this question by rewriting the phase~\eqref{eq:phase-taylor-expansion} with a small inclination angle $\theta\equiv\cos^{-1}(\vu*r_\rmc\vdot\vu*r)$,
    \begin{equation}
        \sigma_\pm-\vu*r\vdot\vb*X_\pm=\frac{1}{2}\theta^2\sigma_\pm-\frac{1}{2}\vb*\delta\vdot\ddot{\vb*X}_\pm\sigma_\pm^2+\frac{1}{6}|\ddot{\vb*X}_\pm|^2\sigma_\pm^3+\cdots,
    \end{equation}
    where $\vb*\delta\equiv\vu*r-\vu*r_\rmc$ is a 3-vector pointing from the cusp direction to the line of sight, with magnitude $|\vb*\delta|=\sqrt{2-2\cos\theta}\simeq\theta$.
The first two terms here are the ones that cause the rapid oscillations that spoil the power-law frequency scaling.
We can therefore estimate the angle $\theta$ at which this power-law scaling fails by setting the first and third terms equal to each other, and setting $\sigma_\pm=(\uppi f|\ddot{\vb*X}_\pm|^2/6)^{-1/3}$ such that the dummy variable $u$ is equal to unity (since we saw that the integral is dominated by $|u|\lesssim1$).
This defines the cusp's \emph{beaming angle},
    \begin{equation}
    \label{eq:beam-angle}
        \theta_\rmb(f)\simeq\frac{2^{2/3}}{3^{1/6}}(|f|\ell)^{-1/3},
    \end{equation}
    which is necessarily very small, since we have $|f|\ell\gg1$.
(Note that we have set the effective mode numbers $\bar{n}_\pm$ to unity here---these can increase the size of the beam like $\theta_\rmb\sim\bar{n}_\pm^{1/3}$ for very wiggly strings.)

Our final expression for the cusp waveform is therefore
    \begin{equation}
    \label{eq:cusp-waveform}
        \tilde{h}_\rmc(f,\vu*r)\simeq A_\rmc\frac{G\mu\ell^{2/3}}{r|f|^{4/3}}\Theta(\vu*r_\rmc\vdot\vu*r-\cos\theta_\rmb)\Theta(|f|-2/\ell),
    \end{equation}
    where $\Theta(x)$ is the Heaviside step function.
We approximate the beam with a sharp, step-function cutoff, as the gradual onset of the exponential frequency damping is somewhat difficult to calculate precisely.
We also include a second step function, to remind us that this expression should not be applied below the fundamental frequency $2/\ell$.
Strictly speaking we should only use equation~\eqref{eq:cusp-waveform} in the high-frequency regime, but in practice we expect that this \emph{under}estimates the low-frequency GW emission, so we can safely use it as a conservative proxy for the GW emission across the whole frequency range (particularly as our results for the corresponding GW memory are dominated by the high-frequency emission).

Using equation~\eqref{eq:cusp-waveform}, we can calculate the total GW energy radiated by the cusp.
It is useful to normalise this against the total energy of the loop, $\mu\ell$, to give the dimensionless energy spectrum
    \begin{equation}
    \label{eq:dimensionless-energy}
        \epsilon(f,\vu*r)\equiv\frac{1}{\mu\ell}\frac{\dd{E_\gw}}{\dd{(\ln f)}\dd[2]{\vu*r}}=\frac{\uppi r^2f^3}{4G\mu\ell}\qty(|\tilde{h}(f,\vu*r)|^2+|\tilde{h}(-f,\vu*r)|^2).
    \end{equation}
For cusps, this is given by
    \begin{equation}
    \label{eq:cusp-energy-spectrum}
        \epsilon_\rmc(f,\vu*r)\simeq\frac{\uppi}{2}A_\rmc^2G\mu(f\ell)^{1/3}\Theta(\vu*r_\rmc\vdot\vu*r-\cos\theta_\rmb)\Theta(f-2/\ell),
    \end{equation}
    which diverges at high frequencies for observers in the beaming direction, $\vu*r=\vu*r_\rmc$.
This divergence is \enquote{hidden} by the beaming angle $\theta_\rmb$, which shrinks fast enough with frequency to ensure that the total radiated energy is finite.
We can see this by defining the isotropically-averaged energy spectrum,
    \begin{equation}
    \label{eq:dimensionless-energy-isotropic}
        \bar{\epsilon}(f)\equiv\int_{S^2}\dd[2]{\vu*r}\epsilon(f,\vu*r),
    \end{equation}
    which gives us an extra factor of $\uppi\theta_\rmb^2$, so that
    \begin{equation}
    \label{eq:cusp-isotropic-energy}
        \bar{\epsilon}_\rmc(f)\simeq(2/3)^{1/3}(\uppi A_\rmc)^2G\mu(f\ell)^{-1/3}\Theta(f-2/\ell).
    \end{equation}
Integrating over frequency, we define the total fraction of the loop's energy that is radiated as
    \begin{equation}
    \label{eq:total-energy}
        \mathcal{E}\equiv\int_0^\infty\frac{\dd{f}}{f}\bar{\epsilon}(f),
    \end{equation}
    which ends up being of order $G\mu$ for cusps,
    \begin{equation}
    \label{eq:cusp-total-energy-primary}
        \mathcal{E}_\rmc\simeq3^{2/3}(\uppi A_\rmc)^2G\mu\ll1.
    \end{equation}
As we will see below, however, the high-frequency divergence in equation~\eqref{eq:cusp-energy-spectrum} can still cause problems elsewhere due to the nonlinear nature of GR.
Indeed, this divergence lies at the heart of the divergent behaviour we will encounter in the nonlinear memory waveforms.

\subsubsection{Kinks}
Now let us consider the kink case.
Suppose one of the left/right mode velocities $\dot{\vb*X}_\pm$ jumps discontinuously from $\vu*n_1$ to $\vu*n_2$ at some value of $\sigma_\pm$.
The oscillatory integral is then dominated by this jump, giving~\cite{Damour:2001bk}
    \begin{equation}
    \label{eq:kink-left-right-integral}
        \mathcal{I}^i_\pm\simeq\frac{\rmi}{\uppi f}\qty(\frac{\hat{n}_2^i}{1-\vu*n_2\vdot\vu*r}-\frac{\hat{n}_1^i}{1-\vu*n_1\vdot\vu*r}).
    \end{equation}
While this expression diverges if either $\vu*n_1$ or $\vu*n_2$ points along the line of sight $\vu*r$, we always project onto the transverse plane by contracting with the polarisation tensors $e^A_{ij}$, meaning that the GW strain is always finite.
Replacing one copy of the stationary-phase integral~\eqref{eq:stationary-phase-leading-order} with this kink integral, and neglecting a factor of order unity which depends on the 3-vectors $\vu*n_1$ and $\vu*n_2$, we therefore obtain the kink waveform
    \begin{equation}
    \label{eq:kink-waveform}
        \tilde{h}_\rmk(f,\vu*r)\simeq A_\rmk\frac{G\mu\ell^{1/3}}{r|f|^{5/3}}\Theta(\vu*r_\rmk\vdot\vu*r-\cos\theta_\rmb)\Theta(|f|-2/\ell),
    \end{equation}
    where $A_\rmk\equiv\sqrt{A_\rmc}/\uppi\lesssim0.2076$.
The corresponding energy spectrum is
    \begin{equation}
    \label{eq:kink-energy-spectrum}
        \epsilon_\rmk(f,\vu*r)\simeq\frac{\uppi}{2}A_\rmk^2G\mu(f\ell)^{-1/3}\Theta(\vu*r_\rmk\vdot\vu*r-\cos\theta_\rmb)\Theta(f-2/\ell),
    \end{equation}
    which, unlike the cusp spectrum, converges at high frequencies, even when the beam points exactly along the line of sight, $\vu*r=\vu*r_\rmk$.

We see that the kink waveform is remarkably similar to the cusp waveform, in that it is also linearly polarised and is also characterised by a power-law falloff in frequency, although the $\sim f^{-5/3}$ falloff here is slightly steeper than the $\sim f^{-4/3}$ we found for cusps.
The beaming angle $\theta_\rmb$ is the same as before, as it is still set by the range of directions $\vu*r$ in which the stationary-phase approximation holds.

One very important difference is that kinks are \emph{persistent} features of cosmic string loops, as opposed to cusps, which are \emph{transient}; this is because cusps are associated with a single value of both $\sigma_+$ and $\sigma_-$ at which $\vb*X_+=\vb*X_-$, which determines a fixed time $t=(\sigma_++\sigma_-)/2$, while kinks have only one of the two coordinates $\sigma_\pm$ fixed, with the corresponding locus in $(t,\sigma)$ describing the propagation of the kink around the loop over time.
As a result, the cusp waveform is associated with a single beaming direction $\vu*r_\rmc$, while the kink waveform is associated with a one-dimensional \enquote{fan} of directions defined by the kink's propagation direction as it travels around the loop.
The unit vector $\vu*r_\rmk$ that appears in equation~\eqref{eq:kink-waveform} is therefore defined as the direction in this \enquote{fan} which passes closest to the observer's line of sight $\vu*r$.
Here we assume that the length of the curve traced out by this fan on the unit sphere is comparable to the sphere's circumference, $\sim2\uppi$.
As a result, integrating equation~\eqref{eq:kink-energy-spectrum} over the sphere picks up a factor of $4\uppi\theta_\rmb$ rather than $\uppi\theta_\rmb^2$, giving
    \begin{equation}
        \bar{\epsilon}_\rmk(f)\simeq\frac{2^{5/3}}{3^{1/6}}(\uppi A_\rmk)^2G\mu(f\ell)^{-2/3}\Theta(f-2/\ell),\qquad\mathcal{E}_\rmk\simeq3^{5/6}(\uppi A_\rmk)^2G\mu.
    \end{equation}

Note that we could repeat this calculation to find the GWs radiated by milder discontinuities in the loop, due to jumps in higher-order derivatives of $\vb*X_\pm$.
However, for each order of the derivative we expect the waveform to be reduced by a factor of $\sim1/(|f|\ell)\ll1$, so that these features are less important than kinks in the high-frequency regime we are interested in.

\subsubsection{Kink-kink collisions}
Finally, the waveform for a kink-kink collision is given by setting both the integrals $\mathcal{I}^i_\pm$ equal to equation~\eqref{eq:kink-left-right-integral}.
This gives
    \begin{equation}
    \label{eq:kink-kink-collision-waveform}
        \tilde{h}_\mathrm{kk}(f,\vu*r)\simeq\frac{G\mu}{r\uppi^2f^2}\Theta(|f|-2/\ell).
    \end{equation}
Since the left-moving kink is associated with a fixed value of $\sigma_+$, and the right-moving kink with a fixed value of $\sigma_-$, their collision occurs at a single point $(\sigma_+,\sigma_-)$ and is therefore \emph{transient} like a cusp, in contrast with the persistent GW emission from each kink individually.
As mentioned previously, the resulting emission is perfectly isotropic, as there is no stationary-phase requirement and therefore no beaming direction.
As a result, kink-kink collisions produce no nonlinear memory, since the angular integral in equation~\eqref{eq:nonlinear-memory} has no TT part for this isotropic emission.
For this reason, kink-kink collisions are unimportant for much of the rest of this chapter.

For completeness, we include the energy spectrum associated with the waveform~\eqref{eq:kink-kink-collision-waveform},
    \begin{equation}
        \epsilon_\mathrm{kk}(f,\vu*r)=\frac{1}{4\uppi}\bar{\epsilon}_\mathrm{kk}(f)\simeq\frac{G\mu}{2\uppi^3f\ell},\qquad\mathcal{E}_\mathrm{kk}\simeq\frac{G\mu}{\uppi^2}.
    \end{equation}

\subsubsection{The gravitational-wave background from loops}
We can calculate the GWB spectrum generated by cusps, kinks, and kink-kink collisions on a cosmic population of loops using the \enquote{Phinney formula} from section~\ref{sec:modelling-stochastic-backgrounds}.
Of the two key ingredients that go into this formula, we have already calculated one: the energy spectra of each GW emission mechanism.
The other ingredient is the comoving rate density $\mathcal{R}$, which we can factorise into the comoving number density of loops, $n$, and the mean GW burst rate from each loop.
(Note that these are all functions of the loop length $\ell$.)
Since we know that each loop's oscillation period is given by $\ell/2$, we can write the latter in terms of the mean number of cusps, kinks, and kink-kink collisions per oscillation, which we denote by $N_i$, $i\in\{\rmc,\rmk,\mathrm{kk}\}$.
Determining these numbers for a realistic loop network is somewhat challenging; the number of cusps per oscillation $N_\rmc$ is thought to be of order unity, while the number of kinks $N_\rmk$ could in principle be much larger.
The number of kink-kink collisions is always $N_\mathrm{kk}=N_\rmk^2/4$, since each of the $N_\rmk/2$ left-moving kinks will collide with all of the $N_\rmk/2$ right-moving kinks once per oscillation period.
Putting all this together, the Phinney formula~\eqref{eq:phinney-formula-parameterised} gives
    \begin{equation}
    \label{eq:gwb-cosmic-strings}
        \Omega(f)=\frac{16\uppi G\mu}{3H_0^2}\sum_{i\in\{\rmc,\rmk,\mathrm{kk}\}}N_i\int\dd{t}a(t)\int\dd{\ell}n(\ell,t)\bar{\epsilon}_i(f/a(t)),
    \end{equation}
    where we have replaced the redshift integral with an integral over cosmic time $t$, with $a(t)$ the FLRW scale factor.

The loop network usually evolves toward a scaling solution~\cite{Bennett:1987vf}, such that the number density is given by
    \begin{equation}
        n(\ell,t)=\frac{a^3(t)}{t^4}\mathcal{F}(\gamma),
    \end{equation}
    where the dimensionless function $\mathcal{F}$ depends on the loop length and on cosmic time only through their dimensionless ratio $\gamma\equiv\ell/t$.
There are three widely-used models for the distribution function:
    \begin{enumerate}
        \item The original \enquote{one-scale} model of \citet{Vilenkin:2000jqa} (see also \citet{Siemens:2006vk});
        \item The model of \citet{Blanco-Pillado:2013qja}, which is calibrated to numerical simulations;
        \item The model of \citet{Ringeval:2005kr} (see also \citet{Lorenz:2010sm}), which is calibrated to a different set of simulations, and includes additional modelling of the effects of backreaction on the loops.
    \end{enumerate}
Following \citet{Auclair:2019wcv}, we refer to these as \enquote{model 1}, \enquote{model 2}, and \enquote{model 3} respectively.
Model 1 is widely considered obsolete, as it is incompatible with both of the main sets of Nambu-Goto network simulations~\cite{Ringeval:2005kr,Blanco-Pillado:2013qja}; however, we include it here for completeness.

\section{Nonlinear memory waveforms}
\label{sec:nonlinear-memory}

Now that we have derived the standard cusp and kink waveforms, equations~\eqref{eq:cusp-waveform} and~\eqref{eq:kink-waveform}, we are almost ready to calculate the nonlinear memory associated with each of these.
Before we start, we develop here some useful expressions---based on the nonlinear memory formula in equation~\eqref{eq:nonlinear-memory}---which will allow us to carry out these calculations entirely in the frequency domain.

We work in terms of the complex GW strain $h(t,\vb*r)=h_+-\rmi h_\times$ throughout.
We distinguish between the primary, oscillatory strain signal (sourced at linear order) and the additional strain due to the nonlinear memory effect by writing these as $h^{(0)}$ and $h^{(1)}$ respectively, reserving $h^{(n)}$ with $n\ge2$ for the \enquote{memory of the memory} and other higher-order memory contributions, which we discuss in sections~\ref{sec:cusp-higher-order} and~\ref{sec:kink-higher-order}.
The leading nonlinear memory correction term can be written in gauge-invariant form as~\cite{Thorne:1992sdb,Favata:2010zu}
    \begin{equation}
    \label{eq:memory-flux}
        h^{(1)}(t,\vb*r)=\frac{2G}{r}\int_{-\infty}^t\dd{t'}\int_{S^2}\dd[2]{\vu*r'}\frac{(e^{+,ij}-\rmi e^{\times,ij})\hat{r}'_i\hat{r}'_j}{1-\vu*r\vdot\vu*r'}\frac{\dd{E^{(0)}_\gw}}{\dd{t'}\dd[2]{\vu*r'}},
    \end{equation}
    which reduces to equation~\eqref{eq:nonlinear-memory} in the limit $t\to+\infty$.
Since the integral is over the (non-negative) GW energy flux from the source, $\dd{E^{(0)}_\gw}/\dd{t}\dd[2]{\vu*r}$, we generically obtain a nonzero memory correction which grows monotonically with time while the source is \enquote{on}.
By projecting onto the GW polarisation tensors we ensure that only the TT part of the angular integral contributes; this TT part vanishes if the emitted flux is exactly isotropic, but is generically non-vanishing for anisotropic emission.

Note that equation~\eqref{eq:memory-flux} corresponds to just one of many quadratic terms that appear on the right-hand side of the relaxed Einstein field equation~\cite{Pati:2000vt,Favata:2008yd}.
While this term has received particular attention in the literature due to its pleasing intuitive interpretation as the nonlinear gravitational counterpart to the linear GW memory effect, the other quadratic terms will also give nonlinear corrections to the GWs emitted by cosmic strings.
Our results here therefore represent just one particular nonlinear contribution to these GW signals.
Since we argue below that nonlinear effects are important near cusps on large cosmic string loops, it would be interesting to explore these additional contributions further in future work.
However, we note that there are two features of the nonlinear memory term that make it particularly important here, and justify our focus on this one term.
The first is that it is \emph{hereditary}, i.e., it depends on the entire history of the source; as we will see, this has the effect of transferring power from very high frequencies down to the frequencies we can observe.
The second is that it is \emph{radiative}, i.e., it determines the GW signal measured by distant observers, rather than just affecting the spacetime near to the source (indeed, Christodoulou's original expression for the nonlinear memory effect~\cite{Christodoulou:1991cr} was derived entirely in terms of gauge-invariant quantities at null infinity, and can be very cleanly identified with equation~\eqref{eq:memory-flux} above).

Returning to equation~\eqref{eq:memory-flux}, we can insert the expressions for the polarisation tensors from equation~\eqref{eq:polarisation-tensors} to write
    \begin{equation}
        (e^{+,ij}-\rmi e^{\times,ij})\hat{r}'_i\hat{r}'_j=[(\vu*\theta-\rmi\vu*\phi)\vdot\vu*r']^2.
    \end{equation}
We can also rewrite the energy flux~\eqref{eq:memory-flux} in terms of the linear GW strain of the source using equation~\eqref{eq:gw-energy-flux}.
The leading GW memory term then becomes
    \begin{equation}
    \label{eq:memory-strain}
        h^{(1)}(t,\vb*r)=\int_{-\infty}^t\frac{\dd{t'}}{2r}\int_{\vu*r'}|r\dot{h}^{(0)}(t',\vb*r')|^2,
    \end{equation}
    where we have written the angular integral as
    \begin{equation}
    \label{eq:integral-shorthand}
        \int_{\vu*r'}[\cdots]\equiv\int_{S^2}\frac{\dd[2]{\vu*r'}}{4\uppi}\frac{[(\vu*\theta-\rmi\vu*\phi)\vdot\vu*r']^2}{1-\vu*r\vdot\vu*r'}[\cdots],
    \end{equation}
    for brevity.
Throughout we refer to the oscillatory strain $h^{(0)}$ on the right-hand side that sources the memory as the \enquote{primary} GW emission.

\subsection{Late-time memory}

One drawback of equation~\eqref{eq:memory-strain} is that it is written in terms of the time-domain primary strain, whereas the cosmic string waveforms that we want to consider are much more naturally expressed in the frequency domain.
This problem disappears when we consider the \emph{late-time memory}, i.e. the total strain offset caused by the cusp,
    \begin{equation}
        \Updelta h^{(1)}\equiv\lim_{t\to\infty}h^{(1)}(t),
    \end{equation}
    as the time integral in equation~\eqref{eq:memory-strain} is then over the entire real line, and we can thus use Parseval's theorem to write
    \begin{equation}
    \label{eq:late-time-memory}
        \Updelta h^{(1)}(\vb*r)=\frac{2\uppi^2}{r}\int_\mathbb{R}\dd{f}\int_{\vu*r'}|rf\tilde{h}^{(0)}(f,\vb*r')|^2,
    \end{equation}
    where $\tilde{h}^{(0)}$ is the Fourier transform of the primary strain signal, and the extra factor of frequency comes from the time derivative on the strain.

\subsection{Frequency-domain memory waveforms}

The late-time memory~\eqref{eq:late-time-memory} is useful for giving a sense of the total size of the memory effect, but in many cases is not directly observable.
For example, the test masses in ground-based GW interferometers like LIGO/Virgo are not freely-falling in the plane of the interferometer arms; they are acted upon by feedback control systems at low frequencies to mitigate seismic noise~\cite{Saulson:1995zi}.
These low-frequency forces mean that the test masses cannot sustain a permanent displacement after the GW has passed.
However, the \enquote{ramping up} of the memory signal from zero at early times to $\Updelta h^{(1)}$ at late times can be measured if it contains power in the sensitive frequency band of the interferometer.

We are therefore interested in calculating the full memory signal in the frequency domain.
The simplest way of doing this is to use the result\footnote{%
One can show this by noting that $\int_{-\infty}^t\dd{t'}g(t')$ is just the convolution of $g(t)$ with the Heaviside step function $\Theta(t)$.
Since $\mathcal{F}[\Theta]=(1/2)\delta(f)-\rmi/(2\uppi f)$, the result follows from the convolution theorem, $\mathcal{F}[\Theta*g]=\mathcal{F}[\Theta]\mathcal{F}[g]$.}
    \begin{equation}
    \label{eq:step-function-fourier}
        \mathcal{F}\qty[\int_{-\infty}^t\dd{t'}g(t')]=\frac{1}{2}\tilde{g}(0)\delta(f)-\frac{\rmi}{2\uppi f}\tilde{g}(f)
    \end{equation}
    for a general function $g(t)$, where $\mathcal{F}[\cdots]$ denotes the Fourier transform, and $\tilde{g}=\mathcal{F}[g]$.
Applying this to equation~\eqref{eq:memory-strain}, we obtain
    \begin{equation}
        \tilde{h}^{(1)}(f)=\frac{1}{2}\Updelta h^{(1)}\delta(f)-\frac{\rmi}{2\uppi f}\int_\mathbb{R}\frac{\dd{t}}{2r}\int_{\vu*r'}\rme^{-2\uppi\rmi ft}|r\dot{h}^{(0)}|^2.
    \end{equation}
We can neglect the term proportional to $\delta(f)$, as this only contributes at $f=0$, and we are interested here in frequencies which are accessible to GW experiments (this zero-frequency term is still captured in equation~\eqref{eq:late-time-memory}).
By replacing each factor of $\dot{h}^{(0)}$ with its Fourier transform and massaging the resulting expression, we find
    \begin{equation}
    \label{eq:memory-frequency-domain}
        \tilde{h}^{(1)}(f)=-\frac{\rmi\uppi}{rf}\int_\mathbb{R}\dd{f'}\int_{\vu*r'}f'(f'-f)r^2\tilde{h}^{(0)}(f',\vu*r')\tilde{h}^{(0)*}(f'-f,\vu*r').
    \end{equation}
This is the simplest way of calculating the frequency-domain memory using only the primary frequency-domain signal $\tilde{h}^{(0)}$.

\section{Memory from cusps}
\label{sec:cusps}

We are now ready to calculate the nonlinear memory from cusps, inserting the primary waveform~\eqref{eq:cusp-waveform} into equations~\eqref{eq:late-time-memory} and~\eqref{eq:memory-frequency-domain} to obtain the late-time memory and frequency-domain waveform, and then iterating this process to obtain higher-order memory corrections.
(This iteration process is conceptually similar to the procedure described by \citet{Talbot:2018sgr,Khera:2020mcz} to incorporate higher-order memory effects in BBH waveform models.)

\subsection{Beaming effects}

The anisotropic beaming of the GWs from cusps is what gives rise to a nonzero memory effect (due to its nonzero TT projection), and is captured in the angular integral $\int_{\vu*r'}\Theta(\vu*r_\rmc\vdot\vu*r'-\cos\theta_\rmb)$, where we are using the shorthand~\eqref{eq:integral-shorthand}.
To compute this integral, it is convenient to define polar coordinates $\vu*r'=(\theta',\phi')$ such that the North pole $\theta'=0$ coincides with the centre of the beam $\vu*r_\rmc$.
The integrand then only has support for $\theta'\in[0,\theta_\rmb]$, so that we obtain
    \begin{align}
    \begin{split}
    \label{eq:spherical-integral-cusp}
        \int_{\vu*r'}\Theta(\vu*r_\rmc\vdot\vu*r'-\cos\theta_\rmb)&=\int_0^{\theta_\rmb}\dd{\theta'}\frac{\sin\theta'}{2}\int_0^{2\uppi}\frac{\dd{\phi'}}{2\uppi}\frac{[(\vu*\theta-\rmi\vu*\phi)\vdot\vu*r']^2}{1-\vu*r\vdot\vu*r'}\\
        &=\frac{1+\cos I}{1-\cos I}\qty[\frac{1}{2}(1-\cos\theta_\rmb)\cos\theta_\rmb-\frac{1}{4}\cos I\sin^2\theta_\rmb],
    \end{split}
    \end{align}
    where $I\equiv\cos^{-1}\vu*r_\rmc\vdot\vu*r$ is the inclination of the beam to the observer's line of sight.
In the high-frequency regime where the primary cusp and kink waveforms are valid, the beam angle is very small, so we expand equation~\eqref{eq:spherical-integral-cusp} to leading order in $\theta_\rmb$ to obtain
    \begin{equation}
    \label{eq:spherical-integral-cusp-result-small-beam}
        \int_{\vu*r'}\Theta(\vu*r_\rmc\vdot\vu*r'-\cos\theta_\rmb)\simeq\frac{\theta_\rmb^2}{4}(1+\cos I).
    \end{equation}

There are a few remarks worth making about equation~\eqref{eq:spherical-integral-cusp-result-small-beam}.
First, we note that it assumes $I\gg\theta_\rmb$; when instead the inclination is much smaller than the beaming angle, $I\ll\theta_\rmb$, the integral~\eqref{eq:spherical-integral-cusp} drops to zero, as the TT part of the angular emission vanishes when the beam is aligned with the line of sight.
The fact that the result~\eqref{eq:spherical-integral-cusp-result-small-beam} is purely real, despite the integrand being complex, shows that the memory strain is linearly polarised, much like the primary cusp and kink waveforms.
Geometrically, the $\theta_\rmb^2/4$ factor represents the fraction of the sphere taken up by the beam, while the $(1+\cos I)$ factor shows how the strength of the memory effect varies with inclination.
In particular, we notice that the memory strain is nonzero when the observer lies outside of the beam, $I>\theta_\rmb$.
In fact, the observed memory strain vanishes only when the beam is face-on ($I\ll\theta_\rmb$) or face-off ($I=\uppi$), as in both cases the angular pattern of the primary GW emission is isotropic around the line of sight.

It is interesting to note that this angular pattern---a broad $\sim(1+\cos I)$ distribution, except at very small inclinations where the memory signal drops to zero---is exactly the same as that of the \emph{linear} GW memory generated by the ejection of an ultrarelativistic \enquote{blob} of matter from a massive object along a fixed axis~\cite{Segalis:2001ns}.
Intuitively, this makes complete sense: the setup here is essentially the same, except that the \enquote{blob} is replaced by a burst of GWs.

\subsection{Late-time memory}

Now that we have computed the angular integral~\eqref{eq:spherical-integral-cusp}, it is straightforward to obtain the late-time memory from the cusp.
Inserting equations~\eqref{eq:cusp-waveform} and~\eqref{eq:beam-angle} into equation~\eqref{eq:late-time-memory} and integrating over frequency, we find
    \begin{equation}
    \label{eq:cusp-late-time-memory-result}
        \Updelta h^{(1)}_\rmc=2\times3^{2/3}(\uppi A_\rmc G\mu)^2(1+\cos I)\ell/r.
    \end{equation}
Inserting the numerical factors, this has a maximum value of $\Updelta h^{(1)}_\rmc\approx14.86\times(G\mu)^2\ell/r$ for nearly-face-on cusps $I\gtrsim0$, and smoothly tapers to zero for face-off cusps $I=\uppi$.
We note that equation~\eqref{eq:cusp-late-time-memory-result} should be taken with a pinch of salt, as it is sensitive to the low-frequency regime where the primary waveform is less accurate; nonetheless, we expect this to give a reasonable estimate of the magnitude of the memory effect.

\subsection{Full waveform}
\label{sec:cusp-full-waveform}

\begin{figure}[t!]
    \includegraphics[width=\textwidth]{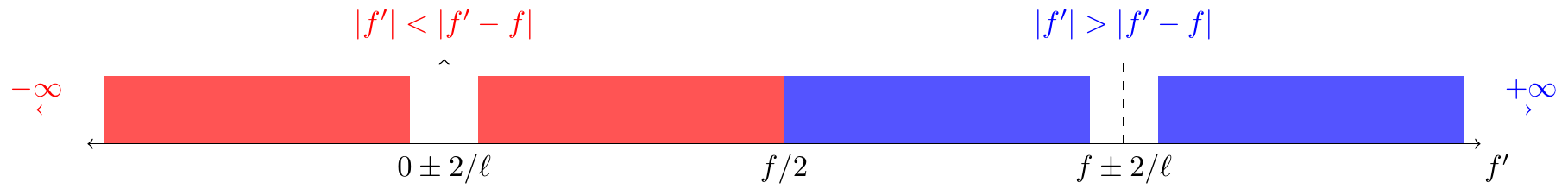}
    \caption{%
    Schematic illustration of the different contributions to $\tilde{h}^{(1)}_\rmc(f)$ from the integral over $f'$ in equation~\eqref{eq:memory-frequency-domain}.
    By introducing a dimensionless dummy variable $u\equiv f'/|f|$ we obtain the two integrals shown in equation~\eqref{eq:cusp-all-frequencies}, one corresponding to the finite interior region $2/\ell<f'<f-2/\ell$, and the other corresponding to the two semi-infinite exterior regions $f'<-2/\ell$ and $f'>f+2/\ell$.
    }
    \label{fig:fprime}
\end{figure}

We now calculate the full frequency-domain cusp memory waveform by inserting equations~\eqref{eq:cusp-waveform} and~\eqref{eq:beam-angle} into equation~\eqref{eq:memory-frequency-domain}.
In doing so, we must be careful to correctly account for the behaviour of the two frequency arguments $f'$ and $f'-f$; in particular, the integrand is only nonzero when both of these arguments have magnitude greater than $2/\ell$, and the size of the beam angle $\theta_\rmb$ must always be set by whichever of the two arguments has the greater magnitude (as this corresponds to a smaller, more restrictive beam).
These different contributions are illustrated in figure~\ref{fig:fprime} for the case where $f$ is positive.
Assuming that $|f|>4/\ell$, we find
    \begin{equation}
    \label{eq:cusp-all-frequencies}
        \tilde{h}^{(1)}_\rmc(f)=-\frac{\rmi\Updelta h^{(1)}_\rmc}{2^{2/3}3\uppi\ell^{1/3}f|f|^{1/3}}\qty[\int_{2/|f|\ell}^\infty\frac{\dd{u}}{u^{1/3}(1+u)}-\int_{2/|f|\ell}^{1/2}\frac{\dd{u}}{u^{1/3}(1-u)}],
    \end{equation}
    where $u$ is a dimensionless dummy variable.
We can simplify this by taking $|f|\gg4/\ell$, as this is the regime where the primary waveforms are valid; in this high-frequency limit, both integrals can be evaluated analytically.
Inserting equation~\eqref{eq:cusp-late-time-memory-result}, the final result is
    \begin{equation}
    \label{eq:cusp-result}
        \tilde{h}^{(1)}_\rmc(f)\simeq-\rmi B_c\frac{(G\mu)^2\ell^{2/3}}{rf|f|^{1/3}}(1+\cos I)\Theta(|f|-2/\ell),
    \end{equation}
    with a numerical prefactor,
    \begin{equation}
        B_\rmc\equiv\uppi A_\rmc^2(3/2)^{2/3}\qty[\frac{4\uppi}{3\sqrt{3}}-{}_2F_1(\tfrac{2}{3},\tfrac{2}{3};\tfrac{5}{3};-1)]\approx1.191,
    \end{equation}
    where ${}_2F_1$ is a hypergeometric function.
While we assumed $|f|>4/\ell$ in order to obtain equation~\eqref{eq:cusp-all-frequencies}, we have checked that our final expression~\eqref{eq:cusp-result} underestimates the true memory signal in the region $2/\ell<|f|<4/\ell$ (assuming the primary signal is accurate at these frequencies, which we expect to be the case to within an order of magnitude if one ignores the low-frequency motion not associated with the cusp), so we can safely leave the low-frequency cutoff at $2/\ell$ as in the primary waveform.
The simple expression~\eqref{eq:cusp-result} thus gives a conservative but generally accurate model of the time-varying part of the cusp memory waveform at all frequencies greater that the fundamental mode of the loop, while the zero-frequency offset is described by equation~\eqref{eq:cusp-late-time-memory-result}.

\begin{figure}[t!]
    \begin{center}
        \includegraphics[width=0.6\textwidth]{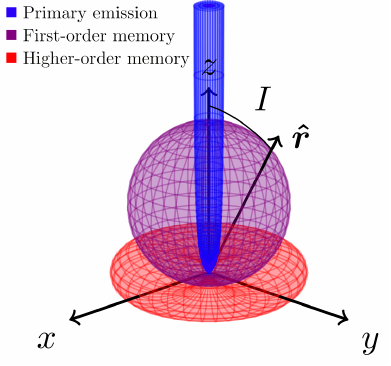}
    \end{center}
    \caption{%
    A cartoon illustration of the angular distribution of the energy radiated by a cusp.
    The primary emission (blue) is concentrated in a narrow beam of width $\theta_\rmb\sim(f\ell)^{-1/3}$, while the first-order memory emission (violet) is proportional to $(1+\cos I)^2$, and the second-order memory emission (red) is proportional to $\sin^4I$.
    }
    \label{fig:cusp-memory-angular-patterns}
\end{figure}

Comparing equation~\eqref{eq:cusp-result} with the primary cusp waveform~\eqref{eq:cusp-waveform}, we see that they are remarkably similar to each other, with both being given by the same simple frequency power law $\sim f^{-4/3}$ and the same dependence on the loop length $\ell$ (the latter being required by dimensional arguments).
There are, however, some important differences:
    \begin{enumerate}
        \item The memory GW emission has broad support on the sphere, while the primary waveform only has support inside a narrow beam (see figure~\ref{fig:cusp-memory-angular-patterns}).
        \item The memory waveform is suppressed by an additional power of $G\mu$ (this makes intuitive sense, given that it is a nonlinear effect sourced by the primary GW emission).
        \item The numerical constant in front of the memory waveform, $B_\rmc$, is an order of magnitude larger than that in front of the primary waveform, $A_\rmc$.
    \end{enumerate}
The first point is particularly crucial, as it lies at the heart of the divergent behaviour that we investigate in sections~\ref{sec:energy-divergence} and~\ref{sec:cusp-higher-order}.

\subsection{Time-domain waveform near the arrival time}

\begin{figure}[t!]
    \begin{center}
        \includegraphics[width=0.7\textwidth]{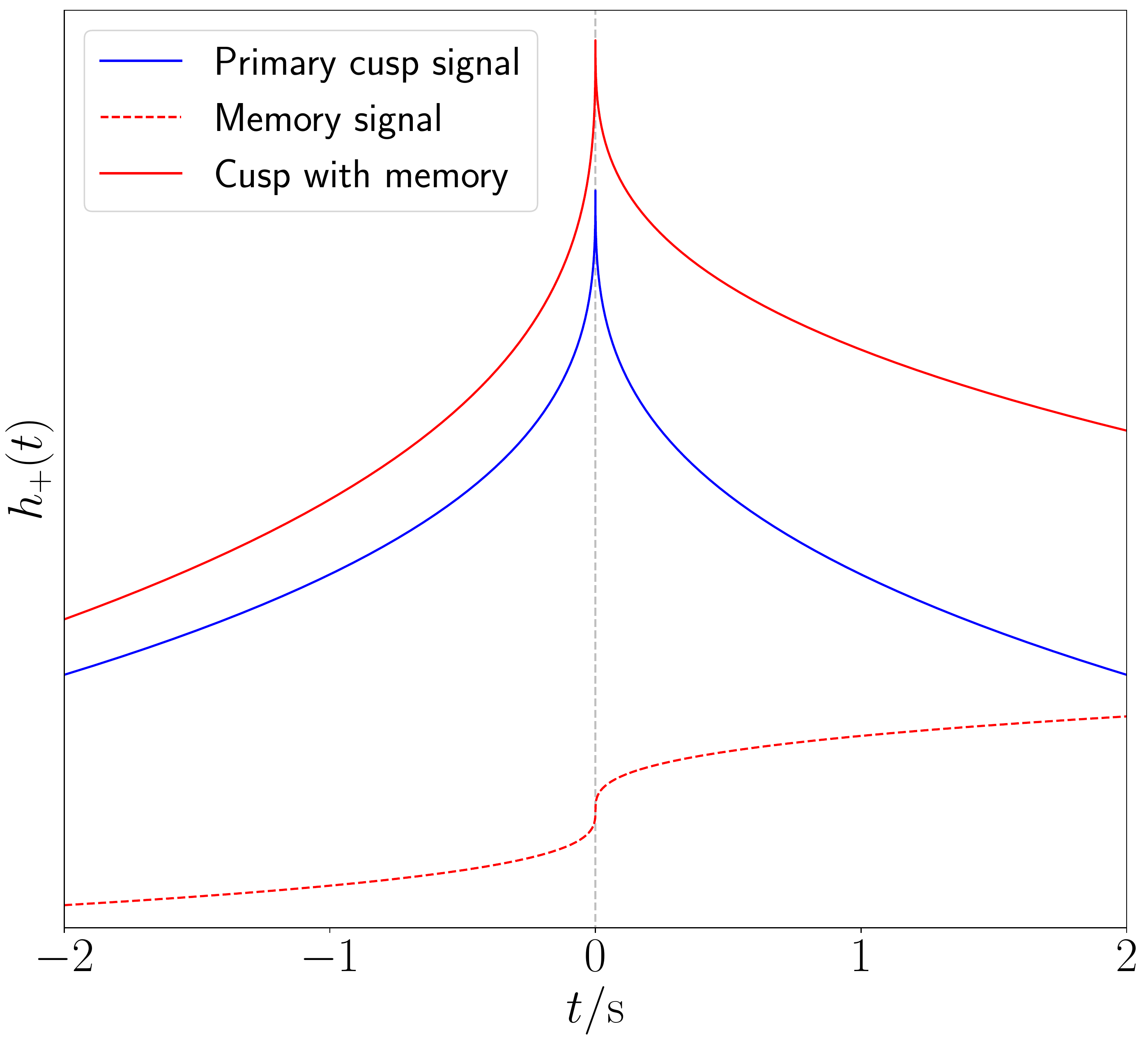}
    \end{center}
    \caption{%
    The time-domain GW strain $h_\rmc(t)$ from a cusp, with and without the leading-order memory contribution~\eqref{eq:cusp-time-domain-approx}.
    The memory is exaggerated by a factor of $\sim1/G\mu$ here to make it visible.
    For small inclinations $I<\theta_\rmb$ the observer lies within the cusp's beam, and sees the memory superimposed on the primary cusp signal (solid red line).
    If the inclination is very small, $I\ll\theta_\rmb$, then the memory vanishes and only the primary cusp signal is observable (solid blue line).
    In most cases however, the observer lies outside of the beam, and only the memory is observable (red dashed line).
    Note that the higher-order memory contributions (order $n\ge2$) are \emph{not} shown here; these would look like step functions in the time domain, with height that either diverges rapidly with $n$ (\enquote{large} loops, $\ell\gtrsim\delta/(G\mu)^3$) or converges so rapidly that the contribution to the total signal is negligible (\enquote{small} loops, $\ell\lesssim\delta/(G\mu)^3$).
    }
    \label{fig:cusp-td-waveform}
\end{figure}

While we have focused on deriving the memory signal in the frequency domain, we can inverse-Fourier-transform equation~\eqref{eq:cusp-result} to obtain a simple closed-form expression in the time domain,
    \begin{equation}
    \label{eq:cusp-time-domain-approx}
        h^{(1)}_\rmc(t)-h^{(1)}_\rmc(t_0)\simeq-2^{5/3}3\uppi B_\rmc\frac{(G\mu)^2}{r}(1+\cos I)(t-t_0)\qty[1-\frac{\Gamma(2/3)}{(4\uppi|t-t_0|/\ell)^{2/3}}].
    \end{equation}
(Here we have re-introduced the time of arrival of the primary cusp signal, $t_0$, rather than setting it to zero.)
This time-domain waveform is shown in figure~\ref{fig:cusp-td-waveform}.
Note that this is real, which means that the signal is linearly polarised, just like the primary waveform~\eqref{eq:cusp-waveform}.

It is important to note that since equation~\eqref{eq:cusp-result} is only valid for high frequencies $f\gg2/\ell$, equation~\eqref{eq:cusp-time-domain-approx} must only be valid for a short duration $|t-t_0|\ll\ell$ around the arrival time.
However, for typical loop sizes this \enquote{short duration} is actually much longer than the relevant observational timescale (typically a few seconds), so the approximation in equation~\eqref{eq:cusp-time-domain-approx} may be a useful one.

\subsection{Radiated energy}
\label{sec:radiated-energy}

Inserting the memory waveform~\eqref{eq:cusp-result} into equation~\eqref{eq:dimensionless-energy}, we find the radiated energy spectrum
    \begin{equation}
        \epsilon^{(1)}_\rmc(f,\vu*r)\simeq\frac{\uppi}{2}B_\rmc^2(G\mu)^3(f\ell)^{1/3}(1+\cos I)^2\Theta(f-2/\ell).
    \end{equation}
The fact that $\tilde{h}^{(0)}_\rmc$ is purely real while $\tilde{h}^{(1)}_\rmc$ is purely imaginary means that there is no coherent cross-energy between the two contributions, so the total energy is just $\epsilon^{(0)}_\rmc+\epsilon^{(1)}_\rmc$.
For observers in the beaming direction the energy spectra of the primary and memory signals scale in the exact same way with frequency, but with a much smaller coefficient for the memory signal.
However, the picture changes drastically when integrating the spectra over the sphere to compute the total emission, as the primary emission is then suppressed by a factor of $\theta_\rmb^2\sim(f\ell)^{-2/3}$, while the memory waveform gives
    \begin{equation}
    \label{eq:cusp-memory-isotropic-energy}
        \bar{\epsilon}^{(1)}_\rmc(f)\simeq\frac{8}{3}(\uppi B_\rmc)^2(G\mu)^3(f\ell)^{1/3}\Theta(f-2/\ell).
    \end{equation}
We see that the isotropic energy spectrum due to the memory emission is blue-tilted (i.e. grows with frequency), while the primary spectrum~\eqref{eq:cusp-isotropic-energy} is red-tilted.
This means that the memory emission dominates at very high frequencies,
    \begin{equation}
        f>\frac{3}{16\ell}\qty(\frac{A_\rmc}{B_\rmc G\mu})^3\approx8\times10^{22}\,\mathrm{Hz}\times\qty(\frac{\ell}{\mathrm{pc}})^{-1}\qty(\frac{G\mu}{10^{-11}})^{-3}.
    \end{equation}

\subsection{Ultraviolet divergence of the radiated energy}
\label{sec:energy-divergence}

The total fraction of the loop's energy radiated by the cusp memory can be found by integrating over frequency, as in equation~\eqref{eq:total-energy}.
For the primary signal, we saw in section~\ref{sec:cusps-kinks} that this gives a small total energy fraction of order $G\mu$.
For the memory signal, however, the integral diverges.
This is clearly unphysical, and shows a breakdown in the validity of equation~\eqref{eq:cusp-result}.
Note however that this breakdown is \emph{not} in the low-frequency regime where we know that equation~\eqref{eq:cusp-result} is inaccurate; rather, the integral has an ultraviolet divergence that goes like $\sim f^{1/3}$ at high frequencies $f\to\infty$.
Instead, we can understand this divergence as a breakdown of the Nambu-Goto approximation for the loop dynamics.

\begin{figure}[t!]
    \begin{center}
        \includegraphics[width=0.7\textwidth]{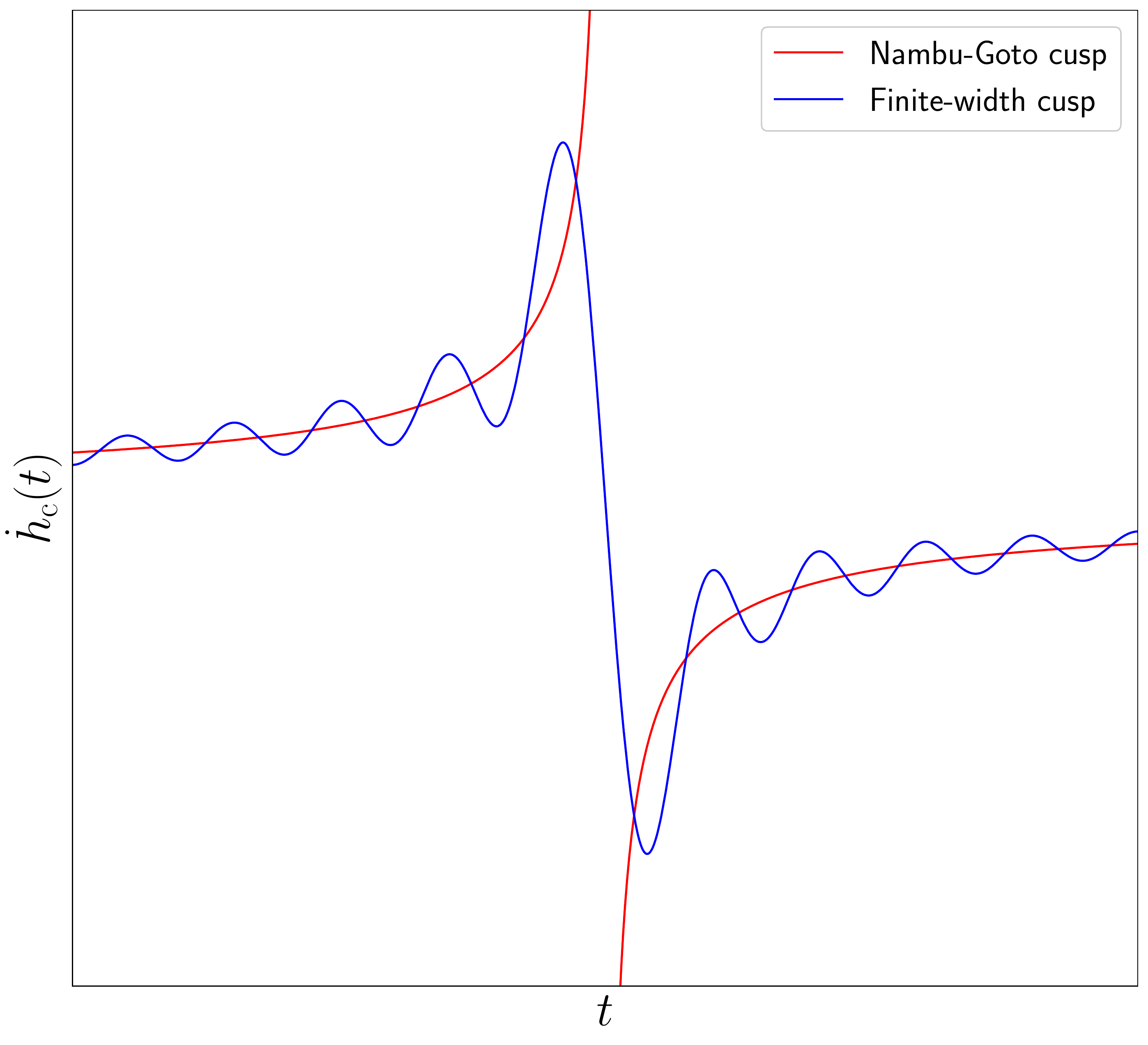}
    \end{center}
    \caption{%
    The time derivative of the cusp strain signal close to the peak, $|t-t_0|\ll\ell$, for an observer in the beaming direction, $I=0$.
    This diverges for the standard Nambu-Goto waveform~\eqref{eq:cusp-waveform}, ruining the validity of the Nambu-Goto approximation, and causing a divergence in the energy radiated by the first-order memory.
    By introducing a frequency cutoff due to the finite string width, $f<1/\delta$, we see that the derivative becomes finite and continuous, and the first-order memory divergence is regularised.
    }
    \label{fig:hdot-cusp}
\end{figure}

In section~\ref{sec:nambu-goto}, we justified the use of the Nambu-Goto action by arguing that possible curvature corrections $\hbar\kappa/\mu$ should generically be very small for macroscopic loops, typically of order $(\delta/\ell)^2$.
However, as we saw in section~\ref{sec:flat-space-eoms}, the Nambu-Goto action generically predicts the formation of cusps, where the GW strain looks locally like $h(t)\sim|t-t_0|^{1/3}$.
As was already recognised by \citet{Damour:2001bk}, the curvature associated with this scales as $\kappa\propto\ddot{h}\propto|t-t_0|^{-5/3}$, which diverges at the peak of the cusp, clearly ruining the validity of the Nambu-Goto approximation.
This problem does not manifest itself in the primary cusp waveform, since the beaming angle $\theta_\rmb$ decreases fast enough with frequency to ensure the total radiated energy is finite, cf. equation~\eqref{eq:cusp-total-energy-primary}.
However, the energy flux in the centre of the beam is still divergent, and as we have shown here, this sources further divergences due to the nonlinear nature of gravity.
Since the memory effect is not beamed, there is nothing suppressing it at high frequencies.

The simplest way to regularise this divergence is to impose an ultraviolet cutoff at some high energy scale, for which the only obvious candidate is the string width scale.
By truncating the frequency-domain cusp waveform at $f\sim1/\delta$, the cusp is smoothed out on timescales $|t-t_0|\sim\delta$, and the curvature reaches a finite maximum value that scales like $\kappa\propto\delta^{-5/3}$.
This smoothing is shown explicitly in figure~\ref{fig:hdot-cusp}, where we see that the time derivative of the strain diverges in the Nambu-Goto case, but is finite and continuous if a finite width is introduced.
(In this heuristic setup, it is not immediately clear whether or not the smoothing afforded by a finite string width is strong enough to prevent higher-curvature terms in the action from becoming important near the cusp; we return to this point later.)

We therefore consider only GWs with frequency $f<1/\delta$.
This has no impact on the primary waveform, since $1/\delta\approx10^{38}\,\mathrm{Hz}\times\qty(G\mu/10^{-11})^{1/2}$, which is beyond the reach of any current or planned GW experiments.
However, the cutoff \emph{does} impact the GW memory.
For example, setting an upper limit of $f=1/\delta$ in the integral~\eqref{eq:total-energy}, we find a finite value for the energy radiated by the cusp memory,
    \begin{equation}
    \label{eq:cusp-total-energy-memory}
        \mathcal{E}^{(1)}_\rmc\simeq8(\uppi B_\rmc)^2(G\mu)^3(\ell/\delta)^{1/3}\approx2\times10^{-16}\times\qty(\frac{G\mu}{10^{-11}})^{19/6}\qty(\frac{\ell}{\mathrm{pc}})^{1/3},
    \end{equation}
    where we have used $\delta\approx\ell_\Pl/\sqrt{G\mu}$ in the second expression.
This shows that the energy radiated due to the first-order memory effect is smaller than that from the primary emission for observationally-allowed values of the string tension, $G\mu\lesssim10^{-11}$.
However, the factor of $(\ell/\delta)^{1/3}\gg1$ in equation~\eqref{eq:cusp-total-energy-memory} is concerning.
Based on equation~\eqref{eq:cusp-total-energy-memory}, cusps on GUT-scale strings ($G\mu=10^{-6}$) would radiate far more energy through the memory effect than through the primary GWs; for $\ell\gtrsim0.4\,\mathrm{pc}$ they would radiate more than the entire energy of the loop.
Such a situation would be unphysical, and would indicate the breakdown of the validity of the primary cusp waveform.
One might argue that this is not an issue, since GUT-scale Nambu-Goto strings are already ruled out by observations; however, we show below that accounting for higher-order memory effects exacerbates the problem, leading to similar unphysical results even for much lower string tensions.

We should note that the cutoff imposed here is rather ad-hoc, and is done in a way that is agnostic to the underlying microphysics of the string.
In reality, the divergence will be resolved in a way which may depend on the microphysics, and the GW memory observables may be sensitive to this; indeed, this is implied by the fact that equation~\eqref{eq:cusp-total-energy-memory} depends directly on the string width $\delta$.

The fact that equation~\eqref{eq:cusp-total-energy-memory} depends on the loop length $\ell$ distinguishes it from the equivalent expression for the primary signal, equation~\eqref{eq:cusp-total-energy-primary}, in which we found that a fixed fraction of the loop's energy (roughly $\sim G\mu$) was radiated by the cusp regardless of the loop length.
The reason for the difference here is that we have introduced a new lengthscale to the problem, the string width $\delta$, so on dimensional grounds we can expect powers of $\ell/\delta$ to appear.
The reason we have a \emph{positive} power of $\ell/\delta$ here is that, as we can see from equation~\eqref{eq:cusp-memory-isotropic-energy}, the energy radiated by the cusp memory at any fixed frequency $f$ is larger for larger loops, scaling like $\propto\ell^{1/3}$.
In equation~\eqref{eq:cusp-total-energy-memory} we are essentially calculating the energy radiated at the string-width frequency $1/\delta$, and thus end up with a factor of $(\ell/\delta)^{1/3}$.
This is a foretaste of one of our key findings in this chapter, which is that the cusp memory diverges only for loops greater than some critical length.

\subsection{Second-order memory}
\label{sec:cusp-higher-order}

The memory waveform~\eqref{eq:cusp-result} describes the spacetime curvature generated by the energy-momentum of the primary GWs from the cusp.
However, these memory GWs themselves carry energy-momentum, and will in turn act as a source of their own GW memory (see reference~\cite{Talbot:2018sgr} for a discussion of this effect in the context of binary black hole coalescences).
We refer to this \enquote{memory of the memory} here as the \emph{second-order memory effect}.

The calculation of this second-order memory contribution is straightforward for cusps, simply substituting equation~\eqref{eq:cusp-result} into the right-hand side of equation~\eqref{eq:memory-frequency-domain} and following all of the same steps as before.
The key difference is in the angular integral, $\int_{\vu*r'}(1+\cos I')^2=\tfrac{1}{6}\sin^2I$, which, due to the broader emission of the first-order memory signal, is not suppressed by a factor of $\theta_\rmb^2$; cf. equation~\eqref{eq:spherical-integral-cusp-result-small-beam}.
The resulting expressions for the late-time memory and frequency-domain waveform are
    \begin{align}
    \begin{split}
        \Updelta h^{(2)}_\rmc&=\frac{2(\uppi B_\rmc)^2}{3r}(G\mu)^4\ell^{4/3}\sin^2I\int_{2/\ell}^\infty\frac{\dd{f}}{f^{2/3}},\\
        \tilde{h}^{(2)}_\rmc(f)&=-\frac{\rmi\uppi B_\rmc^2}{3rf}(G\mu)^4\ell^{4/3}\sin^2I\qty[\int_{2/\ell}^\infty\frac{\dd{f'}}{f'^{1/3}(f'+|f|)^{1/3}}+\int_{2/\ell}^{|f|/2}\frac{\dd{f'}}{f'^{1/3}(|f|-f')^{1/3}}],
    \end{split}
    \end{align}
    both of which contain integrals which diverge due to high-frequency contributions from the first-order memory.
Introducing the $f<1/\delta$ cutoff from section~\ref{sec:energy-divergence} once again, we find to leading order in $\delta$,
    \begin{equation}
    \label{eq:cusp-2nd-order-mem}
        \Updelta h^{(2)}_\rmc\simeq\frac{2\uppi^2\ell}{r}B_\rmc^2(G\mu)^4(\ell/\delta)^{1/3}\sin^2I,\qquad\tilde{h}^{(2)}_\rmc(f)\simeq-\frac{\rmi\Updelta h^{(2)}_\rmc}{2\uppi f}\Theta(1/\delta-|f|)\Theta(|f|-2/\ell).
    \end{equation}
This is interesting in that it departs from the the $\sim f^{-4/3}$ scaling of both the primary waveform~\eqref{eq:cusp-waveform} and the first-order memory waveform~\eqref{eq:cusp-result}; the second-order memory instead has a slower decay with frequency, due to the lack of beaming in the first-order memory.
In fact, for intermediate frequencies $2/\ell<|f|<1/\delta$ the second-order memory is identical to the Fourier transform of a Heaviside step function of height $\Updelta h^{(2)}_\rmc$.
This means that on timescales much shorter than the loop oscillation period $\ell/2$ and much longer than the light-crossing time of the loop width $\delta$, the second-order memory signal looks like a step function in the time domain; this is because the signal is dominated by high frequencies $f\lesssim1/\delta$, and therefore \enquote{switches on} in a very short time interval\footnote{%
This picture is closely related to the zero-frequency limit (ZFL) method for calculating radiation from high-energy gravitational scattering~\cite{Smarr:1977fy,Turner:1978jj,Bontz:1979zfl,Wagoner:1979dd}, which leverages the fact that the scattering is effectively instantaneous compared to the GW frequencies of interest.}, $|t-t_0|\lesssim\delta$.

The waveform~\eqref{eq:cusp-2nd-order-mem} can be used to calculate a second-order correction to the energy radiated by the cusp,
    \begin{align}
    \begin{split}
    \label{eq:cusp-energy-memory-2nd-order}
        \epsilon^{(2)}_\rmc&=\frac{\uppi r^2f^3}{2G\mu\ell}\qty(|\tilde{h}^{(0)}_\rmc+\tilde{h}^{(1)}_\rmc+\tilde{h}^{(2)}_\rmc|^2-|\tilde{h}^{(0)}_\rmc+\tilde{h}^{(1)}_\rmc|^2)=\frac{\uppi r^2f^3}{2G\mu\ell}\qty(|\tilde{h}^{(2)}_\rmc|^2+2\tilde{h}^{(1)}_\rmc\tilde{h}^{(2)*}_\rmc)\\
        &=\qty[\frac{\uppi^3}{2}B_\rmc^4(G\mu)^7(\ell/\delta)^{2/3}f\ell\sin^4I+\uppi^2B_\rmc^3(G\mu)^5(\ell/\delta)^{1/3}(f\ell)^{2/3}\sin^2I(1+\cos I)]\\
        &\qquad\times\Theta(1/\delta-|f|)\Theta(|f|-2/\ell).
    \end{split}
    \end{align}
We see that, unlike for the first-order correction, there are now nonzero cross-terms due to $\tilde{h}^{(1)}_\rmc$ and $\tilde{h}^{(2)}_\rmc$ being exactly in phase with each other, resulting in two new contributions to the energy.
Integrating over frequency and emission direction, the total energy is given by
    \begin{equation}
    \label{eq:cusp-total-energy-memory-2nd-order}
        \mathcal{E}^{(2)}_\rmc=\frac{16}{15}(\uppi B_\rmc)^4(G\mu)^7(\ell/\delta)^{5/3}+4(\uppi B_\rmc)^3(G\mu)^5(\ell/\delta).
    \end{equation}
Comparing with the corresponding first-order result~\eqref{eq:cusp-total-energy-memory} we see a clear pattern emerging, where the second term in equation~\eqref{eq:cusp-total-energy-memory-2nd-order} is multiplied by a factor of $\sim\uppi B_\rmc(G\mu)^2(\ell/\delta)^{2/3}$ compared to the first-order energy, and multiplying by the same factor again gives the first term in equation~\eqref{eq:cusp-total-energy-memory-2nd-order}.
This factor is greater than unity so long as the loop length $\ell$ is larger than
    \begin{equation}
    \label{eq:ell-star}
        \ell_*\equiv\frac{\delta}{(\uppi B_\rmc)^{3/2}(G\mu)^{3}}\approx0.7\,\mathrm{km}\times\qty(\frac{G\mu}{10^{-11}})^{-7/2},
    \end{equation}
    which applies to all macroscopically-large loops.

One would naively expect successive memory corrections for a generic GW source to become less and less important at higher order; the fact that they become \emph{more} important for such a large class of cosmologically-relevant cosmic string loops is surprising, and suggests that something unphysical is happening.
Indeed, if we plug numerical values into equation~\eqref{eq:cusp-total-energy-memory-2nd-order}, we find that loops larger than about $0.1\,\mathrm{pc}\times(G\mu/10^{-11})^{-47/10}$ have $\mathcal{E}^{(2)}_\rmc$ greater than unity, meaning that they radiate away more than the total energy of the loop.
This represents a significant worsening of the issue we identified at the end of section~\ref{sec:energy-divergence}.
The problem only gets worse when we go beyond second-order corrections, as we demonstrate below.

\subsection{Higher-order memory, and another divergence}
\label{sec:higher-order-divergence}

Iterating the procedure described above to calculate the third-order GW memory, it is straightforward to show that it obeys the same step-function-like relation that we found at second order,
    \begin{equation}
        \tilde{h}^{(3)}_\rmc(f)=-\frac{\rmi\Updelta h^{(3)}_\rmc}{2\uppi f}\Theta(1/\delta-|f|)\Theta(|f|-2/\ell),
    \end{equation}
    with the late-time memory given by
    \begin{equation}
        \Updelta h^{(3)}_\rmc=\frac{2\uppi^2}{r}\int_\mathbb{R}\dd{f}\int_{\vu*r'}r^2f^2\qty(|\tilde{h}^{(2)}_\rmc|^2+2\tilde{h}^{(1)}_\rmc\tilde{h}^{(2)*}_\rmc),
    \end{equation}
    where we have made sure to include both of the second-order energy contributions as source terms for the third-order memory.
Upon integration, this becomes
    \begin{equation}
        \Updelta h^{(3)}_\rmc=-\frac{4\delta}{5r}(\ell/\ell_*)^{8/3}\sin^2I\qty(1-\frac{1}{3}\cos^2I)-\frac{2\delta}{r}(\ell/\ell_*)^2\sin^2I\qty(1+\frac{3}{5}\cos I),
    \end{equation}
    where $\ell_*$ is the $G\mu$-dependent lengthscale defined in equation~\eqref{eq:ell-star}.

The third-order memory clearly scales differently for loops with $\ell\gg\ell_*$ (which we call \enquote{large loops}) compared to those with $\ell\ll\ell_*$ (which we call \enquote{small loops}).
When going to fourth order and beyond, we obtain an increasing number of cross-terms in the energy at each order (starting with $|\tilde{h}^{(n)}_\rmc|^2$, then $2\tilde{h}^{(n-1)}_\rmc\tilde{h}^{(n)*}_\rmc$, $2\tilde{h}^{(n-2)}_\rmc\tilde{h}^{(n)*}_\rmc$, and so on, down to $2\tilde{h}^{(1)}_\rmc\tilde{h}^{(n)*}_\rmc$), resulting in a proliferation of terms in the resulting memory expressions, each with a different power of $\ell/\ell_*$.
It is therefore much simpler to treat small and large loops separately, and focus on the leading power of $\ell/\ell_*$ in each case.
This results in a very economical formula for iterating the memory calculation, valid for all $n\ge2$,
    \begin{equation}
    \label{eq:iterate-memory}
        \Updelta h^{(n)}_\rmc=
        \begin{cases}
            \displaystyle\frac{r}{\delta}\int_{\vu*r'}|\Updelta h^{(n-1)}_\rmc|^2, & \text{for}\;\ell\gg\ell_*\\[8pt]
            \displaystyle\frac{6\uppi r}{\delta^2}\int_{\vu*r'}\Updelta h^{(n-1)}_\rmc|\tilde{h}^{(1)}_\rmc(1/\delta)|, & \text{for}\;\ell\ll\ell_*
        \end{cases}
    \end{equation}
    where we have evaluated the frequency integral in each case, leaving just the angular integral.
The frequency-domain waveform is then given by the same quasi-step-function form as before,
    \begin{equation}
        \tilde{h}^{(n)}_\rmc=-\frac{\rmi\Updelta h^{(n)}_\rmc}{2\uppi f}\Theta(1/\delta-|f|)\Theta(|f|-2/\ell),
    \end{equation}
    and the corresponding energy spectra are given by
    \begin{equation}
        \epsilon^{(n)}_\rmc=
        \begin{cases}
            \displaystyle\frac{r^2f}{8\uppi G\mu\ell}|\Updelta h^{(n)}_\rmc|^2, & \text{for}\;\ell\gg\ell_*\\[8pt]
            \displaystyle\frac{r^2f^2}{2G\mu\ell}\Updelta h^{(n)}_\rmc|\tilde{h}^{(1)}_\rmc|, & \text{for}\;\ell\ll\ell_*
        \end{cases}
    \end{equation}

Solving equation~\eqref{eq:iterate-memory} iteratively with equation~\eqref{eq:cusp-2nd-order-mem} as an input, we find for all $n\ge2$,
    \begin{equation}
    \label{eq:cusp-higher-order-final}
        \Updelta h^{(n)}_\rmc=
        \begin{cases}
            \displaystyle\frac{\delta}{r}(\ell/\ell_*)^{2^n/3}L_n(I), & \text{for}\;\ell\gg\ell_*\\[8pt]
            \displaystyle\frac{\delta}{r}(\ell/\ell_*)^{2n/3}S_n(I), & \text{for}\;\ell\ll\ell_*
        \end{cases}
    \end{equation}
    where $L_n(I)$ and $S_n(I)$ are polynomials in $\cos I$ which describe the angular pattern of the memory for large and small loops, respectively---these are described in detail in appendix~\ref{sec:iota-polynomials}.
For small loops, all memory effects are subdominant compared to the primary emission, and equation~\eqref{eq:cusp-higher-order-final} gives a convergent geometric series.
For large loops, on the other hand, the memory emission becomes stronger at each order, and equation~\eqref{eq:cusp-higher-order-final} gives a lacunary series which diverges extremely quickly.
One might hope that the polynomials $L_n(I)$ decrease in magnitude fast enough to counteract the divergence, but we find empirically in appendix~\ref{sec:iota-polynomials} that $|L_n(I)|\approx5(2/5)^{2^{n-2}}\sin^2I$, so the series diverges as long as $\ell\gtrsim(5/2)^3\ell_*\approx1\,\text{km}\times(G\mu/10^{-11})^{-7/2}$, as shown in figure~\ref{fig:cusp-energy}.
We discuss the cause of this divergence in appendix~\ref{sec:rhdot}, and argue that it is caused by a trans-Planckian GW energy flux from the cusp.

\subsection{Memory from cusp collapse}
\label{sec:cusp-collapse-memory}

The results of the previous section show that, even with an ultraviolet cutoff in place at the scale of the string width, the standard cusp waveform~\eqref{eq:cusp-waveform} leads to a divergence for all \enquote{large} loops with length $\ell\gtrsim\delta/(G\mu)^3$.
The fact that the divergence appears in observable, gauge-independent quantities (the memory strain and, therefore, the radiated energy) means that something unphysical must be going on.
Since the only inputs to our calculation are the cusp waveform~\eqref{eq:cusp-waveform} and the GW memory formula~\eqref{eq:memory-flux}, at least one of these two ingredients must break down for cusps on loops of this size.

In order to track down the cause of the divergence, let us list the assumptions that go into equations~\eqref{eq:cusp-waveform} and~\eqref{eq:memory-flux}:
    \begin{enumerate}
        \item The GW frequency is assumed to be much greater than the fundamental mode of the loop, $f\gg2/\ell$.
        This is because the waveform~\eqref{eq:cusp-waveform} is derived using the universal behaviour of the loop on scales $\ll\ell$ near the cusp.
        \item The loop's dynamics are assumed to follow the Nambu-Goto action~\eqref{eq:nambu-goto-action} on lengthscales larger than the loop width $\delta$.
        (We have imposed a cutoff that effaces scales below this, in order to regularise the first divergence we encountered in section~\ref{sec:energy-divergence}.)
        \item The loop is assumed to evolve according to the flat-space equations of motion~\eqref{eq:flat-space-eoms-1}--\eqref{eq:flat-space-eoms-3}; i.e. gravitational backreaction is assumed to be negligible.
        \item The GWs generated by the loop are assumed to be well-described by linear perturbations on a flat background.
    \end{enumerate}

As mentioned earlier, the first assumption cannot be the source of the problem, as the divergence is associated with very high frequencies near the string width scale.

The second assumption is robust so long as ($i$) the higher-order curvature terms in the worldsheet Lagrangian~\eqref{eq:worldsheet-lagrangian} are negligible, and ($ii$) the strings are created through the breaking of a local gauge symmetry, so that the underlying field theory does not give rise to long-range interactions (i.e. we are not considering global strings, which would instead be described by the Kalb-Ramond action~\cite{Vilenkin:2000jqa}).
As mentioned in section~\ref{sec:energy-divergence}, while introducing a finite string width prevents the curvature from diverging, it does not necessarily guarantee that the higher-order curvature terms are negligible.
In principle the memory divergence identified here could be cured by departures from the Nambu-Goto action near the cusp.
However, these departures would have to take place on scales much greater than the string width, which seems very difficult to achieve.

By elimination, it seems that the problem is mostly likely due to assumptions 3 and 4: i.e., that the flat-space description of the cusp's dynamics and GW generation is inconsistent.
Indeed, we can trace the divergence back to the fact that the integrated GW energy flux diverges in the centre of the cusp's beam, $\int\dd{(\ln f)}\epsilon^{(0)}_\rmc(f,\vu*r_\rmc)\to\infty$, which already suggests that a flat-space description is insufficient.
This divergent flux is hidden somewhat by the narrowness of the beam, $\theta_\rmb\sim(f\ell)^{-1/3}$, which ensures that $\mathcal{E}^{(0)}_\rmc$ is finite, but we have shown that the GW memory inherits and amplifies the divergence.

One possible resolution is that the gravitational backreaction of the loop on itself could smooth out the cusp on scales much larger than $\delta$.
However, there is large body of literature on cosmic string backreaction~\cite{Thompson:1988yj,Quashnock:1990wv,Copeland:1990qu,Battye:1994qa,Buonanno:1998is,Carter:1998ix,Wachter:2016hgi,Wachter:2016rwc,Blanco-Pillado:2018ael,Chernoff:2018evo,Blanco-Pillado:2019nto} which indicates that the backreaction timescale is $\sim\ell/(G\mu)$, much longer than the loop's oscillation period.
While it is true that these studies all assume linearised gravity, and are thus likely to underestimate the magnitude of the effect near cusps, it is still hard to see how the loop could backreact fast enough to smooth out the cusp on relatively large scales before the peak of the signal.

All of this suggests that we need some strong-gravity mechanism which acts on a very short timescale while the cusp is forming, and suppresses the cusp's GW emission at frequencies far below the cutoff, $f\ll1/\delta$.
In section~\ref{sec:cusp-collapse}, we propose exactly such a mechanism: we argue that when cusps form on sufficiently large cosmic string loops, they source such extreme spacetime curvature that a small portion of the loop could collapse to form a black hole at a time $\sim G\mu\ell$ before the peak of the cusp emission.
We will see that, remarkably, the loops for which this \enquote{cusp collapse} process is predicted to occur are those with length $\ell\gtrsim\delta/(G\mu)^3$---exactly the same loops for which the higher-order memory divergence occurs.

\begin{figure}[p!]
    \begin{center}
        \includegraphics[width=0.75\textwidth]{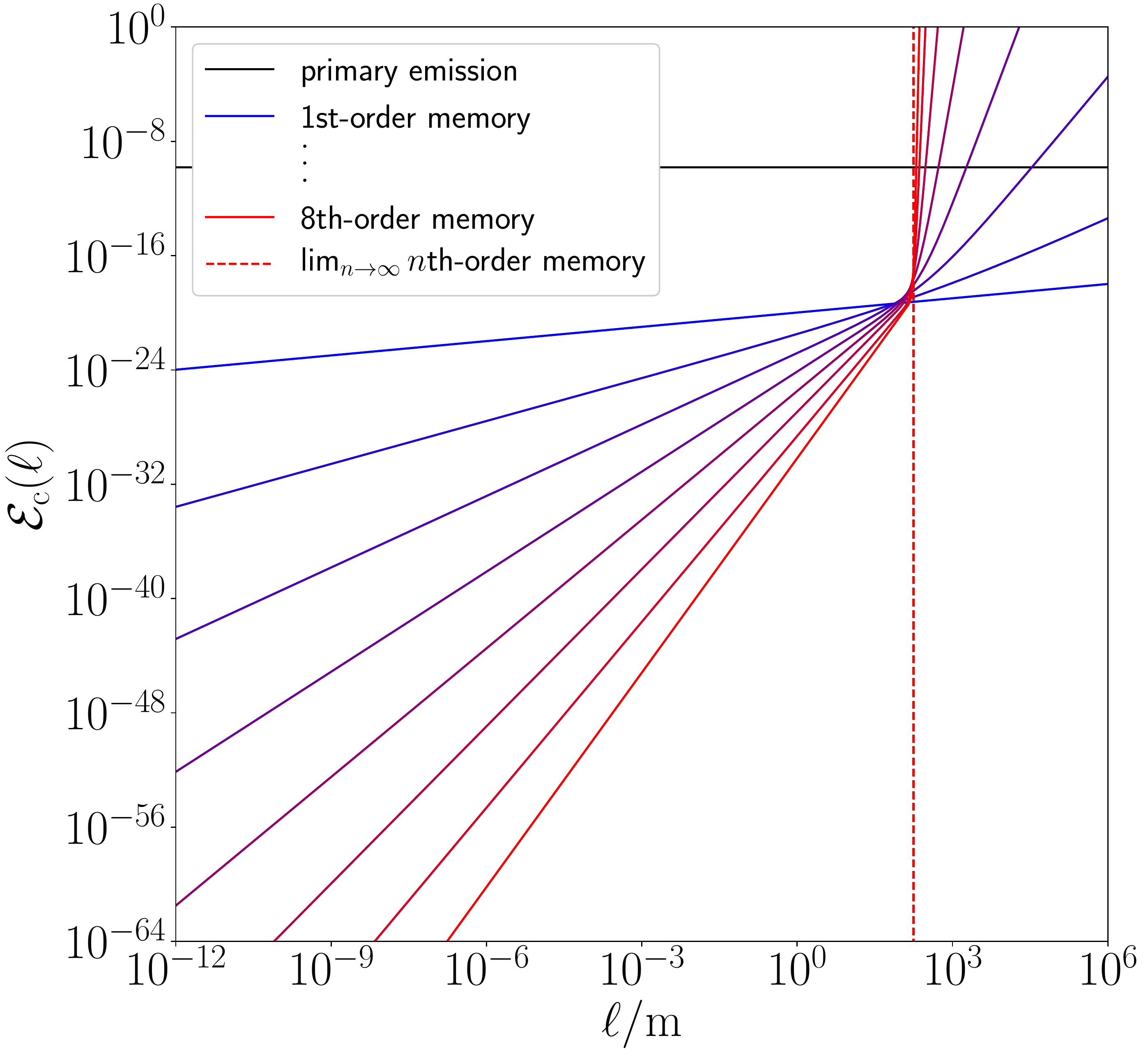}
    \end{center}
    \begin{center}
        \includegraphics[width=0.75\textwidth]{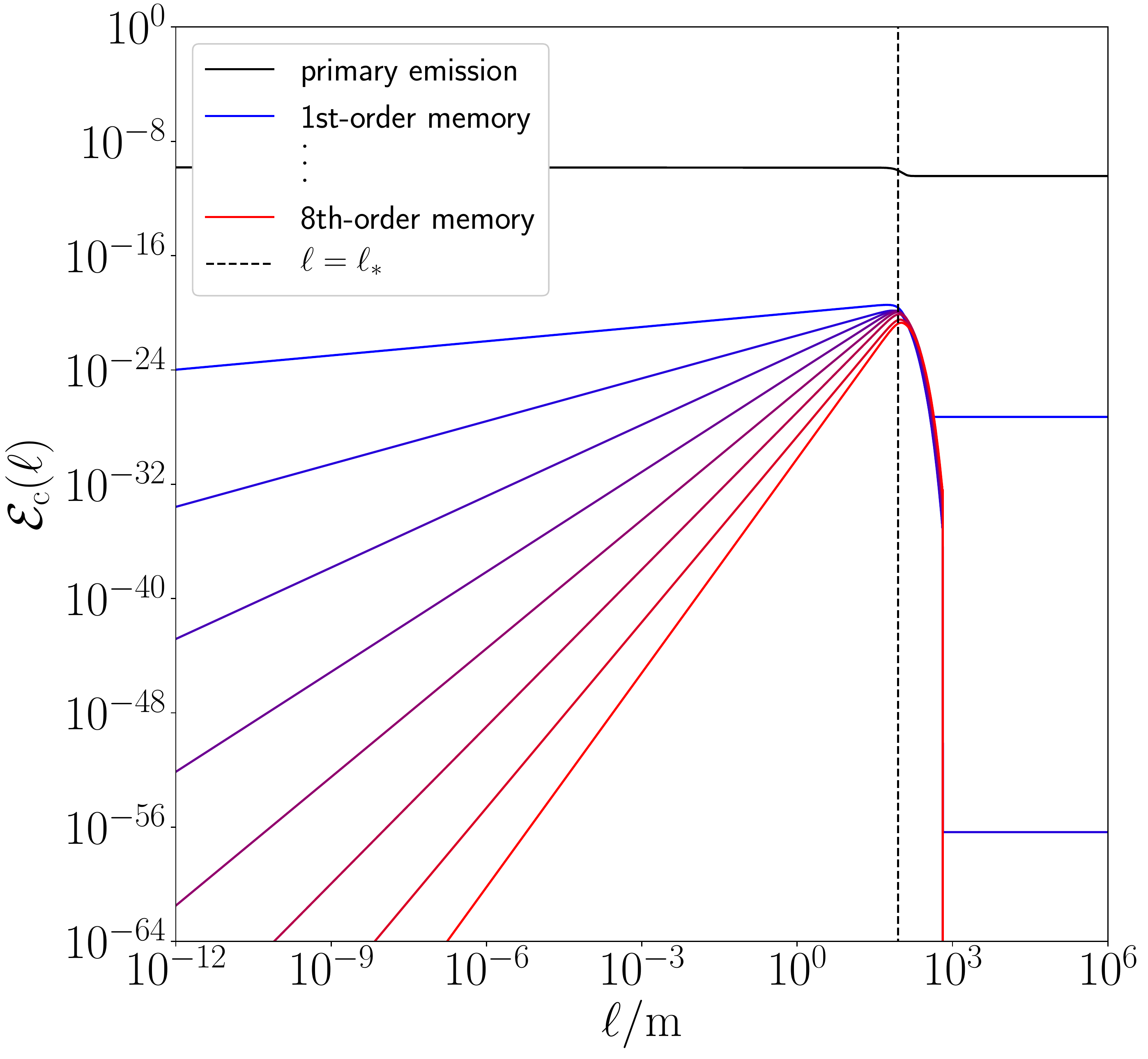}
    \end{center}
    \caption{%
    The fractional energy radiated by a cusp at different orders in the memory expansion as a function of loop length, with $G\mu=10^{-11}$.
    The top panel shows the standard cusp case, clearly illustrating the divergence at $\ell\gtrsim\ell_*\approx90\,\mathrm{m}$.
    The bottom panel shows the cusp collapse case, for which the radiated energy at each order drops to a small, $\ell$-independent value for $\ell\gtrsim\ell_*$, curing the divergence.
    }
    \label{fig:cusp-energy}
\end{figure}

It is difficult to calculate the precise GW signal associated with cusp collapse, but as we will show later in section~\ref{sec:cusp-collapse-gws}, there are two main qualitative differences it introduces compared to the standard cusp waveform: ($i$) the Fourier transform of the primary strain signal, $\tilde{h}^{(0)}_\rmc$, is reduced by a factor $\approx1/2$, as it only receives contributions from the half of the signal at times before the peak; ($ii$) more importantly, there is a loss of power at frequencies $f\gtrsim1/(G\mu\ell)$, due to the truncation immediately before the peak.
Both of these effects influence the corresponding GW memory signal.
We can account for ($i$) by multiplying the strain at each order in the memory expansion by the appropriate power of $1/2$, and can approximate the effect of ($ii$) by introducing a sharp cutoff at frequency $f=1/(G\mu\ell)$, which is equivalent to replacing $\delta\to G\mu\ell$ in all previous expressions.
The critical lengthscale~\eqref{eq:ell-star} which previously marked the onset of the divergence then becomes
    \begin{equation}
        \ell_*\to\frac{\ell}{(\uppi B_\rmc)^{3/2}(G\mu)^2}\gg\ell,
    \end{equation}
    which means that we are always in the \enquote{small loop} regime, $\ell\ll\ell_*$; higher-order memory corrections are suppressed by powers of $G\mu$, and the divergence is avoided completely, as shown in figure~\ref{fig:cusp-energy}.
Note that the cusp collapse process is only predicted to take place for loops with $\ell\gtrsim\delta/(G\mu)^3$, and that the GW memory from smaller loops is still described by the results given above, with $\ell_*$ given by equation~\eqref{eq:ell-star}.

More explicitly, if the divergence is indeed cured by invoking cusp collapse, then the GW observables from the cusp are as follows: the primary waveform is
    \begin{equation}
        \tilde{h}^{(0)}_\rmc\simeq A_\rmc\frac{G\mu\ell^{2/3}}{r|f|^{4/3}}\Theta(\vu*r\vdot\vu*r_\rmc-\cos\theta_\rmb)\Theta(|f|-2/\ell)
        \begin{cases}
            (1/2)\Theta(1/G\mu\ell-|f|), & \text{for}\;\ell\gg\ell_*\\
            \Theta(1/\delta-|f|), & \text{for}\;\ell\ll\ell_*
        \end{cases}
    \end{equation}
    with $\ell_*$ given by~\eqref{eq:ell-star}; the first-order memory waveform is
    \begin{equation}
        \tilde{h}^{(1)}_\rmc\simeq-\rmi B_\rmc\frac{(G\mu)^2\ell^{2/3}}{rf|f|^{1/3}}(1+\cos I)\Theta(|f|-2/\ell)
        \begin{cases}
            (1/4)\Theta(1/G\mu\ell-|f|), & \text{for}\;\ell\gg\ell_*\\
            \Theta(1/\delta-|f|), & \text{for}\;\ell\ll\ell_*
        \end{cases}
    \end{equation}
    with the corresponding late-time memory given by
    \begin{equation}
        \Updelta h^{(1)}_\rmc\approx2\times3^{2/3}(\uppi A_\rmc G\mu)^2(1+\cos I)\frac{\ell}{r}
        \begin{cases}
            1/4, & \text{for}\;\ell\gg\ell_*\\
            1, & \text{for}\;\ell\ll\ell_*
        \end{cases}
    \end{equation}
    and the $n$th-order memory waveforms for $n\ge2$ are all step-function-like,
    \begin{equation}
        \tilde{h}^{(n)}_\rmc\approx-\frac{\rmi\Updelta h^{(n)}_\rmc}{2\uppi f}\Theta(|f|-2/\ell)
        \begin{cases}
            \Theta(1/G\mu\ell-|f|), & \text{for}\;\ell\gg\ell_*\\
            \Theta(1/\delta-|f|), & \text{for}\;\ell\ll\ell_*
        \end{cases}
    \end{equation}
    with height given by
    \begin{equation}
        \Updelta h^{(n)}_\rmc\approx
        \begin{cases}
            \displaystyle\frac{\ell}{r}\qty(\frac{\uppi B_\rmc}{4})^{2^{n-1}}(G\mu)^{1+\frac{2^{n+1}}{3}}L_n(I), & \text{for}\;\ell\gg\ell_*\\
            \displaystyle\frac{\delta}{r}(\ell/\ell_*)^{2n/3}S_n(I), & \text{for}\;\ell\ll\ell_*
        \end{cases}
    \end{equation}

It is interesting to note that loops with length just below the cusp-collapse threshold, $\ell\lesssim\ell_*$, emit more energy through GW memory than those above the threshold.
We can see this already in the primary emission, where the total energy emission is
    \begin{equation}
    \label{eq:cusp-collapse-total-energy-0th-order}
        \mathcal{E}^{(0)}_\rmc\approx3^{2/3}(\uppi A_\rmc)^2G\mu\times
        \begin{cases}
            1/4, & \text{for}\;\ell\gg\ell_*\\
            1, & \text{for}\;\ell\ll\ell_*
        \end{cases}
    \end{equation}
    with only cusps above the collapse threshold being subject to the $1/4$ reduction in GW power due to the truncation of the signal.
However, the difference becomes more significant in the memory emission,
    \begin{equation}
    \label{eq:cusp-collapse-total-energy-1st-order}
        \mathcal{E}^{(1)}_\rmc\approx
        \begin{cases}
            \displaystyle\frac{1}{2}(\uppi B_\rmc)^2(G\mu)^{8/3}, & \text{for}\;\ell\gg\ell_*\\[8pt]
            \displaystyle8(\uppi B_\rmc)^{3/2}(G\mu)^2(\ell/\ell_*), & \text{for}\;\ell\ll\ell_*
        \end{cases}
    \end{equation}
    where there is an extra power of $(G\mu)^{2/3}$ for cusps above the collapse threshold, compared to those just below it.
This pattern continues for higher-order memory, where the total radiated energy for all $n\ge2$ is given by
    \begin{equation}
    \label{eq:cusp-collapse-total-energy-nth-order}
        \mathcal{E}^{(n)}_\rmc\approx
        \begin{cases}
            \displaystyle\frac{1}{2}\qty(\frac{\uppi B_\rmc}{4})^{2^n}(G\mu)^{2^{n+2}/3}\int_{S^2}\frac{\dd[2]{\vu*r}}{4\uppi}|L_n(I)|^2, & \text{for}\;\ell\gg\ell_*\\[8pt]
            \displaystyle\frac{1}{2}(\uppi B_\rmc)^{3/2}(G\mu)^2(\ell/\ell_*)^{(2n-1)/3}\int_{S^2}\frac{\dd[2]{\vu*r}}{4\uppi}6(1+\cos I)S_n(I), & \text{for}\;\ell\ll\ell_*
        \end{cases}
    \end{equation}

In section~\ref{sec:cusp-collapse-gws}, we will also discuss a further GW signature associated with cusp collapse: the high-frequency quasi-normal ringing of the newly-formed black hole after the collapse.
We neglect this effect here however, as it is hard to say anything concrete about the phase evolution and angular pattern of this additional GW emission, and this prevents us from calculating the associated memory signal.
It would be interesting to revisit this contribution to the memory if and when more detailed phase-coherent cusp collapse waveform models become available.

\section{Memory from kinks}
\label{sec:kinks}

As we saw in section~\ref{sec:cusps-kinks}, kinks are persistent features of cosmic string loops (unlike cusps, which are transient): they propagate around the loop at the speed of light, continuously emitting GWs in a beam which traces out a one-dimensional \enquote{fan} of directions, like a lighthouse.
This fact is usually unimportant when computing the primary GW emission from kinks, since the beam only overlaps with the observer's line of sight for a small fraction of the loop oscillation time, meaning that kinks are effectively transient sources for any given observer.
However, in order to calculate a GW memory signal, we need to know the primary GW flux in all directions over the entire history of the source, which for kinks means specifying the beaming direction as a function of time.

We consider here the simplest case, where the beam of the kink traces out a great circle on the sphere at a constant rate.
We choose our polar coordinates such that this circle lies in the equatorial plane, $\theta_\rmk=\uppi/2$.
The azimuthal direction of the kink's beam is then given by $\phi_\rmk=4\uppi\varsigma t/\ell$ with $\varsigma=\pm1$, where the two different signs correspond to left- and right-moving kinks respectively.\footnote{%
The definition of whether a given kink is left- or right-moving is somewhat arbitrary here.
For concreteness, we call kinks with $\varsigma=+1$ \enquote{left-moving}; these move anti-clockwise around the loop when viewed from the North pole $\theta=0$.
Conversely, kinks with $\varsigma=-1$ are called \enquote{right-moving}; these move clockwise when viewed from $\theta=0$.}
From equation~\eqref{eq:kink-waveform} we then have
    \begin{equation}
    \label{eq:kink-phase}
        \tilde{h}^{(0)}_\rmk(f,\vu*r)\simeq A_\rmk\frac{G\mu\ell^{1/3}}{r|f|^{5/3}}\rme^{-\rmi\varsigma\phi f\ell/2}\Theta(\theta_\rmb-|I|)\Theta(|f|-2/\ell),
    \end{equation}
    where the inclination $I\equiv\uppi/2-\theta$ describes the angle between the observer's line of sight and the closest point on the equatorial plane, and takes values $I\in[-\uppi/2,\uppi/2]$ (with positive/negative values corresponding to the observer being above/below the plane).
Note that we have picked up a phase factor $\rme^{-\rmi\varsigma\phi f\ell/2}$ to account for the time at which the kink passes closest to the line of sight.
As we show below, this direction-dependent phase ultimately leads to a strong suppression of the kink memory signal compared to the cusp case.

\subsection{Beaming effects}

As in the cusp case, the first step in calculating the GW memory signal is to compute the angular integral that captures the beaming effects of the kink, $\int_{\vu*r'}\rme^{-\rmi\varsigma\phi'f\ell/2}\Theta(\theta_\rmb-|I|)$.
Here the small circle of radius $\theta_\rmb$ around the North pole that we considered for cusps has been replaced with a narrow band of half-width $\theta_\rmb$ around the equator, and we have included the $\phi$-dependent phase factor from equation~\eqref{eq:kink-phase}.
It is straightforward to integrate out the zenith angle $\theta'$ if we keep only the leading-order term in $\theta_\rmb$; this gives
    \begin{equation}
    \label{eq:kink-angular-integral}
        \int_{\vu*r'}\rme^{-\rmi\varsigma\phi'f\ell/2}\Theta(\theta_\rmb-|I|)\simeq\theta_\rmb K_{f\ell/2}(-\varsigma I),
    \end{equation}
    where we have defined a family of azimuthal integrals,
    \begin{equation}
    \label{eq:K_n-definition}
        K_n(I)\equiv\int_0^{2\uppi}\frac{\dd{\phi}}{2\uppi}\rme^{\rmi n\phi}\frac{(\sin I\cos\phi-\rmi\sin\phi)^2}{1-\cos I\cos\phi}.
    \end{equation}
Notice that the argument $n$ is always an integer, as the GW frequency is always an integer multiple of the loop's fundamental mode $2/\ell$ (although we often ignore this when taking the continuum limit at high frequencies).
Computing the integral equation~\eqref{eq:K_n-definition} for general $n$ therefore corresponds to finding the Fourier spectrum of a complicated nonlinear function of $\phi$; we perform this calculation explicitly in appendix~\ref{sec:K_n}.

\subsection{Late-time memory}

Using equation~\eqref{eq:late-time-memory}, along with the expression for the angular integral $K_0(I)$ from equation~\eqref{eq:K_n-final}, we find that the late-time memory from the kink is given by
    \begin{equation}
        \Updelta h^{(1)}_\rmk=\frac{3^{5/6}\ell}{r}(\uppi A_\rmk G\mu)^2\frac{4\sin|I|+\cos2I-3}{\cos^2I}.
    \end{equation}
Inserting numerical values, the strongest effect is $\Updelta h^{(1)}_\rmk\approx-2.125\times(G\mu)^2\ell/r$ when the observer lies in the plane of the kink $I=0$, an order of magnitude smaller than the maximum cusp memory.
The late-time kink memory decreases smoothly to zero as one approaches the poles $I\to\pm\uppi/2$.

\subsection{Full waveform}

\begin{figure}[t!]
    \begin{center}
        \includegraphics[width=0.6\textwidth]{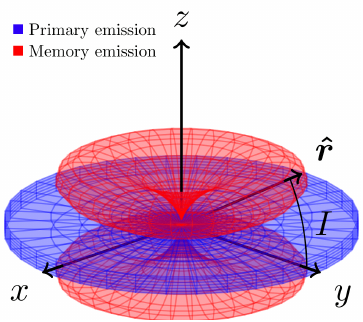}
    \end{center}
    \caption{%
    A cartoon illustration of the angular distribution of the energy radiated by a kink.
    The primary emission (blue) is concentrated in a narrow fan of half-width $\theta_\rmb\sim(f\ell)^{-1/3}$ around the plane of the kink, while the memory emission (red) is concentrated in two lobes either side of this plane, which are exponentially suppressed as one approaches either of the directions normal to the plane.
    }
    \label{fig:kink-memory-angular-patterns}
\end{figure}

The calculation here is very similar to the cusp case in section~\ref{sec:cusp-full-waveform}.
Inserting equation~\eqref{eq:kink-phase} into equation~\eqref{eq:memory-frequency-domain}, and taking care to enforce the frequency cutoffs and to account for the two competing beam angles $\theta_\rmb(f')$ and $\theta_\rmb(f'-f)$, we obtain
    \begin{equation}
        \tilde{h}_\rmk^{(1)}(f)\simeq-\frac{\rmi2^{5/3}\uppi(A_\rmk G\mu)^2\ell^{1/3}}{3^{1/6}rf|f|^{2/3}}K_{f\ell/2}(-\varsigma I)\qty[\int_{2/|f|\ell}^\infty\frac{\dd{u}}{u^{2/3}(1+u)}-\int_{2/|f|\ell}^{1/2}\frac{\dd{u}}{u^{2/3}(1-u)}]
    \end{equation}
    for $|f|>4/\ell$.
(This is analogous to equation~\eqref{eq:cusp-all-frequencies} from the cusp case.)
Taking the limit $|f|\gg2/\ell$, we can evaluate the integrals analytically to find
    \begin{align}
    \begin{split}
    \label{eq:kink-result}
        \tilde{h}^{(1)}_\rmk(f)&\simeq-\rmi B_\rmk\frac{(G\mu)^2\ell^{1/3}}{rf|f|^{2/3}}K_{f\ell/2}(-\varsigma I)\Theta(|f|-4/\ell),\\
        B_\rmk&\equiv12^{5/6}\uppi A_\rmk^2\qty[\frac{2\uppi}{3\sqrt{3}}-{}_2F_1(\tfrac{1}{3},\tfrac{1}{3};\tfrac{4}{3};-1)]\approx0.2915,
    \end{split}
    \end{align}
    where we have set the frequency cutoff at double the fundamental mode $2/\ell$ due to the different behaviour of the angular integral $K_n(I)$ for $n=1$ compared to $n\ge2$.

This waveform~\eqref{eq:kink-result} shares many features with both the cusp memory waveform~\eqref{eq:cusp-result} and the primary kink waveform~\eqref{eq:kink-phase}.
The most important difference from both of those waveforms is the dependence on inclination, which here is a function of frequency.
Using the results derived in appendix~\ref{sec:K_n}, we can rewrite equation~\eqref{eq:kink-result} as
    \begin{equation}
    \label{eq:kink-result-cases}
        \tilde{h}^{(1)}_\rmk(f)\simeq-\rmi B_\rmk\frac{(G\mu)^2\ell^{1/3}}{rf|f|^{2/3}}\frac{4\sin|I|}{\cos^2I}
        \begin{cases}
            \displaystyle\qty(\frac{\cos I}{1+\sin|I|})^{f\ell/2}\Theta(f-4/\ell), & \text{for}\:\varsigma I<0\\
            \displaystyle\qty(\frac{\cos I}{1+\sin|I|})^{-f\ell/2}\Theta(-f-4/\ell), & \text{for}\:\varsigma I>0
        \end{cases}
    \end{equation}
    meaning that the memory signal from left-moving kinks contains only negative frequencies above the equatorial plane and only positive frequencies below the plane, and vice versa for right-moving kinks.

For high frequencies $|n|\gg1$ the angular integral $K_n(I)$ has a maximum value of $\simeq4/(\rme|n|)$ at inclination $I\simeq1/n$.
This means that the kink memory signal is only observable very close to the plane of the kink (but \emph{not} in the plane, where it vanishes; see figure~\ref{fig:kink-memory-angular-patterns}), and is suppressed by an extra power of frequency compared to the primary signal, $\tilde{h}^{(1)}_\rmk\sim f^{-8/3}$.

\subsection{Time-domain waveform near the arrival time}

As with the cusp case, we can reverse-Fourier-transform equation~\eqref{eq:kink-result-cases} to find the time-domain memory strain around time of arrival of the primary kink signal, $|t-t_0|\ll\ell$.
Unlike the cusp case, we obtain a signal which is not linearly polarised, but contains both $+$ and $\times$ polarisation content,
    \begin{align}
    \begin{split}
    \label{eq:kink-memory-time-domain}
        h^{(1)}_{\rmk,+}(t)-h^{(1)}_{\rmk,+}(t_0)&\simeq\frac{16\uppi B_\rmk(G\mu)^2}{2^{1/3}r}(t-t_0)\frac{\sin|I|}{\cos^2I}E_{2/3}\qty(2\ln\frac{1+\sin|I|}{\cos I}),\\
        h^{(1)}_{\rmk,\times}(t)-h^{(1)}_{\rmk,\times}(t_0)&\simeq\frac{64\uppi^2B_\rmk(G\mu)^2}{2^{1/3}\ell r}(t-t_0)^2\frac{\sin\varsigma I}{\cos^2I}E_{-1/3}\qty(2\ln\frac{1+\sin|I|}{\cos I}),
    \end{split}
    \end{align}
    where we have used the generalised exponential integral function, $E_n(z)\equiv\int_1^\infty\dd{x}\rme^{-zx}x^{-n}$.
The $\times$-polarised component is suppressed by an additional factor of $(t-t_0)/\ell\ll1$, meaning that the signal is still approximately $+$-polarised.

An important difference with respect to the cusp case is that since kinks are long-lived rather than transient, their memory signal is not concentrated around a particular arrival time, but can in principle be observed at \emph{all} times.
One can easily obtain expressions analogous to equation~\eqref{eq:kink-memory-time-domain} at any point in the kink's periodic motion by substituting in the appropriate time when evaluating the reverse Fourier transform; in general this leads to a mixing between the $+$- and $\times$-polarisation modes.

\subsection{Radiated energy}

As in the cusp case, we are interested in the total energy radiated by the kink, and how the memory emission adds to this.
Using equation~\eqref{eq:dimensionless-energy}, we find that the dimensionless energy spectrum for the first-order memory is
    \begin{equation}
        \epsilon^{(1)}_\rmk(f,\vu*r)\simeq4\uppi B_\rmk^2\frac{(G\mu)^3}{(f\ell)^{1/3}}\frac{\sin^2I\cos^{f\ell-4}I}{(1+\sin|I|)^{f\ell}}\Theta(f-4/\ell).
    \end{equation}
Recall that when integrating over the sphere, the primary spectrum~\eqref{eq:kink-energy-spectrum} is suppressed by a factor of the beaming angle $\theta_\rmb\sim(f\ell)^{-1/3}$.
The spherical integral for the memory contribution can be evaluated explicitly to give a much stronger suppression,
    \begin{equation}
        \int_{S^2}\dd[2]{\vu*r}\frac{\sin^2I\cos^{f\ell-4}I}{(1+\sin|I|)^{f\ell}}=\frac{8\uppi}{f\ell(f\ell-2)(f\ell+2)}\simeq8\uppi/(f\ell)^3,
    \end{equation}
    so that at high frequencies, the isotropic energy spectrum is approximately
    \begin{equation}
        \bar{\epsilon}^{(1)}_\rmk(f)\simeq32(\uppi B_\rmk)^2\frac{(G\mu)^3}{(f\ell)^{10/3}}\Theta(f-4/\ell).
    \end{equation}
We see that the angular pattern of the kink memory strongly suppresses the isotropic spectrum at high frequencies, meaning that unlike in the cusp case, there is no frequency range where the memory contribution dominates, and the total radiated energy converges without imposing an ultraviolet cutoff,
    \begin{equation}
    \label{eq:kink-total-energy}
        \mathcal{E}^{(1)}_\rmk\simeq\frac{3(\uppi B_\rmk)^2}{2^{8/3}\times5}(G\mu)^3.
    \end{equation}

\subsection{Higher-order memory}
\label{sec:kink-higher-order}

As with the cusp case, we can iterate the memory calculation with our 1st-order memory waveform~\eqref{eq:kink-result} as an input to calculate the 2nd-order memory effect (the \enquote{memory of the memory}).
This involves calculating the angular integral
    \begin{equation}
        \int_{\vu*r'}\rme^{-\rmi\varsigma\phi'f\ell/2}K_{f'\ell/2}(-\varsigma I')K_{(f'-f)\ell/2}(-\varsigma I'),
    \end{equation}
    which in the high-frequency regime $|f'|,|f'-f|\gg2/\ell$ is well-approximated by
    \begin{equation}
        \simeq\frac{16K_{f\ell/2}(-\varsigma I)}{(f'\ell-f\ell/2)^3}\qty[\Theta(f'-4/\ell)\Theta(f'-f-4/\ell)-\Theta(-f'-4/\ell)\Theta(-f'+f-4/\ell)].
    \end{equation}
With this result to hand, the remaining steps are very similar to the calculations for the 1st-order memory described above, yielding
    \begin{align}
    \begin{split}
        \tilde{h}^{(2)}_\rmk(f)&\simeq-\rmi\frac{512\uppi B_\rmk^2(G\mu)^4}{rf|f|^{10/3}\ell^{7/3}}K_{f\ell/2}(-\varsigma I)\Theta(|f|-4/\ell)\int_{4/|f|\ell}^\infty\frac{\dd{u}}{u^{2/3}(u+1)^{2/3}(u+1/2)^3}\\
        &\approx-\rmi\frac{1666\times(G\mu)^4}{rf|f|^{10/3}\ell^{7/3}}K_{f\ell/2}(-\varsigma I)\Theta(|f|-4/\ell).
    \end{split}
    \end{align}
The fact that this has the exact same dependence on the inclination $I$ as the 1st-order memory makes it straightforward to iterate the process to higher orders.
Doing this, we find that for all $n\ge2$, the kink GW memory is given schematically by
    \begin{equation}
        \tilde{h}^{(n)}_\rmk(f)\sim\frac{-\rmi(G\mu)^{2n}f\ell^3}{r(|f|\ell)^{8n/3}}K_{f\ell/2}(-\varsigma I)\Theta(|f|-4/\ell),
    \end{equation}
    multiplied by some numerical constant (for which there does not seem to be a simple expression for all $n$).

The situation here is drastically different from the cusp case.
For cusps we saw that each successive order in the memory was suppressed by larger powers of $G\mu\ll1$, but also enhanced by larger powers of $\ell/\delta\gg1$, and that in certain situations the latter would dominate, causing a divergence.
For kinks, we see instead that each order in the memory is not only suppressed by powers of $G\mu$, but is further suppressed by a factor of $(|f|\ell)^{-8/3}\ll1$ each time.
The signal falls off quickly enough with frequency that higher-order memory contributions are not sensitive to the string-width scale, and no factors of $\ell/\delta$ appear.
This means that there is no situation where the kink memory diverges, and that the higher-order contributions are negligible in any observational scenario.
This lack of divergence is in agreement with the cusp-collapse mechanism we have invoked as a possible resolution for the cusp divergence, as we will see below that kinks are not predicted to form PBHs.

\subsection{Caveats of our approach}

We have only considered the simplest case where kink's beam traverses a fixed plane at a constant rate, in order to make detailed analytical calculations feasible.
This situation is highly idealised, and is not representative of the loops one would find in a cosmological loop network, which would typically contain structure on scales smaller than the loop length $\ell$, causing the path of the beam to vary on those scales.
We expect that such structure is only likely to make a significant \emph{qualitative} difference to our results if it is on a scale corresponding to the GW frequency of interest.
We note that small-scale structure on loops is expected to be damped over time through gravitational backreaction~\cite{Quashnock:1990wv}, so our simple treatment here is not likely to be too unreasonable.
In any case, we do not expect such considerations to change the main conclusion of this section: that memory from kinks is highly suppressed compared to the cusp case.

We have also neglected the fact that kinks always appear in pairs on loops, with one left-mover for every right-mover.
A realistic loop is likely to have several pairs of kinks, with each of these sourcing a GW memory signal like the one calculated here.
Since the kinks travel around the loop at the same average rate, it is possible that the superposition of their memory signals could give rise to interesting coherent effects.
However, regardless of whether the kink memory signals are coherent or not, the GW energy flux will still be of the same order of magnitude, so our main conclusions are unaffected.

\section{Primordial black holes from cusp collapse}
\label{sec:cusp-collapse}

We saw in section~\ref{sec:cusps} that the standard Nambu-Goto cusp waveform~\eqref{eq:cusp-waveform} leads to a divergence in the total GW strain when accounting for all higher-order memory terms.
In section~\ref{sec:cusp-collapse-memory}, we argued that this is due to a breakdown in the weak-field assumption used to derive the waveform in the first place.
In order to resolve the divergence, there must be some form of strong-gravity mechanism which truncates the high-frequency GW emission.

In this section, we explore one tentative candidate for such a mechanism: that the GW emission is suppressed by the formation of a black hole event horizon enclosing the cusp.
Aside from curing the cusp memory divergence, this presents a novel formation channel for primordial black holes.
We present two heuristic arguments for why this might happen, both based on the \emph{hoop conjecture} (which we will define below).
These arguments are not mathematically rigorous, and are subject to some open conceptual issues around how exactly the hoop conjecture applies in highly dynamical and relativistic situations such as cusps~\cite{Blanco-Pillado:2021klh}.
Ultimately we will require fully general-relativistic calculations to gain a complete understanding of strong-gravity effects near cusps, probably in the form of numerical relativity simulations.
With these caveats in mind, we explore this \enquote{cusp-collapse} scenario as one possible resolution of the nonlinear memory divergence identified above, with particular emphasis on potential observational signatures that could allow us to test this proposal.

\subsection{The hoop conjecture}
\label{sec:hoop-conjecture}

The hoop conjecture, first formulated by \citet{Thorne:1972ji}, is a powerful diagnostic for the formation of BH horizons in GR.
The conjecture states that \enquote{horizons form when, and only when, a mass $M$ gets compacted into a region whose circumference in every direction is $\mathcal{C}\le4\uppi GM$}~\cite{Misner:1974qy}.
In other words, if a sphere containing mass $M$ fits inside its own Schwarzschild radius $r_\mathrm{S}\equiv2GM$, it forms a black hole.
This conjecture is intentionally somewhat vaguely defined in several ways; for example, there is no unambiguous way to assign a mass to the gravitational field inside the sphere.\footnote{%
    Doing so would require a \emph{quasi-local} measure of gravitational mass in GR, which is challenging for reasons we touched on in section~\ref{sec:gw-energy}.}
For the present situation, we include only the mass due to the matter fields, $M_\mathrm{sphere}\equiv\int_{\mathcal{B}_r}\dd[3]{\vb*x}T_{00}(t,\vb*x)$, where $\mathcal{B}_r$ is a ball of radius $r$, and the mass is a function of time and of the centre of the ball.
The hoop conjecture then predicts BH formation if
    \begin{equation}
    \label{eq:hoop-condition}
        \frac{2GM_\mathrm{sphere}}{r}\ge1.
    \end{equation}
    which we refer to as the \enquote{hoop condition}.

\begin{figure}[t!]
    \begin{center}
        \includegraphics[width=0.7\textwidth]{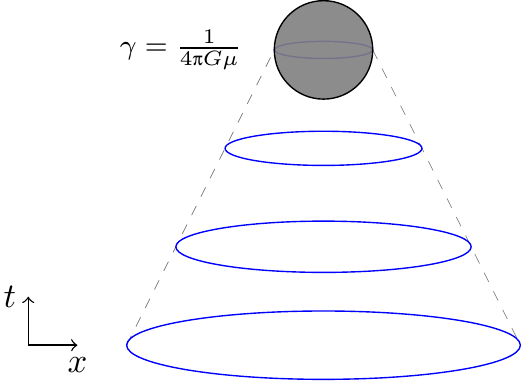}
    \end{center}
    \caption{%
    A collapsing circular cosmic string loop (in blue) forms a PBH (in grey) once its Lorentz factor satisfies equation~\eqref{eq:gamma-condition-circular}.}
    \label{fig:circular-collapse}
\end{figure}

We are interested in cosmic string loops which lead to PBH formation, i.e., solutions to equations~\eqref{eq:flat-space-eoms-1}--\eqref{eq:flat-space-eoms-3} which satisfy the hoop condition~\eqref{eq:hoop-condition} at some point in their evolution.
The simplest example is a circular loop, which contracts at an accelerating rate until the entire loop is compact enough to form a PBH, as illustrated in figure~\ref{fig:circular-collapse}.
This occurs within a single loop oscillation period, and results in a PBH of mass $M_\mathrm{pbh}\sim M_\mathrm{loop}=\mu\ell$, which is smaller than the original loop by a factor of $G\mu$.
One can show that the horizon forms only once the loop's Lorentz factor satisfies
    \begin{equation}
    \label{eq:gamma-condition-circular}
        \gamma=\frac{1}{4\uppi G\mu}.
    \end{equation}
Since we know that $G\mu\ll1$, this corresponds to an ultrarelativistic contraction velocity $v\simeq1-8\uppi^2(G\mu)^2$, and we can understand the PBH formation as being due to relativistic length contraction.
This mechanism for PBH formation from (quasi)circular loops has been studied extensively in the literature~\cite{Hawking:1987bn,Polnarev:1988dh,Hawking:1990tx,Garriga:1992nm,Caldwell:1993kv,Garriga:1993gj,Caldwell:1995fu,MacGibbon:1997pu,Helfer:2018qgv,James-Turner:2019ssu,Aurrekoetxea:2020tuw}.
However, circular collapse is only possible if all three components of the loop's angular momentum are smaller than those of a typical loop by a factor of $\sim G\mu$~\cite{Vilenkin:2000jqa}.
Circular collapse is thus finely-tuned, and only a very small fraction of the cosmic loop population is expected to collapse in this way.

This naturally leads one to ask whether generic (i.e. noncircular) loops can form PBHs.
It is easy to see intuitively that a Lorentz factor of order $\gamma\sim(G\mu)^{-1}$ like that in equation~\eqref{eq:gamma-condition-circular} is a necessary condition for PBH formation, even for noncircular loops.
Suppose we want to form a PBH which contains some fraction $f$ of the loop's mass, $M_\mathrm{pbh}=f\mu\ell$, corresponding to a $\sigma$ interval of $\Updelta\sigma=f\ell$.
This length of string must be compacted into a region of diameter $\lesssim2r_\mathrm{S}=4GM_\mathrm{pbh}$.
The ratio between this lengthscale and the corresponding $\sigma$ interval is related to the loop's tangent vector,
    \begin{equation}
        |\vb*X'|\approx\frac{|\Updelta\vb*X|}{\Updelta\sigma}\lesssim\frac{4GM_\mathrm{pbh}}{f\ell}=4G\mu.
    \end{equation}
We can relate the tangent vector to the loop's dynamics by rearranging equation~\eqref{eq:flat-space-eoms-2} to get
    \begin{equation}
    \label{eq:tangent-vector-gamma}
        |\vb*X'|=\sqrt{1-|\dot{\vb*X}|^2}=\frac{1}{\gamma},
    \end{equation}
    which shows that the hoop condition is generically satisfied if part of the loop has a large enough Lorentz factor,
    \begin{equation}
    \label{eq:gamma-condition}
        \gamma\gtrsim\frac{1}{4G\mu}.
    \end{equation}
This is not a sharp bound, just an order-of-magnitude estimate.
The corresponding (exact) inequality~\eqref{eq:gamma-condition-circular} for circular loops agrees to within a factor of $\uppi$.
We expect equation~\eqref{eq:gamma-condition} to have a similar level of accuracy for generic loop configurations.

\begin{figure}[t!]
    \begin{center}
        \includegraphics[width=0.75\textwidth]{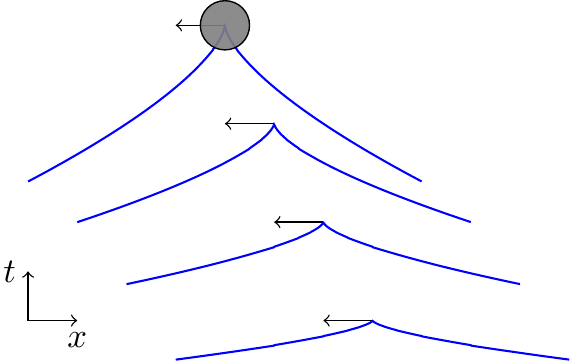}
    \end{center}
    \caption{%
    A segment of a cosmic string loop (in blue) becomes more compact as it develops a cusp.
    Once it satisfies the hoop condition~\eqref{eq:hoop-condition} it collapses to form a PBH (in grey).}
    \label{fig:cusp-collapse}
\end{figure}

Note that in the above argument we have \emph{not} assumed that the entire loop must be moving at such high velocities, only some fraction $f$ of it.
This is in contrast with the literature on (quasi)circular loop collapse, which has looked exclusively at cases where \emph{all} of the loop's mass ends up behind the PBH horizon.\footnote{%
    In fact, \citet{Polnarev:1988dh} briefly mention the possibility of forming a PBH from just part of the loop, but do not discuss this idea in any detail.}
The argument sketched above therefore suggests a change in focus: rather than looking at PBH formation \emph{from} loops, we should be concerned with PBH formation \emph{on} loops.

\subsection{The cusp-collapse mechanism}
\label{sec:cusp-collapse-mechanism}

Now consider what happens for cusps.
Since these correspond to a point on the loop moving at the speed of light, the corresponding Lorentz factor diverges, instantaneously compacting a finite fraction of the loop's mass into an infinitesimally small region.
Equation~\eqref{eq:gamma-condition} then suggests that cusps should therefore lead to some fraction of the loop's mass being enclosed behind a horizon, as illustrated in figure~\ref{fig:cusp-collapse}.

We can look at this idea in more detail by considering the behaviour of solutions to the flat-space EoMs~\eqref{eq:flat-space-eoms-1}--\eqref{eq:flat-space-eoms-3} near a cusp, as we did in section~\ref{sec:cusps-kinks}.
Using equation~\eqref{eq:cusp-xpm-taylor-expansions}, we see that the position and velocity of the loop near the cusp at time $t=0$ are given by
    \begin{align}
    \begin{split}
    \label{eq:solution-near-cusp}
        \vb*X_0(\sigma)&\equiv\frac{1}{2}[\vb*X_+(\sigma)+\vb*X_-(-\sigma)]=\frac{1}{2}\ddot{\vb*X}\sigma^2+\order*{\sigma^3},\\
        \dot{\vb*X}_0(\sigma)&\equiv\frac{1}{2}[\dot{\vb*X}_+(\sigma)+\dot{\vb*X}_-(-\sigma)]=\vu*r_\rmc+\frac{1}{2}\dddot{\vb*X}\sigma^2+\order*{\sigma^3},
    \end{split}
    \end{align}
    so that the distance from the cusp as a function of $\sigma$ is given by
    \begin{equation}
    \label{eq:distance-from-cusp}
        r_0(\sigma)=\sqrt{\vb*X_0\vdot\vb*X_0}=\frac{1}{2}|\ddot{\vb*X}|\sigma^2+\order*{\sigma^3}.
    \end{equation}
We see that the fact that $\dot{\vb*X}_+=\dot{\vb*X}_-=\vu*r_\rmc$ at the cusp means that there is no term of order $\sigma$ in equation~\eqref{eq:distance-from-cusp}, and the distance grows much more slowly for small $\sigma$ than it would on a non-cuspy part of the loop; this is the crucial ingredient for fulfilling the hoop condition.

Consider now a sphere of radius $r\ll\ell$, centred at the cusp.
We see from equation~\eqref{eq:distance-from-cusp} that the portion of the loop contained in the sphere is given by $-\sigma_*\le\sigma\le\sigma_*$, where $\sigma_*\ll\ell$ is defined by $r=r_0(\sigma_*)$, such that $\sigma_*=\sqrt{2r/|\ddot{\vb*X}|}$.
We thus see that the mass contained in the sphere is
    \begin{equation}
        M_\mathrm{sphere}=\mu\int_{-\sigma_*}^{+\sigma_*}\dd{\sigma}=2\mu\sigma_*=\qty(\frac{8\mu^2r}{|\ddot{\vb*X}|})^{1/2}.
    \end{equation}
The hoop condition~\eqref{eq:hoop-condition} is therefore satisfied if $r|\ddot{\vb*X}|\le32(G\mu)^2$, with the limiting PBH mass being $M_\mathrm{pbh}=16G\mu^2/|\ddot{\vb*X}|$.
The fact that this depends on the cusp's acceleration $|\ddot{\vb*X}|$ rather than its velocity may seem surprising at first, but we can understand this intuitively by using equations~\eqref{eq:flat-space-eoms-1} and~\eqref{eq:tangent-vector-gamma} to write $|\ddot{\vb*X}|=|\vb*X''|=\dv*{(1/\gamma)}{\sigma}$.
The acceleration therefore tells us about the rate of change of the Lorentz factor along the loop, and thus controls the size of the region that satisfies the hoop condition, which sets the PBH mass.

We therefore find that, assuming the hoop conjecture can be applied in this simple way, cusps are predicted to form PBHs with mass
    \begin{equation}
    \label{eq:M_PBH}
        M_\mathrm{pbh}=\frac{8}{\uppi\bar{n}}G\mu^2\ell\approx G\mu M_\mathrm{loop},
    \end{equation}
    which are a factor of $G\mu$ smaller than those formed from circular collapse.
(Here we have used the effective mode number $\bar{n}\equiv\ell|\ddot{\vb*X}|/(2\uppi)$ from section~\ref{sec:cusps-kinks}.)

\subsection{Properties of the PBHs}
\label{sec:pbh-properties}

\begin{figure}[t!]
    \begin{center}
        \includegraphics[width=0.6\textwidth]{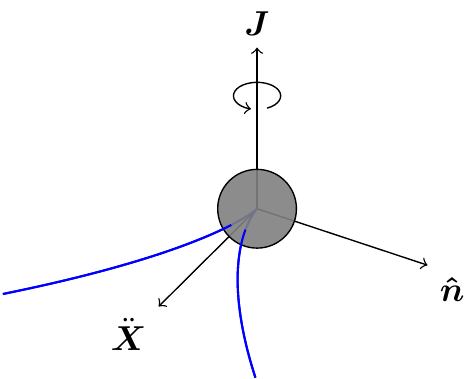}
    \end{center}
    \caption{%
        An illustration of the PBH (in grey) immediately after formation.
        The cusp acceleration $\ddot{\vb*X}$, cusp velocity $\vu*r_\rmc$, and PBH angular momentum $\vb*J$ are all orthogonal to each other.
        The cosmic string (in blue) punctures the horizon at two points separated by a small angle $\sim G\mu$, with its cusp hidden behind the horizon.}
    \label{fig:bh-cusp}
\end{figure}

We can estimate the properties of the PBHs formed through cusp collapse by assuming that all of the energy-momentum inside the sphere of radius $2GM$ at time $t=0$ is trapped behind the horizon.
Using equation~\eqref{eq:conformal-gauge-energy-momentum}, the PBH's linear and angular momenta are then given by
    \begin{align}
    \begin{split}
        P^i&=\int_{\mathcal{B}_r}\dd[3]{\vb*x}T^{0i}(0,\vb*x)=\mu\int_{-\sigma_*}^{+\sigma_*}\dd{\sigma}\dot{X}_0^i,\\
        J^i&=\int_{\mathcal{B}_r}\dd[3]{\vb*x}\tensor{\varepsilon}{^i_{jk}}x^jT^{0k}(0,\vb*x)=\mu\int_{-\sigma_*}^{+\sigma_*}\dd{\sigma}\tensor{\varepsilon}{^i_{jk}}X_0^j\dot{X}_0^k,
    \end{split}
    \end{align}
    where $\varepsilon_{ijk}$ is the Levi-Civita symbol.
Inserting the leading-order terms from equation~\eqref{eq:solution-near-cusp}, and using $M=2\mu\sigma_*=16G\mu^2/|\ddot{\vb*X}|$, we find
    \begin{equation}
        \vb*P=M\qty[\vu*r_\rmc+\frac{32(G\mu)^2}{3}\frac{\dddot{\vb*X}}{|\ddot{\vb*X}|^2}],\qquad \vb*J=\frac{2GM^2}{3}\frac{\ddot{\vb*X}}{|\ddot{\vb*X}|}\times\vu*r_\rmc.
    \end{equation}
Thus we see that immediately after formation, the PBH is moving in the cusp direction $\vu*r_\rmc$ with an ultrarelativistic velocity $v=|\vb*P|/M\approx1$.
In fact, the PBH's Lorentz factor is of the same order of magnitude as our estimate~\eqref{eq:gamma-condition},
    \begin{equation}
        \gamma\le\sqrt{\frac{3}{128}}(G\mu)^{-1},
    \end{equation}
    where we have used equation~\eqref{eq:left-right-constraints-at-cusp} and the triangle inequality $|\ddot{\vb*X}_+|^2+|\ddot{\vb*X}_-|^2\ge|\ddot{\vb*X}_++\ddot{\vb*X}_-|^2$.
We also see that the PBH is spinning around an axis orthogonal to both the cusp's velocity $\vu*r_\rmc$ and its acceleration $\ddot{\vb*X}$, as illustrated in figure~\ref{fig:bh-cusp}, with a dimensionless spin parameter,
    \begin{equation}
        \chi\equiv\frac{|\vb*J|}{GM^2}=2/3,
    \end{equation}
    that is two-thirds of the extremal Kerr value $\chi=1$.
This large spin is a direct consequence of the orthogonality of the cusp's velocity and acceleration, as enforced by equation~\eqref{eq:left-right-constraints-at-cusp}.
It is perhaps surprising at first that we obtain such a specific prediction for the spin, but we can understand this as a consequence of the universality of the string solution near the cusp, as discussed in section~\ref{sec:cusps-kinks}.
We should point out, however, that this prediction does not account for the mass and angular momentum radiated during the collapse; we return to this point in section~\ref{sec:radiation} below.

Note that the mass $M$ includes the kinetic energy of the PBH, which is large due to its ultrarelativistic velocity.
The PBH's rest mass is given by
    \begin{equation}
    \label{eq:rest-mass}
        m\equiv\sqrt{M^2-P^2}=\frac{M}{\gamma}\approx(G\mu)^2\mu\ell,
    \end{equation}
    which is smaller than the total mass of the loop by a factor of $\sim(G\mu)^2$.
This means that the PBH radius is
    \begin{equation}
    \label{eq:pbh-radius}
        r\sim(G\mu)^3\ell,
    \end{equation}
    i.e., the size of the PBH scales linearly with the size of the loop, as expected from the fact that the PBH's mass is a fixed fraction of that of the loop (c.f. equation~\eqref{eq:M_PBH}).
By using the Nambu-Goto approximation, we have assumed throughout that the cosmic strings have zero width, effacing any physics which occurs on lengthscales smaller than the string width $\delta\sim\sqrt{\hbar/\mu}$.
We therefore expect our results to hold only if the PBH radius is larger than $\delta$, which implies
    \begin{equation}
    \label{eq:ell-min-cusp-collapse}
        \ell\gtrsim\frac{\delta}{(G\mu)^3}=\frac{\ell_*}{(G\mu)^2}\approx5.1\,\mathrm{km}\,\times\qty(\frac{G\mu}{10^{-11}})^{-7/2},
    \end{equation}
    and corresponds to a minimum PBH rest mass of
    \begin{equation}
    \label{eq:M-min}
        m_\mathrm{min}\approx\frac{\delta}{G}\approx\frac{m_\Pl}{\sqrt{G\mu}}\approx6.9\,\mathrm{g}\,\times\qty(\frac{G\mu}{10^{-11}})^{-1/2},
    \end{equation}
    where we recall that $m_\Pl/\sqrt{G\mu}$ is the energy scale at which the cosmic strings are formed.
We thus see that there are three different classes of loops, corresponding to three broad ranges of loop lengths: loops smaller than $\delta/(G\mu)$ are driven by field-theoretic effects (as we mentioned in section~\ref{sec:nambu-goto}), loops larger than $\delta/(G\mu)^3$ are able to form PBHs from cusps, and loops inbetween are unchanged compared to the standard treatment in the Nambu-Goto literature.
(These different regimes are summarised in figure~\ref{fig:scales}.)
Remarkably, this cusp-collapse regime $\ell\gtrsim\delta/(G\mu)^3$ corresponds exactly to the \enquote{large-loop} regime in which the nonlinear memory from the cusp diverges, as we showed in section~\ref{sec:cusps}.
This suggests that we are along the right lines in searching for a strong-gravity resolution to this divergence.

The loop punctures the PBH horizon at two points, corresponding to $(t,\sigma)=(0,\pm\sigma_*)$.
Using the expansion~\eqref{eq:solution-near-cusp} for $\vb*X_0(\pm\sigma_*)$, we see that both points lie very near to the $\ddot{\vb*X}$ axis, which is in the PBH's equatorial plane.
By continuing the expansion to at least $\order*{\sigma^3}$, one can show that
    \begin{equation}
        \cos\theta_*\equiv\frac{\vb*X_0(\sigma_*)\vdot\vb*X_0(-\sigma_*)}{|\vb*X_0(\sigma_*)||\vb*X_0(-\sigma_*)|}=1+\order*{\sigma_*^2},
    \end{equation}
    which implies that the angle between the two puncture points is $\theta_*\sim\sigma_*/\ell\sim G\mu$, and thus that both points are very close to the equatorial plane.
(This is important for our discussion of the subsequent dynamics of the loop near the PBH in Section~\ref{sec:loop-pbh-dynamics}.)
Accounting for the finite string width $\delta$, we see that these two puncture points are so close that it is possible for the loop to self-intersect at the PBH horizon.
Using simple trigonometry, the separation between the puncture points on the horizon is roughly
    \begin{equation}
    \label{eq:puncture-separation}
        2Gm_\mathrm{pbh}\tan\frac{\theta_*}{2}\sim(G\mu)^4\ell,
    \end{equation}
    so the loop self-intersects at the horizon if this separation is smaller than the string width $\delta$, which occurs if
    \begin{equation}
    \label{eq:puncture-intersection}
        \ell\lesssim\frac{\delta}{(G\mu)^4}=\frac{\ell_*}{(G\mu)^3}\approx1.7\times10^{-2}\,\mathrm{pc}\times\qty(\frac{G\mu}{10^{-11}})^{-9/2}.
    \end{equation}
In this case, one would expect the string to intercommute near the horizon, meaning that the PBH would be immediately chopped off from the loop at formation.

It is worth pointing out that we could have found the exact same PBH radius~\eqref{eq:pbh-radius} in a much quicker way if we had applied the hoop condition to the GW energy density radiated by the cusp, rather than the energy density of the loop itself.
To see this, write the total GW flux through a sphere of radius $r$ around the cusp as
    \begin{equation}
        M_\gw(r)=\mu\ell\int_{2\uppi/r}^\infty\frac{\dd{f}}{f}\bar{\epsilon}_\rmc(f)\sim G\mu^2\ell^{2/3}r^{1/3},
    \end{equation}
    where we have made sure to only include GWs with wavelengths smaller than the sphere, so that their energy density~\eqref{eq:gw-energy-flux} is well-defined.
This expression then satisfies the hoop condition $2GM_\gw/r=1$ for a sphere of size $r\sim(G\mu)^3\ell$, perfectly matching what we found in equation~\eqref{eq:pbh-radius}.
Indeed, recalling our finding from section~\ref{sec:cusps-kinks} that GWs of frequency $f$ are generated by a region $|\sigma|\sim\ell^{2/3}f^{-1/3}$ surrounding the cusp, we see that substituting in $f\sim2\uppi/r\sim(G\mu)^{-3}/\ell$ gives $|\sigma|\sim G\mu\ell\sim\sigma_*$, so that the string segment radiating these GWs corresponds exactly to the region that collapses according to our arguments above.

\subsection{Other points on the loop: pseudocusps and kinks}
\label{sec:pseudocusp-collapse}

Consider now a generic loop segment at some time $t=0$, whose configuration is locally described by
    \begin{equation}
    \label{eq:loop-configuration}
        \vb*X_\pm(\sigma_\pm)=\vu*r_\pm\sigma_\pm+\frac{1}{2}\ddot{\vb*X}_\pm\sigma_\pm^2+\order*{\sigma_\pm^3}.
    \end{equation}
If $\vu*r_+=\vu*r_-$, the point $\sigma=0$ is a cusp with a divergent Lorentz factor $\gamma\to\infty$, which forms a PBH as described above.
However, our simple argument in equation~\eqref{eq:gamma-condition} suggests that a divergent Lorentz factor is not necessary for PBH formation; all we need is $\gamma=\order{1/G\mu}$.
We therefore expect PBH formation in situations where $\theta\equiv\cos^{-1}\vu*r_+\vdot\vu*r_-\simeq|\vu*r_+-\vu*r_-|$ is nonzero, so long as $\theta$ is small enough (note that this is very reminiscent of our discussion around the beaming angle for the cusp GW signal in section~\ref{sec:cusps-kinks}).
We refer to points on the loop where $\theta$ is small but nonzero as \enquote{pseudocusps}.

In order to estimate how small $\theta$ must be, we generalise equation~\eqref{eq:distance-from-cusp} to give the distance from the pseudocusp for small $\sigma$ at time $t=0$,
    \begin{equation}
    \label{eq:distance-from-pseudocusp}
        r_0(\sigma)\approx\frac{1}{2}\theta|\sigma|+\frac{1}{2}|\ddot{\vb*X}|\sigma^2,
    \end{equation}
    where we have used the constraints~\eqref{eq:left-right-constraints-at-cusp}, and have approximated $(\vu*r_+-\vu*r_-)\vdot\ddot{\vb*X}\approx\theta|\ddot{\vb*X}|$.
Repeating the arguments of section~\ref{sec:cusp-collapse-mechanism}, we find that a PBH forms so long as $\theta<8G\mu$, with the corresponding mass given by $M=2\mu(8G\mu-\theta)/|\ddot{\vb*X}|$.
The loop velocity at $\sigma=0$ is $v\simeq1-\theta^2/8$, so we can translate the bound $\theta<8G\mu$ into a bound on the pseudocusp velocity.
Doing so, we see that pseudocusps collapse to form PBHs so long as their Lorentz factor obeys
    \begin{equation}
    \label{eq:gamma-condition-pseudocusp}
        \gamma\ge\frac{1}{4G\mu},
    \end{equation}
    in agreement with our simple estimate~\eqref{eq:gamma-condition}.
Since pseudocusps occur even more generically on loops than cusps do~\cite{Stott:2016loe}, this result further enhances the PBH formation rate.

Note that in writing the Taylor expansion~\eqref{eq:loop-configuration} we have assumed that the loop is smooth in the neighbourhood of $\sigma=0$, which precludes any discontinuities in the loop's tangent vector, i.e., kinks.
However, it is easy to convince oneself that kinks do not contribute to PBH formation.
A kink at some $\sigma_\mathrm{k}$ near $\sigma=0$ would make equation~\eqref{eq:distance-from-pseudocusp} a piecewise smooth function, with different coefficients for each order in $\sigma$ on either side of the kink.
Generically these coefficients are of the same order of magnitude on both sides of the kink; there is nothing about the kink which forces the $\order{\sigma}$ term in equation~\eqref{eq:distance-from-pseudocusp} to be small, which is what we require for PBH formation.
As we have shown above, the smallness of this term is uniquely associated with a large Lorentz factor, and therefore with (pseudo)cusps.

Of course, this last argument depends strongly on the Nambu-Goto approximation; in a full field-theoretic setting one would expect kinks to carry gradient energy, which may be sufficiently concentrated to satisfy the hoop condition.
However, one would only expect the gradient energy to be large in a region of size comparable to the string width $\delta$, meaning the resulting PBH masses would be near the minimal mass $m_\mathrm{min}\sim\delta/G$ from equation~\eqref{eq:M-min}.
For kinks, as for cusps, we can trust the Nambu-Goto approximation so long as we consider PBHs with mass $m\gg m_\mathrm{min}$.

\subsection{Timescale of cusp collapse}
\label{sec:backreaction}

\begin{figure}[p!]
\thisfloatpagestyle{empty}
\begin{singlespacing}
\begin{center}
\begin{tikzpicture}[scale=12pt]
    \draw[->] (0,0.025) -- (0,0.915);
    \draw[-]  (0.01,0.875) -- (-0.01,0.875) node[anchor=east]{$\ell/(G\mu)^{3/2}$};
    \node[right] at (0.01,0.875) {$
        \begin{cases}
            \text{timescale for free-fall into PBH [equation~\eqref{eq:free-fall-time}]}
        \end{cases}$};
    \draw[-]  (0.01,0.74) -- (-0.01,0.74) node[anchor=east]{$\ell/(G\mu)$};
    \node[right] at (0.01,0.74) {$
        \begin{cases}
            \text{loop decay timescale [equation~\eqref{eq:decay-time}]}\\
            \text{linear backreaction timescale (section~\ref{sec:backreaction})}
        \end{cases}$};
    \draw[-]  (0.01,0.605) -- (-0.01,0.605) node[anchor=east]{$\ell$};
    \node[right] at (0.01,0.605) {$
        \begin{cases}
            \text{loop size (section~\ref{sec:nambu-goto})}\\
            \text{loop oscillation period (section~\ref{sec:nambu-goto})}\\
            \text{circular collapse timescale (section~\ref{sec:hoop-conjecture})}
        \end{cases}$};
    \draw[-]  (0.01,0.47) -- (-0.01,0.47) node[anchor=east]{$G\mu\ell$};
    \node[right] at (0.01,0.47) {$
        \begin{cases}
            \text{cusp collapse timescale [equation~\eqref{eq:collapse-timescale}]}\\
            \text{circular-collapse PBH size (section~\ref{sec:hoop-conjecture})}
        \end{cases}$};
    \draw[-]  (0.01,0.335) -- (-0.01,0.335) node[anchor=east]{$(G\mu)^2\ell$};
    \node[right] at (0.01,0.335) {$
        \begin{cases}
            \text{cusp-collapse energy scale [equation~\eqref{eq:M_PBH}]}\\
            \text{cusp GW emission energy scale [equation~\eqref{eq:gw-radiation-fraction}]}
        \end{cases}$};
    \draw[-]  (0.01,0.2) -- (-0.01,0.2) node[anchor=east]{$(G\mu)^3\ell$};
    \node[right] at (0.01,0.2) {$
        \begin{cases}
            \text{cusp-collapse PBH size [equation~\eqref{eq:pbh-radius}]}\\
            \text{cusp-collapse QNM frequency [equation~\eqref{eq:qnm-omega}]}\\
            \text{frame-dragging due to PBH spin (section~\ref{sec:loop-pbh-dynamics})}
        \end{cases}$};
    \draw[-]  (0.01,0.065) -- (-0.01,0.065) node[anchor=east]{$(G\mu)^4\ell$};
    \node[right] at (0.01,0.065) {$
        \begin{cases}
            \text{horizon puncture-point separation [equation~\eqref{eq:puncture-separation}]}
        \end{cases}$};
    \draw[->] (0,-0.655) -- (0,-0.035);
    \draw[-]  (0.01,-0.075) -- (-0.01,-0.075) node[anchor=east]{$\delta/(G\mu)^4$};
    \node[right] at (0.01,-0.075) {$
        \begin{cases}
            \text{no loop self-intersection at horizon [equation~\eqref{eq:puncture-intersection}]}
        \end{cases}$};
    \draw[-]  (0.01,-0.21) -- (-0.01,-0.21) node[anchor=east]{$\delta/(G\mu)^3$};
    \node[right] at (0.01,-0.21) {$
        \begin{cases}
            \text{cusp-collapse PBHs can form [equation~\eqref{eq:ell-min-cusp-collapse}]}
        \end{cases}$};
    \draw[-]  (0.01,-0.345) -- (-0.01,-0.345) node[anchor=east]{$\delta/(G\mu)$};
    \node[right] at (0.01,-0.345) {$
        \begin{cases}
            \text{Nambu-Goto approximation valid [equation~\eqref{eq:ell-min}]}\\
            \text{particle radiation negligible (section~\ref{sec:nambu-goto})}\\
            \text{circular-collapse PBHs can form (section~\ref{sec:hoop-conjecture})}
        \end{cases}$};
    \draw[-]  (0.01,-0.48) -- (-0.01,-0.48) node[anchor=east]{$\delta$};
    \node[right] at (0.01,-0.48) {$
        \begin{cases}
            \text{string width (section~\ref{sec:nambu-goto})}\\
            \text{string formation scale }m_\Pl/\sqrt{G\mu}\text{ (section~\ref{sec:cosmic-strings})}\\
            \text{minimum PBH size [equation~\eqref{eq:M-min}]}
        \end{cases}$};
    \draw[-]  (0.01,-0.615) -- (-0.01,-0.615) node[anchor=east]{$(G\mu)^{1/2}\delta$};
    \node[right] at (0.01,-0.615) {$
        \begin{cases}
            \text{Planck length }\ell_\Pl\text{ (section~\ref{sec:nambu-goto})}
        \end{cases}$};
\end{tikzpicture}
\end{center}
\end{singlespacing}
    \caption{%
    A summary of the different scales in the cusp-collapse problem, including references to the equation or section where they first appear.
    In the Nambu-Goto approximation there are only two dimensionful quantities: the string tension $\mu$ and the loop length $\ell$.
    Since the string tension usually appears in the dimensionless combination $G\mu$, all of the system's time- and length-scales can be written as $(G\mu)^p\times\ell$ for some power $p$.
    The fact that $G\mu\ll1$ means that there is a strong hierarchy between these scales.
    Going beyond Nambu-Goto introduces another dimensionful quantity with its own hierarchy of scales: the string width $\delta\approx(\mu/\hbar)^{-1/2}$.
    (Note that many of the scales associated with $\delta$ here are lower limits on the loop size $\ell$; e.g., $\delta/(G\mu)$ is the smallest loop size for which the Nambu-Goto approximation is valid.)
    These two sets of scales are shifted relative to each other depending on $\ell$.
    More complicated combinations of $\ell$ and $\delta$ are of course possible; e.g., the evaporation timescales for cusp-collapse and circular-collapse PBHs are $(G\mu)^8\ell^3/\delta^2$ and $(G\mu)^2\ell^3/\delta^2$ respectively.}
    \label{fig:scales}
\end{figure}

One of the main assumptions we have made so far is that the loop is described by the flat-space equations of motion~\eqref{eq:flat-space-eoms-1}--\eqref{eq:flat-space-eoms-3} right up to the instant of PBH formation.
A more complete analysis would account for the gravitational backreaction of the loop on its own dynamics, which one would expect to suppress the cusp.
One might worry whether this suppression is strong enough to prevent the PBH from forming.

Significant evidence against this worry comes from the extensive literature on cosmic-string backreaction~\cite{Thompson:1988yj,Quashnock:1990wv,Copeland:1990qu,Battye:1994qa,Buonanno:1998is,Carter:1998ix,Wachter:2016hgi,Wachter:2016rwc,Blanco-Pillado:2018ael,Chernoff:2018evo,Blanco-Pillado:2019nto}, in which numerous different approaches (both analytical and numerical) have repeatedly shown that backreaction does not prevent cusps from forming.
There is general agreement that cusps are suppressed to some degree by backreaction, but that this suppression occurs gradually over many loop oscillation periods, on a timescale of order the loop decay time
    \begin{equation}
    \label{eq:decay-time}
        t_\mathrm{decay}\sim\frac{\ell}{G\mu}.
    \end{equation}
A serious problem with this argument is that essentially all of the existing work on string backreaction has been done in the weak-field limit, treating the string's gravity as a linear perturbation on the background spacetime.
This linearised approach is clearly unable to answer questions about strong-gravity effects, such as whether or not PBH formation takes place.

One piece of evidence we can consider is the timescale on which the PBH formation occurs in the scenario described above.
The velocity of the string point $\sigma=0$ at times near to the cusp, $|t|\ll\ell$, can be written as
    \begin{equation}
        \dot{\vb*X}_\mathrm{c}(t)=\vu*r_\rmc+t\ddot{\vb*X}+\frac{1}{2}t^2\dddot{\vb*X}+\order*{t^3},
    \end{equation}
    with the corresponding Lorentz factor given by
    \begin{equation}
    \label{eq:gamma-cusp}
        \gamma_\mathrm{c}(t)\simeq\frac{2}{|t|}|\ddot{\vb*X}_+-\ddot{\vb*X}_-|^{-1}\approx\frac{\ell}{\uppi\bar{n}|t|},
    \end{equation}
    where we have used the constraints~\eqref{eq:left-right-constraints-at-cusp}, and the last equality generally holds to within an order of magnitude.
Since PBH formation is associated with the Lorentz factor growing above a certain threshold~\eqref{eq:gamma-condition-pseudocusp}, we can estimate the associated timescale by setting equation~\eqref{eq:gamma-cusp} equal to this threshold, giving
    \begin{equation}
    \label{eq:collapse-timescale}
        \Updelta t_\mathrm{pbh}\sim G\mu\ell.
    \end{equation}
This shows that the PBH is formed on an extremely short timescale: shorter than the loop oscillation period by a factor of $G\mu$, and shorter than the timescale for linear backreaction by a factor of $(G\mu)^2$.
(See figure~\ref{fig:scales} for an overview of the different time- and length-scales in the loop-PBH system.)
Even if nonlinear backreaction is in principle strong enough to prevent the cusp from forming, it seems unlikely that it can act on a short enough timescale to do so, meaning that backreaction seems unlikely to be able prevent PBH formation.\footnote{%
    See also \citet{Thompson:1988yj}, who examines backreaction on cusps and argues geometrically that cusps form \emph{\enquote{no matter how strong the gravitational field near a cusp}}.}

Equation~\eqref{eq:collapse-timescale} is a very important result for us, as it shows that we should expect the collapse process to occur some short but finite time before the peak of the cusp GW signal at $t=0$, thereby suppressing the GW emission at frequencies $f\gtrsim1/\Updelta t$.
As we showed in section~\ref{sec:cusp-collapse-memory}, this loss of power at very high frequencies cures the nonlinear memory divergence that would otherwise occur for cusps.

\subsection{Radiation from the collapse}
\label{sec:radiation}

Our analysis thus far has neglected the effects of gravitational radiation during the collapse.
Radiation is likely to be important, as the collapse is ultrarelativistic and highly nonspherical.
In general, one would expect the final mass, linear momentum, and angular momentum of the PBH to be of the form
    \begin{equation}
    \label{eq:radiation-efficiency}
        M=M_0(1-\epsilon_M),\quad P=P_0(1-\epsilon_P),\quad J=J_0(1-\epsilon_J),
    \end{equation}
    where a zero subscript denotes the na\"{i}ve, zero-radiation quantity calculated above, and the $\epsilon_i$ are three numbers between zero and unity, describing the efficiency with which each quantity is radiated away.
In the context of circular loop collapse, Hawking~\cite{Hawking:1990tx} calculated a theoretical upper bound on $\epsilon_M$ of $1-\sqrt{1/2}\approx29\%$ by explicitly constructing a marginally outer trapped surface in the spacetime of the collapsing loop and requiring that this surface be enclosed by the event horizon of the final PBH.
This argument depends heavily on the circular symmetry of the loop, and no such construction seems possible in our case.

Despite the lack of symmetries here, we can make some interesting statements by requiring that the final PBH spin $\chi=J/(GM^2)$ be less than or equal to unity; otherwise the PBH would be \enquote{overspun} to reveal a naked singularity, violating cosmic censorship~\cite{Penrose:1973um,Wald:1997wa}.
Since $\chi_0=2/3$, we can write
    \begin{equation}
    \label{eq:spin-bound}
        \chi=\frac{2}{3}\frac{(1-\epsilon_J)}{(1-\epsilon_M)^2}\le1.
    \end{equation}
We see that, so long as $\epsilon_J\lesssim2\epsilon_M$, the final spin parameter of the PBH is larger than the na\"{i}ve value $2/3$, which shows that the upper bound~\eqref{eq:spin-bound} is likely to be useful.
In general, we expect $\epsilon_J\lesssim\epsilon_M$; see e.g. \citet{Durrer:1989zi}, in which the rate at which loops radiate angular momentum is shown to be typically an order of magnitude smaller than the rate at which they radiate mass.
If $\epsilon_J=\epsilon_M$, then equation~\eqref{eq:spin-bound} gives $\epsilon_M\le1/3\approx33\%$.
In the limit where $\epsilon_J\to0$, the bound is even stronger, $\epsilon_M\le1-\sqrt{2/3}\approx18\%$.
Since we expect $0<\epsilon_J<\epsilon_M$, the true upper bound for cusp collapse is likely to lie somewhere between these two extremes.

Interestingly, numerical relativity simulations of circular loop collapse performed by \citet{Helfer:2018qgv} found $\epsilon_M\lesssim2\%$, well below Hawking's bound (see also \citet{Aurrekoetxea:2020tuw}).
The authors suggest that this is due to the symmetry of the circular collapse, which means the horizon is nearly spherical when it first forms, suppressing the total radiation.
The initial horizon in our case is likely to be highly distorted, meaning that $\epsilon_M$ is likely to be closer to its upper bound.
Of course, it is possible that cusp collapse radiates angular momentum much more efficiently than is typical for loops as a whole, in which case $\epsilon_J$ could be larger than calculated by \citet{Durrer:1989zi}.
It would then be possible for $\epsilon_M$ to be larger than the rough bounds we found above.

We note in passing that radiation of linear momentum ($\epsilon_P>0$) would lead to a \enquote{rocket effect}~\cite{Hogan:1984is,Vachaspati:1984gt}, in which the loop's centre of mass is given a kick in the opposite direction to the radiation.
However, even if this process is maximally efficient, the radiated momentum is at most $P_0\approx G\mu M_\mathrm{loop}$, so the maximum kick is $v\approx G\mu$.
This pales in comparison to the rms velocity of points on the loop, $v_\mathrm{rms}=\sqrt{1/2}$ that we found in equation~\eqref{eq:loop-v-rms}, so the effect is of negligible interest; radiation from elsewhere on the loop quickly cancels out the kick.

\subsection{Dynamics of the loop-PBH system}
\label{sec:loop-pbh-dynamics}

Once formed, the PBH is inextricably linked to the surrounding string; the portion of the string enclosed behind the horizon cannot escape, and the portion outside the horizon is topologically forbidden from detaching itself.
Since the mass of our PBHs is smaller than that of their parent loops by a factor of $G\mu$, we expect the loop dynamics to be largely unaffected by the presence of the PBH, at least on timescales $\sim\ell$.
In particular, this means that despite its ultrarelativistic velocity, the PBH cannot drag the rest of the loop along with it---instead, we expect the loop's tension to act on the PBH to decelerate it, and for the loop to continue oscillating in essentially the same motion as before.
This could mean that cusp-collapse PBHs do not trace the DM distribution, as their parent loops could easily drag them out of DM haloes.
(This possibility was also pointed out by \citet{Vilenkin:2018zol}, albeit for a different PBH formation scenario.)

Most cusp-collapse PBHs are very small, and decay rapidly through Hawking radiation~\cite{Hawking:1974sw}.
In particular, the evaporation timescale~\eqref{eq:pbh-evaporation-timescale} for the minimum mass~\eqref{eq:M-min} is $\approx t_\Pl(m_\mathrm{pbh}/m_\Pl)^3\approx10^{-27}\,\mathrm{s}\times(G\mu/10^{-11})^{-3/2}$.
It is unclear what effect the loop has on the evaporation process, and vice versa.
The PBH cannot maintain its mass at the minimum value in equation~\eqref{eq:M-min} by accreting the loop, since this would correspond to the loop losing mass at a rate
    \begin{equation}
        \dv{m}{t}\approx\qty(\frac{m_\Pl}{m_\mathrm{pbh}})^2\frac{m_\Pl}{t_\Pl}\approx\mu\approx\frac{M_\mathrm{loop}}{\ell},
    \end{equation}
    i.e. the loop would have to lose all of its mass within a single oscillation period.
This seems very unlikely, given the limited gravitational influence of the PBH---the timescale for an object to free-fall from a distance $\sim\ell$ into the PBH is
    \begin{equation}
    \label{eq:free-fall-time}
        t_\mathrm{ff}\sim\frac{\ell^{3/2}}{\sqrt{Gm_\mathrm{pbh}}}\approx\frac{\ell}{(G\mu)^{3/2}},
    \end{equation}
    so even neglecting the loop's kinetic energy, it would take many oscillation periods for it to be accreted.
We are therefore forced to allow the PBH to decay to sizes smaller than the loop width.
It is hard to envisage a way for the topologically-stable field configuration around the string to be disrupted by the PBH evaporation, so the most likely outcome seems to be that the PBH simply vanishes from the loop.\footnote{%
    \citet{Bonjour:1998rf} and \citet{Gregory:2013xca} found that Abelian-Higgs string-BH systems can exhibit interesting \enquote{flux expulsion} effects when the BH is smaller than the string width; however, these results are only valid for extremal Kerr and Reissner-Nordstr{\"o}m BHs, and it is not clear whether they have any bearing on our sub-extremal PBHs, or on the evaporation process.}

PBHs with rest mass larger than $m_*\approx5\times10^{14}\,\mathrm{g}\approx3\times10^{-19}M_\odot$ evaporate very slowly, and lose a negligible fraction of their mass within a Hubble time~\cite{Carr:2020gox}.
It is therefore interesting to consider how these non-evaporating PBHs interact with their parent loops on cosmological timescales.
For the simplest case of an infinitely long straight string, explicit solutions for the metric of a BH threaded by a cosmic string have been constructed for the Nambu-Goto case~\cite{Aryal:1986sz} and for Abelian-Higgs strings~\cite{Achucarro:1995nu,Chamblin:1997gk,Bonjour:1998rf,Dehghani:2001nz,Ghezelbash:2001pq,Gregory:2013xca}.
For solutions where the BH is rotating, the string is assumed to be aligned with the spin axis.
In each case the solution is static, and the string represents a form of stable long-range hair on the BH.
Since the string is static, its only gravitational effect is to induce a conical singularity along its axis, with a deficit angle $\sim G\mu$~\cite{Vilenkin:1981zs,Garfinkle:1985hr}.
This deficit angle means that the string-BH solution is not asymptotically flat, which explains how it evades the no-hair theorem~\cite{Ruffini:1971bza,Bekenstein:1996pn}.
The deficit angle can also modify the BH's quasi-normal mode spectrum~\cite{Chen:2008zzv,Cheung:2020dxo}.

The relevance of these results is somewhat limited in our case, as the string emanating from the PBH is not static, but continues to oscillate relativistically.
Perhaps even more importantly, the string is not locally aligned with the PBH's spin axis, so does not puncture the PBH at its poles like the cases studied in this literature~\cite{Aryal:1986sz,Achucarro:1995nu,Chamblin:1997gk,Bonjour:1998rf,Dehghani:2001nz,Ghezelbash:2001pq,Gregory:2013xca}.
Instead, due to the geometry of the cusp, the two points where the string punctures the horizon lie in---or very close to---the equatorial plane, and are separated by a small angle $\sim G\mu$ (as we showed in section~\ref{sec:pbh-properties}).
Being in the equatorial plane, one would expect relativistic frame-dragging to pull the string into a spiral configuration around the PBH spin axis on scales $\sim Gm_\mathrm{pbh}$.
(On larger scales $\sim\ell$, the string tension easily overcomes the frame-dragging forces.)
This spiralling of the string around the PBH, combined with the very small separation between the two points at which it punctures the horizon, makes it seem likely that the string intersects itself near to the PBH.
The PBH would thus be chopped off from the rest of the loop, leaving it with only a small segment of string still attached, which it would rapidly accrete.
For sufficiently small loops, we have shown in equation~\eqref{eq:puncture-intersection} that the PBH is likely to be immediately chopped off at the moment of formation.

It is interesting to ask whether two PBHs connected to the same loop could have a greater chance of merging due to the loop dynamics; a similar effect has been demonstrated for the annihilation of monopole-antimonopole pairs connected by strings (so-called \enquote{cosmic necklaces})~\cite{Berezinsky:1997td,Siemens:2000ty}.
However, the two PBHs would likely be separated by a distance $\sim\ell$ much larger than their size, so based on the discussion above we would expect the PBHs to be chopped off before the loop has the chance to pull them together.

There are clearly many uncertainties in how cusp-collapse PBHs affect the loop network, but our very rough arguments here suggest that small PBHs rapidly evaporate to leave the loop essentially unchanged (although its dynamics are affected by the radiation), while large PBHs are likely to be cut off from loop by string self-intersections on small lengthscales $\sim Gm_\mathrm{pbh}\sim(G\mu)^3\ell$.

\section{Observational consequences}
\label{sec:observational-consequences}

Having derived the nonlinear GW memory waveforms emitted by cusps and kinks, and having investigated PBH formation as a potential mechanism for curing the memory divergence we encountered for cusps, we are now ready to ask what implications these results have for observations.
We begin by looking at the impact that cusp collapse has on the \emph{primary} GW emission from cosmic strings, and how this changes existing bounds on $G\mu$, before exploring the detection prospects of the memory GW emission.
We then investigate the mass spectrum of the PBHs formed through cusp collapse, showing that the evaporation of these PBHs leads to novel bounds on $G\mu$ (albeit weaker than those coming from GW observations), and arguing that they form a unique BH population which could act as an observational \enquote{smoking gun} for cosmic strings.

\subsection{Consequences of cusp collapse for GW searches}
\label{sec:cusp-collapse-gws}

\begin{figure}[t!]
    \begin{center}
        \includegraphics[width=0.72\textwidth]{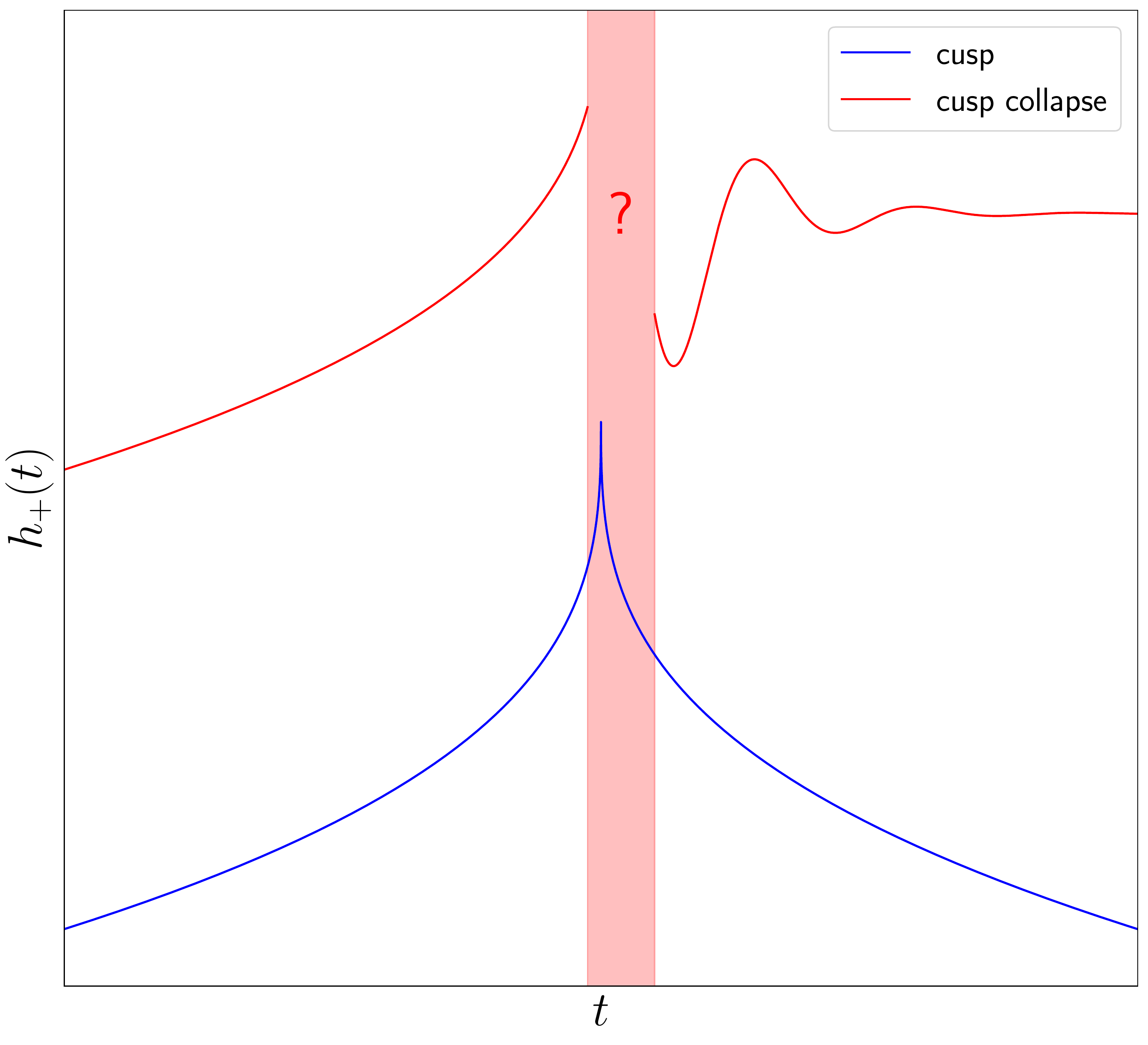}
    \end{center}
    \caption{%
    A heuristic illustration of the different GW signals from collapsing and non-collapsing cusps.
    Shown in blue is the standard time-domain cusp waveform~\eqref{eq:cusp-waveform}, which is symmetric around the peak.
    The cusp collapse waveform, in red, is truncated just before the peak, and eventually culminates in the QNM ringing of the final PBH.
    The uncertain period inbetween is denoted with a question mark.
    Note that the QNM frequency~\eqref{eq:qnm-omega} is much higher than depicted here, and that this figure is only for illustrative purposes.}
    \label{fig:waveform}
\end{figure}

As a first approximation, we can model the effects of cusp collapse by transforming the standard waveform~\eqref{eq:cusp-waveform} to the time domain, and truncating it at some time $t_\mathrm{pbh}<t_0$ when the horizon forms.
In the limit $t_\mathrm{pbh}\to t_0$ where the PBH forms at the peak of the cusp signal, the resulting frequency-domain waveform is exactly half of the standard one, with the other half corresponding to the truncated part of the signal at $t\ge t_0$.
Based on the discussion around equation~\eqref{eq:collapse-timescale}, we expect $t_0-t_\mathrm{pbh}\sim G\mu\ell\ll\ell/2$, so the waveform is truncated slightly before $t_0$, as shown in figure~\ref{fig:waveform}.
This leads to a loss of power at frequencies above $f_\mathrm{pbh}\sim1/(G\mu\ell)$, which cures the memory divergence, as we saw in section~\ref{sec:cusp-collapse-memory}.

A further contribution to the signal comes from the quasinormal ringing of the PBH.
We can describe this very approximately by including only the leading-order $(\ell,m,n)=(2,2,0)$ quasinormal mode (QNM), writing
    \begin{equation}
        h(t)\approx C\frac{Gm_\mathrm{pbh}}{r}\exp[\rmi(\omega t+\phi)-t/\tau],
    \end{equation}
    where $C$ and $\phi$ are unknown real constants.
Using the fitting formulae in \citet{Berti:2005ys}, we take the real and imaginary parts of the $(2,2,0)$ QNM for a PBH with spin $\chi=2/3$ as
    \begin{equation}
        \omega\approx0.5214/(Gm_\mathrm{pbh}),\quad1/\tau\approx0.1715/(Gm_\mathrm{pbh}).
    \end{equation}
Since $Gm_\mathrm{pbh}\approx(G\mu)^3\ell$, we see that the ringdown signal is associated with extremely high frequencies,
    \begin{equation}
    \label{eq:qnm-omega}
        \omega\approx5\times10^{24}\,\mathrm{Hz}\times\qty(\frac{\ell}{\mathrm{pc}})^{-1}\qty(\frac{G\mu}{10^{-11}})^{-3}.
    \end{equation}

Our ignorance about the exact details of the collapse means that we cannot hope to construct an accurate phase-coherent waveform like those found by \citet{Aurrekoetxea:2020tuw} for circular loop collapse.
However, by accounting for the truncation of the cusp signal and the PBH ringdown, we can obtain a reasonable first approximation to the (logarithmic, one-sided) GW energy spectrum,
    \begin{equation}
    \label{eq:cusp-collapse-energy-spectrum}
        \epsilon_\gw\approx\frac{(2/3)^{1/3}}{4}\frac{(\uppi A_\rmc)^2G\mu}{(f\ell)^{1/3}}\Theta(1-G\mu\ell f)+\frac{C^2\ell}{4}(\uppi f)^3(G\mu)^5\qty[\mathcal{L}^2\qty(2\uppi f;\omega,\tfrac{1}{\tau})+\mathcal{L}^2\qty(2\uppi f;-\omega,\tfrac{1}{\tau})].
    \end{equation}
The first term here is reduced by a factor $1/4$ compared to equation~\eqref{eq:cusp-energy-spectrum} (due to a factor $1/2$ in each power of the strain) and is truncated at $f\sim1/(G\mu\ell)$, while the second part is the ringdown contribution, written in terms of the Lorentzian
    \begin{equation}
        \mathcal{L}(x;x_0,\gamma)\equiv\frac{\gamma/\uppi}{\gamma^2+(x-x_0)^2}.
    \end{equation}

We can fix the constant $C$ by setting the total energy radiated by the ringdown term equal to $G\mu\epsilon_M$, where $\epsilon_M$ is the collapse radiation efficiency introduced in equation~\eqref{eq:radiation-efficiency}, and the factor $G\mu$ translates between the loop's mass and the PBH's relativistic mass (i.e. rest mass plus kinetic energy), with $\epsilon_M$ defined as a fraction of the latter.
This gives
    \begin{equation}
        C=\sqrt{\frac{64\uppi\epsilon_M}{1+\omega^2\tau^2}\frac{\tau/Gm_\mathrm{pbh}}{G\mu}}\approx10.70\times(\epsilon_M/G\mu)^{1/2}.
    \end{equation}
We assume a value of $\epsilon_M=10\%$, which is consistent with the upper bounds we found in section~\ref{sec:radiation}, and is comparable to the mass-radiation fraction found in numerical simulations of other ultrarelativistic, strong-gravity phenomena~\cite{Sperhake:2008ga,East:2012mb}.
The total fraction of the loop's mass radiated by the cusp is approximately
    \begin{equation}
    \label{eq:gw-radiation-fraction}
        \int_{2/\ell}^\infty\frac{\dd{f}}{f}\epsilon_\gw\approx
        \begin{cases}
            14.9\,G\mu & \text{cusp},\\
            (3.71+\epsilon_M)\,G\mu & \text{cusp collapse},
        \end{cases}
    \end{equation}
    which shows that the radiation from collapsing cusps is comparable to, but strictly less than, that from non-collapsing cusps.

\begin{figure}[t!]
    \begin{center}
        \includegraphics[width=0.7\textwidth]{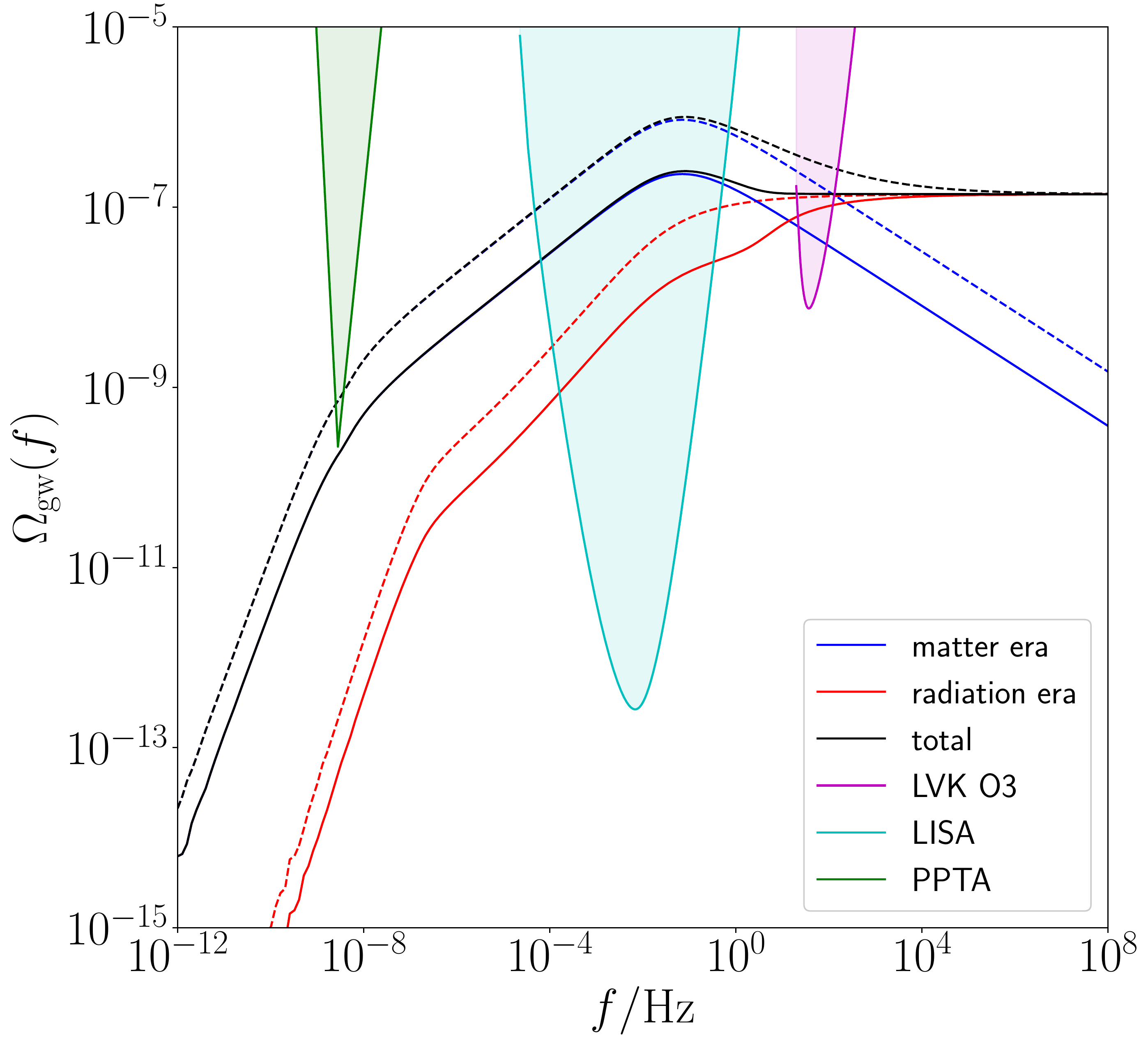}
    \end{center}
    \caption{%
    The GWB spectrum from cusps on cosmic string loops.
    Solid lines include the effects of cusp collapse using equation~\eqref{eq:cusp-collapse-energy-spectrum}, while dotted lines correspond to the standard case without collapse~\eqref{eq:cusp-energy-spectrum}.
    The magenta, cyan, and green curves show the power-law integrated sensitivities of LIGO/Virgo's third observing run~\cite{Abbott:2021kbb,Abbott:2021pi,Abbott:2021nrg}, the Parkes PTA~\cite{Shannon:2015ect,Verbiest:2016vem}, and LISA~\cite{Caprini:2019pxz,Smith:2019wny}, respectively.
    We use \enquote{model 3} of the loop network~\cite{Ringeval:2005kr,Lorenz:2010sm} with $G\mu=3\times10^{-11}$, illustrating how the PPTA bound is weakened due to cusp collapse.
    At high frequencies the spectra with and without cusp collapse become identical; the frequency at which this changeover occurs decreases for smaller values of $G\mu$, meaning that the LIGO/Virgo bound on model 3 is the same in both cases.}
    \label{fig:Omega_gw-pbh}
\end{figure}

We can account for the effect of cusp collapse on the GWB from cosmic string loops by replacing the GW energy spectrum used in equation~\eqref{eq:gwb-cosmic-strings} with that given here in equation~\eqref{eq:cusp-collapse-energy-spectrum}.
A representative example of the resulting GWB spectrum is shown in figure~\ref{fig:Omega_gw-pbh}.
At low frequencies $\Omega_\gw$ is reduced by a factor of $1/4$ compared to the standard spectrum, which relaxes the constraints on $G\mu$ coming from LIGO/Virgo~\cite{Abbott:2019vic} and from PTAs~\cite{Shannon:2015ect,Lasky:2015lej,Verbiest:2016vem,Blanco-Pillado:2017rnf}.
At high frequencies this factor $1/4$ difference vanishes, as the signal is dominated by loops which are too small to undergo cusp collapse.
The changeover between these two regimes depends on the value of $G\mu$, as this sets the size of the smallest loops that can undergo cusp collapse through equation~\eqref{eq:ell-min-cusp-collapse}; for smaller values of $G\mu$ the changeover happens at lower frequencies.
At very high frequencies the QNM emission from PBHs forming in the matter era gives rise to a strong peak, but this is dwarfed by the radiation-era plateau, making it unobservable (regardless, these frequencies are well beyond the reach of any current or planned GW experiment).

\subsection{Detection prospects for the nonlinear memory}
\label{sec:detection-prospects}

Having calculated the nonlinear GW memory waveforms associated with cusps and kinks, it is natural to ask whether these signals are detectable with current or future GW observatories.
Clearly the divergent behaviour diagnosed for cusps in section~\ref{sec:cusp-higher-order} could, in principle, have important observational implications, depending on how the divergence is regulated.
For the purposes of this section we assume the divergence is resolved along the lines of the cusp-collapse scenario described in section~\ref{sec:cusp-collapse}.
We show below that, under this assumption, the cusp and kink memory signals are suppressed so strongly that they are well beyond the reach of GW observatories, even future third-generation interferometers like Einstein Telescope (ET)~\cite{Punturo:2010zz} and Cosmic Explorer (CE)~\cite{Reitze:2019iox}.

We start by calculating the expected detection horizons for individual bursts of GW memory from cusps and kinks---i.e. the maximum distance at which a burst can be detected, on average.
We assume a matched-filter search, such that the optimal root-mean-square SNR (averaging over sky location and polarisation angle) for a frequency-domain waveform $\tilde{h}(f)$ which is isotropically averaged over the source inclination is given by~\cite{Maggiore:2007zz}
    \begin{equation}
        \rho_\mathrm{rms}=\qty[\sum_I\frac{4}{5}\sin^2\alpha_I\int_0^\infty\dd{f}\frac{|\tilde{h}(f)|^2}{P_I(f)}]^{1/2}.
    \end{equation}
The sum here is over different GW detectors, with $P_I(f)$ representing the noise power spectral density (PSD) of detector $I$, and $\alpha_I$ the opening angle between the two interferometer arms (this angle enters through the detector's response function; LIGO, Virgo, and CE have an opening angle of $\uppi/2$, while each of ET's three interferometers has an opening angle of $\uppi/3$).
It is convenient to rewrite this in terms of the fractional energy spectrum, using equation~\eqref{eq:dimensionless-energy} to give
    \begin{equation}
    \label{eq:snr-rms}
        \rho_\mathrm{rms}=\qty[\sum_I\frac{2G\mu\ell}{5\uppi^2r^2}\sin^2\alpha_I\int_0^\infty\dd{f}\frac{\bar{\epsilon}(f_\mathrm{s})}{f_\mathrm{s}^3P_I(f)}]^{1/2},
    \end{equation}
    where $r(z)$ is now the comoving distance to the source, and $f_\mathrm{s}=(1+z)f$ is the source-frame frequency, with $z$ the redshift.
We assume that any cosmic string signal with $\rho_\mathrm{rms}\ge12$ can be confidently detected; in reality, this threshold depends on the distribution of non-Gaussian noise transients (\enquote{glitches}) in the network, and how closely these are able to mimic the waveforms of interest, but we ignore these details here.

Even assuming an optimistic third-generation GW detector network consisting of Einstein Telescope plus two Cosmic Explorers,\footnote{%
    For ET we use the \enquote{ET-D} noise PSD~\cite{Hild:2010id}, while for CE we use the \enquote{CE-2} noise PSD~\cite{Hall:2020dps}.%
} we find that given current constraints on the string tension ($G\mu\lesssim10^{-11}$), the detection horizon for a cusp memory signal is at most $\approx2\,\mathrm{pc}$.
This corresponds to a negligibly small detection rate, as very few cosmic strings are expected within a volume of this size.
For kink memory the result is even more pessimistic, with a detection horizon of $\approx0.008\,\mathrm{AU}$ (a few times larger than the Earth-Moon distance).
Larger values of the string tension $G\mu$ would boost the detectability of these memory bursts, but would be in conflict with existing observational results.

\begin{figure}[t!]
    \begin{center}
        \includegraphics[width=0.7\textwidth]{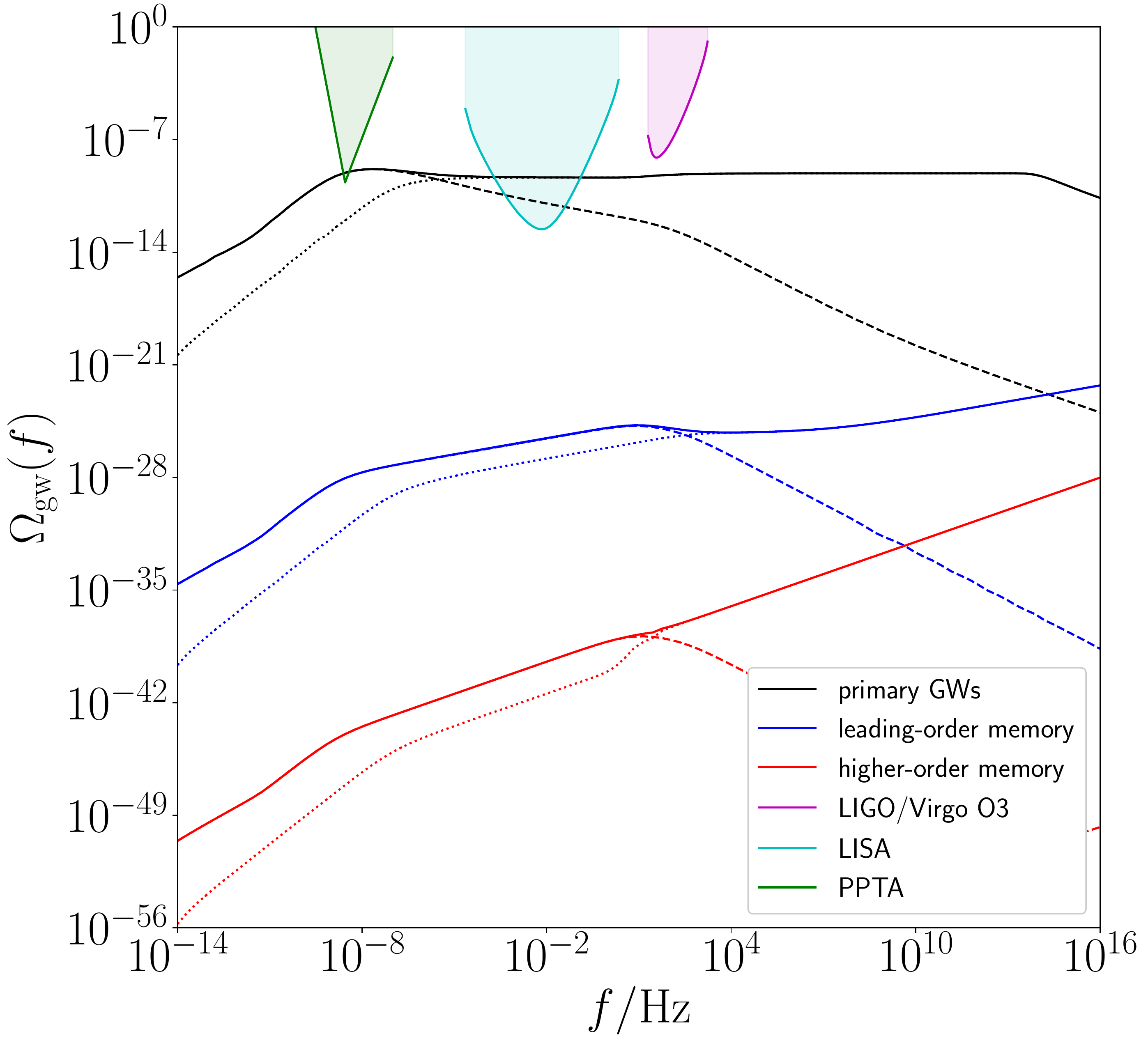}
    \end{center}
    \caption{%
    Contributions to the GWB energy spectrum $\Omega_\gw(f)$ from cosmic strings at different orders in the memory expansion, assuming that the memory divergence is resolved through the cusp-collapse scenario.
    We see that the memory effect is negligible compared to the primary emission.
    Here we assume \enquote{model 2}~\cite{Blanco-Pillado:2013qja,Blanco-Pillado:2017rnf} of the string network, and a string tension of $G\mu=8.1\times10^{-11}$ (this is the largest string tension allowed by current constraints in the cusp-collapse scenario~\cite{Jenkins:2020ctp}; smaller tensions suppress the memory effect even further).
    The dashed/dotted curves show the contributions from the matter/radiation era, respectively, with the solid curves showing the combined spectra.
    A distinct change in all three of the radiation-era spectra is visible at around $f\approx10\,\mathrm{Hz}$; at frequencies above this, the spectra are increasingly dominated by small loops which do not undergo cusp-collapse, and this is why the memory effect becomes more prominent (though still undetectable).
    The sensitivity curves are the same as in figure~\ref{fig:Omega_gw-pbh}.
    }
    \label{fig:Omega_gw-memory}
\end{figure}

In figure~\ref{fig:Omega_gw-memory} we show the GWB spectrum from a particular model of the loop network, including the contributions from first-order and higher-order GW memory from cusps and kinks.
We see that while the primary GWB reaches a plateau at high frequencies, the memory contributions to the GWB grow with frequency above $\approx10\,\mathrm{Hz}$, which makes sense given the slower fall-off of the cusp memory emission at high frequency compared to the primary cusp and kink signals.
However, each order in memory is suppressed by an additional factor of $(G\mu)^2$, and this suppression is strong enough to render the memory contribution unobservable at all frequencies.

\subsection{The PBH mass spectrum}
\label{sec:mass-spectrum}

We argued in section~\ref{sec:flat-space-eoms} that cosmic string loops typically form $N_\rmc\approx1$ cusp per oscillation period, which means that cusp-collapse PBHs are continuously created by the loop network, from the very early Universe to the present day, resulting in a broad distribution of PBH masses.
This contrasts sharply with the standard PBH formation scenario we discussed in section~\ref{sec:primordial-black-holes}, where the collapse typically occurs at a single early epoch, resulting in a monochromatic PBH mass spectrum.

We can write the comoving number density of cusp-collapse PBHs with rest mass between $m$ and $m+\dd{m}$ as
    \begin{equation}
    \label{eq:n-pbh-cusp-collapse}
        n_\mathrm{pbh}(m,t)\dd{m}=\int_0^t\dd{t'}\frac{2N_\rmc}{\ell_m}n_\mathrm{loop}(\ell_m,t')\dd{\ell_m},
    \end{equation}
    where $n_\mathrm{loop}(\ell,t)=(a^3/t^4)\mathcal{F}(\gamma)$ is loop distribution function we encountered in section~\ref{sec:cusps-kinks}, and $\ell_m\approx Gm/(G\mu)^3$ is the loop length required to form a PBH of mass $m$.
We assume $N_\rmc=1$, consistent with much of the literature on cosmic string phenomenology, particularly regarding GW searches~\cite{Abbott:2017mem,Auclair:2019wcv}.
The factor of $2/\ell_m$ here accounts for the oscillation period of the loop which forms the PBH.

Since $\ell_m$ corresponds to a fixed physical (rather than comoving) scale, PBH production only begins once this scale has entered the horizon.
This happens at cosmic time
    \begin{equation}
    \label{eq:collapse-time}
        t_i(m)\approx\ell_m\approx16\,\mathrm{Gyr}\times\frac{m}{10^{-10}M_\odot}\times\qty(\frac{G\mu}{10^{-11}})^{-3},
    \end{equation}
    which means that the largest PBHs form at the present day, with mass
    \begin{equation}
    \label{eq:M_max}
        m_\mathrm{max}\approx0.88\times10^{-10}\,M_\odot\times\qty(\frac{G\mu}{10^{-11}})^3.
    \end{equation}
Coincidentally, this corresponds to the \enquote{sublunar} mass range---one of the few regimes where there are no constraints preventing PBHs from making up the totality of DM~\cite{Carr:2020gox,Carr:2020xqk} (see figure~\ref{fig:pbh-bounds}).
However, the majority of cusp-collapse PBHs have masses much smaller than this.

\begin{figure}[t!]
    \begin{center}
        \includegraphics[width=0.7\textwidth]{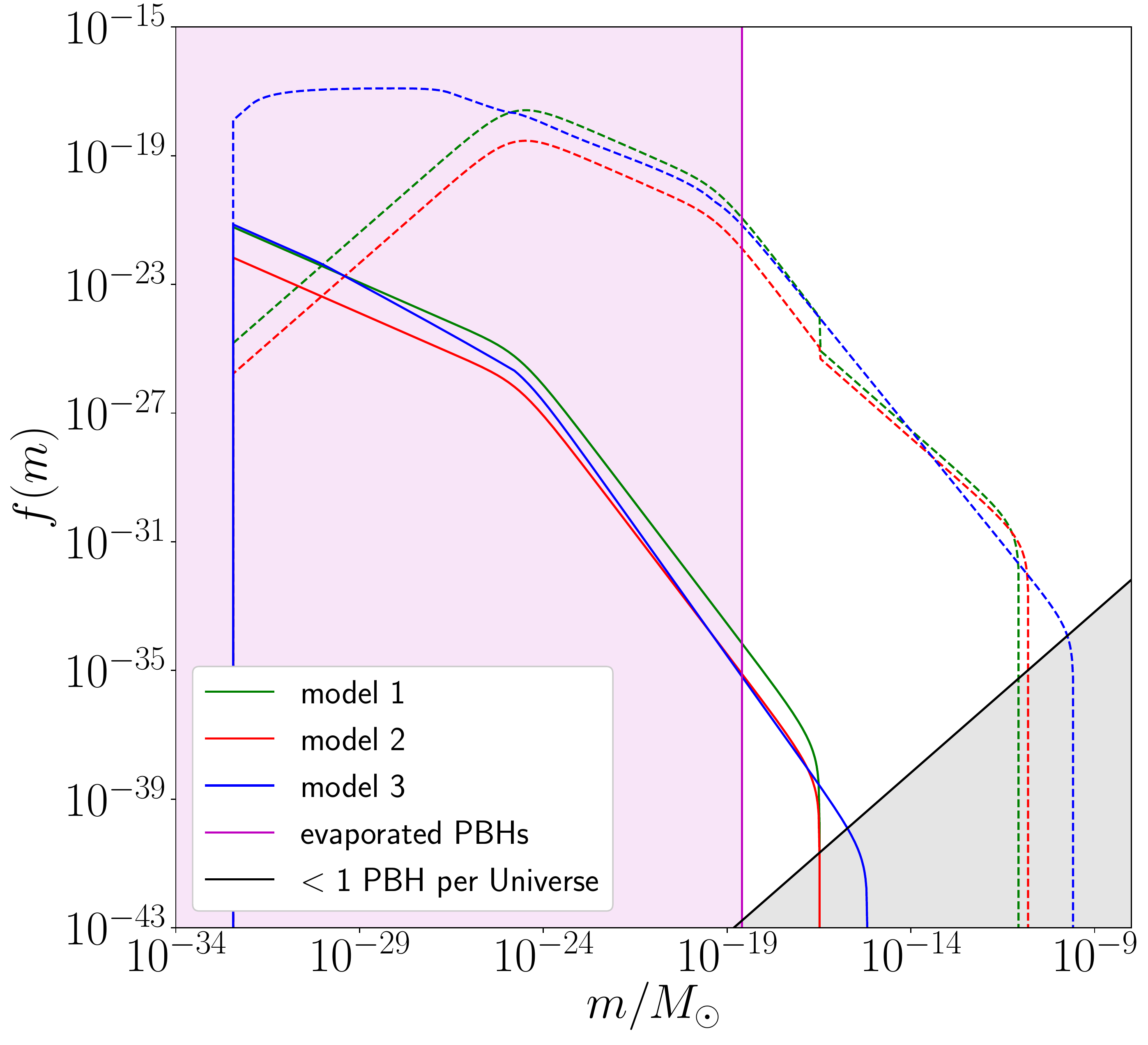}
    \end{center}
    \caption{%
    The present-day PBH mass spectrum \eqref{eq:mass-spectrum} for the three standard models of the cosmic string loop network with $G\mu=10^{-11}$.
    The solid and dashed lines correspond to PBHs formed in the radiation and matter eras respectively.
    The cutoffs at small and large masses are given by equations~\eqref{eq:M-min} and~\eqref{eq:M_max}.
    The magenta region represents PBHs which have evaporated by the present day.
    The grey region corresponds to there being less than one PBH of that mass in the observable Universe.}
    \label{fig:mass-spectra-Gmu-1e-11}
\end{figure}

Inserting equation~\eqref{eq:n-pbh-cusp-collapse} into the expression~\eqref{eq:cdm-mass-fraction} we introduced in section~\ref{sec:primordial-black-holes}, we find that the mass spectrum of cusp-collapse PBHs is
    \begin{equation}
    \label{eq:mass-spectrum}
        f(m)=\frac{2N_\rmc}{\rho_\mathrm{cdm}}\int_0^{t_0}\dd{t}\frac{a^3(t)}{t^4}\mathcal{F}\qty(\frac{Gm}{(G\mu)^3t}),
    \end{equation}
    which is shown in figure~\ref{fig:mass-spectra-Gmu-1e-11} for a representative value of $G\mu$.
In general the integral in equation~\eqref{eq:mass-spectrum} is broken into two parts, corresponding to the different scaling solutions in the matter and radiation eras.
Note that equation~\eqref{eq:mass-spectrum} includes only the energy density due to the rest mass of the PBHs; their kinetic energy at formation is larger by a factor of $1/(G\mu)$, though this will eventually be redshifted away.
This large kinetic energy will likely lead to interesting phenomenology and constraints which are not captured by traditional PBH analyses, as these generally assume the PBHs are formed with negligible peculiar velocities.
We plan to explore this in future work.

Equation~\eqref{eq:mass-spectrum} does not include evaporation due to Hawking radiation, and is therefore only valid for masses greater than $m_*\approx5\times10^{14}\,\mathrm{g}\approx3\times10^{-19}M_\odot$, with PBHs lighter than this evaporating in less than a Hubble time~\cite{Carr:2020gox}.
Nonetheless, the form of $f(m)$ for masses below $m_*$ can be useful for deriving constraints on the overall mass spectrum due to evaporation effects.
For small masses in the radiation era, equation~\eqref{eq:mass-spectrum} approaches a time-independent power law $f(m)=f_*(m/m_*)^{-1/2}$, with $f_*$ a constant depending on $G\mu$ and on the network model.
The negative exponent means that the mass spectrum is dominated by very small PBHs, and that the strongest constraints on cusp collapse come from their evaporation.
In fact, this is the same power law as the mass spectrum resulting from the collapse of circular loops~\cite{MacGibbon:1997pu,James-Turner:2019ssu}, but with a different pre-factor.
In the circular collapse case, the pre-factor depends on the fraction of circular loops, which is unknown; in our case, the pre-factor depends only on $G\mu$, which we can therefore constrain directly.
Using the most up-to-date constraints from \citet{James-Turner:2019ssu}, we find\footnote{%
    These constraints are phrased in terms of a normalisation constant $c_\mathrm{string}$, with the CMB constraint giving $c_\mathrm{string}<2\times10^{-12}$.
    This can be translated to our mass spectrum using $c_\mathrm{string}=2f_*\Omega_\mathrm{cdm}$.}
    \begin{equation}
    \label{eq:Gmu-bounds-CMB}
        G\mu\lesssim
        \begin{cases}
            6.0\times10^{-7} & \text{for model 1,}\\[-4pt]
            1.2\times10^{-6} & \text{for model 2,}\\[-4pt]
            9.3\times10^{-7} & \text{for model 3,}
        \end{cases}
    \end{equation}
    which in turn gives a constraint on the total fraction of DM made up by cusp-collapse PBHs,
    \begin{equation}
        \frac{\rho_\mathrm{pbh}}{\rho_\mathrm{cdm}}\equiv\int_{m_*}^\infty\dd{m}\frac{f(m)}{m}\lesssim
        \begin{cases}
            2.0\times10^{-10} & \text{for model 1,}\\[-4pt]
            7.4\times10^{-11} & \text{for model 2,}\\[-4pt]
            8.0\times10^{-10} & \text{for model 3.}
        \end{cases}
    \end{equation}
The constraint~\eqref{eq:Gmu-bounds-CMB} on $G\mu$ is comparable to those set by CMB analyses~\cite{Ade:2013xla,McEwen:2016aob}, and is almost independent of the network model.
This constraint is set by the damping of small-scale CMB anisotropies due to PBHs decaying at recombination~\cite{Zhang:2007zzh,Carr:2009jm}, and is orders of magnitude stronger than the $\gamma$-ray constraint~\cite{James-Turner:2019ssu}.
It is likely to become more stringent with future CMB missions, and with similar analyses from upcoming $21\,\mathrm{cm}$ intensity-mapping experiments~\cite{Stocker:2018avm,Lucca:2019rxf}.
For now, however, it is unable to compete with the constraints on the order of $G\mu\lesssim10^{-11}$ coming from GW observations.

\subsection{A unique BH population}
\label{sec:populations}

\begin{figure}[t!]
    \begin{center}
        \includegraphics[width=0.75\textwidth]{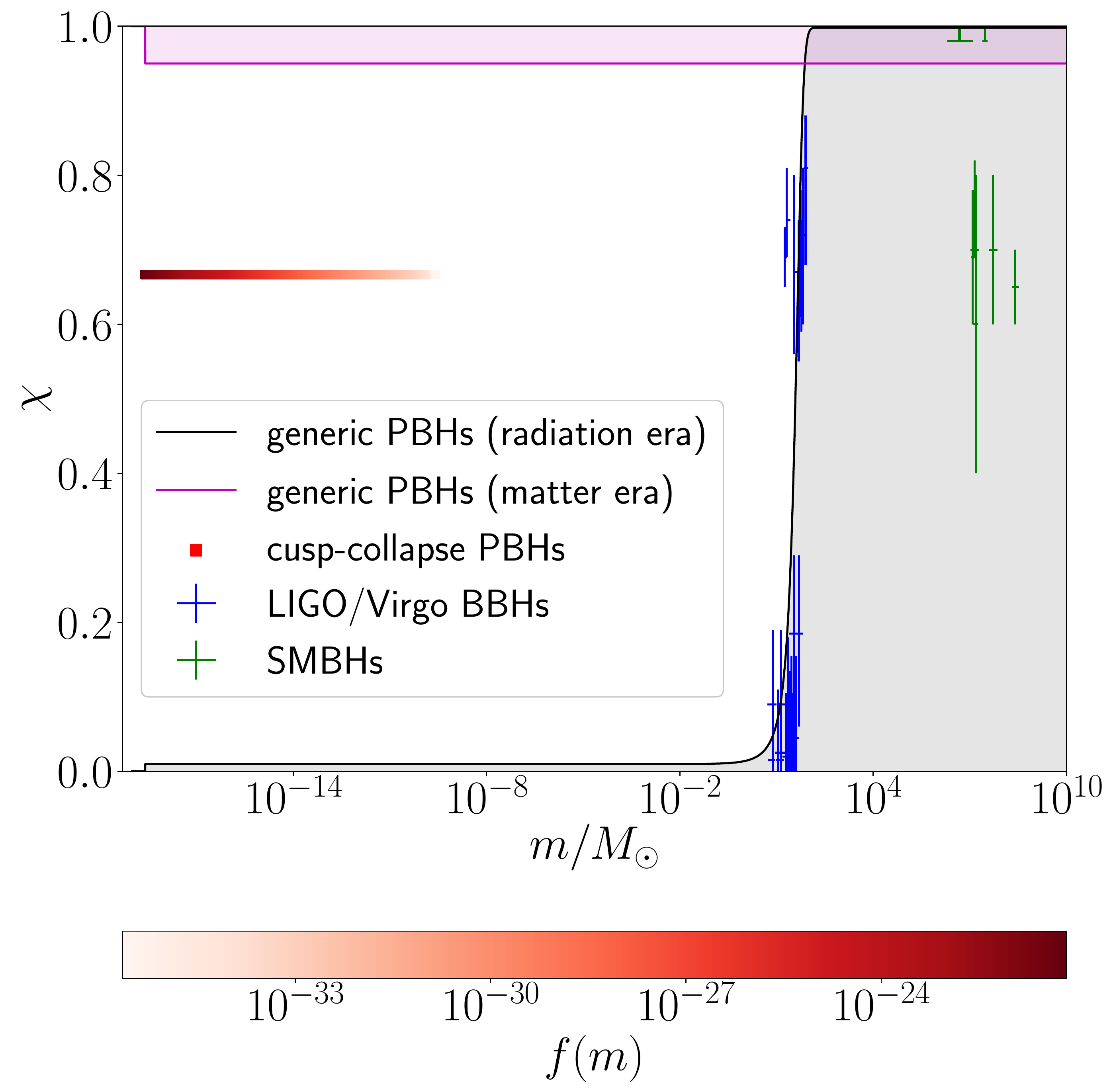}
    \end{center}
    \caption{%
    The location of various primordial and astrophysical BH populations in the Regge plane (BH mass-spin parameter space).
    Blue crosses show the initial and final BHs for each of the ten binary BH mergers in LIGO/Virgo's first GW Transient Catalogue (GWTC-1)~\cite{Abbott:2018mvr}; the spin distribution is noticeably bimodal, with initial BHs having low spins $\chi\lesssim0.2$ and final BHs have large spins $\chi\approx0.7$.
    (The spins of the initial BHs are not confidently measured in GWTC-1, so the spin values and uncertainties plotted here are merely heuristic and are estimated from the inspiralling binary's effective aligned spin $\chi_\mathrm{eff}$.)
    Green crosses show the SMBHs catalogued in \citet{Brenneman:2011wz}.
    The grey region shows the expected parameter space for \enquote{conventional} PBHs formed from overdensities collapsing during radiation domination, including the effects of accretion as calculated in \citet{DeLuca:2020bjf}.
    The magenta region shows a possible population of near-extremal PBHs formed during a period of early matter domination.
    The red region shows cusp-collapse PBHs, shaded according to their mass spectrum for $G\mu=10^{-11}$, cf.~figure~\ref{fig:mass-spectra-Gmu-1e-11}.
    All of the PBH populations are cut off at $m_*\approx3\times10^{-19}M_\odot$, due to evaporation through Hawking radiation.}
    \label{fig:regge-plane}
\end{figure}

We seen in section~\ref{sec:pbh-properties} that cusp-collapse PBHs are universally formed with dimensionless spins of $\chi=2/3$, regardless of the loop size $\ell$ or the string tension $G\mu$.\footnote{%
Note that this is the \enquote{na\"{i}ve} zero-radiation value, and that a fully general-relativistic calculation would likely give a different value for the final spin.
However, our argument in this section still holds, provided that the true value of $\chi$ is significantly larger than zero and less than unity.}
For sufficiently large rest masses $m\gg m_*$, this initial spin value is not affected by Hawking radiation, and survives to the present day~\cite{Arbey:2019jmj}.
This is interesting because it means that cusp-collapse PBHs occupy a unique region of the \enquote{Regge plane}~\cite{Arvanitaki:2010sy,Berti:2015itd,Brito:2017zvb} (i.e., the BH mass-spin parameter space), as we illustrate in Fig~\ref{fig:regge-plane}.

Spins of $\chi\approx2/3$ are common amongst astrophysical BHs.
In particular, this is a very natural value for BHs formed from binary mergers like those observed by LIGO/Virgo~\cite{Abbott:2018mvr}; the majority of such binaries are approximately equal-mass with small initial spins~\cite{Abbott:2018jsj,Fishbach:2019bbm}, which correspond to final spins of $\chi\approx0.687$~\cite{Rezzolla:2007rz}.
SMBHs in active galactic nuclei are also observed to have large spins $\chi\gtrsim0.6$, due to accretion and prior mergers~\cite{Brenneman:2011wz}.
However, such astrophysical processes are unable to create subsolar-mass BHs, which dominate the cusp-collapse PBH mass spectrum for realistic values of $G\mu$, cf. figure~\ref{fig:mass-spectra-Gmu-1e-11}.

On the other hand, subsolar masses are generally possible in other PBH formation mechanisms, but these mechanisms are unable to generate spins $\chi\approx2/3$ like those resulting from cusp collapse.
\enquote{Conventional} PBHs formed from collapsing overdensities during radiation domination are typically born with small spins of order $\chi\sim0.01$~\cite{Chiba:2017rvs,Mirbabayi:2019uph,DeLuca:2019buf}.
These initially low-spinning PBHs can acquire large spins through accretion, saturating the Thorne bound $\chi\approx0.998$~\cite{Thorne:1974ve}, but this process only takes place within a Hubble time if the PBHs in question are sufficiently massive, $m\gtrsim50\,M_\odot$~\cite{DeLuca:2020bjf}.
Subsolar-mass PBHs are extremely inefficient at accreting matter, and remain essentially non-spinning.

Subsolar-mass PBHs can have large spins if they form from collapsing overdensities during a hypothetical period of early matter domination~\cite{Polnarev:1986bi,Carr:2017edp,Harada:2017fjm,Kuhnel:2019zbc}, as radiation pressure is then unable to dissipate angular momentum during the collapse.
However, these PBHs are expected to have near-extremal spins $\chi\approx1$, which are easily distinguishable from the $\chi=2/3$ prediction of cusp collapse.

We therefore see that any observation of a subsolar-mass BH with a large (but non-extremal) spin $\chi\approx2/3$ would be incompatible with any of the other BH formation mechanisms mentioned here, and would be a \enquote{smoking gun} signature of cusp collapse, and of cosmic strings more generally.

\section{Summary and outlook}
\label{sec:cosmic-strings-summary}

In this chapter, we have thoroughly explored the nonlinear GW memory associated with cusps and kinks on Nambu-Goto cosmic strings, deriving detailed analytical waveform models for the memory GWs, including the \enquote{memory of the memory} and other higher-order memory effects.
These are among the first memory observables computed for a cosmological source of GWs, with previous literature having focused almost entirely on astrophysical sources.

The leading-order cusp memory waveform~\eqref{eq:cusp-result} that we have found is strikingly similar to the primary GW signal from the cusp~\eqref{eq:cusp-waveform}, with the same characteristic $\sim f^{-4/3}$ frequency power-law.
However, one very important difference is that this memory signal is emitted in all directions, unlike the primary signal which is confined to a narrow beam of width $\theta_\rmb\sim f^{-1/3}$.
As a result, we find that the total GW energy radiated by the cusp memory \emph{diverges} for a Nambu-Goto (i.e., zero-width) loop.
This divergence can be regularised by introducing a cutoff at the scale of the string width $\delta\sim\ell_\Pl/\sqrt{G\mu}$, but this then introduces powers of $\ell/\delta\gg1$ to the higher-order memory terms, causing the sum of all memory contributions to diverge for loops of length $\ell\gtrsim\delta/(G\mu)^3$.

In section~\ref{sec:cusp-collapse-memory} we have argued that the most plausible explanation for this unphysical memory divergence is the assumption that the spacetime containing the cosmic string loop is well-described by a flat background with linear perturbations.
We have suggested that some strong-gravity mechanism must kick in on loops of length $\ell\gtrsim\delta/(G\mu)^3$ to suppress the high-frequency GW emission from cusps, thereby curing the divergence.
In section~\ref{sec:cusp-collapse}, we have investigated exactly such a mechanism: the collapse of a small cosmic string segment near the cusp to form a PBH.
Remarkably, this cusp-collapse process is predicted to occur for all loops of length $\ell\gtrsim\delta/(G\mu)^3$---exactly the same loops for which we have diagnosed the memory divergence.
We have shown explicitly that if cusp collapse does indeed occur for these loops, the corresponding GW memory is strongly suppressed, and the divergence is cured.

We have also calculated the memory emission associated with kinks, and have shown that this is suppressed due to interference between GWs emitted by the kink at different points in its history.
There is thus no situation in which the kink memory signal diverges; this accords with the cusp-collapse description, as kinks are not predicted to form PBHs in that scenario.

In section~\ref{sec:observational-consequences} we have investigated the observational consequences of our results, focusing on the implications for the GW background.
We find that by requiring the string tension to agree with existing observational bounds, and by invoking cusp collapse to prevent the memory from diverging, the resulting memory signal is very strongly suppressed, and is not likely to be detected by any current or upcoming GW observatories.
Meanwhile, the primary signal from cusps is reduced by a factor of $\approx1/4$ at low frequencies due to cusp collapse, slightly weakening existing bounds on the string tension.
Of course, if cusp collapse does not occur, then it is possible that large loops could source much stronger memory signals; however, one would then need an alternative mechanism for resolving the cusp divergence.

One promising avenue for testing these ideas lies in observing the PBHs themselves.
By calculating the angular momentum of the string segment captured behind the horizon, we have shown that cusp-collapse PBHs are highly spinning, with dimensionless spin parameter equal to two-thirds of the extremal Kerr value, $\chi=2/3$.
This spin is a universal property of the formation mechanism, and is independent of the loop size $\ell$ and string tension $G\mu$.
To the best of our knowledge, cusp collapse is the only known primordial or astrophysical mechanism for generating subsolar-mass BHs with large but sub-extremal spins.
The observation of such a BH would therefore be a \enquote{smoking gun} signal of cusp collapse, and of cosmic strings more generally.
The fact that cusp-collapse PBHs are born with ultrarelativistic velocities will undoubtedly also give rise to some interesting phenomenology, and may allow us to place new constraints on their abundance, aside from those in the standard PBH literature.
It would also be very interesting to calculate the merger rate of cusp-collapse PBH binaries, as well as the corresponding GWB spectrum~\cite{Mandic:2016lcn,Bartolo:2016ami,Clesse:2016ajp,Wang:2016ana,Raidal:2017mfl,Raidal:2018bbj}, as consistency with LIGO/Virgo observations (in particular the subsolar-mass search~\cite{Abbott:2005pf,Abbott:2018oah,Magee:2018opb,Abbott:2019qbw}) would provide another independent constraint on the string tension; we leave this for future work.

Our work in this chapter demonstrates the importance of considering the nonlinear memory associated with a broader class of GW sources than just compact binaries; we have shown that the memory effect is interesting not just from an observational point of view, but also as a tool for sharpening our theoretical understanding and modelling of said GW sources.
As an unexpected by-product of our analysis, we have shown that the standard Nambu-Goto description of cusps is unphysical, and that strong gravity effects (possibly including PBH formation) could play an important r\^ole in a more complete understanding of their dynamics.
This motivates further work to better understand cusps in full GR, including nonlinear gravitational effects beyond those considered here, and thereby understand whether cusp collapse does indeed take place, with the ultimate goal of computing reliable waveform predictions for GW observatories.
It would also be very interesting to study the \emph{linear} memory associated with cusps, and to understand how this contributes to the total GW emission; however, such a study would require us to go beyond the Nambu-Goto approach adopted here.
More generally, our results motivate a broader examination of the nonlinear memory effect in GW astronomy and cosmology, to see what other surprises may be in store.

%% file: chapters/binary-resonance.tex
\chapter{Binary systems as dynamical gravitational-wave detectors}
\label{chap:binary-resonance}

\noindent
In section~\ref{sec:gw-detection} we discussed several current and near-future experimental efforts to detect gravitational waves.
These experiments each hold enormous scientific potential, and together will probe a dizzying range of exotic astrophysical and cosmological phenomena throughout the history of the Universe, including those we covered in section~\ref{sec:gw-sources}.
However, we saw how practical and technical limitations restrict the sensitivity of each experiment to a narrow frequency band, leaving broad swathes of the GW frequency spectrum essentially unexplored, as shown in figure~\ref{fig:gwb-constraints}.
These gaps in the GW spectrum could contain signals that are inaccessible to any current or planned GW observatory.
Perhaps the clearest example is the stochastic GW signal expected from a first-order phase transition: as we discussed in section~\ref{sec:phase-transitions}, this signal is sharply peaked and could, for a broad range of the physical parameter space, be missed by all of the GW experiments we have described.
FOPTs aside, there is also the distinct possibility of these unexplored frequency bands containing completely unexpected signals, beyond just those models that have been proposed in the literature.
Discovering one of these \enquote{unknown unknowns} could revolutionise our understanding of the Universe.

\begin{figure}[t!]
    \includegraphics[width=\textwidth]{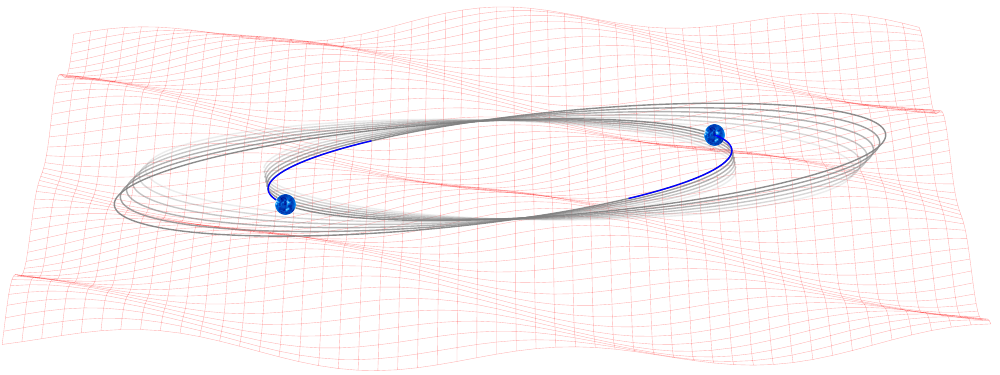}
    \caption{%
    Cartoon illustration of the binary resonance effect.
    Stochastic fluctuations in the background spacetime geometry due to incoming GWs perturb the trajectories of two orbiting masses, causing cumulative changes to their orbital elements.
    }
    \label{fig:cartoon}
\end{figure}

This problem motivates us to explore alternative experimental techniques that can bridge the observational gaps that exist in the GW frequency spectrum.
In this chapter we explore one such possibility, which is to search for GWs by studying their influence on binary systems.
Rather than searching for oscillations in the proper distance between the test masses in an interferometer (as in LIGO/Virgo/KAGRA, etc.), or between pulsars and the Earth (as in PTAs), we consider here the GW-induced oscillations between two freely-falling astronomical bodies in a gravitationally-bound orbit.
While these oscillations are extremely challenging to observe \emph{directly} for any realistic binary, we show below that they can leave lasting imprints on the binary's orbit---particularly if they occur at an integer multiple of the binary's orbital frequency, as this causes the perturbations to be resonantly amplified (as we will see below).
For persistent GW signals, these imprints accumulate over time, eventually giving rise to observable deviations which can be used to infer the GW amplitude, turning the binary into a \emph{dynamical GW detector} (see figure~\ref{fig:cartoon}).
This idea has a long history~\cite{Bertotti:1973clo,Misner:1974qy,Rudenko:1975tes,Mashhoon:1978res,Turner:1979inf,Futamase:1979gu,Mashhoon:1981cos,Linet:1982xc,Nelson:1982pk,Chicone:1995nn,Chicone:1996yc,Chicone:1996pm,Chicone:1999sg,Iorio:2011eu,Iorio:2021cxt}, and has been used to search for GWs with the orbit of the binary pulsar B1913+16~\cite{Hui:2012yp}.
Similar ideas have also been used to search for orbital changes induced by coherently oscillating ultralight DM fields~\cite{Blas:2016ddr,LopezNacir:2018epg,Armaleo:2019gil,Armaleo:2020yml,Desjacques:2020fdi,Blas:2019hxz}.
Nonetheless, this \enquote{binary resonance} effect has received relatively little attention in the GW community, and has not yet been exploited to its full potential.

In this chapter we develop, from first principles, a new formalism for calculating the GW-induced evolution of a generic binary system.
While we could, in principle, use binary resonance to search for any GW signal morphology, there are two practical considerations which lead us to narrow our scope here:
    \begin{enumerate}
        \item We require the GW signal to be \emph{persistent}, so that the GW-induced perturbations to the binary's orbit can accumulate over time and eventually reach a detectable level.
        \item We neglect \emph{narrowband} signals whose power is concentrated in a relatively small range of frequencies, since the probability of the signal's frequency band overlapping with one of the resonant frequencies of the binary is then too small for the searches to have much impact.
    \end{enumerate}
As we saw in section~\ref{sec:gwb-intro}, the Universe provides us with a persistent, broadband GW signal that satisfies these conditions: the stochastic GW background.
We therefore focus on the response of a binary system to the GWB.

The stochastic nature of the signal means that we cannot deterministically calculate the evolution of a binary coupled to the GWB; instead we treat each of the binary's orbital elements as a time-dependent random variable, and study how the statistics of the orbital perturbations evolve over time.
Our key result is a secularly-averaged \emph{Fokker-Planck equation} (FPE), which improves upon previous results by fully specifying the evolution of the joint probability distribution for all six orbital elements.
(Compare with \citet{Hui:2012yp}, who calculated only the variance of the distribution, and focused on just one orbital element: the period.)
By comparing solutions of this FPE with high-precision orbital data from various binary systems, one can place stringent new constraints on the GWB at frequencies that are inaccessible to all other current and future GW observatories.

The remainder of this chapter is structured as follows.
In section~\ref{sec:dynamics} we give a brief, self-contained overview of Keplerian orbits, orbital perturbations, and the formalism of osculating orbital elements.
This section is a review of standard results, focusing only on the tools we need to tackle the binary resonance problem; for more details on the various points we discuss, we refer the reader to, e.g., \citet{Brouwer:1961mcm}, \citet{Murray:2000ssd}, or \citet{Poisson:2014gr}.
In section~\ref{sec:gw-resonance} we specialise this formalism to perturbations from the GWB, and develop a FPE for the orbital elements, expressing the coefficients of the equation in terms of GW transfer functions.
In section~\ref{sec:KM} we write these coefficients explicitly as functions of the orbital elements, and briefly discuss their properties (the technical details of this calculation are given in appendix~\ref{app:kramers-moyal-derivation}).
In section~\ref{sec:circular} we obtain some exact late-time results for the case where the binary's eccentricity is held fixed at zero, which greatly simplifies the FPE.
In section~\ref{sec:full-fpe} we tackle the more complicated general-eccentricity case, consider practical approaches for solving the FPE on observational timescales, and present some example results for the Hulse-Taylor binary pulsar B1913+16.
In section~\ref{sec:observations} we develop the necessary statistical formalism to search for GWB-induced orbital perturbations in observational data, discussing how to compute upper limits and sensitivity forecasts, and how to apply these tools to pulsar timing and laser-ranging experiments.
In section~\ref{sec:microhertz} we use this formalism to calculate forecast constraints on the GWB spectrum from our methods.
We show that binary resonance searches can fill the \enquote{$\upmu$Hz gap} between LISA and PTAs, probing a unique region of the FOPT parameter space, and potentially helping to shed light on the common process signal detected by NANOGrav.

\section{Binary dynamics}
\label{sec:dynamics}

In this section we introduce the machinery of osculating orbital elements---an extremely useful formalism for describing perturbations to binary orbits.
We start by recalling the basic properties of Keplerian orbits (i.e., orbits of two point masses interacting only through Newtonian gravity), before introducing the equations of motion (EoMs) for the osculating elements, and discussing the most important contribution to these EoMs from relativistic effects.
We also introduce alternative sets of orbital elements that are useful in cases where the eccentricity or inclination of the orbit are small.

\subsection{Keplerian orbits}

\begin{figure}[t!]
    \begin{center}
        \includegraphics[width=0.667\textwidth]{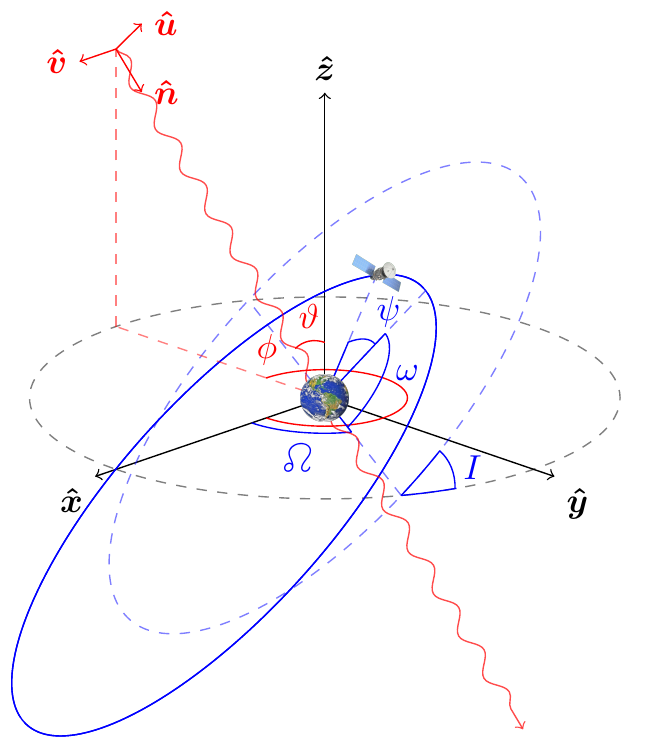}
    \end{center}
    \caption{%
    Schematic diagram of a Keplerian orbit.
    The plane of the orbit (shown in blue) is defined relative to the fixed reference frame $(\vu*x,\vu*y,\vu*z)$ (shown in black) by the inclination and the longitude of ascending node, while the orientation of the orbit within this plane is specified by the argument of pericentre.
    The true anomaly $\psi$ acts as an angular coordinate for the position of the orbit within the orbital plane, measured relative to the pericentre.
    The polarisation tensors describing the effects of an incoming plane GW are described in terms of the basis $(\vu*n,\vu*u,\vu*v)$ (shown in red), which is related to the reference frame by the angles $\vartheta$ and $\phi$.
    (Note that the wavelength of the GW is not shown to scale here---our analysis assumes a wavelength much larger than the size of the orbit.)
    }
    \label{fig:orbital-elements}
\end{figure}

We start with a Keplerian binary, working in cylindrical coordinates $(\vu*r,\vu*\theta,\vu*\ell)$ in a frame with the centre of mass fixed at the origin.
We also introduce a fixed Cartesian reference frame $(\vu*x,\vu*y,\vu*z)$, such that the line of sight of the observer is in the positive $\vu*z$ direction.
The unperturbed EoM for the separation vector $\vb*r$ is
    \begin{equation}
    \label{eq:Keplerian-EoM}
        \ddot{\vb*r}+\frac{GM}{r^2}\vu*r=\vb*0,
    \end{equation}
    where $r\equiv|\vb*r|$ is the radial separation, $\vu*r\equiv\vb*r/r$ is the radial unit vector, and $M\equiv m_1+m_2$ is the total mass.
We define the energy and angular momentum of the binary (in units of the reduced mass $\mu\equiv m_1m_2/M$) by
    \begin{equation}
    \label{eq:e-l-definitions}
        \mathcal{E}=\frac{1}{2}\dot{\vb*r}\vdot\dot{\vb*r}-\frac{GM}{r},\qquad\vb*\ell=\vb*r\cp\dot{\vb*r}=r^2\dot{\theta}\,\vu*\ell.
    \end{equation}
For the total angular momentum we write $\ell\equiv|\vb*\ell|=r^2\dot{\theta}$.
These are all conserved under the Newtonian EoM~\eqref{eq:Keplerian-EoM}, since
    \begin{equation}
    \label{eq:e-l-conservation}
        \dot{\mathcal{E}}=\dot{\vb*r}\vdot\qty(\ddot{\vb*r}+\frac{GM}{r^2}\vu*r)=0,\qquad\dot{\vb*\ell}=\vb*r\cp\ddot{\vb*r}=\vb*r\cp\qty(-\frac{GM}{r^2}\vu*r)=\vb*0,
    \end{equation}
    where we used $\dot{r}=\dot{\vb*r}\vdot\vu*r$.
The fact that $\dot{\vb*\ell}=\vb*0$ means that the binary orbit is confined to a fixed 2D plane, which is specified with respect to the $(\vu*x,\vu*y,\vu*z)$ reference frame by two angles: the inclination $I$, which is the angle between the binary's angular momentum vector $\vb*\ell$ and the observer's line of sight $\vu*z$, and the longitude of ascending node $\asc$, which is the angle between $\vu*x$ and the point where the orbit passes through the reference plane with positive velocity in the $\vu*z$ direction.
(These angles are illustrated in figure~\ref{fig:orbital-elements}.)

We can find the shape of the orbit in this plane by integrating equation~\eqref{eq:Keplerian-EoM}, yielding a family of elliptical solutions
    \begin{equation}
    \label{eq:Kepler-solutions}
        r=\frac{\ell^2/(GM)}{1+e\cos\psi},\qquad\mathcal{E}=-\frac{GM}{2a},\qquad\ell=\sqrt{GMa(1-e^2)},
    \end{equation}
    with the shape of the ellipse described by its semi-major axis $a$ and eccentricity $e$.
Here we have introduced the true anomaly $\psi$ as the angular position of the orbit within the orbital plane, defined such that the pericentre (minimum separation) occurs at $\psi=0$.
The point at which this occurs is defined by the argument of pericentre $\omega\equiv\theta-\psi$, which is measured relative to the ascending node.
Since $\omega$ is constant, the orbit is closed and the motion of the binary is periodic in time, with period $P$ related to the semi-major axis by Kepler's third law,
    \begin{equation}
    \label{eq:kepler-iii}
        \frac{GM}{a^3}=\qty(\frac{2\uppi}{P})^2.
    \end{equation}
In what follows, we work entirely in terms of the period rather than the semi-major axis, as the former is more closely linked to the resonant frequencies of the orbit.

The five constants $(P,e,I,\asc,\omega)$ are almost enough information to specify a particular Keplerian orbit; all that remains is to specify the time at which the binary is at pericentre, $t_0$.
In practice, it is more convenient to replace $t_0$ with the compensated mean anomaly,
    \begin{equation}
    \label{eq:epsilon}
        \eps\equiv\frac{2\uppi}{P}(t-t_0)-\int_0^t\dd{t'}\frac{2\uppi}{P(t')},
    \end{equation}
    as this is more well-behaved when the orbit is perturbed~\cite{Brouwer:1961mcm}.
Note that in the absence of perturbations this reduces to $\eps=-2\uppi t_0/P$.
We call the set $(P,e,I,\asc,\omega,\eps)$ the \emph{orbital elements} of a binary.

\subsection{Perturbations and osculating orbits}
\label{sec:perturbing-force}

Now suppose the binary is acted upon by some small perturbing force~\cite{Burns:1976cm,Murray:2000ssd,Hui:2012yp,Blas:2016ddr}
    \begin{equation}
    \label{eq:perturbing-force}
        \ddot{\vb*r}+\frac{GM}{r^2}\vu*r=\updelta\ddot{\vb*r}=r(\mathcal{F}_r\vu*r+\mathcal{F}_\theta\vu*\theta+\mathcal{F}_\ell\vu*\ell).
    \end{equation}
(The factor of $r$ here is included for later convenience, as the GW perturbations we consider are proportional to the orbital separation.)
Inserting this into equation~\eqref{eq:e-l-conservation} gives
    \begin{equation}
    \label{eq:e-l-evolution}
        \dot{\mathcal{E}}=\dot{\vb*r}\vdot\updelta\ddot{\vb*r}=r\dot{r}\mathcal{F}_r+r^2\dot{\theta}\mathcal{F}_\theta,\qquad\dot{\vb*\ell}=\vb*r\cp\updelta\ddot{\vb*r}=r^2\mathcal{F}_\theta\vu*\ell-r^2\mathcal{F}_\ell\vu*\theta,
    \end{equation}
    so the binary's energy and angular momentum are no longer constant.
As a result, the binary is no longer described by a fixed set of orbital elements.
However, the orbit is still tangent to some Keplerian ellipse at each moment in time.
We therefore define $(P,e,I,\asc,\omega,\eps)$ as functions of time which track the evolution of this tangent ellipse; these are called the \emph{osculating} orbital elements.

By writing the osculating elements as functions of the energy $\mathcal{E}$ and angular momentum $\vb*\ell$ we can translate equation~\eqref{eq:e-l-evolution} into a set of EoMs for the osculating elements.
A full derivation along these lines is given in, e.g., \citet{Burns:1976cm} or \citet{Murray:2000ssd}; here we simply quote the resulting set of equations,
    \begin{align}
    \begin{split}
    \label{eq:Xdot}
        \dot{P}&=\frac{3P^2\gamma}{2\uppi}\qty[\frac{e\sin\psi \mathcal{F}_r}{1+e\cos\psi}+\mathcal{F}_\theta],\qquad\dot{e}=\frac{\dot{P}\gamma^2}{3Pe}-\frac{P\gamma^5\mathcal{F}_\theta}{2\uppi e(1+e\cos\psi)^2},\\
        \dot{I}&=\frac{P\gamma^3\cos\theta \mathcal{F}_\ell}{2\uppi(1+e\cos\psi)^2},\qquad\dot{\asc}=\frac{\tan\theta}{\sin I}\dot{I},\\
        \dot{\omega}&=\frac{P\gamma^3}{2\uppi e}\qty[\frac{(2+e\cos\psi)\sin\psi \mathcal{F}_\theta}{(1+e\cos\psi)^2}-\frac{\cos\psi \mathcal{F}_r}{1+e\cos\psi}]-\cos I\dot{\asc},\\
        \dot{\eps}&=-\frac{P\gamma^4\mathcal{F}_r}{\uppi(1+e\cos\psi)^2}-\gamma(\cos I\dot{\asc}+\dot{\omega}),
    \end{split}
    \end{align}
    where we have defined the dimensionless angular momentum,
    \begin{equation}
        \gamma\equiv\sqrt{1-e^2}=\frac{\ell}{\sqrt{GMa}}.
    \end{equation}
Note that the size and shape of the orbit (determined by $P$ and $e$) is only affected by forces within the plane (i.e., $\mathcal{F}_r$ and $\mathcal{F}_\theta$), while the plane of the orbit (determined by $I$ and $\asc$) is only affected by forces normal to the plane (i.e., $\mathcal{F}_\ell$).
The radial and angular phases of the orbit (determined by $\omega$ and $\eps$) are affected by both.

\subsection{Secular averaging of the perturbations}

The EoMs~\eqref{eq:Xdot} for the osculating elements are somewhat cumbersome to solve, as they each oscillate over the course of an orbital period due to their dependence on the true anomaly $\psi(t)$.
In many situations however (including the present application to GW detection), the perturbing forces $\mathcal{F}_\alpha$ are small, and the resulting changes to the orbital elements therefore occur very slowly, on timescales much longer than the orbital period.
It is then possible to decouple the fast oscillations of the true anomaly from the much more gradual evolution of the orbital elements by calculating a \emph{secular average} of the EoMs,
    \begin{equation}
    \label{eq:secular-average-time}
        \ev*{\dot{X}}_\mathrm{sec}\equiv\int_{t_0}^{t_0+P}\frac{\dd{t}}{P}\dot{X},
    \end{equation}
    where $X\in(P,e,I,\asc,\omega,\eps)$.
The notation $\ev{\cdots}_\mathrm{sec}$ here is chosen to avoid confusion with plain angle brackets $\ev{\cdots}$ we use below to denote an ensemble average of a random variable.
This procedure integrates out the fast oscillations that occur on timescales $\le P$, leaving just the long-timescale dynamics of the system that we are interested in.
Note that we treat the orbital elements themselves as constant inside the integral, which only makes sense in the limit where they are slowly-varying.

It is usually easier to evaluate the secular average if we replace the time integral in equation~\eqref{eq:secular-average-time} with an integral over the true anomaly.
We can do this by noticing that
    \begin{equation}
        \dv{\psi}{t}=\ell/r^2=\frac{2\uppi}{P\gamma^3}(1+e\cos\psi)^2,
    \end{equation}
    where we have used equations~\eqref{eq:e-l-definitions}, \eqref{eq:Kepler-solutions}, and~\eqref{eq:kepler-iii}.
The secular average can therefore be written as
    \begin{equation}
    \label{eq:secular-average-psi}
        \ev*{\dot{X}}_\mathrm{sec}=\int_0^{2\uppi}\frac{\dd{\psi}}{2\uppi}\frac{\gamma^3\dot{X}}{(1+e\cos\psi)^2}.
    \end{equation}
In order to apply this average to the EoMs~\eqref{eq:Xdot}, we need to know the form of the perturbing forces $\mathcal{F}_\alpha$, which in general will vary over the course of the orbit.

\subsection{Secular evolution due to relativistic effects}
\label{sec:relativistic-drift}

In section~\ref{sec:gw-resonance}, we use equation~\eqref{eq:Xdot} to calculate the perturbations to the osculating orbital elements caused by resonance with the GWB.
However, we can use the same set of equations to calculate the perturbations caused by relativistic corrections to the equations of motion, which are particularly important for binaries with short periods.
Following \citet{Poisson:2014gr}, we write the relativistic force components to leading post-Newtonian (PN) order as
    \begin{align}
    \begin{split}
    \label{eq:relativistic-drift}
        \mathcal{F}_r&=\qty(\frac{2\uppi}{P})^2\frac{v_P^2}{\gamma^8}(1+e\cos\psi)^3\qty[3-\eta-e^2(1+3\eta)+e(2-4\eta)\cos\psi+e^2\frac{8-\eta}{2}\sin^2\psi],\\
        \mathcal{F}_\theta&=\qty(\frac{2\uppi}{P})^2\frac{2ev_P^2}{\gamma^8}\sin\psi(1+e\cos\psi)^4(2-\eta),\qquad\mathcal{F}_\ell=0,
    \end{split}
    \end{align}
    where
    \begin{equation}
        v_P\equiv\qty(\frac{2\uppi GM}{P})^{1/3}=\sqrt{\frac{GM}{a}}
    \end{equation}
    is the binary's rms velocity, and we recall from section~\ref{sec:compact-binaries} that $\eta\equiv\mu/M$ is the dimensionless mass ratio.
Inserting these expressions into equation~\eqref{eq:secular-average-psi}, we find that the only non-vanishing secular perturbations to the osculating elements are
    \begin{equation}
    \label{eq:relativistic-drift-conservative}
        \ev*{\dot{\omega}}_\mathrm{sec}=\frac{6\uppi v_P^2}{P\gamma^2},\qquad\ev*{\dot{\eps}}_\mathrm{sec}=\frac{2\uppi v_P^2}{P}\qty(6-7\eta-\frac{15-9\eta}{\gamma}),
    \end{equation}
    where the first line is the famous \emph{perihelion precession}, which provided some of the earliest observational evidence for GR.
These are both 1PN corrections (i.e., order $v_P^2$).

Note that the above implies that the binary's period and eccentricity are conserved at 1PN order.
However, since these are typically the most precisely-measured orbital elements, it is worth including their leading-order relativistic evolution, even if this is at higher PN order and is thus much smaller than the terms in equation~\eqref{eq:relativistic-drift-conservative} for most binaries.
As we saw in section~\ref{sec:compact-binaries}, this leading-order evolution of the period and eccentricity is due to GW radiation reaction, and can be calculated by applying simple energy-balance arguments to the unperturbed Keplerian orbit~\cite{Peters:1963ux}, giving
    \begin{equation}
    \label{eq:relativistic-drift-radiative}
        \ev*{\dot{P}}_\mathrm{sec}=-\frac{192\uppi\eta v_P^5}{5\gamma^7}\qty(1+\tfrac{73}{24}e^2+\tfrac{37}{96}e^4),\qquad\ev*{\dot{e}}_\mathrm{sec}=-\frac{608\uppi\eta v_P^5}{15P\gamma^5}\qty(e+\tfrac{121}{304}e^3).
    \end{equation}
These are just the period and eccentricity decay rates we found in equations~\eqref{eq:period-decay-radiation-reaction} and~\eqref{eq:eccentricity-decay-radiation-reaction}, albeit with slightly different notation here.
We see that both are 2.5PN effects (i.e., order $v_P^5$).
The inclination and longitude of ascending node are both conserved at this PN order.

\subsection{Small-eccentricity and small-inclination orbits}
\label{sec:small-e-I}

For binaries where the eccentricity is small (say, $e\lesssim10^{-3}$) the argument of pericentre $\omega$ becomes poorly-defined, as it becomes increasingly difficult to distinguish the point of closest approach as the orbit becomes increasingly circular.
This in turn means that the compensated mean anomaly $\eps$ becomes poorly-defined, as this is defined relative to the time at which the binary is at pericentre, $t_0$.
These issues can be resolved by defining an alternative set of orbital elements,
    \begin{equation}
    \label{eq:small-e-I}
        \zeta\equiv e\sin\omega,\qquad\kappa\equiv e\cos\omega,\qquad\xi\equiv\omega+\eps.
    \end{equation}
The first two quantities here are sometimes called the \enquote{Laplace-Lagrange eccentric parameters}, while the latter is the compensated mean argument.
We can then describe the orbit of a near-circular binary in terms of the alternative set of osculating elements $(P,\zeta,\kappa,I,\asc,\xi)$~\cite{Murray:2000ssd,Lange:2001rn}.
These evolve according to
    \begin{equation}
        \dot{\zeta}=\dot{e}\sin\omega+\dot{\omega}e\cos\omega,\qquad\dot{\kappa}=\dot{e}\cos\omega-\dot{\omega}e\sin\omega,\qquad\dot{\xi}=\dot{\omega}+\dot{\eps},
    \end{equation}
    with $\dot{e}$, $\dot{\asc}$, $\dot{\omega}$, and $\dot{\eps}$ given by equation~\eqref{eq:Xdot}.
For the relativistic perturbations~\eqref{eq:relativistic-drift}, we thus have
    \begin{equation}
        \ev*{\dot{\zeta}}_\mathrm{sec}=\frac{6\uppi v_P^2}{P}\qty(\kappa-\tfrac{304}{45}\eta v_P^3\zeta),\qquad\ev*{\dot{\kappa}}_\mathrm{sec}=-\frac{6\uppi v_P^2}{P}\qty(\zeta+\tfrac{304}{45}\eta v_P^3\kappa),\qquad\ev*{\dot{\xi}}_\mathrm{sec}=-\frac{4\uppi v_P^2}{P}(3-\eta),
    \end{equation}
    where we have neglected $\order*{e^2}$ terms.
Similarly, $\asc$ is ill-defined for orbits with very small inclination, so in this case we define
    \begin{equation}
        p\equiv I\sin\asc,\qquad q\equiv I\cos\asc,\qquad\lambda\equiv\asc+\xi,
    \end{equation}
    and describe the orbit using $(P,\zeta,\kappa,p,q,\lambda)$.

\section{Resonant gravitational-wave perturbations}
\label{sec:gw-resonance}

In this section we calculate the evolution of the osculating orbital elements of a binary system due to resonance with the GWB.
We start by specifying the perturbing force associated with an incoming plane GW in the limit where the wavelength is much larger than the size of the orbit (i.e., the small-antenna limit).
This allows us to write down a \emph{Langevin equation} describing individual random realisations of the stochastic evolution of the osculating elements.
Using the statistical properties of the GWB, we then derive a secularly-averaged \emph{Fokker-Planck equation} which describes the evolution of the full statistical distribution of the orbital elements over timescales much longer than the binary period.

\subsection{Coupling to the gravitational-wave polarisation modes}

Using the results of section~\ref{sec:gw-test-masses}---in particular, equation~\eqref{eq:test-mass-response}---we can express the response of a binary to an impinging plane GW in the proper detector frame as~\cite{Misner:1974qy,Maggiore:2007zz}
    \begin{equation}
    \label{eq:gw-force}
        \updelta\ddot{r}^i=\frac{1}{2}\ddot{h}^{ij}r_j,
    \end{equation}
    so that the resulting evolution of the binary is described in terms of the perturbing force terms
    \begin{equation}
        \mathcal{F}_r=\frac{1}{2}\ddot{h}_{ij}\hat{r}^i\hat{r}^j,\qquad\mathcal{F}_\theta=\frac{1}{2}\ddot{h}_{ij}\hat{r}^i\hat{\theta}^j,\qquad\mathcal{F}_\ell=\frac{1}{2}\ddot{h}_{ij}\hat{r}^i\hat{\ell}^j,
    \end{equation}
    where $h_{ij}(t)$ is the TT part of the metric perturbation at the position of the binary's centre of mass.
Carrying out a plane-wave decomposition as discussed in section~\ref{sec:polarisation-modes},
    \begin{equation}
        h_{ij}(t)=\int_{S^2}\dd[2]{\vu*n}e_{ij}^A(\vu*n)h_A(t,\vu*n),
    \end{equation}
    these force terms can thus be written as
    \begin{equation}
        \label{eq:force-ddot-h}
        \mathcal{F}_\alpha=\frac{1}{2}\int_{S^2}\dd[2]{\vu*n}e_{ij}^A\hat{r}^i\hat{\alpha}^j\ddot{h}_A,
    \end{equation}
    where $\alpha$ runs over the cylindrical coordinates $(r,\theta,\ell)$.

As we discussed in section~\ref{sec:gw-test-masses}, equation~\eqref{eq:gw-force} is only correct in the limit where the GW wavelength $\lambda$ is much larger than the distance between the two objects in question, which in this case is set by the semi-major axis of the binary's orbit, $a\ll\lambda$.
Using Kepler's third law~\eqref{eq:kepler-iii}, this condition can be rewritten as $fP\ll1/v_P$, where $f$ is the GW frequency.
Since we are interested in GW frequencies which are harmonics of the binary period ($f=n/P$ for some $n\in\mathbb{Z}_+$), this tells us that the analysis below, based on equation~\eqref{eq:gw-force}, is only valid for harmonics that satisfy $n\ll 1/v_P$.
This is not an impediment, since in the cases we are interested in the binary is sub-relativistic, $v_P\ll1$, and as we will see in later sections, the strongest GW contribution to the binary's evolution typically comes from the lowest few harmonics anyway.

\subsection{Langevin formulation}

The stochastic evolution of the binary is described by the evolution equations derived in section~\ref{sec:perturbing-force}, with the perturbing force terms given by equation~\eqref{eq:force-ddot-h}.
All this can be rewritten as a coupled set of nonlinear ordinary differential equations (ODEs),
    \begin{equation}
    \label{eq:langevin}
        \dot{X}_i(\vb*X,t)=V_i(\vb*X)+\Gamma_i(\vb*X,t),
    \end{equation}
    where $X_i$ runs over the set of orbital elements (either $(P,e,I,\asc,\omega,\eps)$, or the small-eccentricity and/or small-inclination alternatives described in section~\ref{sec:small-e-I}), $V$ is the deterministic drift term due to other perturbations (e.g., the relativistic effects described in section~\ref{sec:relativistic-drift}), and $\Gamma$ is the stochastic diffusion term due to resonance with the GWB.
ODEs of this form, in which one of the source terms is a random variable rather than a deterministic function, are called \emph{Langevin equations}~\cite{Risken:1989fpe}.
Rather than possessing a unique, deterministic solution for a given set of initial conditions, Langevin equations are characterised by a large number of distinct solutions, corresponding to different random realisations of the stochastic source term $\Gamma$.

In our case, $\Gamma$ can be written as
    \begin{equation}
    \label{eq:Gamma-def}
        \Gamma_i(\vb*X,t)=\int_{S^2}\dd[2]{\vu*n}T_i^A(X,t,\vu*n)\ddot{h}_A(t,\vu*n),
    \end{equation}
    where the $T_i^A$ are a set of transfer functions describing the coupling between the GWB strain and the orbital elements.
For example, using equations~\eqref{eq:Xdot} and~\eqref{eq:force-ddot-h}, the transfer functions for the period $P$ are
    \begin{equation}
        T_{P}^A=\frac{3P^2\gamma}{4\uppi}\qty(\frac{e\sin\psi}{1+e\cos\psi}\hat{r}^i+\hat{\theta}^i)\hat{r}^je^A_{ij}.
    \end{equation}
However, the exact form of the transfer functions is unimportant for this section---all that matters is that we can write the stochastic term in the form~\eqref{eq:Gamma-def}.
We write out the full set of transfer functions (in terms of their Fourier components) in appendix~\ref{app:kramers-moyal-derivation}.

The transfer functions are explicitly time-dependent via their dependence on trigonometric functions of the true anomaly $\psi(t)$, due to the variation of the binary's response over the course of each orbit.
If we neglect the secular evolution of the orbital elements on long timescales $X/\dot{X}\gg P$, this means that the transfer functions are \emph{periodic} functions of time, with period equal to the binary's orbital period $P$.
This quasi-periodicity allows us to approximate each of the transfer functions as a Fourier series with period $P$,
    \begin{equation}
    \label{eq:transfer-function-sum}
        T^A_i(\vb*X,t,\vu*n)=\sum_{n=-\infty}^{+\infty}\rme^{-2\uppi\rmi nt/P}T^A_{i,n}(\vb*X,\vu*n).
    \end{equation}
We therefore see that the binary's response to an incoming GW is characterised by its harmonic frequencies $f=n/P$.
Over long timescales, these frequencies will change due to the secular evolution of the binary period $P$.
The Fourier modes $T^A_{i,n}$ will also vary on these timescales, as they are functions of the orbital elements.

\subsection{From the Langevin equation to the Fokker-Planck equation}
\label{sec:fokker-planck}

Solving the Langevin equation~\eqref{eq:langevin} gives individual random trajectories of the binary through parameter space, corresponding to different random realisations of the GWB.
However, we are more interested in the ensemble of all possible random trajectories.
We can describe this ensemble in terms of the time-dependent distribution function (DF) for the orbital elements, $W(\vb*X,t)$, which is defined such that the probability of the orbital elements $\vb*X$ belonging to any region $\mathcal{X}$ of parameter space at time $t$ is given by the corresponding integral over the DF,
    \begin{equation}
        \Pr(\vb*X\in\mathcal{X}|t)=\int_{\mathcal{X}}\dd{\vb*X}W(\vb*X,t).
    \end{equation}
This DF can either be interpreted as the probability density function for the stochastic orbital elements of an individual binary, or as the cumulative distribution for the orbital elements of a population of multiple binaries.

\begin{figure}[t!]
    \begin{center}
        \includegraphics[width=0.667\textwidth]{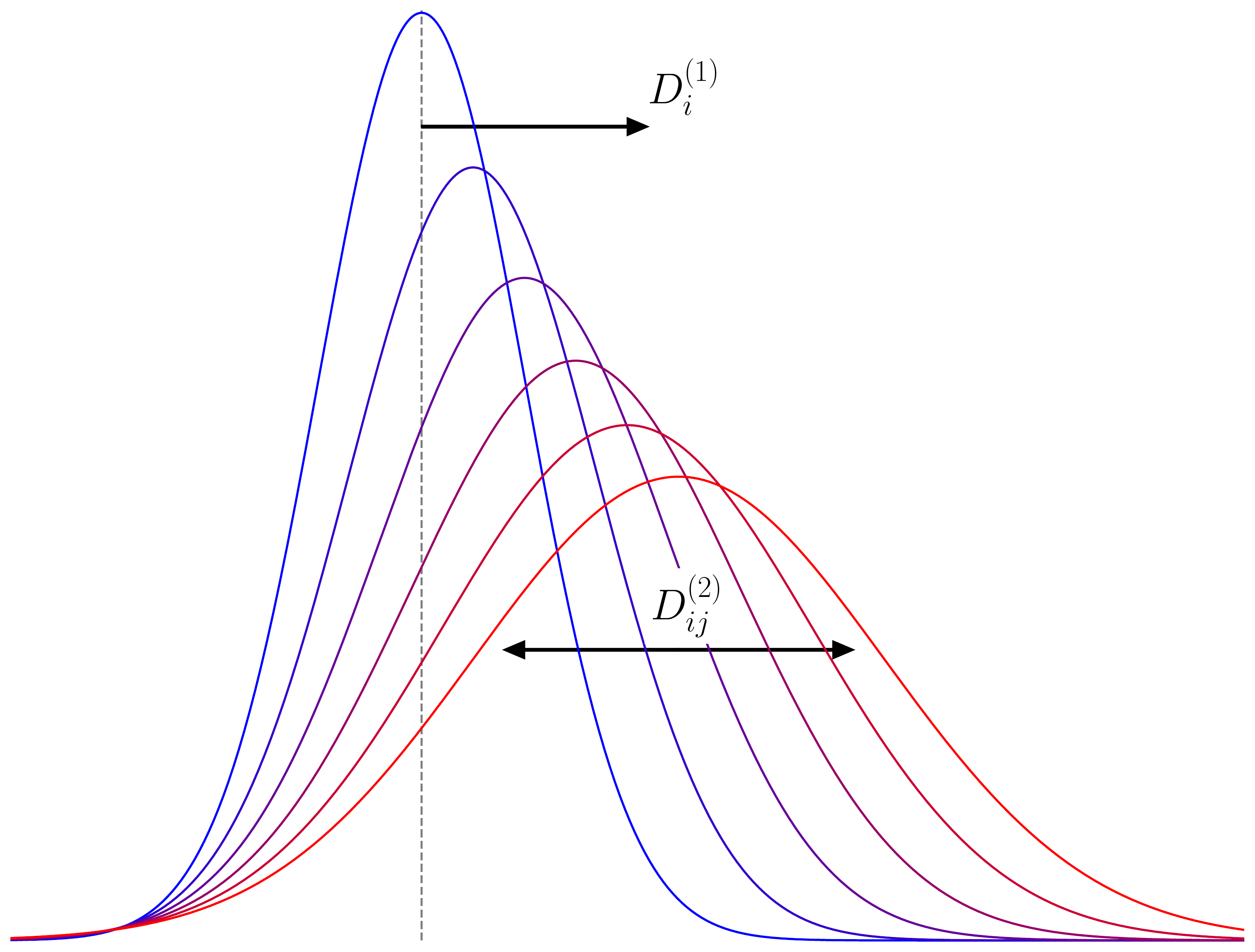}
    \end{center}
    \caption{%
    Heuristic illustration of a distribution function evolving according to the Fokker-Planck equation.
    The drift vector $D^{(1)}_i$ drives the bulk movement of the distribution in one direction, while the diffusion matrix $D^{(2)}_{ij}$ causes the distribution to become broader, as the randomness of the stochastic perturbation propagates into an increasing randomness in the distribution.
    }
    \label{fig:drift-diffusion}
\end{figure}

Formally, we can write down the time evolution of the DF in terms of the Kramers-Moyal (KM) forward expansion~\cite{Risken:1989fpe,Gardiner:2004hsm},
    \begin{equation}
    \label{eq:KM-forward}
        \pdv{W}{t}=\sum_{n=1}^\infty(-)^n\frac{\partial^n}{\partial{X_{i_1}}\partial{X_{i_2}}\cdots\partial{X_{i_n}}}\qty(D^{(n)}_{i_1i_2\cdots i_n}W),
    \end{equation}
    where repeated indices are summed over.
This is determined by the KM coefficients,
    \begin{equation}
        D^{(n)}_{i_1i_2\cdots i_n}(\vb*X,t)\equiv\lim_{\tau\to0}\frac{1}{\tau n!}\ev{\prod_{j=1}^n\qty[X_{i_j}(t+\tau)-X_{i_j}(t)]},
    \end{equation}
    with angle brackets indicating an ensemble average under the distribution $W$ at time $t$.
Since we take the GWB as Gaussian, the KM coefficients for orders $n\ge3$ all vanish,\footnote{%
One could derive this explicitly by using the approach described later in this section to show that $D^{(3)}_{ijk}=0$, as the Pawula theorem~\cite{Pawula:1967zz,Risken:1989fpe} then implies that all higher-order coefficients $n>3$ must vanish in order to guarantee that the DF is normalised.}
leaving just the first two coefficients,
    \begin{align}
    \begin{split}
        D^{(1)}_i&=\lim_{\tau\to0}\frac{1}{\tau}\ev{X_i(t+\tau)-X_i(t)},\\
        D^{(2)}_{ij}&=\lim_{\tau\to0}\frac{1}{2\tau}\ev{[X_i(t+\tau)-X_i(t)][X_j(t+\tau)-X_j(t)]},
    \end{split}
    \end{align}
    which we call the \emph{drift vector} and the \emph{diffusion matrix}, respectively.
The KM forward expansion~\eqref{eq:KM-forward} then becomes the \emph{Fokker-Planck equation} (FPE),
    \begin{equation}
    \label{eq:fpe}
        \pdv{W}{t}=-\partial_i(D^{(1)}_iW)+\partial_i\partial_j(D^{(2)}_{ij}W),
    \end{equation}
    with $\partial_i\equiv\pdv*{}{X_i}$.
The effect of each set of coefficients is illustrated in figure~\ref{fig:drift-diffusion}.

We can calculate the KM coefficients by directly integrating the Langevin equation~\eqref{eq:langevin}, from some initial time $t$ where the orbital elements are \enquote{sharp} (i.e., known exactly rather than randomly distributed), $X_i(t)\equiv x_i$, over some small time interval $\tau$,
    \begin{equation}
    \label{eq:direct-integration}
        X_i(t+\tau)-x_i=\int^{t+\tau}_t\dd{t'}[V_i(\vb*X(t'))+\Gamma_i(\vb*X(t'),t')].
    \end{equation}
(This derivation closely follows that in section~3.3.2 of \citet{Risken:1989fpe}, but is generalised to allow the noise term $\Gamma$ to depend on the system's parameters $X_i$, rather than just being a function of time.)
Both terms under the integral on the right-hand side are random, due to the random spread in the orbital elements for all times $t'>t$.
However, we can express these in terms of the sharp values $x_i$ by Taylor expanding,
    \begin{align}
    \begin{split}
    \label{eq:taylor-expansion}
        V_i(\vb*X(t'))&=V_i(\vb*x)+\partial_jV_i(\vb*x)[X_j(t')-x_j]+\cdots,\\
        \Gamma_i(\vb*X(t'),t')&=\Gamma_i(\vb*x,t')+\partial_j\Gamma_i(\vb*x,t')[X_j(t')-x_j]+\cdots.
    \end{split}
    \end{align}
Wherever $X_i(t')-x_i$ appears on the right-hand side of equation~\eqref{eq:taylor-expansion} we can insert equation~\eqref{eq:direct-integration} and iterate, such that the only orbital elements that appear are the sharp initial values $x_i$.
Then the only randomness is through the stochastic term $\Gamma_i$, whose statistics are completely specified in terms of the GWB moments~\eqref{eq:ddot-h-moments}.

The first few terms in the iterative expansion are
    \begin{align}
    \begin{split}
        X_i(t+\tau)-x_i&=\int_t^{t+\tau}\dd{t'}V_i(\vb*x)+\int_t^{t+\tau}\dd{t'}\partial_jV_i(\vb*x)\int_t^{t'}\dd{t''}[V_j(\vb*x)+\Gamma_j(\vb*x,t'')]+\cdots\\
        &+\int_t^{t+\tau}\dd{t'}\Gamma_i(\vb*x,t')+\int_t^{t+\tau}\dd{t'}\partial_j\Gamma_i(\vb*x,t')\int_t^{t'}\dd{t''}[V_j(\vb*x)+\Gamma_j(\vb*x,t'')]+\cdots.
    \end{split}
    \end{align}
By virtue of the time-independence of the sharp drift term $V_i(\vb*x)$, this immediately simplifies to
    \begin{align}
    \begin{split}
    \label{eq:simplified-expansion}
        X_i(t+\tau)-x_i=\tau V_i(\vb*x)&+\frac{1}{2}\tau^2V_j(\vb*x)\partial_jV_i(\vb*x)+\partial_jV_i(\vb*x)\int_t^{t+\tau}\dd{t'}\int_t^{t'}\dd{t''}\Gamma_j(\vb*x,t'')+\cdots\\
        &+\int_t^{t+\tau}\dd{t'}\Gamma_i(\vb*x,t')+V_j(x)\int_t^{t+\tau}\dd{t'}(t'-t)\partial_j\Gamma_i(\vb*x,t')\\
        &+\int_t^{t+\tau}\dd{t'}\partial_j\Gamma_i(\vb*x,t')\int_t^{t'}\dd{t''}\Gamma_j(\vb*x,t'')+\cdots.
    \end{split}
    \end{align}
Taking the first moment of equation~\eqref{eq:simplified-expansion}, all terms linear in $\Gamma_i$ vanish, and we are left with
    \begin{align}
    \begin{split}
        \ev{X_i(t+\tau)-x_i}&=\tau V_i(\vb*x)+\frac{1}{2}\tau^2V_j(\vb*x)\partial_jV_i(\vb*x)+\cdots\\
        &+\int_t^{t+\tau}\dd{t'}\int_t^{t'}\dd{t''}\ev{\Gamma_j(\vb*x,t'')\partial_j\Gamma_i(\vb*x,t')}+\cdots,
    \end{split}
    \end{align}
so that the drift vector is given by
    \begin{equation}
    \label{eq:KM1}
        D^{(1)}_i=V_i+\lim_{\tau\to0}\frac{1}{\tau}\int_t^{t+\tau}\dd{t'}\int_t^{t'}\dd{t''}\ev{\Gamma_j(\vb*x,t'')\partial_j\Gamma_i(\vb*x,t')}.
    \end{equation}
Similarly, when taking the second moment of equation~\eqref{eq:simplified-expansion} and taking the $\tau\to0$ limit, the only term that survives for the diffusion matrix is
    \begin{equation}
    \label{eq:KM2}
        D^{(2)}_{ij}=\lim_{\tau\to0}\frac{1}{2\tau}\int_t^{t+\tau}\dd{t'}\int_t^{t+\tau}\dd{t''}\ev{\Gamma_i(\vb*x,t')\Gamma_j(\vb*x,t'')}.
    \end{equation}
The fact that both sets of KM coefficients are quadratic in $\Gamma$ makes sense, since this is the smallest power of $\Gamma$ which does not vanish when taking an expectation value.
The difference, however, is that the drift vector has a derivative acting on one factor of $\Gamma$, while the diffusion matrix has no derivatives.
This is because the quadratic-in-$\Gamma$ term appearing in the drift vector corresponds to the second-order term in the Taylor expansion we wrote down in equation~\eqref{eq:taylor-expansion}, while in the diffusion matrix this is simply the product of two first-order Taylor expansion terms.

\subsection{Secular drift and diffusion}

By calculating the KM coefficients from equations~\eqref{eq:KM1} and~\eqref{eq:KM2}, we can now (in principle) obtain the full DF for the orbital elements by integrating the FPE~\eqref{eq:fpe}.
However, in order to calculate the KM coefficients, we must first evaluate the ensemble averages $\ev{\Gamma^2}$ and $\ev{\Gamma\partial\Gamma}$ that appear in the integrands of equations~\eqref{eq:KM1} and~\eqref{eq:KM2} above.
We do this by rewriting these as ensemble averages over the GWB strain.
Using equations~\eqref{eq:Gamma-def} and~\eqref{eq:transfer-function-sum}, we find
    \begin{align}
    \begin{split}
    \label{eq:gamma-2pt-hddot}
        \ev{\Gamma_i(\vb*x,t')\Gamma_j(\vb*x,t'')}&=\sum_{n=-\infty}^{+\infty}\sum_{m=-\infty}^{+\infty}\int_{S^2}\dd[2]{\vu*n}\int_{S^2}\dd[2]{\vu*n'}\rme^{2\uppi\rmi(nt'-mt'')/P}\\
        &\qquad\qquad\qquad\qquad\times T^{A*}_{i,n}(\vb*x,\vu*n)T^{A'}_{j,m}(\vb*x,\vu*n')\ev{\ddot{h}_A(t',\vu*n)\ddot{h}_{A'}(t'',\vu*n')},\\
        \ev{\Gamma_j(\vb*x,t'')\partial_j\Gamma_i(\vb*x,t')}&=\sum_{n=-\infty}^{+\infty}\sum_{m=-\infty}^{+\infty}\int_{S^2}\dd[2]{\vu*n}\int_{S^2}\dd[2]{\vu*n'}\rme^{2\uppi\rmi(nt'-mt'')/P}\\
        &\qquad\qquad\qquad\qquad\times T^{A'}_{j,m}(\vb*x,\vu*n')\partial_jT^{A*}_{i,n}(\vb*x,\vu*n)\ev{\ddot{h}_A(t',\vu*n)\ddot{h}_{A'}(t'',\vu*n')},
    \end{split}
    \end{align}
    so that both terms are determined by the 2-point statistics of the strain.

In section~\ref{sec:gw-density-parameter}, we derived the second moment of the Fourier transform of the strain for a GWB that is Gaussian, stationary, unpolarised, and isotropic, with no nontrivial phase correlations,
    \begin{equation}
        \ev{\tilde{h}_A(f,\vu*n)\tilde{h}^*_{A'}(f',\vu*n')}=\frac{3}{4}H_0^2(2\uppi|f|)^{-3}\Omega(f)\delta_{AA'}\delta(f-f')\delta^{(2)}(\vu*n,\vu*n').
    \end{equation}
Here we are interested in the statistics of the second time derivative of the strain, which is related to the Fourier components by
    \begin{equation}
        \ddot{h}_A=\dv[2]{}{t}\int_{-\infty}^{+\infty}\dd{f}\rme^{-2\uppi\rmi ft}\tilde{h}_A=-\int_{-\infty}^{+\infty}\dd{f}\rme^{-2\uppi\rmi ft}(2\uppi f)^2\tilde{h}_A,
    \end{equation}
    such that
    \begin{equation}
        \label{eq:ddot-h-moments}
        \ev*{\ddot{h}_A(t,\vu*n)\ddot{h}_{A'}(t',\vu*n')}=3\uppi H_0^2\delta_{AA'}\delta(\vu*n,\vu*n')\int_0^\infty\dd{f}\cos[2\uppi f(t-t')]f\Omega(f).
    \end{equation}
Inserting this into equation~\eqref{eq:gamma-2pt-hddot}, we obtain
    \begin{align}
    \begin{split}
    \label{eq:gamma-2nd-moments}
        \ev{\Gamma_i(\vb*x,t')\Gamma_j(\vb*x,t'')}&=3\uppi H_0^2\sum_{n=-\infty}^{+\infty}\sum_{m=-\infty}^{+\infty}\int_{S^2}\dd[2]{\vu*n}T^{A*}_{i,n}T^A_{j,m}\\
        &\qquad\qquad\qquad\qquad\times\int_0^\infty\dd{f}\rme^{2\uppi\rmi(nt'-mt'')/P}\cos[2\uppi f(t'-t'')]f\Omega(f),\\
        \ev{\Gamma_j(\vb*x,t'')\partial_j\Gamma_i(\vb*x,t')}&=3\uppi H_0^2\sum_{n=-\infty}^{+\infty}\sum_{m=-\infty}^{+\infty}\int_{S^2}\dd[2]{\vu*n}T^A_{j,m}\partial_jT^{A*}_{i,n}\\
        &\qquad\qquad\qquad\qquad\times\int_0^\infty\dd{f}\rme^{2\uppi\rmi(nt'-mt'')/P}\cos[2\uppi f(t'-t'')]f\Omega(f).
    \end{split}
    \end{align}

In order to derive the corresponding KM coefficients using equations~\eqref{eq:KM1} and~\eqref{eq:KM2}, we see that we must evaluate two oscillatory time integrals,
    \begin{align}
    \begin{split}
    \label{eq:time-integrals}
        \int_t^{t+\tau}\dd{t'}\int_t^{t+\tau}\dd{t''}&\rme^{-2\uppi\rmi(f-n/P)t'}\rme^{2\uppi\rmi(f-m/P)t''},\\
        \int_t^{t+\tau}\dd{t'}\int_t^{t'}\dd{t''}&\rme^{-2\uppi\rmi(f-n/P)t'}\rme^{2\uppi\rmi(f-m/P)t''},
    \end{split}
    \end{align}
    where we have converted the cosine in equation~\eqref{eq:gamma-2nd-moments} into an exponential.
To do so, we recall that the timescales $\le P$ associated with the binary's resonant frequencies are much shorter than the secular timescales $X/\dot{X}$ we are interested in.\footnote{%
    It would also be interesting to study the dynamics on sub-orbital timescales, as was done by \citet{Rozner:2019gba} and \citet{Desjacques:2020fdi} for the case of a perturbing ultralight dark matter field.
    We leave this for future work.}
Even though we will later take the limit $\tau\to0$, all we really require to derive a FPE valid on secular timescales $\sim X/\dot{X}$ is that $\tau\ll X/\dot{X}$, and since we have $P\ll X/\dot{X}$, we can consistently also demand that $\tau\gg P$.
In this limit, the first integral in equation~\eqref{eq:time-integrals} is approximated by
    \begin{equation}
        \int_t^{t+\tau}\dd{t'}\int_t^{t+\tau}\dd{t''}\rme^{-2\uppi\rmi(f-n/P)t'}\rme^{2\uppi\rmi(f-m/P)t''}\simeq\tau\delta_{mn}\delta(f-n/P).
    \end{equation}
For the second integral, notice that we can write
    \begin{align}
    \begin{split}
        \tau\delta_{mn}\delta(f-n/P)&\simeq\int_t^{t+\tau}\dd{t'}\int_t^{t+\tau}\dd{t''}\rme^{-2\uppi\rmi(f-n/P)t'}\rme^{2\uppi\rmi(f-m/P)t''}\\
        &=\int_t^{t+\tau}\dd{t'}\int_t^{t'}\dd{t''}\rme^{-2\uppi\rmi(f-n/P)t'}\rme^{2\uppi\rmi(f-m/P)t''}\\
        &\qquad+\int_t^{t+\tau}\dd{t'}\int_{t'}^{t+\tau}\dd{t''}\rme^{-2\uppi\rmi(f-n/P)t'}\rme^{2\uppi\rmi(f-m/P)t''},
    \end{split}
    \end{align}
and that in the limit $\tau\gg P$ the latter two terms are approximately equal to each other, so that
    \begin{equation}
        \int_t^{t+\tau}\dd{t'}\int_t^{t'}\dd{t''}\rme^{-2\uppi\rmi(f-n/P)t'}\rme^{2\uppi\rmi(f-m/P)t''}\simeq\frac{1}{2}\tau\delta_{mn}\delta(f-n/P).
    \end{equation}
Plugging this back in to equations~\eqref{eq:KM1},~\eqref{eq:KM2}, and~\eqref{eq:gamma-2nd-moments}, we are able to write the KM coefficients directly in terms of the GW transfer functions,
    \begin{align}
    \begin{split}
    \label{eq:km-final}
        D^{(1)}_i&=V_i+\frac{3\uppi}{2}H_0^2\sum_{n=1}^\infty\int_{S^2}\dd[2]{\vu*n}\frac{n\Omega_n}{P}\Re\qty(T^A_{j,n}\partial_jT^{A*}_{i,n}),\\
        D^{(2)}_{ij}&=\frac{3\uppi}{2}H_0^2\sum_{n=1}^\infty\int_{S^2}\dd[2]{\vu*n}\frac{n\Omega_n}{P}\Re\qty(T^{A*}_{i,n}T^A_{j,n}),
    \end{split}
    \end{align}
    where we have used $T_{i,-n}=T^*_{i,n}$, and where $\Omega_n\equiv\Omega(n/P)$ is the GWB energy density at the binary's $n^\mathrm{th}$ harmonic frequency.
(We always take only the real part when defining the KM coefficients, so for brevity we will leave this implicit from now on.)
Equation~\eqref{eq:km-final} describes the secular drift and diffusion of the orbital elements over timescales much longer than the orbital period $P$.
We neglect the $\ev{\cdots}_\mathrm{sec}$ notation used in section~\ref{sec:dynamics} for brevity, but the interpretation here is exactly the same.

Note that the drift vector in equation~\eqref{eq:km-final} includes not just the expected deterministic drift $V_i$, but also a stochastic term.
As we show below, this \enquote{noise-induced drift} leads to a net evolution of the \emph{mean values} of the orbital elements, and not just their variance.
This is somewhat counter-intuitive, and justifies the careful derivation presented in this section; otherwise, it would be tempting to assume that the GWB affects only the variance of the orbital elements (as was assumed in \citet{Hui:2012yp}, for example).
One can understand this drift as being the result of a \enquote{diffusion gradient} due to the nonlinear coupling between the GWB and the binary: the orbital elements change in response to the GW strain, thereby changing the values of the transfer functions and modifying the response to the strain, with the resulting feedback loop creating a preferred direction in parameter space for the orbital elements to evolve towards.
The partial derivative acting on the transfer function in equation~\eqref{eq:km-final} shows that this term would vanish in the linear case (i.e., where the transfer functions are constant), and therefore that this is a purely nonlinear effect.

We can also understand this stochastic drift as being a consequence of the GWB strain having a \emph{finite correlation time}; i.e., there is some $\tau_\rmc>0$ such that
    \begin{equation}
    \label{eq:correlation-time}
        \ev{\ddot{h}_{ij}(t)\ddot{h}_{ij}(t+\tau)}>0\quad\text{for all}\quad|\tau|<\tau_\rmc,
    \end{equation}
    (with no implied summation over the spatial indices).
Physically speaking, this is because we cannot have arbitrarily high GW frequencies contributing to the GWB spectrum $\Omega(f)$, so equation~\eqref{eq:ddot-h-moments} automatically satisfies equation~\eqref{eq:correlation-time}.
However, since this correlation time is much shorter than the secular timescales we are interested in, we can implicitly take $\tau_\rmc\to0$ at the end of the calculation.
This is the Stratonovich prescription~\cite{Stratonovich:1968mp} for calculating stochastic integrals, which differs from the It\^o prescription~\cite{Ito:1944si} in which the correlation time is assumed to be zero from the start.
Surprisingly, these two prescriptions give different physical predictions when the transfer functions depend on the state of the system, with the stochastic drift term we derived above being present only in the Stratonovich case and not in the It\^o case.
This can present a problem in systems with a white-noise stochastic term, $\ev{\Gamma(t)\Gamma(t')}=\delta(t-t')$, which formally has zero correlation time, as one must determine which of the two prescriptions is more appropriate.
However, we stress that for our case here equation~\eqref{eq:ddot-h-moments} shows that there must be a nonzero correlation time, so the Stratonovich prescription is the only sensible choice.
(We could get equation~\eqref{eq:ddot-h-moments} to be proportional to $\delta(t-t')$ by having $\Omega(f)\propto1/f$ at arbitrarily high frequencies, but this would cause the total GW energy density to diverge.)
The stochastic drift term appearing in equation~\eqref{eq:km-final} is therefore an unambiguous prediction for a binary coupled to the GWB.
For further discussion of the It\^o and Stratonovich prescriptions and the origin of noise-induced drift see, e.g., section~3.3.3 of \citet{Risken:1989fpe}.

\section{Calculating the Kramers-Moyal coefficients}
\label{sec:KM}

\begin{figure}[p!]
    \begin{center}
        \includegraphics[width=0.8\textwidth]{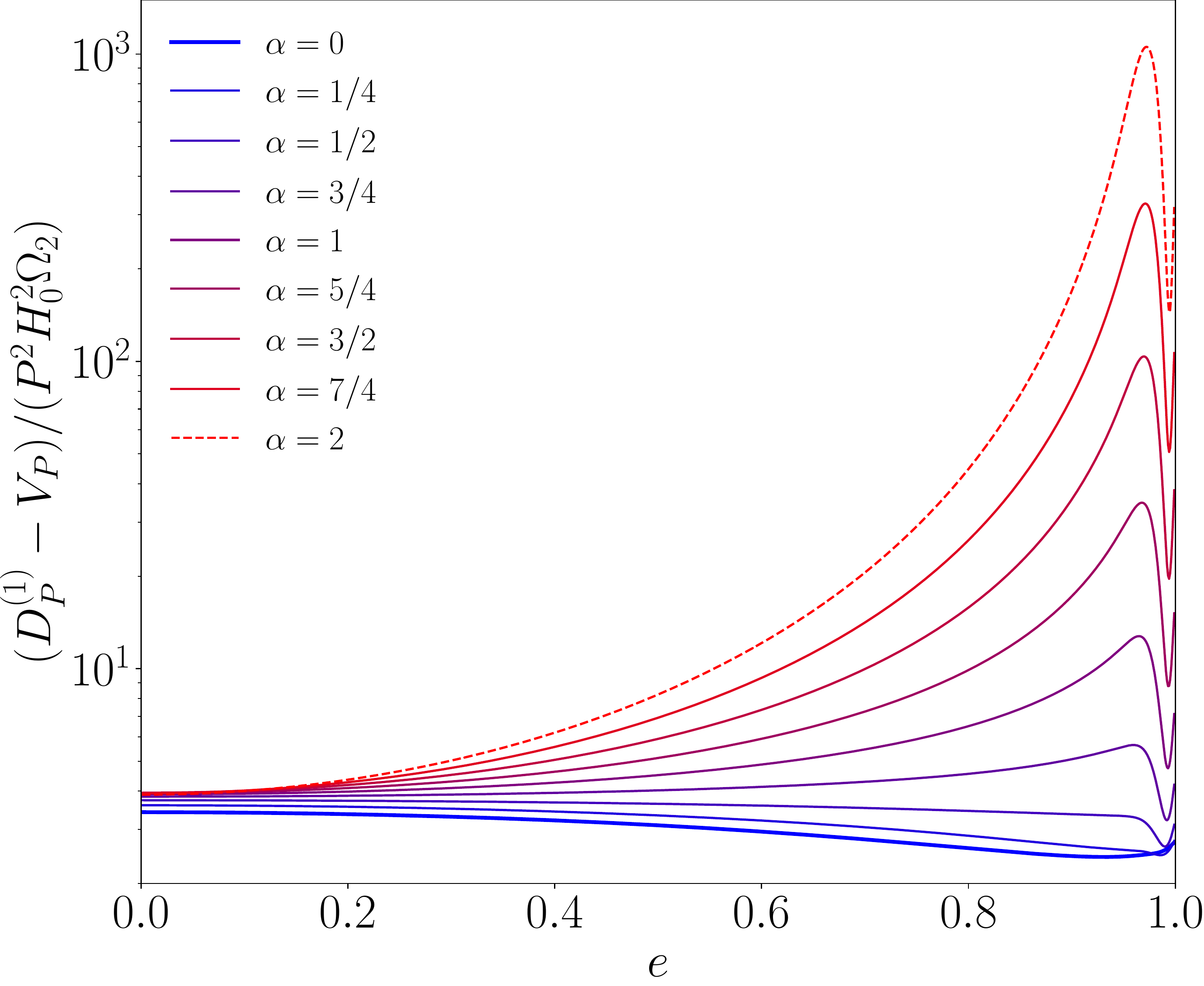}
        \includegraphics[width=0.8\textwidth]{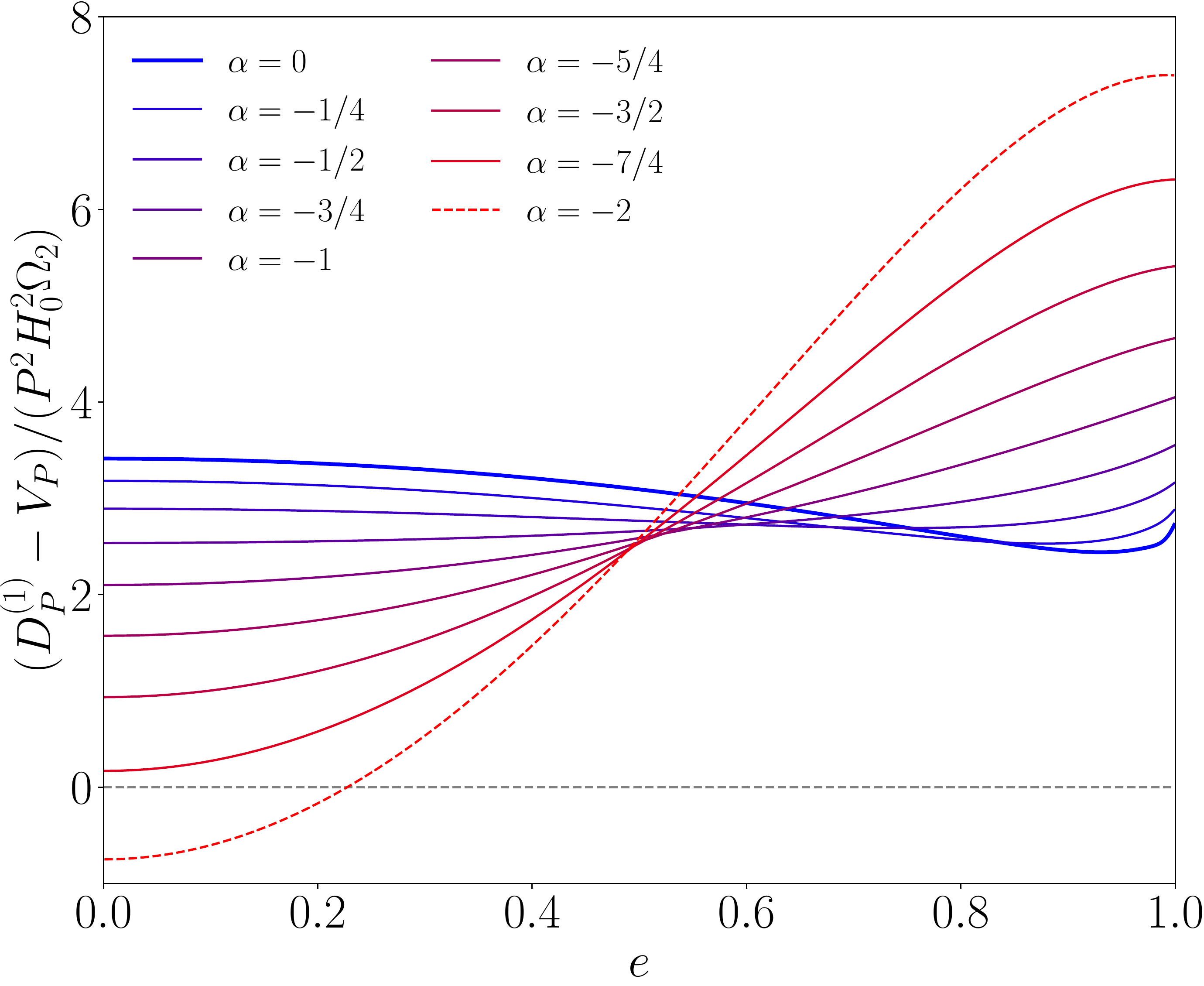}
    \end{center}
    \caption{%
    Stochastic parts of the secular period drift coefficient $D^{(1)}_P$ as a function of eccentricity for various power-law GWB spectra, $\Omega(f)\sim f^\alpha$.
    The top panel shows positive power-law indices, $\alpha=0,\dots,2$, while the bottom panel shows negative indices, $\alpha=0,\dots,-2$.}
    \label{fig:km1-P-ecc}
\end{figure}

In this section we use equation~\eqref{eq:km-final} to derive explicit expressions for the KM coefficients as functions of the binary orbital elements and the GWB spectrum.
We start with the most general expressions, before specialising to the cases of small eccentricity and small inclination.

\subsection{General orbits}

We present here the secular KM coefficients for general eccentricity $e\in(0,1)$.
The details of this calculation are lengthy and unimportant for the final results, so we quote only the final expressions here, with some further details given in appendix~\ref{app:kramers-moyal-derivation}.
In particular, section~\ref{sec:polarisation-tensor-projections} derives the necessary projections of the GW polarisation tensors onto the binary's cylindrical coordinate basis; section~\ref{sec:transfer} uses these projections to write down the transfer functions for all six orbital elements $(P,e,I,\asc,\omega,\eps)$ in terms of functions of the eccentricity called \emph{Hansen coefficients}, which we introduce below; section~\ref{sec:reference-frame} derives the equations describing how each of the KM coefficients transforms under a change of the reference frame, allowing us to select a particular frame which simplifies the calculations; section~\ref{sec:KM-primed-frame} presents the values of the KM coefficients in this frame in terms of the Hansen coefficients; and section~\ref{sec:hansen} writes out all of the necessary Hansen coefficients explicitly as functions of eccentricity.

Putting all of these ingredients together, we write the full set of KM coefficients explicitly in terms of the Hansen coefficients, which are defined as
    \begin{align}
    \begin{split}
    \label{eq:hansen-definition}
        C^{lm}_n(e)&\equiv\ev{\exp(\frac{2\uppi\rmi nt}{P})\frac{\cos m\psi}{(1+e\cos\psi)^l}}_\mathrm{sec},\\
        S^{lm}_n(e)&\equiv\ev{\exp(\frac{2\uppi\rmi nt}{P})\frac{\sin m\psi}{(1+e\cos\psi)^l}}_\mathrm{sec},\\
        E^{lm}_n(e)&\equiv C^{lm}_n(e)+S^{lm}_n(e).
    \end{split}
    \end{align}
The secular diffusion matrix $D^{(2)}_{ij}$ is thus given by
    \begin{align*}
    \begin{split}
        D^{(2)}_{PP}&=\frac{27P^3\gamma^2}{20}\sum_{n=1}^\infty nH_0^2\Omega_n\qty[\qty|E^{02}_n+\frac{e}{2}(E^{11}_n-E^{13}_n)|^2-\frac{(eS^{11}_n)^2}{3}],\\
        D^{(2)}_{Pe}&=\frac{\gamma^2D^{(2)}_{PP}}{3Pe}-\frac{9P^2\gamma^6}{40}\sum_{n=1}^\infty nH_0^2\Omega_nE^{22}_n\qty(\frac{2}{e}E^{02}_n+E^{11}_n-E^{13}_n)^*,\\
        D^{(2)}_{ee}&=\frac{3P\gamma^6}{20e^2}\sum_{n=1}^\infty nH_0^2\Omega_n\qty[\qty|E^{02}_n+\frac{e}{2}(E^{11}_n-E^{13}_n)-\gamma^2E^{22}_n|^2-\frac{(eS^{11}_n)^2}{3}],
    \end{split}
    \end{align*}
    \begin{align}
    \begin{split}
    \label{eq:ecc-diffusion}
        D^{(2)}_{II}&=\frac{3P\gamma^6}{80}\sum_{n=1}^\infty nH_0^2\Omega_n\qty[\qty|E^{20}_n|^2+\qty|E^{22}_n|^2+2\cos2\omega E^{20}_n(E^{22}_n)^*],\\
        D^{(2)}_{I\asc}&=\frac{3P\gamma^6}{40}\frac{\sin2\omega}{\sin I}\sum_{n=1}^\infty nH_0^2\Omega_nE^{20}_n(E^{22}_n)^*,\qquad D^{(2)}_{I\omega}=-\frac{3P\gamma^6}{40}\frac{\sin2\omega}{\tan I}\sum_{n=1}^\infty nH_0^2\Omega_nE^{20}_n(E^{22}_n)^*,\\
        D^{(2)}_{\asc\asc}&=\frac{3P\gamma^6}{80\sin^2I}\sum_{n=1}^\infty nH_0^2\Omega_n\qty[\qty|E^{20}_n|^2+\qty|E^{22}_n|^2-2\cos2\omega E^{20}_n(E^{22}_n)^*],\qquad D^{(2)}_{\asc\omega}=-\cos ID^{(2)}_{\asc\asc},\\
        D^{(2)}_{\omega\omega}&=\cos^2ID^{(2)}_{\asc\asc}\\
        &+\frac{3P\gamma^6}{80e^2}\sum_{n=1}^\infty nH_0^2\Omega_n\qty[\qty|E^{11}_n+E^{13}_n+2E^{21}_n-2E^{23}_n+\frac{e}{2}\qty(E^{20}_n-E^{24}_n)|^2+\frac{4}{3}\qty(C^{11}_n)^2],\\
        D^{(2)}_{\omega\eps}&=-\frac{3P\gamma^7}{80e^2}\sum_{n=1}^\infty nH_0^2\Omega_n\bigg[\qty|E^{11}_n+E^{13}_n+2(E^{21}_n-E^{23}_n)+\frac{e}{2}(E^{20}_n-4E^{22}_n-E^{24}_n)|^2\\
        &\qquad\qquad\qquad\qquad\qquad+\frac{4}{3}C^{11}_n(C^{11}_n-2eC^{20}_n)-4|eE^{22}_n|^2\bigg],\\
        D^{(2)}_{\eps\eps}&=\frac{3P\gamma^8}{80e^2}\sum_{n=1}^\infty nH_0^2\Omega_n\bigg[\qty|E^{11}_n+E^{13}_n+2(E^{21}_n-E^{23}_n)+\frac{e}{2}(E^{20}_n-8E^{22}_n-E^{24}_n)|^2\\
        &\qquad\qquad\qquad\qquad\qquad+\frac{4}{3}(C^{11}_n-2eC^{20}_n)^2\bigg],\\
        D^{(2)}_{PI}&=D^{(2)}_{P\omega}=D^{(2)}_{P\asc}=D^{(2)}_{P\eps}=D^{(2)}_{eI}=D^{(2)}_{e\omega}=D^{(2)}_{e\asc}=D^{(2)}_{e\eps}=D^{(2)}_{I\eps}=D^{(2)}_{\asc\eps}=0,
    \end{split}
    \end{align}
    while the secular drift vector $D^{(1)}_{i}$ is given by
    \begin{align}
    \begin{split}
    \label{eq:ecc-drift}
        D^{(1)}_P&=V_P+\frac{9P^2\gamma^2}{4}\sum_{n=1}^\infty nH_0^2\Omega_n\bigg[\qty|E^{02}_n+\frac{e}{2}(E^{11}_n-E^{13}_n)|^2-\frac{1+4e^2}{15}(S^{11}_n)^2-\frac{e\gamma^2}{15}{S^{11}_n}'S^{11}_n\\
        &\quad+\frac{\gamma^2}{10e}\qty(E^{02}_n+\frac{e}{2}(E^{11}_n-E^{13}_n))\\
        &\qquad\times\qty(3E^{11}_n+E^{13}_n+eE^{20}_n+4E^{21}_n+4eE^{22}_n-4E^{23}_n-eE^{24}_n+2{E^{02}_n}'+e({E^{11}_n}'-{E^{13}_n}'))^*\\
        &\quad-\frac{\gamma^4}{10e}E^{22}_n\qty(E^{11}_n-E^{13}_n+2{E^{02}_n}'+e({E^{11}_n}'-{E^{13}_n}'))^*\bigg],\\
        D^{(1)}_e&=V_e-\frac{P\gamma^6}{20}\sum_{n=1}^\infty nH_0^2\Omega_n\bigg[{S^{11}_n}'S^{11}_n+\frac{3}{e^3}\qty(E^{02}_n+\frac{e}{2}(E^{11}_n-E^{13}_n)-\gamma^2E^{22}_n)\\
        &\quad\times\bigg(E^{02}_n-E^{22}_n-e(E^{11}_n+E^{13}_n+2E^{21}_n-2E^{23}_n+{E^{02}_n}'-\gamma^2{E^{22}_n}')\\
        &\qquad\qquad-\frac{e^2}{2}(E^{20}_n+10E^{22}_n-E^{24}_n+{E^{11}_n}'-{E^{13}_n}')\bigg)^*\bigg],\\
        D^{(1)}_I&=\frac{\sin2I}{2}D^{(2)}_{\asc\asc},\qquad D^{(1)}_\asc=-2\cot ID^{(2)}_{I\asc},\\
        D^{(1)}_\omega&=V_\omega+\frac{3P\gamma^6}{40}\sin2\omega\frac{2-\sin^2I}{\sin^2I}\sum_{n=1}^\infty nH_0^2\Omega_nE^{20}_n(E^{22}_n)^*,\qquad D^{(1)}_\eps=V_\eps.
    \end{split}
    \end{align}
\begin{figure}[p!]
    \begin{center}
        \includegraphics[width=0.8\textwidth]{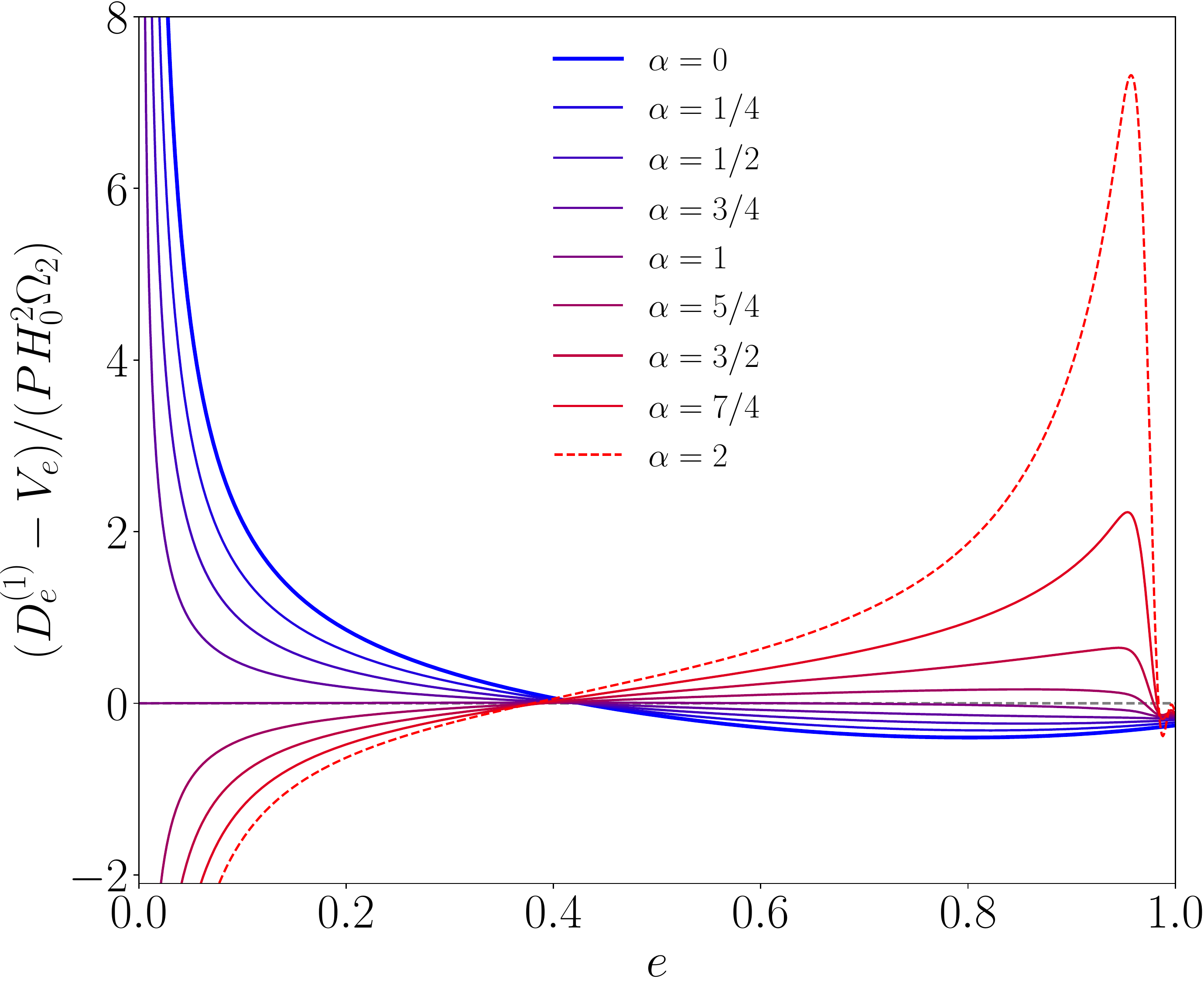}
        \includegraphics[width=0.8\textwidth]{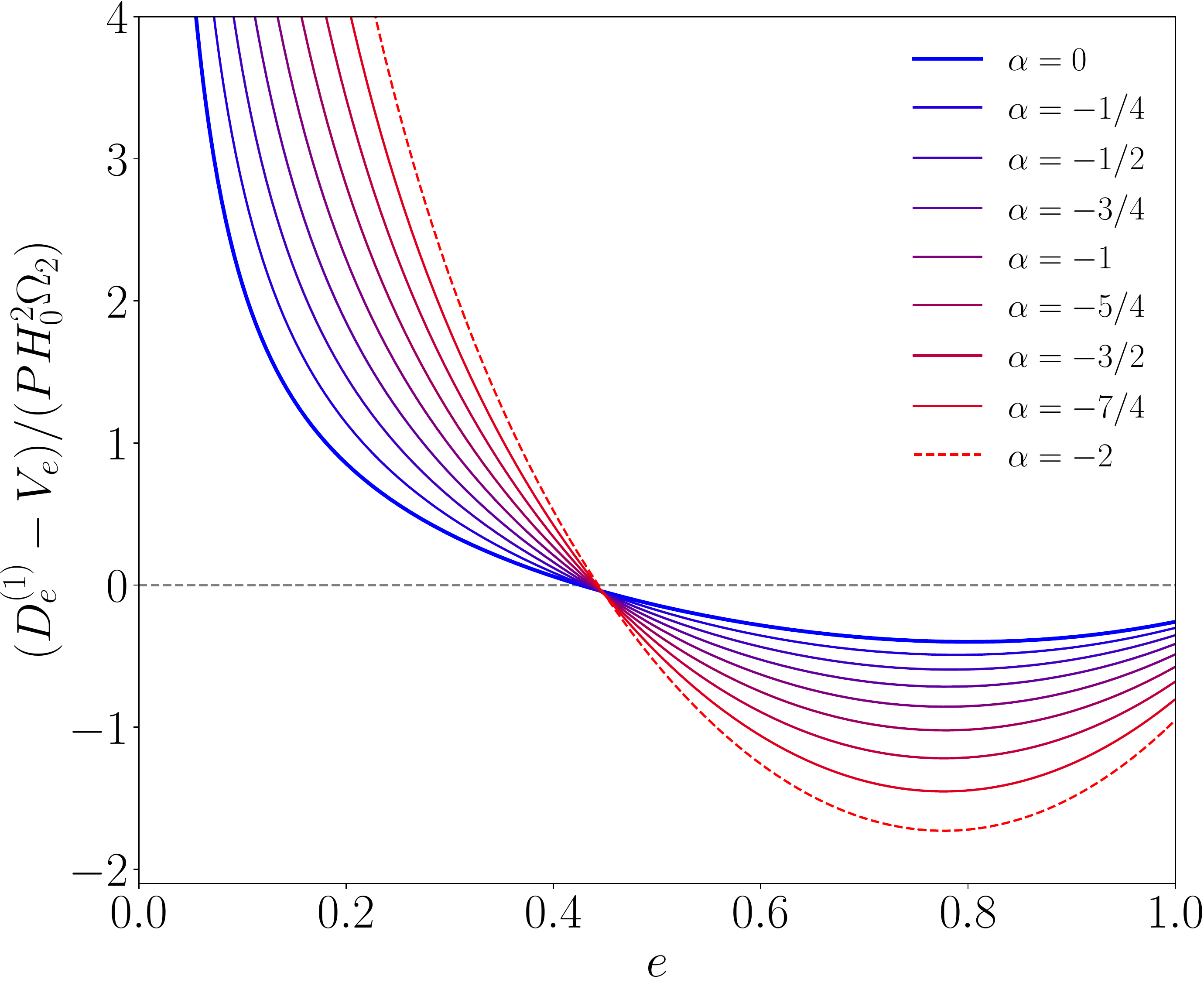}
    \end{center}
    \caption{%
    Same as figure~\ref{fig:km1-P-ecc}, but for the eccentricity drift coefficient $D^{(1)}_e$.}
    \label{fig:km1-e-ecc}
\end{figure}
The deterministic drift terms here are those given in section~\ref{sec:relativistic-drift},
    \begin{align}
    \begin{split}
        V_P&=-\frac{192\uppi\eta v_P^5}{5\gamma^7}\qty(1+\tfrac{73}{24}e^2+\tfrac{37}{96}e^4),\qquad V_e=-\frac{608\uppi\eta v_P^5}{15P\gamma^5}\qty(e+\tfrac{121}{304}e^3),\\
        V_\omega&=\frac{6\uppi v_P^2}{P\gamma^2},\qquad V_\eps=\frac{2\uppi v_P^2}{P}\qty(6-7\eta-\frac{15-9\eta}{\gamma}),
    \end{split}
    \end{align}
    and primes on the Hansen coefficients denote derivatives with respect to eccentricity.

\begin{figure}[p!]
    \begin{center}
        \includegraphics[width=0.8\textwidth]{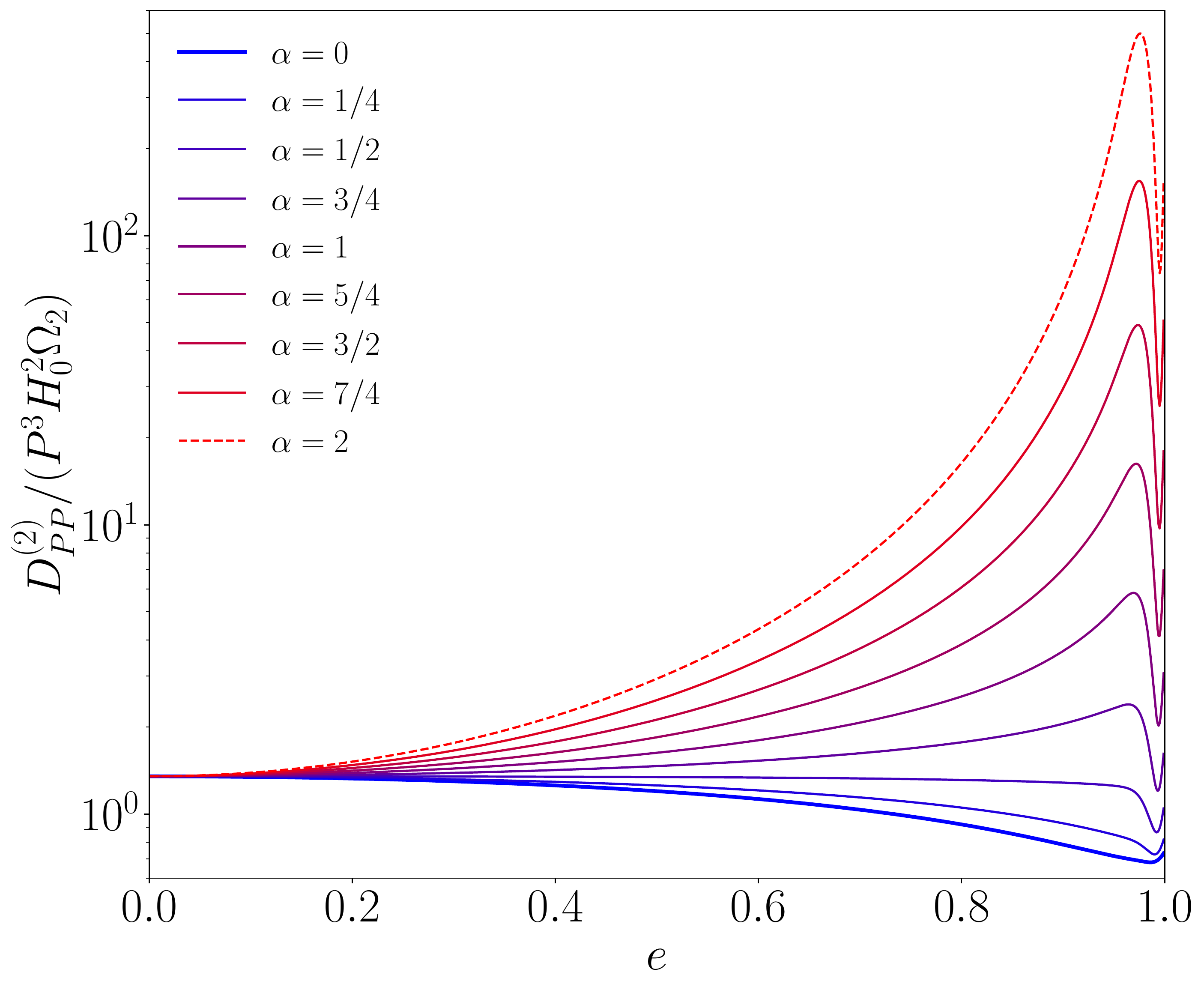}
        \includegraphics[width=0.8\textwidth]{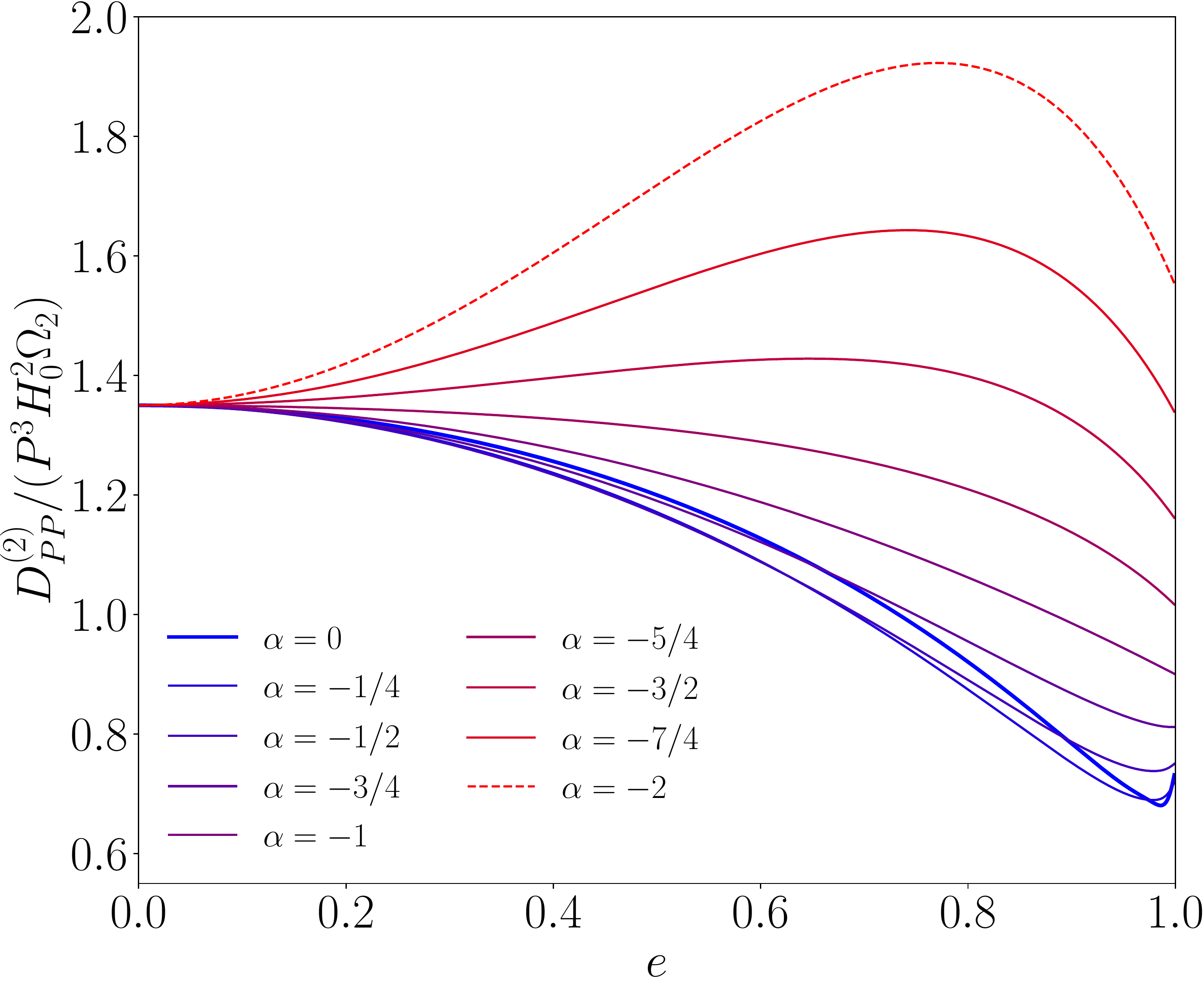}
    \end{center}
    \caption{%
    The secular period-period diffusion coefficients $D^{(2)}_{PP}$ as a function of eccentricity for various power-law GWB spectra, $\Omega(f)\sim f^\alpha$.
    The top panel shows positive power-law indices, $\alpha=0,\dots,2$, while the bottom panel shows negative indices, $\alpha=0,\dots,-2$.}
    \label{fig:km2-PP-ecc}
\end{figure}

\begin{figure}[p!]
    \begin{center}
        \includegraphics[width=0.8\textwidth]{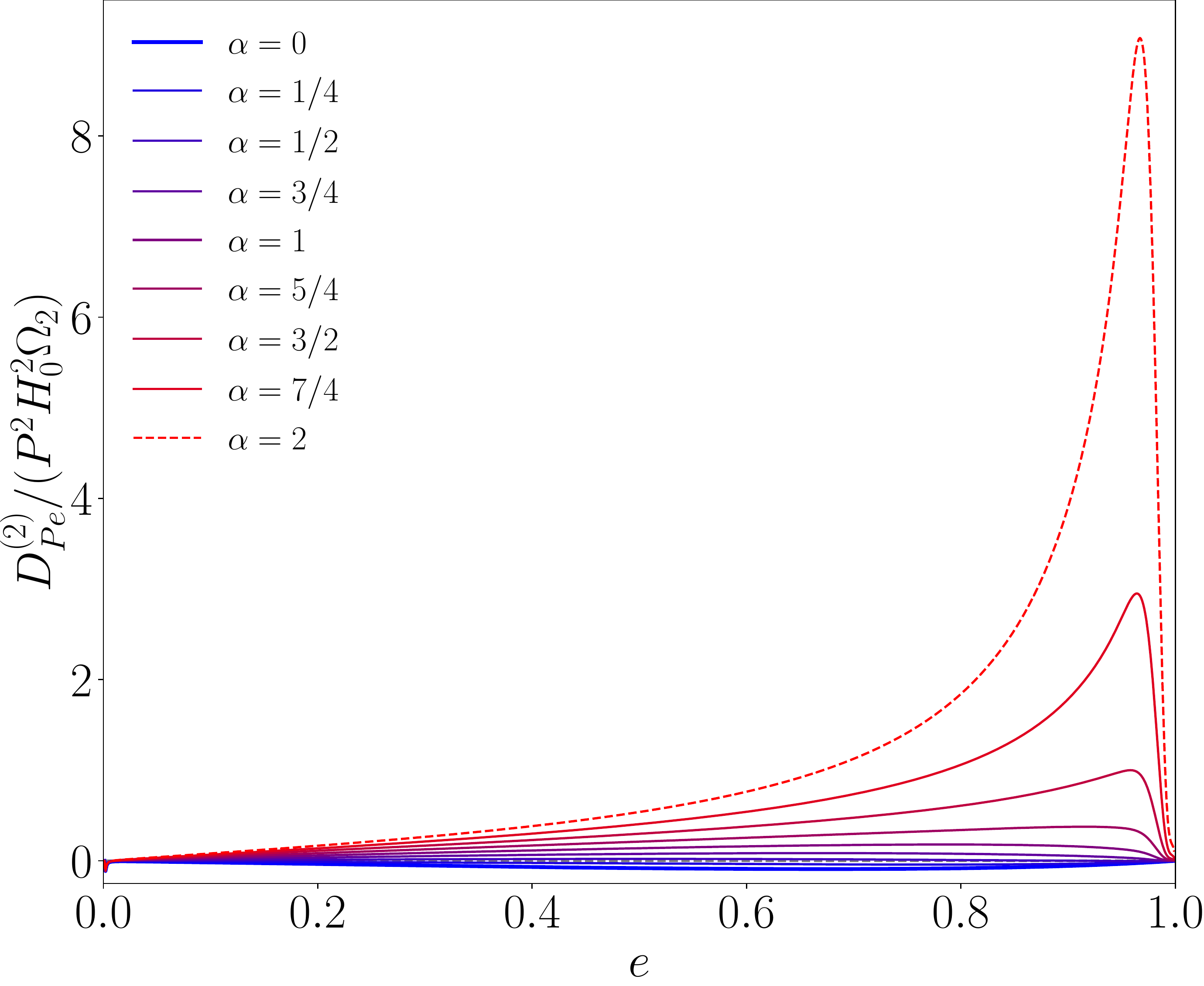}
        \includegraphics[width=0.8\textwidth]{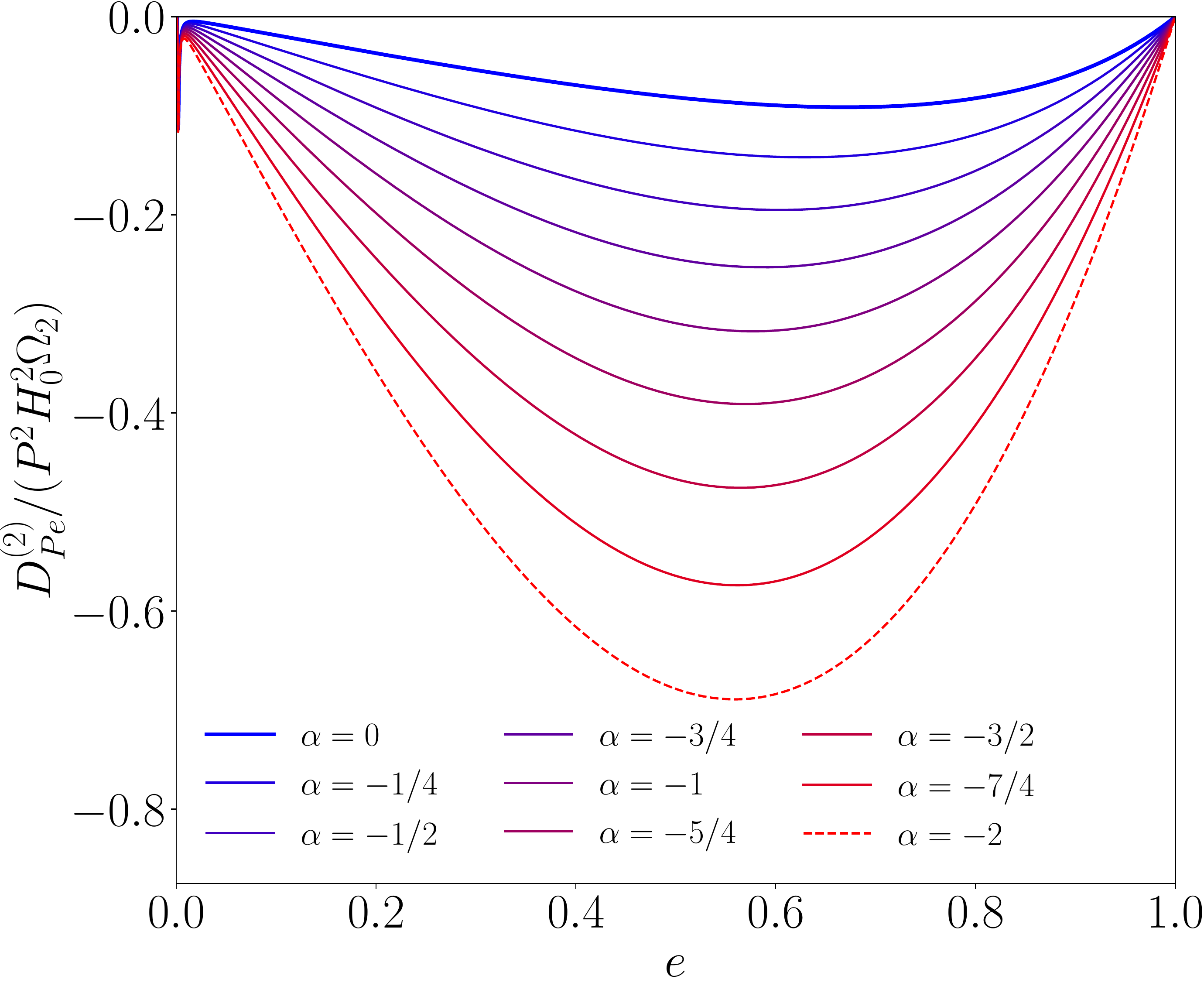}
    \end{center}
    \caption{%
    Same as figure~\ref{fig:km2-PP-ecc}, but for the period-eccentricity diffusion coefficient $D^{(2)}_{Pe}$.}
    \label{fig:km2-Pe-ecc}
\end{figure}

\begin{figure}[p!]
        \begin{center}
            \includegraphics[width=0.8\textwidth]{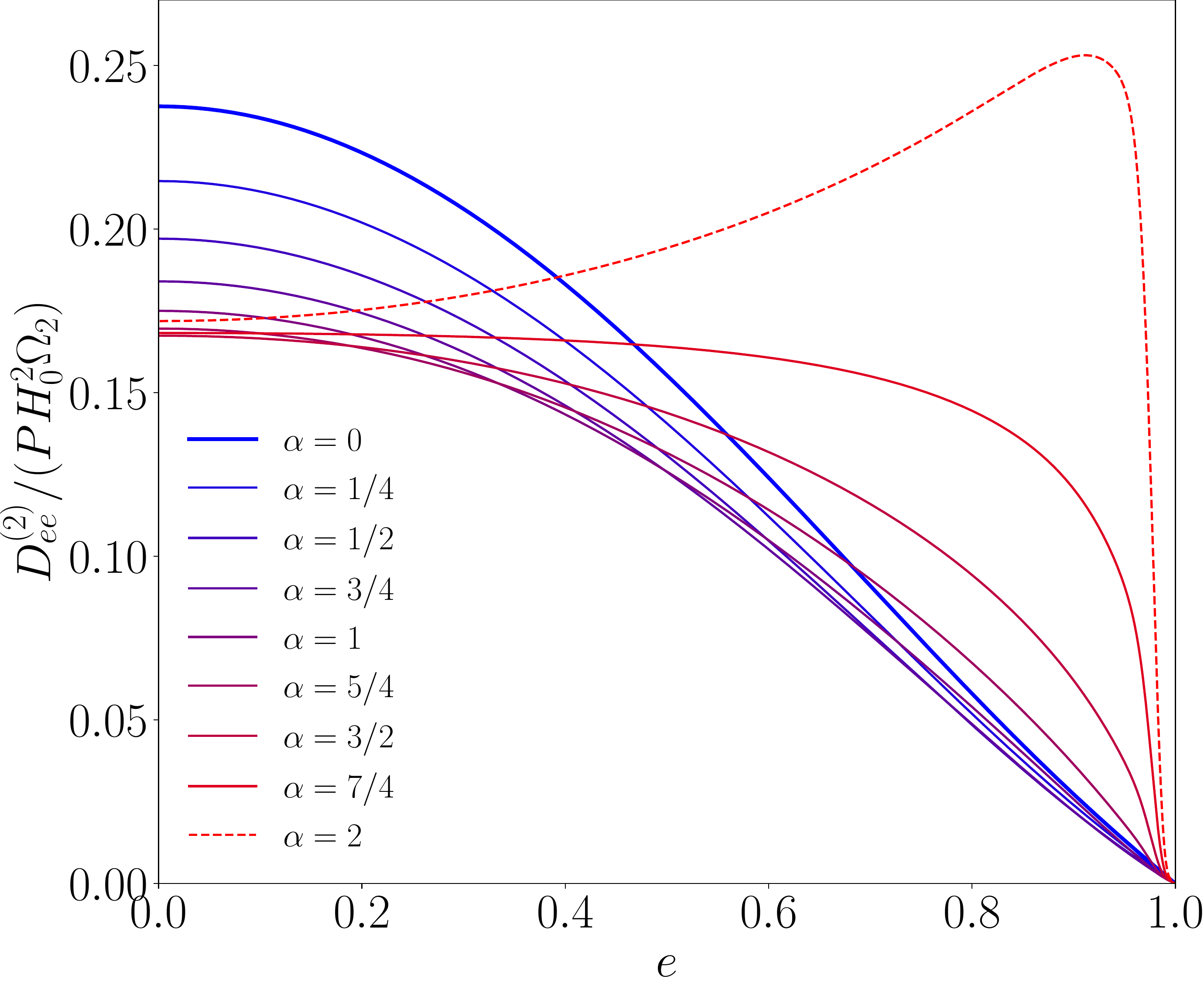}
            \includegraphics[width=0.8\textwidth]{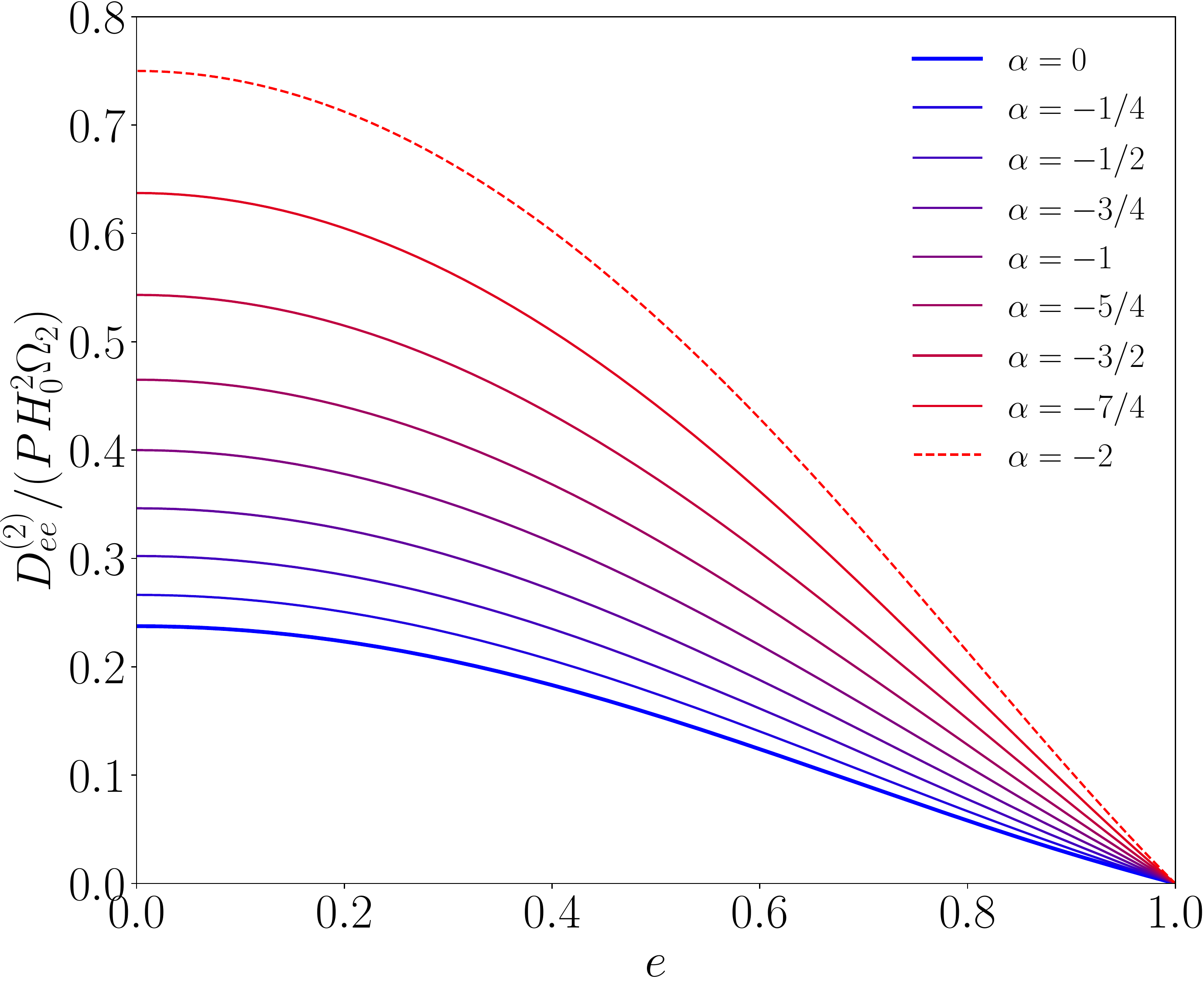}
        \end{center}
        \caption{%
        Same as figure~\ref{fig:km2-PP-ecc}, but for the eccentricity-eccentricity diffusion coefficient $D^{(2)}_{ee}$.}
        \label{fig:km2-ee-ecc}
    \end{figure}

\begin{figure}[p!]
    \begin{center}
        \includegraphics[width=0.8\textwidth]{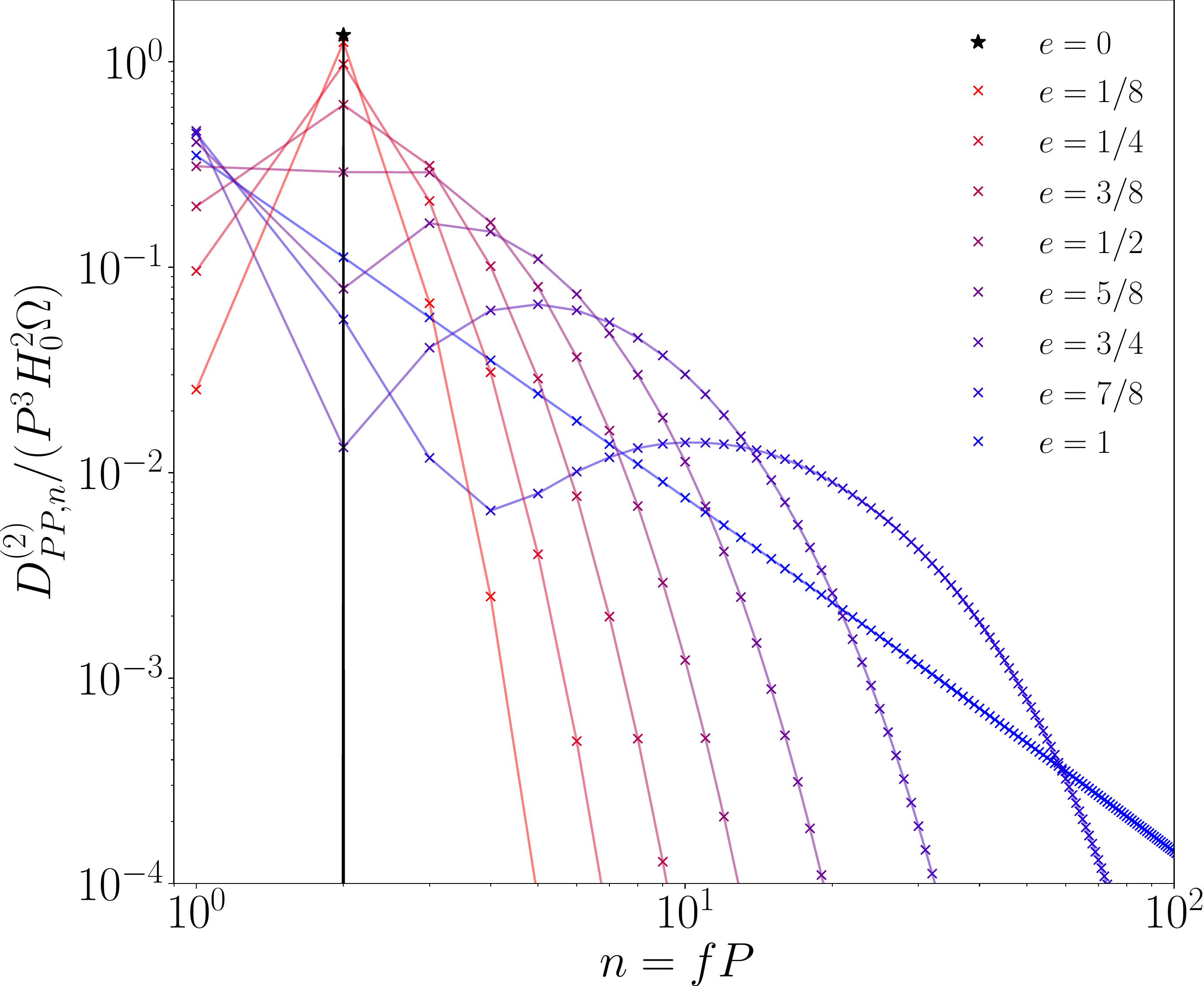}
        \includegraphics[width=0.8\textwidth]{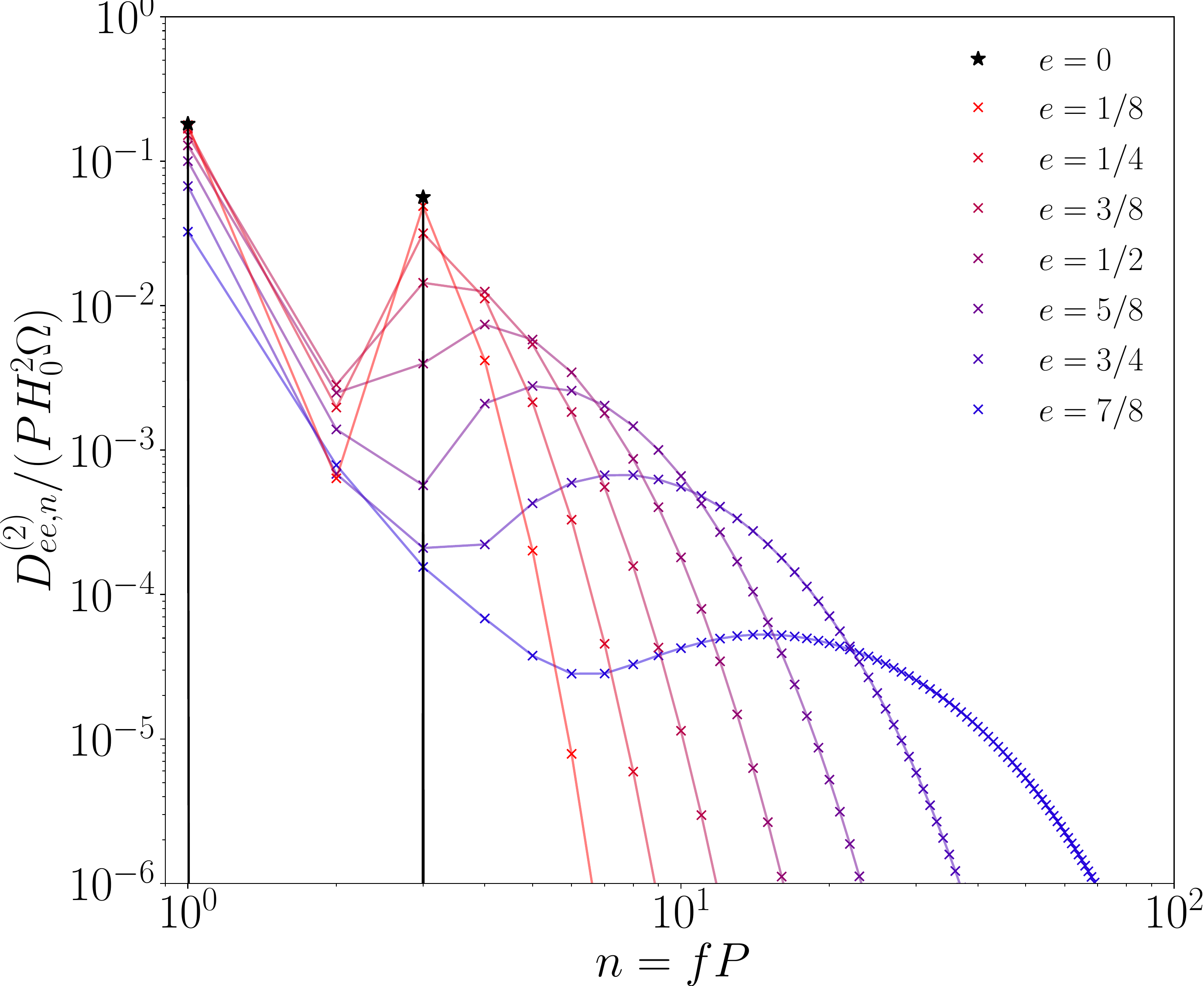}
    \end{center}
    \caption{%
    Contributions to the secular diffusion coefficients $D^{(2)}_{PP}$ (top panel) and $D^{(2)}_{ee}$ (bottom panel) from different harmonic frequencies, for binaries with various eccentricities $e=0,\dots,1$.
    (The subscript \enquote{$n$} here indicates that we have extracted the contribution from the $n^\mathrm{th}$ harmonic.)
    We set $\Omega(f)=\mathrm{constant}$ here, so that each harmonic receives equal weighting; alternative GWB spectra will give different weighting to each harmonic.
    Note that nearly all of the harmonics vanish in the circular case, $e=0$, so that the few harmonics that do contribute appear as vertical lines.
    Note also that $D^{(2)}_{ee}=0$ when $e\to1$.}
    \label{fig:km2-n}
\end{figure}

The values of some of these drift and diffusion coefficients as functions of eccentricity and harmonic frequency are shown in figures~\ref{fig:km1-P-ecc}--\ref{fig:km1-n}; we focus on the coefficients pertaining to the period and eccentricity of the binary, as these are the most important for our envisaged observational applications.
These coefficients, describing the evolution of the size and shape of the orbit, are independent of the remaining four orbital elements $(I,\asc,\omega,\eps)$, which all describe the orientation of the orbit in space.
In figures~\ref{fig:km1-P-ecc}--\ref{fig:km2-ee-ecc} we plot the coefficients as functions of eccentricity, assuming various power-law GWB spectra $\Omega(f)\sim f^\alpha$, $\alpha\in[-2,+2]$.
There are two particularly striking features that are worth mentioning: the first is that $D^{(2)}_{Pe}$ and $D^{(2)}_{ee}$ both vanish as $e\to1$, regardless of the GWB spectrum; and the second is that $D^{(1)}_e$ changes sign at an eccentricity $e\approx0.4$ that is approximately (though not exactly) independent of the GWB spectrum; the significance of this near-universal crossover eccentricity is not immediately clear.

In figures~\ref{fig:km2-n} and~\ref{fig:km1-n} we plot the contributions from each of the binary's harmonic frequencies $f=n/P$.
We see that the evolution of the period is driven by the $n=2$ harmonic for near-circular binaries, with all other harmonics having zero contribution to $D^{(2)}_{PP}$ and $D^{(1)}_P$ in the circular limit $e\to0$.
This mirrors the frequency content of GW \emph{emission} from binaries, which is also dominated by the $n=2$ harmonic for small eccentricity, as we saw in section~\ref{sec:compact-binaries}.
As we discussed there, the significance of $n=2$ can be understood by recognising that advancing a circular orbit in time by $P/2$ (i.e. the inverse of the $n=2$ harmonic) is equivalent to exchanging the positions of the two bodies, resulting in a setup which has the same GW emission and absorption properties.
The eccentricity evolution, on the other hand, is dominated by the $n=1$ and $n=3$ harmonics when $e$ is small.
In all cases, the contribution from higher harmonics becomes stronger for larger eccentricities, as the Fourier spectrum of the binary's response to GWs becomes richer.
(Again, this is the same qualitative pattern that one finds in the case of GW \emph{emission} by the binary.)
In the limit $e\to1$, we see a very simple pattern emerge in which each coefficient approaches a power law in the harmonic number $n$ (except for $D^{(2)}_{ee}$, which vanishes in this limit).

The qualitative link between the GW emission and absorption spectra of binaries noted above is intriguing, and raises the question of whether this relationship can be established more formally.
This might lead to deeper insights into binary--GW interactions, for example by allowing us to prove something akin to a fluctuation-dissipation theorem for this system.
It would also be interesting to make contact with existing results on the scattering of GWs by binaries~\cite{Annulli:2018quj}.
However, on the face of it there are several important differences between the GW absorption and emission processes: for example, the masses of the orbiting bodies are crucial in determining the radiated GW flux, but have no influence at all on GW absorption, since the GW-induced oscillations in the orbital separation are independent of the masses.
(This last statement is a manifestation of the equivalence principle.)
We leave a more thorough exploration of these questions for future work.

Inserting the coefficients in equations~\eqref{eq:ecc-drift} and~\eqref{eq:ecc-diffusion} into equation~\eqref{eq:fpe}, we obtain a FPE which completely describes the secular evolution of a general binary system under GWB resonance; this is the main result of our analysis.
Sections~\ref{sec:circular}, \ref{sec:full-fpe}, and~\ref{sec:observations} explore various strategies for solving this equation, and for using these solutions to place constraints on the GWB spectrum.
Before moving on, we derive some simplified expressions for the KM coefficients in the cases where the eccentricity and/or the inclination are small.

\subsection{Small-eccentricity and small-inclination orbits}
\label{sec:KM-small-e-I}

For binaries with very small eccentricity, we want to recast the results above in terms of the alternative orbital elements $(P,\zeta,\kappa,I,\asc,\xi)$, as defined in equation~\eqref{eq:small-e-I}.
We do this using the coordinate transformation laws for the KM coefficients (see, e.g., section~4.9 of \citet{Risken:1989fpe} for a derivation),
    \begin{equation}
        D^{(1)}_i=\pdv{X_i}{X_{i'}}D^{(1)}_{i'}+\pdv{X_i}{X_{i'}}{X_{j'}}D^{(2)}_{i'j'},\qquad D^{(2)}_{ij}=\pdv{X_i}{X_{i'}}\pdv{X_j}{X_{j'}}D^{(2)}_{i'j'},
    \end{equation}
    where summation over the primed indices is implied.
Neglecting terms of order $e^2\sim\zeta^2\sim\zeta\kappa\sim\kappa^2$, we find the drift coefficients
    \begin{align}
    \begin{split}
    \label{eq:small-e-drift}
        D^{(1)}_P&=V_P+\frac{3P^2}{160}H_0^2(-79\Omega_1+288\Omega_2-27\Omega_3),\\
        D^{(1)}_\zeta&=V_\zeta+\frac{P}{160}\frac{\zeta}{\zeta^2+\kappa^2}H_0^2(25\Omega_1-27\Omega_3),\qquad D^{(1)}_\kappa=V_\kappa+\frac{P}{160}\frac{\kappa}{\zeta^2+\kappa^2}H_0^2(25\Omega_1-27\Omega_3),\\
        D^{(1)}_I&=\frac{3P}{80}H_0^2\Omega_2\cot^2I,\qquad D^{(1)}_\asc=0,\qquad D^{(1)}_\xi=V_\xi,
    \end{split}
    \end{align}
    where the deterministic drift terms are given by
    \begin{equation}
        V_\zeta=\frac{6\uppi v_P^2}{P}\qty(\kappa-\tfrac{304}{45}\eta v_P^3\zeta),\qquad V_\kappa=-\frac{6\uppi v_P^2}{P}\qty(\zeta+\tfrac{304}{45}\eta v_P^3\kappa),\qquad V_\xi=-\frac{4\uppi v_P^2}{P}(3-\eta),
    \end{equation}
    and the diffusion coefficients are
    \begin{align}
    \begin{split}
    \label{eq:small-e-diffusion}
        D^{(2)}_{PP}&=\frac{27P^3}{20}H_0^2\Omega_2,\qquad D^{(2)}_{P\zeta}=-\frac{3P\zeta}{160}H_0^2(25\Omega_1+12\Omega_2-27\Omega_3),\\
        D^{(2)}_{P\kappa}&=-\frac{3P\kappa}{160}H_0^2(25\Omega_1+12\Omega_2-27\Omega_3),\qquad D^{(2)}_{\zeta\zeta}=D^{(2)}_{\kappa\kappa}=\frac{P}{160}H_0^2(29\Omega_1+9\Omega_3),\\
        D^{(2)}_{\zeta\asc}&=-\frac{3P\kappa}{80}H_0^2\Omega_2\frac{\cos I}{\sin^2I},\qquad D^{(2)}_{\zeta\xi}=-\frac{P\kappa}{320}H_0^2\qty[203\Omega_1-12\Omega_2(20+\cot^2I)+63\Omega_3],\\
        D^{(2)}_{\kappa\asc}&=\frac{3P\zeta}{80}H_0^2\Omega_2\frac{\cos I}{\sin^2I},\qquad D^{(2)}_{\kappa\xi}=\frac{P\zeta}{320}H_0^2\qty[203\Omega_1-12\Omega_2(20+\cot^2I)+63\Omega_3],\\
        D^{(2)}_{II}&=\frac{3P}{80}H_0^2\Omega_2,\qquad D^{(2)}_{\asc\asc}=\frac{3P}{80}H_0^2\Omega_2\csc^2I,\qquad D^{(2)}_{\asc\xi}=-\frac{3P}{80}H_0^2\Omega_2\frac{\cos I}{\sin^2I},\\
        D^{(2)}_{\xi\xi}&=\frac{3P}{80}H_0^2\Omega_2(16+\cot^2I),\qquad D^{(2)}_{PI}=D^{(2)}_{P\asc}=D^{(2)}_{P\xi}=D^{(2)}_{\zeta\kappa}=D^{(2)}_{\zeta I}=D^{(2)}_{\kappa I}=D^{(2)}_{I\asc}=D^{(2)}_{I\xi}=0.
    \end{split}
    \end{align}
(We have confirmed that the $D^{(2)}_{PP}$ coefficient here matches the corresponding result in \citet{Hui:2012yp}.)

We see from equation~\eqref{eq:small-e-drift} that the stochastic drift term for the period $P$ is usually positive, and can thus be interpreted physically as describing the softening of the binary due to the absorption of energy from the GWB (the term can become negative if $\Omega_2$ is significantly smaller than $\Omega_1$ and/or $\Omega_3$, but GWB spectra typically vary sufficiently slowly with frequency that this does not occur).
Interestingly, this implies that the net secular drift of the binary period (deterministic plus stochastic) generally changes sign at some critical value of $P$; e.g., for a scale-invariant GWB $\Omega(f)=\mathrm{constant}$, this value is given by
    \begin{equation}
    \label{eq:Pcrit}
        P_\rmc=\qty(\frac{1024\uppi\eta}{91H_0^2\Omega})^{3/11}(2\uppi G\mathcal{M})^{5/11}\approx95\,\mathrm{yr}\times\qty(\frac{\Omega}{10^{-6}}\frac{1/4}{\eta})^{-3/11}\qty(\frac{M}{m_\odot})^{5/11},
    \end{equation}
    which corresponds to a semi-major axis of
    \begin{equation}
        a_\rmc\approx21\,\mathrm{au}\times\qty(\frac{\Omega}{10^{-6}}\frac{1/4}{\eta})^{-2/11}\qty(\frac{M}{m_\odot})^{1/3}.
    \end{equation}
Binaries with $P<P_\rmc$ will decay through GW emission, decreasing their period further, whereas binaries with $P>P_\rmc$ undergo a net softening through GWB absorption, leading to a further increase in their period.
The point $P=P_\rmc$ is thus an unstable fixed point of the Langevin equation for $P$.\footnote{%
    We can understand this instability through a thermodynamic analogy.
    In the absence of the GWB, a binary system radiates GW energy with increasing intensity as it inspirals; its dynamical \enquote{temperature} grows as it loses energy, meaning that it has a \emph{negative heat capacity}---see footnote~\ref{ft:thermo} of chapter~\ref{chap:intro}.
    Similarly, the GWB acts as a heat reservoir and imparts energy to the binary (on average), therefore slowing the orbital motion and decreasing the system's temperature.
    A conventional thermodynamic system with positive heat capacity would equilibriate at a point where the heat loss from GW radiation balanced the heat gain from the GWB.
    Instead, the binary undergoes a runaway increase or decrease in its temperature due to its negative heat capacity.}
Note however that random diffusion due to $D^{(2)}_{PP}$ acts on a similar timescale (see figure~\ref{fig:timescales} and equations~\eqref{eq:drift-timescale} and~\eqref{eq:diffusion-timescale}), and can easily push the system either side of this critical point.

We find a similar phenomenon for the eccentricity.
Transforming back from the Laplace-Lagrange variables $\zeta,\kappa$ for now, equations~\eqref{eq:small-e-drift} and~\eqref{eq:small-e-diffusion} give
    \begin{equation}
        D^{(1)}_e=V_e+\frac{9P}{80e}H_0^2(3\Omega_1-\Omega_3),
    \end{equation}
    so that the (usually positive) stochastic drift diverges as $e\to0$, while the (always negative) deterministic part vanishes as $e\to0$.
The net eccentricity drift thus changes sign at a critical value, just as it does for the period.
For example, assuming a scale-invariant GWB spectrum, this critical value is
    \begin{equation}
        e_\rmc=\sqrt{\frac{27H_0^2\Omega}{4864\uppi\eta v_P^5}}\approx5.9\times10^{-5}\times\qty(\frac{\Omega}{10^{-6}}\frac{1/4}{\eta})^{1/2}\qty(\frac{M}{m_\odot})^{-5/6}\qty(\frac{P}{\mathrm{yr}})^{11/6}.
    \end{equation}
Since $\partial_eD^{(1)}_e<0$ at this point, $e_\rmc$ is a stable fixed point of the corresponding Langevin equation: binaries with larger eccentricity will tend to circularise through GW emission until they reach $e_\rmc$, while binaries with smaller eccentricity will on average have their eccentricity excited through GWB resonance.
This is particularly interesting from an observational point of view, as it suggests that eccentricities smaller than $e_\rmc$ might be less frequently observed in sufficiently old systems, though random diffusion due to $D^{(2)}_{ee}$ can still push systems below this point.

We also see from equation~\eqref{eq:small-e-drift} that the stochastic drift term for the inclination changes sign at $I=\uppi/2$ (this is also true in the general-eccentricity case, see equation~\eqref{eq:ecc-drift}).
Since $\partial_ID^{(1)}_I<0$ at $I=\uppi/2$, this is a stable fixed point of the corresponding Langevin equation; stochastic drift will, on average, drive binaries toward $I=\uppi/2$.
This effect is counteracted, however, by binaries diffusing away from $I=\uppi/2$.
As we show explicitly in section~\ref{sec:circular}, on extremely long timescales these two effects balance each other, leaving an isotropic distribution for the inclination.

In the limit where the binary's inclination is also small, we rewrite the KM coefficients in terms of the orbital elements $(P,\zeta,\kappa,p,q,\lambda)$.
This gives the drift coefficients
    \begin{equation}
    \label{eq:small-e-I-drift}
        D^{(1)}_p=-\frac{Pp}{40}H_0^2\Omega_2,\qquad D^{(1)}_q=-\frac{Pq}{40}H_0^2\Omega_2,
    \end{equation}
    and the diffusion coefficients,
    \begin{align}
    \begin{split}
    \label{eq:small-e-I-diffusion}
        D^{(2)}_{\zeta\lambda}&=-\frac{P\kappa}{320}H_0^2\qty(203\Omega_1-240\Omega_2+63\Omega_3),\qquad D^{(2)}_{\kappa\lambda}=\frac{P\zeta}{320}H_0^2\qty(203\Omega_1-240\Omega_2+63\Omega_3),\\
        D^{(2)}_{pp}&=D^{(2)}_{qq}=\frac{3P}{80}H_0^2\Omega_2,\qquad D^{(2)}_{p\lambda}=\frac{3Pq}{160}H_0^2\Omega_2\qquad D^{(2)}_{q\lambda}=-\frac{3Pp}{160}H_0^2\Omega_2\qquad D^{(2)}_{\lambda\lambda}=\frac{3P}{5}H_0^2\Omega_2,\\
        D^{(2)}_{Pp}&=D^{(2)}_{Pq}=D^{(2)}_{P\lambda}=D^{(2)}_{\zeta\kappa}=D^{(2)}_{\zeta p}=D^{(2)}_{\zeta q}=D^{(2)}_{\kappa p}=D^{(2)}_{\kappa q}=D^{(2)}_{pq}=0,
    \end{split}
    \end{align}
    where the coefficients not listed are identical to those in equations~\eqref{eq:small-e-drift} and~\eqref{eq:small-e-diffusion}.
Note that this implies $D^{(1)}_I=3PH_0^2\Omega_2/(80I)$, which diverges as $I\to0$.
This means that binaries are quickly excited away from zero inclination, similar to what happens for the eccentricity as $e\to0$.

\section{Some exact results for circular binaries}
\label{sec:circular}

We have shown that the osculating orbital elements of a binary coupled to the GWB evolve according to a nonlinear six-dimensional FPE, for which no analytical solution is generally available.
However, in the small-eccentricity limit $e\to0$, we found in equations~\eqref{eq:small-e-drift} and~\eqref{eq:small-e-diffusion} that the drift and diffusion of the binary period $P$ are independent of all of the other orbital elements.
This allows us to treat $P$ separately by solving the one-dimensional FPE
    \begin{equation}
        \pdv{W}{t}=-\pdv{J}{P},
    \end{equation}
    where $W(P,t)$ is now the single-variable DF for $P$, marginalised over the other orbital elements, and where
    \begin{equation}
    \label{eq:prob-current}
        J(P,t)\equiv D^{(1)}W-\partial_P(D^{(2)}W)
    \end{equation}
    is the \emph{probability current}~\cite{Risken:1989fpe,Gardiner:2004hsm}.
(We have suppressed the $P$ subscripts on the drift and diffusion coefficients for this single-variable case.)
In this section, we derive some exact results for this simplified equation.
These results highlight the power of our Fokker-Planck formalism, which allows us to answer these questions about the full shape of the DF in a way that previous analyses are unable to.

\subsection{Quasi-stationary period distribution}

The simplest kind of solution to look for is a stationary (i.e., time-independent) distribution, corresponding to constant probability current throughout the parameter space.
Setting $J=\mathrm{constant}$ in equation~\eqref{eq:prob-current}, we can use the integrating factor (which is defined up to an arbitrary constant multiplicative factor)
    \begin{equation}
    \label{eq:integrating-factor}
        I(P)\equiv\exp(\int\dd{P}\frac{D^{(1)}}{D^{(2)}})
    \end{equation}
    to obtain
    \begin{equation}
    \label{eq:quasi-stationary}
        W=\frac{I(P)}{D^{(2)}}\qty(C-J\int\frac{\dd{P}}{I(P)}),
    \end{equation}
    with $C$ a constant which, for a given value of $J$, is fixed by the normalisation of the DF.
Clearly, the functional form of equation~\eqref{eq:quasi-stationary} depends on the GWB energy spectrum $\Omega(f)$.
As a simple example, consider a scale-invariant GWB, $\Omega(f)=\mathrm{constant}$, which has
    \begin{equation}
        D^{(1)}=V_P+\frac{273}{80}P^2H_0^2\Omega,\qquad D^{(2)}=\frac{27}{20}P^3H_0^2\Omega,\qquad I(P)\propto P^{91/36}\exp[\frac{91}{132}\qty(\frac{P_\rmc}{P})^{11/3}].
    \end{equation}
A full exploration of spectra beyond this simple scale-invariant case is beyond our scope here, but we expect the qualitative results of this section to be reasonably robust to this choice.

\begin{figure}[p!]
    \begin{center}
        \includegraphics[width=0.8\textwidth]{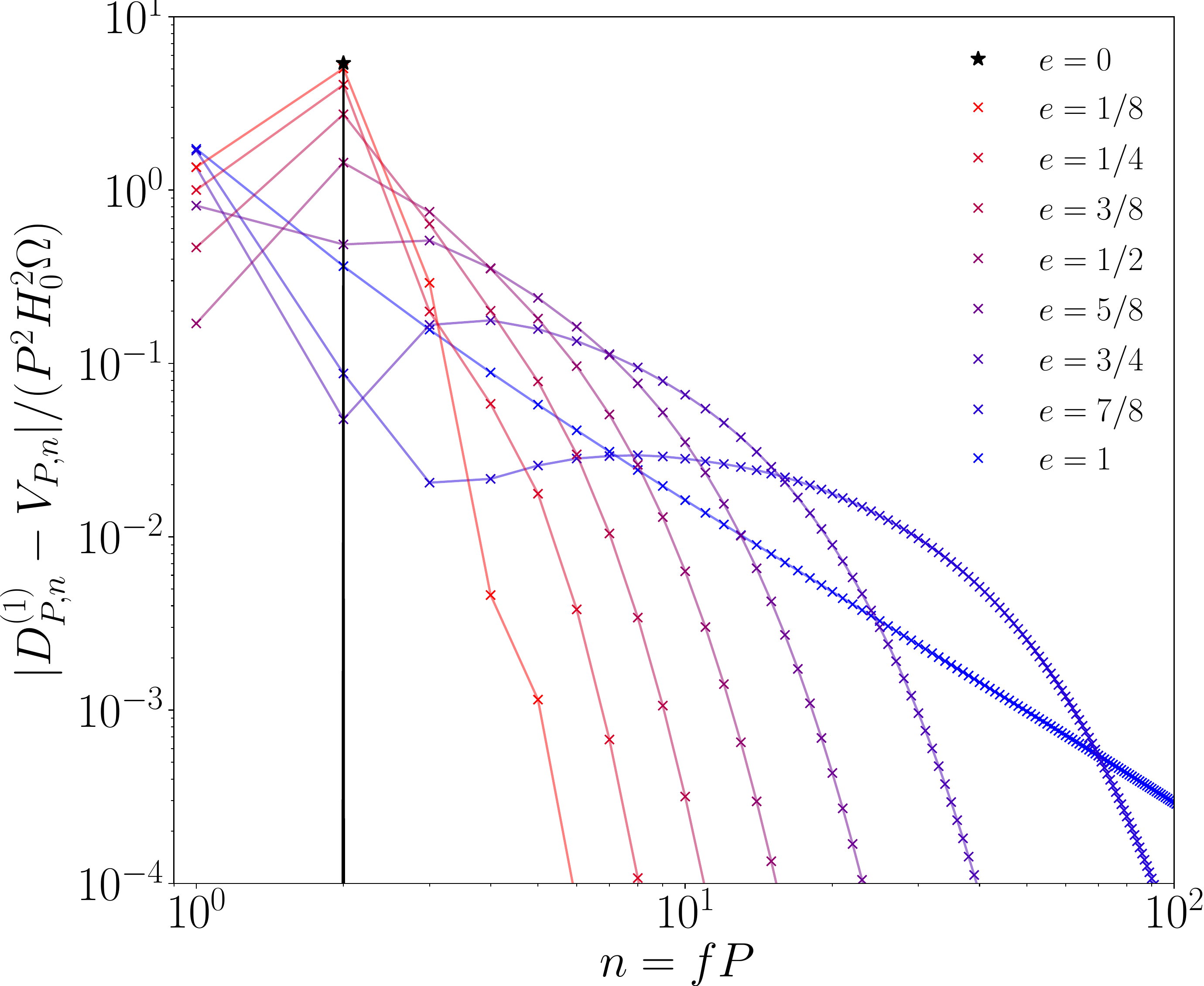}
        \includegraphics[width=0.8\textwidth]{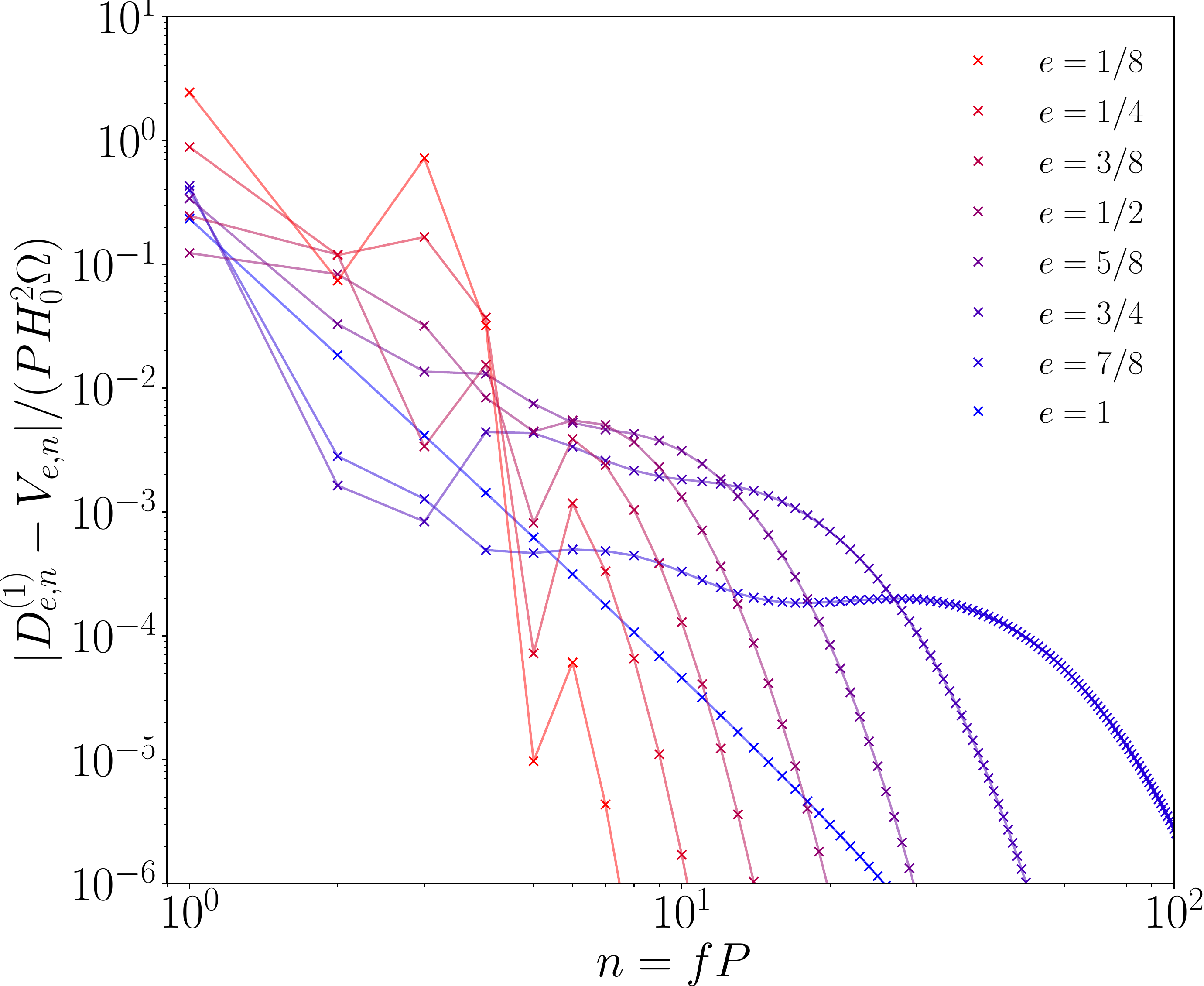}
    \end{center}
    \caption{%
    Contributions to the stochastic parts of the secular drift coefficients $D^{(1)}_P$ (top panel) and $D^{(1)}_e$ (bottom panel) from different harmonic frequencies, for binaries with various eccentricities $e=0,\dots,1$.
    (The subscript \enquote{$n$} here indicates that we have extracted the contribution from the $n^\mathrm{th}$ harmonic.)
    We show the absolute values, as the drift coefficients have both positive and negative contributions.
    Note that $D^{(1)}_e\to+\infty$ as $e\to0$.}
    \label{fig:km1-n}
\end{figure}

We can fix $J$ and $C$ by imposing the appropriate boundary conditions.
For some minimum value of $P$ the binary merges or is tidally disrupted, whereas for some maximum value the binary becomes gravitationally unbound, so at both extremes we require absorbing boundary conditions---i.e., the DF must go to zero at both boundaries.
Systems with absorbing boundary conditions do not admit nonzero stationary solutions~\cite{Gardiner:2004hsm}; formally, the conditional probability of the binary having period $P$ at time $t$, given that it initially had period $P_0$ at time zero, obeys $\lim_{t\to\infty}W(P,t|P_0,0)=0$ across the entire parameter space.
Intuitively, this is because all of the initial probability mass is eventually absorbed by one or other of the boundaries.
We can also understand this in terms of the instability discussed in section~\ref{sec:KM-small-e-I}; binaries either side of the critical period $P_\rmc$ undergo a runaway evolution away from this point, reaching one of the two boundaries within finite time, thus leaving an empty distribution in the limit $t\to\infty$.

\begin{figure}[t!]
    \begin{center}
        \includegraphics[width=0.8\textwidth]{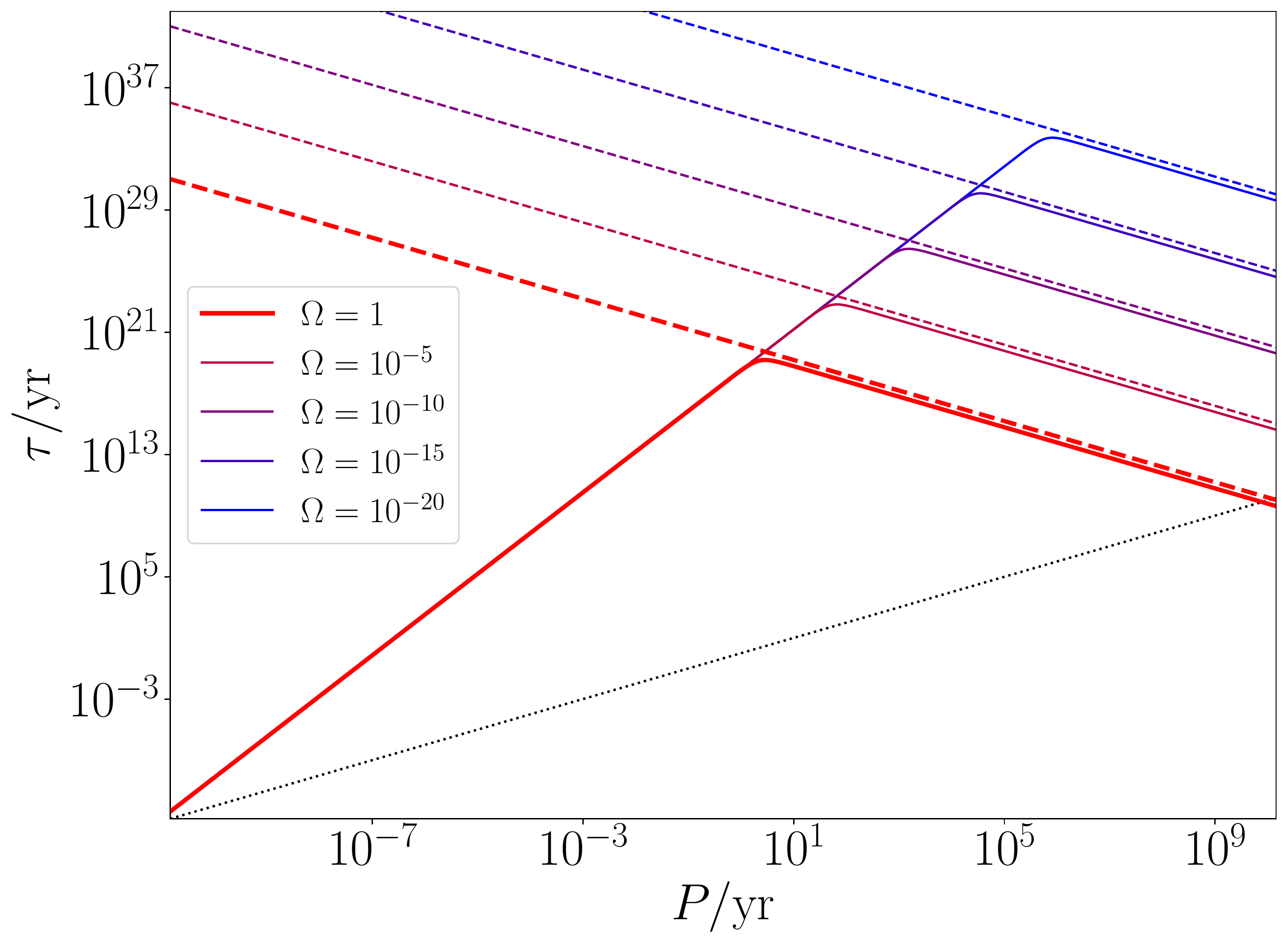}
    \end{center}
    \caption{%
   Evolution timescales for the DF of a circular binary with total mass $M=m_\odot$ immersed in a scale-invariant GWB $\Omega(f)=\mathrm{constant}$.
    Solid curves show the drift timescale~\eqref{eq:drift-timescale}, which is dominated by GW emission at \enquote{short} periods $P\lesssim P_\rmc$ and by GW absorption from the GWB at \enquote{long} periods $P\gtrsim P_\rmc$.
    Dashed curves show the diffusion timescale~\eqref{eq:diffusion-timescale}.
    The dotted black curve shows $\tau=P$; our secular-averaging assumption only holds in the region above this curve.}
    \label{fig:timescales}
\end{figure}

In practice, however, the timescale over which the binary evolves is set by its period, and binaries with short periods will approach stationarity much faster than binaries with long periods.
More concretely, for a flat GWB spectrum we have a drift timescale
    \begin{equation}
    \label{eq:drift-timescale}
        \tau_\mathrm{drift}\equiv\frac{P}{|D^{(1)}|}\simeq
        \begin{cases}
            \displaystyle\frac{80}{273PH_0^2\Omega}, & P\gg P_\rmc, \\[4pt]
            \displaystyle\frac{5P^{8/3}}{192\uppi\eta(2\uppi GM)^{5/3}}, & P\ll P_\rmc,
        \end{cases}
    \end{equation}
    and a diffusion timescale
    \begin{equation}
    \label{eq:diffusion-timescale}
        \tau_\mathrm{diff}\equiv\frac{P^2}{|D^{(2)}|}=\frac{20}{27PH_0^2\Omega}.
    \end{equation}
As shown in figure~\ref{fig:timescales}, the fastest timescale is many orders of magnitude shorter near the lower boundary than it is near the upper boundary.
It may be reasonable, therefore, to look for \enquote{quasi-stationary} solutions, where equilibrium is established near the lower boundary, but where the boundary condition for long periods is neglected.
We achieve this by choosing $J$ and $C$ such that the DF goes to zero at the lower boundary, and such that the DF is normalised, but without enforcing any condition at the upper boundary.

\begin{figure}[t!]
    \begin{center}
        \includegraphics[width=0.8\textwidth]{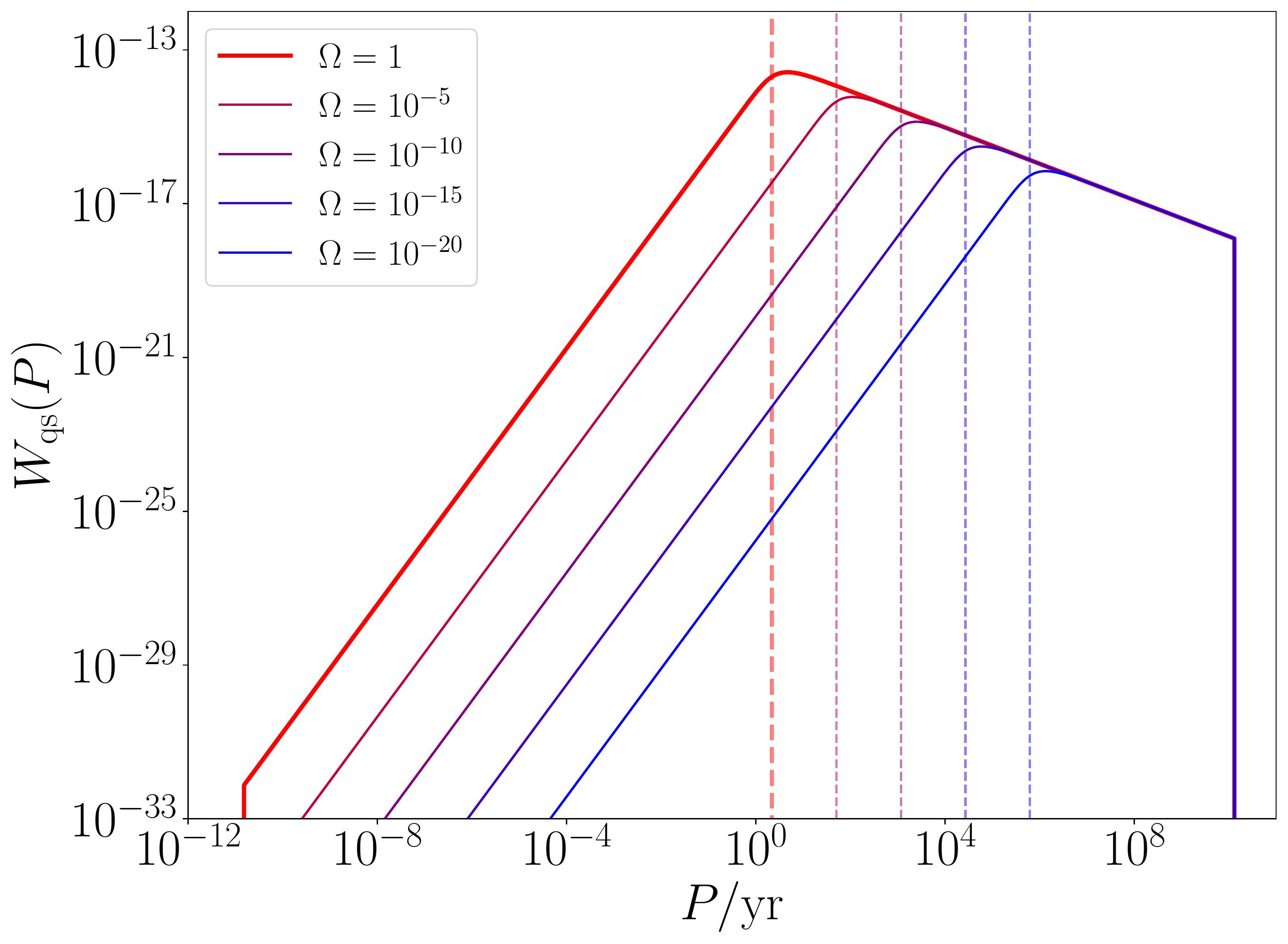}
    \end{center}
    \caption{%
    The quasi-stationary distribution~\eqref{eq:quasi-stationary-solution} for the period $P$ of a binary with mass $M=m_\odot$ coupled to a scale-invariant GWB, $\Omega(f)=\mathrm{constant}$.
    The upper and lower cutoffs are due to the age of the Universe and the binary's ISCO period~\eqref{eq:isco}, respectively.
    The dashed vertical lines indicate the value of $P_\rmc$ for each $\Omega$, as defined by equation~\eqref{eq:Pcrit}---this is roughly the peak of the distribution in each case.
    Binaries with periods $P<P_\rmc$ decay deterministically through GW emission and are removed from the distribution, whereas binaries with $P>P_\rmc$ are supported against decay by resonant absorption of the GWB.
    As we lower the GWB intensity $\Omega$, the resonance becomes weaker, and the binaries must have longer periods to avoid decay.}
    \label{fig:quasi-stationary-solution}
\end{figure}

As a simple example, consider a scale-invariant GWB spectrum $\Omega(f)=\mathrm{constant}$, and fix the DF to be zero at the period corresponding to the binary's innermost stable circular orbit,
    \begin{equation}
    \label{eq:isco}
        P_\mathrm{ISCO}\equiv6^{3/2}\times2\uppi GM,
    \end{equation}
    with the binary assumed to merge at periods shorter than this.
The corresponding quasi-stationary distribution is then given in terms of the dimensionless variable $\varrho\equiv P/P_\mathrm{ISCO}$ by
    \begin{equation}
    \label{eq:quasi-stationary-solution}
        W_\mathrm{qs}\propto\exp(\frac{\lambda}{\varrho^{11/3}})\qty[\frac{E_{77/132}(\lambda\varrho^{-11/3})}{\varrho^2}-\frac{E_{77/132}(\lambda)}{\varrho^{17/36}}],
    \end{equation}
    where $E_n(z)\equiv\int_1^\infty\dd{t}\rme^{-zt}t^{-n}$ is the exponential integral function, and
    \begin{equation}
        \lambda\equiv\frac{91}{132}\qty(\frac{P_\rmc}{P_\mathrm{ISCO}})^{11/3}=\frac{\sqrt{2}\eta(GMH_0)^{-2}}{8019\sqrt{3}\uppi\Omega}\gg1
    \end{equation}
    is a dimensionless constant which quantifies the strength of the deterministic drift $V_P$ relative to the secular diffusion $D^{(2)}$.
While the functional form of equation~\eqref{eq:quasi-stationary-solution} is somewhat opaque, one can show that it approximately interpolates between $W_\mathrm{qs}\sim P^{5/3}$ for $P_\mathrm{ISCO}\ll P\ll P_\rmc$ and $W_\mathrm{qs}\sim P^{-17/36}$ for $P\gg P_\rmc$ (this broken-power-law behaviour is clearly seen in figure~\ref{fig:quasi-stationary-solution}).
We normalise equation~\eqref{eq:quasi-stationary-solution} by integrating up to some maximum period, which we choose to be the age of the Universe.
By substituting equation~\eqref{eq:quasi-stationary-solution} into equation~\eqref{eq:prob-current} and evaluating at $\varrho=1$ where $W=0$ and $\pdv*{W}{P}>0$, we see that the probability current is strictly negative, $J<0$.
Since we specified that $J=\mathrm{constant}$, this means there is a uniform net flow towards shorter periods throughout the parameter space, regardless of how soft the binary is.

For hard binaries with $P<P_\rmc$ it is obvious that we should have $J<0$, as the negative deterministic drift is the most important effect and quickly drives the binary toward merger.
For soft binaries with $P>P_\rmc$ the negative probability current is less immediately obvious; it shows us that, while there is a net drift $D^{(1)}$ pushing these binaries towards longer periods, on average they are nonetheless expected to flow towards shorter periods.
We can understand this somewhat counter-intuitive behaviour by including diffusion as well as drift effects.
Indeed, note that by equation~\eqref{eq:prob-current}, the condition $J<0$ is equivalent to
    \begin{equation}
        \pdv{(\ln D^{(2)}W)}{(\ln P)}>\frac{PD^{(1)}}{D^{(2)}}=\frac{91}{36}\qty[1-(P/P_\rmc)^{-11/3}],
    \end{equation}
    where the RHS tends to a positive constant value in the $P\gg P_\rmc$ region we are interested in.
What this is telling us is that, despite the positive drift coefficient, we can still have a negative probability current if and only if the diffusion coefficient $D^{(2)}$ grows sufficiently quickly with $P$, as this makes it sufficiently likely for the binary's random walk to wander below $P_\rmc$ and then rapidly approach short periods through GW emission.
Interestingly, we find that for the quasi-stationary distribution we have $\pdv*{(\ln D^{(2)}W_\mathrm{qs})}{(\ln P)}\simeq91/36$, so this diffusive effect is \emph{only just} strong enough to cause a net negative probability current.

We can repeat this process for any given GWB spectrum to write down a corresponding quasi-stationary solution for the period of a circular binary.
However, the time taken to relax to this distribution is extremely long for typical GWB spectra (see figure~\ref{fig:timescales}), so these solutions are physically uninteresting in most cases.
Besides, the assumption of a perfectly circular binary is overly simplistic, as we have shown in section~\ref{sec:KM-small-e-I} that the eccentricity distribution relaxes away from zero on shorter timescales.
Nonetheless, the approach in this section is still useful for building intuitive understanding of the dynamics of the full DF, and may be useful for, e.g., studies of the orbital element distributions of old populations of binaries, where the full shape of the distribution is vitally important.
Finding quasi-stationary solutions for the full multivariate FPE is much more challenging, and for eccentric binaries there is no guarantee that such a solution with the appropriate lower boundary condition even exists.

\subsection{Mean coalescence time}

For any one-dimensional FPE, it is possible to write down an explicit formula for the \emph{mean first passage time} at either of its boundaries---i.e., the average time taken for an individual random trajectory to reach that boundary, as a function of the initial position~\cite{Gardiner:2004hsm}.
In our case, this is a useful tool for understanding how the presence of a particular GWB spectrum impacts upon the eventual fate of a binary system.
Applying this to our lower absorbing boundary at the ISCO, we thus have the mean coalescence time of a binary coupled to the GWB, as a function of its initial period $P_i$,
    \begin{equation}
        \ev{\tau(P_i)}=\int_{P_\mathrm{ISCO}}^{P_i}\dd{P}\int_P^{P_\mathrm{max}}\dd{P'}\frac{I(P')}{I(P)D^{(2)}(P')},
    \end{equation}
    where $I(P)$ is the integrating factor defined in equation~\eqref{eq:integrating-factor}.

Returning to the example of a scale-invariant GWB, this becomes
    \begin{equation}
    \label{eq:coalescence-time}
        \ev{\tau(P_i)}=\int_1^{\varrho_i}\frac{\dd{\varrho}}{\varrho^{91/36}}\int_{\varrho}^{\varrho_\mathrm{max}}\frac{\dd{\varrho'}}{\varrho'^{17/36}}\frac{20\,\rme^{\lambda(\varrho'^{-11/3}-\varrho^{-11/3})}}{27P_\mathrm{ISCO}H_0^2\Omega}.
    \end{equation}
This double integral is challenging to evaluate in general.
However, we can easily verify equation~\eqref{eq:coalescence-time} by showing that it reproduces the standard expression for the coalescence time due to deterministic GW emission in cases where the GWB resonance is weak.
Setting $P_i\ll P_\rmc$, we can safely take the limit $\lambda\to\infty$ and extract the leading-order and next-to-leading-order terms,
    \begin{equation}
        \ev{\tau(P_i)}\simeq\frac{405GM}{16\eta}\qty(\frac{P_i}{P_\mathrm{ISCO}})^{8/3}\qty[1-\frac{96}{91}\rme^{-\frac{91}{132}(P_\rmc/P_i)^{11/3}}\frac{P_\mathrm{max}^{151/36}}{P_i^{19/36}P_\rmc^{11/3}}].
    \end{equation}
The leading-order term agrees with the deterministic coalescence time one finds by integrating $V_P$, as expected.
There is a small negative contribution from the next-to-leading-order term, which indicates that GWB resonance slightly speeds up the coalescence in this regime.
We can understand this by noticing that the term associated with $D^{(2)}$ in equation~\eqref{eq:prob-current} is always negative for the quasi-stationary distribution, so that diffusion always has a net negative contribution to the probability current and thus, on average, always help drive binaries towards merger.

\subsection{Including the remaining orbital elements}

Having found a quasi-stationary period distribution, it is now relatively easy to obtain stationary distributions for the remaining orbital elements $(I,\asc,\xi)$, so long as we hold $\zeta$ and $\kappa$ fixed at zero.
To do so, we write the full FPE as
    \begin{equation}
        \pdv{W}{t}=-\partial_iJ_i,\qquad J_i\equiv D^{(1)}_iW-\partial_j(D^{(2)}_{ij}W),
    \end{equation}
    where $W$ is now interpreted as the multivariate DF over the four orbital elements $(P,I,\asc,\xi)$, and the four corresponding probability currents are
    \begin{align}
    \begin{split}
    \label{eq:probability-currents}
        J_P&=D^{(1)}_PW-\partial_P(D^{(2)}_{PP}W),\qquad J_I=D^{(1)}_IW-D^{(2)}_{II}\partial_IW,\\
        J_\asc&=-D^{(2)}_{\asc\asc}\partial_\asc W-D^{(2)}_{\asc\xi}\partial_\xi W,\qquad J_\xi=V_\xi W-D^{(2)}_{\asc\xi}\partial_\asc W-D^{(2)}_{\xi\xi}\partial_\xi W.
    \end{split}
    \end{align}
We have used equations~\eqref{eq:small-e-drift} and~\eqref{eq:small-e-diffusion} to simplify these expressions, in particular using the independence of the diffusion terms from most of the orbital elements to take them outside the partial derivatives.

Let us first consider the inclination $I$.
By definition, this is constrained to lie in the interval $[0,\uppi]$.
Unlike for the period $P$, a binary reaching one of the boundaries of this interval is not removed from the distribution; instead of absorbing boundary conditions, we have \emph{reflecting} boundary conditions, i.e., the probability current $J_I$ vanishes at both boundaries.
However, if the distribution is stationary then $J_I=\mathrm{constant}$, so the current must vanish everywhere on $[0,\uppi]$.
Setting $J_I=0$ in equation~\eqref{eq:probability-currents} gives
    \begin{equation}
        \partial_I\ln W=D^{(1)}_I/D^{(2)}_{II}=\cot I.
    \end{equation}
Integrating this we find $W\propto\sin I$, which corresponds to an isotropic distribution (since $\cos I$ is uniformly distributed).
This makes intuitive sense: on long timescales, GWB resonance causes the binary to \enquote{forget} its initial orbital plane, such that the resulting stationary distribution is spherically symmetric.
This also agrees with our finding in section~\ref{sec:KM-small-e-I} that $D^{(1)}_I=0$ at $I=\uppi/2$, which is the mean inclination of an isotropic distribution.

For $\asc$ and $\xi$ it is natural to impose periodic boundary conditions for the DF and for the probability currents,
    \begin{align}
    \begin{split}
        W(P,I,\asc,\xi,t)&=W(P,I,\asc+2\uppi,\xi,t)=W(P,I,\asc,\xi+2\uppi,t),\\
        J_i(P,I,\asc,\xi,t)&=J_i(P,I,\asc+2\uppi,\xi,t)=J_i(P,I,\asc,\xi+2\uppi,t).
    \end{split}
    \end{align}
Stationarity requires $\partial_\asc J_\asc=\partial_\xi J_\xi=0$.
By inspection, we see that this is achieved if $\partial_\asc W=\partial_\xi W=0$, so that the DF depends only on $P$ and $I$.
This corresponds to $\asc$ and $\xi$ being uniformly distributed, which also satisfies the periodic boundary conditions; $\asc$ then has zero probability current, while $\xi$ has a uniform negative current due to the deterministic drift $V_\xi$, which depends only on the period.

\section{Solving the full Fokker-Planck equation}
\label{sec:full-fpe}

We now consider the full FPE for all six orbital elements $(P,e,I,\asc,\omega,\eps)$.
Allowing nonzero eccentricity $e>0$ leads to much more complicated KM coefficients, and means that the results of section~\ref{sec:circular} are no longer applicable.
Nonetheless, we can use the fact that diffusion of the orbital elements due to the GWB takes place on very long timescales $\tau_\mathrm{diff}\sim1/(PH_0^2\Omega)\gg P$ (see figure~\ref{fig:timescales}).
This allows us to develop some useful approximate solution schemes for much shorter observational timescales.

\subsection{Perturbative short-time solution}

The FPE can be written as an operator equation
    \begin{equation}
        \pdv{W}{t}=L_\mathrm{FP}W,\qquad L_\mathrm{FP}(\vb*X)\equiv-\partial_iD^{(1)}_i+\partial_i\partial_jD^{(2)}_{ij}.
    \end{equation}
This has the formal solution\footnote{%
    Here we take advantage of the time-independence of the secular KM coefficients.
    For time-dependent coefficients, one would need to instead construct a time-ordered Dyson series~\cite{Risken:1989fpe}, though this gives the same result for short times.} $W(\vb*X,t)=\exp(tL_\mathrm{FP})W(\vb*X,0)$, which for short times $t\ll\tau_\mathrm{diff}$ can be expanded as
    \begin{equation}
    \label{eq:short-time-expansion}
        W(\vb*X,t)=\qty[1+tL_\mathrm{FP}(\vb*X,0)+\order{t/\tau_\mathrm{diff}}^2]W(\vb*X,0),
    \end{equation}
    where the right-hand side depends only on data at time zero.

Suppose that at time zero the binary's orbital elements take on the \enquote{sharp} values $x_i$.
The initial condition for the DF is then $W(\vb*X,0)=\delta^{(6)}(\vb*X-\vb*x)$.
By using a Fourier representation of the delta function, we can evaluate equation~\eqref{eq:short-time-expansion} to find~\cite{Risken:1989fpe}
    \begin{equation}
    \label{eq:short-time-solution}
        W(\vb*X,t)=\frac{1}{\sqrt{\det4\uppi tD^{(2)}}}\exp{-\frac{\qty[D^{(2)}]^{-1}_{ij}}{4t}\qty(X_i-x_i-D^{(1)}_it)\qty(X_j-x_j-D^{(1)}_jt)}+\order{\frac{t}{\tau_\mathrm{diff}}}^2,
    \end{equation}
    i.e., on short timescales, the DF is a multivariate Gaussian with mean $x_i+D^{(1)}_it$ and covariance matrix $2tD^{(2)}_{ij}$.
Here $[D^{(2)}]^{-1}_{ij}$ represents the elements of the inverse of the diffusion matrix, and both the drift vector and diffusion matrix are evaluated at $(\vb*X,t)=(\vb*x,0)$.

\subsection{Evolution of moments of the orbital elements}

The short-time expansion shows that on observational timescales the DF of the orbital elements is approximately Gaussian, and is therefore completely characterised by its first two moments: the mean and the covariance matrix, which we write as
    \begin{equation}
        \bar{X}_i\equiv\ev{X_i},\qquad C_{ij}\equiv\mathrm{Cov}[X_i,X_j]=\ev{\qty(X_i-\bar{X}_i)\qty(X_j-\bar{X}_j)}.
    \end{equation}
It is therefore useful to take moments of the FPE to find the time evolution of these quantities, rather than attempting to calculate the full time-dependent DF.
In doing so, we can calculate the backreaction of perturbations on the evolution of the binary, obtaining corrections to the linear growth found in equation~\eqref{eq:short-time-solution}.

The first moment of the FPE gives
    \begin{equation}
        \dv{\bar{X}_i}{t}=\pdv{}{t}\int\dd{\vb*X}X_iW=\int\dd{\vb*X}X_i\pdv{W}{t}=-\int\dd{\vb*X}X_i\partial_j\qty(D^{(1)}_jW)+\int\dd{\vb*X}X_i\partial_j\partial_k\qty(D^{(2)}_{jk}W).
    \end{equation}
We integrate by parts, and assume that the DF falls off fast enough that all boundary terms vanish, leaving
    \begin{equation}
    \label{eq:mean-evolution}
        \dv{\bar{X}_i}{t}=\int\dd{\vb*X}D^{(1)}_iW=\ev{D^{(1)}_i}.
    \end{equation}
This fall-off assumption is very reasonable here, as the diffusion rate is extremely small for realistic binaries, so the DF will only have support very near to the mean value.
Doing the same for second moment gives
    \begin{align}
    \begin{split}
        \dv{t}\ev{X_iX_j}&=\int\dd{\vb*X}X_iX_j\pdv{W}{t}=-\int\dd{\vb*X}X_iX_j\partial_k\qty(D^{(1)}_kW)+\int\dd{\vb*X}X_iX_j\partial_k\partial_\ell\qty(D^{(2)}_{k\ell}W)\\
        &=\int\dd{\vb*X}\qty(X_iD^{(1)}_j+X_jD^{(1)}_i)W+2\int\dd{\vb*X}D^{(2)}_{ij}W=\ev{X_iD^{(1)}_j}+\ev{X_jD^{(1)}_i}+2\ev{D^{(2)}_{ij}}.
    \end{split}
    \end{align}
We can combine these to give the evolution equation for the covariance matrix,
    \begin{align}
    \begin{split}
    \label{eq:covariance-evolution}
        \dv{C_{ij}}{t}&=\dv{t}(\ev{X_iX_j}-\bar{X}_i\bar{X}_j)=\ev{X_iD^{(1)}_j}+\ev{X_jD^{(1)}_i}+2\ev{D^{(2)}_{ij}}-\ev{X_i}\ev{D^{(1)}_j}-\ev{X_j}\ev{D^{(1)}_i}\\
        &=\mathrm{Cov}\qty[X_i,D^{(1)}_j]+\mathrm{Cov}\qty[X_j,D^{(1)}_i]+2\ev{D^{(2)}_{ij}}.
    \end{split}
    \end{align}

\subsection{The slow-diffusion approximation}

Equations~\eqref{eq:mean-evolution} and~\eqref{eq:covariance-evolution} fully describe the evolution of the mean and covariance of the orbital elements.
However, they are given in terms of ensemble averages over nonlinear functions of the orbital elements, which we cannot perform without knowing the full DF.
Even if we were to evaluate them approximately by assuming a Gaussian distribution, the resulting expressions would be very cumbersome.

In the case where the variance is small and any given orbital element $X_i$ is \enquote{close} to its mean value $\ev{X_i}$ (in a probabilistic sense), one can instead Taylor expand an arbitrary function of the elements around the mean,
    \begin{equation}
        f(\vb*X)=f(\bar{\vb*X})+\qty(X_i-\bar{X}_i)\partial_if(\bar{\vb*X})+\frac{1}{2}\qty(X_i-\bar{X}_i)\qty(X_j-\bar{X}_j)\partial_i\partial_jf(\bar{\vb*X})+\cdots,
    \end{equation}
    so that the mean of the function is approximated by
    \begin{equation}
    \label{eq:small-var-approx}
        \ev{f(\vb*X)}\simeq f(\bar{\vb*X})+\frac{1}{2}C_{ij}\partial_i\partial_jf(\bar{\vb*X}).
    \end{equation}
(Note that the first-order term in the expansion vanishes when taking the mean, as the first central moment is identically zero.)

\begin{figure}[p!]
    \thisfloatpagestyle{empty}
    \begin{center}
        \includegraphics[height=0.9\textwidth,angle=-90,origin=c]{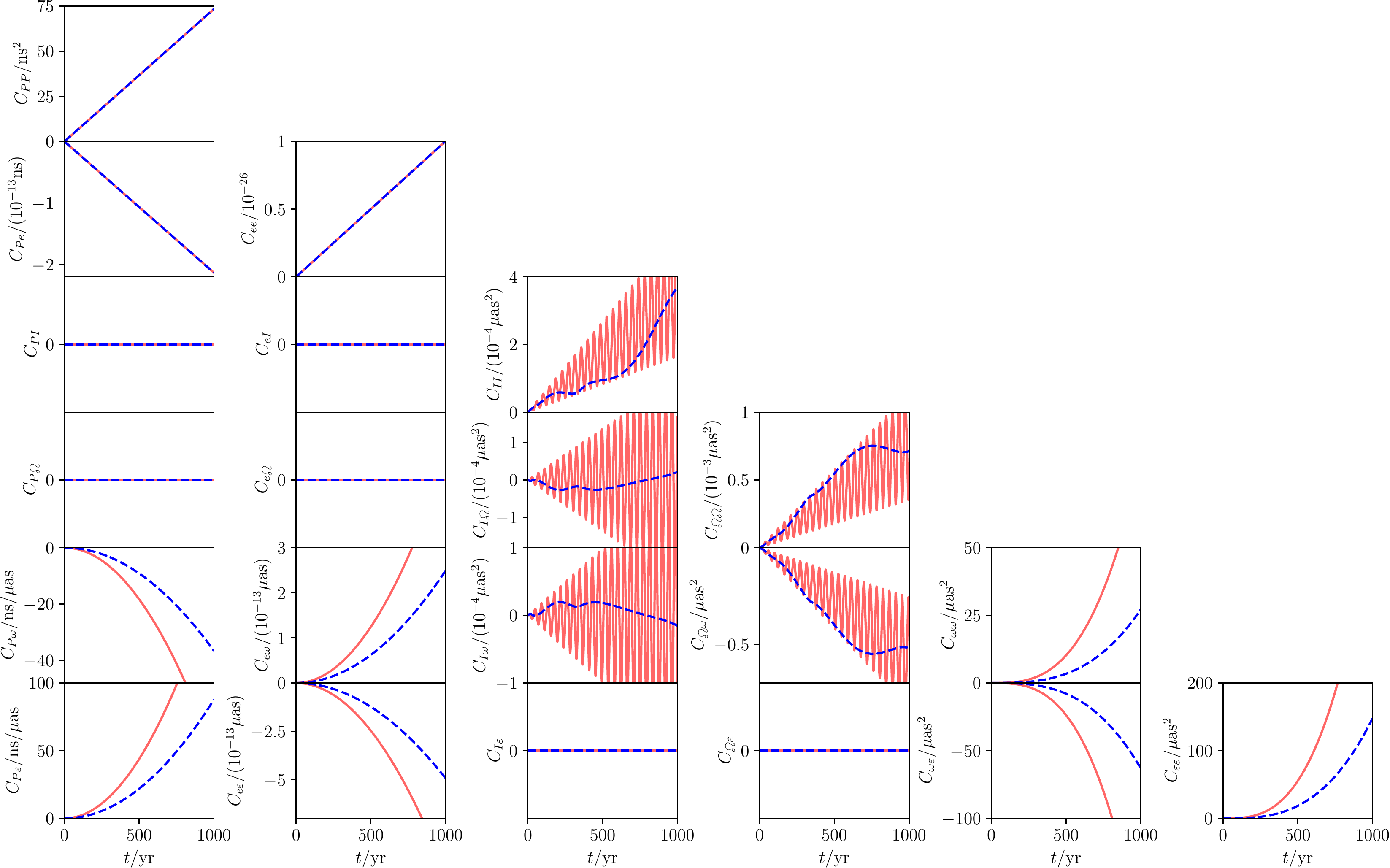}
    \end{center}
    \caption{%
    The covariance matrix $C_{ij}(t)$ of the orbital elements of the binary pulsar B1913+16 (the Hulse-Taylor system) over a $1\,\mathrm{kyr}$ interval, assuming a scale-invariant GWB spectrum $\Omega(f)=10^{-5}$, and including the first 400 harmonics.
    The blue dashed curves show numerical solutions of the evolution equations~\eqref{eq:slow-diffusion-evolution-equations}, while the pale red curves show the naive solution $t\times\dv*{C_{ij}}{t}$, which is exact only when $\dv*{C_{ij}}{t}=\mathrm{constant}$.}
    \label{fig:HT_cov_t}
\end{figure}

We can justify using equation~\eqref{eq:small-var-approx} by noting that the diffusion matrix calculated in section~\ref{sec:KM} is very small in most physical situations.
To keep track of how this smallness propagates into the evolution equations, we introduce a formal small parameter $\epsilon$ (which we will later set to unity), writing $D^{(2)}_{ij}\to\epsilon D^{(2)}_{ij}$.
For sharp initial conditions $\bar{X}_i=x_i$, $C_{ij}=0$, we see from equation\eqref{eq:covariance-evolution} that $C_{ij}=\order{\epsilon}$, as the $\mathrm{Cov}[\vb*X,D^{(1)}]$ terms are initially zero.
We therefore also write $C_{ij}\to\epsilon C_{ij}$.
We thus see that equation~\eqref{eq:small-var-approx} is justified if we neglect terms of order $\epsilon^2$.
We call this the \emph{slow-diffusion} approximation, as it relies on the fact that the timescale $\tau_\mathrm{diff}$ over which the covariance grows is long compared to the observation time.
Note that the stochastic contribution to the drift vector is generally of the same order as the diffusion matrix, so we also write
    \begin{equation}
        D^{(1)}_i=V_i+\epsilon\updelta D^{(1)}_i,
    \end{equation}
    where the stochastic term $\updelta D^{(1)}_i$ is suppressed by a factor of $\epsilon$.

Applying this approximation to equations~\eqref{eq:mean-evolution} and~\eqref{eq:covariance-evolution}, we obtain the moment evolution equations to first order in $\epsilon$,
    \begin{align}
    \begin{split}
    \label{eq:moment-evolution-slow-diffusion}
        \dv{\bar{X}_i}{t}&=V_i+\epsilon\updelta D^{(1)}_i+\frac{1}{2}\epsilon C_{jk}\partial_j\partial_kV_i+\order*{\epsilon^2},\\
        \epsilon\dv{C_{ij}}{t}&=2\epsilon D^{(2)}_{ij}+\epsilon C_{ik}\partial_kV_j+\epsilon C_{jk}\partial_kV_i+\order*{\epsilon^2},
    \end{split}
    \end{align}
    where the drift vector and diffusion matrix are both evaluated at $\bar{\vb*X}$.

We see that the evolution equation~\eqref{eq:moment-evolution-slow-diffusion} for the mean orbital elements includes both $\order{\epsilon^0}$ and $\order{\epsilon^1}$ terms.
For numerical reasons, it is convenient to separate these.
We therefore write the mean orbital elements as
    \begin{equation}
        \bar{\vb*X}(t)=\bar{\vb*X}_0(t)+\epsilon\updelta\bar{\vb*X}(t),
    \end{equation}
    where $\bar{\vb*X}_0$ represents the values the elements would take in the absence of the GWB, which obey the deterministic evolution equation
    \begin{equation}
        \dv{\bar{X}_{0,i}}{t}=V_i(\bar{\vb*X}_0).
    \end{equation}
(Note that this separation of deterministic and stochastic parts of the drift is always possible, regardless of the size of the deterministic term.)
Meanwhile, $\updelta\bar{\vb*X}$ represents the deviation in the mean due to GWB resonance, and evolves according to
    \begin{equation}
        \epsilon\dv{\updelta\bar{X}_i}{t}=V_i(\bar{\vb*X})-V_i(\bar{\vb*X}_0)+\epsilon\updelta D^{(1)}_i(\bar{\vb*X})+\frac{1}{2}\epsilon C_{jk}\partial_j\partial_kV_i(\bar{\vb*X})+\order*{\epsilon^2}.
    \end{equation}
We can Taylor-expand the terms that are evaluated at $\bar{\vb*X}=\bar{\vb*X}_0+\epsilon\updelta\bar{\vb*X}$ to give
    \begin{equation}
        \epsilon\dv{\updelta\bar{X}_i}{t}=\epsilon\updelta\bar{X}_j\partial_jV_i(\bar{\vb*X}_0)+\epsilon\updelta D^{(1)}_i(\bar{\vb*X}_0)+\frac{1}{2}\epsilon C_{jk}\partial_j\partial_kV_i(\bar{\vb*X}_0)+\order*{\epsilon^2}.
    \end{equation}

With the appropriate terms identified, we can send $\epsilon\to1$.
Our full set of evolution equations, to leading order in the slow-diffusion approximation, reads
    \begin{align}
    \begin{split}
    \label{eq:slow-diffusion-evolution-equations}
        \dv{\bar{X}_{0,i}}{t}&=V_i,\\
        \dv{\updelta\bar{X}_i}{t}&\simeq\updelta D^{(1)}_i+\updelta\bar{X}_j\partial_jV_i+\frac{1}{2}C_{jk}\partial_j\partial_kV_i,\\
        \dv{C_{ij}}{t}&\simeq2D^{(2)}_{ij}+C_{ik}\partial_kV_j+C_{jk}\partial_kV_i,
    \end{split}
    \end{align}
    with all terms evaluated at the deterministic mean elements $\bar{\vb*X}_0$.
By writing the FPE in this form, we have replaced a six-dimensional, second-order PDE with 33 coupled, one-dimensional, first-order ODEs: six each for the deterministic mean elements $\bar{X}_{0,i}$ and the perturbations $\updelta\bar{X}_i$, with the remaining 21 coming from the independent components of the $6\times6$ symmetric matrix $C_{ij}$.
In figures~\ref{fig:HT_cov_t} and~\ref{fig:HT_corner} and in section~\ref{sec:microhertz} below, we explore the behaviour of these equations numerically with a purpose-built Python code \texttt{gwresonance},\footnote{\url{https://github.com/alex-c-jenkins/gw-resonance}} which uses a fifth-order Runge-Kutta method, as implemented in the \texttt{scipy.integrate} library~\cite{Virtanen:2020sci}.

\begin{figure}[t!]
    \includegraphics[width=\textwidth]{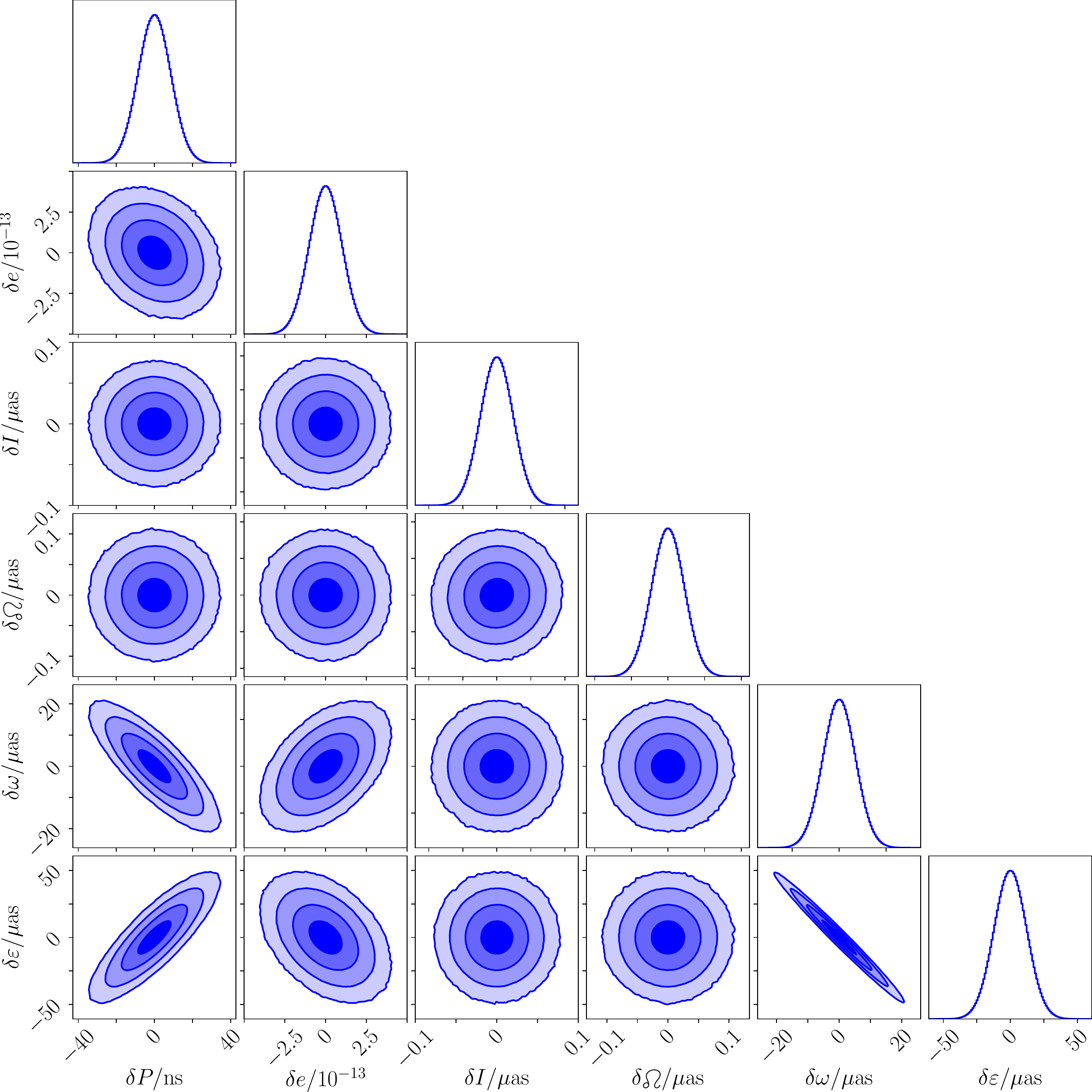}
    \caption{%
    Corner plot showing the distribution of the orbital elements of B1913+16 at the end of the $1\,\mathrm{kyr}$ numerical integration shown in figure~\ref{fig:HT_cov_t}.}
    \label{fig:HT_corner}
\end{figure}

It is interesting to note that, starting from sharp initial conditions, some elements of the covariance matrix remain fixed at zero under equation~\eqref{eq:slow-diffusion-evolution-equations}:
    \begin{equation}
    \label{eq:zero-cov}
        C_{PI}=C_{P\asc}=C_{eI}=C_{e\asc}=C_{I\eps}=C_{\asc\eps}=0,
    \end{equation}
    i.e., stochastic perturbations to these pairs of orbital elements remain statistically uncorrelated at first order in the slow-diffusion approximation.
This is due to the vanishing of the corresponding elements of the diffusion matrix $D^{(2)}_{ij}$ in equation~\eqref{eq:ecc-diffusion}, as well as the fact that $V_I=V_\asc=0$.
However, all other elements of the covariance matrix generically grow over time.

In figure~\ref{fig:HT_cov_t}, we show an example of an integration of the evolution equations~\eqref{eq:slow-diffusion-evolution-equations} for the Hulse-Taylor binary pulsar B1913+16~\cite{Hulse:1974eb}.
We see that, in this example, the period-eccentricity sector of the covariance matrix grows linearly with time,
    \begin{equation}
        C_{PP}\sim C_{Pe}\sim C_{ee}\sim t,
    \end{equation}
    indicating that the right-hand side of equation~\eqref{eq:slow-diffusion-evolution-equations} is dominated by the $D^{(2)}$ term for these components.
In contrast, many of the components involving the argument of pericentre and compensated mean anomaly (specifically $C_{P\omega}$, $C_{P\eps}$, $C_{e\omega}$, $C_{e\eps}$, $C_{\omega\omega}$, $C_{\omega\eps}$, and $C_{\eps\eps}$) have relatively smaller values for the diffusion matrix, and are instead driven by the $C\partial V$ terms, leading them to grow like $\sim t^2$ (since the components of the covariance matrix sourcing them grow like $C\sim t$).
The components $C_{II}$, $C_{I\asc}$, $C_{I\omega}$, $C_{\asc\asc}$, and $C_{\asc\omega}$ evolve more erratically; this is due to the presence of $\sin2\omega$ and $\cos2\omega$ terms in the corresponding components of the diffusion matrix (see equation~\eqref{eq:ecc-diffusion}), which oscillate on a timescale $\uppi/\ev{\dot{\omega}}_\mathrm{sec}\approx43\,\mathrm{yr}$ due to the perihelion precession of the Hulse-Taylor system~\cite{Weisberg:2010zz,Weisberg:2016jye}.
Finally, the remaining six components of the covariance matrix are zero throughout the integration time, as expected from equation~\eqref{eq:zero-cov}.

In figure~\ref{fig:HT_corner} we show the distribution of the orbital elements at the end of the integration in figure~\ref{fig:HT_cov_t}.
We see that the orbital elements are, on the whole, weakly correlated with each other, with the notable exceptions of the pairs $(P,\omega)$, $(P,e)$, and particularly $(\omega,\eps)$, which are highly covariant.
Note that these distributions include the overall offset due to stochastic drift, but that this is less important than diffusion in this case, and is therefore harder to distinguish by eye.

\subsection{Growth rate of non-Gaussianity}

By using equation~\eqref{eq:slow-diffusion-evolution-equations} we have neglected all higher-order moments of the distribution, and thus fail to capture any non-Gaussianities in the DF.
We can measure the departure from Gaussianity by tracking the evolution of the third central moment,
    \begin{equation}
        S_{ijk}\equiv\ev{(X_i-\bar{X}_i)(X_j-\bar{X}_j)(X_k-\bar{X}_k)},
    \end{equation}
    as this vanishes identically for a Gaussian distribution.
In particular, for each individual orbital element $X_i$ we have, to leading order in the slow-diffusion approximation,
    \begin{equation}
    \label{eq:skewness-evolution}
        \dot{S}_{iii}\simeq-3\bar{X}_iC_{ij}\partial_jV_i\, ,
    \end{equation}
    (with no summation over $i$).
Unlike the equations for the first two moments, the right-hand side of equation~\eqref{eq:skewness-evolution} is initially zero---non-Gaussianity is only sourced once the covariance has had a chance to grow.
We also see that non-Gaussianity can only be sourced at leading order for orbital elements which have a nonzero deterministic drift $V_i$; the inclination $I$ and longitude of ascending node $\asc$ remain Gaussian at leading order.
This justifies our assumption that the distribution is Gaussian on observational timescales.
However, studies of, e.g., the properties of old binary populations will require us to drop this assumption and solve for the full DF.
The results of section~\ref{sec:circular} are an important first step in studying cases like these.

\section{Statistical formalism for gravitational-wave searches}
\label{sec:observations}

In this section we describe how to estimate the sensitivity of a given binary to a generic GWB spectrum using two different high-precision probes of binary dynamics: timing of millisecond pulsars, and laser ranging experiments.
This will allow us to compute forecasts for the upper limits that each observational probe will be able to place on various GWB spectra.
To do so, we develop a novel likelihood-ratio statistic which accounts for the fact that not only the \emph{measured} orbital elements, but also the \emph{true} orbital elements, are random variables.

\subsection{Distribution of the observed orbital elements}

Consider a set of observations of a binary system, which are broken up into intervals much longer than the orbital period, but much shorter than the orbital diffusion timescale.
These observations give us a discrete series of measurements of the orbital elements $\vb*X(t)$, one for each interval, with the intervals being labelled by $t$.
(It is necessary to split up the data into intervals, since each high-precision measurement of the orbital elements requires a large number of individual data points.)
We assume these measurements are made with Gaussian noise and zero bias.
The measured values $\vu*X(t)$ are thus drawn from a multivariate Gaussian distribution centred on the true values $\vb*X(t)$, with log-likelihood $\mathcal{L}=\ln p$ given by
    \begin{equation}
        -2\mathcal{L}(\vu*X|\vb*X)=\sum_t\qty[\ln\det2\uppi\mathsf{M}+(\vu*X-\vb*X)^\mathsf{T}\mathsf{M}^{-1}(\vu*X-\vb*X)],
    \end{equation}
    where the measurement noise is described by the covariance matrix $\mathsf{M}$.
The form of the covariance matrix depends on how sensitive the timing residuals are to each orbital element---we discuss this further below.

The true orbital elements $\vb*X(t)$ are themselves random due to the uncertainty caused by GWB resonance.
As discussed in section~\ref{sec:full-fpe}, we can usually approximate the distribution of the elements as Gaussian (at least on observational timescales), so that the log-likelihood reads
    \begin{equation}
        -2\mathcal{L}(\vb*X|\Omega)=\sum_t\qty[\ln\det2\uppi\mathsf{C}+(\vb*X-\bar{\vb*X})^\mathsf{T}\mathsf{C}^{-1}(\vb*X-\bar{\vb*X})],
    \end{equation}
    where the mean values $\bar{\vb*X}(t)$ and covariance matrix $\mathsf{C}(t)$ both depend on the GWB spectrum $\Omega(f)$, and are computed as described in section~\ref{sec:full-fpe}.

We can marginalise over the unknown \enquote{true} elements $\vb*X(t)$ to obtain a likelihood function for the measured elements $\vu*X(t)$ for a given GWB spectrum,
    \begin{equation}
    \label{eq:likelihood}
        -2\mathcal{L}(\vu*X|\Omega)\equiv-2\ln\qty[\int\dd{\vb*X}p(\vu*X|\vb*X)p(\vb*X|\Omega)]=\sum_t\qty[\ln\det2\uppi\mathsf{N}+(\vu*X-\bar{\vb*X})^\mathsf{T}\mathsf{N}^{-1}(\vu*X-\bar{\vb*X})],
    \end{equation}
    where $\mathsf{N}\equiv\mathsf{M}+\mathsf{C}$ is the combined covariance matrix, incorporating the measurement uncertainty as well as the stochasticity of the orbital elements.

\subsection{Likelihood-ratio test}

Given a set of measured orbital elements $\vu*X$, we can phrase the GWB detection problem as a hypothesis test, where
    \begin{itemize}
        \item $\mathcal{H}_0$ is the \emph{null hypothesis}, that there is no GWB present, $\Omega=0$;
        \item $\mathcal{H}_\Omega$ is the \emph{alternative hypothesis}, that there is a GWB present, $\Omega\ne0$.
    \end{itemize}
The simplest version of this problem is when we are searching for a GWB with a fixed spectral shape (e.g., a power law), in which case the only unknown is a single parameter setting the overall amplitude, which we denote $\Omega$.
(Concretely, this parameter $\Omega$ should then refer to the amplitude of the GWB at some fixed reference frequency, c.f. equation~\eqref{eq:power-law-gwb}.)

A very natural way of carrying out such a hypothesis test is by using the log-likelihood-ratio statistic
    \begin{equation}
        \Lambda(\vu*X)\equiv2\max_{\Omega>0}\qty[\mathcal{L}(\vu*X|\Omega)-\mathcal{L}(\vu*X|0)],
    \end{equation}
    i.e., we compare the likelihood~\eqref{eq:likelihood} in the case of the null hypothesis $\Omega=0$ with the maximum value of the likelihood as a function of $\Omega$ in the case of the alternative hypothesis.
Since the maximum value of the likelihood over the entire range of values of $\Omega$ is always greater than or equal to its value at $\Omega=0$, we see that $\Lambda\ge0$.

Using equation~\eqref{eq:likelihood}, along with the fact that $\mathsf{N}=\mathsf{M}$ when $\Omega=0$, we find
    \begin{align}
    \label{eq:detection-statistic}
        \Lambda(\vu*X)=\max_{\Omega>0}\sum_t\qty[(\vu*X-\bar{\vb*X}_0)^\mathsf{T}\mathsf{M}^{-1}(\vu*X-\bar{\vb*X}_0)-(\vu*X-\bar{\vb*X})^\mathsf{T}\mathsf{N}^{-1}(\vu*X-\bar{\vb*X})-\ln\det(\mathsf{I}+\mathsf{M}^{-1}\mathsf{C})],
    \end{align}
    where $\bar{\vb*X}_0$ is the value of $\bar{\vb*X}$ when $\Omega=0$, and where we have used
    \begin{equation}
        \frac{\det\mathsf{N}}{\det\mathsf{M}}=\det\mathsf{M}^{-1}\mathsf{N}=\det(\mathsf{I}+\mathsf{M}^{-1}\mathsf{C}).
    \end{equation}
Given a set of measurements $\vu*X(t)$, we can thus compute the likelihood ratio statistic using equation~\eqref{eq:detection-statistic} by solving the FPE to find $\bar{\vb*X}(t)$ and $\mathsf{C}(t)$ for a large number of possible of values of $\Omega$ in order to maximise the likelihood.
If the resulting value of $\Lambda$ is large enough, then we reject the null hypothesis and claim a detection of the GWB.

An obvious question is:
What value of $\Lambda$ is \enquote{large enough}?
To define the detection threshold, we need to know the distribution of $\Lambda$ in the case where $\mathcal{H}_0$ is true.
Comparing the observed value of $\Lambda$ with this null distribution then allows us to directly infer the statistical significance of the results.
Since $\Lambda$ is a complicated function of the data $\vu*X$, it is difficult to find an exact distribution, even though we have fully specified how the data are distributed.
However, in the limit where we have a large number of measurements (i.e., the data cover a large number of time intervals $t$), Wilks' theorem tells us that $\Lambda$ follows a chi-squared distribution~\cite{Casella:2002stat}, $\lim_{n_t\to\infty}\Lambda\sim\chi^2_1$, where $n_t$ is the number of time segments.
(In this case the chi-squared distribution has one degree of freedom, as $\mathcal{H}_\Omega$ has one more free parameter than $\mathcal{H}_0$; we could imagine using a more complicated model for the GWB with $k$ parameters, in which case the appropriate distribution would be $\chi^2_k$.)
In this limit, we can therefore set the threshold for detecting the GWB at a given confidence level according to the corresponding $p$-value of $\chi^2_1$; e.g., a detection with 95\% confidence would require $\Lambda\ge3.841$.

\subsection{Sensitivity forecasts}

We can also use the likelihood-ratio statistic discussed above to estimate the sensitivity of future observing campaigns to different GWB spectra.
To do so, we simply compute the expected value of $\Lambda$ under the GWB hypothesis $\mathcal{H}_\Omega$, and find the smallest value of $\Omega$ for which this expectation surpasses the detection threshold---this tells us the weakest GWB that we can expect to detect with a given set of observations.

Let us use $\ev{\cdots}_\Omega$ to denote an expectation value under $\mathcal{H}_\Omega$.
By definition, we have
    \begin{align}
        \ev{\hat{X}_i}_\Omega=\bar{X}_i,\qquad\ev{\qty(\hat{X}_i-\bar{X}_i)\qty(\hat{X}_j-\bar{X}_j)}_\Omega=\mathsf{N}_{ij},
    \end{align}
    where the mean vector $\bar{\vb*X}$ and covariance matrix $\mathsf{C}$ here are computed using the true underlying value of $\Omega$.
In principle, is different from the value $\hat{\Omega}$ that maximises the likelihood, and it is the latter which determines the values for $\bar{\vb*X}$ and $\mathsf{C}$ that appear in the statistic $\Lambda$.
However, in the $n_t\to\infty$ limit discussed above we have $\hat{\Omega}\to\Omega$ (in statistics parlance, the maximum-likelihood estimator is efficient~\cite{Casella:2002stat}), so the two are interchangeable.
It is thus straightforward to show that, in this limit, the expected value of $\Lambda$ under $\mathcal{H}_\Omega$ is
    \begin{align}
    \begin{split}
    \label{eq:expected-lambda}
        \ev{\Lambda}_\Omega&=\sum_t\qty[\updelta\bar{\vb*X}^\mathsf{T}\mathsf{M}^{-1}\updelta\bar{\vb*X}+\tr\mathsf{M}^{-1}\mathsf{C}-\ln\det(\mathsf{I}+\mathsf{M}^{-1}\mathsf{C})].
    \end{split}
    \end{align}
We see that the stochastic drift in the orbital elements appears quadratically here, which means that it typically contributes less to the detectability of the GWB than diffusion, which enters linearly through the covariance matrix.

Once we have specified the covariance matrix $\mathsf{M}$ for our observations, we can compute equation~\eqref{eq:expected-lambda} and find the smallest value of $\Omega$ that we can expect to detect.

\subsection{Application to pulsar timing}

While the formalism we have developed above is applicable to a very broad class of astrophysical binary systems, one of the main applications is in the case where one member of the binary is a millisecond pulsar (MSP)~\cite{Lorimer:2008se}.
Analysis of the timing data from this MSP then allows us to determine its orbit with incredible precision, with uncertainties as small as a few parts per billion in some cases.
These precision measurements allow us to search for the very small stochastic perturbations to the orbit described above.
(The same principle has been used to set novel constraints on ultralight dark matter, due to its resonant effects on binary pulsar orbits~\cite{Blas:2016ddr,LopezNacir:2018epg,Blas:2019hxz,Armaleo:2019gil,Armaleo:2020yml,Desjacques:2020fdi}.)
Here we follow the approach of \citet{Blandford:1976pt} (see also \citet{Epstein:1977pn} and \citet{Damour:1986pn}) to estimate the covariance matrix $\mathsf{M}$ for the orbital elements of a binary pulsar.

For each of the $n_t$ observation intervals used to construct the likelihood ratio statistic above, one must observe some number $n_\obs$ of pulse arrival times, compare these arrival times with a timing formula for the binary, and thereby infer the binary's orbital elements at that time from the timing residuals.
In the limit $n_\obs\to\infty$, the resulting covariance matrix $\mathsf{M}$ describing the uncertainty in the orbital elements is given by the inverse of the \emph{Fisher matrix},
    \begin{equation}
        \mathsf{M}=\mathsf{F}^{-1},\qquad\mathsf{F}_{ij}\equiv-\ev{\partial_i\partial_j\mathcal{L}(\vb*t|\vb*X)},
    \end{equation}
    where $\mathcal{L}$ is the log-likelihood describing the distribution of arrival times $\vb*t=(t_1,t_2,\ldots,t_{n_\obs})$ as a function of the orbital elements $\vb*X$, and the angle brackets here denote an expectation value under that distribution.
We make the standard assumptions that the times-of-arrival (ToAs) of the pulses form a set of uncorrelated Gaussian random variables with constant variance $\sigma^2$ (i.e., the timing noise is independent of the binary's orbit), and with mean values given by the timing formula,
    \begin{equation}
        \ev{t_a}\equiv\mathcal{T}_a(\vb*X),\qquad\mathrm{Cov}[t_a,t_b]\equiv\delta_{ab}\sigma^2,
    \end{equation}
    so that the log-likelihood is given by
    \begin{equation}
        -2\mathcal{L}(\vb*t|\vb*X)=\sum_{a=1}^{n_\obs}\ln2\uppi\sigma^2+\frac{1}{\sigma^2}\qty(t_a-\mathcal{T}_a)^2.
    \end{equation}
The Fisher matrix for a likelihood of this form (as derived in, e.g., \citet{Tegmark:1996bz}) is
    \begin{equation}
    \label{eq:fisher-sum}
        \mathsf{F}_{ij}\equiv\frac{1}{\sigma^2}\sum_{a=1}^{n_\obs}\pdv{\mathcal{T}_a}{X_i}\pdv{\mathcal{T}_a}{X_j}.
    \end{equation}

To evaluate the Fisher matrix, we therefore need the derivatives of the timing formula $\mathcal{T}$ with respect to each of the orbital elements.
Using the standard Blandford-Teukolsky timing formula~\cite{Blandford:1976pt}, we write these as
    \begin{align}
    \begin{split}
    \label{eq:blandford-teukolsky}
        \pdv{\mathcal{T}}{P}&=\frac{v_P\sin I}{1+m_1/m_2}\frac{E-e\sin E}{2\uppi(1-e\cos E)}(\sin\omega\sin E-\gamma\cos\omega\cos E),\\
        \pdv{\mathcal{T}}{e}&=-\frac{P}{2\uppi}\frac{v_P\sin I}{1+m_1/m_2}\qty[\sin\omega(1+\sin^2E)+\frac{\cos\omega\sin E}{\gamma}(e-\gamma^2\cos E)],\\
        \pdv{\mathcal{T}}{I}&=\frac{P}{2\uppi}\frac{v_P\cos I}{1+m_1/m_2}\qty[\sin\omega(\cos E-e)+\gamma\cos\omega\sin E],\\
        \pdv{\mathcal{T}}{\asc}&=0,\\
        \pdv{\mathcal{T}}{\omega}&=\frac{P}{2\uppi}\frac{v_P\sin I}{1+m_1/m_2}\qty[\cos\omega(\cos E-e)-\gamma\sin\omega\sin E],\\
        \pdv{\mathcal{T}}{\eps}&=-\frac{P}{2\uppi}\frac{v_P\sin I}{1+m_1/m_2}\frac{\sin\omega\sin E-\gamma\cos\omega\cos E}{1-e\cos E},
    \end{split}
    \end{align}
    where the \emph{eccentric anomaly} $E(t)$ is defined by
    \begin{equation}
    \label{eq:eccentric-anomaly}
        r=a(1-e\cos E),\qquad\cos\psi=\frac{\cos E-e}{1-e\cos E},
    \end{equation}
    and acts as an alternative to the true anomaly $\psi$ as a way of parameterising the orbital ellipse.
We see that the timing formula does not depend on the longitude of ascending node, $\pdv*{\mathcal{T}}{\asc}=0$, which means that $\asc$ cannot be determined with pulsar timing (physically this is because $\asc$ corresponds to a rotation around the line-of-sight axis, and thus does not affect the observable motion parallel to the line of sight).

In the limit of many ToAs, $n_\obs\to\infty$, and assuming that these are distributed uniformly in time, we can replace the sum in equation~\eqref{eq:fisher-sum} with an integral to obtain
    \begin{equation}
    \label{eq:fisher-integral}
        \mathsf{F}_{ij}\simeq\frac{n_\obs}{T_\obs\sigma^2}\int_0^{T_\obs}\dd{t}\pdv{\mathcal{T}}{X_i}\pdv{\mathcal{T}}{X_j}\simeq\frac{n_\obs}{T_\obs\sigma^2}\frac{P}{2\uppi}\int_0^{2\uppi T_\obs/P}\dd{E}(1-e\cos E)\pdv{\mathcal{T}}{X_i}\pdv{\mathcal{T}}{X_j},
    \end{equation}
    where $T_\obs$ is the time interval over which the $n_\obs$ pulse measurements are made, and where we have used Kepler's equation,
    \begin{equation}
    \label{eq:keplers-equation}
        \dv{E}{t}=\frac{2\uppi/P}{1-e\cos E}.
    \end{equation}
For simplicity, we assume that $T_\obs$ is an integer multiple of the binary period $P$, and begins when the binary is at pericentre; if this is not the case, then the following formulae contain additional phase factors which do not affect the order of magnitude of the results.

We find that it is convenient to define
    \begin{equation}
        \mathsf{F}_{ij}=\frac{n_\obs P^2}{16\uppi^2\sigma^2}\qty(\frac{v_P\sin I}{1+m_1/m_2})^2\;\tilde{\mathsf{F}}_{ij},
    \end{equation}
    pulling out some factors which appear in all of the Fisher matrix elements.
Inserting the derivatives~\eqref{eq:blandford-teukolsky} into equation~\eqref{eq:fisher-integral}, we find that this factorised form of the Fisher matrix is given by

    \begin{align}
    \begin{split}
    \label{eq:msp-fisher-matrix}
        \tilde{\mathsf{F}}_{PP}&=\frac{8e}{P^2}(1+\tfrac{1}{4}e^2)+\frac{\cos2\omega}{P^2}f_1(e)+\frac{2\uppi T_\obs}{P^3}\sin2\omega f_2(e)+\frac{8\uppi^2T_\obs^2}{3P^4}\qty[1-\tfrac{1}{4}\cos2\omega f_3(e)],\\
        \tilde{\mathsf{F}}_{Pe}&=\frac{4}{P}\qty(1+\tfrac{3}{8}e)\qty(1-\tfrac{1}{6}e^2)+\frac{\uppi T_\obs e}{P^2\gamma}\sin2\omega-\frac{8\cos2\omega}{3P}\qty(1+\tfrac{3}{4}e+\tfrac{1}{6}e^2+\tfrac{1}{16}e^3),\\
        \tilde{\mathsf{F}}_{PI}&=\frac{\cot I}{P}\qty[2e\qty(1+\tfrac{1}{4}e)+\cos2\omega\qty(1-2e-\tfrac{3}{2}e^2)],\\
        \tilde{\mathsf{F}}_{P\omega}&=-\frac{\sin2\omega}{P}(1-2e-\tfrac{3}{2}e^2)-\frac{2\uppi T_\obs\gamma}{P^2},\\
        \tilde{\mathsf{F}}_{P\eps}&=-\frac{\sin2\omega}{P}f_2(e)-\frac{2\uppi T_\obs}{P^2}\qty[1-\tfrac{1}{4}\cos2\omega f_3(e)],\\
        \tilde{\mathsf{F}}_{ee}&=\frac{5}{\gamma^2}\qty[1-\tfrac{3}{4}e^2-\tfrac{1}{20}e^4-\tfrac{9}{10}\cos2\omega\qty(1-\tfrac{23}{18}e^2+\tfrac{1}{18}e^4)],\\
        \tilde{\mathsf{F}}_{eI}&=3e\cot I\qty[1+\tfrac{1}{12}e^2-\tfrac{11}{6}\cos2\omega\qty(1-\tfrac{1}{22}e^2)],\\
        \tilde{\mathsf{F}}_{e\omega}&=\tfrac{11}{2}e\sin2\omega\qty(1-\tfrac{1}{22}e^2),\qquad\tilde{\mathsf{F}}_{e\eps}=\frac{e}{\gamma}\sin2\omega,\\
        \tilde{\mathsf{F}}_{II}&=2\cot^2I\qty(1+\tfrac{3}{2}e^2-\tfrac{5}{2}e^2\cos2\omega),\qquad\tilde{\mathsf{F}}_{I\omega}=5e^2\cot I\sin2\omega,\qquad\tilde{\mathsf{F}}_{I\eps}=0,\\
        \tilde{\mathsf{F}}_{\omega\omega}&=2+3e^2+5e^2\cos2\omega,\qquad\tilde{\mathsf{F}}_{\omega\eps}=2\gamma,\qquad\tilde{\mathsf{F}}_{\eps\eps}=4\frac{\sin^2\omega+\gamma\cos^2\omega}{1+\gamma},
    \end{split}
    \end{align}
    where the $f_i(e)$ are functions of eccentricity, which are given to $\order{e^{14}}$ by
    \begin{align}
    \begin{split}
        f_1(e)&\simeq1+\tfrac{8}{9}e-\tfrac{3}{16}e^2-\tfrac{448}{225}e^3-\tfrac{175}{288}e^4-\tfrac{11584}{11025}e^5-\tfrac{2975}{9216}e^6-\tfrac{67264}{99225}e^7-\tfrac{96733}{460800}e^8\\
        &-\tfrac{5818432}{12006225}e^9-\tfrac{278579}{1843200}e^{10}-\tfrac{149726912}{405810405}e^{11}-\tfrac{20910823}{180633600}e^{12},\\
        f_2(e)&\simeq1+\tfrac{4}{3}e+\tfrac{5}{8}e^2+\tfrac{2}{15}e^3+\tfrac{1}{48}e^4+\tfrac{1}{210}e^5-\tfrac{29}{768}e^6-\tfrac{31}{1260}e^7-\tfrac{359}{7680}e^8\\
        &-\tfrac{3559}{110880}e^9-\tfrac{469}{10240}e^{10}-\tfrac{19087}{576576}e^{11}-\tfrac{6099}{143360}e^{12},\\
        f_3(e)&\simeq e^2+\tfrac{1}{2}e^4+\tfrac{5}{16}e^6+\tfrac{7}{32}e^8+\tfrac{21}{128}e^{10}+\tfrac{33}{256}e^{12}.
    \end{split}
    \end{align}
We have checked numerically that these expressions are accurate to within $\approx2\%$ even at large eccentricity, $e=0.95$; for smaller eccentricities, the accuracy is even better.

For binaries with small eccentricity ($e\lesssim10^{-3}$), we can change to the alternative orbital elements $(P,\zeta,\kappa,I,\asc,\xi)$ to find
    \begin{align}
    \begin{split}
        \tilde{\mathsf{F}}_{PP}&=\frac{1}{P^2}\qty[-\frac{\zeta^2-\kappa^2}{\zeta^2+\kappa^2}+\frac{4\uppi T_\obs}{P}\frac{\zeta\kappa}{\zeta^2+\kappa^2}+\frac{8\uppi^2T_\obs^2}{3P^2}],\qquad\tilde{\mathsf{F}}_{P\zeta}=\frac{4\zeta}{3P}\frac{5\zeta^2+6\kappa^2}{(\zeta^2+\kappa^2)^{3/2}},\\
        \tilde{\mathsf{F}}_{P\kappa}&=\frac{4\kappa}{3P}\frac{\kappa^2}{(\zeta^2+\kappa^2)^{3/2}},\qquad\tilde{\mathsf{F}}_{PI}=-\frac{\cot I}{P}\frac{\zeta^2-\kappa^2}{\zeta^2+\kappa^2},\qquad\tilde{\mathsf{F}}_{P\xi}=-\frac{2}{P}\qty(\frac{\zeta\kappa}{\zeta^2+\kappa^2}+\frac{\uppi T_\obs}{P}),\\
        \tilde{\mathsf{F}}_{\zeta\zeta}&=\frac{19}{2},\qquad\tilde{\mathsf{F}}_{\zeta\kappa}=\frac{3}{4}\zeta\kappa,\qquad\tilde{\mathsf{F}}_{\zeta I}=\frac{1}{2}\zeta\cot I\qty(15+\frac{2\zeta^2}{\zeta^2+\kappa^2}),\qquad\tilde{\mathsf{F}}_{\zeta\xi}=2\kappa\frac{\zeta^2-\kappa^2}{\zeta^2+\kappa^2},\\
        \tilde{\mathsf{F}}_{\kappa\kappa}&=\frac{1}{2},\qquad\tilde{\mathsf{F}}_{\kappa I}=-\frac{1}{2}\kappa\cot I\qty(5-\frac{2\zeta^2}{\zeta^2+\kappa^2}),\qquad\tilde{\mathsf{F}}_{\kappa\xi}=\zeta\qty(\frac{5}{2}-\frac{\zeta^2}{\zeta^2+\kappa^2}),\\
        \tilde{\mathsf{F}}_{II}&=2\cot^2I,\qquad\tilde{\mathsf{F}}_{I\xi}=0,\qquad\tilde{\mathsf{F}}_{\xi\xi}=2.
    \end{split}
    \end{align}

Assuming some values for the observation interval $T_\obs$, number of ToAs $n_\obs$, and rms timing noise $\sigma$, we can thus use the Fisher matrix elements given above to calculate the expected likelihood ratio~\eqref{eq:expected-lambda} for a given GWB spectrum, and therefore compute upper limit forecasts.

\subsection{Application to laser ranging}

Another extremely precise observational probe of binary dynamics is \emph{laser ranging} (LR), in which laser pulses are fired at retroreflectors on bodies orbiting the Earth; typically the Moon (Lunar Laser Ranging, LLR)~\cite{Murphy:2013qya} or artificial satellites (Satellite Laser Ranging, SLR)~\cite{Ciufolini:2016ntr}.
By measuring the round-trip times of these pulses, the size of the orbit can be tracked over time with millimetre precision.

We estimate the sensitivity of a generic LR experiment in a very similar way to our treatment of pulsar timing above.
Individual range measurements are assumed to be unbiased, with uncorrelated Gaussian noise of variance $\sigma^2$.
We write their mean value at time $t$ in terms of the eccentric anomaly~\eqref{eq:eccentric-anomaly}, and define the Fisher matrix in the limit of many uniformly-spaced observations analogously to equation~\eqref{eq:fisher-integral},
    \begin{equation}
    \label{eq:fisher-lr}
        \mathsf{F}_{ij}\simeq\frac{n_\obs}{T_\obs\sigma^2}\frac{P}{2\uppi}\int_0^{2\uppi T_\obs/P}\dd{E}(1-e\cos E)\pdv{r}{X_i}\pdv{r}{X_j}.
    \end{equation}
The necessary partial derivatives are given by
    \begin{equation}
        \label{eq:derivatives-lr}
        \pdv{r}{P}=\frac{2a}{3P}(1-e\cos E)-\frac{ae\sin E}{P}\frac{E-e\sin E}{1-e\cos E},\qquad\pdv{r}{e}=a\frac{e-\cos E}{1-e\cos E},\qquad\pdv{r}{\eps}=\frac{ae\sin E}{1-e\cos E},
    \end{equation}
    where we have used the time-integrated form of equation~\eqref{eq:keplers-equation},
    \begin{equation}
        E-e\sin E=\frac{2\uppi t}{P}+\eps.
    \end{equation}
Note that we have neglected the three angles describing the orientation of the orbital plane in space, $(I,\asc,\omega)$, as these are not directly measured by the round-trip times of the laser pulses.

Inserting equation~\eqref{eq:derivatives-lr} into equation~\eqref{eq:fisher-lr}, we obtain the Fisher matrix,
    \begin{align}
    \begin{split}
    \label{eq:lr-fisher-matrix}
        \mathsf{F}_{PP}&\simeq\frac{4n_\obs a^2}{9P^2\sigma^2}\bigg[\frac{3\uppi^2T_\obs^2}{2P^2}(e^2+\tfrac{1}{4}e^4)+1+3e+\tfrac{27}{16}e^2+4e^3+\tfrac{315}{256}e^4\bigg],\\
        \mathsf{F}_{Pe}&\simeq\frac{3n_\obs a^2}{4P\sigma^2}\qty(e+\tfrac{8}{9}e^2+\tfrac{5}{8}e^3+\tfrac{8}{45}e^4),\qquad\mathsf{F}_{P\eps}\simeq-\frac{\uppi n_\obs T_\obs a^2}{2P^2\sigma^2}\qty(e^2+\tfrac{1}{4}e^4),\\
        \mathsf{F}_{ee}&=\frac{n_\obs a^2}{\sigma^2}\qty(2-\gamma-\frac{1-\gamma}{e^2}),\qquad\mathsf{F}_{\eps\eps}=\frac{n_\obs a^2}{\sigma^2}\qty(1-\gamma).
    \end{split}
    \end{align}
We have neglected $\order{e^6}$ terms for the first three entries here, since the Moon and the artificial satellites that are typically tracked with SLR have small eccentricities (e.g., the Lunar eccentricity is $e\approx0.055$).

\section[Bridging the $\mu$Hz gap]{Bridging the $\upmu$Hz gap}
\label{sec:microhertz}

As we argued at the beginning of this chapter, the various gaps between the sensitive frequency bands of current and near-future GW experiments present a serious problem for GW astronomy, as failing to explore these gaps could mean missing out on discovering any number of exciting signals hiding at these frequencies.
A particularly large and well-known gap in the GW landscape occurs at roughly $10^{-7}$--$10^{-4}$~Hz, between the sensitive bands of pulsar timing arrays and future space-based interferometers such as LISA (see figure~\ref{fig:gwb-constraints}).
Accessing these frequencies is challenging, requiring \enquote{detectors} of astronomical scale that are nonetheless sensitive to the subtle effects of GWs.
For example, one proposal is to construct an extremely long-baseline space-based interferometer, roughly the size of the Earth's orbit around the Sun (the $\upmu$Ares concept~\cite{Sesana:2019vho}); however, such ideas are currently very futuristic.
Instead, we can take advantage of the fact that these $\upmu\mathrm{Hz}$ frequencies correspond to the orbital frequencies of binaries with periods ranging from days to years, allowing us to fill this gap with binary resonance searches.
For shorter periods on the order of hours, we can also begin to explore the LISA band in the decade before LISA flies.

In this section, we use the formalism developed in the preceding sections to explore the GWB constraints that are possible with current and near-future observational data.
Our main results are based on three different high-precision probes of binary orbital dynamics:
    \begin{description}
        \item[MSP:] Timing of binary millisecond pulsars (MSPs), with periods between $P\approx1.5\,\mathrm{hr}$ and $P\approx5.3\,\mathrm{yr}$~\cite{Manchester:2004bp};
        \item[LLR:] Laser-ranging measurements of the Moon's orbit around the Earth ($P\approx27\,\mathrm{days}$)~\cite{Murphy:2013qya};
        \item[SLR:] Laser-ranging measurements of the orbits of artificial satellites around the Earth, in particular the LAGEOS-1 satellite ($P\approx3.8\,\mathrm{hr}$)~\cite{Ciufolini:2016ntr}, as this has been regularly producing laser-ranging data for longer than any other satellite mission.
    \end{description}

To get a sense of how strong we can expect our forecast constraints to be, it is instructive to carry out a back-of-the-envelope calculation in which the rms perturbation to the orbital period after time $T$ is $\sigma_P=\sqrt{2TD^{(2)}_{PP}}$.
Taking the LLR case as an example, for a SGWB intensity $\Omega_\mathrm{gw}=10^{-5}$ and an observation period of $T=15\,\mathrm{yr}$, this gives $\sigma_P\sim1\,\upmu\mathrm{s}$.
This corresponds to a rms perturbation to the semi-major axis of $\sigma_a=(2a/3P)\sigma_P\sim0.1\,\mathrm{mm}$.
Given that each LLR `normal point' measurement determines the Earth-Moon distance to within $\sim3\,\mathrm{mm}$, we see that a campaign of $\sim1000$ such measurements should be capable of detecting this signal (assuming that the total measurement uncertainty scales like the inverse square root of the number of independent measurements).

To obtain our forecasts, we numerically evolve the evolution equations~\eqref{eq:slow-diffusion-evolution-equations} for the first and second moments of the orbital elements of each of these systems using our Python code \texttt{gwresonance}.
We start all integrations from sharp initial conditions, and include the first 400 harmonics when dynamically computing the KM coefficients (although the evolution is almost always dominated by the first three harmonics; c.f. figure~\ref{fig:combs}).
By integrating over the duration of a given observing campaign, and specifying the precision with which a given experiment can measure the orbital elements (via the Fisher matrix; either equation~\eqref{eq:msp-fisher-matrix} for LR or equation~\eqref{eq:lr-fisher-matrix} for MSPs), we can compute the expectation value of the likelihood-ratio statistic~\eqref{eq:expected-lambda}, $\ev{\Lambda}_\Omega$.
We thus estimate the detection threshold for a given experiment and for a given GWB power-law index $\alpha$ by finding the smallest GWB amplitude such that $\ev{\Lambda}_\Omega\ge3.841$, using a numerical root-finding procedure.
We then iterate this procedure over different power-law indices, $\alpha=-10,-9.75,-9.5,\ldots,+10$, to construct the PI curves for these searches~\cite{Thrane:2013oya}, as defined in section~\ref{sec:stochastic-searches}.
These curves represent the sensitivity of each binary to the GWB, under the assumption that the GWB spectrum is reasonably well-modelled as a power law with $|\alpha|\le10$ in that binary's sensitive frequency band.

For each of our binary resonance probes (MSP, LLR, and SLR), we calculate two PI curves: one which reflects the data currently available in 2021, and one which should be achievable by 2038, by which time LISA is expected to have completed its nominal 4-year mission~\cite{Amaro-Seoane:2017drz}.
Before we present the results of these calculations, we first discuss the details of how we model the sensitivity of each set of observations, some of the details of which are given in table~\ref{tab:binary-resonance-data}.

\subsection{Binary pulsars}

\begin{table}[p!]
    \begin{center}
    \begin{singlespacing}
        \rotatebox{-90}{
        \begin{tabular}{l l l l l l l l}
            & $P/\mathrm{day}$ & $e$ & $I/\mathrm{rad}$ & $\omega/\mathrm{rad}$ & $m_1/m_\odot$ & $m_2/m_\odot$ & $T_\obs/\mathrm{yr}$\\
            \hline
            \textbf{Binary pulsars} & & & & & & & \\
            J0737$-$3039A~\cite{Kramer:2006nb} & $\phantom{01234}0.1023$ & $0.08778$ & $1.548$ & $3.574$ & $1.338$ & $\phantom{1}1.249$ & $18$\\
            B1913$+$16~\cite{Weisberg:2016jye} & $\phantom{01234}0.3230$ & $0.6171$ & $0.8223$ & $5.106$ & $1.440$ & $\phantom{1}1.389$ & $46$\\
            B2127$+$11C~\cite{Jacoby:2006dy} & $\phantom{01234}0.3353$ & $0.6814$ & $1.047$ & $6.027$ & $1.358$ & $\phantom{1}1.354$ & $31$\\
            B1534$+$12~\cite{Fonseca:2014qla} & $\phantom{01234}0.4207$ & $0.2737$ & $1.347$ & $4.945$ & $1.333$ & $\phantom{1}1.346$ & $30$\\
            J1829$+$2456~\cite{Haniewicz:2020jro} & $\phantom{01234}1.176$ & $0.1391$ & $1.32$ & $4.013$ & $1.306$ & $\phantom{1}1.569$ & $17$\\
            J1439$-$5501~\cite{Lorimer:2008se} & $\phantom{01234}2.118$ & $0.00004985$ & $1.047$ & $4.831$ & $1.299$ & $\phantom{1}1.376$ & $17$\\
            B2303$+$46~\cite{vanKerkwijk:1999xj} & $\phantom{0123}12.34$ & $0.6584$ & $1.047$ & $0.6122$ & $1.16$ & $\phantom{1}1.37$ & $36$\\
            J0045$-$7319~\cite{Kaspi:1996pul} & $\phantom{0123}51.17$ & $0.8079$ & $0.77$ & $2.012$ & $1.4$ & $\phantom{1}8.8$ & $30$\\
            J1740$-$3052~\cite{Madsen:2012rs} & $\phantom{012}231.0$ & $0.5789$ & $0.93$ & $3.118$ & $1.4$ & $20.0$ & $20$\\
            B1259$-$63~\cite{Miller-Jones:2018waj} & $\phantom{01}1237.0$ & $0.8699$ & $2.69$ & $2.420$ & $1.35$ & $\phantom{1}4.140$ & $29$\\
            J1638$-$4725~\cite{Manchester:2004bp} & $\phantom{01}1941.0$ & $0.955$ & $1.047$ & $1.545$ & $1.35$ & $\phantom{1}8.079$ & $15$\\
            \hline
            \textbf{Laser ranging} & & & & & & & \\
            LLR (APOLLO)~\cite{Murphy:2013qya} & $\phantom{0123}27.32$ & $0.0549$ & $0.0899$ & --- & $3.003\times10^{-6}$ & $3.692\times10^{-8}$ & $15$\\
            SLR (LAGEOS-1)~\cite{Ciufolini:2016ntr,ilrs:lageos} & $\phantom{01234}0.1563$ & $0.0045$ & $1.917$ & --- & $3.003\times10^{-6}$ & $2.047\times10^{-28}$ & $46$\\
            \hline
            \textbf{Solar system}~\cite{Murray:2000ssd,jpl:sbdb} & & & & & & & \\
            Mercury & $\phantom{0123}87.97$ & $0.2056$ & $0.1108$ & --- & $1.0$ & $1.660\times10^{-7}$ & $4.5\times10^9$ \\
            Venus & $\phantom{012}224.7$ & $0.006773$ & $0.03760$ & --- & $1.0$ & $2.448\times10^{-6}$ & $4.5\times10^9$ \\
            Earth & $\phantom{012}365.3$ & $0.01671$ & $0.02743$ & --- & $1.0$ & $3.003\times10^{-6}$ & $4.5\times10^9$ \\
            Mars & $\phantom{012}687.0$ & $0.09341$ & $0.02847$ & --- & $1.0$ & $3.227\times10^{-7}$ & $4.5\times10^9$ \\
            Jupiter & $\phantom{01}4333.0$ & $0.04839$ & $0.005619$ & --- & $1.0$ & $9.545\times10^{-4}$ & $4.5\times10^9$ \\
            Saturn & $\phantom{0}10760.0$ & $0.05415$ & $0.01615$ & --- & $1.0$ & $2.858\times10^{-4}$ & $4.5\times10^9$ \\
            Uranus & $\phantom{0}30710.0$ & $0.04717$ & $0.01736$ & --- & $1.0$ & $4.365\times10^{-5}$ & $4.5\times10^9$ \\
            Neptune & $\phantom{0}60220.0$ & $0.008586$ & $0.01284$ & --- & $1.0$ & $5.150\times10^{-5}$ & $4.5\times10^9$ \\
            Pluto & $\phantom{0}90610.0$ & $0.2488$ & $0.2715$ & --- & $1.0$ & $6.385\times10^{-9}$ & $4.5\times10^9$ \\
            KBO 79360 Sila-Nunam & $105200.0$ & $0.007505$ & $0.03939$ & --- & $1.0$ & $5.451\times10^{-12}$ & $4.5\times10^9$ \\
            KBO 523678 & $112200.0$ & $0.007206$ & $0.06268$ & --- & $1.0$ & --- & $4.5\times10^9$
        \end{tabular}
        }
    \end{singlespacing}
    \end{center}
    \caption{%
        Table of masses and orbital elements of the binary systems used to generate our results in figures~\ref{fig:all-constraints}--\ref{fig:solar-system}.
        We show a maximum of four significant digits, though in many cases these quantities have been measured with much greater precision.
        For the binary pulsars, $m_1$ and $m_2$ refer to the masses of the pulsar and its companion respectively, while for the other binaries they are chosen so that $m_1\ge m_2$.}
    \label{tab:binary-resonance-data}
\end{table}

We extract the orbital elements of 322 binary MSPs from the ATNF pulsar catalogue\footnote{\url{https://www.atnf.csiro.au/research/pulsar/psrcat/}}~\cite{Manchester:2004bp}, discarding 106 due to incomplete information, as well as the extremely wide binary J2032+4127, whose 46~yr period~\cite{Lyne:2015oua} means that the system has completed less than one complete orbit since its discovery in 2009~\cite{Abdo:2010en}.
For the remaining 215 MSPs, we extract the period $P$, eccentricity $e$, and argument of pericentre $\omega$; for near-circular systems $e\le10^{-3}$ the latter two are replaced by the Laplace-Lagrange parameters $\zeta$, $\kappa$, as these are more numerically stable when $e$ is very small.
The strongest GW constraints typically come from binaries with longer periods, although the sensitivity also depends on the eccentricity and argument of pericentre in a more complicated way.

The inclinations of binary MSPs are generally poorly-determined due to a degeneracy with the (often unknown) masses of the pulsar and its companion.
For most of the 215 systems, we assume a pulsar mass of $m_p=1.35\,m_\odot$ and an inclination of $I=\uppi/3$, as this corresponds to the median value of the companion mass $m_c$, which we extract from the catalogue.
These assumptions are necessary to calculate our predictions for the orbital evolution, since the two masses set the deterministic decay of the period and eccentricity through GW radiation; in practice one would marginalise over the uncertainties in the masses, and propagate this uncertainty in to the model for the orbital evolution.
In order to refine our results, we replace these assumed values with more accurate mass and inclination determinations from the literature for the following MSPs, which produce the best GWB bounds from our sample: J0737$-$3039A~\cite{Kramer:2006nb} (the double pulsar), B1913$+$16~\cite{Weisberg:2016jye} (the Hulse-Taylor system), B2127$+$11C~\cite{Jacoby:2006dy}, B1534$+$12~\cite{Fonseca:2014qla}, J1829$+$2456~\cite{Haniewicz:2020jro}, J1439$-$5501~\cite{Lorimer:2008se}, B2303$+$46~\cite{vanKerkwijk:1999xj}, J0045$-$7319~\cite{Kaspi:1996pul}, J1903$+$0327~\cite{Freire:2010tf}, J1740$-$3052~\cite{Madsen:2012rs}, and B1259$-$63~\cite{Miller-Jones:2018waj}.

Using these orbital elements and masses, we integrate the moment evolution equations~\eqref{eq:slow-diffusion-evolution-equations} from sharp initial conditions, with the initial time set to the year in which each system was discovered.
With these details specified, the GWB sensitivity is then set by the number of ToAs per observing interval, $n_\obs$, and the rms timing noise $\sigma$ associated with each ToA.
We assume each ToA corresponds to a 10-minute pulse integration time.
For our 2021 sensitivity curves, we assume each system is monitored for two weeks every year, with ToAs being gathered for two hours every day within this period; this corresponds to the data cadence for B1913+16~\cite{Hui:2012yp}, and gives 168 ToAs per year.
We further assume $\sigma=1\,\upmu\mathrm{s}$.
For our 2038 sensitivity curves, we assume an observing campaign of 365 ToAs per year (i.e., 10 minutes of observations per pulsar per day) with $\sigma=80\,\mathrm{ns}$, which is the forecast 10-minute ToA uncertainty of next-generation radio telescopes like the SKA~\cite{Liu:2011cka}.

It is important to note that our 2038 bounds are based only on known pulsars.
However, the SKA and other future radio telescopes are expected to discover large numbers of new pulsars~\cite{Janssen:2014dka}, some of which may be in binaries with orbits that are particularly sensitive probes of GWB resonance.
We make no assumptions about these as-yet undiscovered pulsars, meaning that our 2038 bounds are conservative in this sense.

\subsection{Laser ranging experiments}

For our LLR results we use the Lunar orbital elements, Lunar mass, and Earth mass tabulated in \citet{Murray:2000ssd}.
We base our 2021 sensitivity calculations on the APOLLO experiment, which has been observing since 2006, collecting roughly 260 \enquote{normal point} range measurements per year with a rms uncertainty of $\sigma\approx3\,\mathrm{mm}$~\cite{Murphy:2013qya}.
For our 2038 sensitivity curve, we assume an observation campaign which collects 1040 normal points per year (four times the current level) with an order-of-magnitude improvement in precision, $\sigma=0.3\,\mathrm{mm}$ (this would likely require the installation of new retroreflectors on the Lunar surface~\cite{Murphy:2013qya}, as the degradation of the existing reflectors is currently the main impediment to LLR sensitivity improvements).
We emphasise that including only the APOLLO experiment represents a conservative estimate of LLR sensitivity, as this excludes other experiments which have been collecting LLR data since 1969 (albeit with much less precision than the APOLLO data).

For our SLR results we focus on the LAGEOS-I satellite, with a start date of 1976, and using the satellite mass and orbital elements tabulated on the International Laser Ranging Service LAGEOS webpage.\footnote{\url{https://ilrs.gsfc.nasa.gov/missions/satellite_missions/current_missions/lag1_general.html}}
We assume that 50,000 normal points are collected per year for our 2021 sensitivity curve~\cite{ilrs:lageos}, rising to 200,000 per year by 2038 (again, a factor of four increase), and assume the same normal point uncertainties as for LLR in both cases.

We note that any futuristic GW mission in the solar system focusing on the frequency band of interest here may face the challenge of modelling the gravity gradient noise from asteroids, as recently pointed out by \citet{Fedderke:2020yfy}.
However, this is several orders of magnitude too small to affect the forecasts we present here.
Nonetheless, this is indicative of the kinds of systematic uncertainties that are likely to impact our proposed laser-ranging GW searches.
In principle, any non-GW-induced evolution in the Moon's orbit can be modelled as part of the deterministic drift $V_i$, but this is likely to include nuisance terms which will need to be marginalised over, potentially reducing the sensitivity of the search.
For now we neglect these potential systematics, and leave a more detailed study of their impact on laser-ranging GW searches for future work.

\subsection{Results}

\begin{figure}[t!]
    \includegraphics[width=\textwidth]{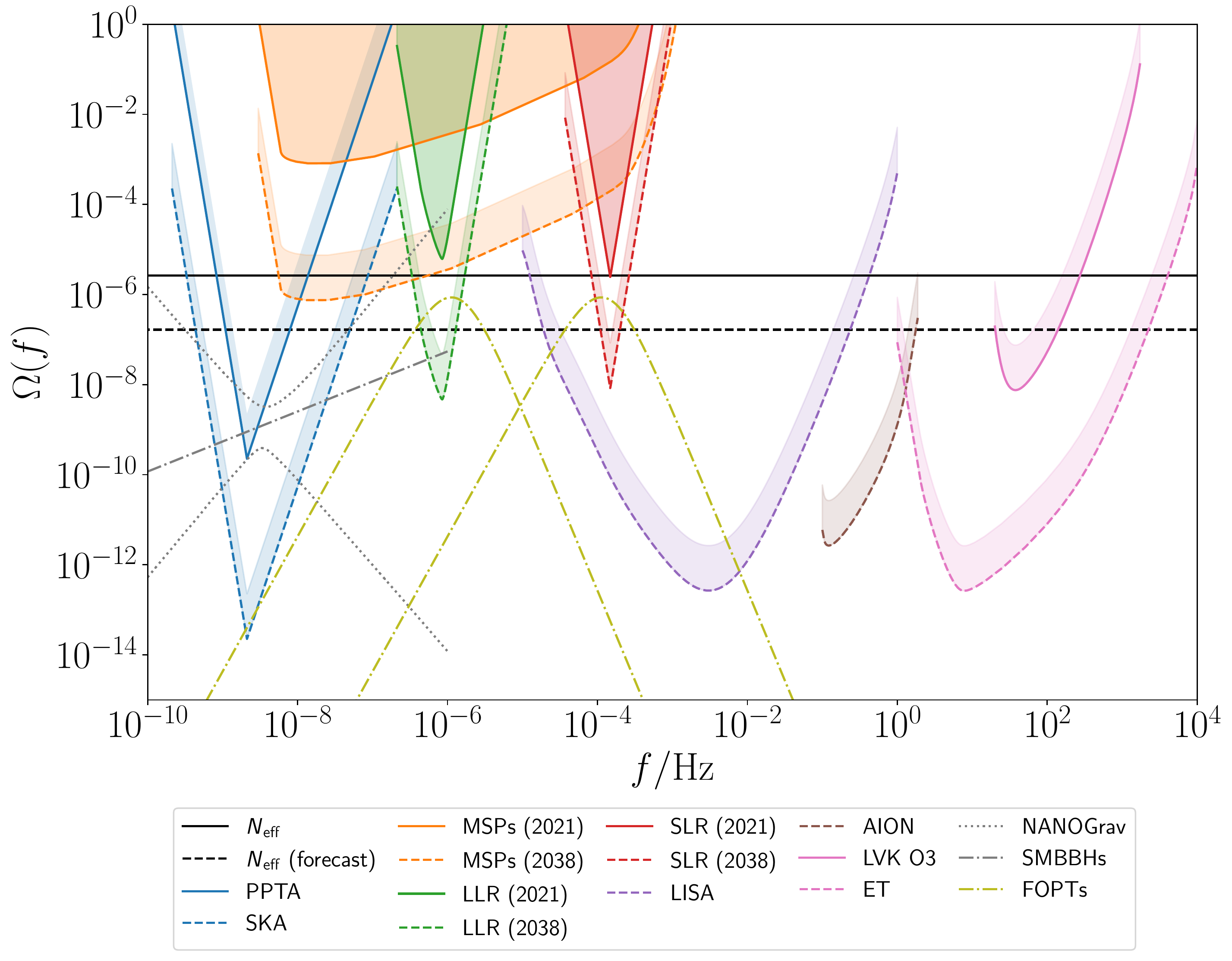}
    \caption{%
    Power-law integrated sensitivity curves of current and near-future GW experiments.
    Solid curves indicate existing results from the LIGO/Virgo/KAGRA Collaboration (LVK)~\cite{Abbott:2021kbb}, pulsar timing by the Parkes PTA~\cite{Lasky:2015lej}, and indirect constraints from $N_\mathrm{eff}$~\cite{Pagano:2015hma}, as well as expected present-day sensitivities of binary resonance searches with binary millisecond pulsars (MSPs), Lunar laser ranging (LLR), and satellite laser ranging (SLR), which are presented for the first time here.
    Dashed curves indicate our binary resonance forecast sensitivities for 2038, along with expected bounds from Einstein Telescope (ET)~\cite{Punturo:2010zz}, LISA~\cite{Amaro-Seoane:2017drz}, the Square Kilometre Array (SKA)~\cite{Janssen:2014dka}, and the proposed km-scale atom interferometer AION~\cite{Badurina:2019hst}, as well as improved $N_\mathrm{eff}$ constraints~\cite{Pagano:2015hma}.
    The grey dotted curves indicate a range of possible signals associated with the common process (CP) detected by NANOGrav~\cite{Arzoumanian:2020vkk}, while the overlaid dash-dotted curve shows the median inferred amplitude for the NANOGrav CP when assuming a $\Omega\sim f^{2/3}$ spectrum, as expected for inspiralling supermassive binary black holes (SMBBHs).
    The yellow dash-dotted curves show two first-order phase transition (FOPT) spectra at temperatures $T_*=2\,\mathrm{GeV}$ and $200\,\mathrm{GeV}$, peaking at $f\approx1\,\upmu\mathrm{Hz}$ and $\approx100\,\upmu\mathrm{Hz}$ respectively.}
    \label{fig:all-constraints}
\end{figure}

Our binary resonance PI curves are shown in figure~\ref{fig:all-constraints}, alongside the sensitivities of the current and near-future GW experiments discussed in section~\ref{sec:gw-detection}.
There are various other constraints at lower frequencies not shown here, including those from CMB B-mode searches~\cite{Ade:2015tva,Namikawa:2019tax} and CMB spectral distortions~\cite{Kite:2020uix}, as well as potential future constraints in the frequency band we are interested in, e.g. from astrometry~\cite{Moore:2017ity,Darling:2018hmc,Wang:2020pmf,Garcia-Bellido:2021zgu,Aoyama:2021xhj}, helioseismology~\cite{Lopes:2015pca}, modulation of \enquote{continuous-wave} GW signals~\cite{Bustamante-Rosell:2021daj}, the $\upmu$Ares concept~\cite{Sesana:2019vho}, the moon's normal modes~\cite{Harms:2020mma,Jani:2020gnz}, and high-cadence PTA observations~\cite{Perera:2018pts,Wang:2020hfh}.
However, all these constraints are either very futuristic, not applicable to stochastic GW signals, or not strong enough to be competitive with our forecasts.

\begin{figure}[t!]
    \begin{center}
        \includegraphics[width=0.667\textwidth]{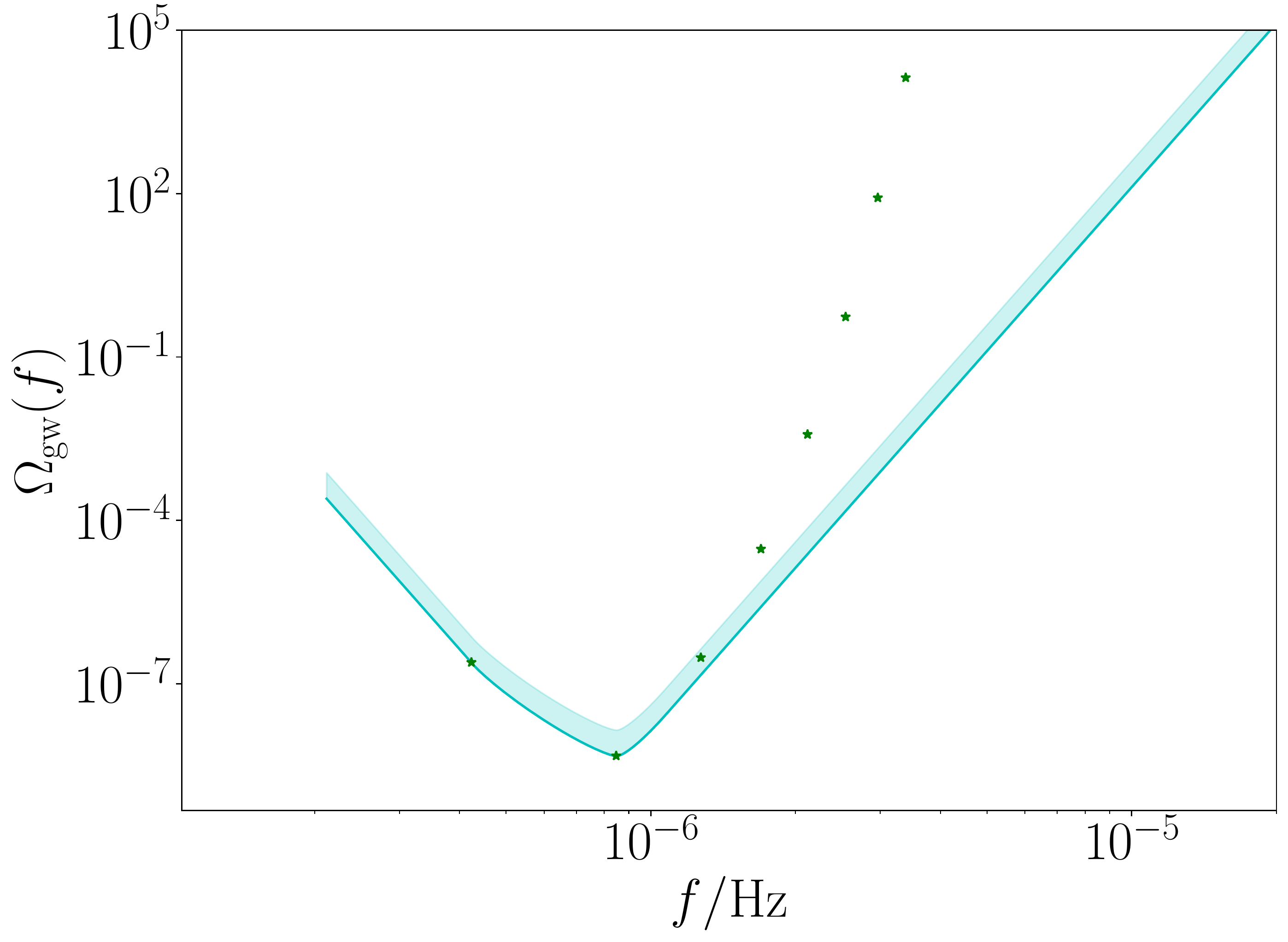}
    \end{center}
    \begin{center}
        \includegraphics[width=0.667\textwidth]{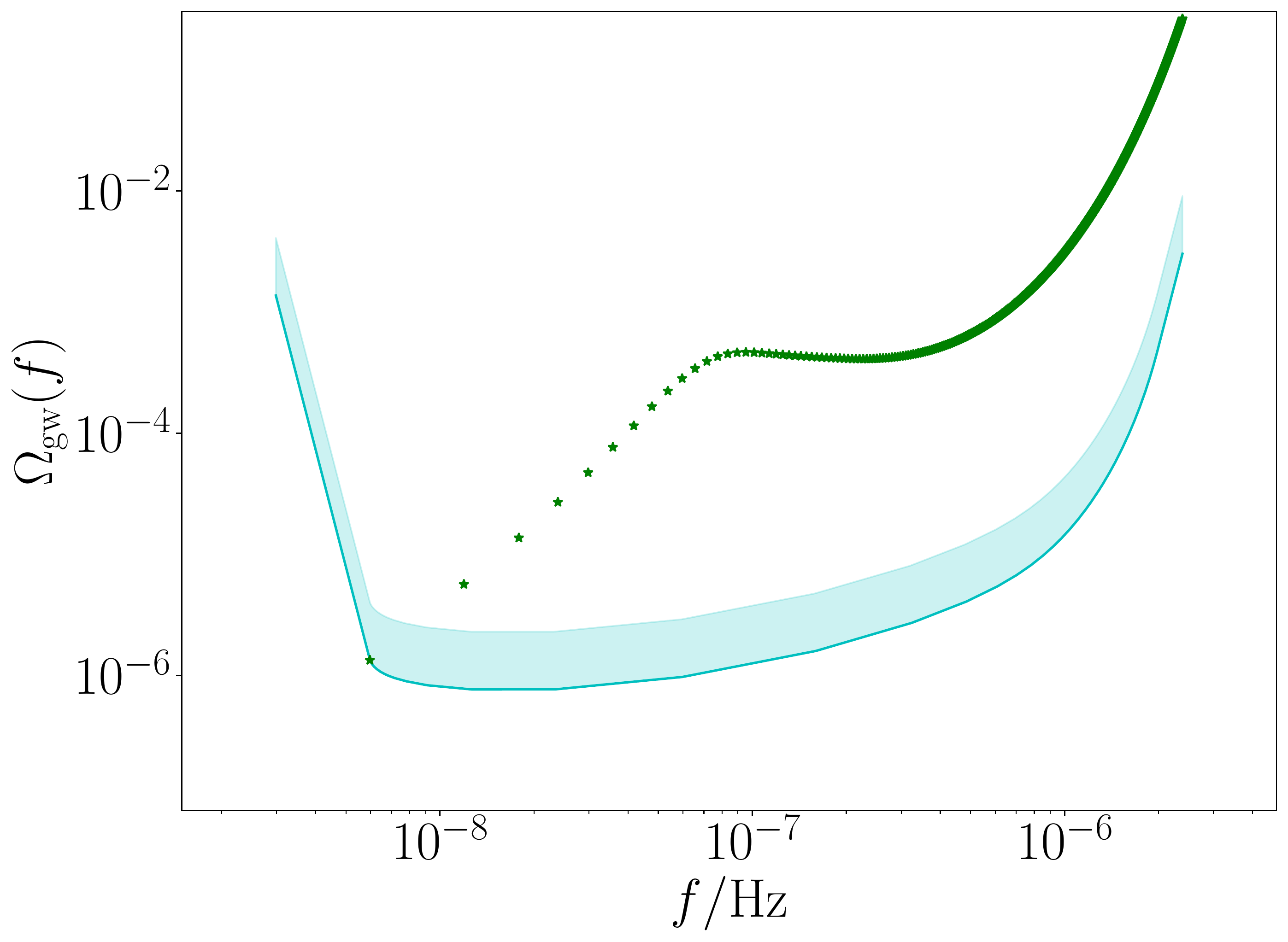}
    \end{center}
    \caption{%
    Comparison of the continuous PI curves (cyan) and discrete frequency \enquote{comb} constraints (green points) for two binary systems at 2038 sensitivity: the Earth-Moon system in the top panel, and the binary pulsar J1638-4725 in the bottom panel.
    The mismatch between the discrete harmonics and the PI curves is partly due to the collective constraining power of the large number of harmonics at high frequencies (where the linearly-spaced harmonics become more finely spaced on a logarithmic scale), and partly due to the range of power-law indices considered (e.g., the APOLLO constraints degrade faster than $\Omega\sim f^{10}$ for harmonics $n>3$).
    }
    \label{fig:combs}
\end{figure}

Figure~\ref{fig:combs} shows a couple of examples of how the shapes of these PI curves depend on the \enquote{comb} of independent constraints at each of the binary's resonant frequencies.
For low-eccentricity cases such as the Earth-Moon system ($e\approx0.055$), the $n=2$ harmonic is by far the most sensitive, giving a PI curve which is sharply peaked at this frequency.
On the other hand, high-eccentricity systems such as the binary pulsar J1638-4725 ($e\approx0.955$) can have sensitivity out to harmonics of order $n\sim100$ or more, giving much broader PI curves.

We find that laser-ranging experiments are already able to place cosmologically relevant bounds with present data; LLR has an expected sensitivity of $\Omega(f)\ge6.2\times10^{-6}$ at $f=0.85\,\upmu\mathrm{Hz}$ ($95\%$ confidence upper limit), while SLR with the LAGEOS satellite is sensitive to $\Omega(f)\ge2.4\times10^{-6}$ at $f=0.15\,\mathrm{mHz}$.
These represent by far the most sensitive direct GWB searches to date in the broad frequency band between ground-based interferometers at $f\gtrsim10\,\mathrm{Hz}$ and PTAs at $f\sim\mathrm{nHz}$, a full three orders of magnitude stronger than existing constraints from the Cassini spacecraft~\cite{Armstrong:2003ay} and the Earth's normal modes~\cite{Coughlin:2014xua}, and are competitive with the indirect, integrated $N_\mathrm{eff}$ constraint we discussed in section~\ref{sec:cosmological-gwb-bounds} which currently sets $\int\dd{(\ln f)}\Omega(f)\le2.6\times10^{-6}$ for $f\gtrsim10^{-15}$~\cite{Pagano:2015hma}.
By the end of the LISA mission in 2038, we expect these bounds to improve to $\Omega(f)\ge4.8\times10^{-9}$ for LLR and $\Omega(f)\ge8.3\times10^{-9}$ for SLR, significantly better than the $N_\mathrm{eff}$ constraint, which is expected to reach $\int\dd{(\ln f)}\Omega(f)\le1.7\times10^{-7}$ by that time~\cite{Pagano:2015hma}.

The frequencies $f=0.85\,\upmu\mathrm{Hz}$ and $f=0.15\,\mathrm{mHz}$ mentioned above correspond to the $n=2$ harmonics of the Earth-Moon and Earth-LAGEOS systems, respectively.
The corresponding sensitivity curves are strongly peaked in both cases, since the coupling to the $n=2$ harmonic is by far the strongest for low-eccentricity orbits like that of the Moon ($e\approx0.055$) and LAGEOS ($e\approx0.0045$), as we saw in section~\ref{sec:KM-small-e-I}.
The next most sensitive frequency in both cases is the $n=1$ harmonic, which is sensitive to $\Omega(f)\ge3.2\times10^{-4}$ for LLR and $\Omega(f)\ge2.2\times10^{-2}$ for SLR at present, improving to $\Omega(f)\ge2.5\times10^{-7}$ and $\Omega(f)\ge7.5\times10^{-5}$ respectively by 2038.
(See figure~\ref{fig:combs} for the individual sensitivities of each harmonic of the Earth-Moon system.)

\begin{figure}[t!]
    \begin{center}
        \includegraphics[width=0.667\textwidth]{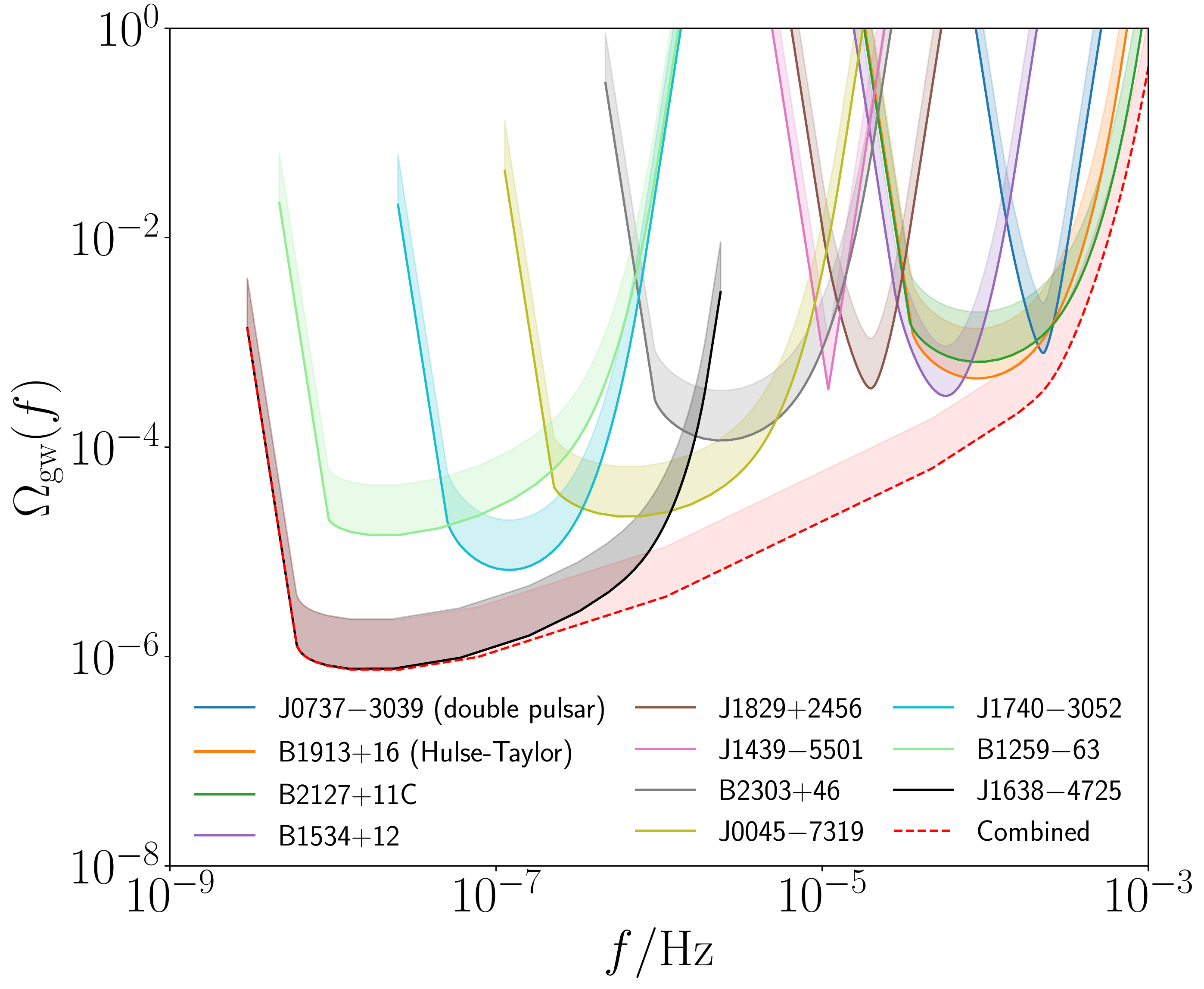}
    \end{center}
    \caption{%
    GWB PI curves from 11 binary pulsar systems at 2038 sensitivity.
    The red dashed curve shows the combined bound from these systems plus 204 others in the ATNF catalogue~\cite{Manchester:2004bp}, and corresponds to the red dashed curve in figure~\ref{fig:all-constraints}.
    The assorted shapes of the curves shown here depend on the binary orbital parameters (particularly the eccentricity), and illustrate the utility of our formalism in accurately capturing the response of each system to the GWB.}
    \label{fig:binary-pulsars-2038}
\end{figure}

While binary pulsars are not able to compete with the laser-ranging experiments in terms of sheer sensitivity, they do provide useful bounds across a much wider frequency band, spanning nearly five decades in frequency from $\approx6\,\mathrm{nHz}$ up to $\approx0.2\,\mathrm{mHz}$.
This is partly due to the broad range of orbital periods of various systems, and partly to the large eccentricities of many of these binaries, which gives them sensitivity to much higher harmonics.
The overall binary pulsar sensitivity curves shown in figure~\ref{fig:all-constraints} are computed by combining the overlapping PI curves of all 215 binaries that we extract from the ATNF pulsar catalogue~\cite{Manchester:2004bp}, as illustrated in figure~\ref{fig:binary-pulsars-2038}.
The most stringent sensitivity from this combined curve is $\Omega(f)\ge8.2\times10^{-4}$ at frequencies $f=14$--$25\,\mathrm{nHz}$ with present data, expected to reach $\Omega(f)\ge7.5\times10^{-7}$ by 2038.
These forecasts are much stronger than the upper limit set by \citet{Hui:2012yp}, which relied exclusively on data from the Hulse-Taylor system B1913+16; this makes sense in light of figure~\ref{fig:binary-pulsars-2038}, which shows us that this system does not produce particularly strong constraints compared to some of the other binaries in our sample.

\subsection{Implications for phase transitions and NANOGrav}

Figure~\ref{fig:all-constraints} also shows various potential GWB signals around the $\upmu$Hz band probed by our proposed binary resonance searches.
The most important to mention here are the phase transition spectra that we first discussed in section~\ref{sec:phase-transitions}, partly because FOPTs are a robust prediction of many well-motivated extensions to the Standard Model, and partly because the spectral shape of a FOPT signal highlights the constraining power of binary resonance searches.
While binary resonance probes are not competitive with GW interferometers and PTAs in searching for GWB spectra which are roughly flat over many decades in frequency (e.g., from cosmic strings), they can prove extremely useful for spectra that are confined to a narrow frequency band.
Here we focus on the GWB signal due to sound waves in the plasma, as described by equations~\eqref{eq:fopt-gwb-spectrum}--\eqref{eq:fopt-peak-intensity}, which is thought to be the dominant contribution for most FOPTs~\cite{Caprini:2019egz}.
As we mentioned previously, one also generally expects contributions from bubble wall collisions and from turbulent flows in the plasma.
These can only enhance the spectra we have investigated, meaning that our results here should be viewed as conservative bounds.

\begin{figure}[t!]
    \begin{center}
        \includegraphics[width=0.75\textwidth]{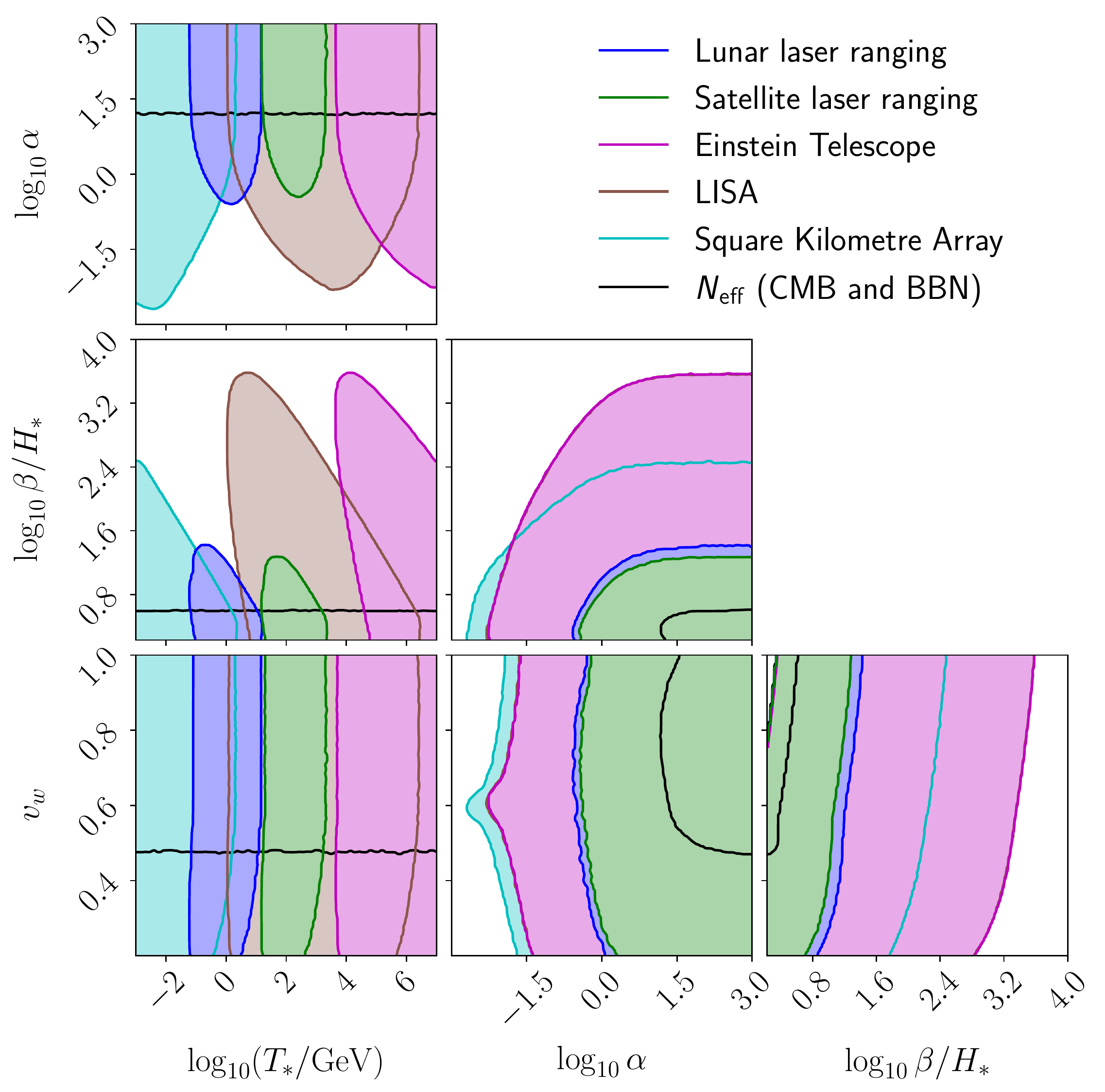}
    \end{center}
    \caption{%
    Exclusion regions of the FOPT parameter space for various GWB searches at 2038 sensitivity.
    Here $T_*$ is the temperature at which the FOPT occurs, $\alpha$ is the energy density released by the FOPT in units of the radiation density at the transition epoch, $\beta/H_*$ is the inverse duration of the transition in units of the Hubble rate at the transition epoch, and $v_w$ is the bubble wall velocity.
    }
    \label{fig:fopt}
\end{figure}

In figure~\ref{fig:fopt} we perform a scan over the FOPT parameters $(T_*,\alpha,\beta/H_*,v_w)$ for transitions occurring between $T_*=10^{-3}\,\mathrm{GeV}$ and $10^7\,\mathrm{GeV}$, identifying regions of parameter space where the corresponding GWB signal is expected to be detected by binary resonance searches and other GW probes at 2038 sensitivity.
We use equations~\eqref{eq:fopt-gwb-spectrum}--\eqref{eq:fopt-peak-intensity}, subject to the requirement that the mean bubble separation,
    \begin{equation}
    \label{eq:bubble-separation}
        R_*=\frac{(8\uppi)^{1/3}}{\beta}\max(v_w,c_\mathrm{s}),
    \end{equation}
    is smaller than the Hubble scale $1/H_*$ (with $c_\mathrm{s}=1/\sqrt{3}$ the speed of sound in the plasma).
For the efficiency parameter $\kappa$ which appears in equation~\eqref{eq:fopt-peak-intensity}, we use the fitting functions in the appendix of \citet{Espinosa:2010hh}, while for the sound wave lifetime we take~\cite{Ellis:2020awk}
    \begin{equation}
        \tau_\mathrm{sw}=R_*\times\qty(\frac{3\kappa}{4}\frac{\alpha}{1+\alpha})^{-1/2}.
    \end{equation}
In order to compute the $N_\mathrm{eff}$ constraints, we use the integrated form of this spectrum,
    \begin{equation}
        \int_{-\infty}^{+\infty}\dd{(\ln f)}\Omega(f)=\frac{343\sqrt{7/3}}{360}\Omega(f_*)\approx1.46\Omega(f_*).
    \end{equation}
(Strictly speaking this is an overestimate, as it includes frequencies $f\lesssim10^{-15}\,\mathrm{Hz}$ that do not contribute to $N_\mathrm{eff}$; however, this has negligible effect on the results in practice.)

We use the Markov chain Monte-Carlo sampler \texttt{emcee}\footnote{\url{https://emcee.readthedocs.io/en/stable/}}~\cite{ForemanMackey:2012ig} to explore the FOPT parameter space, using the following priors (and discarding any samples for which the mean bubble separation~\eqref{eq:bubble-separation} is larger than the horizon, $R_*H_*>1$).
    \begin{itemize}
        \item Transition temperature $T_*$: log-uniform in $[10^{-3},10^7]$ GeV.
        \item Transition strength $\alpha$: log-uniform in $[10^{-3},10^3]$.
        \item Inverse duration $\beta/H_*$: log-uniform in $[10^0,10^4]$.
        \item Bubble wall velocity $v_w$: uniform in $[0.2,1]$.
    \end{itemize}

The resulting exclusion regions in figure~\ref{fig:fopt} show FOPTs which can be detected at $\ge95\%$ confidence.
We find that LLR and SLR are able to probe significant regions of the FOPT parameter space at $T_*\sim\mathrm{GeV}$ and $\sim100\,\mathrm{GeV}$ respectively.
While SLR is less sensitive than LISA and will provide only complementary information, LLR will probe a region of the parameter space that is not accessible by any other planned GW experiment, thus providing a unique and valuable contribution to the search for phase transitions in the early Universe.
FOPTs are only one example of a strongly-peaked GWB spectrum, but they demonstrate that binary resonance searches (and LLR in particular) have unique GW discovery potential.

Another potential GWB signal shown in figure~\ref{fig:all-constraints} is the stochastic common process detected by NANOGrav in their 12.5-year PTA dataset~\cite{Arzoumanian:2020vkk}, which we discussed in section~\ref{sec:pulsar-timing}.
Assuming that the signal seen by NANOGrav is indeed due to GWs, and that its spectrum can be extrapolated into the $\upmu$Hz band, we find that present-day LLR data are able to probe some of the steeper spectra allowed by the NANOGrav data (roughly $\Omega\sim f^{1.8}$), which could correspond to a strongly blue-tilted inflationary tensor spectrum~\cite{Vagnozzi:2020gtf,Kuroyanagi:2020sfw}.\footnote{Such spectra can avoid the existing LVK and $N_\mathrm{eff}$ constraints if one allows for a nonstandard thermal history~\cite{Kuroyanagi:2020sfw}.}
If instead we assume that the NANOGrav signal follows the $\Omega\sim f^{2/3}$ scaling expected from inspiralling SMBBHs, we find that the spectrum should be detectable with 2038 LLR data.
This provides further motivation for the binary resonance searches we propose, showing that LLR can probe the nature of GW signals detected in the nHz band by NANOGrav and other PTAs.\footnote{%
    It is also possible for the spectrum to drop off before reaching $\upmu$Hz frequencies, so that one cannot naively extrapolate the spectrum from NANOGrav frequencies as we have here; however, in this scenario a non-detection of the GWB with LLR would reveal this drop-off, and would thus still provide valuable observational information.}

\subsection{Solar system bounds}

All of the binary resonance searches discussed above rely on precision measurements of orbital elements over observational timescales of years to decades.
However, our theoretical framework can also be used to study the GWB-induced evolution of binaries on much longer timescales, e.g. the evolution of planetary orbits since the formation of the Solar System $\sim4.5\,\mathrm{Gyr}$ ago.
This amplifies the size of the effect we are interested in, as the deviations in the orbital elements typically grow like the square root of the elapsed time.
However, this also entails a loss of precision, as the initial values of the orbital elements are unknown.

\begin{figure}[t!]
    \begin{center}
        \includegraphics[width=0.667\textwidth]{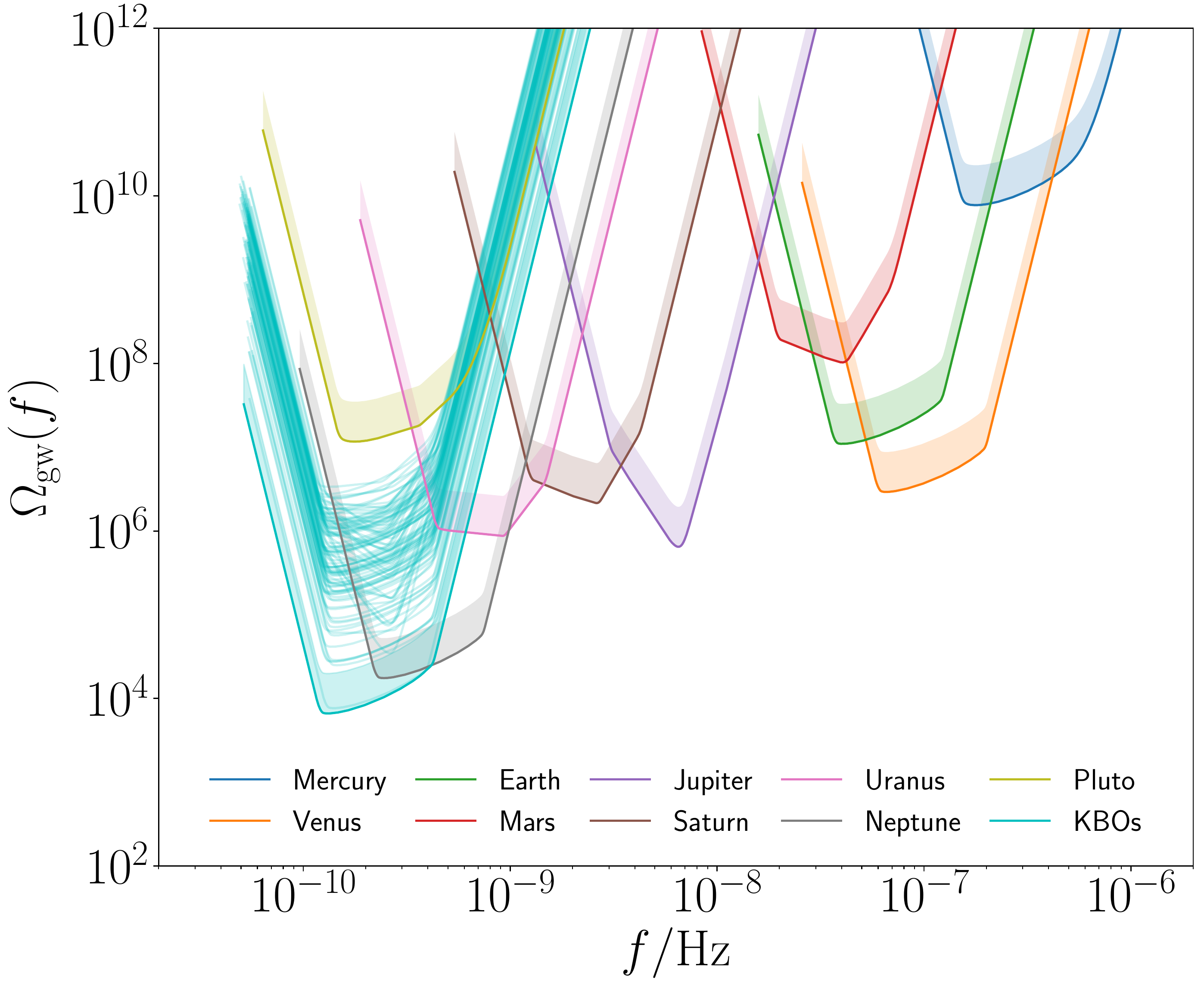}
    \end{center}
    \caption{%
    GWB PI curves inferred from the present-day orbital elements of various Solar System bodies.
    The faint cyan curves show constraints from 110 individual KBOs from the NASA/JPL Small-Body Database, while the solid cyan curve shows the combined KBO constraint.
    }
    \label{fig:solar-system}
\end{figure}

In figure~\ref{fig:solar-system} we show GWB constraints from the observed orbital elements of the eight Solar System planets, along with the dwarf planet Pluto and 110 classical Kuiper Belt Objects (KBOs).
We find that these are all orders of magnitude weaker than the precision binary resonance constraints possible with binary pulsars and laser ranging, with the strongest limit of $\Omega\le6.6\times10^3$ at $f=0.13\,\mathrm{nHz}$ coming from 523678 (2013 XB$_{26}$), a classical KBO on a very low-eccentricity orbit~\cite{jpl:sbdb}.

To produce these constraints, we integrate the evolution equations~\eqref{eq:slow-diffusion-evolution-equations} over the age of the solar system ($\sim4.5~\mathrm{Gyr}$), and compare the present-day periods, eccentricities, and inclinations of various solar system bodies to the rms changes in each of these predicted due to binary resonance,
    \begin{equation}
    \label{eq:rms-x}
        \sigma_i=\sqrt{\updelta\bar{X}_i^2+C_{ii}},
    \end{equation}
    (no summation over the repeated index).
Since the GWB tends to drive binaries towards longer periods, higher eccentricities, and larger inclinations, we can infer an upper limit on the GWB intensity by requiring equation~\eqref{eq:rms-x} to be less than the present-day values of each of these quantities.
In doing so, we account for the redshifting of GWs over cosmological timescales, setting
    \begin{equation}
        \Omega=\Omega_0\times(1+z)^4,\qquad f=f_0\times(1+z),
    \end{equation}
    with \enquote{0} subscripts denoting the present-day values that we place bounds on.
Since the solar system formed at redshift $z\approx0.41$, this can affect the final bounds by roughly a factor of $(1+z)^4\approx3.9$.
The redshifting of the GW frequency also broadens the shape of the resulting PI curve.

We extract the present-day orbital elements and masses of the eight planets and Pluto from \citet{Murray:2000ssd}, as well as those of 110 dynamically cold \enquote{classical} KBOs from the NASA/JPL Small-Body Database~\cite{jpl:sbdb}.\footnote{\url{https://ssd.jpl.nasa.gov/sbdb.cgi}}
The individual PI curves of the KBOs are combined to give an overall PI curve for the Kuiper Belt constraint, which is dominated by 523678 (2013 XB26) at low frequencies, and by 79360 Sila-Nunam (1997 CS$_{29}$) at high frequencies, primarily due to their low eccentricities $e\approx0.007$.

\section{Summary and outlook}
\label{sec:binary-resonance-summary}

In this chapter we have developed a powerful new formalism for calculating the statistical evolution of binary systems coupled to the GWB, deriving a secular FPE which captures the full probability distribution of all six orbital elements on timescales much longer than the binary period.
The KM coefficients describing this FPE, given in equations~\eqref{eq:ecc-drift} and~\eqref{eq:ecc-diffusion}, and illustrated in figures~\ref{fig:km1-P-ecc}--\ref{fig:km1-n}, encapsulate the rich dynamical structure that arises from the interactions between tensorial GW perturbations and elliptical orbits.

The full FPE is a six-dimensional nonlinear PDE, and it is therefore challenging to find exact solutions.
Nonetheless, we have extracted some qualitative features of the late-time behaviour in section~\ref{sec:circular} by fixing the eccentricity to zero.
This analysis illustrates one of the key advantages of our formalism over previous approaches: its ability to capture the full shape of the DF.
We find that, while the stochastic drift effect due to the SGWB tends to increase the energy of the binary and counteract its orbital decay through GW emission, diffusion tends to have the opposite effect, so that the net influence of SGWB resonance is to drive the binary towards merger.
At the same time, the GWB perturbations act to erase any memory of the binary's initial configuration in space, driving the orbit towards an isotropic distribution in the inclination $I$, and uniform distributions in the other angular variables.
We also find, however, that there is a strong drive towards larger eccentricities in the $e\to0$ limit, so these results with $e=0$ must be taken with a pinch of salt.

We have also developed, in section~\ref{sec:full-fpe}, a practical approach for numerically integrating the full FPE with general eccentricity $e\in(0,1)$ on observational timescales, taking advantage of the fact that these are typically much shorter than the diffusion timescale.
The resulting set of equations~\eqref{eq:slow-diffusion-evolution-equations} are illustrated in figures~\ref{fig:HT_cov_t} and~\ref{fig:HT_corner}, using the binary pulsar B1913+16 as an example.
Combined with the statistical tools developed in section~\ref{sec:observations}, this allows us to calculate GWB sensitivity curves for probes of binary dynamics such as pulsar timing and Lunar/satellite laser ranging.
We have presented these results in section~\ref{sec:microhertz}, showing that our methods can improve upon existing bounds in the $\upmu$Hz frequency band by several orders of magnitude, thereby constraining unique regions of the FOPT parameter space, and possibly shedding light on the nature of the NANOGrav common process signal.

These results motivate further work to develop binary resonance into a precision tool for GW astronomy.
The most important task will be to develop the necessary data analysis pipelines to conduct GW searches with pulsar-timing and laser-ranging data, in a way that fully utilises the theoretical developments in this work.
However, there also many possible avenues for developing our theoretical formalism further.
One important problem is to develop practical numerical integration schemes that go beyond the approach in section~\ref{sec:full-fpe} and capture non-Gaussian features in the DF, thereby taking full advantage of the Fokker-Planck approach (this will be particularly important for studies of populations of binaries~\cite{Barr:2016vxv,Hui:2018mkc}).
It would also be interesting to relax some of our assumptions, for example abandoning the secular-averaging approach and attempting to capture the evolution of binaries within a single orbital period, or perhaps relaxing some of the usual assumptions about the GW strain statistics to develop searches for GWBs that are non-Gaussian, anisotropic, or have nonstandard polarisation content.
One could also explore the sensitivity of binaries to narrowband sources, using a similar approach to \citet{Blas:2019hxz} to consider GW frequencies between the binary's resonant frequencies.
There is also no reason to restrict ourselves to just binaries; in future work, we plan to consider the GWB-driven evolution of other gravitationally-bound systems such as hierarchical triples or many-body systems such as galaxies and globular clusters.
Finally, our work could be extended even further by considering other stochastic perturbing fields which may exist in the Universe, such as ultralight scalars \cite{Foster:2017hbq,Kalaydzhyan:2018zsx,Centers:2019dyn,Dror:2021nyr}.


%% file: chapters/conclusion.tex
\chapter{Conclusion}
\label{chap:conclusion}

\epigraph{%
    \enquote{I may not have gone where I intended to go,\\
    but I think I have ended up where I needed to be.}
    }{Douglas Adams}


The goal of this thesis has been to investigate some of the many ways in which gravitational-wave observations can give us fascinating new insights into the fundamental laws that govern the Universe; from the statistical distribution of matter on the largest scales, down to the particles and interactions that make up the fabric of the cosmos.
In particular, we have focused on three exciting frontiers for GW cosmology: ($i$) probing late-time cosmology using anisotropies in the astrophysical GW background, ($ii$) exploring the intersection between nonlinear gravity and particle physics beyond the Standard Model with cosmic strings, and ($iii$) searching for first-order phase transitions and other cosmological GW signals with binary resonance.

We began in chapter~\ref{chap:anisotropies} by motivating the importance of treating the AGWB as anisotropic, bringing it into line with other cosmological observables such as the CMB, and thereby giving us access to novel information that cannot be extracted from the isotropic component alone.
The underlying hope here is that, since the compact binaries that make up the AGWB signal are formed from stellar evolution, they can act as tracers of the distribution of galaxies and matter on large scales, possibly revealing cosmological information complementary to that provided by existing observational probes.
With this aim in mind, we have developed a model for the angular power spectrum of the AGWB, combining well-motivated astrophysical recipes for computing the CBC rate density with a large and detailed mock galaxy catalogue based on the Millennium $N$-body simulation.
We find that the AGWB has strong angular fluctuations, ranging from a few percent on the largest angular scales up to nearly 100\% at higher multipoles $\ell\sim100$.
While these anisotropies are well below the sensitivity of present GW searches, we have shown that they are likely to be detected by next-generation experiments such as Einstein Telescope and Cosmic Explorer, providing a direct measurement of the statistical clustering of GW sources.
Unfortunately, the information that we will be able to extract from such a detection is likely to be limited by the presence of \emph{shot noise}, which we have shown is a natural consequence of the AGWB consisting of a finite number of transient signals.
Though we have developed an optimal data-analysis method for inferring the true angular power spectrum in the presence of shot noise, the uncertainty associated with these analyses is still much larger than the cosmic variance limit, making it difficult to extract precise cosmological information from the signal.

There are, however, several exciting roads ahead for anisotropic AGWB searches.
One approach---which we have already highlighted in section~\ref{sec:anisotropies-summary}---is to cross-correlate the AGWB with other cosmological probes, which has the dual advantage of boosting the detectability of the signal and reducing the amplitude of the shot noise.
While there have been several preliminary studies into such cross-correlation searches, there is still plenty of work to be done, both in modelling the expected signal and its dependence on various astrophysical and cosmological factors, and implementing efficient statistical methods for inferring the cross-correlation spectrum from GW strain data.
This future work may include going beyond two-point statistics such as the angular power spectrum, and trying to capture the full non-Gaussian statistics of the AGWB, perhaps by computing the bispectrum and trispectrum.
Another exciting possibility is to search for the \emph{kinematic dipole} in the AGWB (and, indeed, in other stochastic backgrounds).
As we showed in section~\ref{sec:inhomogeneous}, this dipole encodes information about the frequency spectra of the sources contributing to the background, so that, if measured, it could prove extremely useful in discriminating between different possible sources of a given GWB signal.

In chapter~\ref{chap:cosmic-strings}, we investigated the GW emission from Nambu-Goto cosmic strings, an extremely promising and well-motivated prediction of many models of particle physics beyond the Standard Model.
In particular, we calculated the \emph{nonlinear GW memory} associated with cusps and kinks on Nambu-Goto loops, i.e., the hereditary GW signal sourced by the energy-momentum of the initial cusp and kink GW emission.
If observed, this nonlinear memory would provide a fascinating test of the fundamental field-theoretic properties of general relativity, as well as probing the nonlinear gravitational dynamics of the cosmic strings themselves.
We have obtained simple analytical waveforms for the nonlinear memory emitted by cusps and kinks, including the \enquote{memory of the memory} and other higher-order effects.
Surprisingly, while the memory from kinks is strongly suppressed, the memory from cusps \emph{diverges} for sufficiently large loop sizes.
This divergence is unphysical, and, we have argued, indicates a breakdown in the weak-field treatment of Nambu-Goto loop dynamics that is ubiquitous in the literature, requiring some kind of strong-gravity mechanism to reduce the high-frequency GW emission from the cusp and thereby resolve the divergence.

In section~\ref{sec:cusp-collapse}, we have argued that one possible tentative resolution to the cusp divergence would be for the offending cusps to collapse and form primordial black holes.
The formation of an event horizon truncates the cusp's GW spectrum before the signal reaches its peak, which, as we have explicitly demonstrated, suppresses the cusp's memory emission and cures the divergence.
Using heuristic arguments based on the hoop conjecture, we show that such a situation arises very naturally for the exact same range of loop lengths that give rise to the memory divergence.
If these cusp-collapse PBHs do indeed form, we show that they are likely to be born with very small masses, large (but non-extremal) spins, and relativistic velocities, making them very distinct from all other known populations of BHs (both astrophysical and primordial).
Thus, while the nonlinear memory is not itself observable in the cusp-collapse scenario, the PBHs formed this way could act as a \enquote{smoking gun} signature of cosmic strings.

The most pressing next steps associated with this work are clearly to ascertain whether or not cusp collapse is the correct picture for resolving the nonlinear memory divergence, and if not, whether the nonlinear memory could in fact leave a substantial observational imprint on GW observations or not.
This will likely require numerical relativity simulations, as the relativistic velocities and lack of isometries associated with cusps makes detailed analytical treatments very challenging without resorting to the weak-field approximation.
Another interesting avenue to pursue would be to calculate the \emph{linear} GW memory associated with matter radiation from cusps, though this would require going beyond the simple Nambu-Goto approximation we have adopted here.

In chapter~\ref{chap:binary-resonance}, the final of our three forays into GW cosmology, we have developed an exciting new GW search method based on monitoring the orbits of binary systems.
The key principle is that impinging GWs with frequency equal to an integer multiple of the binary's orbital frequency interact resonantly with the system, leaving a lasting imprint on the size, shape, and orientation of the orbit.
Stochastic GW backgrounds are ideal targets for these searches, as they are \emph{persistent}, thus giving rise to perturbations in the orbital elements that grow over time, and \emph{broadband}, meaning that they are much more likely to overlap with a given binary's resonant frequencies than a narrowband signal.
However, due to the phase incoherence of the GWB, we are unable to deterministically model the corresponding evolution of the orbital elements over time.
We therefore develop, from first principles, a Fokker-Planck equation that tracks the time-dependent probability distribution of each of the orbital elements.

In section~\ref{sec:microhertz}, we have applied these theoretical tools to forecast the sensitivity of various observational probes of binary dynamics to the GWB, focusing on the timing of binary millisecond pulsars and laser ranging of the Moon and artificial satellites.
We have shown that these searches can reach impressive sensitivity levels in the \enquote{$\upmu$Hz gap} between the sensitivities of LISA and pulsar timing arrays, down to the level of $\Omega\sim10^{-8}$ by the end of LISA's nominal 4-year mission in the late 2030s.
This level of sensitivity would allow us to search for a range of exciting GW signals in this thus-far-unexplored frequency range; for example, ruling out a unique region of the parameter space for first-order phase transitions that cannot be reached by other planned GW experiments, as well as shedding some light on the potential GWB signal seen by NANOGrav in their 12.5-year dataset.

There are numerous exciting research directions that these binary resonance searches could take us in.
Perhaps the most important is to develop full data analysis pipelines for analysing pulsar-timing and laser-ranging data, allowing us to set upper limits on (or make detections of!) the GWB spectrum.
It would also be very interesting to extend our formalism to study how other gravitationally-bound systems, such as galaxies or stellar triples, are perturbed by the GWB.

Clearly, each of the three core topics in this thesis is ripe for further theoretical and observational development as we step further into the era of GW astronomy.
As existing detectors are upgraded, and next-generation instruments begin to come online, we can look forward to an ever-growing number of detected GW signals, and perhaps, eventually, the first signs of exotic new physics.
While it is not yet clear exactly what this fast-approaching treasure trove of GW data will teach us about cosmology and fundamental physics, we can be confident that there \emph{are} exciting discoveries in store, just as there have been every time humans have used a new tool for looking at the night sky.

%% file: chapters/appendices.tex
\begin{appendices}
\chapter{Directional GWB sensitivity forecast for the Einstein Telescope}
\label{sec:et}

In section~\ref{sec:agwb-sky} (and, in particular, figure~\ref{fig:C_ell-cbcs}), we showed that the lowest multipoles of the AGWB angular power spectrum are likely to be detected by the Einstein Telescope, a third-generation ground-based GW interferometer.
This statement depends purely on the detector geometry and expected noise spectrum of the ET proposal, and is relatively straightforward to derive.
However, since, to my knowledge, this calculation has not been presented explicitly anywhere in the literature, I include it here for completeness.
We focus purely on the directional GWB searches that are possible by cross-correlating data between each of ET's three interferometers (ignoring for now the potential issue of correlated noise sources), and use the formalism of \citet{Thrane:2009fp} to derive the corresponding Fisher matrix for the $C_\ell$'s.
In practice, ET will likely be joined by other third-generation observatories such as Cosmic Explorer~\cite{Hall:2020dps,Reitze:2019iox}, and it should therefore be possible to increase the sensitivity further (both in terms of amplitude and in terms of the number of accessible multipoles) by conducting cross-correlation searches between these different detectors.
However, the resulting sensitivity then depends on the relative locations and orientations of the detectors, which are not yet known, so we focus for now on searches using only ET data.

The \emph{directional} overlap reduction function between two Michelson interferometers operating in the small-antenna limit $f\ll1/L$ (where $L$ is the arm length) is given by~\cite{Romano:2016dpx}
    \begin{equation}
        \gamma_{IJ}(f,\vu*r)=\frac{1}{2}\mathcal{D}_I^A(\vu*r)\mathcal{D}_J^A(\vu*r)\cos[2\uppi f\vu*r\vdot\upDelta\vb*x],
    \end{equation}
    which is just the integrand of the isotropic ORF~\eqref{eq:orf} we encountered in section~\ref{sec:stochastic-searches}, $\Gamma_{IJ}(f)=\int_{S^2}\dd[2]{\vu*r}/(4\uppi)\gamma_{IJ}(f,\vu*r)$.
The detector response functions $\mathcal{D}_I^A$ encode the coupling of interferometer $I$ to a plane wave with polarisation $A$ arriving from the direction $\vu*r$.
These are given in terms of the unit vectors $\vu*u_I$ and $\vu*v_I$ that point along each of the arms,
    \begin{equation}
        \mathcal{D}_I^A=\frac{1}{2}e^A_{ij}(\hat{u}^i_I\hat{u}^j_I-\hat{v}^i_I\hat{v}^j_I),
    \end{equation}
    where $e^A_{ij}$ are the polarisation tensors~\eqref{eq:polarisation-tensors}, which are given in terms of the standard spherical polar unit vectors $\vu*\theta$, $\vu*\phi$,
    \begin{equation}
        e^+_{ij}=\hat{\theta}_i\hat{\theta}_j-\hat{\phi}_i\hat{\phi}_j,\qquad e^\times_{ij}=2\hat{\theta}_{(i}\hat{\phi}_{j)}.
    \end{equation}
For ET, we have three co-located (i.e. $\upDelta\vb*x=0$) interferometers, each having two arms which make up two sides of an equilateral triangle~\cite{Punturo:2010zz}.
We can therefore write
    \begin{equation}
        \mathcal{D}_1^A=\frac{1}{2}e^A_{ij}(\hat{u}^i_1\hat{u}^j_1-\hat{u}^i_2\hat{u}^j_2),\qquad \mathcal{D}_2^A=\frac{1}{2}e^A_{ij}(\hat{u}^i_2\hat{u}^j_2-\hat{u}^i_3\hat{u}^j_3),\qquad \mathcal{D}_3^A=\frac{1}{2}e^A_{ij}(\hat{u}^i_3\hat{u}^j_3-\hat{u}^i_1\hat{u}^j_1),
    \end{equation}
    and define a Cartesian coordinate frame in which the arm vectors are given by
    \begin{equation}
        \vu*u_1=(1,0,0),\qquad\vu*u_2=-\frac{1}{2}(1,\sqrt{3},0),\qquad\vu*u_3=-\frac{1}{2}(1,-\sqrt{3},0),
    \end{equation}
In the same Cartesian basis, we have
    \begin{equation}
        \vu*\theta=(\cos\theta\cos\phi,\cos\theta\sin\phi,-\sin\theta),\qquad\vu*\phi=(-\sin\phi,\cos\phi,0).
    \end{equation}
The three overlap reduction functions are therefore
    \begin{align}
    \begin{split}
    \label{eq:orfs}
        \gamma_{12}&=-\frac{3}{128}(1+\cos^2\theta)^{2}\Re{1+2\rme^{4\rmi\phi+2\uppi\rmi/3}}-\frac{3}{32}\cos^2\theta\Re{1-2\rme^{4\rmi\phi+2\uppi\rmi/3}},\\
        \gamma_{13}&=-\frac{3}{128}(1+\cos^2\theta)^2\Re{1+2\rme^{4\rmi\phi}}-\frac{3}{32}\cos^2\theta\Re{1-2\rme^{4\rmi\phi}},\\
        \gamma_{23}&=-\frac{3}{128}(1+\cos^2\theta)^{2}\Re{1+2\rme^{4\rmi\phi-2\uppi\rmi/3}}-\frac{3}{32}\cos^2\theta\Re{1-2\rme^{4\rmi\phi-2\uppi\rmi/3}}.
    \end{split}
    \end{align}
Note that the fact that the detectors are co-located makes the ORFs frequency-independent (up until the point where the small-antenna approximation breaks down, $f\approx1/L\approx1/(10\,\mathrm{km})\approx30\,\mathrm{kHz}$).
We are interested in ET's response to different spherical harmonics of the GWB, so we decompose equation~\eqref{eq:orfs} into spherical harmonic components, $[\gamma_{IJ}]_{\ell m}\equiv\int_{S^2}\dd[2]{\vu*n}Y_{\ell m}(\vu*n)\gamma_{IJ}(\vu*n)$.
The resulting nonzero components are shown in table~\ref{tab:orfs}.
We see that each ET pair is only sensitive to four independent spherical harmonic components of the GWB: those corresponding to $(\ell,m)$ values of $(0,0)$, $(2,0)$, $(4,0)$, and $(4,\pm4)$.

\begin{table}[t!]
    \begin{center}
    \begin{tabular}{c | c c c}
        $(\ell,m)$ & $[\gamma_{12}]_{\ell m}$ & $[\gamma_{13}]_{\ell m}$ & $[\gamma_{23}]_{\ell m}$ \\ [0.5ex]
        \hline
        $(0,0)$ & $-\dfrac{3\sqrt{\uppi}}{20}$ & $-\dfrac{3\sqrt{\uppi}}{20}$ & $-\dfrac{3\sqrt{\uppi}}{20}$ \\ [2ex]
        $(2,0)$ & $-\dfrac{3}{14}\sqrt{\dfrac{\uppi}{5}}$ & $-\dfrac{3}{14}\sqrt{\dfrac{\uppi}{5}}$ & $-\dfrac{3}{14}\sqrt{\dfrac{\uppi}{5}}$ \\ [2ex]
        $(4,0)$ & $-\dfrac{\sqrt{\uppi}}{280}$ & $-\dfrac{\sqrt{\uppi}}{280}$ & $-\dfrac{\sqrt{\uppi}}{280}$ \\ [2ex]
        $(4,\pm4)$ & $-\dfrac{\rme^{\mp2\uppi\rmi/3}}{4}\sqrt{\dfrac{\uppi}{70}}$ & $-\dfrac{1}{4}\sqrt{\dfrac{\uppi}{70}}$ & $-\dfrac{\rme^{\pm2\uppi\rmi/3}}{4}\sqrt{\dfrac{\uppi}{70}}$
    \end{tabular}
    \caption{
        Non-zero spherical harmonic components for each of the three ORFs.
    }
    \label{tab:orfs}
    \end{center}
\end{table}

Given these ORF spherical harmonic components, we can calculate the Fisher matrix~\cite{Romano:2016dpx,Thrane:2009fp} for a power-law AGWB $\Omega\sim f^{2/3}$,
    \begin{equation}
        \Gamma_{\ell m,\ell'm'}=\qty(\frac{3H_0^2}{2\uppi^2f_\mathrm{ref}^3})^2\sum_{I>J}\sum_{t,f}[\gamma^*_{IJ}]_{\ell m}(f,t)[\gamma_{IJ}]_{\ell'm'}(f,t)\frac{(f/f_\mathrm{ref})^{-14/3}}{P_I(f,t)P_J(f,t)},
    \end{equation}
    where $f_\mathrm{ref}$ is the reference frequency at which the angular power spectrum is estimated, $P_I$ is the noise PSD in interferometer $I$, the first sum is over detector pairs, and the second sum is over frequency bins and time intervals.
For concreteness, we take the time intervals as having length $\tau=100\,\mathrm{s}$, so that we have a frequency resolution of $1/\tau=0.01\,\mathrm{Hz}$; this means that the frequency-bin sum is over $f_n=n/\tau$, with $n\in\mathbb{Z}^+$.
As we have seen above, the ORFs are time- and frequency-independent.
We further assume that the noise PSDs are time-independent (i.e., stationary).
The Fisher matrix then simplifies to
    \begin{equation}
        \Gamma_{\ell m,\ell'm'}=\qty(\frac{3H_0^2}{2\uppi^2f_\mathrm{ref}^3})^2\frac{T}{\tau}\sum_{IJ}[\gamma^*_{IJ}]_{\ell m}[\gamma_{IJ}]_{\ell'm'}\sum_{n=1}^\infty\frac{(n/\tau f_\mathrm{ref})^{-14/3}}{P_I(n/\tau)P_J(n/\tau)}.
    \end{equation}
Using the design ET-D noise PSD curve~\cite{Hild:2010id} with a reference frequency $f_\mathrm{ref}=25\,\mathrm{Hz}$, this becomes
    \begin{equation}
        \Gamma_{\ell m,\ell'm'}=\qty(2.509\times10^{25}\times\frac{T}{1\,\mathrm{year}})\sum_{IJ}[\gamma^*_{IJ}]_{\ell m}[\gamma_{IJ}]_{\ell'm'}.
    \end{equation}
The resulting uncertainty on the angular power spectrum is given in terms of the inverse Fisher matrix $\Gamma^{-1}$ by~\cite{Thrane:2009fp}
    \begin{equation}
        \mathrm{Var}[C_\ell]=\frac{2}{(2\ell+1)^2}\sum_{m,m'}\qty|(\Gamma^{-1})_{\ell m,\ell m'}|^2,
    \end{equation}
    with the corresponding 95\% confidence upper limits (assuming a half-Gaussian distribution, with no support below zero),
    \begin{equation}
        \mathrm{UL}_{95\%}=\erf^{-1}(0.95)\sqrt{2\mathrm{Var}[C_\ell]}\approx1.960\sqrt{\mathrm{Var}[C_\ell]}.
    \end{equation}

In practice, the Fisher matrix is singular due to \enquote{blind-spots} in the detector response, and cannot be inverted.
We instead use the Moore-Penrose pseudoinverse, which effectively ignores the spherical harmonics that cannot be measured, and provides the unique minimum-chi-squared solution to the deconvolution problem.
The resulting upper limits are shown in table~\ref{tab:ULs} and in figure~\ref{fig:C_ell-cbcs}.

\begin{table}[t!]
    \begin{center}
    \begin{tabular}{c | c c c}
        & $\ell=0$ & $\ell=2$ & $\ell=4$ \\ [0.5ex]
        \hline
        $\mathrm{UL}_{95\%}$ & $2.625\times10^{-25}$ & $2.143\times10^{-26}$ & $2.917\times10^{-24}$
    \end{tabular}
    \caption{
        Resulting 95\% confidence upper limits on the $C_\ell$'s from ET alone.
    }
    \label{tab:ULs}
    \end{center}
\end{table}

\chapter{Compound Poisson shot noise model}
\label{sec:compound-poisson}

In section~\ref{sec:calc-shot-power} we introduced the compound Poisson random variable $\Lambda\equiv\sum_{i=1}^N\lambda_i$ as a model of the CBC event count in a given spatial volume element $\updelta V$, where the number of galaxies $N\sim\mathrm{Pois}[\bar{n}\updelta V]$ and the number of CBCs in each galaxy $\lambda_i\sim\mathrm{Pois}\qty[R\tau_\mathrm{s}]$ are both Poisson random variables.\footnote{%
    A more sophisticated approach would account for the statistical properties of haloes~\cite{Cooray:2002dia}.
    However, this simple assumption is sufficient for the calculation here.}
(We recall that $\bar{n}$ is the mean galaxy number density, $R$ is the CBC event rate per galaxy, and $\tau_\mathrm{s}=\tau/(1+z)$ is the source-frame time interval corresponding to an observation time $\tau$.)
In this appendix we derive a few of the relevant statistical properties of this model.
We focus initially on a single point in the galaxy/CBC parameter space, $\vb*\zeta$, so that the event rates per galaxy are all i.i.d.

\section{Cumulants of the distribution}

We can characterise the distribution of $\Lambda$ by its cumulants,
    \begin{equation}
        \kappa_n\equiv\left.\dv[n]{K_\Lambda\qty(x)}{x}\right|_{x=0},
    \end{equation}
    where $K_\Lambda(x)$ is the cumulant-generating function (CGF),
    \begin{equation}
    \label{eq:Lambda-cgf}
        K_\Lambda(x)\equiv\ln\ev{\rme^{\Lambda x}}.
    \end{equation}
The first two cumulants, $\kappa_1,\kappa_2$, are just the mean and the variance, while higher cumulants $\kappa_n$ represent the \enquote{connected components} of the $n$th moments of the distribution, and vanish if the distribution is Gaussian.
We can always write each of the moments $\ev*{\Lambda^n}$ as a polynomial in the cumulants; for example, the fourth moment is
    \begin{equation}
    \label{eq:fourth-moment-cumulants}
        \ev*{\Lambda^4}=\kappa_1^4+6\kappa_2\kappa_1^2+3\kappa_2^2+4\kappa_3\kappa_1+\kappa_4.
    \end{equation}

We can find the full set of cumulants for $\Lambda$ by evaluating the CGF~\eqref{eq:Lambda-cgf} explicitly.
As a warm-up exercise before we do this, we start by finding the CGF for each of the per-galaxy CBC event counts, $\lambda\sim\mathrm{Pois}[\bar{\lambda}]$,
    \begin{equation}
        K_\lambda(x)\equiv\ln\ev{\rme^{\lambda x}}=\ln\qty[\sum_{k=0}^\infty\Pr(\lambda=k)\rme^{kx}]=\ln\qty[\sum_{k=0}^\infty\frac{\bar{\lambda}^k\rme^{-\bar{\lambda}}}{k!}\rme^{kx}]=\ln\rme^{\bar{\lambda}(\rme^x-1)}=\bar{\lambda}(\rme^x-1).
    \end{equation}
(Similarly, $K_N(x)=\bar{N}(\rme^x-1)$, as $N$ is also Poisson-distributed.)
Note that this has the property that all of the cumulants are equal, $\kappa_1=\kappa_2=\dots=\bar{\lambda}$ (in a sense, this is the defining feature of the Poisson distribution).
The CGF for the full compound distribution is then given by a very similar calculation,
    \begin{align}
    \begin{split}
    \label{eq:Lambda-cgf-explicit}
        K_\Lambda(x)&=\ln\qty[\sum_{k=0}^\infty\Pr(N=k)\ev{\prod_{i=1}^k\rme^{\lambda_ix}}]=\ln\qty[\sum_{k=0}^\infty\Pr(N=k)\ev{\rme^{\lambda x}}^k]\\
        &=\ln\qty[\sum_{k=0}^\infty\frac{\bar{N}^k\rme^{-\bar{N}}}{k!}\rme^{kK_\lambda(x)}]=\ln\rme^{\bar{N}(\rme^{K_\lambda(x)}-1)}=\bar{N}\qty(\rme^{K_\lambda(x)}-1)=K_N(K_\lambda(x)),
    \end{split}
    \end{align}
    where in the second equality we have used the fact that the $\lambda$'s are i.i.d.
It is straightforward to take derivatives of equation~\eqref{eq:Lambda-cgf-explicit} to obtain the cumulants; the first four are
    \begin{equation}
        \kappa_1=\bar{N}\bar{\lambda}\equiv\bar{\Lambda},\qquad\kappa_2=\bar{\Lambda}(1+\bar{\lambda})\equiv\Var[\Lambda],\qquad\kappa_3=\bar{\Lambda}(1+3\bar{\lambda}+\bar{\lambda}^2),\qquad\kappa_4=\bar{\Lambda}(1+7\bar{\lambda}+6\bar{\lambda}^2+\bar{\lambda}^3),
    \end{equation}
    where $\bar{N}=\bar{n}\updelta V$ and $\bar{\lambda}=R\tau_\mathrm{s}$.
Note that the expression for the variance here corresponds to equation~\eqref{eq:var-Lambda}, with the two terms representing temporal and spatial shot noise, respectively.
We can simplify things significantly by noting that the CBC rate per galaxy is typically on the order of $R\sim\mathrm{Myr}^{-1}$, while the source-frame observing time $\tau_\mathrm{s}$ is at most a few years, so we have $\bar{\lambda}\sim10^{-6}$.
This means that, to a very good approximation, we can take
    \begin{equation}
        \kappa_1\simeq\kappa_2\simeq\kappa_3\simeq\kappa_4\simeq\dots\simeq\bar{\Lambda},
    \end{equation}
    i.e., $\Lambda$ is approximately Poisson-distributed.

\section{Fourth moment of the noisy spherical harmonic components}

In order to calculate the variance of our improved $C_\ell$ estimator~\eqref{eq:new-estimator} in the presence of shot noise, we need to evaluate the fourth moment of the noisy SHCs, $\Omega_{\ell m}^i$.
We can do this by generalising our results above to a set of distinct volume elements observed at different times $\updelta V_i$, whose total CBC counts $\Lambda_i$ are i.i.d..
To do so, we start with equation~\eqref{eq:fourth-moment-cumulants} and insert a Kronecker symbol $\delta_{ij\cdots}$ into each term in the sum, enforcing the statistical independence of each of the different volume elements,
    \begin{align}
    \begin{split}
        \ev*{\Lambda_i\Lambda_j\Lambda_k\Lambda_l}_S&=\kappa_1^4+\kappa_2\kappa_1^2\qty(\delta_{ij}+\delta_{ik}+\delta_{il}+\delta_{jk}+\delta_{jl}+\delta_{kl})\\
        &+\kappa_2^2\qty(\delta_{ij}\delta_{kl}+\delta_{ik}\delta_{jl}+\delta_{il}\delta_{jk})+\kappa_3\kappa_1\qty(\delta_{ijk}+\delta_{ijl}+\delta_{ikl}+\delta_{jkl})+\kappa_4\delta_{ijkl}.
    \end{split}
    \end{align}
Here $\delta_{ijk}$ is a generalisation of the Kronecker delta $\delta_{ij}$, which is equal to unity if $i=j=k$ and vanishes otherwise (and similarly for $\delta_{ijkl}$).
Note that we have re-introduced the notation for the shot noise average $\ev{\dots}_S$ from section~\ref{sec:shot}, as it is important to distinguish between this and the cosmological average $\ev{\dots}_\Omega$ when calculating the variance of the $C_\ell$ estimator under both ensembles.

Rewriting in terms of the comoving CBC rate density, $\mathcal{R}\qty(\vb*r,\vb*\zeta)\equiv nR=\lim_{\updelta V\to0}\Lambda/(\tau_\mathrm{s}\updelta V)$, and introducing $\vb*\zeta$ to represent the parameters of the CBC (masses, spins, \dots) and of the galaxy (star formation rate, metallicity, \dots), we obtain
    \begin{align}
    \begin{split}
    \label{eq:R-4pt}
        \ev*{\mathcal{R}_i\mathcal{R}_j\mathcal{R}_k\mathcal{R}_l}_S&=\bar{\mathcal{R}}_i\bar{\mathcal{R}}_j\bar{\mathcal{R}}_k\bar{\mathcal{R}}_l+\qty[\bar{\mathcal{R}}_i\bar{\mathcal{R}}_j\bar{\mathcal{R}}_k(1+\bar{\lambda}_k)\frac{\delta_{kl}}{\tau_{s,k}}+5\,\mathrm{perms.}]\\
        &+\qty[\bar{\mathcal{R}}_i\bar{\mathcal{R}}_k(1+\bar{\lambda}_i)(1+\bar{\lambda}_k)\frac{\delta_{ij}}{\tau_{s,i}}\frac{\delta_{kl}}{\tau_{s,k}}+2\,\mathrm{perms.}]\\
        &+\qty[\bar{\mathcal{R}}_i\bar{\mathcal{R}}_j(1+3\bar{\lambda}_j+\bar{\lambda}_j^2)\frac{\delta_{jkl}}{\tau_{s,j}^2}+3\,\mathrm{perms.}]+\bar{\mathcal{R}}_i(1+7\bar{\lambda}_i+6\bar{\lambda}_i^2+\bar{\lambda}_i^3)\frac{\delta_{ijkl}}{\tau_{s,i}^3},
    \end{split}
    \end{align}
    where here the Kronecker symbols are shorthand for
    \begin{equation}
    \label{eq:delta-shorthand}
        \delta_{ij}\to\delta_{ij}\delta^3(\vb*r_i-\vb*r_j)\delta(\vb*\zeta_i,\vb*\zeta_j),
    \end{equation}
    i.e., two GW sources must coincide in space, in time, and in parameter space, in order to contribute to the shot noise.

Now, using equation~\eqref{eq:omega-definition} to write the SHCs in terms of the CBC rate density, we find
    \begin{align}
    \begin{split}
        &\ev{\Omega_{\ell m}^i\Omega_{\ell m}^{j*}\Omega_{\ell m'}^{i'*}\Omega_{\ell m'}^{j'}}_S\\
        &\quad=\qty(\frac{2G}{3})^4\int\dd[3]{\vb*r_i}\dd[3]{\vb*r_j}\dd[3]{\vb*r_{i'}}\dd[3]{\vb*r_{j'}}Y_{\ell m}^{i*}Y_{\ell m}^j Y_{\ell m'}^{i'}Y_{\ell m'}^{j'*}(1+z_i)^{-2}(1+z_j)^{-2}(1+z_{i'})^{-2}(1+z_{j'})^{-2}\\
        &\quad\times\qty(\frac{r_H^4}{r_ir_jr_{i'}r_{j'}})^2\int\dd{\vb*\zeta_i}\dd{\vb*\zeta_j}\dd{\vb*\zeta_{i'}}\dd{\vb*\zeta_{j'}}\dv{E_i}{(\ln f_{\mathrm{s},i})}\dv{E_j}{(\ln f_{\mathrm{s},j})}\dv{E_{i'}}{(\ln f_{\mathrm{s},i'})}\dv{E_{j'}}{(\ln f_{\mathrm{s},j'})}\ev{\mathcal{R}_i\mathcal{R}_j\mathcal{R}_{i'}\mathcal{R}_{j'}}_S,
    \end{split}
    \end{align}
    where $Y_{\ell m}^i$ is shorthand for $Y_{\ell m}(\vu*r_i)$, etc.
Using \eqref{eq:R-4pt} and the properties of the spherical harmonics, we get
    \begin{align}
    \begin{split}
    \label{eq:Omega-4pt}
        &\ev{\Omega_{\ell m}^i\Omega_{\ell m}^{j*}\Omega_{\ell m'}^{i'*}\Omega_{\ell m'}^{j'}}_S\\
        &=|\Omega_{\ell m}|^2|\Omega_{\ell m'}|^2+\mathcal{W}_\tau|\Omega_{\ell m}|^2\qty[\delta_{i'j'}+\delta_{m,-m'}(\delta_{i'j}+\delta_{ij'})]+\mathcal{W}_\tau|\Omega_{\ell m'}|^2\qty[\delta_{ij}+\delta_{mm'}(\delta_{ii'}+\delta_{jj'})]\\
        &+\mathcal{W}^2_\tau(\delta_{ij}\delta_{i'j'}+\delta_{mm'}\delta_{ii'}\delta_{jj'}+\delta_{m,-m'}\delta_{ij'}\delta_{i'j})+c^{(3)}_{\ell m}\mathcal{X}_\tau\Omega_{\ell0}\delta_{m'0}(\delta_{ii'j}+\delta_{ijj'})\\
        &+c^{(3)}_{\ell m'}\mathcal{X}_\tau\Omega_{\ell0}\delta_{m0}(\delta_{ii'j'}+\delta_{i'jj'})+c^{(4)}_{\ell m m'}\mathcal{Y}_\tau\delta_{ii'jj'},
    \end{split}
    \end{align}
    where the shorthand~\eqref{eq:delta-shorthand} is now no longer used.
The coefficients on the 3- and 4-point terms are
    \begin{align}
    \begin{split}
        c^{(3)}_{\ell m}&\equiv(-)^m\frac{(2\ell+1)^{3/2}}{\sqrt{4\uppi}}
        \begin{pmatrix}
            \ell & \ell & \ell \\
            0 & 0 & 0
        \end{pmatrix}
        \begin{pmatrix}
            \ell & \ell & \ell \\
            0 & m & -m
        \end{pmatrix},\\
        c^{(4)}_{\ell m m'}&\equiv\sum_{\ell'=0}^{2\ell}(-)^{m+m'}\frac{(2\ell'+1)(2\ell+1)^2}{4\uppi}
        \begin{pmatrix}
            \ell & \ell & \ell' \\
            -m & m & 0
        \end{pmatrix}
        \begin{pmatrix}
            \ell & \ell & \ell' \\
            m' & -m' & 0
        \end{pmatrix}
        \begin{pmatrix}
            \ell & \ell & \ell' \\
            0 & 0 & 0
        \end{pmatrix}^2,
    \end{split}
    \end{align}
    and are just combinations of Wigner $3j$ symbols.
In addition to the 2-point shot noise power $\mathcal{W}_\tau$ we defined before, we now have 3-point and 4-point terms that appear,
    \begin{align}
    \begin{split}
        \mathcal{W}_\tau&\equiv\qty(\frac{2G}{3})^2\frac{r_H^4}{\tau}\int\frac{\dd{r}}{(1+z)^3r^2}\int\dd{\vb*\zeta}\bar{\mathcal{R}}\qty(1+\bar{\lambda})\qty[\dv{E}{(\ln f_\mathrm{s})}]^2,\\
        \mathcal{X}_\tau&\equiv\qty(\frac{2G}{3})^3\frac{r_H^6}{\tau^2}\int\frac{\dd{r}}{(1+z)^4r^4}\int\dd{\vb*\zeta}\bar{\mathcal{R}}(1+3\bar{\lambda}+\bar{\lambda}^2)\qty[\dv{E}{(\ln f_\mathrm{s})}]^3,\\
        \mathcal{Y}_\tau&\equiv\qty(\frac{2G}{3})^4\frac{r_H^8}{\tau^3}\int\frac{\dd{r}}{(1+z)^5r^6}\int\dd{\vb*\zeta}\bar{\mathcal{R}}(1+7\bar{\lambda}+6\bar{\lambda}^2+\bar{\lambda}^3)\qty[\dv{E}{(\ln f_\mathrm{s})}]^4,
    \end{split}
    \end{align}
    where $\bar{\lambda}(r,\vb*\zeta)=\bar{\mathcal{R}}\tau_\mathrm{s}/\bar{n}\ll1$.
(Note that the first line here is just a different way of writing equation~\eqref{eq:W-main}.)

Taking the cosmological average of \eqref{eq:Omega-4pt} and subtracting the 2nd moments, we therefore find
    \begin{align}
    \begin{split}
        \Cov\qty[\Omega_{\ell m}^i\Omega_{\ell m}^{j*},\Omega_{\ell m'}^{i'}\Omega_{\ell m'}^{j'*}]_{S,\Omega}&=\delta_{mm'}\qty(C_\ell+\delta_{ii'}\mathcal{W}_\tau)\qty(C_\ell+\delta_{jj'}\mathcal{W}_\tau)\\
        &+\delta_{m,-m'}\qty(C_\ell+\delta_{ij'}\mathcal{W}_\tau)\qty(C_\ell+\delta_{i'j}\mathcal{W}_\tau)+c^{(4)}_{\ell mm'}\mathcal{Y}_\tau\delta_{ii'jj'}.
    \end{split}
    \end{align}
Note that the term proportional to $\mathcal{X}_\tau$ has vanished, due to the SHCs having zero mean for $\ell>0$ (recall that we do not attempt to estimate $C_0$, due to the degeneracy with the unknown mean intensity $\bar{\Omega}$).
The term proportional to $\mathcal{Y}_\tau$ is associated with the fourth cumulant $\kappa_4$, and is therefore a sign of the non-Gaussian nature of the shot noise fluctuations (as $\kappa_n=0$ for all $n>2$ in the Gaussian case).
However, this term does not end up contributing to the variance of our improved estimator $\hat{C}_\ell$, as this includes only off-diagonal pairs $i\ne j$ and $i'\ne j'$ by design.

\chapter{Higher-order memory from cosmic strings}
\label{app:angular-integrals}

\section{Understanding the origin of the higher-order memory divergence}
\label{sec:rhdot}

Here we derive a condition~\eqref{eq:divergence} on a generic GW signal $h(t)$ which, we argue heuristically, is necessary for the associated nonlinear memory to diverge.
We find this condition by considering a simple toy model in which we can tune the \enquote{sharpness} of the signal to find where the divergence sets in.
We then show that this condition predicts the divergence for cusps on \enquote{large} cosmic string loops, while also providing an explanation for why the memory from compact binaries never diverges.

\subsubsection{Gaussian pulse as a toy model}
Our goal here is to derive a condition on how \enquote{sharp} a putative GW signal must be to give rise to a divergent nonlinear memory expansion.
To investigate this, we use a toy model in which the primary GW is a Gaussian pulse which reaches our detector at retarded time $t=0$,
    \begin{equation}
    \label{eq:pulse}
        h^{(0)}(t)=\frac{A}{r}\exp(-\frac{t^2}{2\sigma^2}).
    \end{equation}
This has amplitude $A$ and width $\sigma$, both of which have dimensions of length.
We ignore the polarisation and angular pattern of the pulse, though these can play an important role (e.g., for the case of kinks, as discussed in section~\ref{sec:kinks} above), and focus on the pulse's behaviour as a function of time.
The Fourier transform of equation~\eqref{eq:pulse} is
    \begin{equation}
        \tilde{h}^{(0)}(f)=\frac{A}{r}\sqrt{2\uppi}\sigma\exp(-\frac{(2\uppi f\sigma)^2}{2}),
    \end{equation}
    so we see that the pulse has a characteristic frequency of $\sim1/\sigma$.
The pulse is smooth and infinitely differentiable for all finite $\sigma$; however, by making $\sigma$ small, we can make the pulse arbitrarily \enquote{sharp} (equivalently, we can make the characteristic frequency arbitrarily large).
This sharpness can be quantified by the maximum value of the time derivative of the strain,
    \begin{equation}
        \max_t|r\dot{h}^{(0)}(t)|=\rme^{-1/2}A/\sigma\sim[\text{amplitude}]\times[\text{characteristic frequency}],
    \end{equation}
    i.e. if the dimensionless ratio $A/\sigma$ is much greater than unity the pulse is \enquote{sharp}, and if $A/\sigma$ is much less than unity the pulse is \enquote{soft}.

The pulse's first-order memory can be written as
    \begin{equation}
        r\dot{h}^{(1)}=C|r\dot{h}^{(0)}|^2=C\qty(\frac{At}{\sigma^2})^2\exp(-\frac{t^2}{\sigma^2}),
    \end{equation}
    where $C$ is some constant arising from the angular integral, which we assume to be $\order{1}$ for all orders in the memory expansion.
(This assumption can easily be violated: e.g., if the integrand has vanishing TT component then $C=0$.)
We immediately see that the memory signal is larger than the primary GW signal near the arrival time ($|t|\sim\sigma$) if and only if $A/\sigma\gg1$, or equivalently,
    \begin{equation}
    \label{eq:divergence}
        \max_t|r\dot{h}(t)|\gg1.
    \end{equation}
Using equation~\eqref{eq:gw-energy-flux}, we see that this is equivalent to $\max_t|\dot{E}_\gw|\gg1/G=m_\Pl/t_\Pl$, i.e. the memory is greater than the primary signal if the GW energy flux is trans-Planckian.

Moving on to the second-order memory, there are two contributions: the self-energy of the first-order memory, and its cross-energy with the primary signal,
    \begin{align}
    \begin{split}
        r\dot{h}^{(2)}&=Cr^2(|\dot{h}^{(1)}+\dot{h}^{(0)}|^2-|\dot{h}^{(0)}|^2)=C|r\dot{h}^{(1)}|^2+2Cr^2\dot{h}^{(1)}\dot{h}^{(0)}\\
        &=C^3\qty(\frac{At}{\sigma^2})^4\exp(-\frac{2t^2}{\sigma^2})-2C^2\qty(\frac{At}{\sigma^2})^3\exp(-\frac{3t^2}{2\sigma^2}).
    \end{split}
    \end{align}
The general pattern is straightforward from here.
Treating the \enquote{sharp} regime $A/\sigma\gg1$ and the \enquote{soft} regime $A/\sigma\ll1$ separately (analogous to the separation between the \enquote{large loop} and \enquote{small loop} regimes for cusps), we find for all $n\ge1$,
    \begin{equation}
        r\dot{h}^{(n)}\simeq
        \begin{cases}
            \displaystyle\frac{1}{C}\qty[\frac{CAt}{\sigma^2}\exp(-\frac{t^2}{2\sigma^2})]^{2^n}, & A/\sigma\gg1\\[10pt]
            \displaystyle\frac{1}{4C}\qty[-\frac{2CAt}{\sigma^2}\exp(-\frac{t^2}{2\sigma^2})]^{n+1}, & A/\sigma\ll1
        \end{cases}
    \end{equation}
    which shows that the memory expansion diverges if equation~\eqref{eq:divergence} holds.

Of course, there are many ways in which this argument could fail, some of which we have already mentioned (in particular, if $C\ll1$).
However, the general takeaway is that if the time derivative of a GW strain signal is large, then a memory divergence \emph{might} occur.
For $\max_t|r\dot{h}|\ll1$ on the other hand, it seems very likely that the memory expansion must always converge.
If this is indeed the case, then we could interpret equation~\eqref{eq:divergence} as a necessary, but not sufficient, condition for the memory divergence.

\subsubsection{Application to cosmic strings}
How does this fit into our results for cosmic string cusps?
Neglecting numerical constants, the time-domain cusp waveform looks like
    \begin{equation}
        h_\rmc^{(0)}(t)\sim-\frac{G\mu}{r}\ell^{2/3}|t|^{1/3}+\text{constant},
    \end{equation}
    so the time derivative \emph{diverges} at $t=0$ due to the absolute value function.
If we introduce a cutoff at the string width scale $\delta$, we find that $\max_t|r\dot{h}_\rmc^{(0)}|\sim G\mu(\ell/\delta)^{2/3}$.
Naively applying the condition~\eqref{eq:divergence}, we would thus expect the cusp memory signal to diverge for loops of length $\ell\gtrsim\delta/(G\mu)^{3/2}$.
However, this does \emph{not} happen, for the simple reason that there is a small factor $C\ll1$ arising from the angular integral, and as mentioned above this causes the condition~\eqref{eq:divergence} to fail.

We can circumvent this issue by considering the first-order memory signal from the cusp as the source of the divergence: this should give $C=\order{1}$ as it has a broad emission pattern, and is not concentrated into a narrow beam.
Indeed, using equation~\eqref{eq:cusp-time-domain-approx} we find that the maximum time derivative for the first-order cusp memory signal is $\max_t|r\dot{h}_\rmc^{(1)}|\sim(G\mu)^2(\ell/\delta)^{2/3}$.
Applying the condition~\eqref{eq:divergence} again, we find that the cusp memory signal should diverge if $\ell\gtrsim\delta/(G\mu)^3$, which is exactly what we find from the careful analysis in section~\ref{sec:higher-order-divergence}.
This supports the idea that equation~\eqref{eq:divergence} gives a necessary condition for the memory divergence, which is generally applicable if the factor $C$ arising from the angular integral is not too small.

\subsubsection{Application to compact binaries}
Here we use a very simple heuristic analysis to show that $|r\dot{h}(t)|$ is at most $\order{1}$ for compact binary coalescences, so that the memory divergence condition~\eqref{eq:divergence} never holds.
We neglect numerical factors throughout.

The GW signal from a CBC can be written as $h(t)=\mathcal{A}(t)\rme^{\rmi\phi(t)}$, where the leading-order Newtonian contributions to the amplitude and phase are~\cite{Maggiore:2007zz}
    \begin{equation}
        \mathcal{A}\sim\frac{(G\mathcal{M})^{5/4}}{r\tau^{1/4}},\qquad\phi\sim\qty(\frac{\tau}{G\mathcal{M}})^{5/8},
    \end{equation}
    with $\tau$ the time until coalescence and $\mathcal{M}\equiv\eta^{3/5}M$ the chirp mass, where $\eta\le1/4$ is the dimensionless mass ratio.
The time derivative of the strain is therefore
    \begin{equation}
        |r\dot{h}(t)|=\sqrt{|r\dot{\mathcal{A}}|^2+|r\dot{\phi}\mathcal{A}|^2}\sim\sqrt{\qty(\frac{G\mathcal{M}}{\tau})^{5/2}+\qty(\frac{G\mathcal{M}}{\tau})^{5/4}}.
    \end{equation}
Formally, this diverges in the Newtonian analysis as $\tau\to0$.
However, introducing a cutoff at the ISCO radius truncates the signal at $\tau_\mathrm{min}\sim\eta^{-8/5}G\mathcal{M}$, so that $\max_t|r\dot{h}(t)|\sim\sqrt{\eta^2+\eta^4}$.
Since $\eta$ is at most $\order{1}$, we see that the condition~\eqref{eq:divergence} is never met.
This makes complete sense, given that we know that the memory from CBC signals cannot diverge.
As a by-product, this expression leads us to conjecture that equal-mass binaries ($\eta=1/4$) should give rise to stronger higher-order memory effects than extreme mass-ratio inspirals ($\eta\ll1$).

\begin{figure}[t!]
    \begin{center}
        \includegraphics[width=0.667\textwidth]{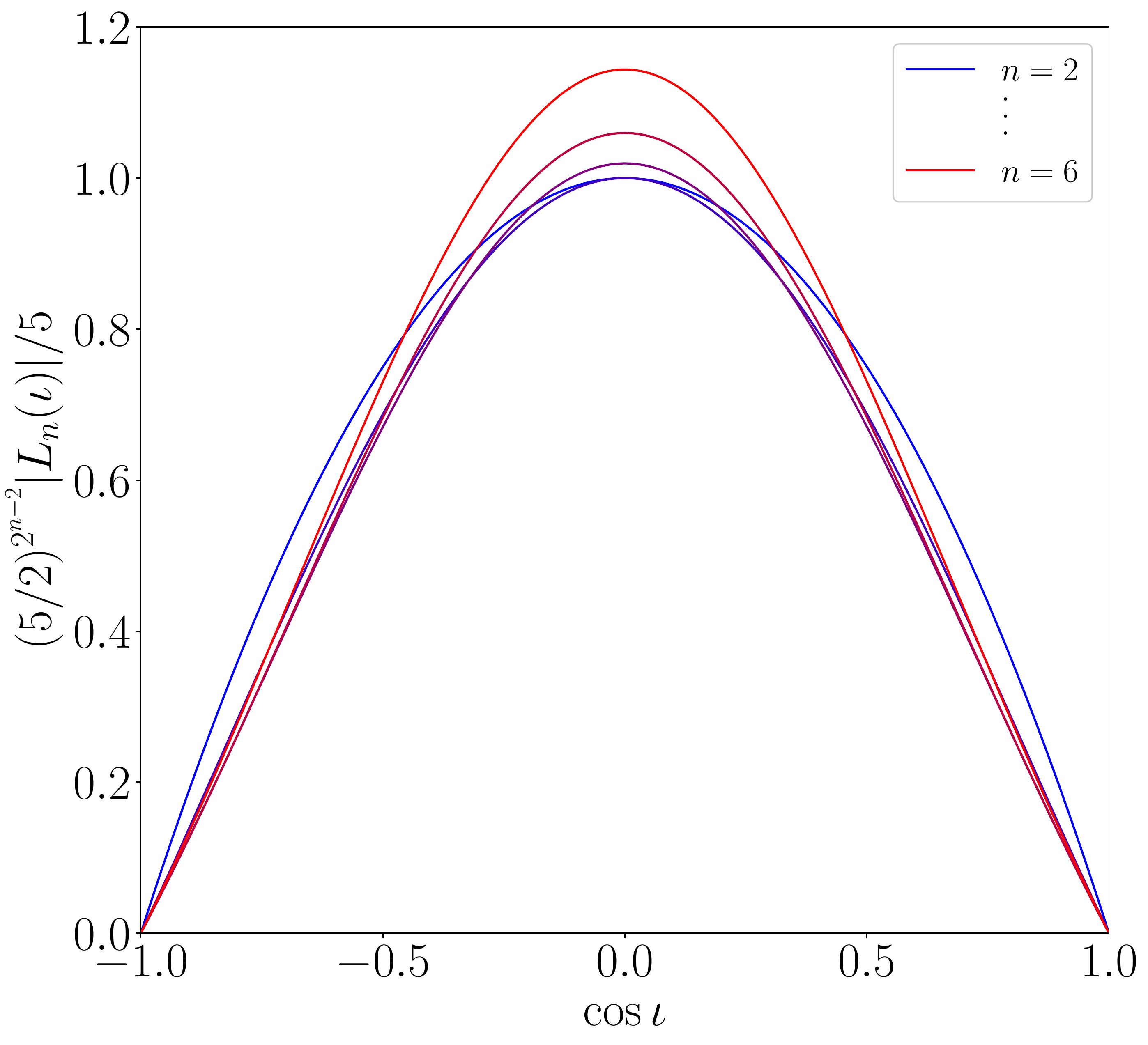}
    \end{center}
    \begin{center}
        \includegraphics[width=0.667\textwidth]{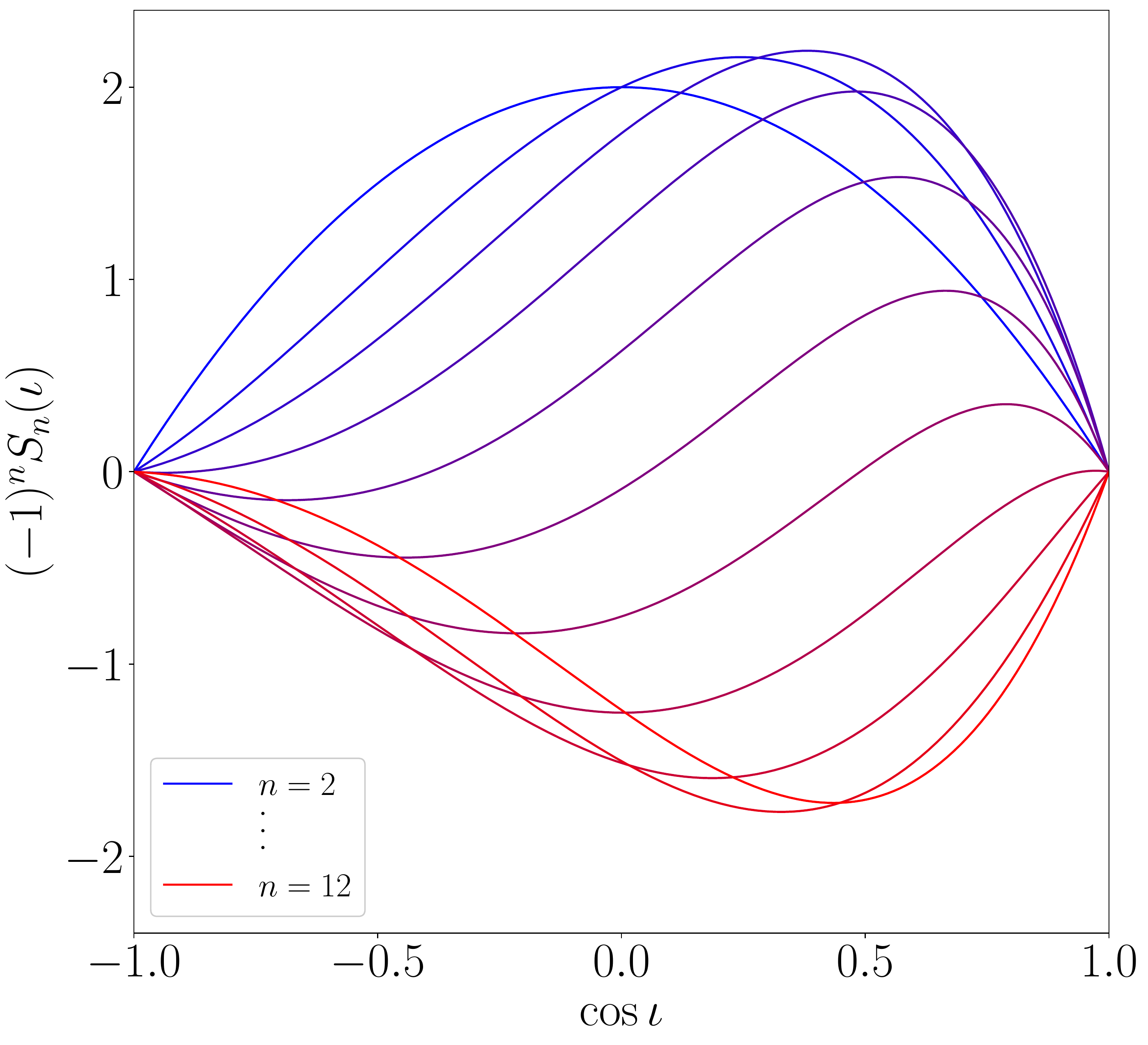}
    \end{center}
    \caption{%
    Functions describing the strength of the $n$th-order cusp memory signal as a function of inclination, with $L_n(I)$ representing \enquote{large} loops $\ell\gg\ell_*$, and $S_n(I)$ representing \enquote{small} loops $\ell\ll\ell_*$.
    }
    \label{fig:L_n-S_n}
\end{figure}

\section{Angular integrals for higher-order cusp memory}
\label{sec:iota-polynomials}

Here we derive the angular patterns of the higher-order cusp memory in the \enquote{large-loop} ($\ell\gg\ell_*$) and \enquote{small-loop} ($\ell\ll\ell_*$) limits, which are described by the functions $L_n(I)$ and $S_n(I)$ respectively, with $I$ the inclination between the cusp beaming direction and the observer's line of sight.
These are defined by inserting the $n$th-order memory formula~\eqref{eq:cusp-higher-order-final} into the iterative relation~\eqref{eq:iterate-memory}, which gives
    \begin{equation}
    \label{eq:iota-polynomials-iterative}
        L_{n+1}(I)=\int_{\vu*r'}|L_n(I')|^2,\qquad S_{n+1}(I)=\int_{\vu*r'}6(1+\cos I')S_n(I'),
    \end{equation}
    for all $n\ge2$, with $L_2(I)=S_2(I)=2\sin^2I$.
(The integral symbol $\int_{\vu*r'}$ is defined in equation~\eqref{eq:integral-shorthand}.)

We find that $L_n$ and $S_n$ can be written as polynomials in $\cos I$ (of order $2^{n-1}$ and $n$, respectively).
The iterative process~\eqref{eq:iota-polynomials-iterative} can therefore be carried out by evaluating a simpler family of integrals, $C_n(I)\equiv\int_{\vu*r'}\cos^nI'$.
To do so, we choose our $\vu*r'=(\theta',\phi')$ coordinates such that $\theta'$ is zero along the line of sight, so that $\cos\theta'\equiv\vu*r\vdot\vu*r'$, with the cusp beaming direction defined relative to the line of sight by $\vu*r_\rmc\vdot\vu*r=\cos I$.
We then have $\cos I'\equiv\vu*r_\rmc\vdot\vu*r'=\cos\theta'\cos I-\cos\phi'\sin\theta'\sin I$, and equation~\eqref{eq:integral-shorthand} becomes
    \begin{equation}
        \int_{\vu*r'}=\int_{S^2}\frac{\dd[2]{\vu*r'}}{4\uppi}(1+\cos\theta')\rme^{-2\rmi\phi'},
    \end{equation}
    so that we can use a binomial expansion of $\cos^nI'=(\cos\theta'\cos I-\cos\phi'\sin\theta'\sin I)^n$ to write
    \begin{equation}
        C_n(I)=\sum_{k=0}^n
        \begin{pmatrix}
            n \\ k
        \end{pmatrix}
        (-\sin I)^k\cos^{n-k}I\int_{-1}^{+1}\frac{\dd{x}}{2}(1+x)x^{n-k}(1-x^2)^{k/2}\int_0^{2\uppi}\frac{\dd{\phi'}}{2\uppi}\rme^{-2\rmi\phi'}\cos^k\phi',
    \end{equation}
    where we have set $x=\cos\theta'$.
Using the Beta function identity
    \begin{equation}
        \mathrm{B}(a,b)\equiv\int_0^1\dd{x}x^{a-1}(1-x)^{b-1}=\frac{\Gamma(a)\Gamma(b)}{\Gamma(a+b)},
    \end{equation}
    we then obtain, for all $n\ge0$,
    \begin{align}
    \begin{split}
        C_{2n}(I)&=\sum_{k=0}^n\frac{\sqrt{\uppi}(n-k)(2n)!\cos^{2k}I\sin^{2(n-k)}I}{2^{2n+1}k!(n-k+1)!\Gamma(n+\tfrac{3}{2})},\\
        C_{2n+1}(I)&=\sum_{k=0}^n\frac{\sqrt{\uppi}(n-k)(2n+1)!\cos^{2k+1}I\sin^{2(n-k)}I}{2^{2n+2}k!(n-k+1)!\Gamma(n+\tfrac{5}{2})}.
    \end{split}
    \end{align}
Calculating each iteration of equation~\eqref{eq:iota-polynomials-iterative} is then reduced to writing down the appropriate linear combination of $C_n(I)$ for different $n$.

\begin{table}[p!]
    \begin{center}
    \begin{tabular}{c | l}
        $n$ & $S_n(I)$ \\
        \hline
        $2$ & $2\sin^2I$ \\[12pt]
        $3$ & $-\dfrac{2\sin^2I}{5}(5+3\cos I)$ \\[12pt]
        $4$ & $\dfrac{2\sin^2I}{25}(25+24\cos I+3\cos2I)$ \\[12pt]
        $5$ & $-\dfrac{2\sin^2I}{175}(154+201\cos I+42\cos2I+3\cos3I)$ \\[12pt]
        $6$ & $\dfrac{\sin^2I}{12250}(15575+28032\cos I+7932\cos2I+960\cos3I+45\cos4I)$ \\[12pt]
        $7$ & $-\dfrac{\sin^2I}{245000}[466422\cos I+172824\cos2I+5(29702+5619\cos3I+450\cos4I+15\cos5I)]$ \\[12pt]
        $8$ & $\dfrac{\sin^2I}{245000}\big(-30194+299024\cos I+157771\cos2I$ \\
        & $\qquad\qquad+32552\cos3I+3490\cos4I+200\cos5I+5\cos6I\big)$ \\[12pt]
        $9$ & $-\dfrac{\sin^2I}{13475000}\big[4934669\cos I+6432382\cos2I+5\big(-2137388+347047\cos3I$ \\
        & $\qquad\qquad\qquad\qquad\qquad\qquad\qquad+46508\cos4I+3565\cos5I+154\cos6I+3\cos7I\big)\big]$ \\[12pt]
        $10$ & $\dfrac{\sin^2I}{10375750000}\big[-5257311104\cos I+2472686192\cos2I$ \\
        & $\qquad\qquad+5\big(214599584\cos3I+36187180\cos4I$ \\
        & $\qquad\qquad-7(383512899-491360\cos5I-28400\cos6I-960\cos7I-15\cos8I)\big)\big]$ \\[12pt]
        $11$ & $-\dfrac{\sin^2I}{1888386500000}\big[-2359095296230\cos I-55089058544\cos2I+116827655852\cos3I$ \\
        & $\qquad\qquad+27832319320\cos4I+3261899900\cos5I$ \\
        & $\qquad\qquad+49(-59685302066+4749680\cos6I+214995\cos7I+5850\cos8I+75\cos9I)\big]$ \\[12pt]
        $12$ & $\dfrac{\sin^2I}{122745122500000}\big[-213800187213104\cos I-34153675430026\cos2I$ \\
        & $\qquad\qquad\qquad+5\big(-37491558286234+292405153376\cos3I$ \\
        & $\qquad\qquad\qquad+238091184664\cos4I+37145938400\cos5I+3229010267\cos6I$ \\
        & $\qquad\qquad\qquad+179962104\cos7I+6563970\cos8I+147000\cos9I+1575\cos10I\big)\big]$
    \end{tabular}
    \end{center}
    \caption{\label{tab:S_n}%
        The first few small-loop angular functions, as defined by equation~\eqref{eq:iota-polynomials-iterative} with $S_2(I)=2\sin^2I$.
        Note that $S_n(I)$ is a polynomial in $\cos I$ of order $n$.}
\end{table}

\begin{table}[p!]
    \begin{center}
    \rotatebox{-90}{%
    \begin{tabular}{c | l}
        $n$ & $L_n(I)$ \\
        \hline
        $2$ & $2\sin^2I$ \\[12pt]
        $3$ & $-\dfrac{2\sin^2I}{15}(5-\cos2I)$ \\[12pt]
        $4$ & $\dfrac{\sin^2I}{141750}[3737\cos2I-262\cos4I+7(-2070+\cos6I)]$ \\[12pt]
        $5$ & $\dfrac{\sin^2I}{123092512042500000}\bigg[91170878299511\cos2I-8844505203863\cos4I+627703754313\cos6I-31676232034\cos8I$ \\
        & $\qquad\qquad\qquad\qquad\qquad\qquad+1106354755\cos10I+1001(-326445712830-24241\cos12I+245\cos14I)\bigg]$ \\[12pt]
        $6$ & $\dfrac{\sin^2I}{4375937836474054406594550954572130000000000000000}\bigg[2337436144549410554623113364688535605014856\cos2I$ \\
        & $\qquad\qquad\qquad-252383762462012251730735377642370477964415\cos4I+22676810284676710047197528844734128748260\cos6I$ \\
        & $\qquad\qquad\qquad-1739064548768558706903275538411759890046\cos8I+114707816611411474747133898122922244740\cos10I$ \\
        & $\qquad\qquad\qquad-6538676998608577029462451915641765109\cos12I+322110616165743776447552901776475040\cos14I$ \\
        & $\qquad\qquad\qquad-13638911209793535304662793111510580\cos16I+490882191972236519729403447127344\cos18I$ \\
        & $\qquad\qquad\qquad+1463(-10081167364753857595829662265\cos20I+ 246246861424248720285801892\cos22I$ \\
        & $\qquad\qquad\qquad+1495(-3145080481324598891074\cos24I+44031911264492955804\cos26I$ \\
        & $\qquad\qquad\qquad+6525(-569906751862925826959924173050076-61436527080929\cos28I+271199768840\cos30I)))\bigg]$ \\
    \end{tabular}
    }
    \end{center}
    \caption{\label{tab:L_n}%
        The first few large-loop angular functions, as defined by equation~\eqref{eq:iota-polynomials-iterative} with $L_2(I)=2\sin^2I$.
        Note that $L_n(I)$ is a polynomial in $\cos I$ of order $2^{n-1}$, such that the number of terms grows exponentially with $n$.
        Despite this apparent complexity, we find that these formulae can be approximated by the simple expression $|L_n(I)|\approx5(2/5)^{2^{n-2}}\sin^2I$.}
\end{table}

The resulting expressions for $S_n(I)$ and $L_n(I)$ for the first few $n\ge2$ are shown in Tables~\ref{tab:S_n} and~\ref{tab:L_n} respectively.
We find empirically that the large-loop angular functions can be approximated by $|L_n(I)|\approx5(2/5)^{2^{n-2}}\sin^2I$ with an accuracy of $\sim10\%$, as is illustrated in the top panel of figure~\ref{fig:L_n-S_n}.
The small-loop angular functions $S_n(I)$ do not seem to follow such a simple pattern.

\section{Angular integrals for kink memory}
\label{sec:K_n}

Here we compute the integral $K_n(I)$ defined in equation~\eqref{eq:K_n-definition} for all angular modes $n\in\mathbb{Z}$.
Note that while the integrand is complex, the integral itself is always real, as the imaginary part of the integrand is an odd function of $\phi$.
Note also that we can focus on non-negative $n\ge0$ by exploiting the symmetry property $K_{-n}(I)=K_n(-I)$.

\begin{figure}[t!]
    \begin{center}
        \includegraphics[width=0.667\textwidth]{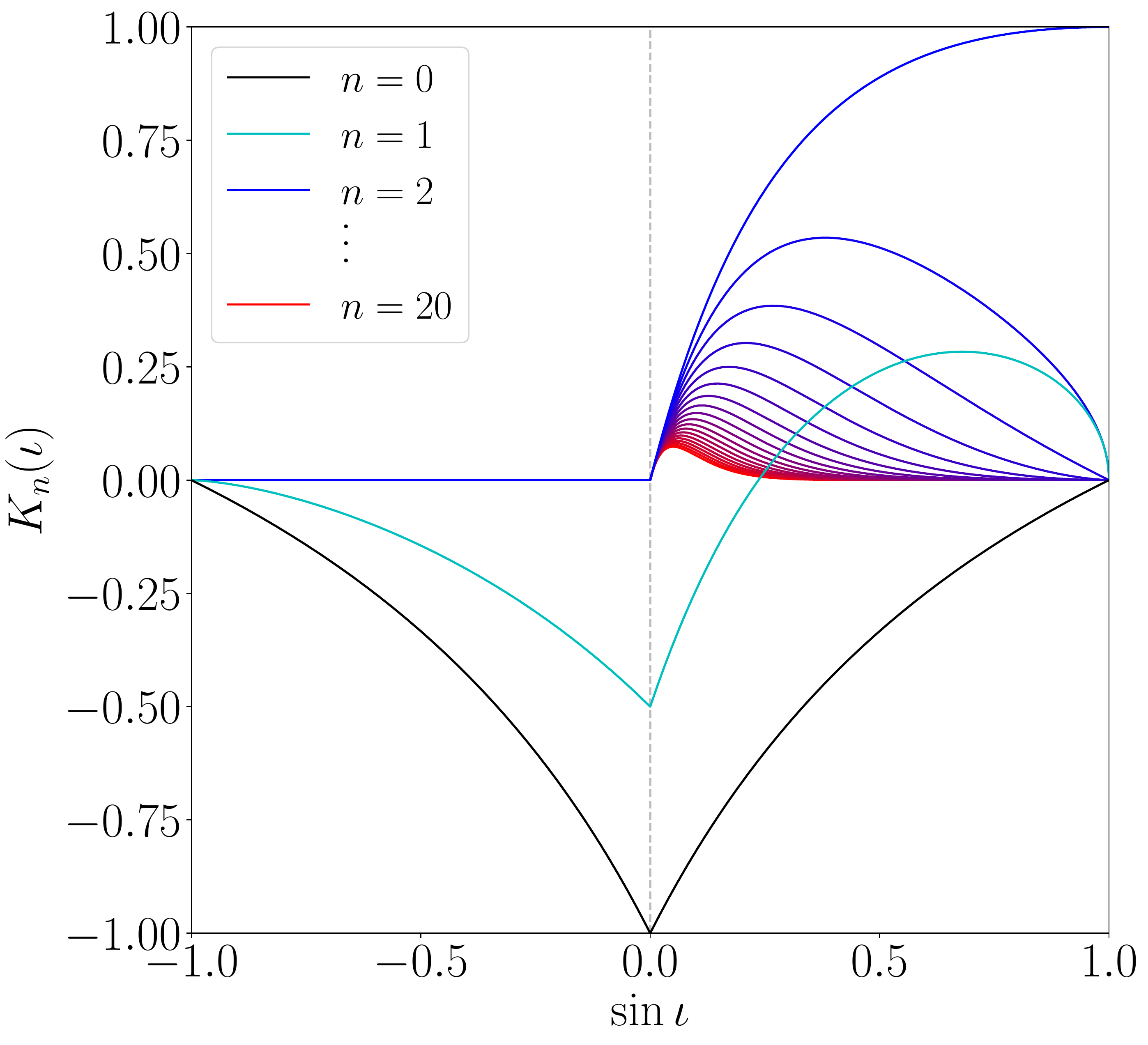}
    \end{center}
    \caption{%
    The angular integral $K_n(I)$ for non-negative $n$, as given by equation~\eqref{eq:K_n-final}.
    The corresponding curves for negative $n$ are obtained by reflecting around $I=0$.
    }
    \label{fig:kink-angular-integral}
\end{figure}

For some special values of $I$ we can obtain the full spectrum immediately from equation~\eqref{eq:K_n-definition},
    \begin{equation}
    \label{eq:K_n-special-values}
        K_n(0)=-\delta_{n,0}-\frac{1}{2}(\delta_{n,1}+\delta_{n,-1}),\qquad K_n(\pm\uppi/2)=\delta_{n,\pm2}.
    \end{equation}
For general values $I\in[-\uppi/2,+\uppi/2]$, we proceed by expanding the denominator of equation~\eqref{eq:K_n-definition} using the geometric series and the binomial theorem,
    \begin{equation}
    \label{eq:K_n-denominator-series}
        \frac{1}{1-\cos I\cos\phi}=\sum_{k=0}^\infty\cos^kI\cos^k\phi=\sum_{k=0}^\infty\sum_{m=0}^k
        \begin{pmatrix}
            k \\ m
        \end{pmatrix}
        \frac{\cos^kI}{2^k}\rme^{-\rmi(k-2m)\phi}.
    \end{equation}
This converges everywhere on $I\in[-\uppi/2,+\uppi/2]$ except for $I=0$, in which case we use equation~\eqref{eq:K_n-special-values} instead.
We can then integrate term-by-term to extract the contribution from each order in the series.
The result, valid for all $n\ge0$, is
    \begin{align}
    \begin{split}
        K_n(I)&=\sum_{k=\max(0,2-n)}^\infty
        \begin{pmatrix}
            2k+n-2 \\ k
        \end{pmatrix}
        \qty(\frac{\cos I}{2})^{2k+n-2}\qty(\frac{1+\sin I}{2})^2\\
        &-2\sum_{k=0}^\infty
        \begin{pmatrix}
            2k+n \\ k
        \end{pmatrix}
        \qty(\frac{\cos I}{2})^{2k+n+2}+\sum_{k=0}^\infty
        \begin{pmatrix}
            2k+n+2 \\ k
        \end{pmatrix}
        \qty(\frac{\cos I}{2})^{2k+n+2}\qty(\frac{1-\sin I}{2})^2.
    \end{split}
    \end{align}
The first sum must be carried out separately for the three cases $n=0$, $n=1$, and $n>1$, yielding the following expressions:
    \begin{align}
    \begin{split}
    \label{eq:K_n-final}
        K_0(I)&=\frac{4\sin|I|+\cos2I-3}{2\cos^2I},\\
        K_1(I)&=\frac{\sin|I|-1}{2\sin|I|}\qty[\cos I+(\sin I-1)\frac{1-\sin|I|+\sin I(1+2\sin I+\sin|I|)}{\cos^3I}],\\
        K_n(I)&=
        \begin{cases}
            \dfrac{4\sin I}{\cos^2I}\qty(\dfrac{\cos I}{1+\sin I})^n, & I>0\\
            0, & I\le0
        \end{cases},\qquad\text{for }n>1.
    \end{split}
    \end{align}
These clearly agree with equation~\eqref{eq:K_n-special-values} for $I=\pm\uppi/2$.
Note that for all $n$, the integral $K_n(I)$ is not differentiable at $I=0$; this is related to the fact that the series in equation~\eqref{eq:K_n-denominator-series} diverges at $I=0$.
However, despite this formal divergence, we see that equation~\eqref{eq:K_n-final} agrees with equation~\eqref{eq:K_n-special-values} in the limit $I\to0$, whether this limit is taken from above or from below.
We therefore use equation~\eqref{eq:K_n-final} over the full domain $I\in[-\uppi/2,+\uppi/2]$.
The resulting curves for the first few non-negative $n$ are illustrated in figure~\ref{fig:kink-angular-integral}.

For $n\ge2$, the integral only has support for $I>0$, and peaks at an inclination $I_*$ given by $I_*=\sin^{-1}[(n/2)-\sqrt{(n/2)^2-1}]$.
When calculating the kink memory signal we are interested in the high-frequency regime, which corresponds to large $n$.
In the limit $n\to\infty$, we have $I_*\simeq1/n$ and $K_n(I_*)\simeq4/(\rme n)$, meaning that the memory signal is strongly suppressed, and is only observable very close to (but strictly outside of) the plane of the kink.

\chapter{Derivation of the secular Kramers-Moyal coefficients}
\label{app:kramers-moyal-derivation}

\section{Polarisation tensors in the binary's coordinate frame}
\label{sec:polarisation-tensor-projections}

In order to describe the GW polarisation tensors $e^A_{ij}$, we introduce the orthonormal frame $(\vu*u,\vu*v,\vu*n)$, where $\vu*n$ is the GW propagation direction (see figure~\ref{fig:orbital-elements}).
We want to find the components of these basis vectors in the frame of the binary, $(\vu*r,\vu*\theta,\vu*\ell)$, as this determines the binary's response to the GW.
First, we transform from the GW frame to the fixed reference frame $(\vu*x,\vu*y,\vu*z)$ by applying the standard rotations with respect to the zenith $\vartheta$ and azimuth $\phi$,
    \begin{equation}
        \begin{pmatrix}
            x \\[-2pt]
            y \\[-2pt]
            z
        \end{pmatrix}
        =\mathsf{R}_\phi\,\mathsf{R}_\vartheta\,
        \begin{pmatrix}
            u \\[-2pt]
            v \\[-2pt]
            n
        \end{pmatrix},\qquad
        \mathsf{R}_\vartheta=
        \begin{pmatrix}
            \cos\vartheta & 0 & \sin\vartheta \\[-2pt]
            0 & 1 & 0 \\[-2pt]
            -\sin\vartheta & 0 & \cos\vartheta
        \end{pmatrix},\quad
        \mathsf{R}_\phi=
        \begin{pmatrix}
            \cos\phi & -\sin\phi & 0 \\[-2pt]
            \sin\phi & \cos\phi & 0 \\[-2pt]
            0 & 0 & 1
        \end{pmatrix}.
    \end{equation}
This reference frame is transformed to the binary frame $(\vu*r,\vu*\theta,\vu*\ell)$ with three further rotations, which specify the inclination $I$, the longitude of ascending node $\asc$, and the argument of the binary in the orbital plane $\theta=\psi+\omega$~\cite{Murray:2000ssd},
    \begin{equation}
        \begin{pmatrix}
            r \\[-2pt]
            \theta \\[-2pt]
            \ell
        \end{pmatrix}
        =\mathsf{R}_\theta\,\mathsf{R}_I\,\mathsf{R}_\asc\,
        \begin{pmatrix}
            x \\[-2pt]
            y \\[-2pt]
            z
        \end{pmatrix},
    \end{equation}
    where
    \begin{equation}
        \mathsf{R}_\asc=
        \begin{pmatrix}
            \cos\asc & \sin\asc & 0 \\[-2pt]
            -\sin\asc & \cos\asc & 0 \\[-2pt]
            0 & 0 & 1
        \end{pmatrix},\qquad
        \mathsf{R}_I=
        \begin{pmatrix}
            1 & 0 & 0 \\[-2pt]
            0 & \cos I & \sin I \\[-2pt]
            0 & -\sin I & \cos I
        \end{pmatrix},\qquad
        \mathsf{R}_\theta=
        \begin{pmatrix}
            \cos\theta & \sin\theta & 0 \\[-2pt]
            -\sin\theta & \cos\theta & 0 \\[-2pt]
            0 & 0 & 1
        \end{pmatrix}.
    \end{equation}
We thus obtain the desired relationship between the binary frame and the GW frame by applying all five rotations,
    \begin{equation}
        \begin{pmatrix}
            r \\[-2pt]
            \theta \\[-2pt]
            \ell
        \end{pmatrix}
        =\mathsf{R}_\theta\,\mathsf{R}_I\,\mathsf{R}_\asc\,\mathsf{R}_\phi\,\mathsf{R}_\vartheta\,
        \begin{pmatrix}
            u \\[-2pt]
            v \\[-2pt]
            n
        \end{pmatrix}.
    \end{equation}

The various contractions with the polarisation tensors are then given by
    \begin{align}
    \begin{split}
    \label{eq:polarisation-tensor-contractions}
        e_{ij}^+\hat{r}^i\hat{r}^j&=\qty[\sin\vartheta\sin I\sin\theta-\cos\vartheta(\cos\varphi\cos\theta+\sin\varphi\cos I\sin\theta)]^2-\qty[\cos\varphi\cos I\sin\theta-\sin\varphi\cos\theta]^2,\\
        e_{ij}^\times\hat{r}^i\hat{r}^j&=2(\cos\varphi\cos I\sin\theta-\sin\varphi\cos\theta)[\cos\vartheta(\cos\varphi\cos\theta+\sin\varphi\cos I\sin\theta)-\sin\vartheta\sin I\sin\theta],\\
        e_{ij}^+\hat{r}^i\hat{\theta}^j&=(\sin\varphi\cos\theta-\cos\varphi\cos I\sin\theta)(\cos\varphi\cos I\cos\theta+\sin\varphi\sin\theta)\\
        &-[\cos\vartheta\cos\varphi\sin\theta+(\sin\vartheta\sin I-\cos\vartheta\sin\varphi\cos I)\cos\theta]\\
        &\qquad\times[\cos\vartheta(\cos\varphi\cos\theta+\sin\varphi\cos I\sin\theta)-\sin\vartheta\sin I\sin\theta],\\
        e_{ij}^\times\hat{r}^i\hat{\theta}^j&=\cos\vartheta\cos2\varphi\cos I\cos2\theta+\sin\vartheta\sin I(\sin\varphi\cos2\theta-\cos\varphi\cos I\sin2\theta)\\
        &+\frac{1}{2}\cos\vartheta\sin2\varphi(1+\cos^2I)\sin2\theta,\\
        e_{ij}^+\hat{r}^i\hat{\ell}^j&=\cos\varphi\sin I(\cos\varphi\cos I\sin\theta-\sin\varphi\cos\theta)\\
        &-(\sin\vartheta\cos I+\cos\vartheta\sin\varphi\sin I)\qty[\cos\vartheta(\cos\varphi\cos\theta+\sin\varphi\cos I\sin\theta)-\sin\vartheta\sin I\sin\theta],\\
        e_{ij}^\times\hat{r}^i\hat{\ell}^j&=\sin\vartheta(\sin\varphi\cos I\cos\theta-\cos\varphi\cos2I\sin\theta)-\cos\vartheta\sin I(\cos2\varphi\cos\theta+\sin2\varphi\cos I\sin\theta),
    \end{split}
    \end{align}
    where we define $\varphi\equiv\phi-\asc$.
We recall that $\vartheta,\phi$ are the spherical coordinates of the incoming plane GW, $\asc$ is the longitude of ascending node, and $\theta\equiv\psi+\omega$ is the orbital argument with respect to the ascending node, with $\psi$ the true anomaly and $\omega$ the argument of pericentre; these are all illustrated in figure~\ref{fig:orbital-elements}.

\section{Transfer functions}
\label{sec:transfer}

The Fourier components of the transfer functions are defined in terms of the polarisation tensor contractions discussed in section~\ref{sec:polarisation-tensor-projections} by
    \begin{align}
    \begin{split}
    \label{eq:transfer-functions}
        T^A_{P,n}&=\frac{3P^2\gamma}{4\uppi}\ev{\frac{e\sin\psi}{1+e\cos\psi}e_{ij}^A\hat{r}_i\hat{r}_j+e_{ij}^A\hat{r}_i\hat{\theta}_j}_n,\qquad T^A_{e,n}=\frac{\gamma^2}{3Pe}T^A_{P,n}-\frac{P\gamma^5}{4\uppi e}\ev{\frac{e_{ij}^A\hat{r}_i\hat{\theta}_j}{(1+e\cos\psi)^2}}_n,\\
        T^A_{I,n}&=\frac{P\gamma^3}{4\uppi}\ev{\frac{\cos\theta}{(1+e\cos\psi)^2}e_{ij}^A\hat{r}_i\hat{\ell}_j}_n,\qquad T^A_{\asc,n}=\frac{P\gamma^3}{4\uppi\sin I}\ev{\frac{\sin\theta}{(1+e\cos\psi)^2}e_{ij}^A\hat{r}_i\hat{\ell}_j}_n,\\
        T^A_{\omega,n}&=\frac{P\gamma^3}{4\uppi e}\ev{\frac{\sin\psi(2+e\cos\psi)}{(1+e\cos\psi)^2}e_{ij}^A\hat{r}_i\hat{\theta}_j-\frac{\cos\psi e_{ij}^A\hat{r}_i\hat{r}_j}{1+e\cos\psi}}_n-\cos I\,T^A_{\asc,n},\\
        T^A_{\eps,n}&=-\frac{P\gamma^4}{2\uppi}\ev{\frac{e^A_{ij}\hat{r}_i\hat{r}_j}{(1+e\cos\psi)^2}}_n-\gamma\cos I\,T^A_{\asc,n}-\gamma T^A_{\omega,n},
    \end{split}
    \end{align}
    where we have introduced the secular averaging operation
    \begin{equation}
    \label{eq:secular-average}
        \ev{\cdots}_n\equiv\int_{t_0}^{t_0+P}\frac{\dd{t}}{P}\exp(\frac{2\uppi\rmi nt}{P})(\cdots),
    \end{equation}
    which extracts the $n^\mathrm{th}$-order Fourier coefficient of a given function of the true anomaly $\psi(t)$, holding the orbital elements fixed (as they vary over much longer timescales).
The subscript $n$ distinguishes this from the ensemble average $\ev{\cdots}$, and from the secular average $\ev{\cdots}_\mathrm{sec}$ (with the latter being equivalent to $\ev{\cdots}_n$ for $n=0$).

The Fourier components arising in the eccentric transfer functions can be expressed in terms of \emph{Hansen coefficients}, $C^{lm}_n$, $S^{lm}_n$~\cite{Brumberg:1995atcm}.
These are functions of eccentricity that have been used for centuries in celestial mechanics to describe Keplerian motion.
We define them here by
    \begin{equation}
    \label{eq:hansen-definition-appendix}
        C^{lm}_n(e)=\ev{\frac{\cos m\psi}{(1+e\cos\psi)^l}}_n,\qquad S^{lm}_n(e)=\ev{\frac{\sin m\psi}{(1+e\cos\psi)^l}}_n,
    \end{equation}
    with explicit expressions for particular sets of $(l,m)$ given in section~\ref{sec:hansen}.

Inserting equation~\eqref{eq:polarisation-tensor-contractions} into equation~\eqref{eq:transfer-functions} and using equation~\eqref{eq:hansen-definition-appendix}, we can thus write the transfer functions as linear combinations of the Hansen coefficients.
These are then the input to computing the KM coefficients.

\section{Transforming the reference frame}
\label{sec:reference-frame}

The polarisation tensor contractions in equation~\eqref{eq:polarisation-tensor-contractions} are completely general and apply to any choice of reference frame.
However, the corresponding expressions for the transfer functions are very lengthy, making it prohibitively difficult to calculate the KM coefficients.
We circumvent this difficulty by choosing a particular reference frame in which the transfer functions are much simpler, before transforming back to a general reference frame.

Given two reference frames, $(\vu*x,\vu*y,\vu*z)$ and $(\vu*{x}{}',\vu*{y}{}',\vu*{z}{}')$, we have two corresponding sets of orbital elements for the binary, $X=(P,e,I,\asc,\omega,\eps)$ and $X'=(P,e,I',\asc',\omega',\eps)$ (the period, eccentricity, and compensated mean anomaly are the same in both frames, as they do not depend on the orientation of the binary in space).
The KM coefficients for the unprimed elements are given in terms of those for the primed ones by (see, e.g., section~4.9 of \citet{Risken:1989fpe} for a derivation)
    \begin{equation}
        D^{(1)}_i=\pdv{X_i}{X_{i'}}D^{(1)}_{i'}+\pdv{X_i}{X_{i'}}{X_{j'}}D^{(2)}_{i'j'},\qquad D^{(2)}_{ij}=\pdv{X_i}{X_{i'}}\pdv{X_j}{X_{j'}}D^{(2)}_{i'j'},
    \end{equation}
    where primed indices run over the primed orbital elements.
We therefore require the first and second partial derivatives of the unprimed elements with respect to the primed ones.
These can be deduced from the following relations, which are derived from section~2.8 of \citet{Murray:2000ssd},
    \begin{align}
    \begin{split}
    \label{eq:orbit-angles-transformation}
        \cos I&=s_{I'}\sin I'\sin\asc'\vu*{x}{}'\vdot\vu*z-s_{I'}\sin I'\cos\asc'\vu*{y}{}'\vdot\vu*z+\cos I'\vu*{z}{}'\vdot\vu*z,\\
        \sin I\sin\asc&=s_{I'}\sin I'\sin\asc'\vu*{x}{}'\vdot\vu*x-s_{I'}\sin I'\cos\asc'\vu*{y}{}'\vdot\vu*x+\cos I'\vu*{z}{}'\vdot\vu*x,\\
        \sin I\sin\omega&=(\cos\asc'\cos\omega'-\cos I'\sin\asc'\sin\omega')\vu*{x}{}'\vdot\vu*z\\
        &+(\sin\asc'\cos\omega'+\cos I'\cos\asc'\sin\omega')\vu*{y}{}'\vdot\vu*z+\sin I'\sin\omega\vu*{z}{}'\vdot\vu*z,
    \end{split}
    \end{align}
    where
    \begin{equation}
        s_{I'}\equiv
        \begin{cases}
            +1 & \text{if }\cos I'>0,\\
            -1 & \text{if }\cos I'<0.
        \end{cases}
    \end{equation}

We are now free to specify the primed frame such that the KM coefficients are easier to compute.
One particularly simple choice is to choose $(\vu*{x}{}',\vu*{y}{}',\vu*{z}{}')$ such that $I'=\uppi/4$ and $\asc'=\omega'=0$.
(It may seem at first that $I'=0$ is a simpler choice, as the reference frame then coincides with the binary's frame.
However, there is a coordinate singularity associated with $I'=0$ which makes some of the associated coefficients poorly-behaved.
Taking $I'=\uppi/4$ is much easier, particularly since we then have $\sin I'=\cos I'$, which simplifies many of the resulting expressions.)

Having specified the primed frame, we require the first and second derivatives to transform back to the unprimed frame, which is relevant for a general observer.
Using equation~\eqref{eq:orbit-angles-transformation}, we find that the nonzero derivatives, evaluated at $(I',\asc',\omega')=(\tfrac{\uppi}{4},0,0)$, are given by
    \begin{align}
    \begin{split}
        \pdv{I}{I'}&=\cos\omega,\quad\pdv{I}{\asc'}=-\frac{\sin\omega}{\sqrt{2}},\qquad\pdv{\asc}{I'}=\frac{\sin\omega}{\sin I},\quad\pdv{\asc}{\asc'}=\frac{\cos\omega}{\sqrt{2}\sin I},\qquad\pdv{\omega}{I'}=-\frac{\sin\omega}{\tan I},\\
        \pdv{\omega}{\asc'}&=\frac{1}{\sqrt{2}}\qty(1-\frac{\cos\omega}{\tan I}),\quad\pdv{\omega}{\omega'}=1,\qquad\pdv[2]{I}{{I'}}=\frac{\sin^2\omega}{\tan I},\quad\pdv{I}{I'}{\asc'}=\frac{\sin\omega}{\sqrt{2}}\qty(\frac{\cos\omega}{\tan I}-1),\\
        \pdv[2]{I}{{\asc'}}&=\frac{\cos\omega}{2}\qty(\frac{\cos\omega}{\tan I}-1),\qquad\pdv[2]{\asc}{{I'}}=-\frac{\sin2\omega}{\sin I\tan I},\quad\pdv{\asc}{I'}{\asc'}=\frac{\cos\omega-\cos2\omega\cot I}{\sqrt{2}\sin I},\\
        \pdv[2]{\asc}{{\asc'}}&=\frac{\sin\omega}{\sin I}\qty(\frac{\cos\omega}{\tan I}-\frac{1}{2}),\qquad\pdv[2]{\omega}{{I'}}=\sin2\omega\frac{2-\sin^2I}{2\sin^2I},\\
        \pdv{\omega}{I'}{\asc'}&=\frac{1}{2\sqrt{2}}\qty(\cos2\omega\frac{2-\sin^2I}{\sin^2I}-\frac{2\cos\omega}{\tan I}-1),\quad\pdv[2]{\omega}{{\asc'}}=\frac{\sin\omega}{\tan I}\qty(\frac{1}{2}-\cos\omega\frac{2-\sin^2I}{\sin2I}).
    \end{split}
    \end{align}
It is straightforward to confirm that these are well-behaved and take on the appropriate values when $(I,\asc,\omega)\to(I',\asc',\omega')$.

We thus find that the unprimed drift coefficients are given in terms of those in the primed frame by
    \begin{align}
    \begin{split}
    \label{eq:drift-transformation}
        D^{(1)}_I&=\cos\omega\qty[D^{(1)}_{I'}+\qty(\frac{\cos\omega}{\tan I}-1)\frac{D^{(2)}_{\asc'\asc'}}{2}]+\frac{\sin^2\omega}{\tan I}D^{(2)}_{I'I'},\\
        D^{(1)}_\asc&=\frac{1}{\sin I}\qty[\sin\omega D^{(1)}_{I'}-\frac{\sin2\omega}{\tan I}D^{(2)}_{I'I'}+\sin\omega\qty(\frac{\cos\omega}{\tan I}-\frac{1}{2})D^{(2)}_{\asc'\asc'}],\\
        D^{(1)}_\omega&=\frac{\sin\omega}{\sin I}\qty[-\cos ID^{(1)}_{I'}+\cos\omega\frac{2-\sin^2I}{\sin I}D^{(2)}_{I'I'}+\qty(\cos I-\cos\omega\frac{2-\sin^2I}{\sin I})\frac{D^{(2)}_{\asc'\asc'}}{2}],
    \end{split}
    \end{align}
    with the diffusion coefficients given by
    \begin{align}
    \begin{split}
    \label{eq:diffusion-transformation}
        D^{(2)}_{II}&=\cos^2\omega D^{(2)}_{I'I'}+\frac{\sin^2\omega}{2}D^{(2)}_{\asc'\asc'},\qquad D^{(2)}_{I\asc}=\frac{\sin2\omega}{2\sin I}\qty(D^{(2)}_{I'I'}-\frac{1}{2}D^{(2)}_{\asc'\asc'}),\\
        D^{(2)}_{I\omega}&=\sin\omega\qty[-\frac{\cos\omega}{\tan I}D^{(2)}_{I'I'}+\frac{1}{2}\qty(\frac{\cos\omega}{\tan I}-1)D^{(2)}_{\asc'\asc'}-\frac{1}{\sqrt{2}}D^{(2)}_{\asc'\omega'}],\\
        D^{(2)}_{\asc\asc}&=\frac{1}{\sin^2I}\qty(\sin^2\omega D^{(2)}_{I'I'}+\frac{\cos^2\omega}{2}D^{(2)}_{\asc'\asc'}),\\
        D^{(2)}_{\asc\omega}&=\frac{1}{\sin I}\qty[-\frac{\sin^2\omega}{\tan I}D^{(2)}_{I'I'}+\frac{\cos\omega}{2}\qty(1-\frac{\cos\omega}{\tan I})D^{(2)}_{\asc'\asc'}+\frac{\cos\omega}{\sqrt{2}}D^{(2)}_{\asc'\omega'}],\\
        D^{(2)}_{\omega\omega}&=\frac{\sin^2\omega}{\tan^2I}D^{(2)}_{I'I'}+\frac{1}{2}\qty(1-\frac{\cos\omega}{\tan I})^2D^{(2)}_{\asc'\asc'}+\sqrt{2}\qty(1-\frac{\cos\omega}{\tan I})D^{(2)}_{\asc'\omega'}+D^{(2)}_{\omega'\omega'},\\
        D^{(2)}_{\omega\eps}&=D^{(2)}_{\omega'\eps},\qquad D^{(2)}_{XI}=D^{(2)}_{X\asc}=D^{(2)}_{X\omega}=D^{(2)}_{X\eps}=D^{(2)}_{I\eps}=D^{(2)}_{\asc\eps}=0,
    \end{split}
    \end{align}
    where $X$ here stands for $P$ or $e$.

\section{Kramers-Moyal coefficients in the primed frame}
\label{sec:KM-primed-frame}

Here we give the KM coefficients evaluated in the primed coordinate frame $(I',\asc',\omega')=(\tfrac{\uppi}{4},0,0)$.
To calculate these, we first evaluate the polarisation tensor contractions from section~\ref{sec:polarisation-tensor-projections} in the primed frame, and insert these into equation~\eqref{eq:transfer-functions} to find the appropriate GW transfer functions, expressing the secular averages in terms of Hansen coefficients that are listed below in section~\ref{sec:hansen}.
These secular transfer functions are then inserted into equation~\eqref{eq:km-final}, integrating over the GW propagation direction $\vu*n=(\vartheta,\phi)$ to obtain the KM coefficients.
The resulting expressions for the drift vector are
    \begin{align}
    \begin{split}
        D^{(1)}_P&=V_P+\frac{9P^2\gamma^2}{80}\sum_{n=1}^\infty nH_0^2\Omega_n\bigg[2E^{02}_n\qty(10eE^{02}_n+(3+7e^2)E^{11}_n+(1-11e^2)E^{13}_n-2e\gamma^2E^{22}_n)^*\\
        &\qquad\qquad\qquad+E^{11}_n\qty(e(3+2e^2)E^{11}_n-2e(1+4e^2)E^{13}_n-2\gamma^2E^{22}_n)^*-e(1-6e^2)\qty|E^{13}_n|^2\\
        &\qquad\qquad\qquad+8\gamma^2\qty(E^{02}_n+\frac{e}{2}(E^{11}_n-E^{13}_n))\qty(E^{21}_n-E^{23}_n+\frac{e}{4}(E^{20}_n+6E^{22}_n-E^{24}_n))^*\\
        &\qquad\qquad\qquad+4\gamma^2\qty(E^{02}_n+\frac{e}{2}(E^{11}_n-E^{13}_n)-\gamma^2E^{22}_n)\qty({E^{02}_n}'+\frac{e}{2}(E^{11}_n-E^{13}_n)')^*\\
        &\qquad\qquad\qquad+2\gamma^2E^{13}_n\qty(E^{22}_n)^*-\frac{4e}{3}S^{11}_n\qty((1+4e^2)S^{11}_n+e\gamma^2{S^{11}_n}')\bigg],\\
        D^{(1)}_e&=V_e+\frac{3P\gamma^6}{20e^2}\sum_{n=1}^\infty nH_0^2\Omega_n\bigg[\frac{1}{e}\qty|E^{02}_n+\frac{e}{2}(E^{11}_n-E^{13}_n)|^2+\frac{\gamma^4}{e}(S^{22}_n)^2-\frac{e^2}{3}{S^{11}_n}'S^{11}_n\\
        &+\qty(E^{02}_n+\frac{e}{2}(E^{11}_n-E^{13}_n)-\gamma^2E^{22}_n)\bigg(-\frac{2}{e}E^{02}_n+\frac{1}{2}(E^{11}_n+3E^{13}_n)+2(E^{21}_n-E^{23}_n)+{E^{02}_n}'\\
        &\qquad\qquad\qquad\qquad\qquad-\gamma^2{E^{22}_n}'+\frac{e}{2}(E^{20}_n+12E^{22}_n-E^{24}_n+(E^{11}_n-E^{13}_n)')\bigg)^*\bigg],\\
        D^{(1)}_{I'}&=\frac{1}{2}D^{(2)}_{\asc'\asc'},\qquad D^{(1)}_{\asc'}=0,\qquad D^{(1)}_{\omega'}=V_{\omega'},\qquad D^{(1)}_{\eps}=V_\eps,
    \end{split}
    \end{align}
    while the nonzero components of the diffusion matrix are
    \begin{align}
    \begin{split}
        D^{(2)}_{PP}&=\frac{27P^3\gamma^2}{20}\sum_{n=1}^\infty nH_0^2\Omega_n\qty[\qty|E^{02}_n+\frac{e}{2}(E^{11}_n-E^{13}_n)|^2-\frac{(eS^{11}_n)^2}{3}],\\
        D^{(2)}_{Pe}&=\frac{\gamma^2D^{(2)}_{PP}}{3Pe}-\frac{9P^2\gamma^6}{40}\sum_{n=1}^\infty nH_0^2\Omega_nE^{22}_n\qty(\frac{2}{e}E^{02}_n+E^{11}_n-E^{13}_n)^*,\\
        D^{(2)}_{ee}&=\frac{3P\gamma^6}{20e^2}\sum_{n=1}^\infty nH_0^2\Omega_n\qty[\qty|E^{02}_n+\frac{e}{2}(E^{11}_n-E^{13}_n)-\gamma^2E^{22}_n|^2-\frac{(eS^{11}_n)^2}{3}],\\
        D^{(2)}_{I'I'}&=\frac{3P\gamma^6}{80}\sum_{n=1}^\infty nH_0^2\Omega_n\qty|E^{20}_n+E^{22}_n|^2,\qquad D^{(2)}_{\asc'\asc'}=-\sqrt{2}D^{(2)}_{\asc'\omega'}=\frac{3P\gamma^6}{40}\sum_{n=1}^\infty nH_0^2\Omega_n\qty|E^{20}_n-E^{22}_n|^2,\\
        D^{(2)}_{\omega'\omega'}&=\frac{1}{2}D^{(2)}_{\asc'\asc'}+\frac{3P\gamma^6}{80e^2}\sum_{n=1}^\infty nH_0^2\Omega_n\qty[\qty|E^{11}_n+E^{13}_n+2E^{21}_n-2E^{23}_n+\frac{e}{2}\qty(E^{20}_n-E^{24}_n)|^2+\frac{4}{3}\qty(C^{11}_n)^2],\\
        D^{(2)}_{\omega'\eps}&=-\frac{3P\gamma^7}{80e^2}\sum_{n=1}^\infty nH_0^2\Omega_n\bigg[\qty|E^{11}_n+E^{13}_n+2(E^{21}_n-E^{23}_n)+\frac{e}{2}(E^{20}_n-4E^{22}_n-E^{24}_n)|^2\\
        &\qquad\qquad\qquad\qquad\qquad+\frac{4}{3}C^{11}_n(C^{11}_n-2eC^{20}_n)-4e^2|E^{22}_n|^2\bigg],\\
        D^{(2)}_{\eps\eps}&=\frac{3P\gamma^8}{80e^2}\sum_{n=1}^\infty nH_0^2\Omega_n\qty[\qty|E^{11}_n+E^{13}_n+2(E^{21}_n-E^{23}_n)+\frac{e}{2}(E^{20}_n-8E^{22}_n-E^{24}_n)|^2+\frac{4}{3}(C^{11}_n-2eC^{20}_n)^2].
    \end{split}
    \end{align}
These KM coefficients can be transformed back to a general reference frame using equations~\eqref{eq:drift-transformation} and~\eqref{eq:diffusion-transformation}, resulting in the expressions given in section~\ref{sec:KM}.

\section{Hansen coefficients}
\label{sec:hansen}

Using various formulae given in \citet{Brumberg:1995atcm}, we write the Hansen coefficients as
    \begin{align}
    \begin{split}
        \label{eq:hansen-general}
        &C^{lm}_n(e)=\frac{{}_2F_1(-m-l-1,m-l-1;1;\beta^2)}{2\gamma^{2l}(1+\beta^2)^{l+1}}[J_{n-m}(ne)+J_{n+m}(ne)]\\
        &+\sum_{k=1}^\infty\bigg\{\frac{(m-l-1)_k\beta^k}{2k!\gamma^{2l}(1+\beta^2)^{l+1}}{}_2F_1(-m-l-1,k+m-l-1;k+1;\beta^2)[J_{n-m-k}(ne)+J_{n+m+k}(ne)]\\
        &\qquad+\frac{(-m-l-1)_k\beta^k}{2k!\gamma^{2l}(1+\beta^2)^{l+1}}{}_2F_1(k-l-m-1,-l+m-1;k+1;\beta^2)[J_{n-m+k}(ne)+J_{n+m-k}(ne)]\bigg\},\\
        &S^{lm}_n(e)=\frac{\rmi\,{}_2F_1(-m-l-1,m-l-1;1;\beta^2)}{2\gamma^{2l}(1+\beta^2)^{l+1}}[J_{n-m}(ne)-J_{n+m}(ne)]\\
        &+\sum_{k=1}^\infty\bigg\{\frac{\rmi(m-l-1)_k\beta^k}{2k!\gamma^{2l}(1+\beta^2)^{l+1}}{}_2F_1(-m-l-1,k+m-l-1;k+1;\beta^2)[J_{n-m-k}(ne)-J_{n+m+k}(ne)]\\
        &\qquad+\frac{\rmi(-m-l-1)_k\beta^k}{2k!\gamma^{2l}(1+\beta^2)^{l+1}}{}_2F_1(k-m-l-1,m-l-1;k+1;\beta^2)[J_{n-m+k}(ne)-J_{n+m-k}(ne)]\bigg\},
    \end{split}
    \end{align}
    where we define the expansion variable $\beta\equiv e/(1+\gamma)$, and where $(\cdots)_k$ is a rising Pochhammer symbol, defined by $(n)_k\equiv n(n+1)(n+2)\cdots(n+k-1)$, while ${}_2F_1(a,b;c;z)$ is a hypergeometric function, and $J_n(z)$ is a Bessel function of the first kind.
We see that in the circular case $e=0$ we have $\beta=0$, ${}_2F_1(a,b;c;0)=1$, and $J_n(0)=\delta_{n,0}$, and these expressions in equation~\eqref{eq:hansen-general} simplify to $C^{lm}_n=\tfrac{1}{2}(\delta_{n,m}+\delta_{n,-m})$ and $S^{lm}_n=\tfrac{\rmi}{2}(\delta_{n,m}-\delta_{n,-m})$, which can be confirmed by directly integrating equation~\eqref{eq:hansen-definition-appendix}.
Note also that for general eccentricity $e\in(0,1)$, from the definition of the Pochhammer symbol, the sums over $k$ in equation~\eqref{eq:hansen-general} terminate if and only if $m\le l+1$, otherwise the corresponding Hansen coefficients will have an infinite number of terms.

Using equation~\eqref{eq:hansen-general}, we can directly compute all the Hansen coefficients that appear in equations~\eqref{eq:ecc-drift} and~\eqref{eq:ecc-diffusion}, obtaining the cosine coefficients
    \begin{align}
    \begin{split}
        \label{eq:hansen-cosine-explicit}
        C^{02}_n&=-\qty[\frac{1+\gamma-\frac{2}{3}e^2(1+\frac{1}{4}\gamma)}{(1+\gamma)^4}]12\gamma^2J_n+\sum_{k=0}^\infty\frac{\beta^k(1-\beta^2)^3}{2(1+\beta^2)}(J_{n-k-2}+J_{n+k+2}),\\
        C^{11}_n&=\frac{1}{2n\gamma^2}(J_{n-1}-J_{n+1}),\\
        C^{13}_n&=-\qty[\frac{1+\gamma-\frac{9}{8}e^2(1+\frac{5}{9}\gamma)+\frac{1}{5}e^4}{\beta\gamma^2(1+\gamma)^6}]20e^4J_n+\sum_{k=0}^\infty\frac{\beta^k(1-\beta^2)^5}{2(1+\beta^2)^2\gamma^2}(J_{n-k-3}+J_{n+k+3})\\
        &-\qty[\frac{1-\gamma-\frac{9}{4}e^2(1-\frac{7}{9}\gamma)+\frac{3}{4}e^4(1-\frac{5}{3}\gamma)}{e^3\beta\gamma^2}]2(J_{n-1}+J_{n+1})\\
        &-\qty[\frac{1-\gamma-3e^2(1-\frac{5}{6}\gamma)+3e^4(1-\frac{5}{8}\gamma)-e^6(1-\frac{15}{32}\gamma)}{e^5\gamma^2}]8(J_{n-2}+J_{n+2}),\\
        C^{20}_n&=-\qty[\frac{n^2(1-\frac{1}{2}e^2)-3n+2}{n^4e^2\gamma^4}]8J_{n-2}+\qty(\frac{n-1}{n^3e\gamma^4})4J_{n-3},\\
        C^{21}_n&=\qty[\frac{n^3(1-\tfrac{7}{4}e^2+\tfrac{3}{4}e^4)-n^2(1-\tfrac{7}{2}e^2+e^4)-4n(1+\tfrac{1}{2}e^2)+4}{n^4e^3\gamma^4}]4J_{n-2}\\
        &-\qty[\frac{n^2(1-\tfrac{3}{2}e^2+\tfrac{1}{2}e^4)+n(1+e^2)-2}{n^3e^2\gamma^4}]2J_{n-3},\\
        C^{22}_n&=-\qty[\frac{n^3(1-\frac{7}{4}e^2+\frac{3}{4}e^4)-2n^2(1-\frac{13}{8}e^2+\frac{7}{16}e^4)-n(1+\frac{1}{2}e^2)+2-e^2}{n^4e^4\gamma^4}]16J_{n-2}\\
        &+\qty[\frac{n^2(1-\frac{3}{2}e^2+\frac{1}{2}e^4)+\frac{1}{2}ne^2-1+\frac{1}{2}e^2}{n^3e^3\gamma^4}]8J_{n-3},\\
        C^{23}_n&=\qty[\frac{n^2\gamma^6+3n(1-\frac{5}{4}e^2+\frac{1}{4}e^4)+2-\frac{3}{2}e^2}{n^2e^3\gamma^4}]4J_n-\qty(\frac{1-\frac{5}{4}e^2+\frac{1}{4}e^4}{n^2e^2\gamma^4})12J_{n-1},\\
        C^{24}_n&=\frac{8\gamma^3}{(1+\gamma)^4}(J_{n-4}+J_{n+4})+\sum_{k=0}^\infty\frac{\beta^k(1+\gamma)^3}{16\gamma^4}\bigg[(1-\beta^2)^7(J_{n-k-4}+J_{n+k+4})\\
        &\qquad\qquad\qquad\qquad\qquad\qquad\qquad+\frac{(-7)_k}{k!}{}_2F_1(1,k-7;k+1;\beta^2)(J_{n+k-4}+J_{n-k+4})\bigg],
    \end{split}
    \end{align}
    and the sine coefficients,
    \begin{align}
    \begin{split}
        \label{eq:hansen-sine-explicit}
        S^{02}_n&=-\qty[\frac{1-\gamma-2e^2(1-\frac{3}{4}\gamma)+e^4}{e^3}]2\rmi(J_{n-1}-J_{n+1})+\sum_{k=0}^\infty\frac{\rmi\beta^k(1-\beta^2)^3}{2(1+\beta^2)}(J_{n-k-2}-J_{n+k+2}),\\
        S^{11}_n&=\frac{\rmi J_n}{ne\gamma},\\
        S^{13}_n&=-\qty[\frac{1-3e^2(1-\frac{5}{6}\gamma)+3e^4(1-\frac{5}{8}\gamma)-e^6}{e^4\gamma^2}]4\rmi(J_{n-1}-J_{n+1})\\
        &-\qty[\frac{1-\gamma-3e^2(1-\frac{5}{6}\gamma)+3e^4(1-\frac{5}{8}\gamma)-e^6(1-\frac{15}{32}\gamma)}{e^5\gamma^2}]8\rmi(J_{n-2}-J_{n+2})\\
        &+\sum_{k=0}^\infty\frac{\rmi\beta^k(1-\beta^2)^5}{2(1+\beta^2)^2\gamma^2}(J_{n-k-3}-J_{n+k+3}),\\
        S^{21}_n&=\frac{-\rmi}{n^2e\gamma^3}[(n\gamma^2+2)J_n-2eJ_{n-1}],\\
        S^{22}_n&=\qty[\frac{n^3(1-\frac{5}{4}e^2+\frac{1}{4}e^4)-2n^2(1-\frac{9}{8}e^2)-n(1+e^2)+2}{n^4e^4\gamma^3}]16\rmi J_{n-2}-\qty[\frac{n^3\gamma^2+n^2e^2-n}{n^4e^3\gamma^3}]8\rmi J_{n-3},\\
        S^{23}_n&=\frac{15e^2}{8\gamma^3}\rmi(J_{n-1}-J_{n+1})-\frac{3e}{2\gamma^3}\rmi(J_{n-2}-J_{n+2})+\frac{1-\frac{1}{4}e^2}{2\gamma^3}\rmi(J_{n-3}-J_{n+3}),\\
        S^{24}_n&=\frac{8\rmi\gamma^3}{(1+\gamma)^4}(J_{n-4}-J_{n+4})\\
        &+\sum_{k=0}^\infty\frac{\beta^k(1+\gamma)^3}{16\gamma^4}\qty[(1-\beta^2)^7(J_{n-k-4}-J_{n+k+4})+\frac{(-7)_k}{k!}{}_2F_1(1,k-7;k+1;\beta^2)(J_{n+k-4}-J_{n-k+4})].
    \end{split}
    \end{align}
Note that all the Bessel functions appearing in equations~\eqref{eq:hansen-cosine-explicit} and~\eqref{eq:hansen-sine-explicit} have their argument equal to $ne$, even if they are not of order $n$; we suppress this argument for brevity.

\end{appendices}